\documentclass{pasj00}

\def\Umlaut#1{\"{#1}}

\def\commenta{$^*$}
\def\commentb{$^\dagger$}
\def\commentc{$^\ddagger$}
\def\commentd{$^\S$}

\def\submitted{submitted}
\def\inpress{in press}
\def\inprep{in preparation}

\def\arxiv#1{ (arXiv astro-ph/#1)}

\DeclareAbbreviation\AAHam{Astron. Abh. Hamburg. Sternw.}
\DeclareAbbreviation\AARv{Astron. Astrophys. Rev.}
\DeclareAbbreviation\AAS{American Astron. Soc. Meeting Abstracts}
\DeclareAbbreviation\an{Astron. Nachr.}
\DeclareAbbreviation\AcA{Acta Astron.}
\DeclareAbbreviation\Afz{Astrofizika}
\DeclareAbbreviation\AnTok{Tokyo Astron. Obs. Annals, Sec. Ser.}
\DeclareAbbreviation\Ap{Astrophysics}
\DeclareAbbreviation\ARep{Astron. Rep.}
\DeclareAbbreviation\ATel{Astronomer's Telegram}
\DeclareAbbreviation\ATsir{Astron. Tsirk.}
\DeclareAbbreviation\AcApS{Acta Astrophys. Sinica}
\DeclareAbbreviation\AstL{Astron. Lett.}
\DeclareAbbreviation\BaltA{Baltic Astron.}
\DeclareAbbreviation\BASI{Bull. Astron. Soc. India}
\DeclareAbbreviation\BeSN{Be Newslett.}
\DeclareAbbreviation\CBET{CBET}
\DeclareAbbreviation\ChJAA{Chinese J. of Astron. and Astrophys.}
\DeclareAbbreviation\GCN{GRB Coord. Netw. Circ.}
\DeclareAbbreviation\ibvs{IBVS}
\DeclareAbbreviation\JAD{J. Astron. Data}
\DeclareAbbreviation\JAVSO{J. American Assoc. Variable Star Obs.}
\DeclareAbbreviation\JBAA{J. British Astron. Assoc.}
\DeclareAbbreviation\LowOB{Lowell Obs. Bull.}
\DeclareAbbreviation\MitVS{Mitteil. Ver\"{a}nderl. Sterne}
\DeclareAbbreviation\MmSAI{Mem. Soc. Astron. Ita.}
\DeclareAbbreviation\Msngr{Messenger}
\DeclareAbbreviation\NewA{New Astron.}
\DeclareAbbreviation\NewAR{New Astron. Rev.}
\DeclareAbbreviation\OAP{Odessa Astron. Publ.}
\DeclareAbbreviation\Obs{Observatory}
\DeclareAbbreviation\OEJV{Open European J. on Var. Stars}
\DeclareAbbreviation\PASA{Publ. Astron. Soc. Australia}
\DeclareAbbreviation\PAZh{Pis'ma AZh}
\DeclareAbbreviation\PhR{Phys. Rep.}
\DeclareAbbreviation\PVSS{Publ. Variable Stars Sect. R. Astron. Soc. New Zealand}
\DeclareAbbreviation\PZ{Perem. Zvezdy}
\DeclareAbbreviation\PZP{Perem. Zvezdy Pril.}
\DeclareAbbreviation\QJRAS{QJRAS}
\DeclareAbbreviation\RMxAA{Rev. Mexicana Astron. Astrof.}
\DeclareAbbreviation\RvMA{Reviews of Modern Astron.}
\DeclareAbbreviation\SASS{Society for Astronom. Sciences Ann. Symp.}
\DeclareAbbreviation\Sci{Science}
\DeclareAbbreviation\SPIE{SPIE Proc.}
\DeclareAbbreviation\SvA{Soviet Astronomy}
\DeclareAbbreviation\SvAL{Soviet Astronomy Letters}
\DeclareAbbreviation\VeSon{Ver\"{o}ff. Sternw. Sonneberg}
\DeclareAbbreviation\VSOLJBul{VSOLJ Variable Star Bull.}
\DeclareAbbreviation\yCat{VizieR Online Data Catalog}
\DeclareAbbreviation\ZA{Z. Astrophys.}

\def\ASPConf#1#2{ASP Conf. Ser. #1, #2}

\def\PublisherASP{San Francisco: ASP}
\def\PublisherReidel{Dordrecht: D. Reidel Publishing Company}

\newcounter{author}
\def\authorcount#1#2{\refstepcounter{author}\label{#1}
                     \altaffiltext{\ref{#1}}{#2}}

\begin{document}
\SetRunningHead{T. Kato et al.}{Period Variations in SU UMa-Type Dwarf Novae}

\Received{200X/XX/XX}
\Accepted{200X/XX/XX}

\title{Survey of Period Variations of Superhumps in SU UMa-Type Dwarf Novae}

\author{Taichi~\textsc{Kato},\altaffilmark{\ref{affil:Kyoto}*}
        Akira~\textsc{Imada},\altaffilmark{\ref{affil:Imada}}
        Makoto~\textsc{Uemura},\altaffilmark{\ref{affil:Uemura}}
        Daisaku~\textsc{Nogami},\altaffilmark{\ref{affil:HidaKwasan}}
        Hiroyuki~\textsc{Maehara},\altaffilmark{\ref{affil:HidaKwasan}}
        Ryoko~\textsc{Ishioka},\altaffilmark{\ref{affil:Ishioka}}
        Hajime~\textsc{Baba},\altaffilmark{\ref{affil:Baba}}
        Katsura~\textsc{Matsumoto},\altaffilmark{\ref{affil:Matsumoto}}
        Hidetoshi~\textsc{Iwamatsu},\altaffilmark{\ref{affil:Kyoto}}
        Kaori~\textsc{Kubota},\altaffilmark{\ref{affil:Kyoto}}
        Kei~\textsc{Sugiyasu},\altaffilmark{\ref{affil:Kyoto}}
        Yuichi~\textsc{Soejima},\altaffilmark{\ref{affil:Kyoto}}
        Yuuki~\textsc{Moritani},\altaffilmark{\ref{affil:Kyoto}}
        Tomohito~\textsc{Ohshima},\altaffilmark{\ref{affil:Kyoto}}
        Hiroyuki~\textsc{Ohashi},\altaffilmark{\ref{affil:Kyoto}}
        Junpei~\textsc{Tanaka},\altaffilmark{\ref{affil:Kyoto}}
        Mahito~\textsc{Sasada},\altaffilmark{\ref{affil:Uemura}}
        Akira~\textsc{Arai},\altaffilmark{\ref{affil:Uemura}}
        Kazuhiro~\textsc{Nakajima},\altaffilmark{\ref{affil:Njh}}
        Seiichiro~\textsc{Kiyota},\altaffilmark{\ref{affil:Kis}}
        Kenji~\textsc{Tanabe},\altaffilmark{\ref{affil:OUS}}
        Kayuyoshi~\textsc{Imamura},\altaffilmark{\ref{affil:OUS}}
        Nanae~\textsc{Kunitomi},\altaffilmark{\ref{affil:OUS}}
        Kenji~\textsc{Kunihiro},\altaffilmark{\ref{affil:OUS}}
        Hiroki~\textsc{Taguchi},\altaffilmark{\ref{affil:OUS}}
        Mitsuo~\textsc{Koizumi},\altaffilmark{\ref{affil:OUS}}
        Norimi~\textsc{Yamada},\altaffilmark{\ref{affil:OUS}}
        Yuichi~\textsc{Nishi},\altaffilmark{\ref{affil:OUS}}
        Mayumi~\textsc{Kida},\altaffilmark{\ref{affil:OUS}}
        Sawa~\textsc{Tanaka},\altaffilmark{\ref{affil:OUS}}
        Rie~\textsc{Ueoka},\altaffilmark{\ref{affil:OUS}}
        Hideki~\textsc{Yasui},\altaffilmark{\ref{affil:OUS}}
        Koichi~\textsc{Maruoka},\altaffilmark{\ref{affil:OUS}}
        Arne~\textsc{Henden},\altaffilmark{\ref{affil:AAVSO}}
        Arto~\textsc{Oksanen},\altaffilmark{\ref{affil:Nyrola}}
        Marko~\textsc{Moilanen},\altaffilmark{\ref{affil:Nyrola}}
        Petri~\textsc{Tikkanen},\altaffilmark{\ref{affil:Nyrola}}
        Mika~\textsc{Aho},\altaffilmark{\ref{affil:Nyrola}}
        Berto~\textsc{Monard},\altaffilmark{\ref{affil:Monard}}
        Hiroshi~\textsc{Itoh},\altaffilmark{\ref{affil:Ioh}}
        Pavol~A.~\textsc{Dubovsky},\altaffilmark{\ref{affil:Dubovsky}}
        Igor~\textsc{Kudzej},\altaffilmark{\ref{affil:Dubovsky}}
        Radka~\textsc{Dancikova},\altaffilmark{\ref{affil:Dancikova}}
        Tonny~\textsc{Vanmunster},\altaffilmark{\ref{affil:Vanmunster}}
        Jochen~\textsc{Pietz},\altaffilmark{\ref{affil:Pietz}}
        Greg~\textsc{Bolt},\altaffilmark{\ref{affil:Bolt}}
        David~\textsc{Boyd},\altaffilmark{\ref{affil:DavidBoyd}} 
        Peter~\textsc{Nelson},\altaffilmark{\ref{affil:Nelson}}
        Thomas~\textsc{Krajci},\altaffilmark{\ref{affil:Krajci}}
        Lewis~M.~\textsc{Cook},\altaffilmark{\ref{affil:LewCook}}
        Ken'ichi~\textsc{Torii},\altaffilmark{\ref{affil:Torii}}
        Donn~R.~\textsc{Starkey},\altaffilmark{\ref{affil:Starkey}}
        Jeremy~\textsc{Shears},\altaffilmark{\ref{affil:Shears}}
        Lasse-Teist~\textsc{Jensen},\altaffilmark{\ref{affil:Jensen}}
        Gianluca~\textsc{Masi},\altaffilmark{\ref{affil:Masi}} 
        Tom\'{a}\v{s}~\textsc{Hynek},\altaffilmark{\ref{affil:JohannPalisa}}
        Rudolf~\textsc{Nov\'{a}k},\altaffilmark{\ref{affil:Novak}}
        Radek~\textsc{Kocian},\altaffilmark{\ref{affil:JohannPalisa}}
        Luk\'{a}\v{s}~\textsc{Kr\'{a}l},\altaffilmark{\ref{affil:JohannPalisa}}
        Hana~\textsc{Kucakova},\altaffilmark{\ref{affil:JohannPalisa}}
        Marek~\textsc{Kolasa},\altaffilmark{\ref{affil:JohannPalisa}}
        Petr~\textsc{Stastny},\altaffilmark{\ref{affil:JohannPalisa}}
        Bart~\textsc{Staels},\altaffilmark{\ref{affil:AAVSO}}$^,$\altaffilmark{\ref{affil:Staels}}
        Ian~\textsc{Miller},\altaffilmark{\ref{affil:Miller}}
        Yasuo~\textsc{Sano},\altaffilmark{\ref{affil:Sano}}
        Pierre~de~\textsc{Ponthi\`ere},\altaffilmark{\ref{affil:Ponthiere}}
        Atsushi~\textsc{Miyashita},\altaffilmark{\ref{affil:Seikei}}
        Tim~\textsc{Crawford},\altaffilmark{\ref{affil:Crawford}}
        Steve~\textsc{Brady},\altaffilmark{\ref{affil:Brady}}
        Roland~\textsc{Santallo},\altaffilmark{\ref{affil:Santallo}}
        Tom~\textsc{Richards},\altaffilmark{\ref{affil:Richards}}
        Brian~\textsc{Martin},\altaffilmark{\ref{affil:Martin}}
        Denis~\textsc{Buczynski},\altaffilmark{\ref{affil:Buczynski}}
        Michael~\textsc{Richmond},\altaffilmark{\ref{affil:RIT}}
        Jim~\textsc{Kern},\altaffilmark{\ref{affil:RIT}}
        Stacey~\textsc{Davis},\altaffilmark{\ref{affil:RIT}}
        Dustin~\textsc{Crabtree},\altaffilmark{\ref{affil:RIT}}
        Kevin~\textsc{Beaulieu},\altaffilmark{\ref{affil:RIT}}
        Tracy~\textsc{Davis},\altaffilmark{\ref{affil:RIT}}
        Matt~\textsc{Aggleton},\altaffilmark{\ref{affil:RIT}}
        Etienne~\textsc{Morelle},\altaffilmark{\ref{affil:Morelle}}
        Elena~P.~\textsc{Pavlenko},\altaffilmark{\ref{affil:Pavlenko}}
        Maksim~\textsc{Andreev},\altaffilmark{\ref{affil:Terskol}}
        Alexander~\textsc{Baklanov},\altaffilmark{\ref{affil:Pavlenko}}
        Michael~D.~\textsc{Koppelman},\altaffilmark{\ref{affil:Koppelman}}
        Gary~\textsc{Billings},\altaffilmark{\ref{affil:Billings}}
        \v{L}ubom\'{i}r~\textsc{Urban\v{c}ok},\altaffilmark{\ref{affil:Urbancok}}
        Yenal~\textsc{\"Ogmen},\altaffilmark{\ref{affil:Ogmen}}
        Bernard~\textsc{Heathcote},\altaffilmark{\ref{affil:Heathcote}}
        Tomas~L.~\textsc{Gomez},\altaffilmark{\ref{affil:Gomez}}
        Alon~\textsc{Retter},\altaffilmark{\ref{affil:Retter}}
        Krzysztof~\textsc{Mularczyk},\altaffilmark{\ref{affil:Warsaw}}
        Kamil~\textsc{Z{\l}oczewski},\altaffilmark{\ref{affil:NicolausCopernicus}}
        Arkadiusz~\textsc{Olech},\altaffilmark{\ref{affil:NicolausCopernicus}}
        Piotr~\textsc{Kedzierski},\altaffilmark{\ref{affil:Warsaw}}
        Roger~D.~\textsc{Pickard},\altaffilmark{\ref{affil:BAAVSS}}$^,$\altaffilmark{\ref{affil:Pickard}}
        Chris~\textsc{Stockdale},\altaffilmark{\ref{affil:Stockdale}}
        Jani~\textsc{Virtanen},\altaffilmark{\ref{affil:Virtanen}}
        Koichi~\textsc{Morikawa},\altaffilmark{\ref{affil:Morikawa}}
        Franz-Josef~\textsc{Hambsch},\altaffilmark{\ref{affil:GEOS}}$^,$\altaffilmark{\ref{affil:BAV}}$^,$\altaffilmark{\ref{affil:Hambsch}}
        Gordon~\textsc{Garradd},\altaffilmark{\ref{affil:Garradd}}
        Carlo~\textsc{Gualdoni},\altaffilmark{\ref{affil:Gualdoni}}
        Keith~\textsc{Geary},\altaffilmark{\ref{affil:AAVSO}}
        Toshihiro~\textsc{Omodaka},\altaffilmark{\ref{affil:Kagoshima}}
        Nobuyuki~\textsc{Sakai},\altaffilmark{\ref{affil:Kagoshima}}
        Raul~\textsc{Michel},\altaffilmark{\ref{affil:UNAM}}
        A.~A.~\textsc{C\'{a}rdenas},\altaffilmark{\ref{affil:UNAM}}
        Kosmas~D.~\textsc{Gazeas},\altaffilmark{\ref{affil:Athens}}
        Panos~G.~\textsc{Niarchos},\altaffilmark{\ref{affil:Athens}}
        A.~V.~\textsc{Yushchenko},\altaffilmark{\ref{affil:Odessa}}
        Franco~\textsc{Mallia},\altaffilmark{\ref{affil:Mallia}}
        Marco~\textsc{Fiaschi},\altaffilmark{\ref{affil:Fiaschi}}
        Gerry~A.~\textsc{Good},\altaffilmark{\ref{affil:GerryGood}}
        Stan~\textsc{Walker},\altaffilmark{\ref{affil:Walker}}
        Nick~\textsc{James},\altaffilmark{\ref{affil:James}}
        Ken-ichi~\textsc{Douzu},\altaffilmark{\ref{affil:Matsumoto}}
        Wm~Mack~\textsc{Julian}~II,\altaffilmark{\ref{affil:Julian}}
        Neil~D.~\textsc{Butterworth},\altaffilmark{\ref{affil:Butterworth}}
        Sergey~Yu.~\textsc{Shugarov},\altaffilmark{\ref{affil:Sternberg}}$^,$\altaffilmark{\ref{affil:Slovak}}
        Igor~\textsc{Volkov},\altaffilmark{\ref{affil:Sternberg}}$^,$\altaffilmark{\ref{affil:Slovak}}
        Drahomir~\textsc{Chochol},\altaffilmark{\ref{affil:Slovak}}
        Natalia~\textsc{Katysheva},\altaffilmark{\ref{affil:Sternberg}}
        Alexander E.~\textsc{Rosenbush},\altaffilmark{\ref{affil:Rosenbush}}
        Maria~\textsc{Khramtsova},\altaffilmark{\ref{affil:Terskol}}
        Petri~\textsc{Kehusmaa},\altaffilmark{\ref{affil:Kehusmaa}}
        Maciej~\textsc{Reszelski},\altaffilmark{\ref{affil:Reszelski}}
        James~\textsc{Bedient},\altaffilmark{\ref{affil:AAVSO}}
        William~\textsc{Liller},\altaffilmark{\ref{affil:Liller}}
        Grzegorz~\textsc{Pojma\'nski},\altaffilmark{\ref{affil:Pojmanski}}
        Mike~\textsc{Simonsen},\altaffilmark{\ref{affil:Simonsen}}
        Rod~\textsc{Stubbings},\altaffilmark{\ref{affil:Stubbings}}
        Patrick~\textsc{Schmeer},\altaffilmark{\ref{affil:Schmeer}}
        Eddy~\textsc{Muyllaert},\altaffilmark{\ref{affil:VVSBelgium}}
        Timo~\textsc{Kinnunen},\altaffilmark{\ref{affil:Kinnunen}}
        Gary~\textsc{Poyner},\altaffilmark{\ref{affil:Poyner}}
        Jose~\textsc{Ripero},\altaffilmark{\ref{affil:Ripero}}
        Wolfgang~\textsc{Kriebel},\altaffilmark{\ref{affil:BAV}}$^,$\altaffilmark{\ref{affil:Kriebel}}
}

\authorcount{affil:Kyoto}{
     Department of Astronomy, Kyoto University, Kyoto 606-8502}
\email{$^*$tkato@kusastro.kyoto-u.ac.jp}

\authorcount{affil:Imada}{
     Okayama Astrophysical Observatory, National Astronomical
     Observatory of Japan, Asakuchi, Okayama 719-0232}

\authorcount{affil:Uemura}{
     Astrophysical Science Center, Hiroshima University, Kagamiyama, 1-3-1
     Higashi-Hiroshima 739-8526}

\authorcount{affil:HidaKwasan}{
     Kwasan and Hida Observatories, Kyoto University, Yamashina,
     Kyoto 607-8471}

\authorcount{affil:Ishioka}{
     Subaru Telescope, National Astronomical Observatory of Japan, 650
     North A'ohoku Place, Hilo, HI 96720, USA}

\authorcount{affil:Baba}{
     Institute of Space and Astronautical Science, Japan Aerospace
     Exploration Agency, 3-1-1 Yoshinodai, Sagamihara, Kanagawa 229-8510}

\authorcount{affil:Matsumoto}{
     Osaka Kyoiku University, 4-698-1 Asahigaoka, Osaka 582-8582}

\authorcount{affil:Njh}{
     Variable Star Observers League in Japan (VSOLJ),
     124 Isatotyo, Teradani, Kumano, Mie 519-4673}

\authorcount{affil:Kis}{
     VSOLJ, 405-1003 Matsushiro, Tsukuba, Ibaraki 305-0035}

\authorcount{affil:OUS}{
     Department of Biosphere-Geosphere Systems, Faculty of Informatics,
     Okayama University of Science, 1-1 Ridai-cho, Okayama, Okayama 700-0005}

\authorcount{affil:AAVSO}{
     American Association of Variable Star Observers, 49 Bay State Rd.,
     Cambridge, MA 02138, USA}

\authorcount{affil:Nyrola}{
     Nyrola observatory, Jyvaskylan Sirius ry, Vertaalantie
     419, FI-40270 Palokka, Finland}

\authorcount{affil:Monard}{
     Bronberg Observatory, CBA Pretoria, PO Box 11426, Tiegerpoort 0056,
     South Africa}

\authorcount{affil:Ioh}{
     VSOLJ, 1001-105 Nishiterakata, Hachioji, Tokyo 192-0153}

\authorcount{affil:Dubovsky}{
     Vihorlat Observatory, Mierova 4, Humenne, Slovakia}

\authorcount{affil:Dancikova}{
     Phillips Academy Andover, USA}

\authorcount{affil:Vanmunster}{
     Center for Backyard Astrophysics (Belgium), Walhostraat 1A, B-3401,
     Landen, Belgium}

\authorcount{affil:Pietz}{
     Nollenweg 6, 65510 Idstein, Germany}

\authorcount{affil:Bolt}{
     Camberwarra Drive, Craigie, Western Australia 6025, Australia}

\authorcount{affil:DavidBoyd}{
     Silver Lane, West Challow, Wantage, OX12 9TX, UK}

\authorcount{affil:Nelson}{
     RMB 2493, Ellinbank 3820, Australia}

\authorcount{affil:Krajci}{
     Center for Backyard Astrophysics New Mexico, PO Box 1351 Cloudcroft,
     New Mexico 83117, USA}

\authorcount{affil:LewCook}{
     Center for Backyard Astrophysics (Concord), 1730 Helix Ct. Concord,
     California 94518, USA}

\authorcount{affil:Torii}{
     Department of Earth and Space Science, Graduate School of
     Science, Osaka University, 1-1 Machikaneyama-cho, Toyonaka, Osaka
     560-0043}

\authorcount{affil:Starkey}{
     DeKalb Observatory, H63, 2507 County Road 60, Auburn, Indiana 46706, USA}

\authorcount{affil:Shears}{
     ``Pemberton'', School Lane, Bunbury, Tarporley, Cheshire, CW6 9NR, UK}

\authorcount{affil:Jensen}{
     Sondervej 38, DK-8350 Hundslund, Denmark}

\authorcount{affil:Masi}{
     The Virtual Telescope Project, Via Madonna del Loco 47, 03023
     Ceccano (FR), Italy}

\authorcount{affil:JohannPalisa}{
     Observatory and Planetarium of Johann Palisa, VSB -- Technical
     University Ostrava, Trida 17. listopadu 15, Ostrava -- Poruba 708 33, 
     Czech Republic}

\authorcount{affil:Novak}{
     Institute of Computer Science, Faculty of Civil Engineering,
     Brno University of Technology, 602 00 Brno, Czech Republic}

\authorcount{affil:Staels}{
     Center for Backyard Astrophysics (Flanders),
     American Association of Variable Star Observers (AAVSO),
     Alan Guth Observatory, Koningshofbaan 51, Hofstade, Aalst, Belgium}

\authorcount{affil:Miller}{
     Furzehill House, Ilston, Swansea, SA2 7LE, UK}

\authorcount{affil:Sano}{
     VSOLJ, Nishi juni-jou minami 3-1-5, Nayoro, Hokkaido 096-0022}

\authorcount{affil:Ponthiere}{
     American Association of Variable Star Observers (AAVSO),
     15 rue Pr\'e Mathy, 5170 Lesve (Profondeville), Belgium}

\authorcount{affil:Seikei}{
     Seikei Meteorological Observatory, Seikei High School}

\authorcount{affil:Crawford}{
     Arch Cape Observatory, 79916 W. Beach Road, Arch Cape, OR 97102}

\authorcount{affil:Brady}{
     5 Melba Drive, Hudson, NH 03051, USA}

\authorcount{affil:Santallo}{
     Southern Stars Observatory, Po Box 60972, 98702 FAAA TAHITI,
     French Polynesia}

\authorcount{affil:Richards}{
     Woodridge Observatory, 8 Diosma Rd, Eltham, Vic 3095, Australia}

\authorcount{affil:Martin}{
     The King's University College; Center for Backyard Astrophysics
     (Alberta), Edmonton, Alberta, Canada T6B 2H3}

\authorcount{affil:Buczynski}{
     Conder Brow Observatory, Fell Acre, Conder Brow, Little Fell Lane,
     Scotforth, Lancs LA2 0RQ, England}

\authorcount{affil:RIT}{
     Physics Department, Rochester Institute of Technology, Rochester,
     New York 14623, USA}

\authorcount{affil:Morelle}{
     9 rue Vasco de GAMA, 59553 Lauwin Planque, France}

\authorcount{affil:Pavlenko}{
     Crimean Astrophysical Observatory, 98409, Nauchny, Crimea, Ukraine}

\authorcount{affil:Terskol}{
     Institute of Astronomy, Russian Academy of Sciences, 361605 Peak Terskol,
     Kabardino-Balkaria, Russia}

\authorcount{affil:Koppelman}{
     Department of Astronomy, University of Minnesota, MN, USA}

\authorcount{affil:Billings}{
     2320 Cherokee Drive NW, Calgary, Alberta, T2L 0X7 Canada}

\authorcount{affil:Urbancok}{
     \v{S}\'{i}d Astronomical Observatory, \v{S}\'{i}d 303, 986 01,
     The Slovak republic}

\authorcount{affil:Ogmen}{
     Green Island Observatory, Ge\c{c}itkale, Magosa, via Mersin, North Cyprus}

\authorcount{affil:Heathcote}{
     Barfold Observatory, 165 Sievers Lane, Glenhope, 3444, Victoria, Australia}

\authorcount{affil:Gomez}{
     ICMAT (CSIC-UAM-UC3M-UCM), Serrano 113bis, 28006 Madrid, Spain}

\authorcount{affil:Retter}{
     86a/6 Hamaccabim St., PO Box 4264, Shoham, 60850 Israel}

\authorcount{affil:Warsaw}{
     Warsaw University Observatory, Al. Ujazdowskie 4,
     00-478 Warsaw, Poland}

\authorcount{affil:BAAVSS}{
     The British Astronomical Association, Variable Star Section (BAA VSS),
     Burlington House, Piccadilly, London, W1J 0DU, UK}

\authorcount{affil:Pickard}{
     3 The Birches, Shobdon, Leominster, Herefordshire, HR6 9NG, UK}

\authorcount{affil:NicolausCopernicus}{
     Nicolaus Copernicus Astronomical Center, Bartycka 18, 00-716 Warsaw, Poland}

\authorcount{affil:Stockdale}{
     8 Matta Drive, Churchill, Victoria  3842, Australia}

\authorcount{affil:Virtanen}{
     Ollilantie 98, 84880 Ylivieska, Finland}

\authorcount{affil:Morikawa}{
     468-3 Satoyamada, Yakage-cho, Oda-gun, Okayama 714-1213}

\authorcount{affil:GEOS}{
     Groupe Europ\'een d'Observations Stellaires (GEOS),
     23 Parc de Levesville, 28300 Bailleau l'Ev\^eque, France}

\authorcount{affil:BAV}{
     Bundesdeutsche Arbeitsgemeinschaft f\"ur Ver\"aderliche Sterne
     (BAV), Munsterdamm 90, 12169 Berlin, Germany}

\authorcount{affil:Hambsch}{
     Vereniging Voor Sterrenkunde (VVS), Oude Bleken 12, 2400 Mol, Belgium}

\authorcount{affil:Garradd}{
     PO Box 157, NSW 2340, Australia}

\authorcount{affil:Gualdoni}{
     22100 Como, Italy}


\authorcount{affil:Kagoshima}{
     Faculty of Science, Kagoshima University, 1-21-30 Korimoto, Kagoshima,
     Kagoshima 890-0065}

\authorcount{affil:UNAM}{
     Instituto de Astronom\'{\i}a UNAM, Apartado Postal 877, 22800 Ensenada
     B.C., M\'{e}xico}

\authorcount{affil:Athens}{
     Department of Astrophysics, Astronomy and Mechanics, University of
     Athens, Panepistimipolis, GR-157 84, Zografos, Athens, Greece}

\authorcount{affil:Odessa}{
     Department of Astronomy and Astronomical Observatory, Odessa National
     University, Shevchenko Park, 270014 Odessa, Ukraine}

\authorcount{affil:Mallia}{
     Campo Catino Astronomical Observatory, Via dei Siculi, 37,
     04100 Latina, Italy}

\authorcount{affil:Fiaschi}{
     Astronomical Observatory G. Colombo, Via Caltana 242,
     35011 Campodarsego PD, Italy}

\authorcount{affil:GerryGood}{
     Albuquerque, New Mexico, USA}

\authorcount{affil:Walker}{
     Wharemaru Observatory, P.O. Box 13, Awanui 0552, New Zealand}

\authorcount{affil:James}{
     11 Tavistock Road, Chelmsford, Essex CM1 6JL, UK}

\authorcount{affil:Julian}{
     4597 Rockaway Loop, Rio Rancho, NM 87124, USA}

\authorcount{affil:Butterworth}{
     24 Payne Street, Mount Louisa, Queensland 4814, Australia}

\authorcount{affil:Sternberg}{
     Sternberg Astronomical Institute, Moscow University, Universitetsky
     Ave., 13, Moscow 119992, Russia}

\authorcount{affil:Slovak}{
     Astronomical Institute of the Slovak Academy of Sciences, 05960,
     Tatranska Lomnica, the Slovak Republic}

\authorcount{affil:Rosenbush}{
     Main Astronomical Observatory, Golosiiv, Kyiv-127, 03680, Ukraine}

\authorcount{affil:Kehusmaa}{
     Slope Rock Observatory, Uima-altaankatu 19, FIN-05820 Hyvinkaa, Finland}

\authorcount{affil:Reszelski}{
     Al. 1-go Maja 29/4, 64500 Szamotuly, Poland}

\authorcount{affil:Liller}{
     Center for Nova Studies, Casilla 5022, Vi\~{n}a del Mar, Chile}

\authorcount{affil:Pojmanski}{
     Warsaw University Observatory, Al. Ujazdowskie 4, 00-478 Warsaw, Poland}

\authorcount{affil:Simonsen}{
     AAVSO, C. E. Scovil Observatory, 2615 S. Summers Rd., Imlay City,
     Michigan 48444, USA}

\authorcount{affil:Stubbings}{
     Tetoora Observatory, Tetoora Road, Victoria, Australia}

\authorcount{affil:Schmeer}{
     Bischmisheim, Am Probstbaum 10, 66132 Saarbr\"{u}cken, Germany}

\authorcount{affil:VVSBelgium}{
     Vereniging Voor Sterrenkunde (VVS),  Moffelstraat 13 3370
     Boutersem, Belgium}

\authorcount{affil:Kinnunen}{
     Sinirinnantie 16, SF-02660 Espoo, Finland}

\authorcount{affil:Poyner}{
     BAA Variable Star Section, 67 Ellerton Road, Kingstanding,
     Birmingham B44 0QE, UK}

\authorcount{affil:Ripero}{
     President of CAA (Centro Astronomico de Avila) and Variable and SNe
     Group M1, Buenavista 7, Ciudad Sto. Domingo, 28110 Algete/Madrid, Spain}

\authorcount{affil:Kriebel}{
     D-84072 Osterwaal Post Au, Germany}



\maketitle

\begin{abstract}
   We systematically surveyed period variations of superhumps in
SU UMa-type dwarf novae based on newly obtained data and past
publications.  In many systems, the evolution of superhump period are
found to be composed of three distinct stages: early evolutionary
stage with a longer superhump period, middle stage with systematically
varying periods, final stage with a shorter, stable superhump period.
During the middle stage, many systems with superhump periods less than
0.08 d show positive period derivatives.
We present observational characteristics of these stages and
greatly improved statistics.  Contrary to the earlier claim,
we found no clear evidence for variation of period derivatives between
superoutburst of the same object.
We present an interpretation that the lengthening of the superhump period
is a result of outward propagation of the eccentricity wave and
is limited by the radius near the tidal truncation.  We interpret that
late stage superhumps are rejuvenized excitation of 3:1 resonance when
the superhumps in the outer disk is effectively quenched.
The general behavior of period variation, particularly in systems
with short orbital periods, appears to follow the scenario
proposed in \citet{kat08wzsgelateSH}.
We also present an observational summary of WZ Sge-type dwarf novae.
Many of WZ Sge-type dwarf novae showed long-enduring superhumps during
the post-superoutburst stage having periods longer than those during
the main superoutburst.
The period derivatives in WZ Sge-type dwarf novae are found to be
strongly correlated with the fractional superhump excess,
or consequently, mass ratio.
WZ Sge-type dwarf novae with a long-lasting rebrightening or with
multiple rebrightenings tend to have smaller period derivatives and
are excellent candidate for the systems around or after the period
minimum of evolution of cataclysmic variables.
\end{abstract}

\newpage

\section{Introduction}

   Dwarf novae (DNe) are a class of cataclysmic variables (CVs), which are
close binary systems consisting of a white dwarf and a red-dwarf secondary
transferring matter via the Roche-lobe overflow.
SU UMa-type dwarf novae, a subclass of DNe, show superhumps during
their long, bright outbursts (superoutbursts)
[see e.g. \citet{vog80suumastars}; \citet{war85suuma}].
The origin of superhumps is basically understood as a result of
varying tidal dissipation in an eccentric accretion disk, whose
eccentricity is excited by the 3:1 orbital resonance
(\cite{whi88tidal}; \cite{osa89suuma}; \cite{osa96review}).

   Until the mid-1990's, the period of superhumps ($P_{\rm SH}$)
had been considered to decrease during superoutburst
(cf. \cite{war85suuma}), which was explained as a result of
decreasing radius of the accretion disk
during superoutburst \citep{osa85SHexcess}.  In recent years,
the existence of objects with positive period derivatives
($P_{\rm dot} = \dot{P}/P$) of superhumps, particularly among systems
with short orbital periods ($P_{\rm orb}$) has been established 
(see e.g. \cite{kat01hvvir}).  Since the superhump period, or its variation,
is related to the radius of the accretion disk, or propagation of
the eccentricity wave (see e.g. \cite{hir90SHexcess}; \cite{lub91SHa};
\cite{kat98super}), the period variation is expected to provide
diagnostics of the dynamics in the outbursting accretion disk.
A number of pieces of research have been issued in this perspective
(e.g. \cite{uem05tvcrv}; \cite{ima06j0137}; \cite{soe09asas1600}).
In particular, \citet{uem05tvcrv} reported markedly different
$P_{\rm dot}$'s between different superoutbursts of TV Crv.
\citet{uem05tvcrv} proposed an interpretation that this difference
is caused by the different mass (angular momentum) in the accretion disk
at the onset of the superoutburst, following the theory by
\citet{osa03DNoutburst}.
If this is confirmed, $P_{\rm dot}$ is expected to provide
an observational measure of the mass in the disk.

   More recently, long-lasting superhumps with period unexpectedly
($\sim$0.5 \%) longer than superhump periods during the
slowly fading stage of WZ Sge-type superoutbursts have been
established \citep{kat08wzsgelateSH}.
\citet{kat08wzsgelateSH} suggested that
they are superhumps arising from the disk matter outside the
3:1 resonance, or around the tidal truncation.  \citet{kat08wzsgelateSH}
also proposed the transient 2:1 resonance in the outer disk could
regulate the excitation and propagation of the 3:1 resonance,
leading to a novel interpretation of the variety of $P_{\rm dot}$
in different SU UMa-type dwarf novae.

   Motivated by these suggestions, we present a new systematic survey
of $P_{\rm dot}$ in SU UMa-type dwarf novae.  The lack of published
times of maxima in some of  references having been one of the major
obstacles in the research of period variations of superhumps, we present
times of all measured superhumps for potential future analysis.

   In section \ref{sec:obs}, we describe our observation and method
of analysis.  In section \ref{sec:general}, we describe
general properties of period variation in superhumps.  General
discussions are given in section \ref{sec:discussion}.
Section \ref{sec:wzsgestars} is dedicated to WZ Sge-type dwarf novae.
These sections are placed
before section \ref{sec:individual} (individual objects)
because of the large amount of data presented in section
\ref{sec:individual}.  We finally give section \ref{sec:conclusion}
for a summary of new findings.
The names of the objects are sometimes abbreviated in
tables, figures and sections \ref{sec:general} and \ref{sec:discussion};
for original names of these objects, refer to section \ref{sec:individual}.
Alternative designations were sometimes used when the original names
were difficult to abbreviate properly.

\section{Observation and Analysis}\label{sec:obs}

   The data were obtained under campaigns led by the VSNET Collaboration
\citep{VSNET}.  In some objects, we used archival data for published
papers, and the public data from the AAVSO International Database\footnote{
$<$http://www.aavso.org/data/download/$>$.
}
as a supplementary purpose.  The majority of the data were acquired
by time-resolved CCD photometry of with 30 cm-class telescopes, whose
observational details on individual objects will be presented in
separate papers dealing with in-depth analysis and discussion on
individual objects.\footnote{
   During this analysis, it became evident that the KU computer
   lost ntp connection between 2008 May 16 and November 25.
   The times of observations
   during this period have been corrected by correlating with
   other simultaneous observations.  The maximum correction amounted
   to 0.005 d and estimated maximum error of correction 0.001 d.
   The details of the corrections and these effects will be discussed
   in Ohshima et al., in preparation.
   The objects affected were V466 And, VY Aqr,
   KP Cas, V1251 Cyg, V630 Cyg, HO Del, V699 Oph, PV Per,
   UW Tri, DO Vul, NSV 5285, SDSS J1627, OT J0211, OT J0238,
   OT J1631 and OT J1914.
   The maximum uncertainty caused by these corrections were
   0.00001--0.00002 d for periods of V466 And, V1251 Cyg and PV Per,
   and less than 0.00001 d for other objects.
   The maximum uncertainty for $P_{\rm dot}$ was less than
   $1 \times 10^{-5}$ .
}
We generally restricted our analysis to superoutburst
plateau and rapid fading stage.  In a few very well-observed cases,
we dealt with post-superoutburst evolution of superhumps.

\begin{figure}
  \begin{center}
    \FigureFile(88mm,70mm){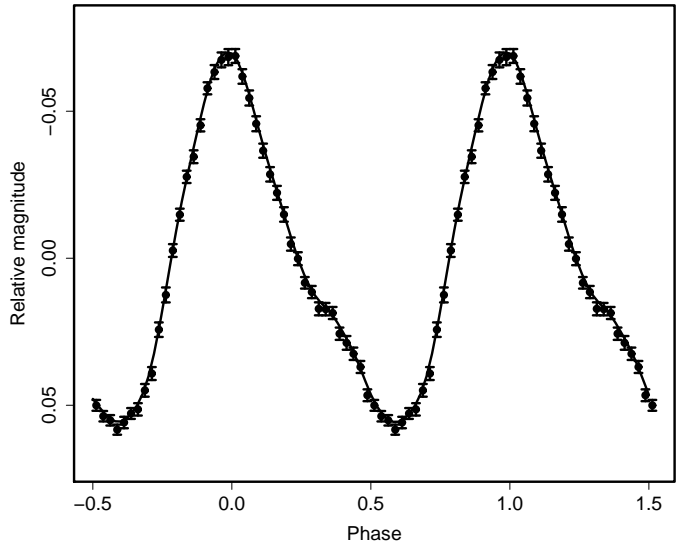}
  \end{center}
  \caption{Template light curve (phase-averaged light curve of superhumps
  in GW Lib).}
  \label{fig:gwsh}
\end{figure}

   After correction for systematic differences between observers, and
subtracting the general trend by fitting low-order (typically three
to five) polynomials, we extracted times of superhump maxima by numerically
fitting a template superhump light curve around the times of observed
maxima.  We did not use the full superhump cycle but generally used
phases $-$0.4 to 0.4 in order to minimize the contamination from
potentially present secondary maxima.  We employed a phase-averaged
(and spline interpolated) light curve of superhumps of GW Lib as the
template, which is one of the best-sampled object among all SU UMa-type
dwarf novae (figure \ref{fig:gwsh}).

This usage of a fixed template has an advantage of much higher
signal-to-noise and thereby higher precision in determining maxima than
eye estimates (typically reducing the scatter by a factor of $\sim$ 5)
or than fitting using lower-quality template light curves prepared for
individual objects.  The usage of a fixed template, however, has
a potential disadvantage of systematic errors caused by the variation
in the superhump profile and the difference of the profile from
the template.
These potential effects have been examined by comparisons between previously
reported times of maxima (referring to the same data) and
those determined in the present work.
No significant systematic differences were found to affect the
determination of $P_{\rm dot}$.
In some cases, comparisons with other authors have yielded significant
constant offsets (individually described in section \ref{sec:individual}),
presumably caused by different methods in extracting maxima.
These offsets were also found to be insensitive to determining $P_{\rm dot}$
after adjustment by constant offsets.

   We generally used Phase Dispersion Minimization (PDM, \cite{PDM})
for determining mean superhump periods described in the text.  The values
determined using linear regressions to times of superhump maxima
can be slightly different from those determined with the PDM.
When segments (in $E$) are shown, these periods were derived from
a linear regression of maxima times of superhumps unless otherwise
noticed.

   Since we mainly focus on period variations of superhumps,
we only present superhump maxima and mostly omit individual light curves
of outbursts, light curves of superhumps and results of PDM analysis
to save space in section \ref{sec:individual}.
Individual $O-C$ diagrams are not usually shown for the same reason;
selected examples of $O-C$ diagrams are summarized in section
\ref{sec:general}.  We, however, tried to include a comparison of
$O-C$ diagrams if different superoutbursts of the same object were
observed, and tried to include the result of period analysis and
the superhump profile if they provide the first solid presentation.
In section \ref{sec:individual}, we also included selected
observations of several superoutbursts not sufficiently covered to determine
$P_{\rm dot}$, if the determination of $P_{\rm SH}$ is meaningful
in itself or if the inclusion improves the statistical quality.
We also included partially observed superoutbursts for completeness,
and this inclusion for future research is justified by a suggestion
that $P_{\rm dot}$ can be measured from a combination of
different superoutbursts (subsection \ref{sec:different}).

   We also calculated superhump periods and derivatives when times of
superhump maxima were available in the literature.  We employed the
same procedure as in the analysis of our own data.  This work
comprises the largest homogeneous survey of variation of superhumps
in SU UMa-type dwarf novae.

\begin{table*}
\caption{List of Superoutbursts.}\label{tab:outobs}
\begin{center}
\begin{tabular}{cccc}
\hline\hline
Subsection & Object & Year & Observers or references\commenta \\
\hline
\ref{obj:foand}    & FO And & 1994 & \citet{kat95foand} \\
\ref{obj:kvand}    & KV And & 1994 & \citet{kat95kvand} \\
                   & KV And & 2002 & Tor, KU, Tan \\
\ref{obj:lland}    & LL And & 1993 & \citet{kat04lland} \\
                   & LL And & 2004 & KU, AAVSO, Mhh, Njh \\
\ref{obj:v402and}  & V402 And & 2005 & Mhh, AAVSO \\
                   & V402 And & 2006 & Mhh \\
\end{tabular}
\end{center}
\end{table*}

\addtocounter{table}{-1}
\begin{table*}
\caption{(continued) List of Superoutbursts.}
\begin{center}
\begin{tabular}{cccc}
\hline\hline
Subsection & Object & Year & Observers or references \\
\hline
\ref{obj:gxcas}    & GX Cas & 1994 & \citet{nog98gxcasv419lyr} \\
                   & GX Cas & 1996 & JEN \\
                   & GX Cas & 1999 & KU \\
                   & GX Cas & 2006 & KU, Njh \\
\ref{obj:htcas}    & HT Cas & 1985 & \citet{zha86htcas} \\
\ref{obj:kpcas}    & KP Cas & 2008 & JSh, Boy, OUS, BSt, JWM, Nov, KU, Mhh, Njh \\
\ref{obj:v452cas}  & V452 Cas & 1999 & KU \\
\hline
\end{tabular}
\end{center}
\end{table*}

\addtocounter{table}{-1}
\begin{table*}
\caption{(continued) List of Superoutbursts.}
\begin{center}
\begin{tabular}{cccc}
\hline\hline
Subsection & Object & Year & Observers or references \\
\hline
\ref{obj:v632cyg}  & V632 Cyg & 2008 & DPV, Njh, AAVSO, VIR, Mhh \\
\ref{obj:v1028cyg} & V1028 Cyg & 1995 & \citet{bab00v1028cyg}, AAVSO \\
                   & V1028 Cyg & 1996 & Oud, AAVSO \\
                   & V1028 Cyg & 1999 & KU, Buc \\
                   & V1028 Cyg & 2001 & Bil \\
                   & V1028 Cyg & 2002 & KU, OUS, Tor, Bil \\
                   & V1028 Cyg & 2004 & Njh, Nyr, DRS \\
\hline
\end{tabular}
\end{center}
\end{table*}

\addtocounter{table}{-1}
\begin{table*}
\caption{(continued) List of Superoutbursts.}
\begin{center}
\begin{tabular}{cccc}
\hline\hline
Subsection & Object & Year & Observers or references \\
\hline
\ref{obj:v844her}  & V844 Her & 1997 & JEN \\
                   & V844 Her & 1999 & \citet{oiz07v844her} \\
                   & V844 Her & 2002 & \citet{oiz07v844her} \\
                   & V844 Her & 2006 & \citet{oiz07v844her} \\
                   & V844 Her & 2008 & KU, DPV, Mhh \\
\ref{obj:v1108her} & V1108 Her & 2004 & AAVSO, VAN, KU, San, RIT, Hen, Kop, COO, Njh, Mhh, DRS, Boy, CTX \\
\ref{obj:ruhor}    & RU Hor & 2003 & MLF, GBo \\
\hline
\end{tabular}
\end{center}
\end{table*}

\addtocounter{table}{-1}
\begin{table*}
\caption{(continued) List of Superoutbursts.}
\begin{center}
\begin{tabular}{cccc}
\hline\hline
Subsection & Object & Year & Observers or references \\
\hline
\ref{obj:v1159ori} & V1159 Ori & 1993 & \citet{pat95v1159ori} \\
                   & V1159 Ori & 2002 & KU, Kis, Tor, Tan, Oud \\
\ref{obj:v344pav}  & V344 Pav & 2004 & \citet{uem04v344pav} \\
\ref{obj:efpeg}    & EF Peg & 1991 & \citet{kat02efpeg}, \citet{how93efpeg} \\
                   & EF Peg & 1997 & KU \\
\ref{obj:v364peg}  & V364 Peg & 2004 & Kra, VAN \\
\ref{obj:v368peg}  & V368 Peg & 2000 & KU, But, Nyr, Kra \\
\hline
\end{tabular}
\end{center}
\end{table*}

\addtocounter{table}{-1}
\begin{table*}
\caption{(continued) List of Superoutbursts.}
\begin{center}
\begin{tabular}{cccc}
\hline\hline
Subsection & Object & Year & Observers or references \\
\hline
\ref{obj:wytri}    & WY Tri & 2000 & \citet{van01wytri}, KU, Nov \\
\ref{obj:suuma}    & SU UMa & 1989 & \citet{uda90suuma} \\
                   & SU UMa & 1999 & KU, Buc, Mhh \\
\ref{obj:swuma}    & SW UMa & 1991 & Oud \\
                   & SW UMa & 1996 & \citet{sem97swuma}, \citet{nog98swuma} \\
                   & SW UMa & 1997 & JEN \\
                   & SW UMa & 2000 & KU, JEN, Nov, Buc, Mar, Pav, Mas, AAVSO \\
\hline
\end{tabular}
\end{center}
\end{table*}

\addtocounter{table}{-1}
\begin{table*}
\caption{(continued) List of Superoutbursts.}
\begin{center}
\begin{tabular}{ccccc}
\hline\hline
Subsection & Object & Year & Observers or references & ID\commentc \\
\hline
\ref{obj:nsv4838}   & NSV 4838 & 2005 & PIE, VAN, Boy & \\
                    & NSV 4838 & 2007 & Mhh, KU, Njh & \\
\ref{obj:nsv5285}   & NSV 5285 & 2008 & KU & \\
\ref{obj:nsv14652}  & NSV 14652 & 2004 & PIE & \\
\ref{obj:j0232}     & 1RXS J0232 & 2007 & Nel, GBo, AAVSO & Pi of the Sky \\
\ref{obj:j0423}     & 1RXS J0423 & 2008 & BXS, JSh, IMi, BSt, DPV, Ioh, Mhh & \\
\ref{obj:j0532}     & 1RXS J0532 & 2005 & KGE, Mhh, DPP, VAN, Nyr, JSh, COO & \citet{ber05j0532} \\
\end{tabular}
\end{center}
\end{table*}

\addtocounter{table}{-1}
\begin{table*}
\caption{(continued) List of Superoutbursts.}
\begin{center}
\begin{tabular}{ccccc}
\hline\hline
Subsection & Object & Year & Observers or reference & ID \\
\hline
\ref{obj:j0557}     & OT J0557 & 2006 & \citet{uem09j0557}, Boy, VAN, Nyr & \citet{klo06j0557cbet777} \\
\ref{obj:j0747}     & OT J0747 & 2008 & Kis, GBo, Njh, Mhh, Nel, BXS, DPP, & Itagaki  \citep{yam08j0747cbet1216}  \\
                    &          &      & JSh, CTX, Ioh, AAVSO & \\
\ref{obj:j0807}     & OT J0807 & 2007 & Mhh, HHO, Kra, Njh, Kis, DPV & Itagaki \\
\ref{obj:j0814}     & OT J0814 & 2008 & URB, DPV, Njh, KU & CSS080409:081419$-$005022 \\
\ref{obj:j0845}     & OT J0845 & 2008 & Njh, Kis, Mhh & Itagaki \citep{yam08j0845cbet1225} \\
\ref{obj:j0902}     & OT J0902 & 2008 & KU, Mhh & CSS080304:090240$+$052501 \\
\ref{obj:j1021}     & OT J1021 & 2006 & \citet{uem08j1021}, AAVSO & \citet{chr06j1021cbet746} \\
\ref{obj:j1026}     & OT J1026 & 2009 & Njh & Itagaki \citep{yam09j1026cbet1644} \\
\ref{obj:j1028}     & OT J1028 & 2009 & GBo, KU, Mhh, Kis & CSS090331:102843$-$081927 \\
\ref{obj:j1112}     & OT J1112 & 2007 & Ioh, GBo, Mhh, Kis & Pi of the Sky \\
\ref{obj:j1300}     & OT J1300 & 2008 & GBo, Mhh, PIE & CSS080702:130030$+$115101 \\
\ref{obj:j1440}     & OT J1440 & 2009 & IMi, Mhh, KU, OUS & CSS090530:144011$+$494734 \\
\ref{obj:j1443}     & OT J1443 & 2009 & Njh, Ioh, Mhh, GBo, KU, Kis & CSS090418:144342$-$175550 \\
\ref{obj:j1631}     & OT J1631 & 2008 & DPV, PIE, Mhh, KU, Njh & CSS080505:163121$+$103134 \\
\ref{obj:j1914}     & OT J1914 & 2008 & Mhh, KU, Njh, DPV, Nyr, AAVSO & Itagaki \citep{yam08j1914cbet1535} \\
\ref{obj:j1959}     & OT J1959 & 2005 & VAN, KU, Njh & \citet{ren05j1959iauc8591} \\
\ref{obj:j2131}     & OT J2131 & 2008 & Mhh, Ioh, SAc & Itagaki \citep{yam08j1631cbet1631} \\
\ref{obj:j2137}     & OT J2137 & 2008 & GBo, Mhh, Njh, DPV, SAc, Kis, Kry & Itagaki \\
\ref{obj:j0222}     & TSS J0222 & 2005 & \citet{ima06tss0222} & \citet{qui05tss0222atel658} \\
\hline
\end{tabular}
\end{center}
\end{table*}

\section{General Properties}\label{sec:general}

\subsection{Distribution of Superhump Periods}

\begin{figure}
  \begin{center}
    \FigureFile(88mm,70mm){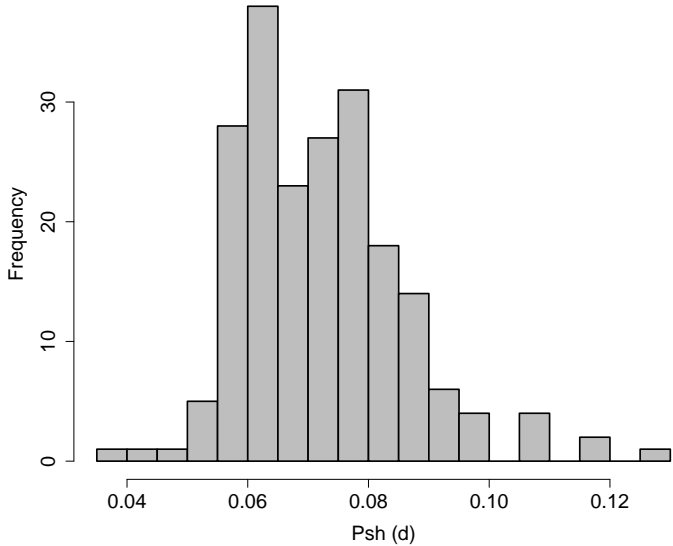}
  \end{center}
  \caption{Distribution of superhump periods in this survey.
  }
  \label{fig:phist}
\end{figure}

   Figure \ref{fig:phist} shows the distribution of superhump periods
in this survey.  With the best statistics ever achieved, we can see
the maximum of the distribution close to $P_{\rm SH} = 0.06$ d
and a monotonous decrease in population towards longer periods.

\subsection{General Tendency in Period Variations}\label{sec:tendency}

\begin{figure}
  \begin{center}
    \FigureFile(88mm,130mm){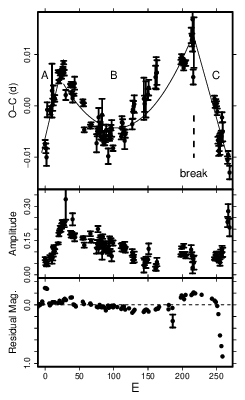}
  \end{center}
  \caption{Representative $O-C$ diagram showing three stages (A--C)
  of $O-C$ variation.  The data were taken from the 2000 superoutburst
  of SW UMa.  (Upper:) $O-C$ diagram.  Three distinct stages
  (A -- evolutionary stage, B -- middle stage, and C -- stage after
  transition to a shorter period) and the location of the period break
  between stages B and C are shown. (Middle): Amplitude of superhumps.
  As shown in \citet{soe09asas1600}, the maximum amplitudes of superhumps
  coincide with transitions between stages (A to B and B to C).
  (Lower:) Deviations from linear decline during the superoutburst
  plateau.  As seen in \citet{soe09asas1600} and \citet{kat03hodel},
  rebrightening during the terminal plateau also corresponds to the
  transition from stage B to C.
  }
  \label{fig:stage}
\end{figure}

   As already demonstrated by several authors (e.g. \cite{ole03ksuma};
\cite{soe09asas1600}), short-$P_{\rm SH}$ SU UMa-type dwarf novae
usually show three distinct stages of period evolution
(figure \ref{fig:stage}):
(A) early stage of superhump evolution having a longer $P_{\rm SH}$,
(B) middle segment with a stabilized period usually with a positive
$P_{\rm dot}$,\footnote{
   This segment occasionally appears to be composed of two linear segments
   forming a ``V''-shaped dip.  Although this could suggest that the
   stage B may not be a continuous entity, we preserve the current
   staging for simplicity and for direct comparison with earlier works.
   Such instances will be individually discussed in section
   \ref{sec:individual}.
} and (C) late stage with a shorter, stable superhump
period.
In well-observed systems, the transitions between stages A and B,
and stages B and C are usually abrupt, associated with discontinuous
period changes.  Although \citet{ole03ksuma} referred these transitions
to decreasing superhump periods, treating as if they are smooth
variations, we adopted the above phenomenological staging because the
transitions are usually discontinuous.

\begin{figure}
  \begin{center}
    \FigureFile(88mm,175mm){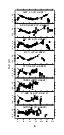}
  \end{center}
  \caption{$O-C$ diagrams of SU UMa-type dwarf novae showing three distinct
  stages.
  }
  \label{fig:ocsamp}
\end{figure}

   Figure \ref{fig:ocsamp} shows $O-C$ diagrams
of representative systems taken from section \ref{sec:individual}
and from literature, in which systems all the three stages were observed.
The figures are arranged in the increasing order of superhump periods
(the periods given in the figures refer to the mean periods during
the stage B).  The thin lines are quadratic fits to the stage B.
Note that the range of cycle counts ($E$) is different between
figures and that the start of stage B was defined to be $E=20$
for better visualization.

   We can see the following general tendency on these figures:
(1) the period derivative during the stage B becomes systematically
smaller with increasing $P_{\rm SH}$, and (2) the duration of the stage B
becomes systematically shorter with increasing $P_{\rm SH}$ or
the fractional superhump excess $\epsilon = P_{\rm SH}/P_{\rm orb}-1$
(figures \ref{fig:bdur}, \ref{fig:bdureps}).

   The last four long-$P_{\rm SH}$ systems (SU UMa, DH Aql, SDSS J1556,
UV Gem) and BZ Cir have nearly zero or negative $P_{\rm dot}$,
but are included in this sequence of figures because they have
all the three distinct stages and because the behavior in these objects
can be understood as a smooth extension of the tendency in
shorter-$P_{\rm SH}$ systems.

   Note also that a few historically controversial systems (V436 Cen:
\cite{sem80v436cen}, and \cite{war83v436cen} for a discussion;
OY Car: \cite{krz85oycarsuper}, and \cite{pat93vyaqr} for a discussion)
can well fit the present general tendency and no anomalies were apparent.

\addtocounter{figure}{-1}
\begin{figure}
  \begin{center}
    \FigureFile(88mm,175mm){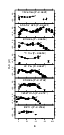}
  \end{center}
  \caption{$O-C$ diagrams of SU UMa-type dwarf novae showing three distinct
  stages (continued).
  }
\end{figure}

\addtocounter{figure}{-1}
\begin{figure}
  \begin{center}
    \FigureFile(88mm,175mm){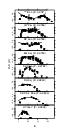}
  \end{center}
  \caption{$O-C$ diagrams of SU UMa-type dwarf novae showing three distinct
  stages (continued).
  }
\end{figure}

\begin{figure}
  \begin{center}
    \FigureFile(88mm,100mm){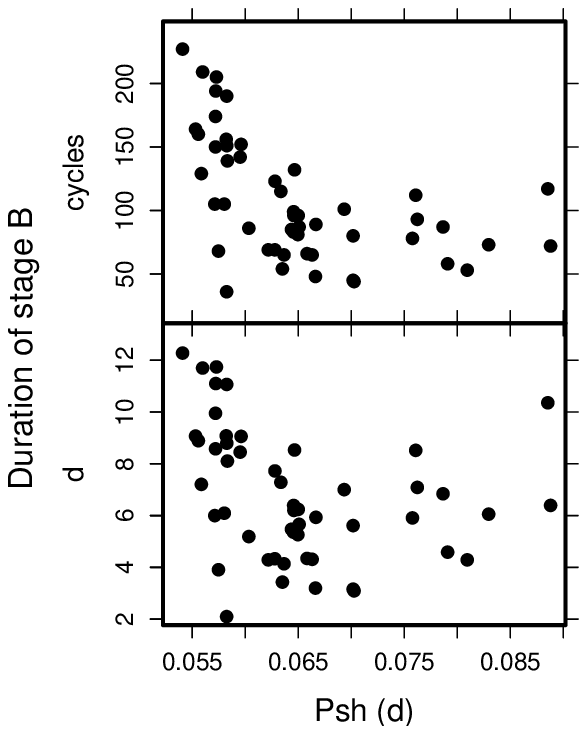}
  \end{center}
  \caption{Duration of stage B.  The duration of stage B decreases with
  increasing $P_{\rm SH}$ both in cycle numbers (upper) and
  time in days (lower).  We used mean $P_{\rm SH}$ during the stage B
  as the representative $P_{\rm SH}$.
  }
  \label{fig:bdur}
\end{figure}

\begin{figure}
  \begin{center}
    \FigureFile(88mm,100mm){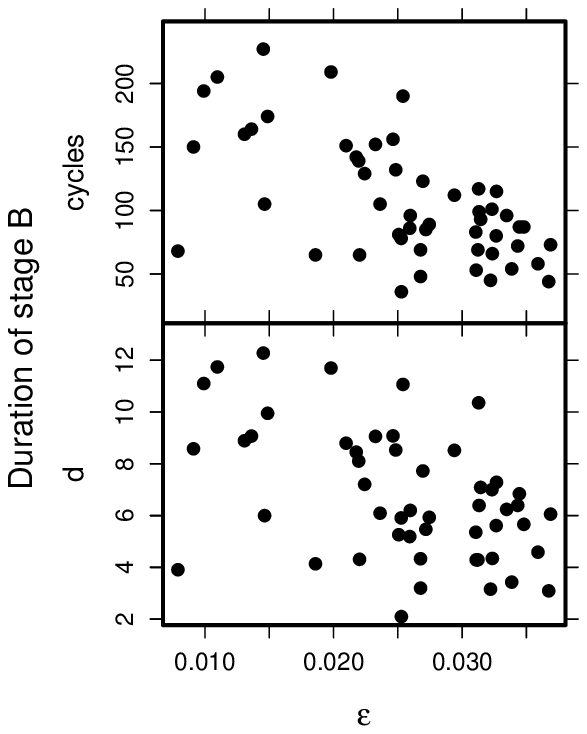}
  \end{center}
  \caption{Duration of stage B.  The duration of stage B decreases with
  increasing $\epsilon$ both in cycle numbers (upper) and
  time in days (lower).  We used mean $P_{\rm SH}$ during the stage B
  for evaluating $\epsilon$.
  }
  \label{fig:bdureps}
\end{figure}

\subsection{Transition to a Shorter Period}

\begin{figure}
  \begin{center}
    \FigureFile(88mm,175mm){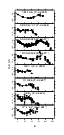}
  \end{center}
  \caption{$O-C$ diagrams of SU UMa-type dwarf novae showing transition
  in the superhump period.
  }
  \label{fig:octrans}
\end{figure}

   Following the stage B, most of well-observed objects
showed a transition to a stage with a shorter $P_{\rm SH}$.
When stage A was not observed or non-existent, this transition on
the $O-C$ diagram appears as a form of ``period break''.  The corresponding
location of this break is shown in figure \ref{fig:stage}.
Figure \ref{fig:octrans} shows $O-C$ diagrams
of selected systems taken from section \ref{sec:individual}
and literature,
in which systems this transition (stage B to stage C) was recorded,
but stage A was not observed.
The durations of the stage B in these systems were not as exactly defined
as in the objects treated in subsection \ref{sec:tendency}.

   Combined with the objects is subsection \ref{sec:tendency},
this transition found to be quite generally, if not always, seen in many
SU UMa-type dwarf novae.  In many well-observed objects, the periods
of superhumps varied little after this transition, in contrast to
the systematic variation seen during the stage B.

\addtocounter{figure}{-1}
\begin{figure}
  \begin{center}
    \FigureFile(88mm,175mm){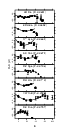}
  \end{center}
  \caption{$O-C$ diagrams of SU UMa-type dwarf novae showing transition
  in the superhump period (continued).
  }
\end{figure}

\subsection{Global Period Derivatives}

   Several authors, including us, have pointed out that $P_{\rm dot}$
in SU UMa-type dwarf novae has a strong correlation with $P_{\rm SH}$
(e.g. \cite{kat01hvvir}; \cite{kat03v877arakktelpucma}; \cite{uem05tvcrv};
\cite{rut07v419lyr}).  These works, however, were based on results from
different segments of $O-C$ diagrams for extracting $P_{\rm dot}$.
On the other hand, \citet{pat93vyaqr} and their descendant papers
calculated $P_{\rm SH}$ from the entire superoutburst (frequently
consisting of stages A--C), and led to a conclusion that almost all
$P_{\rm dot}$'s were negative or zero
(see also a discussion in \cite{ole03ksuma}).

\begin{figure*}
  \begin{center}
    \FigureFile(160mm,140mm){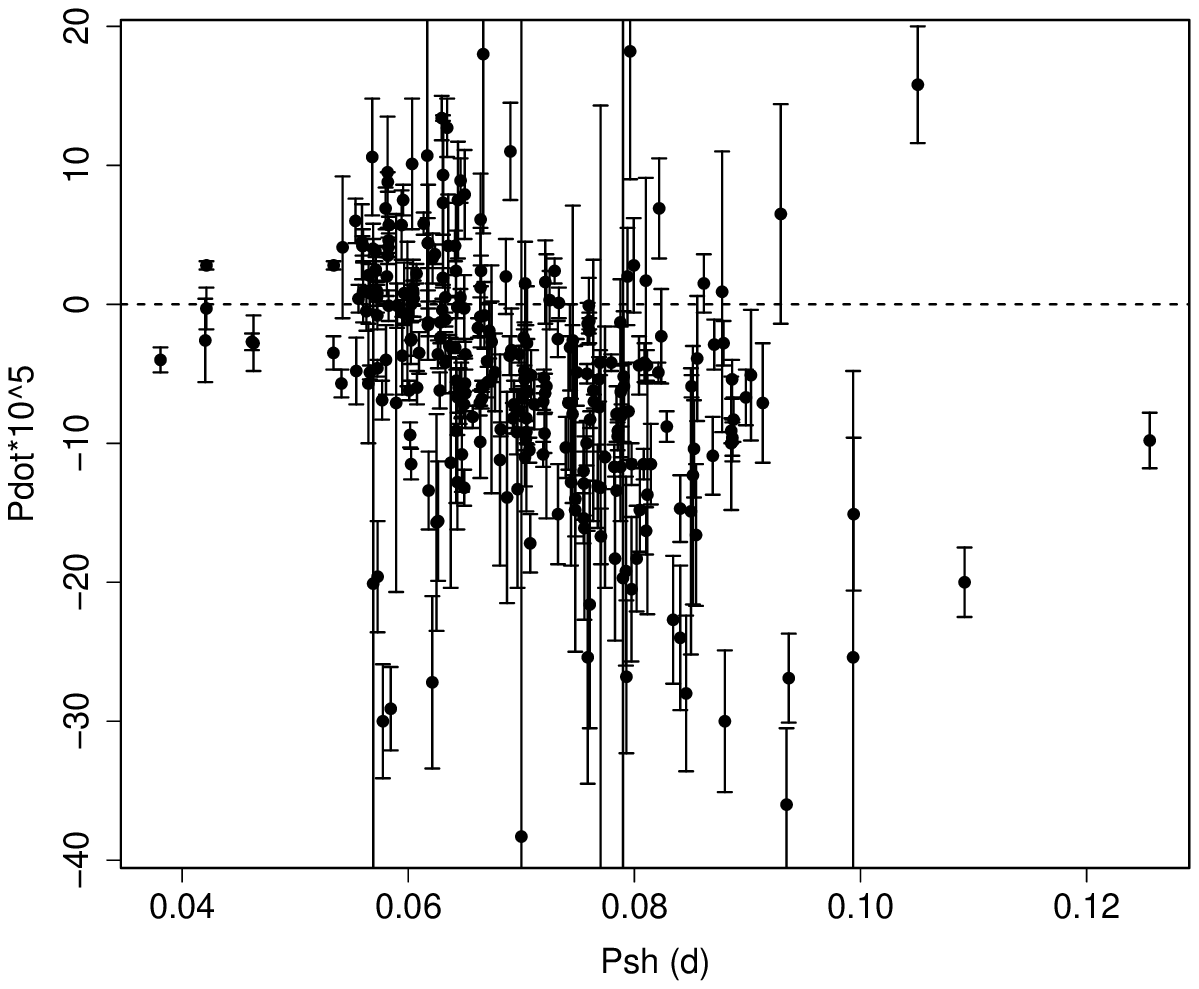}
  \end{center}
  \caption{Globally Determined $P_{\rm dot}$.
  Several objects with extremely negative $P_{\rm dot}$
  (e.g. AX Cap: $-83.0(10.5) \times 10^{-5}$, $P_{\rm SH}$ = 0.1131 d,
   MN Dra: $-165.9(17.7) \times 10^{-5}$, $P_{\rm SH}$ = 0.1077 d,
   NY Ser: $-143.7(7.8) \times 10^{-5}$, 0.1072 d,
   GX Cas: $-66.3(15.2) \times 10^{-5}$, 0.0939 d,
   UV Gem: $-53.4(3.8) \times 10^{-5}$, 0.0931 d)
   are outside this figure.
  }
  \label{fig:global}
\end{figure*}

   We nominally calculated $P_{\rm dot}$ for the entire superoutburst
(restricting to $0 \le E \le 200$ to avoid contaminations from
post-superoutburst variations) and simulated the treatment by
\citet{pat93vyaqr}.  The results presented in figure \ref{fig:global}\footnote{
  Individual values of $P_{\rm dot}$ are not presented because this
  analysis is meaningful only in the context of statistical comparison
  with previous research, and because globally determined $P_{\rm dot}$'s
  on highly structured $O-C$'s are no better than nominal values.
  Better-defined $P_{\rm dot}$ for individual objects are discussed
  in \ref{sec:pdotb} and later (sub)sections.
}
indicate that more than half of systems below $P_{\rm SH} = 0.065$ d
have negative $P_{\rm dot}$.  The presence of systems with positive
$P_{\rm SH}$ and the decreasing trend of $P_{\rm dot}$ with increasing
$P_{\rm SH}$ are already evident from this global determination.

\subsection{Period Derivatives during Stage B}\label{sec:pdotb}

   Since the stages B and C were better studied than the stage A
in many systems, and since they have general properties common to the
majority of superoutbursts, we first describe the stages B and C.

\begin{figure*}
  \begin{center}
    \FigureFile(120mm,80mm){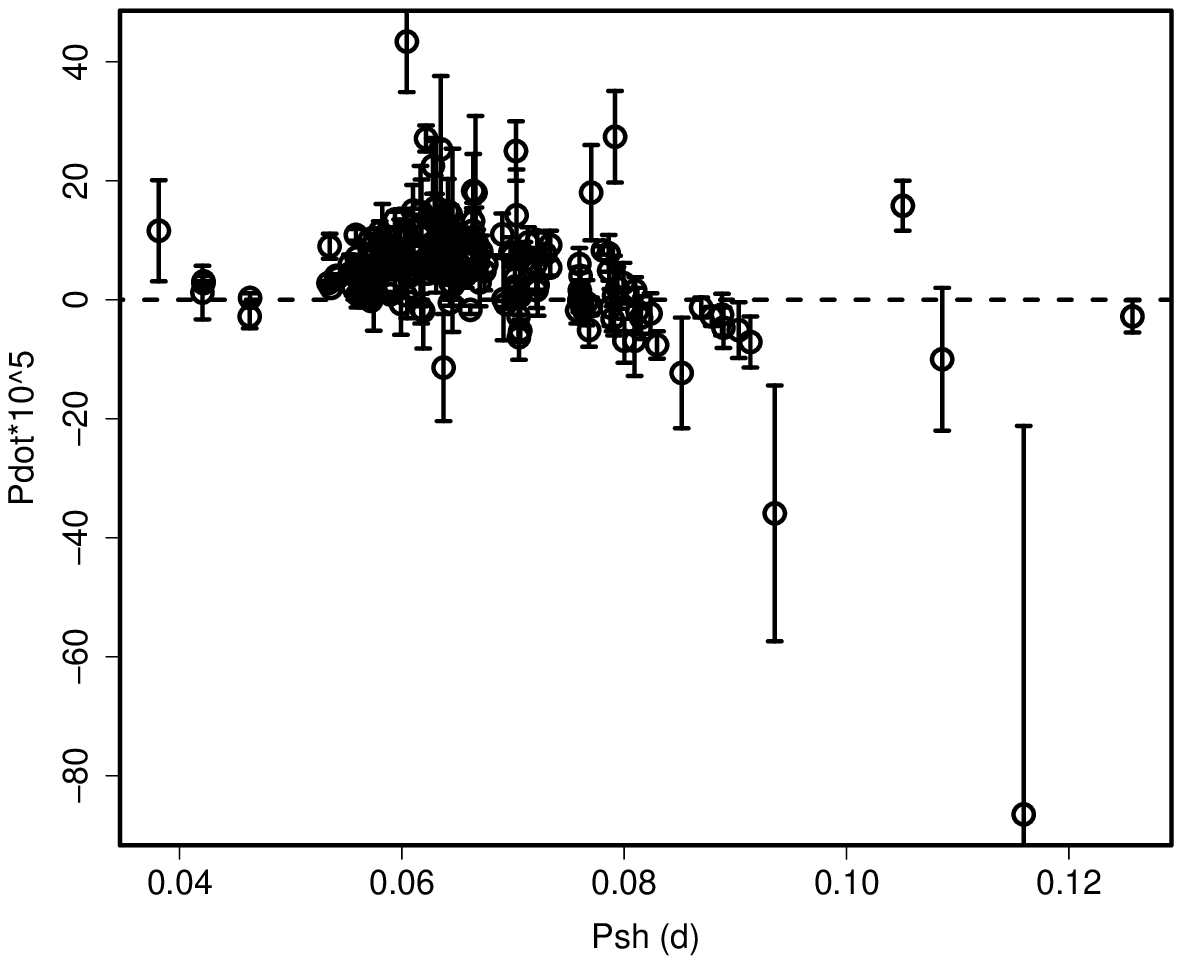}
  \end{center}
  \caption{$P_{\rm dot}$ for stage B versus the mean $P_{\rm SH}$ during
  stage B.
  }
  \label{fig:pdotpsh}
\end{figure*}

\begin{figure*}
  \begin{center}
    \FigureFile(120mm,80mm){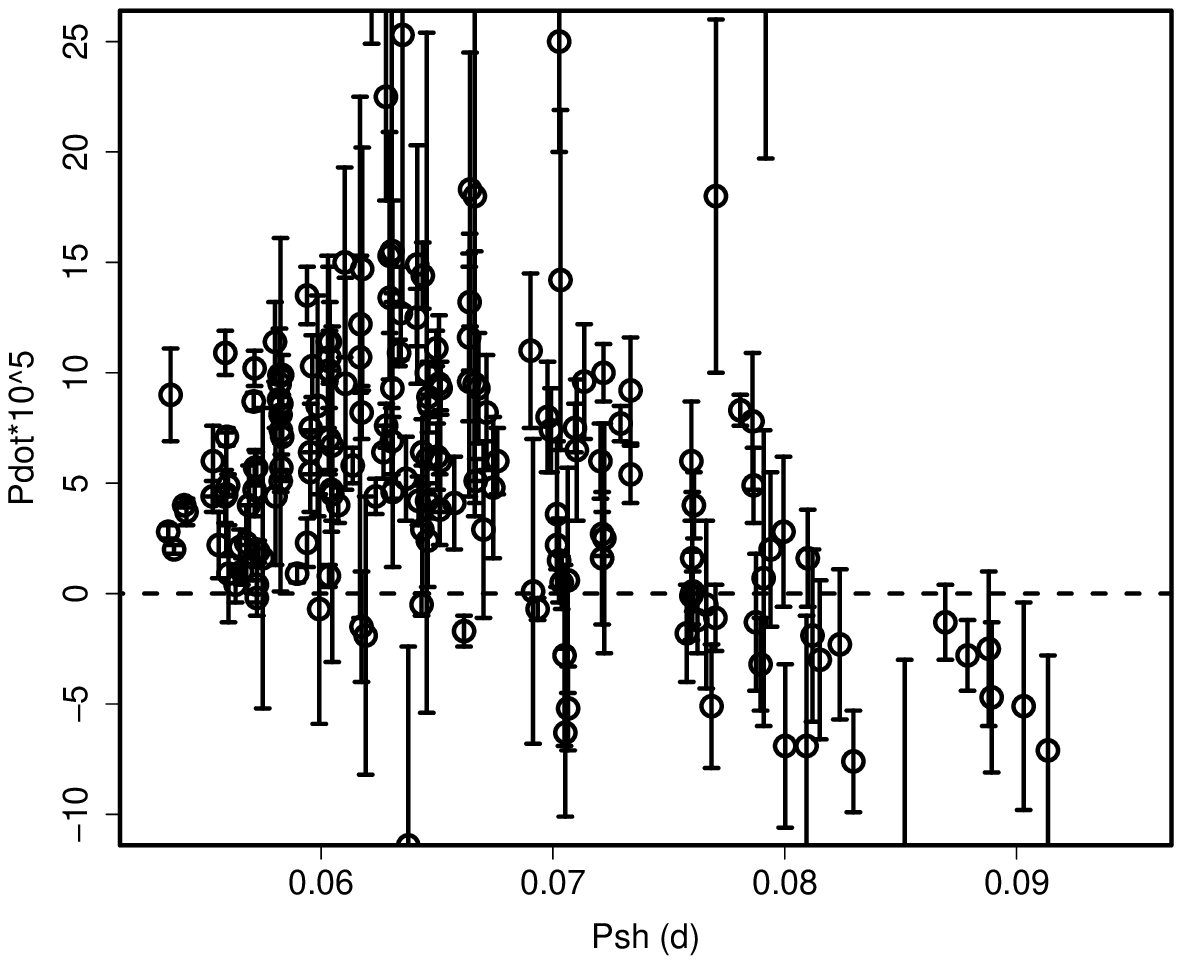}
  \end{center}
  \caption{$P_{\rm dot}$ for stage B versus $P_{\rm SH}$ (enlarged).
  }
  \label{fig:pdotpsh2}
\end{figure*}

   We determined $P_{\rm dot}$ for the stage B.  This treatment corresponds
to the analysis in \citet{kat01hvvir} for short-$P_{\rm SH}$ systems.
The values are listed in table \ref{tab:perlist} as well as other
parameters discussed in subsection \ref{sec:pshstageb}.\footnote{
   The intervals ($E_1$ and $E_2$) for the stages B and C given in the table
   sometimes overlap because of occasional observational ambiguity
   in determining the stages.  The values of $P_{\rm orb}$ are
   taken from \citet{RitterCV7}.
}
The results are shown in figures \ref{fig:pdotpsh} and \ref{fig:pdotpsh2}.
This figure is essentially an improvement of the corresponding figures
presented in \citet{kat01hvvir} and \citet{kat03v877arakktelpucma},
in that the present samples do not include globally determined $P_{\rm dot}$
or locally determined $P_{\rm dot}$ around the transitions (stage A to B
or stage B to C), and in that $P_{\rm dot}$ were (re-)determined
in a homogeneous way from the times of superhump maxima, either published
in the literature or re-examined in this paper.
Note, in particular, that two unusual systems,
V485 Cen and EI Psc, now have more usual $P_{\rm dot}$ in contrast to
\citet{kat01hvvir}.  This was caused by an error in estimating
$P_{\rm dot}$ in the original paper (V485 Cen: \cite{ole97v485cen})
and a combination of two sets of published superhump maxima (EI Psc:
\cite{uem02j2329}; \cite{ski02j2329}).  The figure indicates that
systems with $P_{\rm SH} < 0.08$ d have a general tendency
of a positive $P_{\rm dot}$ during the stage B.

\begin{figure*}
  \begin{center}
    \FigureFile(160mm,140mm){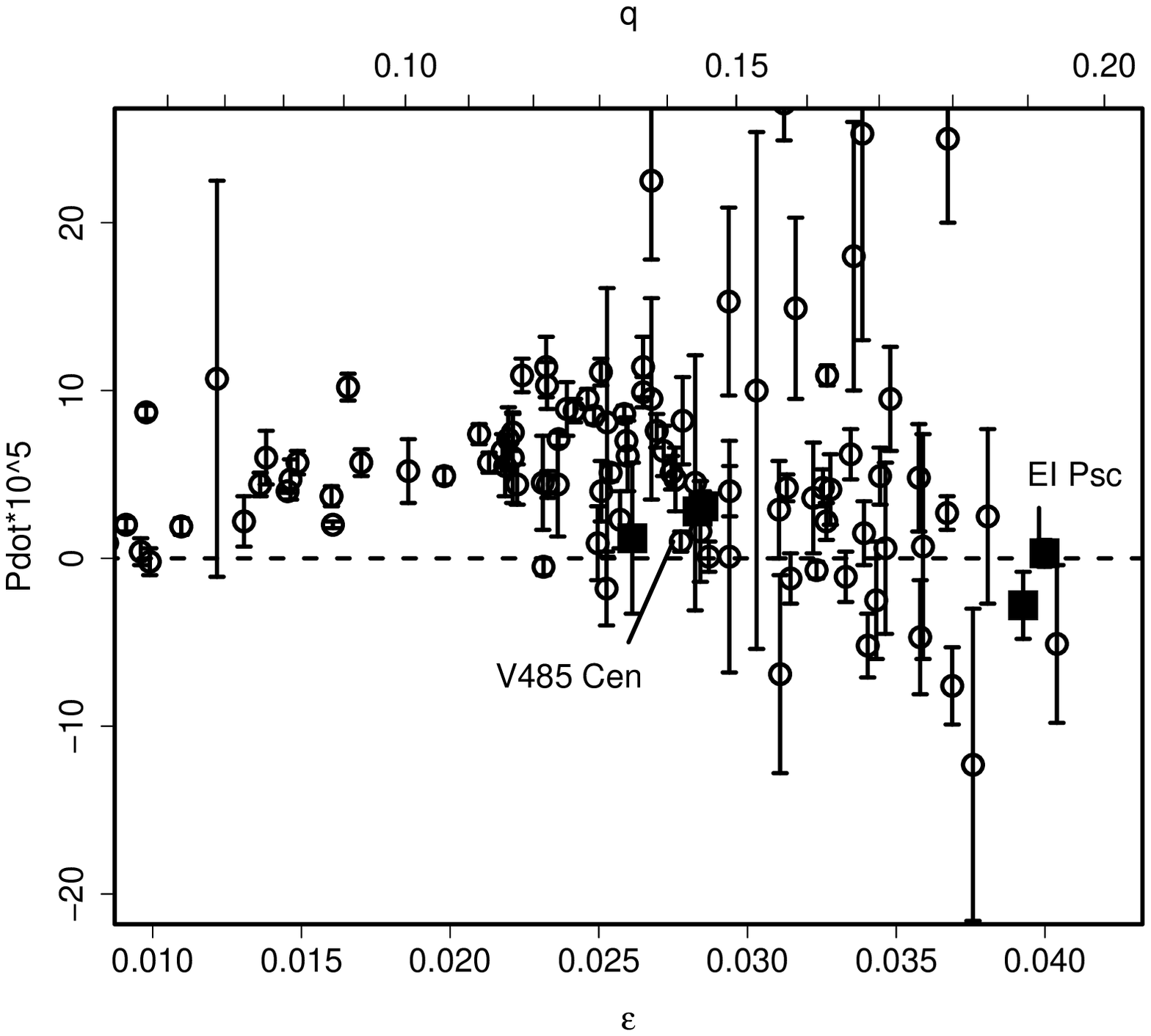}
  \end{center}
  \caption{$P_{\rm dot}$ (for stage B) versus $\epsilon$.
  $P_{\rm dot}$ for stage B has a strong correlation with fractional
  superhump excess ($\epsilon$), which is believed to be an excellent
  measure for $q$.  The $\epsilon$ was determined from the mean
  $P_{\rm SH}$ during the stage B.
  Two systems with unusually short $P_{\rm orb}$
  (filled squares: EI Psc, V485 Cen) follow the same relation as the
  rest of systems.  One exceptionally large-$\epsilon$ object
  (TU Men: $\epsilon$ = 0.073, $P_{\rm dot}$ = $-2.8(2.7) \times 10^{-5}$)
  is located outside this figure.
  }
  \label{fig:pdoteps}
\end{figure*}

   Figure \ref{fig:pdoteps} shows the relation between $P_{\rm dot}$
(for the stage B) versus $\epsilon$.  The period derivative has a strong
correlation with $\epsilon$, which is believed to be an excellent
measure for the mass ratio $q = M_2/M_1$.
It would be worth noting that two systems with
unusually short $P_{\rm orb}$ (filled squares: EI Psc, V485 Cen)
follow the same relation as the rest of systems, suggesting that
$P_{\rm dot}$ is more dependent on $q$ than on $P_{\rm orb}$.
$P_{\rm dot}$ reaches a maximum around $\epsilon$ = 0.025
(equivalent to $q$ = 0.12).

\subsection{Superhump Periods during Stages B and C}\label{sec:pshstageb}

\begin{figure}
  \begin{center}
    \FigureFile(88mm,70mm){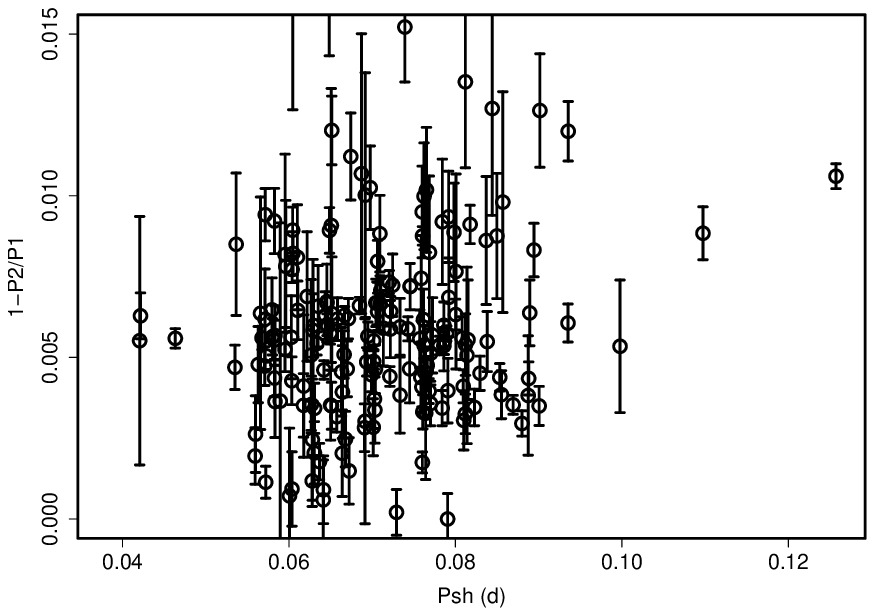}
  \end{center}
  \caption{Fractional decrease of superhump period between stages B and C
  versus $P_{\rm SH}$.
  }
  \label{fig:dpporb}
\end{figure}

   Figure \ref{fig:dpporb} summarizes fractional decrease of the superhump
period between stage B (hereafter period $P_1$) and
stage C (hereafter period $P_2$) versus $P_{\rm SH}$.
The superhump period usually decrease by $\sim$0.5 \% during
the transition from stage B to C.
There appears to be a weak relation between the fractional decrease
and $P_{\rm SH}$: the decrease is larger in longer-$P_{\rm SH}$ systems.

\begin{figure}
  \begin{center}
    \FigureFile(88mm,70mm){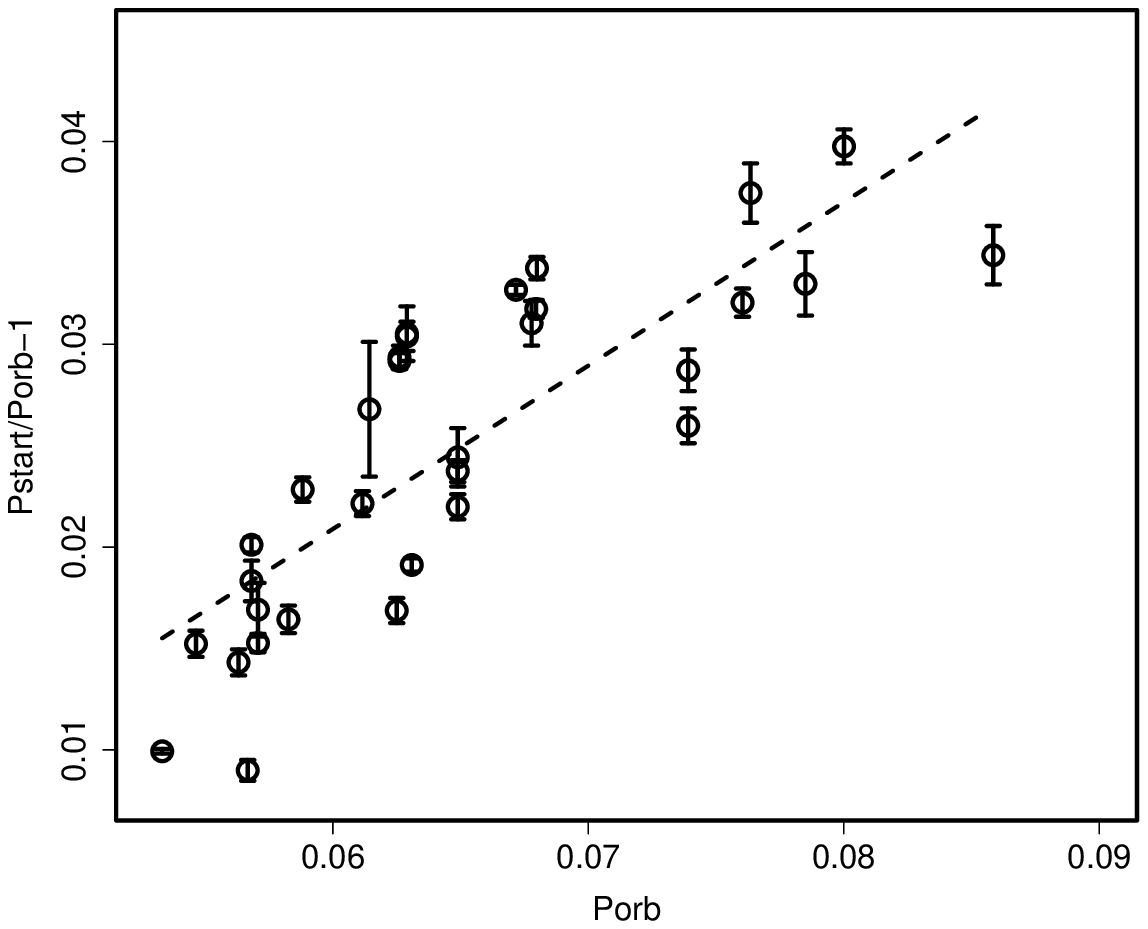}
  \end{center}
  \caption{Fractional superhump excess at the beginning of stage B versus
  mean $P_{\rm orb}$.  The dashed line represents equation
  \ref{equ:pstartporb}.
  }
  \label{fig:pstartporb}
\end{figure}

   Figure \ref{fig:pstartporb} shows the relation between the fractional
superhump excess at the beginning of the stage B (calculated using the
mean $P_{\rm SH}$ and $P_{\rm dot}$) versus $P_{\rm orb}$.
The figure was drawn for
systems with a well-defined stage B (corresponding to subsection
\ref{sec:tendency}) and with a known $P_{\rm orb}$.  The relation
is tighter than the well-known relation between the global $P_{\rm SH}$
and $P_{\rm orb}$ (e.g. \cite{mol92SHexcess}).
A linear regression to the data has yielded the following relation:

\begin{equation}
P_{\rm SH (start)}/P_{\rm orb}-1 = -0.033(6) + 0.87(9) P_{\rm orb}
\label{equ:pstartporb}.
\end{equation}

\begin{figure}
  \begin{center}
    \FigureFile(88mm,70mm){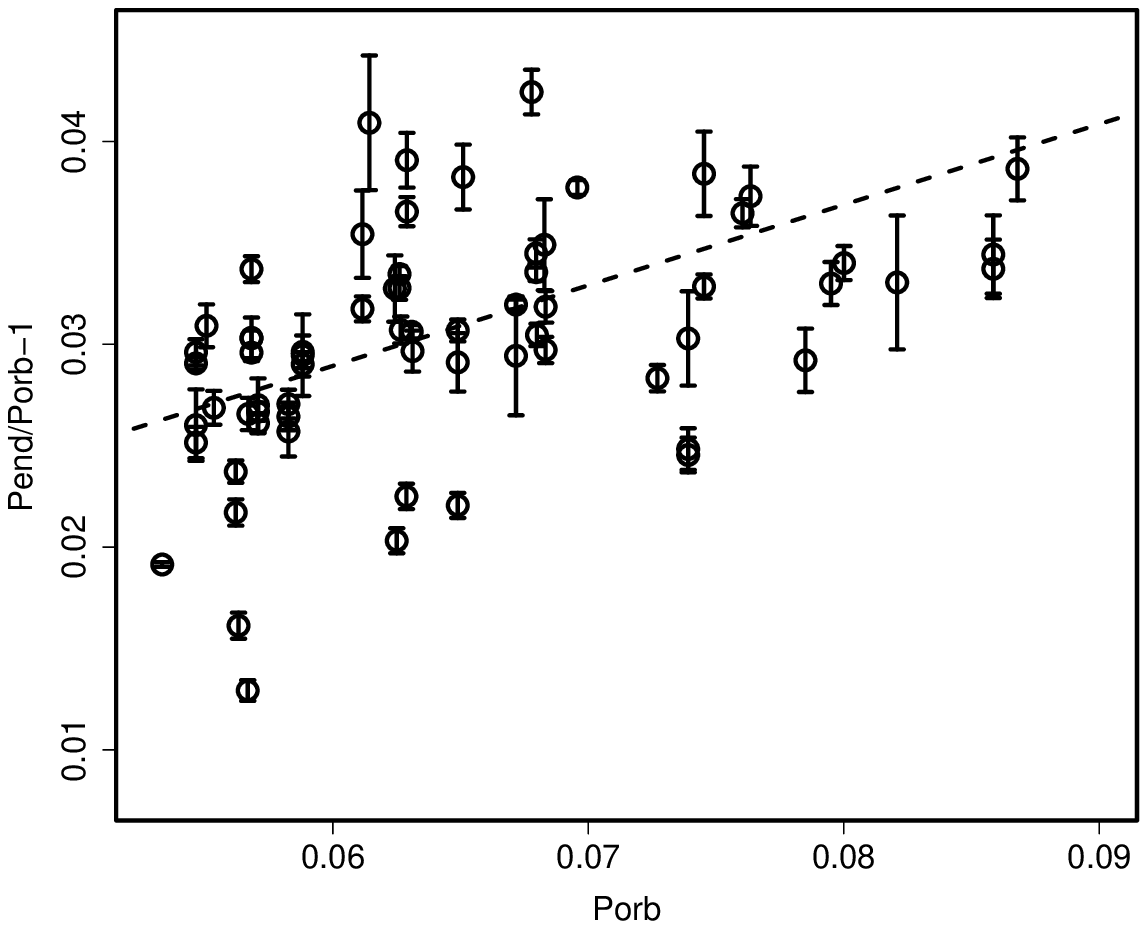}
  \end{center}
  \caption{Fractional superhump excess at the end of the stage B versus
  the mean $P_{\rm orb}$.  The dashed line represents equation
  \ref{equ:pendporb}.  The figure is restricted to the displayed
  range for a comparison with figure \ref{fig:pstartporb}.
  }
  \label{fig:pendporb}
\end{figure}

   Figure \ref{fig:pendporb} shows the relation between the fractional
superhump excess at the end of the stage B, i.e. the longest superhump
period for positive-$P_{\rm dot}$ systems, versus mean $P_{\rm orb}$.
This fractional period excess has, in contrast to one at the
beginning of the stage B, a fairly common value of $\sim$0.03 (slightly
increasing with increasing $P_{\rm orb}$, equation \ref{equ:pendporb})
below the period gap.
The difference in dependence to $P_{\rm orb}$ between
these two periods is striking, and is most prominent at shorter
$P_{\rm orb}$ except extreme WZ Sge-type dwarf novae
(for WZ Sge-type dwarf novae, see description and discussion
in section \ref{sec:wzsgestars}).
This difference appears to determine the $P_{\rm dot}$ -- $P_{\rm SH}$
relation (subsection \ref{sec:pdotb}).

\begin{equation}
P_{\rm SH (end)}/P_{\rm orb}-1 = 0.001(4) + 0.44(6) P_{\rm orb}
\label{equ:pendporb}.
\end{equation}

\begin{figure}
  \begin{center}
    \FigureFile(88mm,70mm){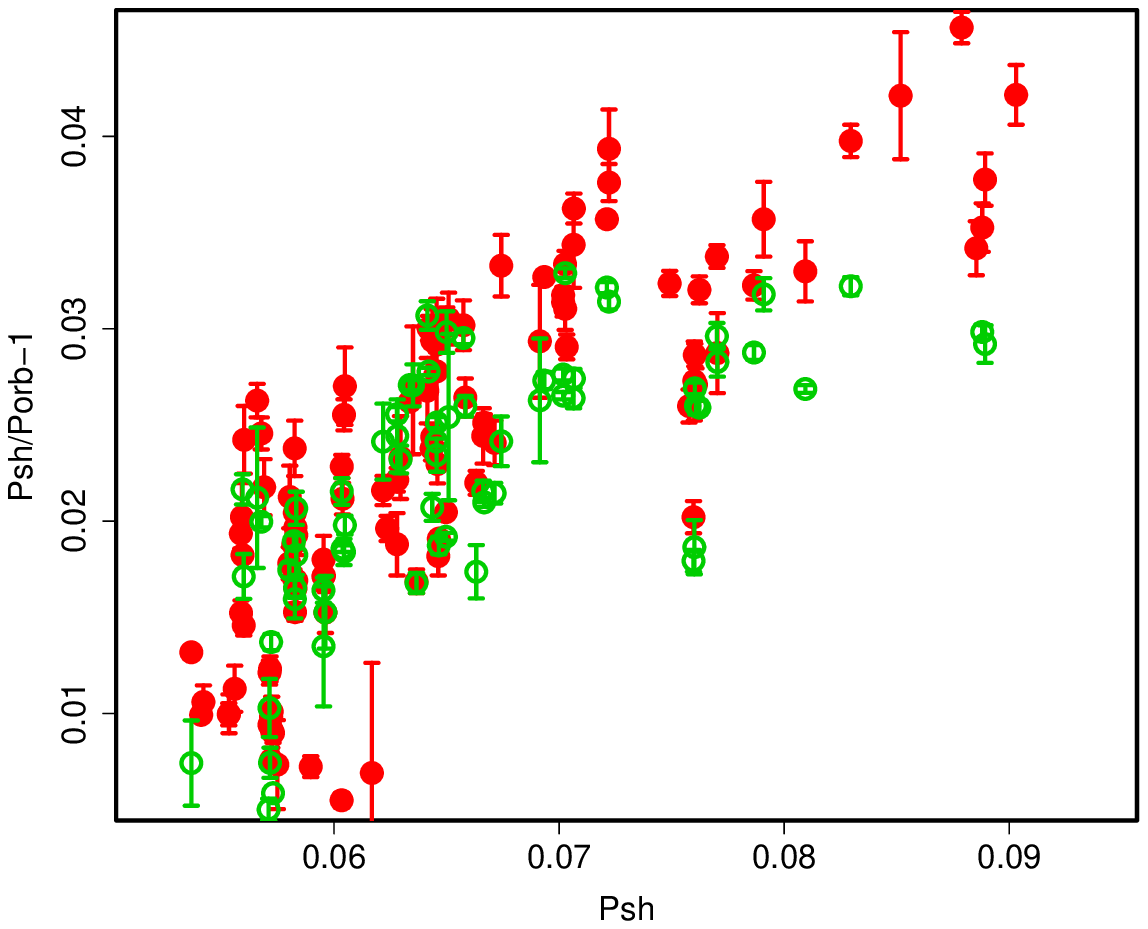}
  \end{center}
  \caption{Comparison of fractional superhump excesses between the stage C
  and the start of the stage B.
  The open and filled circles represent fractional
  superhump excesses in the stage C and at the start of the stage B,
  respectively.
  The superhump excesses during the stage C are almost identical to those
  at the start of the stage B.  The figure is restricted to the displayed
  range for better visibility.
  }
  \label{fig:pstartp2}
\end{figure}

   The superhump excesses (or periods) during the stage C are almost
identical to those at the start of the stage B (figure \ref{fig:pstartp2}).

   For readers' convenience, we also provide relations between
$P_1$ and $P_{\rm orb}$ (equation \ref{equ:p1porb}, the samples are the same
as in figure \ref{fig:pstartporb}) and $P_2$ and $P_{\rm orb}$
(equation \ref{equ:p2porb}).

\begin{equation}
P_1/P_{\rm orb}-1 = -0.017(7) + 0.66(10) P_{\rm orb}
\label{equ:p1porb}.
\end{equation}

\begin{equation}
P_2/P_{\rm orb}-1 = -0.012(4) + 0.56(5) P_{\rm orb}
\label{equ:p2porb}.
\end{equation}

   These equations can be used for estimating $P_{\rm orb}$ (as in
\cite{RitterCV7}) when superhump periods for specific stages are known.
The potential availability of $P_2$ for estimating $P_{\rm orb}$
would provide an excellent alternative to $P_{\rm SH}$ at the start of
the stage B, since the break between the stages B and C is easier to
detect than the start of the stage B, particularly when the superoutburst
is detected during its later course.

   The overall behavior of the stages B and C in positive-$P_{\rm dot}$
systems can be summarized:

\begin{itemize}
\item The superhumps during the stage B start with a short period,
which is well correlated with $P_{\rm orb}$.
\item The superhumps evolve during the stage B toward a longer period,
which commonly has a $\sim$ 3 \% excess to $P_{\rm orb}$.
\item The superhump period return to the initial period during the stage C.
\end{itemize}

\subsection{Superhump Periods during Stage A}

\begin{figure}
  \begin{center}
    \FigureFile(88mm,70mm){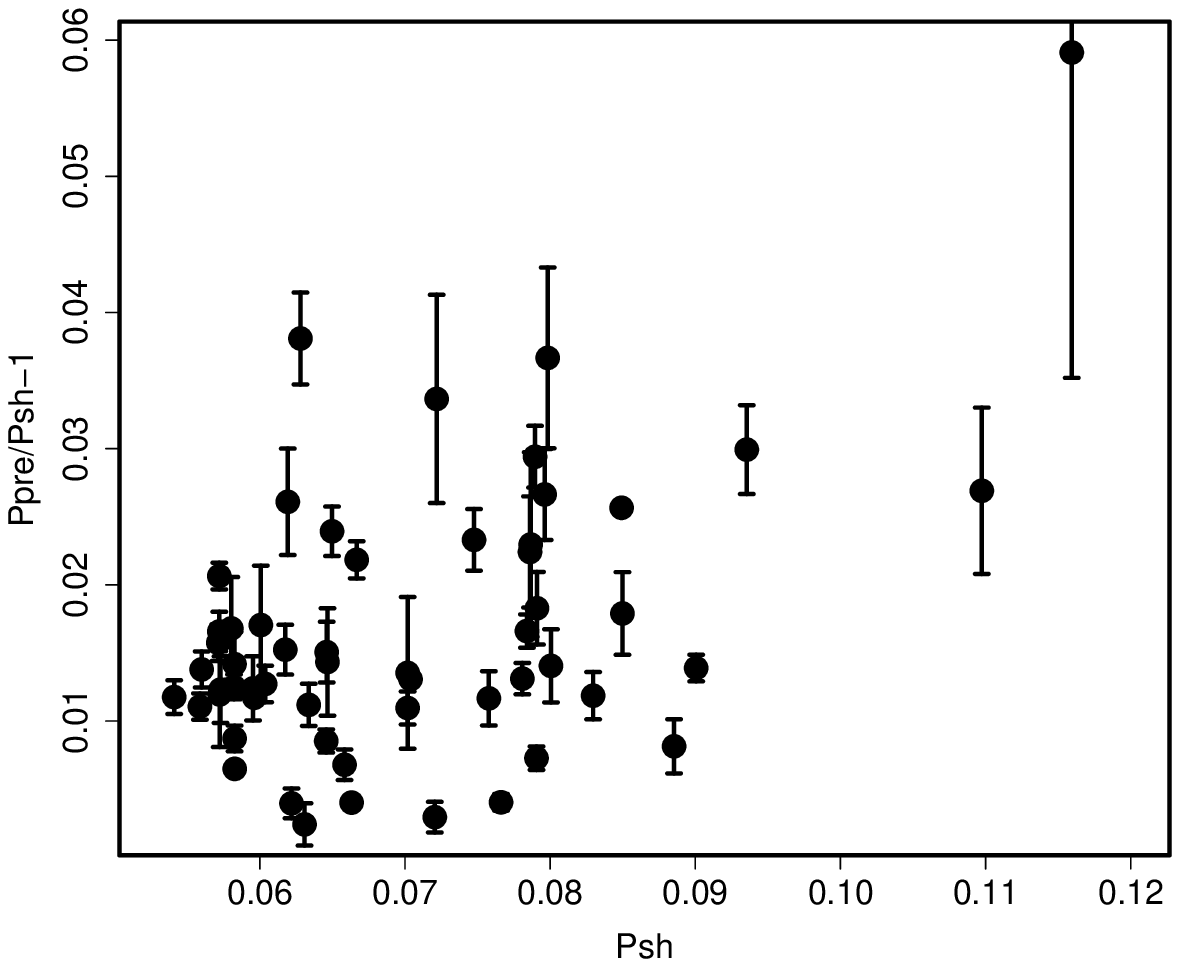}
  \end{center}
  \caption{Superhump periods during the stage A.  Superhumps in this stage
  has a period typically 1.0--1.5 \% longer than the one during the stage B.
  Some systems have still longer periods ($\sim$3 \% longer than the one
  during the stage B).
  }
  \label{fig:ppre}
\end{figure}

   The stage A usually constitutes $\sim$20 superhump cycles.
Table \ref{tab:pera} and figure \ref{fig:ppre} summarize the recorded
superhump periods during the stage A.
Note, however, the periods during this stage were not very
precisely determined because of the shortness of the interval, and
because the amplitudes of superhumps are still small.
Fractional period excesses during this stage to the mean superhump
period during the stage B tend to cluster around 1.0--1.5 \%, with some
exceptional systems having larger ($\sim$3 \%) excesses.

\subsection{Difference Between Different Superoutbursts}\label{sec:different}

  \citet{uem05tvcrv} reported significantly different $P_{\rm dot}$'s
between different superoutbursts of the same object, TV Crv.
Several authors, however, have reported results contrary to this
finding (e.g. \cite{oiz07v844her}; \cite{soe09swuma}; \cite{ohs09qzvir}).

\begin{figure}
  \begin{center}
    \FigureFile(88mm,70mm){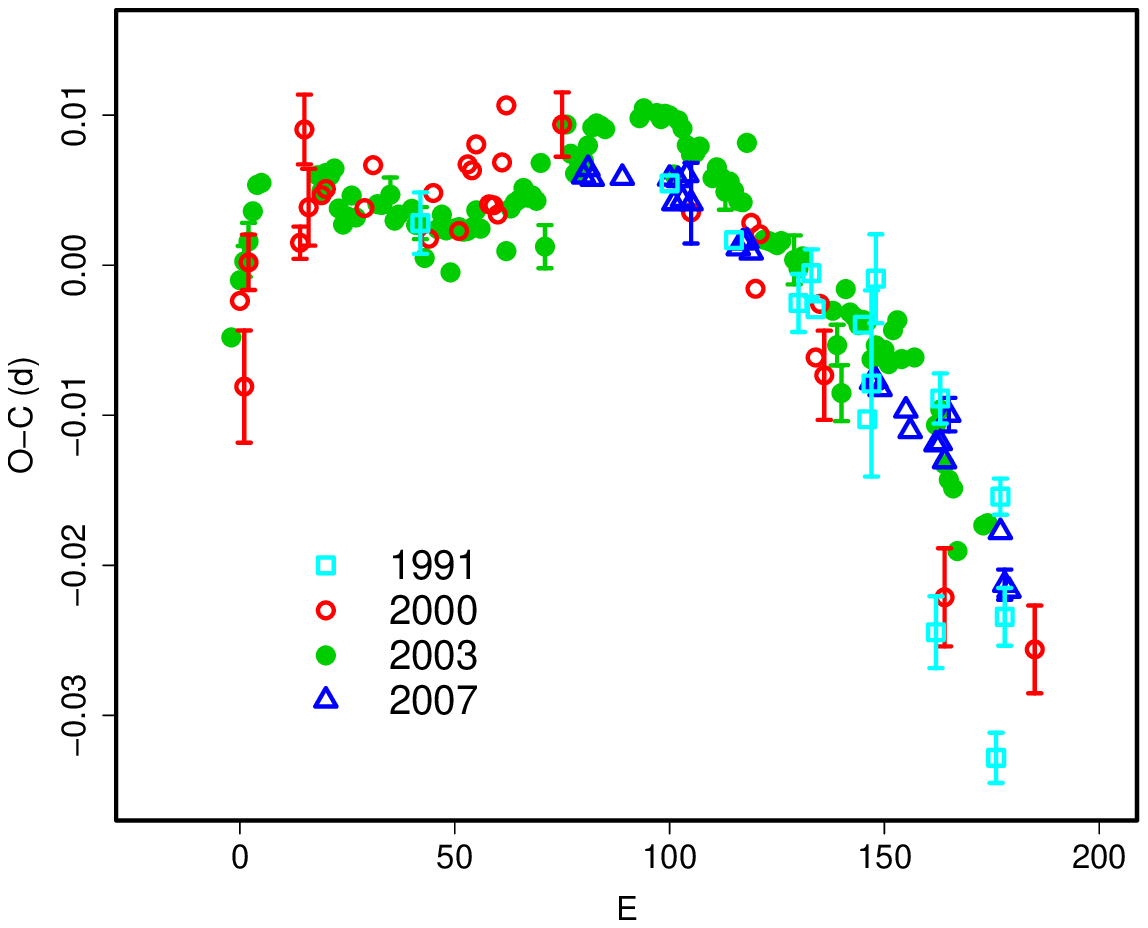}
  \end{center}
  \caption{Comparison of $O-C$ diagrams of UV Per between different
  superoutbursts.
  }
  \label{fig:uvpercomp}
\end{figure}

   We further examined different superoutbursts of the same objects,
and found no convincing evidence for strong variation of
$P_{\rm dot}$ between different superoutbursts.  On the contrary,
the behavior of superhump period in the same object appears to
be similar between different superoutbursts (e.g. figure
\ref{fig:uvpercomp}; figures in section \ref{sec:individual}).
The difference reported in the past was apparently a result of
observation of different stages of superoutbursts (A--C) and the
insufficient coverage of the entire superoutburst.

\begin{figure}
  \begin{center}
    \FigureFile(88mm,70mm){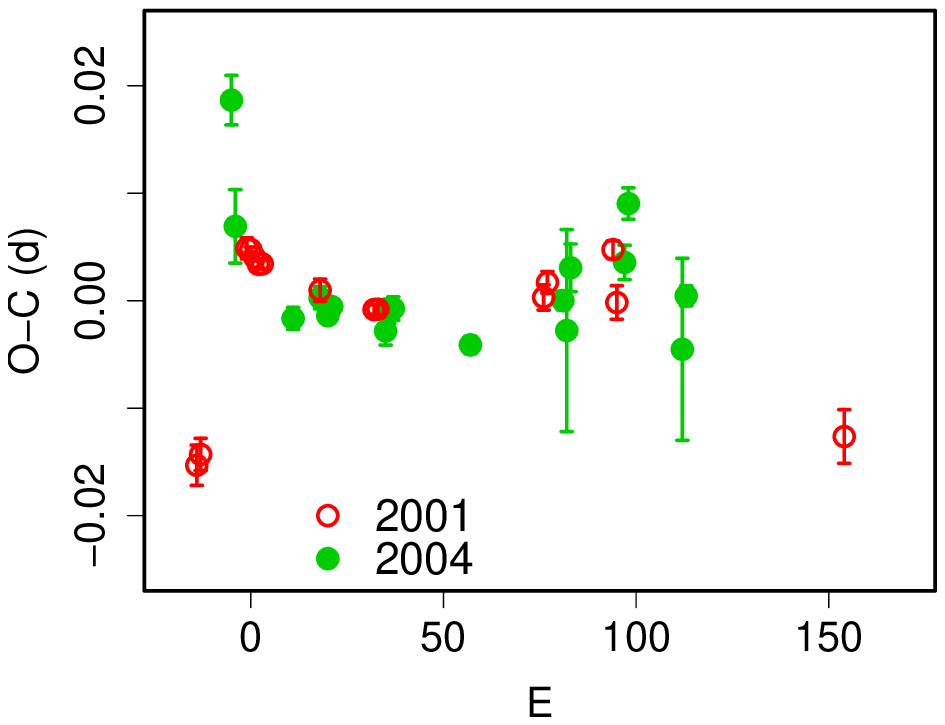}
  \end{center}
  \caption{Comparison of $O-C$ diagrams of TV Crv between 2001 and
  2004 superoutbursts.  $E=0$ corresponds to the start of the stage B.
  }
  \label{fig:tvcrvcomp}
\end{figure}

   A re-examination of the TV Crv case has also shown that the claim
by \citet{uem05tvcrv} was not convincing (figure \ref{fig:tvcrvcomp}).

   During the 2004 superoutburst of V2527 Oph, no anomalous behavior
in $P_{\rm dot}$ was observed even in the presence of a clear
precursor outburst.
Similar situations were observed in GO Com (2003), PU CMa (2008),
AQ Eri (2009), QZ Vir (1993, 2009) and 1RXS J0532 (2005).\footnote{
  The period evolution during the 2008 superoutburst of 1RXS J0423,
  which was associated with a precursor, was slow
  (subsection \ref{sec:j0423}).  It is not clear whether
  the existence of a precursor is responsible in this instance.
}
The proposed relation between the presence of a precursor outburst
and $P_{\rm SH}$ \citep{uem05tvcrv} is not supported by
these instances.

   Although further work is needed to exclude the presence of
different period behavior between different superoutbursts,
the close agreement of the behavior between different superoutbursts
in many objects might be used to construct a combined $O-C$ diagram
and to determine $P_{\rm dot}$ from different superoutbursts even if
observational coverage of each outburst is incomplete.

\section{Discussion}\label{sec:discussion}

\subsection{Existence of Stage B--C Transition}

   In section \ref{sec:general}, we described that most of
well-observed systems show stage B--C transitions.
There are, however, some objects (or superoutbursts) without
prominent stage B--C transitions even though the late stage of
superoutbursts is well observed.  WZ Sge-type dwarf novae with
small $P_{\rm dot}$, in particular, have tendency to lack the
stage B--C transition (see also section \ref{sec:individual}).

\begin{figure*}
  \begin{center}
    \FigureFile(180mm,70mm){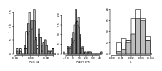}
  \end{center}
  \caption{Existence of stage B--C transition versus $P_{\rm SH}$,
  $P_{\rm dot}$ and $\epsilon$.  The gray color indicates superoutbursts
  with a stage B--C transition.  The existence of stage B--C
  transitions is most strongly correlated with $\epsilon$.
  }
  \label{fig:stagec}
\end{figure*}

   We examined superoutbursts regarding the existence of stage B--C
transitions.  The sample was selected by criteria of ``well-observed''
quality (quality A or B) and observational coverage for at least
50 superhump cycles.
As shown in figure \ref{fig:stagec}, the existence of stage B--C
transitions is most strongly correlated with $\epsilon$.
In systems with a small $\epsilon$ (typically $\epsilon < 0.015$),
only a few superoutbursts showed stage B--C transitions.
The result suggests that the appearance of this transition is
strongly dependent on $q$.

   In systems with $\epsilon > 0.02$, there are some superoutbursts
without a clear transition to the stage C.  The best observed example
might be V844 Her in 2006 \citep{oiz07v844her}.  During this superoutburst,
there was no indication of a transition even after 146 superhump cycles,
when the outburst just entered the rapid decline stage.
In this case, however, the transition may have occurred after the termination
of the observation since a transition was recorded during the rapid fading
and subsequent stage during the 2008 superoutburst of the same object.
The present statistical analysis may be similarly biased toward
the lower detection rate of the transition for systems with long-lasting
stage B, i.e. systems with a shorter $P_{\rm SH}$ or a smaller $\epsilon$
(subsection \ref{sec:tendency}).
Even considering this effect, the apparent rarity of the transition
in small-$P_{\rm dot}$ WZ Sge-type dwarf novae is likely significant,
since these objects were often observed even after the termination of their
superoutbursts.

\begin{table*}
\caption{Superhump Periods and Period Derivatives}\label{tab:perlist}
\begin{center}
\begin{tabular}{cccccccccccccc}
\hline\hline
Object & Year & $P_1$ (d) & err & \multicolumn{2}{c}{$E_1$\commenta} & $P_{\rm dot}$\commentb & err\commentb & $P_2$ (d) & err & \multicolumn{2}{c}{$E_2$\commenta} & $P_{\rm orb}$ (d) & Q\commentc \\
\hline
FO And & 1994 & 0.074554 & 0.000052 & 0 & 14 & -- & -- & 0.074018 & 0.000012 & 13 & 27 & 0.07161 & C \\
KV And & 1994 & 0.074601 & 0.000122 & 0 & 55 & -- & -- & 0.074063 & -- & 55 & 95 & -- & C \\
KV And & 2002 & 0.074501 & 0.000045 & 0 & 41 & -- & -- & 0.074155 & 0.000064 & 41 & 82 & -- & C \\
LL And & 1993 & 0.056900 & 0.000088 & 0 & 56 & -- & -- & -- & -- & -- & -- & 0.055055 & C \\
LL And & 2004 & 0.056583 & 0.000022 & 0 & 290 & 1.0 & 0.6 & 0.056223 & 0.000201 & 290 & 326 & 0.055055 & C \\
V402 And & 2005 & 0.063230 & 0.000058 & 0 & 41 & -- & -- & -- & -- & -- & -- & -- & C \\
V402 And & 2006 & 0.063439 & 0.000062 & 0 & 79 & 12.7 & 2.1 & -- & -- & -- & -- & -- & C \\
\end{tabular}
\end{center}
\end{table*}

\addtocounter{table}{-1}
\begin{table*}
\caption{Superhump Periods and Period Derivatives (continued)}
\begin{center}
\begin{tabular}{cccccccccccccc}
\hline\hline
Object & Year & $P_1$ & err & \multicolumn{2}{c}{$E_1$} & $P_{\rm dot}$ & err & $P_2$ & err & \multicolumn{2}{c}{$E_2$} & $P_{\rm orb}$ & Q \\
\hline
WX Cet & 1998 & 0.059529 & 0.000014 & 15 & 157 & 6.4 & 1.0 & 0.059217 & 0.000038 & 149 & 220 & 0.058261 & A \\
WX Cet & 2001 & 0.059549 & 0.000028 & 0 & 129 & 7.5 & 1.1 & -- & -- & -- & -- & 0.058261 & B \\
WX Cet & 2004 & 0.059534 & 0.000023 & 0 & 137 & 5.5 & 1.8 & 0.059047 & 0.000182 & 136 & 169 & 0.058261 & C \\
Z Cha & 1982 & 0.077252 & 0.000064 & 0 & 38 & -- & -- & 0.076813 & 0.000063 & 38 & 65 & 0.074499 & B \\
RX Cha & 2009 & 0.084921 & 0.000021 & 0 & 34 & -- & -- & -- & -- & -- & -- & -- & C \\
BZ Cir & 2004 & 0.076614 & 0.000019 & 13 & 68 & $-$0.5 & 3.8 & 0.076250 & 0.000010 & 66 & 146 & -- & B \\
PU CMa & 2003 & 0.057962 & 0.000054 & 0 & 51 & -- & -- & 0.057587 & 0.000019 & 51 & 144 & 0.056694 & B \\
\hline
\end{tabular}
\end{center}
\end{table*}

\addtocounter{table}{-1}
\begin{table*}
\caption{Superhump Periods and Period Derivatives (continued)}
\begin{center}
\begin{tabular}{cccccccccccccc}
\hline\hline
Object & Year & $P_1$ & err & \multicolumn{2}{c}{$E_1$} & $P_{\rm dot}$ & err & $P_2$ & err & \multicolumn{2}{c}{$E_2$} & $P_{\rm orb}$ & Q \\
\hline
HO Del & 1994 & 0.064559 & 0.000056 & 0 & 49 & 10.0 & 15.4 & 0.064128 & 0.000054 & 47 & 94 & 0.06266 & C \\
HO Del & 2001 & 0.064280 & 0.000120 & 0 & 2 & -- & -- & -- & -- & -- & -- & 0.06266 & C \\
HO Del & 2008 & 0.064363 & 0.000017 & 11 & 96 & 6.4 & 1.5 & 0.063958 & 0.000044 & 95 & 165 & 0.06266 & B \\
BC Dor & 2003 & 0.068473 & 0.000016 & 0 & 61 & -- & -- & 0.068021 & 0.000007 & 59 & 146 & -- & C \\
CP Dra & 2003 & 0.083698 & 0.000028 & 0 & 15 & -- & -- & 0.082977 & 0.000162 & 36 & 49 & -- & C \\
CP Dra & 2009 & 0.083822 & 0.000073 & 0 & 26 & -- & -- & 0.083362 & 0.000027 & 24 & 97 & -- & B \\
DM Dra & 2003 & 0.075707 & 0.000051 & 0 & 40 & -- & -- & 0.075285 & 0.000044 & 38 & 81 & -- & C \\
\hline
\end{tabular}
\end{center}
\end{table*}

\addtocounter{table}{-1}
\begin{table*}
\caption{Superhump Periods and Period Derivatives (continued)}
\begin{center}
\begin{tabular}{cccccccccccccc}
\hline\hline
Object & Year & $P_1$ & err & \multicolumn{2}{c}{$E_1$} & $P_{\rm dot}$ & err & $P_2$ & err & \multicolumn{2}{c}{$E_2$} & $P_{\rm orb}$ & Q \\
\hline
SX LMi & 1994 & 0.069481 & 0.000017 & 0 & 45 & -- & -- & 0.069088 & 0.000025 & 45 & 118 & 0.06717 & B \\
SX LMi & 2001 & 0.069144 & 0.000033 & 0 & 85 & 0.1 & 6.9 & 0.068935 & 0.000216 & 84 & 113 & 0.06717 & C \\
SX LMi & 2002 & 0.069341 & 0.000004 & 14 & 115 & $-$0.7 & 0.5 & 0.069004 & 0.000030 & 115 & 130 & 0.06717 & C \\
BR Lup & 2003 & 0.082284 & 0.000043 & 0 & 17 & -- & -- & 0.082000 & 0.000018 & 50 & 95 & 0.0795 & C \\
BR Lup & 2004 & -- & -- & -- & -- & -- & -- & 0.082193 & 0.000038 & 0 & 96 & 0.0795 & C \\
AY Lyr & 1987 & 0.075970 & 0.000018 & 0 & 92 & $-$0.1 & 2.0 & -- & -- & -- & -- & -- & B \\
AY Lyr & 2008 & 0.076232 & 0.000099 & 0 & 28 & -- & -- & 0.075471 & 0.000077 & 27 & 54 & -- & B \\
\hline
\end{tabular}
\end{center}
\end{table*}

\addtocounter{table}{-1}
\begin{table*}
\caption{Superhump Periods and Period Derivatives (continued)}
\begin{center}
\begin{tabular}{cccccccccccccc}
\hline\hline
Object & Year & $P_1$ & err & \multicolumn{2}{c}{$E_1$} & $P_{\rm dot}$ & err & $P_2$ & err & \multicolumn{2}{c}{$E_2$} & $P_{\rm orb}$ & Q \\
\hline
VZ Pyx & 1996 & 0.075754 & 0.000012 & 0 & 27 & -- & -- & -- & -- & -- & -- & 0.07332 & C \\
VZ Pyx & 2000 & -- & -- & -- & -- & -- & -- & 0.075492 & 0.000016 & 0 & 94 & 0.07332 & C \\
VZ Pyx & 2004 & 0.075875 & 0.000060 & 0 & 52 & -- & -- & -- & -- & -- & -- & 0.07332 & C \\
VZ Pyx & 2008 & 0.076045 & 0.000021 & 0 & 27 & -- & -- & 0.075379 & 0.000006 & 54 & 80 & 0.07332 & C \\
DV Sco & 2004 & 0.099776 & 0.000202 & 0 & 17 & -- & -- & 0.099243 & 0.000032 & 17 & 53 & -- & C \\
MM Sco & 2002 & 0.061324 & 0.000058 & 0 & 25 & -- & -- & -- & -- & -- & -- & -- & C \\
NY Ser & 1996 & 0.108610 & -- & -- & -- & -- & -- & 0.105677 & 0.000304 & 18 & 37 & -- & C \\
\hline
\end{tabular}
\end{center}
\end{table*}

\addtocounter{table}{-1}
\begin{table*}
\caption{Superhump Periods and Period Derivatives (continued)}
\begin{center}
\begin{tabular}{cccccccccccccc}
\hline\hline
Object & Year & $P_1$ & err & \multicolumn{2}{c}{$E_1$} & $P_{\rm dot}$ & err & $P_2$ & err & \multicolumn{2}{c}{$E_2$} & $P_{\rm orb}$ & Q \\
\hline
IY UMa & 2000 & 0.075776 & 0.000015 & 23 & 101 & $-$1.8 & 2.2 & -- & -- & -- & -- & 0.073909 & A \\
IY UMa & 2002 & 0.076009 & 0.000024 & 0 & 137 & 1.6 & 3.0 & 0.075287 & 0.000104 & 135 & 228 & 0.073909 & C \\
IY UMa & 2004 & 0.076030 & 0.000011 & 0 & 130 & 0.1 & 0.9 & 0.075897 & 0.000022 & 126 & 169 & 0.073909 & C \\
IY UMa & 2006 & 0.076082 & 0.000021 & 42 & 154 & 4.0 & 1.5 & 0.075830 & 0.000034 & 153 & 221 & 0.073909 & C \\
IY UMa & 2007 & 0.075849 & 0.000071 & 0 & 15 & -- & -- & 0.075517 & 0.000043 & 13 & 54 & 0.073909 & C \\
IY UMa & 2009 & 0.076233 & 0.000016 & 30 & 123 & $-$1.2 & 1.5 & 0.075823 & 0.000027 & 122 & 189 & 0.073909 & B \\
KS UMa & 2003 & 0.070179 & 0.000009 & 15 & 95 & 2.2 & 1.1 & 0.069837 & 0.000021 & 95 & 147 & 0.06796 & A \\
\hline
\end{tabular}
\end{center}
\end{table*}

\addtocounter{table}{-1}
\begin{table*}
\caption{Superhump Periods and Period Derivatives (continued)}
\begin{center}
\begin{tabular}{cccccccccccccc}
\hline\hline
Object & Year & $P_1$ & err & \multicolumn{2}{c}{$E_1$} & $P_{\rm dot}$ & err & $P_2$ & err & \multicolumn{2}{c}{$E_2$} & $P_{\rm orb}$ & Q \\
\hline
SDSS J1005 & 2009 & -- & -- & -- & -- & -- & -- & 0.077469 & 0.000021 & 0 & 49 & -- & C \\
SDSS J1100 & 2009 & 0.067569 & 0.000025 & 0 & 64 & -- & -- & -- & -- & -- & -- & -- & C \\
SDSS J1227 & 2007 & 0.064593 & 0.000022 & 33 & 129 & 6.1 & 2.1 & -- & -- & -- & -- & 0.062958 & B \\
SDSS J1524 & 2009 & 0.067136 & 0.000023 & 0 & 89 & 8.2 & 2.6 & 0.066720 & 0.000035 & 88 & 163 & 0.065319 & B \\
SDSS J1556 & 2007 & 0.082961 & 0.000022 & 12 & 85 & $-$7.6 & 2.3 & 0.082587 & 0.000038 & 85 & 145 & 0.08001 & B \\
SDSS J1627 & 2008 & 0.109741 & 0.000087 & 15 & 50 & -- & -- & 0.108771 & 0.000022 & 49 & 150 & -- & A \\
SDSS J1702 & 2005 & 0.105065 & 0.000083 & 0 & 85 & 15.8 & 4.2 & -- & -- & -- & -- & 0.100082 & B \\
\hline
\end{tabular}
\end{center}
\end{table*}

\subsection{Relation between stage C Superhumps and Late Superhumps}

   During the final stage of a superoutburst and the subsequent
post-superoutburst stages, some SU UMa-type dwarf nova have been
reported to exhibit modulations having approximately the same period as
$P_{\rm SH}$, but having a maximum phase $\sim$0.5 offset from those
of usual superhumps.  These modulations have been traditionally called
``late superhumps'' (\cite{hae79lateSH}; \cite{vog83lateSH};
\cite{vanderwoe88lateSH}; \cite{hes92lateSH}).
We, however, could not find very convincing evidence for this
phenomenon in many well-sampled objects
(see e.g. QZ Vir: \cite{ohs09qzvir}).
Instead, there seems to be almost ubiquitous presence of a transition
from the stage B to C associated with a period shortening
(section \ref{sec:pshstageb})
and the continuity of superhump phases in well-observed systems
(see also \cite{ohs09qzvir}).

   This might suggest that at least some of claimed ``late superhumps''
in the literature actually referred to superhumps during the stage C.
The $P_{\rm SH} (= P_2)$ being typically $\sim$ 0.5--1.0 \% shorter than in
earlier stages ($P_1$), a observational gap in $\sim$ 30--50 cycles
($\sim$2--3 d) can result a phase shift of 0.15--0.5, and it may have
been attributed to a $\sim$0.5 phase offset.
Although it would be already difficult to
re-examine historical observations reporting late superhumps, we should
pay attention to this possibility and avoid attributing the term
``late superhumps'' simply because a phase offset is detected.
If this interpretation is indeed the case, the term ``late superhumps''
should better be attributed to superhumps during the stage C ($P_2$).\footnote{
   Note, however, we used ``late superhumps'' for late-stage superhumps
   different from ordinary ones in WZ Sge-type dwarf novae
   \citep{kat08wzsgelateSH}.
}

   There is some evidence of traditional late superhumps in
DT Oct (subsection \ref{sec:dtoct}) and HS Vir (subsection \ref{sec:hsvir}).
It may be that this type of traditional late superhumps is only
observed in systems with a high mass-transfer rate, enabling sufficient
luminosity from the hot spot.

\subsection{Implications of Period Transition in Interpreting Observations}\label{sec:transimplication}

   One of the important consequences of the period transition between stages B
and C in interpreting observations is that this appearance of a new,
stable, period is sometimes confused with the orbital period
(see likely examples,
IX Dra: \cite{ole04ixdra}, OT J102146.4$+$234926: \cite{uem08j1021}).
Photometrically claimed orbital periods during superoutbursts,
especially those giving $\epsilon <1$ \% need to be carefully
re-examined.

   Furthermore, the presence of two distinct periods with fair stability
might be problematic in identifying multiple periodicity by
analyzing power spectra of the entire data (e.g. \cite{pat03suumas}).

   A typical difference of 0.5--1.0\% in $P_{\rm SH}$ between
$P_1$ and $P_2$ corresponds to a difference of 0.03--0.05 in
$q$ \citep{pat05SH}.
This difference could result a systematic error in calibrating
$\epsilon$--$q$ relation, or estimating $q$ depending on the stage
when $P_{\rm SH}$ is measured.  The situation could be worse if the
relation is applied to superhump periods obtained around the termination
of the stage B (subsection \ref{sec:pdotb}, figure \ref{fig:pendporb}).
This issue is further discussed in subsection \ref{sec:epsq}.

\subsection{Minimum Superhump Period and 3:1 Resonance}

\begin{figure}
  \begin{center}
    \FigureFile(88mm,70mm){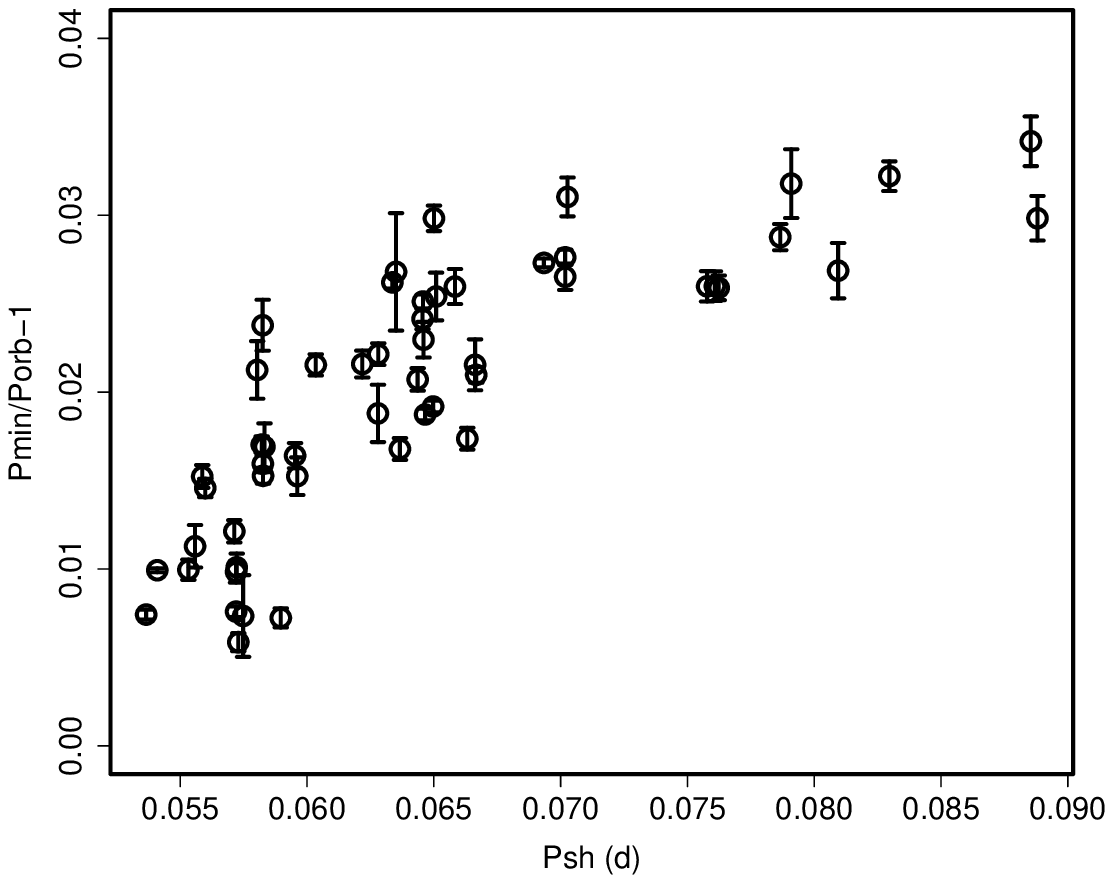}
  \end{center}
  \caption{Relation between the fractional period excess for the
  minimum $P_{\rm SH}$ and $P_{\rm SH}$ ($P_1$).
  }
  \label{fig:pminpsh}
\end{figure}

\begin{figure}
  \begin{center}
    \FigureFile(88mm,70mm){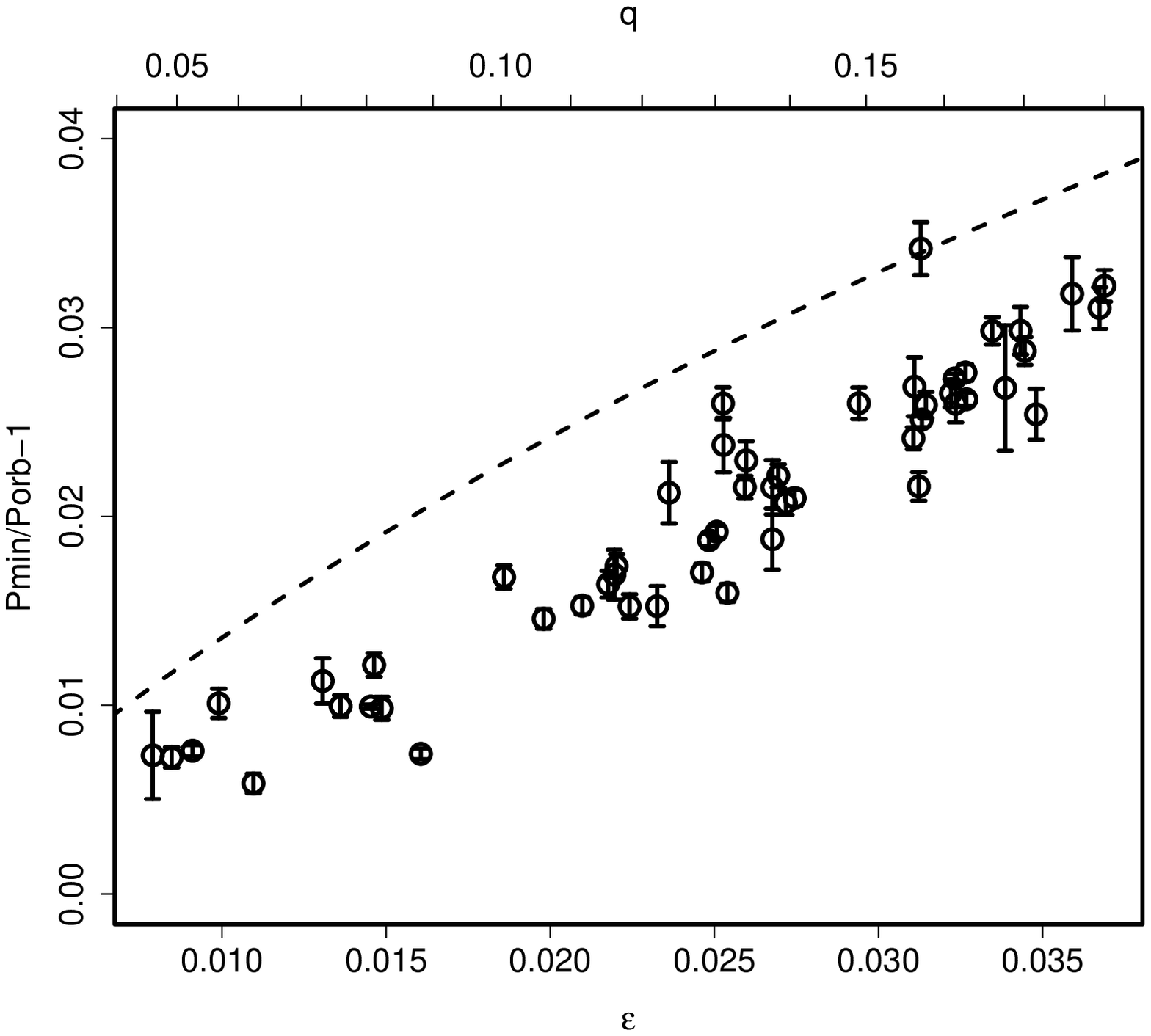}
  \end{center}
  \caption{Relation between the fractional period excess for the
  minimum $P_{\rm SH}$ and $q$, scaled from $P_1$.
  The dashed line represents fractional excess expected for single
  particle dynamical precession rate at 1:3 resonance.
  }
  \label{fig:pmineps}
\end{figure}

\begin{figure}
  \begin{center}
    \FigureFile(88mm,70mm){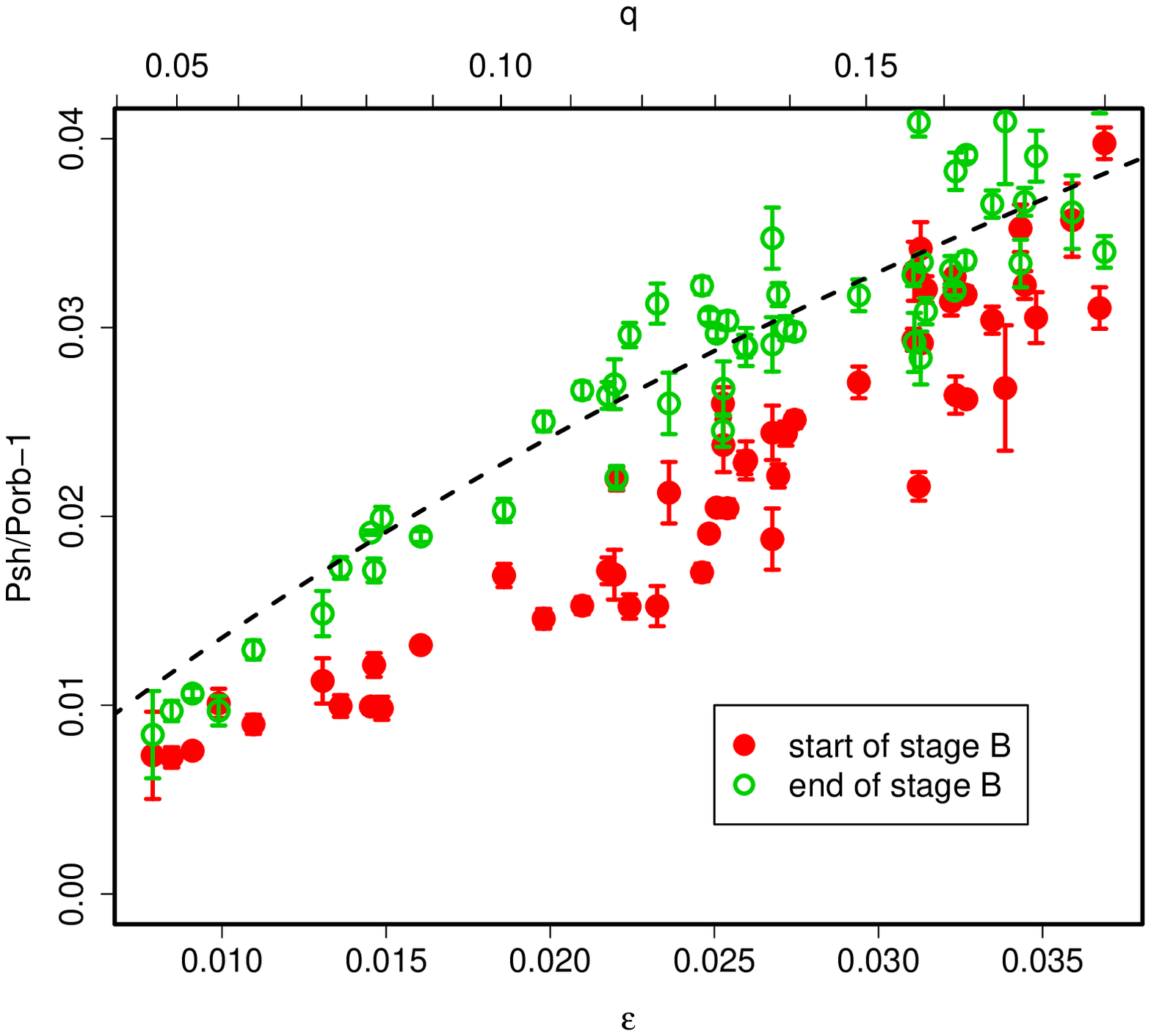}
  \end{center}
  \caption{Relation between the fractional period excess for the
  different epochs of stage B $P_{\rm SH}$ and $q$, scaled from $P_1$.
  The dashed line is the same as in figure \ref{fig:pmineps}.
  }
  \label{fig:pothereps}
\end{figure}

   Among surveyed sets of parameters, we have noticed that the
fractional period excess for minimum $P_{\rm SH}$ of
a given superoutburst (either $P_2$ or $P_{\rm SH}$ at the start of
the stage B for $P_{\rm dot} > 0$ systems) of a given system
is most smoothly and tightly correlated with other system parameters
(figures \ref{fig:pminpsh}, \ref{fig:pmineps}; for a comparison of
other representative $P_{\rm SH}$, see figure \ref{fig:pothereps}).
In figure \ref{fig:pmineps}, we give $\epsilon$ expected for
dynamical precession rate at the 3:1 resonance ($\epsilon_{3:1}$),
using the $\epsilon$--$q$ relation (\cite{pat05SH}, using the updated
one discussed in subsection \ref{sec:epsq}) and angular velocity
of disk precession formulated by \citet{osa85SHexcess}.
The $\epsilon$ for the minimum $P_{\rm SH}$ best parallels the expected 
$\epsilon$ for the 3:1 resonance.  We therefore regard that the
minimum $P_{\rm SH}$ represents the precession at the 3:1 resonance.
This interpretation can naturally explain the ubiquitous presence of
the stage C and the stability of the superhump period during the stage C.
The systematic difference between $\epsilon_{3:1}$ and observed values
is likely attributed to the scaling problem in interpreting hydrodynamical
precession rate of a disk as a whole by single-particle dynamical precession
(see \cite{smi07SH}) rather than the real difference.

   In systems lacking the stage C, such as many of extreme
WZ Sge-type dwarf novae, the $P_{\rm SH}$ appears to always reflect
the precession rate at the 3:1 resonance.
The stability of the $P_{\rm SH}$ in such systems can then be naturally
explained.  In positive $P_{\rm dot}$ systems, $P_{\rm SH}$ at the
start of the stage B is almost identical to $P_2$
(subsection \ref{sec:pshstageb}).  This can be understood as
superhumps excited at the 3:1 resonance quickly dominates at the start
of the stage B in these systems.

\subsection{Maximum Superhump Period and Disk Radius}\label{sec:maxradius}

   By assuming this interpretation and assuming the radial dependence of
precession rate \citep{mur00SHprecession}, we can calculate the disk
radius from $\epsilon$ at other epochs.\footnote{
   In scaling the radius, we used the radius of single-particle
   dynamical 3:1 resonance for simplicity.
   This radius may be systematically too large \citep{smi07SH}.
   Other factors proposed to affect superhump periods include 
   changes in temperature or pressure (\cite{hir93SHperiod};
   \cite{mur98SH}; \cite{mon01SH}; \cite{pea06SH}.
   Since the disk temperature is expected to decrease
   during the decline phase, a slowing effect on the precession due to
   the pressure forces is expected to decrease \citet{mon01SH}.
   This expectation is contrary to the global period decrease generally
   observed, and we regard that the variation in the disk temperature
   is unlikely the primary cause of the period variation.
   We therefore focus on dynamical precession and did not consider
   other effects for simplicity.
}

\begin{figure}
  \begin{center}
    \FigureFile(88mm,70mm){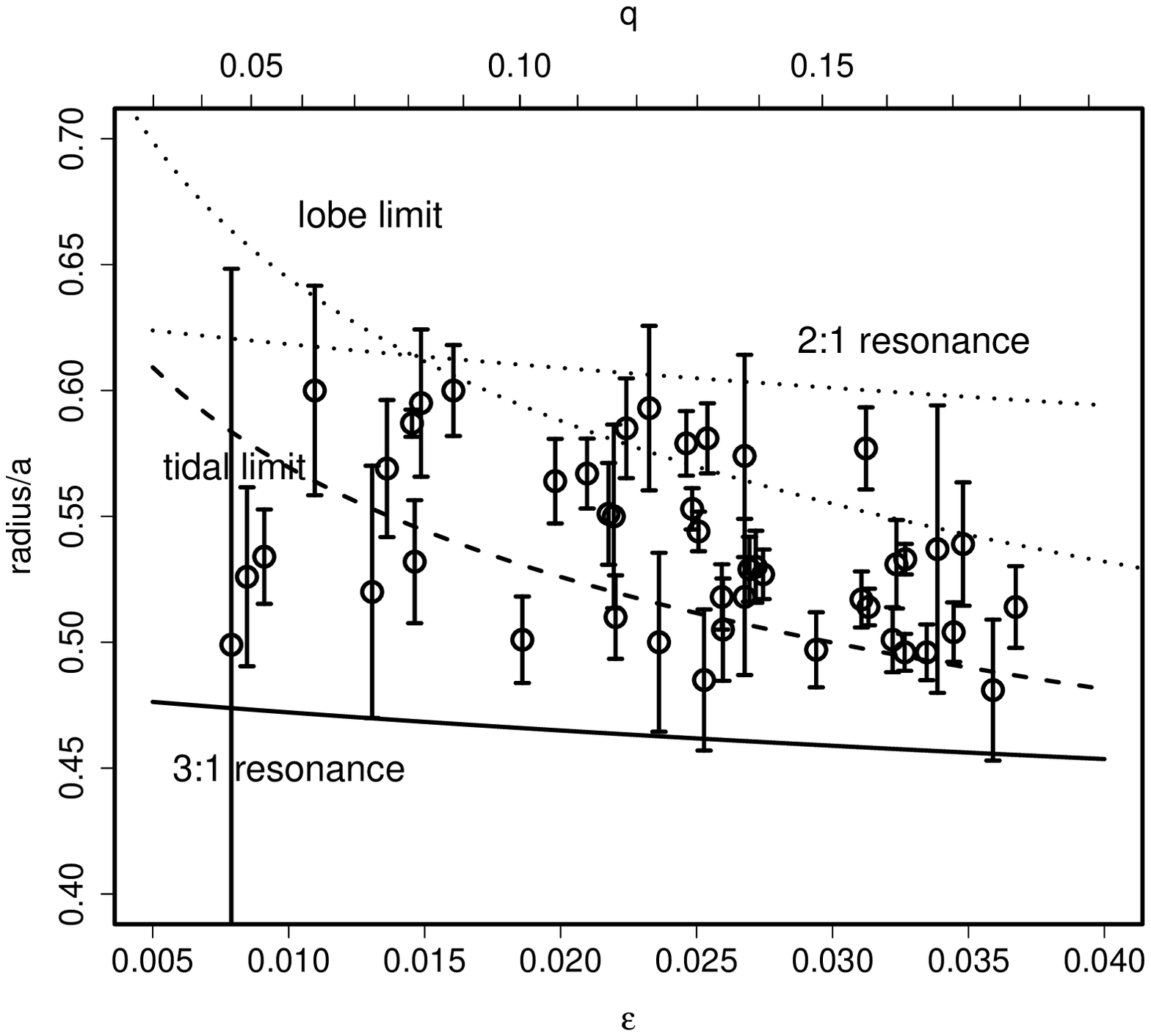}
  \end{center}
  \caption{Disk radius at the end of stage B scaled from ratios of
  $\epsilon$ (for $P_1$) between the end of stage B and
  the minimum $P_{\rm SH}$.
  The locations of various resonances and limits are the same as
  in \citet{kat08wzsgelateSH}.
  }
  \label{fig:pendr}
\end{figure}

   The radii calculated for the end of stage B for systems with
$P_{\rm dot} > 0$, corresponding to the maximum radii, are given in
table \ref{tab:pendr} and figure \ref{fig:pendr}.
The radii at the end of stage B for positive
$P_{\rm dot}$ systems are reasonably situated, considering the errors
and the simplified treatment, around the radii of tidal truncation or slightly
beyond this.  This result can lead to a picture that superhumps are
initially excited at the 3:1 resonance, whose outward propagation
(if there is sufficient matter outside the 3:1 resonance)
is limited by tidal truncation.  This probably determines the maximum
attainable $P_{\rm SH}$ in positive $P_{\rm dot}$ systems.
Since the superhumps usually quickly decay near the end of stage B,
the large dissipation at large radius seems to quickly quench
the eccentricity power.

\begin{table}
\caption{Superhump Periods during Stage A}\label{tab:pera}
\begin{center}
\begin{tabular}{cccc}
\hline\hline
Object & Year & period (d) & err \\
\hline
V455 And & 2007 & 0.05803 & 0.00008 \\
V466 And & 2008 & 0.05815 & 0.00008 \\
DH Aql & 2003 & 0.08079 & 0.00012 \\
VY Aqr & 2008 & 0.06558 & 0.00026 \\
EG Aqr & 2006 & 0.08128 & 0.00018 \\
TT Boo & 2004 & 0.07911 & 0.00009 \\
UZ Boo & 2003 & 0.06354 & 0.00024 \\
AX Cap & 2004 & 0.12279 & 0.00277 \\
GX Cas & 1996 & 0.09690 & 0.00055 \\
V1040 Cen & 2002 & 0.06243 & 0.00007 \\
WX Cet & 1989 & 0.06031 & 0.00003 \\
WX Cet & 1998 & 0.06027 & 0.00014 \\
RX Cha & 2009 & 0.08710 & -- \\
BZ Cir & 2004 & 0.07692 & 0.00005 \\
PU CMa & 2008 & 0.05901 & 0.00022 \\
AL Com & 1995 & 0.05799 & 0.00014 \\
AL Com & 2001 & 0.05791 & 0.00022 \\
GO Com & 2003 & 0.06323 & 0.00010 \\
V632 Cyg & 2008 & 0.06628 & 0.00007 \\
V1028 Cyg & 1995 & 0.06269 & 0.00011 \\
V1113 Cyg & 1994 & 0.07963 & 0.00007 \\
KV Dra & 2009 & 0.06109 & 0.00026 \\
XZ Eri & 2008 & 0.06519 & 0.00021 \\
UV Gem & 2003 & 0.09635 & 0.00030 \\
AW Gem & 1995 & 0.08276 & 0.00053 \\
AW Gem & 2009 & 0.08185 & 0.00038 \\
V844 Her & 2006 & 0.05649 & 0.00005 \\
MM Hya & 2001 & 0.06032 & 0.00025 \\
RZ Leo & 2000 & 0.08046 & 0.00053 \\
GW Lib & 2007 & 0.05473 & 0.00007 \\
V419 Lyr & 2006 & 0.09131 & 0.00009 \\
V585 Lyr & 2003 & 0.06113 & 0.00008 \\
AB Nor & 2002 & 0.08174 & 0.00027 \\
DT Oct & 2003 & 0.07650 & 0.00017 \\
V2527 Oph & 2004 & 0.07226 & 0.00008 \\
V368 Peg & 2005 & 0.07130 & -- \\
UV Per & 2003 & 0.06813 & 0.00009 \\
UV Per & 2007 & 0.06659 & 0.00002 \\
QY Per & 1999 & 0.08037 & 0.00032 \\
WZ Sge & 2001 & 0.05838 & 0.00006 \\
SU UMa & 1999 & 0.08054 & 0.00021 \\
SW UMa & 1991 & 0.05908 & 0.00014 \\
SW UMa & 2000 & 0.05877 & 0.00006 \\
SW UMa & 2006 & 0.05894 & 0.00005 \\
BC UMa & 2003 & 0.06512 & 0.00006 \\
BZ UMa & 2007 & 0.07113 & 0.00039 \\
DV UMa & 2005 & 0.08929 & -- \\
DV UMa & 2007 & 0.08926 & 0.00018 \\
IY UMa & 2000 & 0.07666 & 0.00015 \\
KS UMa & 2003 & 0.07095 & 0.00009 \\
HS Vir & 1996 & 0.08118 & 0.00021 \\
\hline
\end{tabular}
\end{center}
\end{table}

\addtocounter{table}{-1}
\begin{table}
\caption{Superhump Periods during Stage A (continued)}
\begin{center}
\begin{tabular}{cccc}
\hline\hline
Object & Year & period (d) & err \\
\hline
HV Vir & 2002 & 0.05864 & 0.00002 \\
1RXS J0423 & 2008 & 0.07970 & 0.00010 \\
ASAS J0233 & 2006 & 0.05676 & 0.00007 \\
ASAS J1025 & 2006 & 0.06407 & 0.00010 \\
ASAS J1536 & 2004 & 0.06557 & 0.00014 \\
ASAS J1600 & 2005 & 0.06653 & 0.00012 \\
CTCV J0549 & 2006 & 0.08650 & 0.00026 \\
SDSS J1556 & 2007 & 0.08395 & 0.00014 \\
SDSS J1627 & 2008 & 0.11269 & 0.00067 \\
OT J1443 & 2009 & 0.07461 & 0.00055 \\
\hline
\end{tabular}
\end{center}
\end{table}

\begin{table}
\caption{Estimated disk radius at the end of stage B.}\label{tab:pendr}
\begin{center}
\begin{tabular}{ccccc}
\hline\hline
Object & Year & $\epsilon$ & $r$$^a$ & err$^b$ \\
\hline
V1108 Her & 2004 & 0.008 & 0.499 & 0.149 \\
OT J1112 & 2007 & 0.008 & 0.526 & 0.036 \\
WZ Sge & 2001 & 0.009 & 0.534 & 0.019 \\
AL Com & 1995 & 0.011 & 0.600 & 0.042 \\
TSS J0222 & 2005 & 0.013 & 0.520 & 0.050 \\
DI UMa & 2007a & 0.014 & 0.569 & 0.027 \\
GW Lib & 2007 & 0.015 & 0.587 & 0.005 \\
\end{tabular}
\end{center}
\end{table}

\begin{table}
\caption{Estimated disk radius at the start of stage B.}\label{tab:pstartr}
\begin{center}
\begin{tabular}{ccccc}
\hline\hline
Object & Year & $\epsilon$ & $r$$^a$ & err$^b$ \\
\hline
V1108 Her & 2004 & 0.008 & 0.474 & 0.149 \\
OT J1112 & 2007 & 0.008 & 0.474 & 0.036 \\
WZ Sge & 2001 & 0.009 & 0.474 & 0.019 \\
AL Com & 2001 & 0.010 & 0.472 & 0.036 \\
AL Com & 1995 & 0.011 & 0.552 & 0.042 \\
TSS J0222 & 2005 & 0.013 & 0.471 & 0.050 \\
DI UMa & 2007a & 0.014 & 0.472 & 0.027 \\
\end{tabular}
\end{center}
\end{table}

   This picture generally well applies to systems with $\epsilon > 0.02$
(corresponding to $q > 0.11$).  Objects with smaller $\epsilon$ tend to
deviate from this trend.  These objects include extreme WZ Sge-type dwarf
novae (WZ Sge, V455 And, AL Com) while some of
(what are usually regarded as) WZ Sge-type dwarf novae (GW Lib, HV Vir)
have a similar tendency to ordinary SU UMa-type dwarf novae.
The small radii for V436 Cen, UV Per (2007) and others may have been
a result of undersampling of superhump timings; the case for UV Per
is particularly likely because other well-sampled superoutbursts of the
same object generally gave larger radii.

   The difference among WZ Sge-type dwarf novae can be attributed to
the matter left beyond the 3:1 resonance \citep{kat08wzsgelateSH}:
if the 2:1 resonance is strong enough to accrete much of the matter
beyond the radius of 3:1 resonance, the propagation of the eccentricity
wave beyond the 3:1 resonance would not produce a strong superhump
signal with a longer period.  Further observations, however, are
especially needed in these cases whether different types of superoutbursts
(cf. \cite{uem08alcom}) in the same WZ Sge-type object lead to different
behavior of $P_{\rm SH}$.

\subsection{Superhump Period at the Start of Stage B}

\begin{figure}
  \begin{center}
    \FigureFile(88mm,70mm){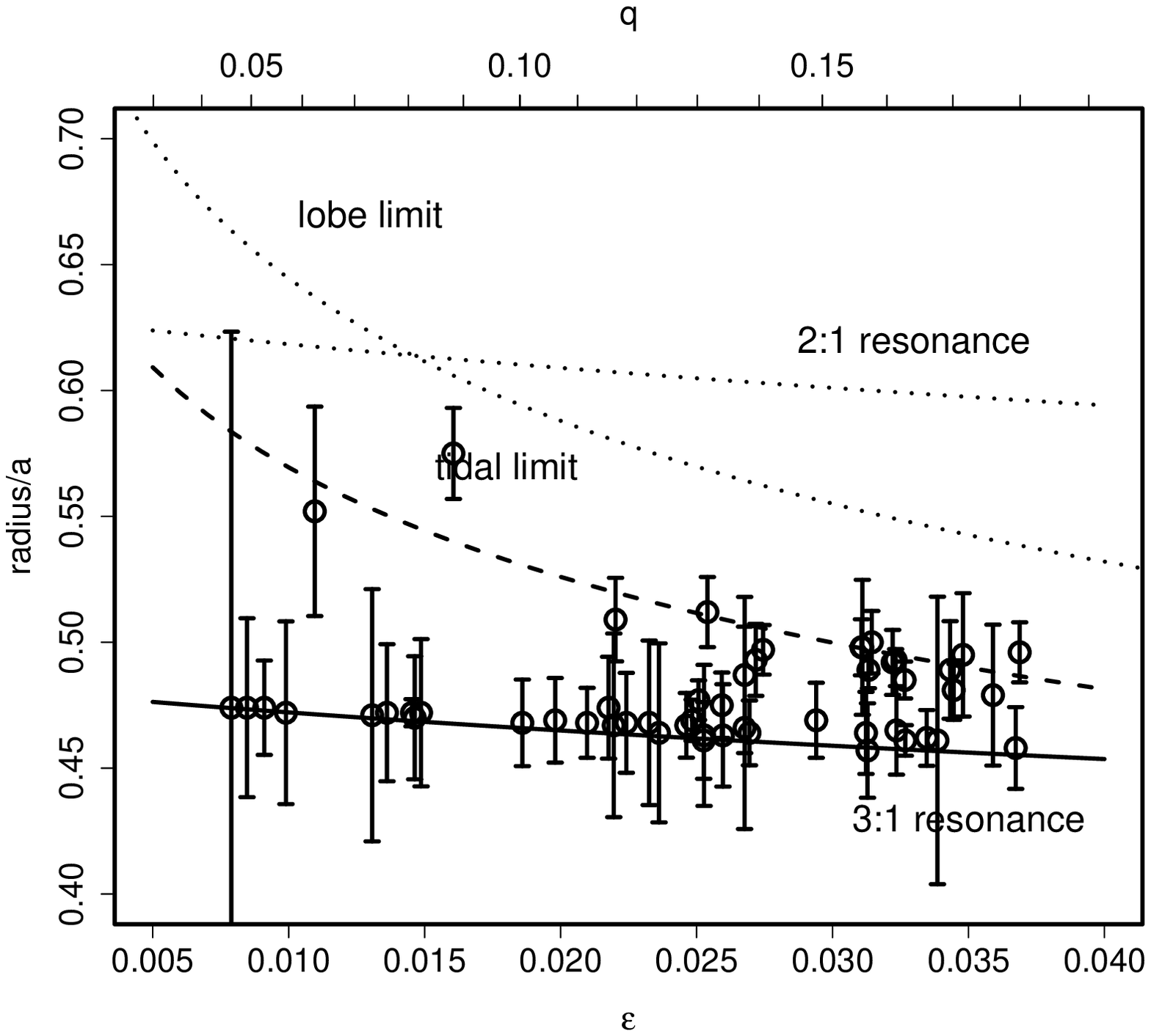}
  \end{center}
  \caption{Disk radius at the start of stage B scaled from ratios of
  $\epsilon$ between the end of the stage B and the minimum $P_{\rm SH}$.
  }
  \label{fig:pstartr}
\end{figure}

   The radii calculated for the start of stage B are given in
table \ref{tab:pstartr} and figure \ref{fig:pstartr}.
In systems with positive $P_{\rm dot}$, these radii match the
supposed 3:1 resonance.  The exceptions, AL Com in 1995 and
OT J0238 in 2008, are likely a result of the poorly determined
stage C superhumps.
In negative $P_{\rm dot}$ systems (generally corresponding to
$\epsilon > 0.025$), the decrease in the $P_{\rm SH}$ can be explained
if superhumps are initially excited slightly outside the 3:1 resonance.
In such systems, the large tidal torque caused by the large $q$ might
enable eccentricity wave originating even outside the 3:1 resonance.

\subsection{Stage C Superhumps in Positive $P_{\rm dot}$ Systems}\label{sec:stagec}

   In our interpretation, the stage C superhumps in positive $P_{\rm dot}$
are regarded as superhumps stably originating from the radius of the
3:1 resonance.  It looks as if superhumps
are newly excited around the radius of 3:1 resonance after the original
superhumps reached a larger radius (limiting radius as discussed in
subsection \ref{sec:maxradius}) and their eccentric power is quenched.
It may be that superhumps can be rejuvenized if the eccentricity
of the original superhumps become sufficiently weak and there is
still sufficient matter around the 3:1 resonance.  Such a condition
could be realized when the matter beyond the 3:1 resonance still
remains after the termination of a superoutburst
(cf. \cite{kat08wzsgelateSH}) and if this matter is efficiently
accreted inward.  The brightening associated with the appearance
(or regrowth) of superhumps at the start of the stage C can be naturally
explained by this accretion and increased dissipation due to
a renewed tidal instability.

\subsection{Stage A Superhumps}

\begin{figure}
  \begin{center}
    \FigureFile(88mm,70mm){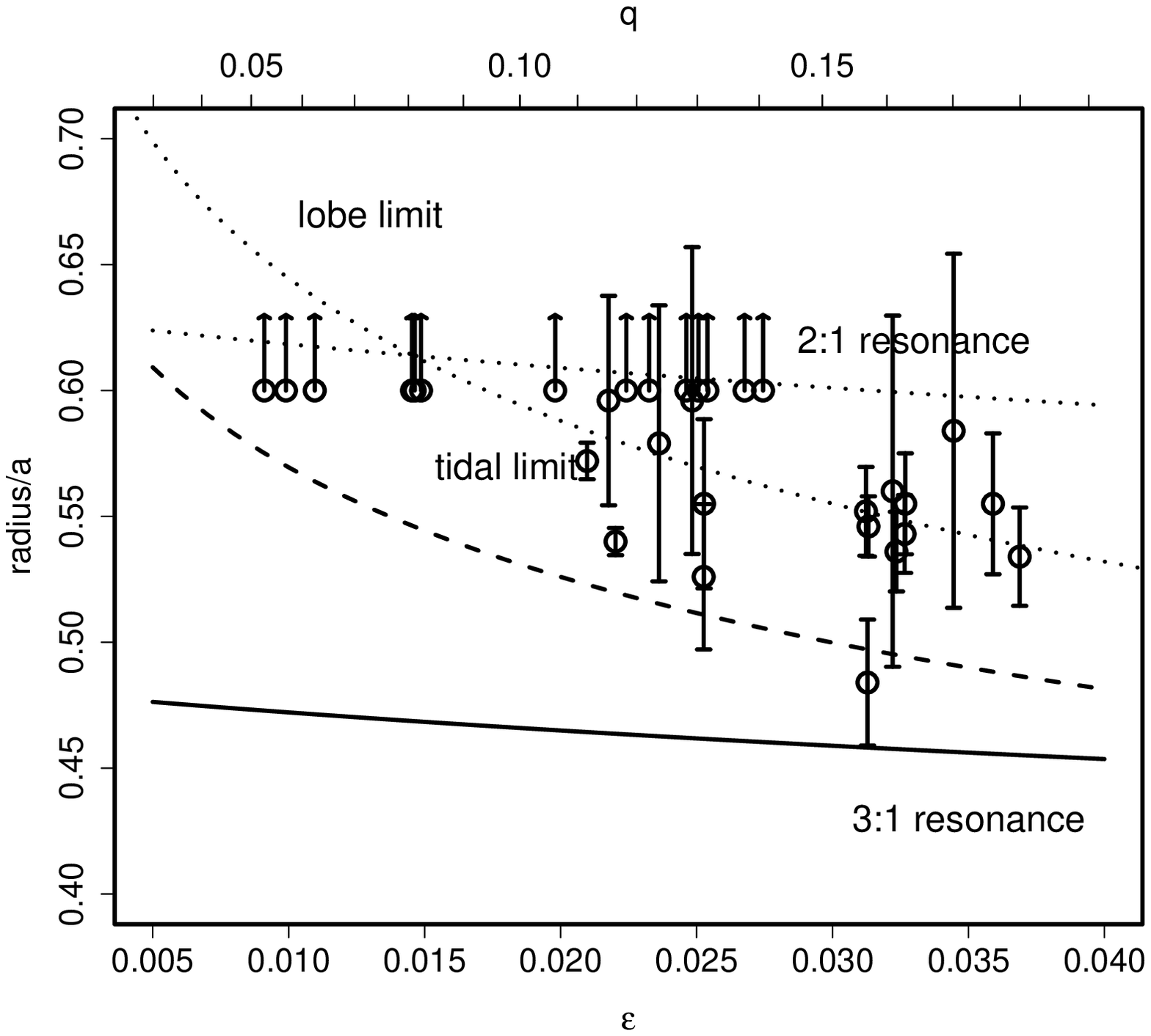}
  \end{center}
  \caption{Disk radius during the stage A scaled from ratios of
  $\epsilon$ between the end of stage B and the minimum $P_{\rm SH}$.
  Upper arrow show lower limits.
  }
  \label{fig:pstagear}
\end{figure}

   We similarly calculated the radii for the start of the stage A
(figure \ref{fig:pstagear}).  In some systems, the fractional superhump
excesses exceed the range in \citet{mur00SHprecession},
and they are shown in lower limits.  The periods of stage A superhumps
can be understood if they originate from the outermost disk.
Since stage A and B superhumps show a smooth transition in phase,
the eccentricity excited during the stage A in the outside the disk
appears to efficiently excite the strong eccentricity at the radius
of the 3:1 resonance.  It may be that the eccentricity invoked during
the stage A can efficiently work as a seed perturbation at the radius
of the 3:1 resonance.
The situation might be the same for the stage B--C transition.

   Although many of very well observed superoutbursts show stage A,
some superoutbursts showed different behavior.  Among them, QZ Vir in 1993
and 1RXS J0532 in 2005 associated with precursor outbursts did not show
long-period superhumps as in usual stage A.  The initial period of
superhumps during the 1993 superoutburst of QZ Vir was close to the
orbital period \citep{kat97tleo}, an exceptional case in this study.
The existence of a prominent precursor in these superoutburst can be
understood as a result of a small disk-mass at the onset of superoutbursts
\citep{osa03DNoutburst}.  In these superoutbursts, the disk mass may
have been so small that virtually no mass was present beyond the
3:1 resonance.

   Note, however, stage A with long-period superhumps was
definitely recorded during superoutbursts of GO Com in 2003
\citep{ima05gocom} and PU CMa in 2008 (subsection \ref{tab:pucma}).
The condition whether stage A appears or not may depend on other factors.

   With smoothed particle hydrodynamics (SPH), \citet{mur98SH} reported
a longer superhump period during the early stage of eccentricity growth.
Although this might correspond to stage A superhumps, the exact
identification should await further investigation.

\subsection{ER UMa Stars}\label{sec:erumastars}

   ER UMa stars are a subgroup of SU UMa-type dwarf novae characterized
by the shortness (19--50 d) of their supercycles
(\cite{kat95eruma}; \cite{rob95eruma}; \cite{mis95PGCV};
\cite{nog95rzlmi}; \cite{osa95eruma}).
It has been demonstrated that at least some of
ER UMa stars show large-amplitude superhumps at the onset of superoutbursts
\citep{kat96erumaSH} and a phase reversal of superhumps during the
early plateau stage \citep{kat03erumaSH}.  \citet{osa03DNoutburst}
interpreted large-amplitude superhumps in the early stage is a consequence
of tidal heating at the outer edge by the continuous presence of tidal
instability, resulting a superoutburst driven by the tidal instability.
The origin of the phase reversal is not yet well understood.
\citet{kat03erumaSH} suspected that a movement of the location of
the strongest tidal dissipation to the opposite direction somehow
happened, while \citet{ole04ixdra} considered a beat between the superhump
and orbital periods.\footnote{
   As discussed in subsection \ref{sec:transimplication} this ``orbital
   period'' likely referred to $P_2$ rather than the true $P_{\rm orb}$.
}

   Due to the complexity in the hump profile and limited availability
of high-quality raw data, we do not discuss on these objects in detail.
An $O-C$ analysis for ER UMa is presented here, and brief discussions
on V1159 Ori and RZ LMi are given in Appendix section \ref{sec:individual}.

\begin{figure}
  \begin{center}
    \FigureFile(88mm,110mm){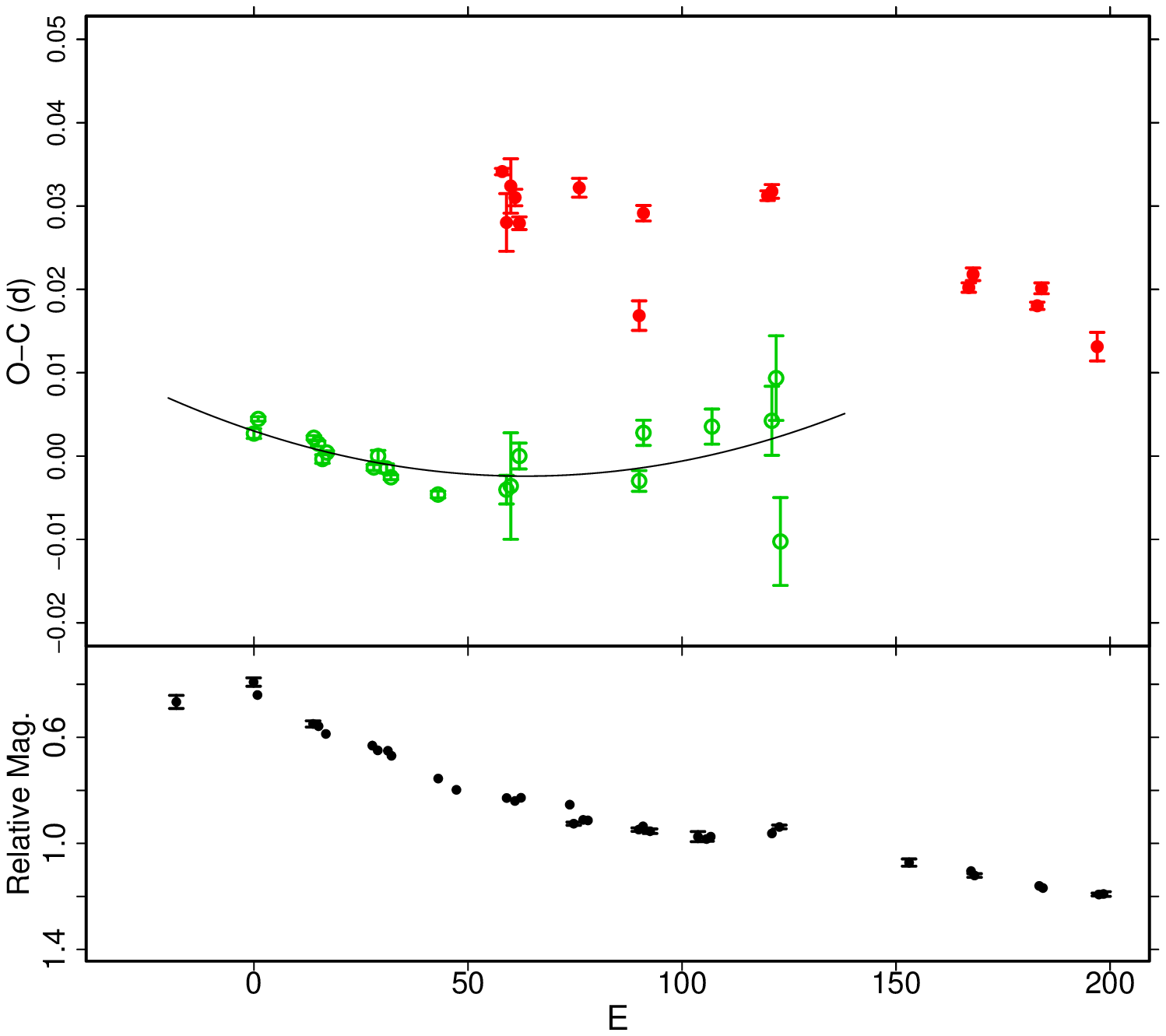}
  \end{center}
  \caption{$O-C$ variation in ER UMa (1995).  (Upper) $O-C$.
  The open and filled circles represent early-stage and later-stage
  superhumps described in \citet{kat03erumaSH}.
  (Lower) Light curve.
  }
  \label{fig:erumahumpall}
\end{figure}

   Upon examination of the data used in
\citet{kat03erumaSH}, we noticed that the early-stage superhumps
can be tracked for a while even after the occurrence of the
reported phase reversal (figure \ref{fig:erumahumpall}, open circles).
These superhumps appear to comfortably follow a positive $P_{\rm dot}$
expected for this $P_{\rm SH}$.  In ER UMa, the stage C appeared to
have started earlier than in ordinary SU UMa-type dwarf novae, and was
observed as a regrowth of superhumps associated with a phase reversal
(figure \ref{fig:erumahumpall}, filled circles).  It may be that
a combination of a large mass-transfer rate from the secondary,
and the small amount of disk matter beyond the 3:1 resonance in ER UMa
stars \citep{osa03DNoutburst} serves a condition enabling early
rejuvenization of superhumps (cf. subsection \ref{sec:stagec}).
It is not known why only ER UMa stars show a $\sim$0.5 phase shift
at the onset of the stage C.  Detailed observations of ER UMa stars
might provide a clue to understanding the nature of the stage B--C
transition.

\begin{figure}
  \begin{center}
    \FigureFile(88mm,70mm){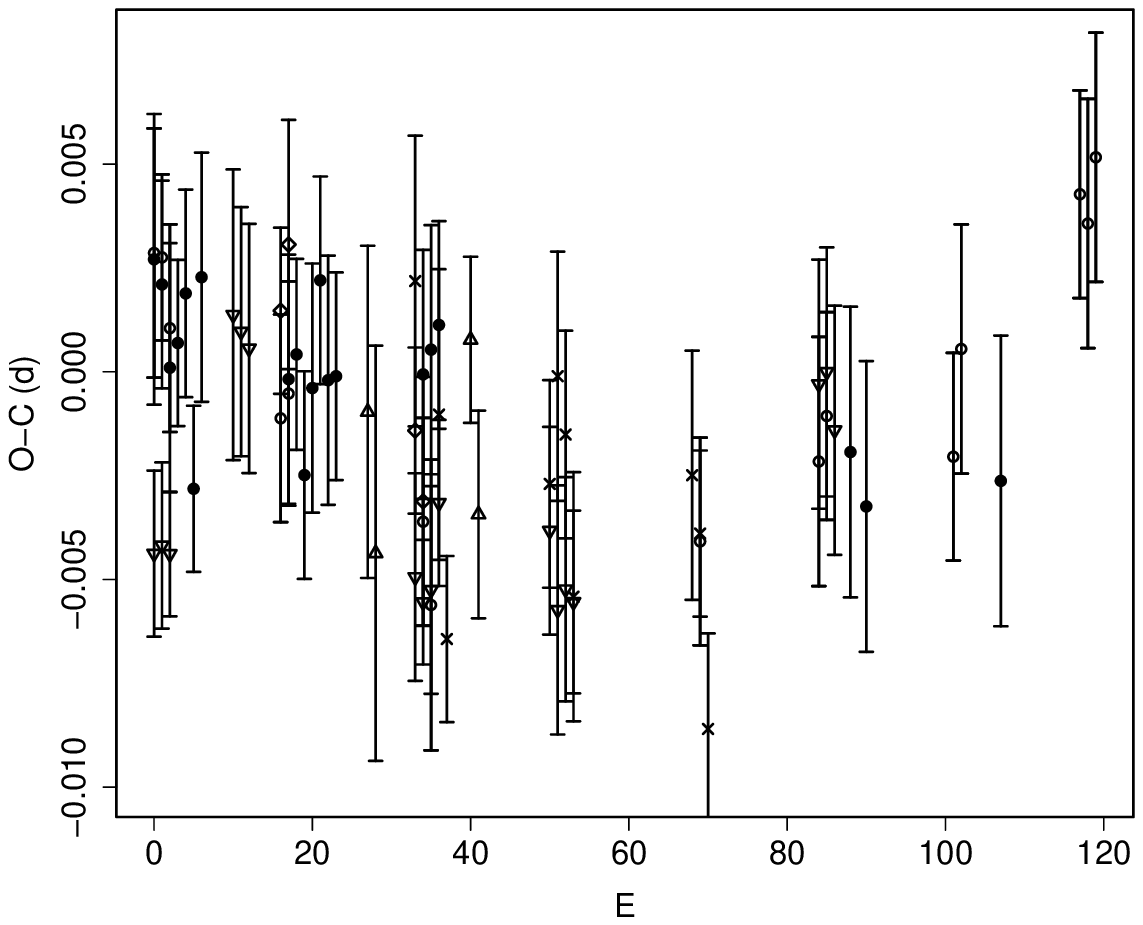}
  \end{center}
  \caption{$O-C$ variation in RZ LMi.  The hump maxima are taken
  from \citet{ole08rzlmi} and are shifted so that the start of
  individual superoutburst corresponds to $E=0$.
  }
  \label{fig:rzlmiocadd}
\end{figure}

   The behavior of superhumps in RZ LMi is still poorly known.
\citet{ole08rzlmi} reported that its superhump periods were almost constant
other than one well-observed, 2004 superoutburst.  \citet{ole08rzlmi}
claimed that the phases of superhumps were even coherent between
different superoutbursts.  We should, however, note that many of
observations by \citet{ole08rzlmi} covered only a few days of
individual superoutbursts, making it difficult to estimate
$P_{\rm dot}$ for individual superoutbursts.  Instead, reported
superhump maxima in \citet{ole08rzlmi} can be reasonably well
expressed by a slightly positive $P_{\rm dot}$, by the same
overlaying method used in subsection \ref{sec:different}
(figure \ref{fig:rzlmiocadd}).  We consider that the $P_{\rm dot}$
observed during the 2004 superoutburst is typical for this object
and the slight difference in $P_{\rm SH}$ between different
superoutbursts \citep{ole08rzlmi} was a result of observation of
different phase of superoutbursts.  This interpretation needs to
be tested by continuous observation throughout different superoutbursts.
It would be intriguing to see whether or not the stage C is present
in RZ LMi (see subsection \ref{sec:rzlmi}),
which might provide a clue why superoutbursts in RZ LMi
are quenched so early (cf. \cite{osa95rzlmi}; \cite{hel01eruma}).

   Most recently, \citet{rut08diuma} reported a positive, but a relatively
small $P_{\rm dot}$ in another ultra-short $P_{\rm SH}$ ER UMa star,
DI UMa.  \citet{rut08diuma} also reported superhump-like variations during
the rising stage but were shifted in phase by $\sim$0.5 $P_{\rm SH}$.
These variations may have been stage A superhumps, and we obtained
a period of 0.0569(2) d by assuming the phase continuity.
The exact identification of their nature should await a further study.
If the superhumps were evolving in period during the rising stage of
DI UMa, as in the stage A in ordinary SU UMa-type dwarf novae, the onset
of tidal instability likely coincides with the ignition of the outburst,
on the contrary to the expectation in \citet{osa03DNoutburst} that
tidal instability triggers ER UMa-type superoutbursts.

\subsection{Long-Period Systems}\label{sec:longp}

\begin{figure}
  \begin{center}
    \FigureFile(88mm,110mm){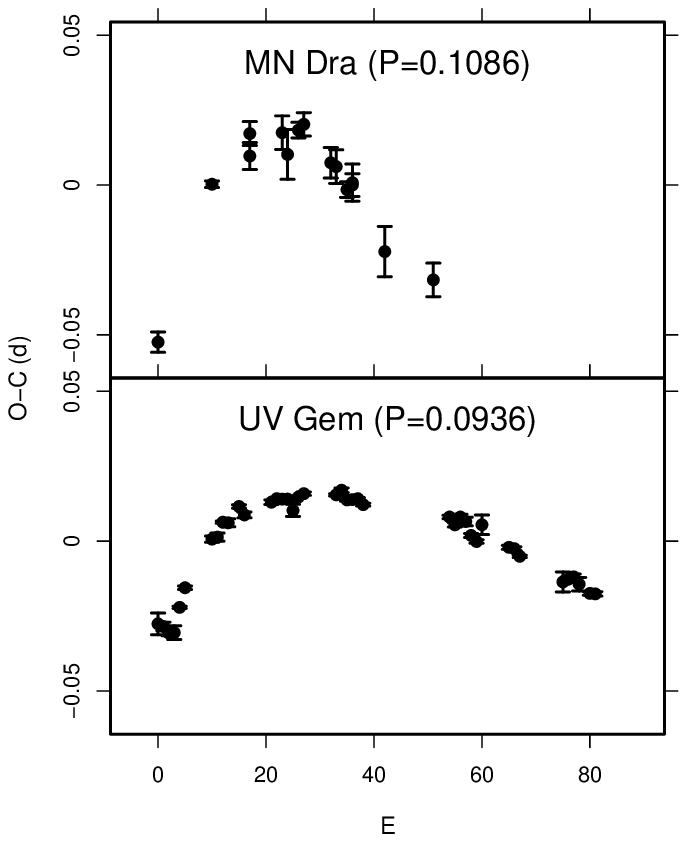}
  \end{center}
  \caption{$O-C$ variations of Long-$P_{\rm SH}$ systems with
  smooth period variations.}
  \label{fig:lp1}
\end{figure}

\begin{figure}
  \begin{center}
    \FigureFile(88mm,110mm){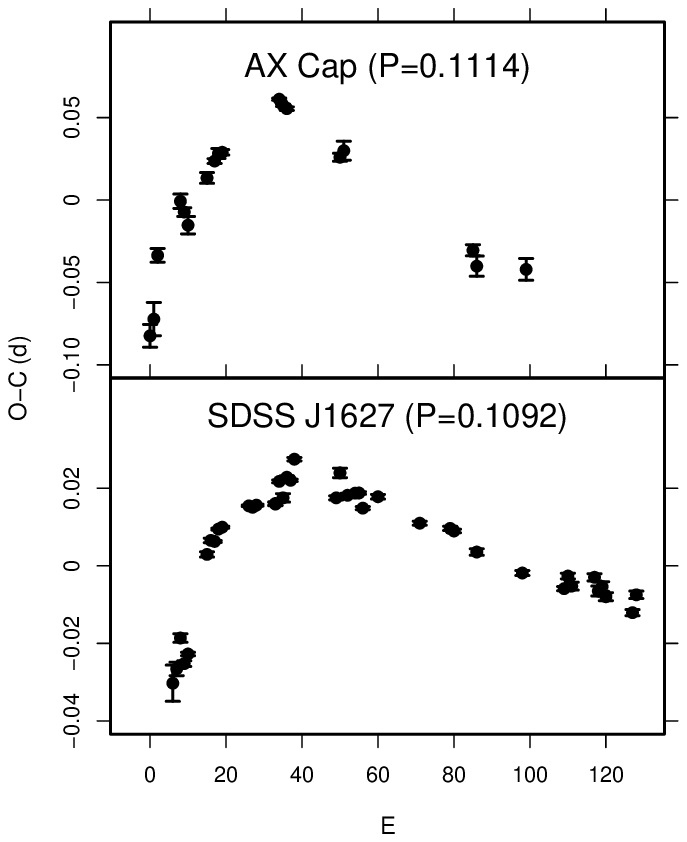}
  \end{center}
  \caption{$O-C$ variations of Long-$P_{\rm SH}$ systems with
  period breaks.}
  \label{fig:lp2}
\end{figure}

   The period variation of superhumps in long-period ($P_{\rm SH}$)
systems appears to vary from system to system.
Some systems, such as MN Dra and UV Gem, show smoothly decreasing
$P_{\rm SH}$ (figure \ref{fig:lp1}), while others, such as
AX Cap and SDSS J1627, show stage transitions (accompanied by
a break in the $O-C$ diagram and a well-defined stage C superhumps
with a fairly constant period) as in short-$P_{\rm SH}$
systems (figure \ref{fig:lp2}).
Note, however, the degree of period variation is strongly
different from system to system.  Although the global pattern of
period variation is similar between AX Cap and SDSS J1627, the amplitude
of $O-C$'s are several times larger in the former system.
There are apparently a class of systems with a much smaller period
variation, such as TU Men, EF Peg, BF Ara and V725 Aql.
Although further confirmation is necessary, SDSS J1702 (and possibly
V725 Aql) even appears to have a positive period derivative.

   The systems with smoothly decreasing $P_{\rm SH}$ look like to have
more frequent normal outbursts than in systems with stage transitions.
The latter class of long-$P_{\rm SH}$ SU UMa-type dwarf novae
seems to somehow mimic short-$P_{\rm SH}$ SU UMa-type dwarf novae
with infrequent superoutbursts.  Whether there is a difference in
$q$ or other system parameters, or whether suppression of normal
outbursts somehow works in the latter class need to be tested
by further observations.

\subsection{Superhumps in Black-Hole X-Ray Transients}\label{sec:BHXT}

   Black-hole X-ray transients (BHXTs)  are known to show superhumps
(cf. \cite{bai92gumus}; \cite{kat95v518per}; \cite{odo96BHXNSH};
\cite{has01BHXNSH}; \cite{uem02j1118}).

\begin{figure}
  \begin{center}
    \FigureFile(88mm,70mm){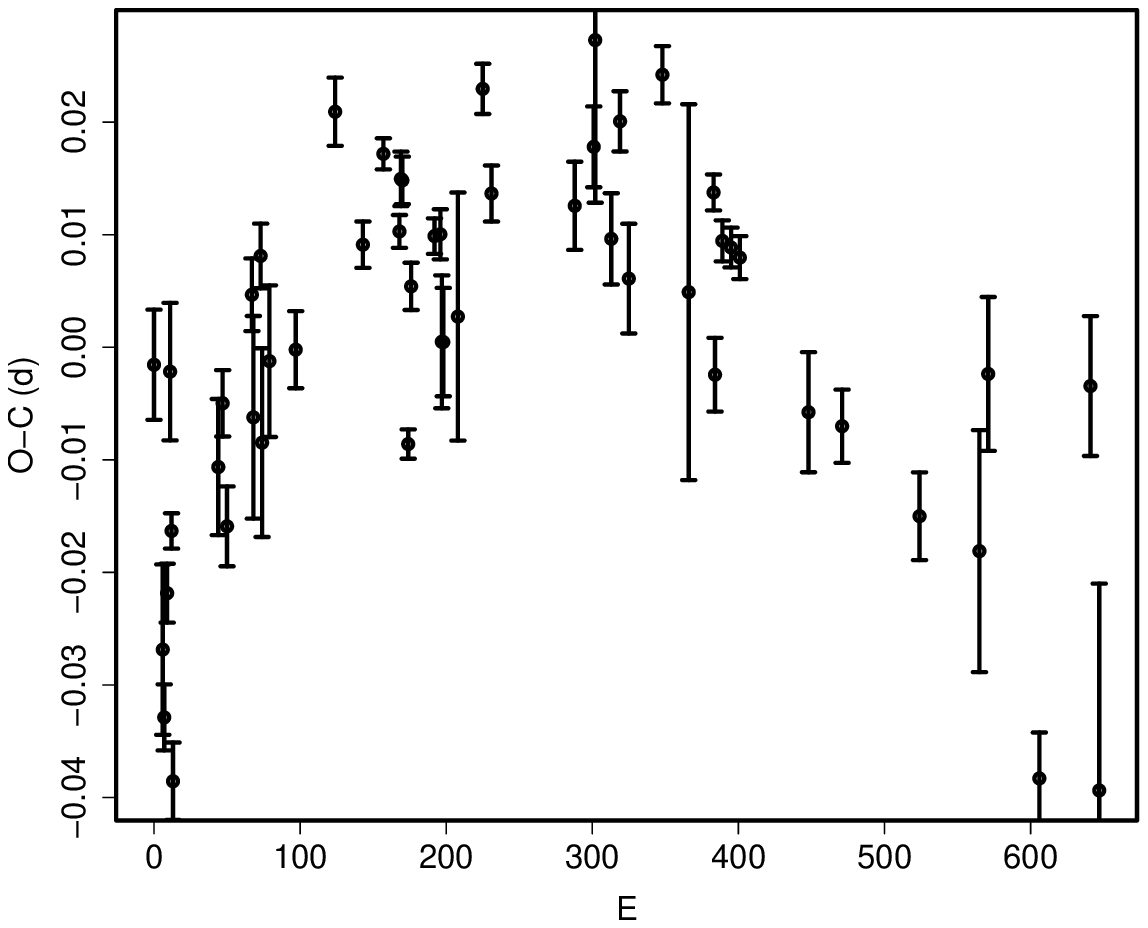}
  \end{center}
  \caption{$O-C$ diagram of KV UMa (=XTE J1118$+$480) during the 2000
  outburst.}
  \label{fig:kvumaoc}
\end{figure}

   KV UMa (=XTE J1118$+$480) is the best studied superhumping system
among BHXTs.  The $O-C$ diagram (figure \ref{fig:kvumaoc}, see
subsection \ref{sec:kvuma} for the data)
closely resemble those of SU UMa-type dwarf novae with intermediate
$P_{\rm orb}$.
The similarity of the $O-C$ variation between SU UMa-type dwarf novae
and a BHXT suggests that the evolution mechanism of superhumps
is similar between these systems.  The degree of period variation,
such as $P_2/P_1-1$ = 0.001 and
global $P_{\rm dot}$ = $-0.43(0.05) \times 10^{-5}$, is an order of
magnitude smaller than those of typical SU UMa-type dwarf novae.
This difference may be attributed to the difference in the emission
mechanism of superhumps between CVs and BHXTs \citep{has01BHXNSH}.
In BHXTs, the outer region of the accretion disk may be efficiently
shadowed by the inner region and may not be sufficiently ionized
for the eccentricity wave to propagate.  A study of period variation
of superhumps in BHXTs is expected to provide additional clue in
understanding the origin of superhumps in these systems and might
serve as a potential tool for studying the structure of the outer
accretion disks in these systems.

   An updated analysis for V518 Per is also presented in subsection
\ref{sec:v518per}.

\subsection{$\epsilon$-$q$ Relation}\label{sec:epsq}

\begin{figure}
  \begin{center}
    \FigureFile(88mm,70mm){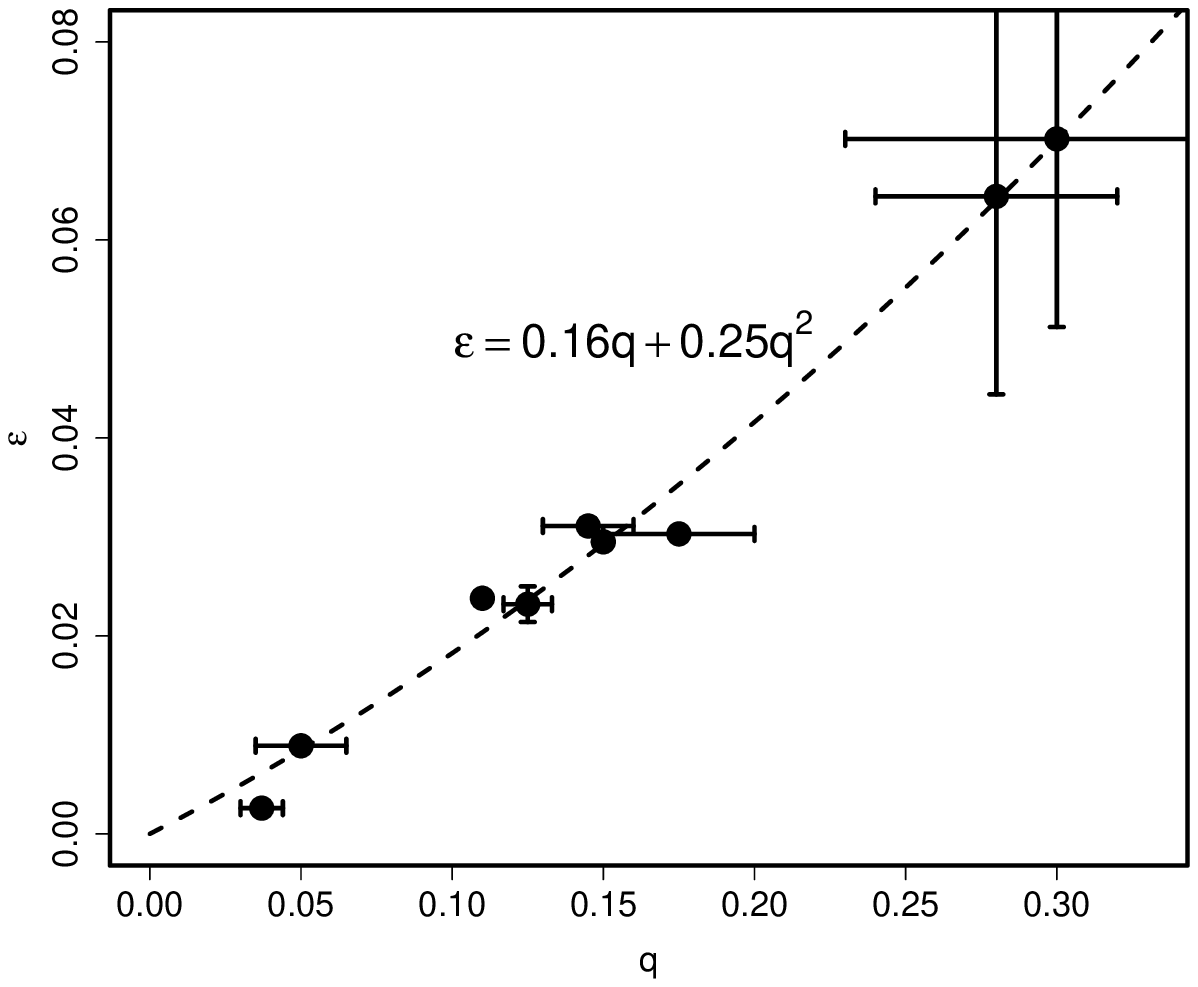}
  \end{center}
  \caption{Fractional superhump excess versus mass-ratio.
  $\epsilon$ denotes fractional superhump excess for the minimum $P_{\rm SH}$.}
  \label{fig:epsq}
\end{figure}

   Since it has become more evident that the shortest $P_{\rm SH}$
(in many cases, this agrees with $P_2$), rather than mean $P_{\rm SH}$,
represents the characteristic $P_{\rm SH}$ for SU UMa-type superoutbursts,
we re-calibrated the $\epsilon$-$q$ relation using the shortest $P_{\rm SH}$
as in the way in \citet{pat05SH}.  The data are given in table
\ref{tab:epsq} (the $q$ and $\epsilon$ for DW UMa and UU Aqr are from
\cite{pat05SH}; $\epsilon$ for other objects are newly determined
in this work).  The updated $\epsilon$-$q$ is shown in equation
\ref{equ:epsq} and figure \ref{fig:epsq}.

\begin{equation}
\epsilon = 0.16(2)q + 0.25(7)q^2
\label{equ:epsq}.
\end{equation}

\begin{table}
\caption{Fractional superhump excess versus mass-ratio}\label{tab:epsq}
\begin{center}
\begin{tabular}{ccc}
\hline\hline
Object & $\epsilon$ & $q$ \\
\hline
KV UMa & 0.0026(2) & 0.037(7) \\
WZ Sge & 0.0089(1) & 0.050(15) \\
XZ Eri & 0.0238(4) & 0.110(2) \\
IY UMa & 0.0238(18) & 0.125(8) \\
Z Cha & 0.0311(8) & 0.145(15) \\
DV UMa & 0.0295(2) & 0.150(1) \\
OU Vir & 0.0303(2) & 0.175(25) \\
DW UMa & 0.0644(20) & 0.28(4) \\
UU Aqr & 0.0702(19) & 0.30(7) \\
\hline
\end{tabular}
\end{center}
\end{table}

\subsection{$\epsilon$-$P_{\rm orb}$ Relation}\label{sec:epsporb}

\begin{figure}
  \begin{center}
    \FigureFile(88mm,70mm){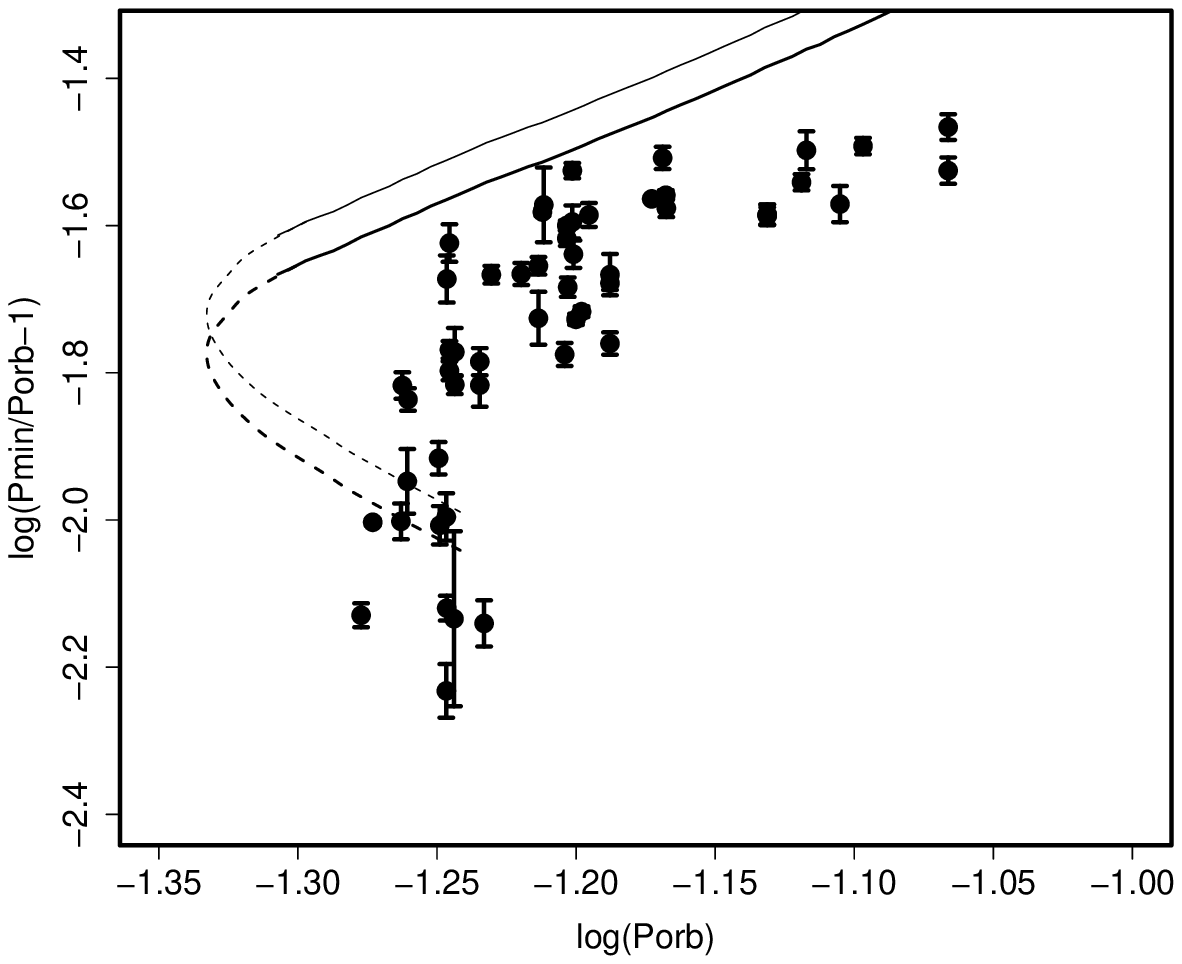}
  \end{center}
  \caption{Fractional superhump excess versus orbital period.
  $\epsilon$ denotes fractional superhump excess for the minimum $P_{\rm SH}$.
  The two set of curves and dashed curves represent predicted $\epsilon$
  for zero-age main-sequence following figure 20 of \citet{pat03suumas};
  the upper (thin) curve represents the relation in \citet{pat03suumas} and
  the lower (thick) curve represents the relation based on the improved
  $\epsilon$-$q$ relation.
  }
  \label{fig:epsporb}
\end{figure}

   The improved relation between $\epsilon$ and $P_{\rm orb}$ is
shown in figure \ref{fig:epsporb}.  The predicted location of
Roche-lobe filling zero-age main sequence is also shown following
\citet{pat03suumas}.  Although the new calibration on the $\epsilon$-$q$
relation seems to slightly improve the deviation between observed and
predicted $\epsilon$, there still remains significant disagreement
between them.  The disagreement is the greatest where the period minimum 
appears to reside: $-1.27 < log(P_{\rm orb}) < -1.25$
($0.053 < P_{\rm orb} < 0.056$ d).

\section{Period Variation of Superhumps in WZ Sge-Type Dwarf Novae}\label{sec:wzsgestars}

\subsection{Late-Stage Superhumps in WZ Sge-Type Dwarf Novae: Case Studies}\label{sec:latestage}

   WZ Sge-type dwarf novae (see e.g. \cite{bai79wzsge}; \cite{dow90wxcet};
\cite{kat01hvvir}) are a subgroup of SU UMa-type dwarf novae
characterized by large-amplitude (typically $\sim$ 8 mag) superoutbursts
with very long (typically $\sim$ 10 yr) recurrence times.

   Some SU UMa-type dwarf novae with long recurrence times, most notably
WZ Sge-type dwarf novae, are known to
exhibit long-enduring superhumps during the late post-superoutburst stage.
We will examine selected special cases (though the discussion
may not be necessarily applicable to general cases) which provide
new insight into the relation of late-stage superhumps and other
periodicities.

\begin{figure*}
  \begin{center}
    \FigureFile(160mm,160mm){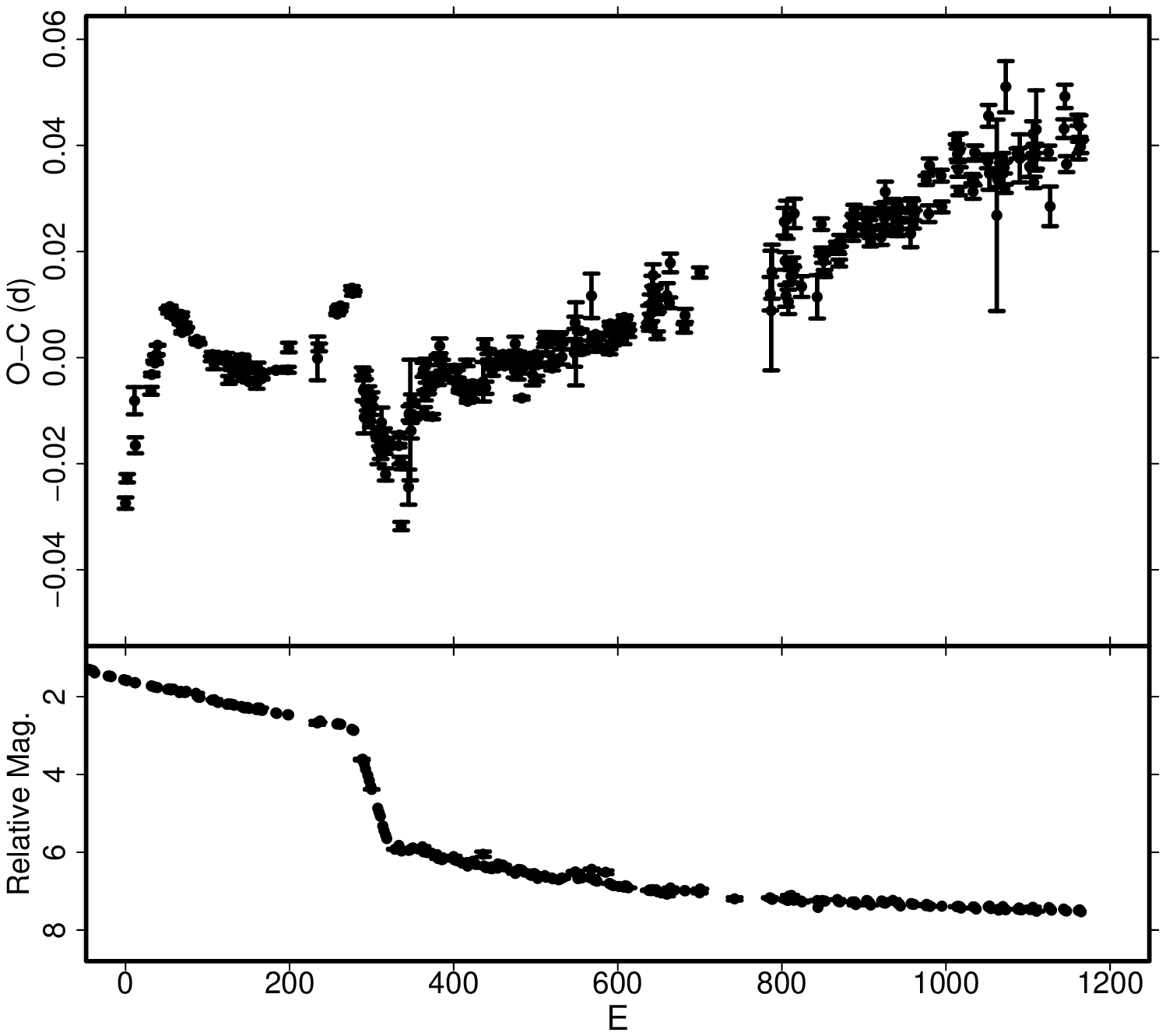}
  \end{center}
  \caption{$O-C$ variation in GW Lib (2007).  (Upper) $O-C$;
  (Lower) Light curve.  The early stage of the superoutburst, when
  ordinary superhump were not observed, is outside ($E < 0$) the figure.
  }
  \label{fig:gwlibhumpall}
\end{figure*}

   The first case is GW Lib in 2007.
During the late post-superoutburst stage, this object showed
very stable superhumps whose period is $\sim$0.5 \% longer than
those of the ordinary superhumps \citep{kat08wzsgelateSH}.
These superhumps during the late post-superoutburst stage appear to be
on a smooth extension of the $O-C$ diagram of the stage B
(figure \ref{fig:gwlibhumpall}).  This suggests
that these superhumps are intrinsically of the same origin, and
the transition to the stage C around the termination of the
superoutburst looks like a disturbance in the $O-C$ diagram.

   This temporary emergence of a new periodicity is in reality
attributed to orbital humps (cf. subsection \ref{sec:gwlib}).
Similar behavior was also recorded in well-observed WZ Sge-type systems
V455 And (subsection \ref{sec:v455and}) and WZ Sge
(subsection \ref{sec:wzsge}).  This phenomenon thus appears
common to many WZ Sge-type dwarf novae, but apparently not very striking
in usual SU UMa-type dwarf novae.  \citet{osa02wzsgehump} presented
an interpretation that the orbital humps observed in WZ Sge-type
superoutbursts can be well reproduced by a projection effect of the
superhump source, rather than by an enhanced hot spot.  Our observation
in SDSS J080434.20$+$510349.2 (hereafter SDSS J0804)
supports this interpretation \citep{kat09j0804}.
There appears to be a condition that this mechanism strongly works
during the late stage of WZ Sge-type superoutbursts.  There also remains
a possibility that a mechanism similar to early superhumps
works in this phase (see subsection \ref{sec:gwlib}).

   Following \citet{kat08wzsgelateSH},
late post-superoutburst superhumps are supposed to originate from
the precessing eccentric disk near the tidal truncation.
This leads to a picture that the eccentric disk continues to slowly
expand after the end of stage B, and finally reaches the tidal truncation
where the period stabilizes.  This picture is a natural extension
of the explanation of ``late superhumps'' in WZ Sge-type dwarf novae
proposed by \citet{kat08wzsgelateSH}.
During the plateau stage when the
disk is still bright enough, newly excited superhumps
(stage C superhumps) can temporarily dominate over the superhump
signal arising from the outer, relatively faint, disk, and
behaves as a temporarily disturbance until the entire disk returns
to the cool state.  This interpretation, however, needs to be
verified by more detailed study and by a comparison with numerical
simulations of superhumps incorporating the thermal instability.

\begin{figure*}
  \begin{center}
    \FigureFile(160mm,160mm){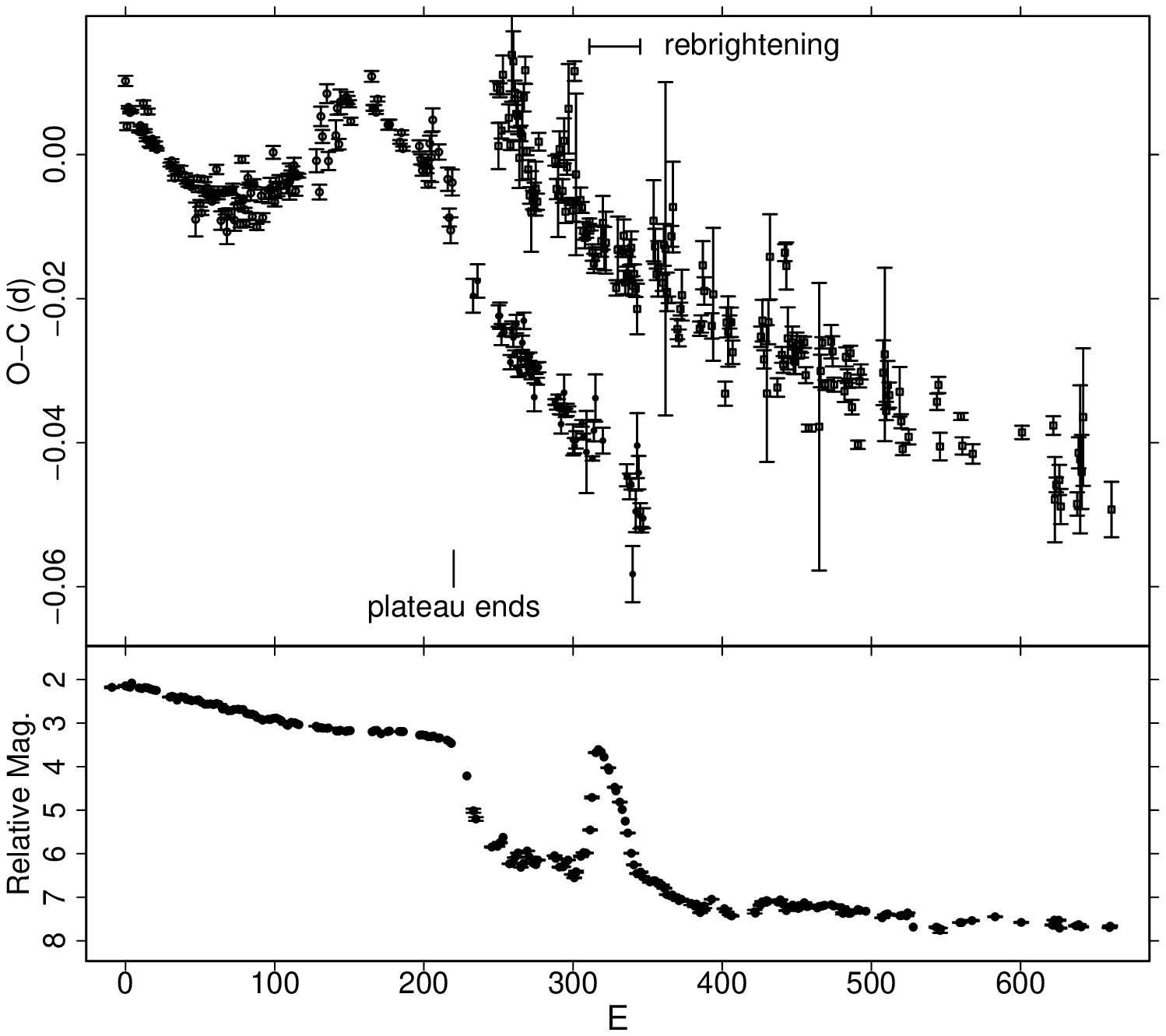}
  \end{center}
  \caption{$O-C$ variation in ASAS J0025 (2004).  (Upper) $O-C$.
  Different symbols refer to humps of different categories
  (see subsection \ref{sec:asas0025});
  (Lower) Light curve.  The earliest stage of the superoutburst
  was not observed.
  }
  \label{fig:asas0025humpall}
\end{figure*}

   The second case is ASAS J002511$+$1217.2
(figure \ref{fig:asas0025humpall}).  Following a typical
stage B--C evolution, the object showed double-humped superhumps
with a shorter period between the end of the superoutburst plateau
and the rebrightening.  Outside this stage, superhumps during the
late post-superoutburst stage are on a smooth extension of the stage C
superhumps (see subsection \ref{sec:asas0025} for details).
Although the situation looks somewhat different from GW Lib,
superhumps during the late post-superoutburst stage appears to
have evolved from the stage C superhumps.  The early post-superoutburst
stage and the rebrightening acted like a disturbance as in GW Lib,
although the emergence of orbital period was not yet confirmed in
this case.
It may be that $m=2$ waves were transiently excited in the inner disk,
and the phenomenological difference from GW Lib may be associated with
the presence of a rebrightening.  It would be worth noting that both
GW Lib and ASAS J002511$+$1217.2 did not show a $\sim$0.5 phase shift
during the late stages.

   We give a summary of late-stage superhumps in WZ Sge-type dwarf novae
in table \ref{tab:latehump}.  The values of late-stage superhumps
($P_{\rm late}$) listed in the table are representative periods.
Since $P_1$ here represents a mean period of stage B, not one
at its beginning, $P_{\rm late}$ can be shorter than $P_1$
in large $P_{\rm dot}$ systems (e.g. ASAS J0025), 
See subsections of individual objects for the details.

\begin{table*}
\caption{Late-stage superhumps in WZ Sge-type dwarf novae.}\label{tab:latehump}
\begin{center}
\begin{tabular}{ccccc}
\hline\hline
Object     & $P_{\rm orb}$ (d) & $P_1$ (d) & $P_{\rm late}$ (d) & source \\
\hline
V455 And   & 0.056309 & 0.057144(11) & 0.057188(6) & this work \\
EG Cnc     & 0.05997  & 0.060337(6)  & 0.06051(2)  & this work, \citet{pat98egcnc} \\
GW Lib     & 0.05332  & 0.054095(10) & 0.054156(1) & this work \\
WZ Sge     & 0.056688 & 0.057204(5)  & 0.057488(14) & this work \\
ASAS J0025 & 0.056540\commenta & 0.057093(12) & 0.056995(3) & this work \\
ASAS J1536 & --       & 0.064602(24) & 0.064729(13) & this work \\
SDSS J0804 & 0.059005 & 0.059539(11) & 0.059659(5) & \citet{kat09j0804} \\
OT J0747   & --       & 0.060750(7)  & 0.060771(3) & this work \\
\hline
  \multicolumn{4}{l}{\commenta candidate $P_{\rm orb}$.} \\
\end{tabular}
\end{center}
\end{table*}

\subsection{Period Variation in WZ Sge-Type Dwarf Novae}

   Although the borderline between WZ Sge-type dwarf novae and ordinary
SU UMa-type dwarf novae is somewhat ambiguous, it has been proposed
that a 2:1 orbital resonance in low-$q$ systems is responsible
for the phenomenon \citep{osa02wzsgehump}.  As already introduced in
\citet{kat08wzsgelateSH}, early superhumps (double-wave humps with
a period close to $P_{\rm orb}$ seen during the earliest stages of
WZ Sge-type superoutbursts; see also \cite{kat02wzsgeESH}) are
considered to be a manifestation of the 2:1 resonance
\citep{osa02wzsgehump}.  By the inferred mechanism, the existence of early
superhumps might a best feature in discriminating WZ Sge-type dwarf
novae from ordinary SU UMa-type dwarf novae (in low-inclination systems,
though, the amplitudes of early superhumps can be too low to detect;
e.g. GW Lib, Imada et al., in preparation).  In this paper, we deal
with objects with early superhumps or objects with very rare
(less than once in several years) and large-amplitude superoutbursts
as WZ Sge-type dwarf novae and analogs.

   \citet{kat08wzsgelateSH} also listed nearly constant to positive
$P_{\rm dot}$ as one of the common properties of WZ Sge-type dwarf novae.
We examine this further in this subsection.

   Table \ref{tab:wztab} summarizes properties of superoutbursts of
WZ Sge-type dwarf novae.  The quiescent magnitudes were mainly taken
from the on-line version of \citet{RitterCV7}, supplemented for
V1251 Cyg (Henden, AAVSO-discussion 14842), V592 Her, HV Vir
(SDSS $g$ values), GW Lib (typical quiescent magnitudes reported
to VSNET).  The maximum magnitudes were mean magnitudes around maximum
from reports to VSNET and other literature; $V$-band measurements are
preferentially used whenever available.  $P_{\rm SH}$ refers to $P_1$.

\begin{table*}
\caption{Parameters of WZ Sge-type superoutbursts.}\label{tab:wztab}
\begin{center}
\begin{tabular}{cccccccccccc}
\hline\hline
Object & Year & $P_{\rm SH}$ & $P_{\rm orb}$ & $P_{\rm dot}$\commenta & err\commenta & $\epsilon$ & Type\commentb & $N_{\rm reb}$\commentc & delay\commentd & Max & Min \\
\hline
LL And & 1993 & 0.056900 & 0.055055 & -- & -- & 0.034 & D? & 0? & -- & 14.3 & 20.0 \\
LL And & 2004 & 0.056583 & 0.055055 & 1.0 & 0.6 & 0.028 & D? & 0? & -- & 12.6 & 20.0 \\
V455 And & 2007 & 0.057133 & 0.056309 & 4.7 & 1.2 & 0.015 & D & 0 & 10 & 8.7 & 16.1 \\
V466 And & 2008 & 0.057203 & 0.056365 & 5.7 & 0.7 & 0.015 & D & 0 & $>$12 & 12.7 & 21.2 \\
UZ Boo & 1994 & 0.061743 & -- & $-$1.5 & 2.5 & -- & B & 2: & 1--5 & 11.7 & 20.4 \\
UZ Boo & 2003 & 0.061922 & -- & $-$1.9 & 6.3 & -- & B & 4 & 3: & 12.8 & 20.4 \\
CG CMa & 1999 & -- & 0.063275 & -- & -- & -- & A/B & 1/2 & -- & 13.7 & [22 \\
\end{tabular}
\end{center}
\end{table*}

   Figure \ref{fig:wzpdoteps} shows the relation between $P_{\rm dot}$
versus $\epsilon$ for WZ Sge-type dwarf novae.
For systems with $\epsilon < 0.026$, $P_{\rm dot}$
is a strong function of $\epsilon$ (equation \ref{equ:wzpdoteps}).
If $\epsilon$ indeed reflects $q$, the low $q$, rather than $P_{\rm orb}$,
is most responsible for smaller $P_{\rm dot}$.  Systems with nearly zero
$P_{\rm dot}$ appear to represent a population with low-mass secondaries.
Combined with figure \ref{fig:wzpdottype}, low-$P_{\rm dot}$ systems with
$P_{\rm SH} < 0.057$ d can be considered as a consequence of terminal
evolution of CVs around the period minimum.  Two long-$P_{\rm SH}$ objects
(OT J1112 and EG Cnc\footnote{
   The $P_{\rm orb}$ has been controversial \citep{kat04egcnc}.  The present
   analysis of $P_{\rm dot}$--$\epsilon$ relation seems to support
   the period identification of \citet{pat98egcnc}.  Accurate determination
   of the period of early superhumps, as well as independent estimates
   of $P_{\rm dot}$ in future superoutbursts is still wanted.
}) are either good candidates for ``period bouncers'',
or the period minimum is broader than had been considered and these
objects are presently reaching the period minimum at these
$P_{\rm orb}$.
We should note, however, this empirical calibration implicitly assume that
all superhumps in WZ Sge-type dwarf novae during the plateau stage B
superhumps.  If some systems show stage C superhump even in this phase,
$P_{\rm dot}$, and hence $q$ might be underestimated (see a discussion
in 1RXS J0232, subsection \ref{sec:j0232}).
Among our sample, 1RXS J0232 is a single candidate for a period
bouncer having a longer superhump period than 0.0603 d (EG Cnc).
The relative lack of promising candidates for period bouncers with
long superhump periods, despite the greatly improved statistics,
should be worth noting.

\begin{equation}
P_{\rm dot} = -0.00002(1) + 0.0040(6) \epsilon
\label{equ:wzpdoteps}.
\end{equation}

   Some object with WZ Sge-type characteristics (early superhumps and
large outburst amplitudes) are present in a range of $\epsilon > 0.026$
(BC UMa, V1251 Cyg, RZ Leo).  These objects do not follow the relation
in equation \ref{equ:wzpdoteps} and appear to have higher $q$.
These object may either consist ``borderline'' WZ Sge-type dwarf
novae (cf. \cite{pat03suumas}), or the existence of a large disk-mass
at the onset of superoutbursts may enable the 2:1 resonance to appear
in some high-$q$ systems.

\subsection{Period Variation versus Outburst Type}\label{sec:wzsgeouttype}

   WZ Sge-type dwarf novae are known to exhibit a wide variety of
outburst morphology, especially in post-outburst rebrightenings
(\cite{kat04egcnc}; \cite{ima06tss0222}).

\begin{figure}
  \begin{center}
    \FigureFile(88mm,160mm){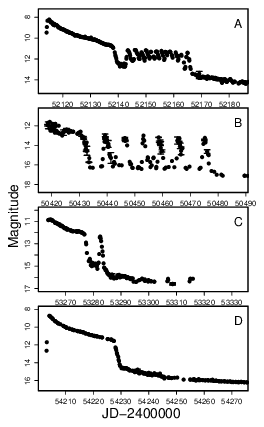}
  \end{center}
  \caption{Types of WZ Sge-type outbursts.  A: WZ Sge (2001),
  B: EG Cnc (1996, data from \cite{kat04egcnc}), C: ASAS J0025 (2004),
  D: GW Lib (2007).
  }
  \label{fig:outtype}
\end{figure}

\begin{figure}
  \begin{center}
    \FigureFile(88mm,70mm){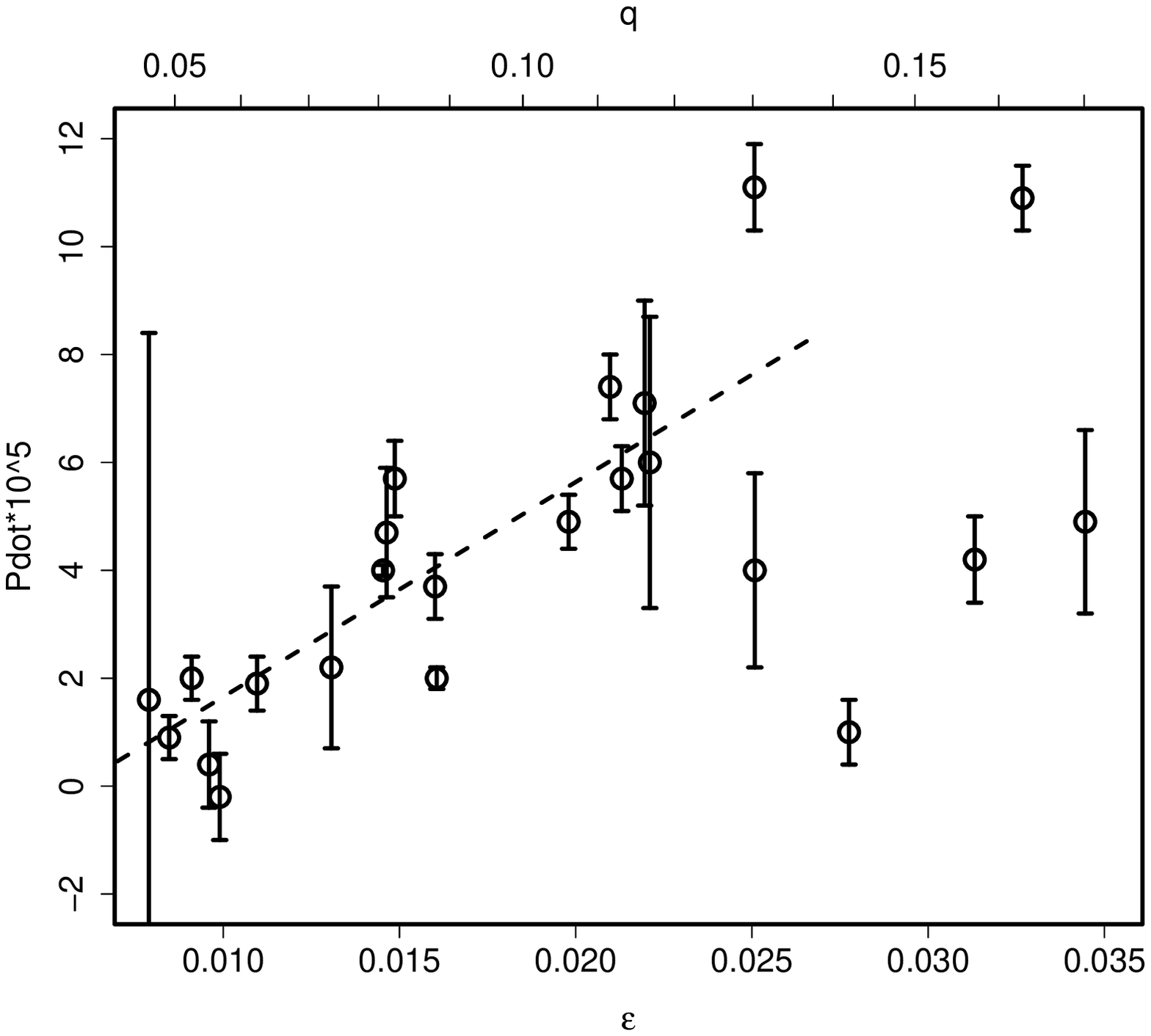}
  \end{center}
  \caption{$P_{\rm dot}$ versus $\epsilon$ for WZ Sge-type
  dwarf novae.  ASAS J0025 was excluded from this figure due to the
  uncertain $P_{\rm orb}$.
  The dashed line represents equation \ref{equ:wzpdoteps}.
  }
  \label{fig:wzpdoteps}
\end{figure}

\begin{figure}
  \begin{center}
    \FigureFile(88mm,70mm){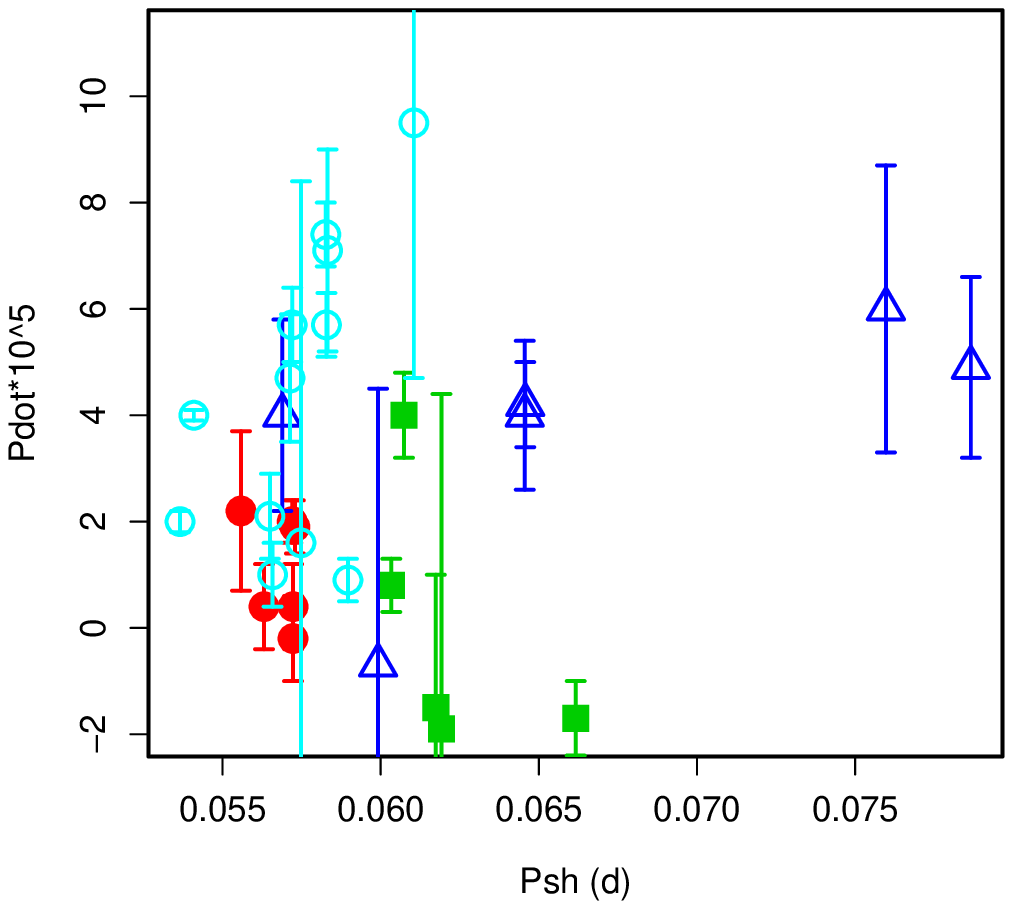}
  \end{center}
  \caption{$P_{\rm dot}$ versus $P_{\rm SH}$ for WZ Sge-type
  dwarf novae.  Symbols represent the type of outburst:
  type-A (filled circles), type-B (filled squares),
  type-C (open triangles), type-D (open circles).
  }
  \label{fig:wzpdottype}
\end{figure}

   Figure \ref{fig:wzpdottype} shows the relation between $P_{\rm dot}$
versus $P_{\rm SH}$ and outburst type, where the nomenclature of
classification is after \citet{ima06tss0222}\footnote{
   Although the original classification \citet{ima06tss0222} was for
   WZ Sge (SU UMa)-type dwarf novae, it should be worth noting that
   these types of rebrightenings sometimes appear in X-ray transients
   \citep{kuu96TOAD}.
} and type-D represents
outbursts without a rebrightening (figure \ref{fig:outtype}).
The $P_{\rm dot}$ tends to decrease with decreasing $P_{\rm SH}$.
There appear to be two populations among WZ Sge-type dwarf novae: systems
with $P_{\rm dot}$ nearly zero ($P_{\rm dot} < +2 \times 10^{-5}$)
and systems with larger $P_{\rm dot}$.

   Type-A outbursts (filled circles; long-duration rebrightening)
are restricted to a region with short $P_{\rm SH}$ and small $P_{\rm dot}$.
Type-B outbursts (filled squares; multiple rebrightenings)
tend to be located in a region with small $P_{\rm dot}$ but with larger
$P_{\rm SH}$ than type-A.
Type-C outbursts (open triangles; single rebrightening) are located
in a region with middle-to-longer $P_{\rm SH}$ and larger $P_{\rm dot}$.
Type-D outbursts (open circles; no rebrightening) tend to have a small
$P_{\rm SH}$ and a various $P_{\rm dot}$.
It should be noted that these classifications are not always the
property unique for each objects, but can be different between
superoutbursts of the same object \citep{uem08alcom}.

   The distinction between type-A and type-D outbursts in short $P_{\rm orb}$
systems may be understood in a scenario presented in
\citet{kat08wzsgelateSH}.  That is, in most extreme WZ Sge-type systems,
the 2:1 resonance can be strong enough to accrete much of the matter
beyond the 3:1 resonance and leave no room for a positive $P_{\rm dot}$.
In less extreme systems, the remnant matter beyond the 3:1 resonance
enables outward propagation of the eccentricity wave and resulting
a positive $P_{\rm dot}$.

   If $P_{\rm dot}$ indeed reflects $q$, the location of type-B
outbursts would indicate that these objects have small $q$, comparable
to those with type-A outbursts, but longer $P_{\rm orb}$.
In these type-B superoutbursts, the intervals between superoutbursts
tend to be shorter than in objects with type-A outbursts,
and the delay in appearance of ordinary superhumps is generally
shorter.  It may be that type-B outbursts are a variety of
type-A outbursts with a smaller disk mass at the onset of the
outbursts.  The presence of a type-B outburst in AL Com with
a possibly fainter maximum \citep{uem08alcom} seems to support
this interpretation.
The presence of low-amplitude outbursts during the 1978 and
2001 superoutbursts of WZ Sge (\cite{pat81wzsge}; \cite{pat02wzsge})
would be a signature of a smooth transition between type-A and type-B
outbursts (see also \cite{osa02wzsgehump}).
The relatively long $P_{\rm orb}$ in type-B objects might suggest
that the binary configuration in these systems is somehow responsible
for an early ignition of a superoutburst than in objects with
type-A outbursts.
Another potential interpretation is that objects with type-B outbursts
have a lower $q$ (cf. \cite{pat98egcnc}) than in other systems.
If this is the case, a smaller tidal torque in low-$q$ systems
might be insufficient to sustain a long-duration type-A rebrightening.
The apparent presence of a higher $\epsilon$ system (SDSS J0804:
\cite{kat09j0804} and \cite{zha08j0804})
among objects with type-B outbursts, however, would indicate that
not all type-B outbursts can be attributed to the low $q$.

   Type-C outbursts are less featured than other types of outbursts;
these outbursts resemble more usual superoutbursts with
a rebrightening frequently seen in a broader spectrum of
SU UMa-type dwarf novae.  A further explanation would be needed why
short-$P_{\rm orb}$ systems have little tendency to show type-C outbursts,
despite the apparent presence of sufficient matter beyond the 3:1
resonance.

\subsection{Delay of Appearance of Superhumps in WZ Sge-Type Dwarf Novae: Relation with Outburst Type}\label{sec:wzsgedelay}

   \citet{kat08wzsgelateSH} suggested that the long delays of appearance
of ordinary superhumps in WZ Sge-type superoutburst can be attributed
to the suppression of the 3:1 resonance by the 2:1 resonance, rather than
these delays reflect the long growth time of the 3:1 resonance in
low-$q$ systems.  The similarity of the duration of the stage A
($\sim$ 20 cycles), which can be considered as the growth time of
superhumps, between SU UMa-type dwarf novae and WZ Sge-type dwarf novae
would also support this interpretation.
We also surveyed these delay times in WZ Sge-type outbursts
for better statistics, and included them in table \ref{tab:wztab}.
In several cases, the delay times could not be well constrained due to
the gap in observations, or due to the apparent delay in detection
of the outburst.  In such cases, the possible ranges of the delay times
are given.  Since the development of superhumps usually takes $\sim$1 d,
the values have $\sim$ 1 d uncertainty even in well-observed systems,
and they may be different from values given in the different literature.

   It is noteworthy that all well-observed type-A and type-D superoutbursts
have longer (6--12 d, or even longer) delay times than in type-C superoutbursts
(typically $\sim$ 5 d).  Many of type-B superoutburst were, unfortunately,
not sampled very well, but they appear to have shorter (1--5 d) delay
times.  These results strengthen the similarity between type-A and type-D
superoutbursts (subsection \ref{sec:wzsgeouttype}).  Following
\citet{kat08wzsgelateSH}, the 2:1 resonance in these outbursts are
strong enough to accrete most of the matter beyond the 3:1 resonance,
and the the small $P_{\rm dot}$ and the lack of type-C rebrightening
may be a natural consequence.
Shorter delay times in type-C superoutbursts and the strongly positive
$P_{\rm dot}$ can be interpreted as a result of a smaller mass and
a smaller effect of the 2:1 resonance, leaving significant amount
of matter beyond the 3:1 resonance \citep{kat08wzsgelateSH}.

   Type-B superoutbursts appear to have intermediate delay times
between type-A/D and ordinary SU UMa-type superoutbursts (1--3 d).
This would suggest that the matter beyond the 3:1 resonance is smaller,
and the 2:1 resonance is weaker than in type-A/D superoutbursts.
The origin of type-B superoutbursts with low $P_{\rm dot}$'s can then
be understood as a consequence of small mass outside the 3:1 resonance
(although the 2:1 resonance still works, the small mass in the outer disk
does not allow sufficient outward propagation of the eccentricity wave),
rather than a consequence of extremely low-$q$ expected for period bouncers.
Further detailed observations of type-B superoutbursts and determination
of $P_{\rm orb}$ would discriminate these possibilities.

\subsection{Delay of Appearance of Superhumps: Comparison between Different Superoutbursts}

   \citet{kat08wzsgelateSH} also suggested superoutbursts with
a different extent are expected to show different delay times.
In the present survey, HV Vir appears to perfectly fit
this expectation.  A fainter superoutburst in 2002 led to a shorter
growth time compared to the 1992 one.  In WX Cet, the delay time ($\ge 4d$)
in the bright superoutburst in 1989 was longer than $\sim$ 2 d
in the 1998 superoutburst (\cite{kat01wxcet}; subsection \ref{sec:wxcet}).
Different superoutbursts of SW UMa (subsection \ref{sec:swuma})
also followed this tendency (see also \cite{soe09swuma}).
Ohshima et al. (in preparation) also suggested that the delay time in
V844 Her during the bright superoutburst in 2008 appears to be longer
than those in other superoutbursts of the same object
(see also subsection \ref{sec:v844her}).
In BC UMa (subsection \ref{sec:bcuma}), the duration of
the stage B was dependent on the extent of the superoutburst.

   In summary, the present survey generally strengthened the expectations
in \citet{kat08wzsgelateSH}.

\section{Individual Objects}\label{sec:individual}

\subsection{FO Andromedae}\label{obj:foand}

   We reanalyzed the data in \citet{kat95foand}.   The times of
superhump maxima are listed in table \ref{tab:foandoc1994}.
This observation covered the late stage of the superoutburst and
most likely caught the stage B--C transition.
The mean periods were 0.07455(5) d for $E \le 14$ (stage B)
and 0.07402(1) d for $13 \le E \le 27$ (stage C).

\begin{table}
\caption{Superhump maxima of FO And (1994).}\label{tab:foandoc1994}
\begin{center}
\begin{tabular}{ccccc}
\hline\hline
$E$ & max$^a$ & error & $O-C^b$ & $N^c$ \\
\hline
0 & 49578.1462 & 0.0008 & $-$0.0033 & 23 \\
13 & 49579.1158 & 0.0006 & 0.0022 & 49 \\
14 & 49579.1896 & 0.0007 & 0.0018 & 31 \\
26 & 49580.0778 & 0.0007 & 0.0001 & 34 \\
27 & 49580.1520 & 0.0006 & 0.0001 & 48 \\
68 & 49583.1917 & 0.0014 & $-$0.0009 & 42 \\
\hline
  \multicolumn{5}{l}{$^{a}$ BJD$-$2400000.} \\
  \multicolumn{5}{l}{$^{b}$ Against $max = 2449578.1495 + 0.074163 E$.} \\
  \multicolumn{5}{l}{$^{c}$ Number of points used to determine the maximum.} \\
\end{tabular}
\end{center}
\end{table}

\subsection{KV Andromedae}\label{obj:kvand}

   KV And was originally reported as a large-amplitude dwarf nova
\citep{kur77kvandkwand}.  \citet{kat94kvand} and \citet{kat95kvand}
reported the detection of superhumps, whose period suggested a more
usual dwarf nova rather than a short-period, WZ Sge-like object.

   We have analyzed two superoutbursts in 1994 (reanalysis of
\cite{kat95kvand}) and 2002.  The results are presented
in tables \ref{tab:kvandoc1994} and \ref{tab:kvandoc2002}.
During both outbursts, the superhump period likely decreased.
The global $P_{\rm dot}$'s were $-12.8(6.0) \times 10^{-5}$
and $-8.2(2.9) \times 10^{-5}$, respectively.  The period changes
can be also interpreted as a result of transition from stage B to C
(see table \ref{tab:perlist}).

\begin{table}
\caption{Superhump maxima of KV And (1994).}\label{tab:kvandoc1994}
\begin{center}
\begin{tabular}{ccccc}
\hline\hline
$E$ & max$^a$ & error & $O-C^b$ & $N^c$ \\
\hline
0 & 49576.2102 & 0.0077 & 0.0012 & 16 \\
1 & 49576.2723 & 0.0022 & $-$0.0110 & 9 \\
27 & 49578.2175 & 0.0011 & $-$0.0002 & 45 \\
28 & 49578.3010 & 0.0017 & 0.0089 & 21 \\
41 & 49579.2609 & 0.0005 & 0.0016 & 49 \\
55 & 49580.3074 & 0.0021 & 0.0065 & 48 \\
95 & 49583.2699 & 0.0017 & $-$0.0069 & 43 \\
\hline
  \multicolumn{5}{l}{$^{a}$ BJD$-$2400000.} \\
  \multicolumn{5}{l}{$^{b}$ Against $max = 2449576.2090 + 0.074398 E$.} \\
  \multicolumn{5}{l}{$^{c}$ Number of points used to determine the maximum.} \\
\end{tabular}
\end{center}
\end{table}

\begin{table}
\caption{Superhump maxima of KV And (2002).}\label{tab:kvandoc2002}
\begin{center}
\begin{tabular}{ccccc}
\hline\hline
$E$ & max$^a$ & error & $O-C^b$ & $N^c$ \\
\hline
0 & 52584.1913 & 0.0005 & $-$0.0030 & 337 \\
1 & 52584.2647 & 0.0005 & $-$0.0040 & 326 \\
2 & 52584.3449 & 0.0014 & 0.0019 & 126 \\
13 & 52585.1613 & 0.0007 & 0.0009 & 346 \\
14 & 52585.2402 & 0.0028 & 0.0055 & 254 \\
27 & 52586.1978 & 0.0033 & $-$0.0030 & 72 \\
40 & 52587.1639 & 0.0020 & $-$0.0029 & 186 \\
41 & 52587.2449 & 0.0041 & 0.0038 & 197 \\
42 & 52587.3158 & 0.0015 & 0.0004 & 58 \\
53 & 52588.1312 & 0.0032 & $-$0.0016 & 135 \\
54 & 52588.2100 & 0.0010 & 0.0028 & 145 \\
55 & 52588.2863 & 0.0016 & 0.0049 & 114 \\
67 & 52589.1712 & 0.0017 & $-$0.0019 & 228 \\
68 & 52589.2474 & 0.0018 & $-$0.0000 & 115 \\
69 & 52589.3196 & 0.0021 & $-$0.0022 & 115 \\
80 & 52590.1372 & 0.0019 & $-$0.0019 & 62 \\
81 & 52590.2109 & 0.0034 & $-$0.0025 & 93 \\
82 & 52590.2907 & 0.0062 & 0.0029 & 95 \\
\hline
  \multicolumn{5}{l}{$^{a}$ BJD$-$2400000.} \\
  \multicolumn{5}{l}{$^{b}$ Against $max = 2452584.1944 + 0.074310 E$.} \\
  \multicolumn{5}{l}{$^{c}$ Number of points used to determine the maximum.} \\
\end{tabular}
\end{center}
\end{table}

\subsection{LL Andromedae}\label{sec:lland}\label{obj:lland}

   LL And is an eruptive object discovered in 1979 \citep{wil79lland}.
Little had been known until its first-ever outburst since the discovery
in 1993, during which \citet{kat04lland} established the SU UMa-type
nature of this object, and reported a superhump period of 0.05697(3) d.
The superhump maxima
determined from these observations are listed in table
\ref{tab:llandoc1993}.  Excluding $E = 37$ with a large error and
a significant deviation in $O-C$, the overall $P_{\rm dot}$ was
$+19.7(17.3) \times 10^{-5}$.

\begin{figure}
  \begin{center}
    \FigureFile(88mm,110mm){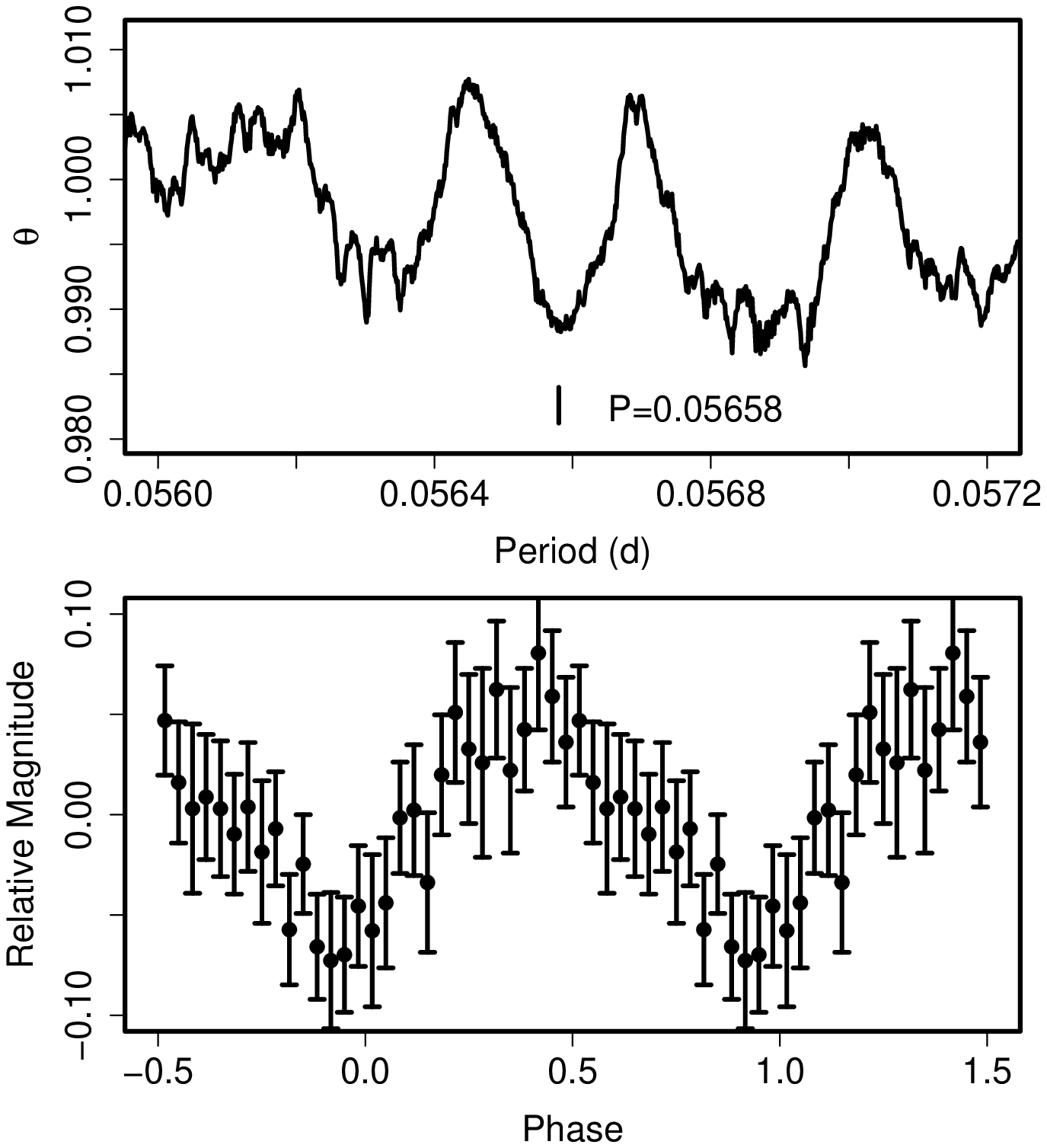}
  \end{center}
  \caption{Superhumps in LL And (2004). (Upper): PDM analysis.
     The selection of the period was based on the 1993 observation.
     (Lower): Phase-averaged profile.}
  \label{fig:llandshpdm}
\end{figure}

   The object underwent another superoutburst in 2004 May--June.
The object was very unfavorably situated for long time-series photometry.
The data were unavoidably taken at a large air-mass, $f(z)$.
We subtracted the first-order atmospheric extinction term, $c f(z)$,
where $c$ was numerically determined for each observer by minimizing
the deviation of the subtracted result from the general fading trend.
With the help of the superhump
period obtained in 1993, we selected the most likely mean superhump
period of 0.05658(2) d with PDM analysis (figure \ref{fig:llandshpdm}).
The times of superhump maxima
determined using this period are given in table \ref{tab:llandoc2004}.
The period and period derivative determined from $0 \leq E \leq 290$
were 0.05658(2) d and $P_{\rm dot}$ = $+1.0(0.6) \times 10^{-5}$,
respectively.  The resultant $\epsilon$ of 2.8 \% is still large for
this $P_{\rm orb}$ (see a discussion in \cite{kat04lland}).

\begin{table}
\caption{Superhump maxima of LL And (1993).}\label{tab:llandoc1993}
\begin{center}
\begin{tabular}{ccccc}
\hline\hline
$E$ & max$^a$ & error & $O-C^b$ & $N^c$ \\
\hline
0 & 49330.9100 & 0.0142 & 0.0028 & 22 \\
1 & 49330.9592 & 0.0058 & $-$0.0048 & 22 \\
36 & 49332.9513 & 0.0011 & $-$0.0043 & 22 \\
37 & 49333.0232 & 0.0056 & 0.0108 & 10 \\
53 & 49333.9219 & 0.0016 & $-$0.0009 & 21 \\
54 & 49333.9759 & 0.0018 & $-$0.0039 & 21 \\
55 & 49334.0375 & 0.0014 & 0.0008 & 20 \\
56 & 49334.0931 & 0.0019 & $-$0.0005 & 20 \\
\hline
  \multicolumn{5}{l}{$^{a}$ BJD$-$2400000.} \\
  \multicolumn{5}{l}{$^{b}$ Against $max = 2449330.9072 + 0.056900 E$.} \\
  \multicolumn{5}{l}{$^{c}$ Number of points used to determine the maximum.} \\
\end{tabular}
\end{center}
\end{table}

\begin{table}
\caption{Superhump maxima of LL And (2004).}\label{tab:llandoc2004}
\begin{center}
\begin{tabular}{ccccc}
\hline\hline
$E$ & max$^a$ & error & $O-C^b$ & $N^c$ \\
\hline
0 & 53152.8407 & 0.0049 & 0.0031 & 55 \\
84 & 53157.5827 & 0.0030 & $-$0.0045 & 51 \\
95 & 53158.2028 & 0.0020 & $-$0.0064 & 71 \\
96 & 53158.2636 & 0.0011 & $-$0.0022 & 115 \\
131 & 53160.2482 & 0.0018 & 0.0034 & 207 \\
149 & 53161.2683 & 0.0016 & 0.0058 & 80 \\
172 & 53162.5636 & 0.0069 & 0.0005 & 47 \\
290 & 53169.2458 & 0.0066 & 0.0107 & 84 \\
308 & 53170.2472 & 0.0036 & $-$0.0057 & 140 \\
325 & 53171.2115 & 0.0043 & $-$0.0026 & 103 \\
326 & 53171.2684 & 0.0134 & $-$0.0022 & 58 \\
\hline
  \multicolumn{5}{l}{$^{a}$ BJD$-$2400000.} \\
  \multicolumn{5}{l}{$^{b}$ Against $max = 2453152.8376 + 0.056543 E$.} \\
  \multicolumn{5}{l}{$^{c}$ Number of points used to determine the maximum.} \\
\end{tabular}
\end{center}
\end{table}

\subsection{V402 Andromedae}\label{obj:v402and}

\begin{figure}
  \begin{center}
    \FigureFile(88mm,70mm){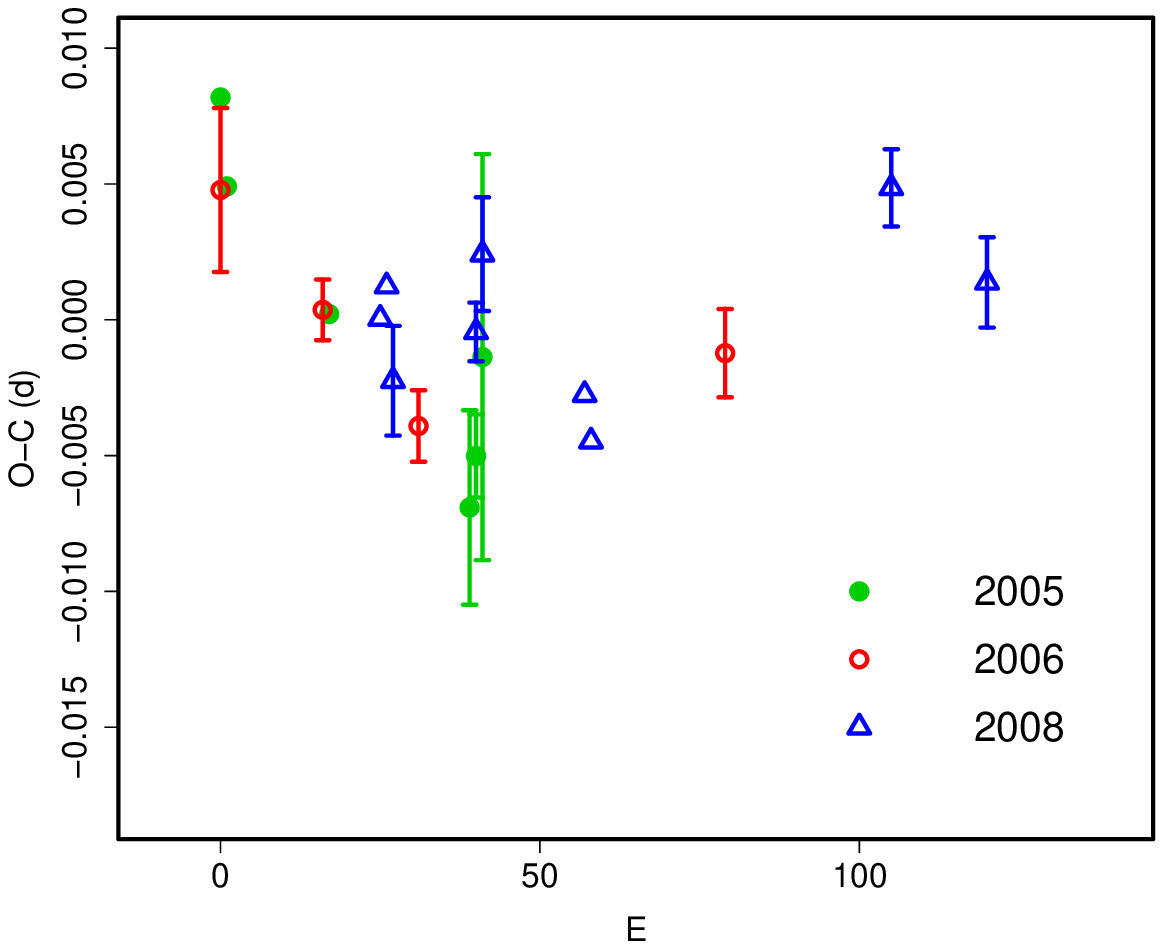}
  \end{center}
  \caption{Comparison of $O-C$ diagrams of V402 And between different
  superoutbursts.  A period of 0.06350 d was used to draw this figure.
  Approximate cycle counts ($E$) after the start of the
  superoutburst were used.
  }
  \label{fig:v402andcomp}
\end{figure}

   V402 And is a dwarf nova discovered by \citet{ant98v1008herv402andv369peg}.
The SU UMa-type nature was confirmed during the 2000 superoutburst
(vsnet-alert 5274).
We analyzed the 2005, 2006 and 2008 superoutbursts (tables
\ref{tab:v402andoc2005}, \ref{tab:v402andoc2006}, \ref{tab:v402andoc2008}).
The 2005 and 2006 superoutbursts were observed during their early stages
and the 2008 one was observed during its middle stage.
The resultant $P_{\rm dot}$ were $+12.7(2.1) \times 10^{-5}$ for the 2006
superoutburst and $+4.2(3.7) \times 10^{-5}$ for the 2008, respectively.
A shorter mean $P_{\rm SH}$ for the 2005 superoutburst during its early
stage is also consistent with the positive $P_{\rm dot}$.
A combined $O-C$ diagram is presented in figure \ref{fig:v402andcomp}.

\begin{table}
\caption{Superhump maxima of V402 And (2005).}\label{tab:v402andoc2005}
\begin{center}
\begin{tabular}{ccccc}
\hline\hline
$E$ & max$^a$ & error & $O-C^b$ & $N^c$ \\
\hline
0 & 53671.0724 & 0.0005 & 0.0020 & 107 \\
1 & 53671.1326 & 0.0006 & $-$0.0010 & 134 \\
17 & 53672.1439 & 0.0010 & $-$0.0014 & 117 \\
39 & 53673.5338 & 0.0036 & $-$0.0026 & 13 \\
40 & 53673.5992 & 0.0015 & $-$0.0004 & 22 \\
41 & 53673.6663 & 0.0075 & 0.0035 & 19 \\
\hline
  \multicolumn{5}{l}{$^{a}$ BJD$-$2400000.} \\
  \multicolumn{5}{l}{$^{b}$ Against $max = 2453671.0704 + 0.063230 E$.} \\
  \multicolumn{5}{l}{$^{c}$ Number of points used to determine the maximum.} \\
\end{tabular}
\end{center}
\end{table}

\begin{table}
\caption{Superhump maxima of V402 And (2006).}\label{tab:v402andoc2006}
\begin{center}
\begin{tabular}{ccccc}
\hline\hline
$E$ & max$^a$ & error & $O-C^b$ & $N^c$ \\
\hline
0 & 53952.2426 & 0.0030 & 0.0029 & 125 \\
16 & 53953.2541 & 0.0011 & $-$0.0006 & 130 \\
31 & 53954.2024 & 0.0013 & $-$0.0039 & 105 \\
79 & 53957.2531 & 0.0016 & 0.0017 & 135 \\
\hline
  \multicolumn{5}{l}{$^{a}$ BJD$-$2400000.} \\
  \multicolumn{5}{l}{$^{b}$ Against $max = 2453952.2397 + 0.063439 E$.} \\
  \multicolumn{5}{l}{$^{c}$ Number of points used to determine the maximum.} \\
\end{tabular}
\end{center}
\end{table}

\begin{table}
\caption{Superhump maxima of V402 And (2008).}\label{tab:v402andoc2008}
\begin{center}
\begin{tabular}{ccccc}
\hline\hline
$E$ & max$^a$ & error & $O-C^b$ & $N^c$ \\
\hline
0 & 54755.0811 & 0.0007 & 0.0010 & 103 \\
1 & 54755.1458 & 0.0006 & 0.0022 & 136 \\
2 & 54755.2058 & 0.0020 & $-$0.0013 & 72 \\
15 & 54756.0331 & 0.0011 & 0.0000 & 98 \\
16 & 54756.0995 & 0.0021 & 0.0029 & 54 \\
32 & 54757.1103 & 0.0006 & $-$0.0028 & 113 \\
33 & 54757.1721 & 0.0009 & $-$0.0046 & 128 \\
80 & 54760.1659 & 0.0014 & 0.0033 & 135 \\
95 & 54761.1149 & 0.0017 & $-$0.0007 & 100 \\
\hline
  \multicolumn{5}{l}{$^{a}$ BJD$-$2400000.} \\
  \multicolumn{5}{l}{$^{b}$ Against $max = 2454755.0801 + 0.063532 E$.} \\
  \multicolumn{5}{l}{$^{c}$ Number of points used to determine the maximum.} \\
\end{tabular}
\end{center}
\end{table}

\subsection{V455 Andromedae}\label{sec:v455and}\label{obj:v455and}

\begin{figure}
  \begin{center}
    \FigureFile(88mm,110mm){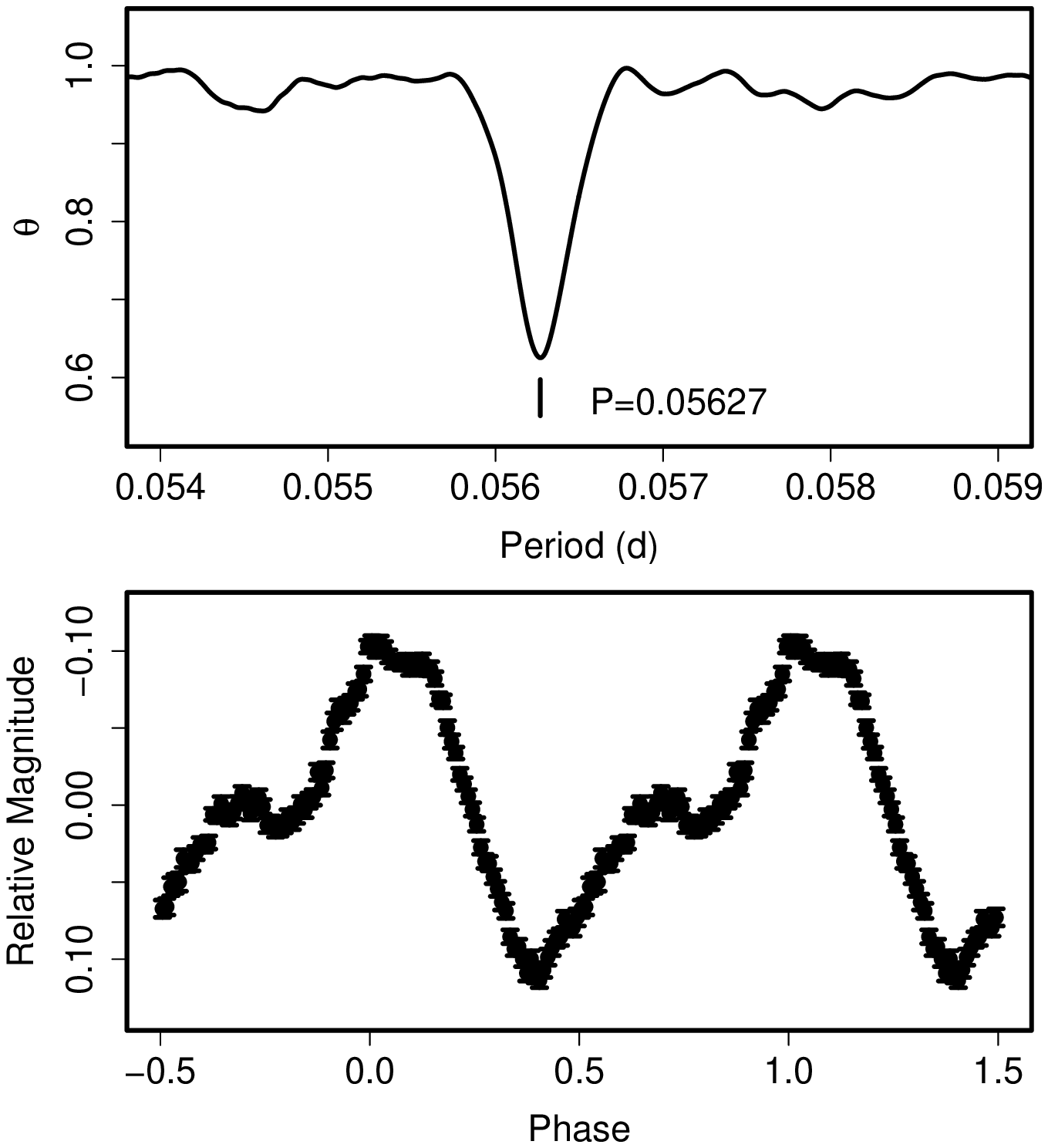}
  \end{center}
  \caption{Early superhumps in V455 And (2007) for BJD 2454349--2454356.
     (Upper): PDM analysis.
     (Lower): Phase-averaged profile.}
  \label{fig:v455andeshpdm}
\end{figure}

\begin{figure}
  \begin{center}
    \FigureFile(88mm,110mm){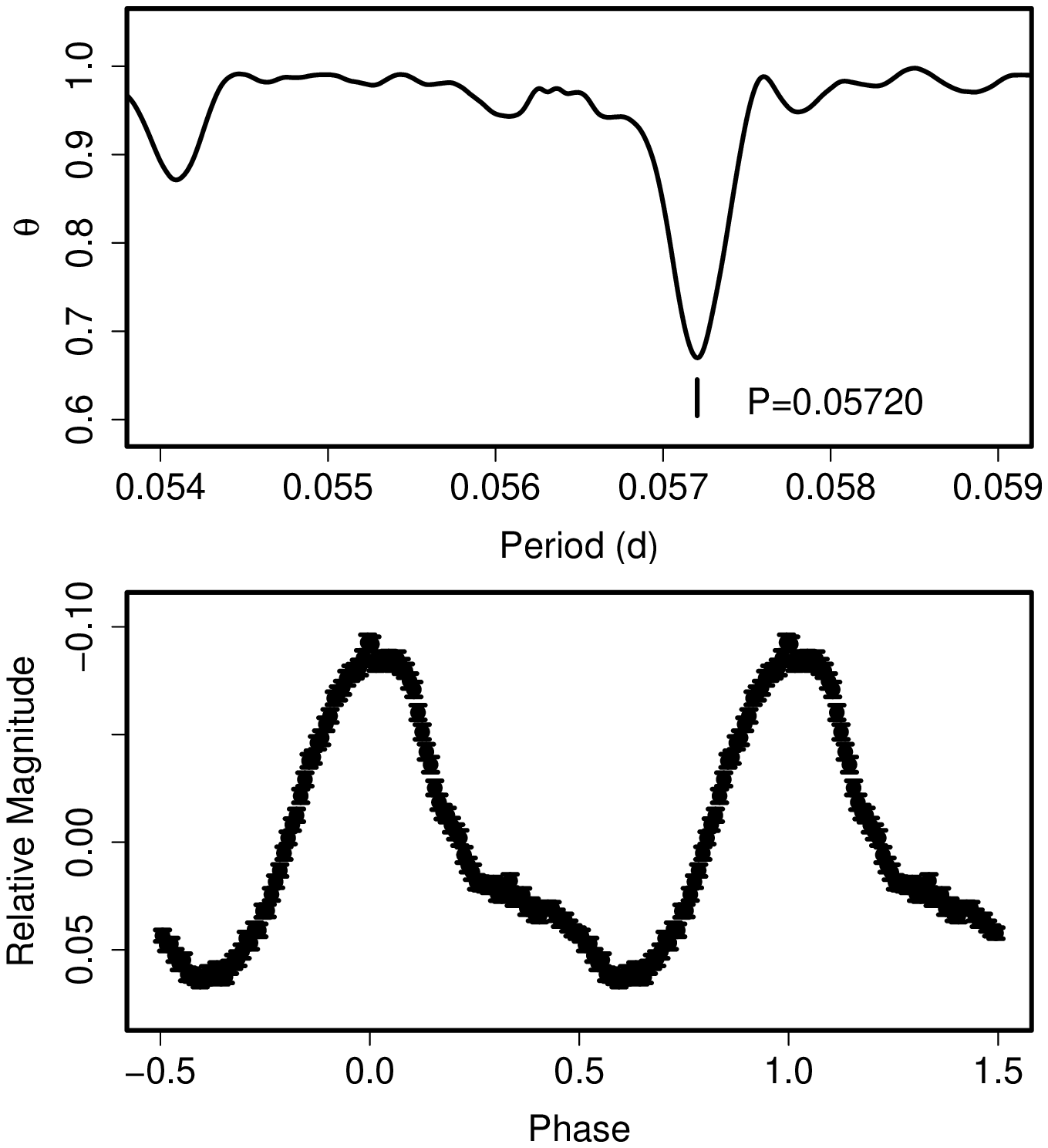}
  \end{center}
  \caption{Ordinary superhumps in V455 And (2007) for BJD 2454357.3--2454366.
     (Upper): PDM analysis.
     (Lower): Phase-averaged profile.}
  \label{fig:v455andshpdm}
\end{figure}

   V455 And = HS 2331$+$3905 \citep{ara05v455and} underwent
a spectacular superoutburst, the first time in its history, in 2007
(H. Maehara, vsnet-alert 9530; \cite{tem07v455andcbet1053}).
Following a rapidly rising stage, the object developed early superhumps
(vsnet-alert 9543) similar to those in WZ Sge.  
After about eleven days, ordinary superhumps appeared
(vsnet-alert 9582, 9584).  Representative mean periods of early and
ordinary superhumps were 0.0562675(18) d (figure \ref{fig:v455andeshpdm})
and 0.0572038(14) d (figure \ref{fig:v455andshpdm}), respectively.

   The maxima times of ordinary superhumps (tables \ref{tab:v455andoc2007})
were obtained after subtracting phase-averaged orbital variations
(mean orbital variations were determined from averages for 3--5 d during the
main outburst and fading stage, 10 d for the post-superoutburst stage).
During BJD 2454356--2454357.3, sporadic humps having a period close to
superhumps were observed in addition to early superhumps.
No apparent superhump signal was detected before this epoch.
For the interval $E \le 20$, clear stage A evolution was observed
with a mean period of 0.05803(8) d (disregarding $E=3$ and $E=11$).
We determined $P_{\rm dot}$ of $+4.7(1.2) \times 10^{-5}$ from
maxima of $23 \le E \le 128$, after which the phases of maxima
coincide with orbital humps and were disregarded (see a discussion in
WZ Sge, subsection \ref{sec:wzsge}).

\begin{figure}
  \begin{center}
    \FigureFile(88mm,70mm){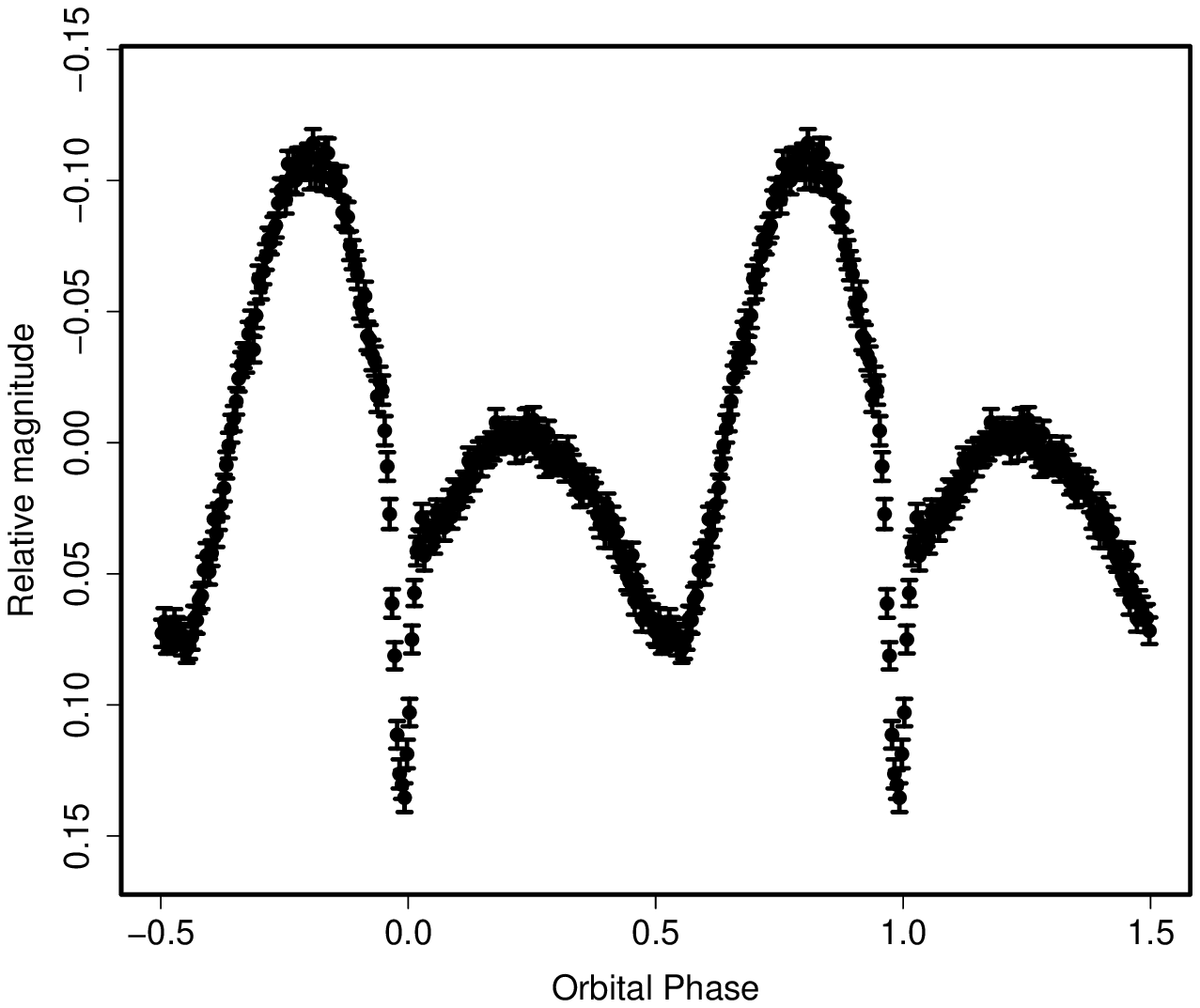}
  \end{center}
  \caption{Averaged orbital light curve of V455 And during the
  post-superoutburst stage.}
  \label{fig:v455postorb}
\end{figure}

\begin{figure}
  \begin{center}
    \FigureFile(88mm,180mm){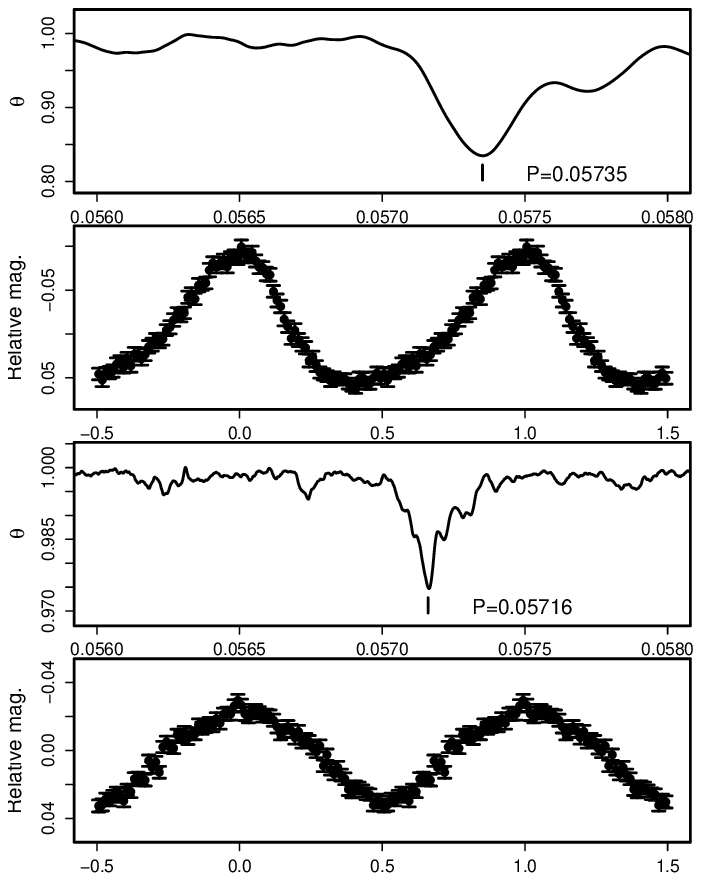}
  \end{center}
  \caption{Period analysis of V455 And (2007) during the post-superoutburst
  stage.  Upper two figures represent the PDM analysis and mean superhump
  profile (after subtracting the orbital variation) before BJD 2454377,
  the epoch of the period change.  Lower two figures represent the
  PDM analysis and mean superhump profile after BJD 2454377.}
  \label{fig:v455postpdm}
\end{figure}

   In contrast to WZ Sge, the orbital variations were so
strong (figure \ref{fig:v455postorb}) that it was practically impossible
to directly extract the times of superhump maxima from the light curve
during the post-superoutburst stage.
We therefore measured the times of superhump maxima during the
this stage after subtracting the orbital light curve
(table \ref{tab:v455andoc2007post}).  A relatively large scatter in
the $O-C$'s was probably a result from the interfering orbital variation.
There was an apparent change in the period around $E=170$.
The mean superhump periods (disregarding maxima coinciding orbital
humps and discrepant ones deviating by more than 0.018 d from the
mean trend) before and after the change were 0.057295(2) d
and 0.057154(1) d, respectively.  These periods were longer than the
$P_{\rm SH}$ during the main superoutburst (cf. \cite{kat08wzsgelateSH}).
Figure \ref{fig:v455postpdm} shows period analysis and mean superhump
profiles during the post-superoutburst stage.

\begin{figure*}
  \begin{center}
    \FigureFile(160mm,160mm){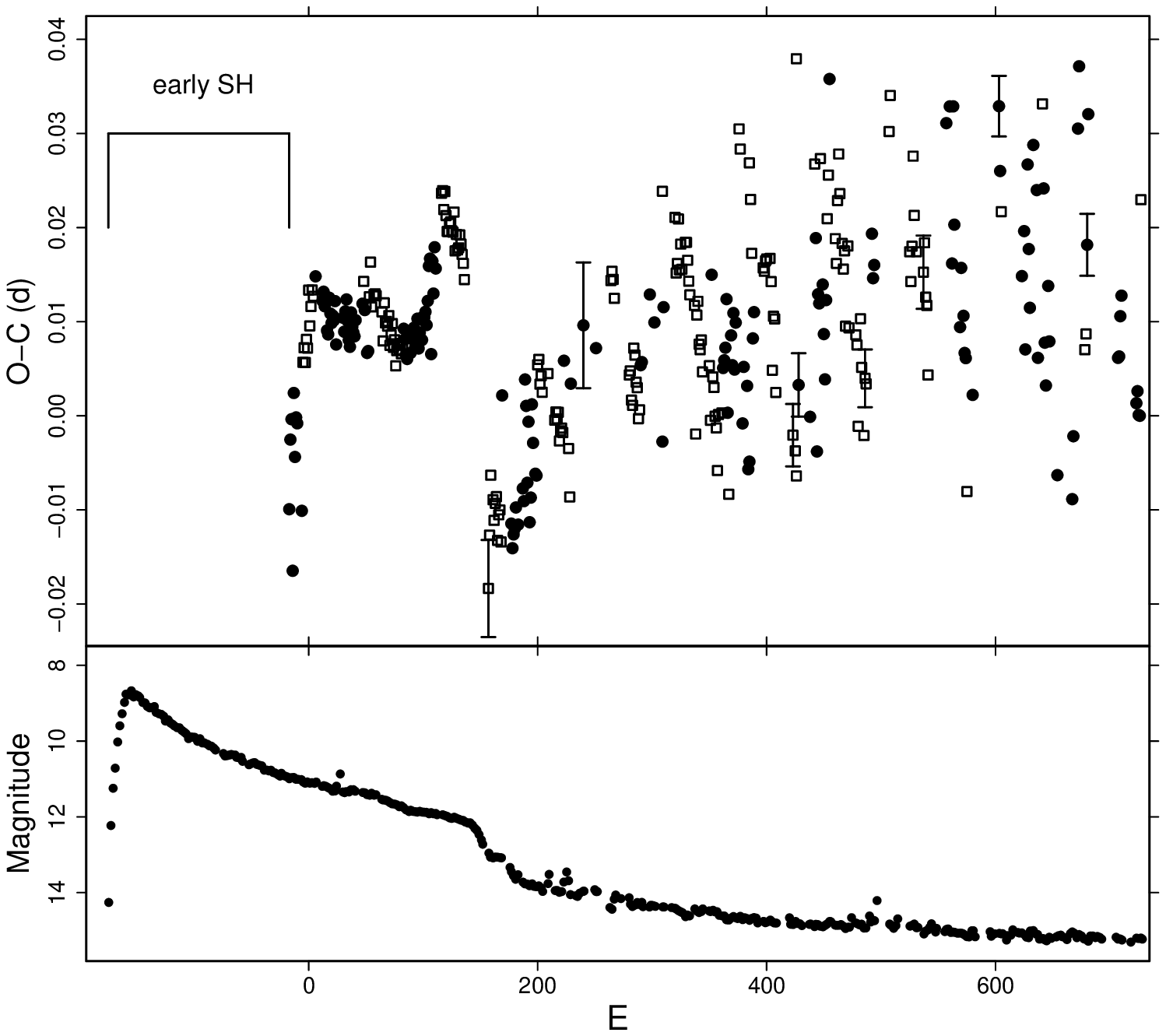}
  \end{center}
  \caption{$O-C$ variation in V455 And (2007).  (Upper) $O-C$.
  Open squares indicate humps coinciding with the phase of orbital humps.
  Filled squares are humps outside the phase of orbital humps.
  We used a period of 0.05714 d for calculating the $O-C$'s.
  The global evolution of the $O-C$ diagram is remarkably similar
  to that of GW Lib (figure \ref{fig:gwlibhumpall}).
  (Lower) Light curve.
  }
  \label{fig:v455humpall}
\end{figure*}

   The overall evolution of $O-C$'s was remarkably similar to that of GW Lib
(figure \ref{fig:v455humpall}; only the first half of the post-superoutburst
stage is shown for better visibility of the general feature).

   A full analysis of the observation will be presented in Maehara et al.,
in preparation.

\begin{table}
\caption{Superhump maxima of V455 And (2007).}\label{tab:v455andoc2007}
\begin{center}
\begin{tabular}{cccccc}
\hline\hline
$E$ & max$^a$ & error & $O-C^b$ & phase$^c$ & $N^d$ \\
\hline
0 & 54357.3037 & 0.0005 & $-$0.0134 & 0.23 & 466 \\
1 & 54357.3682 & 0.0002 & $-$0.0061 & 0.37 & 703 \\
2 & 54357.4275 & 0.0002 & $-$0.0041 & 0.43 & 832 \\
3 & 54357.4686 & 0.0003 & $-$0.0203 & 0.16 & 829 \\
4 & 54357.5446 & 0.0002 & $-$0.0014 & 0.51 & 659 \\
5 & 54357.5949 & 0.0002 & $-$0.0083 & 0.40 & 836 \\
6 & 54357.6563 & 0.0001 & $-$0.0042 & 0.49 & 859 \\
\end{tabular}
\end{center}
\end{table}

\addtocounter{table}{-1}
\begin{table}
\caption{Superhump maxima of V455 And (2007) (continued).}
\begin{center}
\begin{tabular}{cccccc}
\hline\hline
$E$ & max & error & $O-C$ & phase & $N$ \\
\hline
70 & 54361.3261 & 0.0003 & 0.0030 & 0.66 & 542 \\
71 & 54361.3869 & 0.0004 & 0.0065 & 0.74 & 500 \\
72 & 54361.4393 & 0.0003 & 0.0017 & 0.67 & 408 \\
74 & 54361.5549 & 0.0004 & 0.0028 & 0.72 & 288 \\
75 & 54361.6121 & 0.0004 & 0.0028 & 0.74 & 440 \\
76 & 54361.6690 & 0.0006 & 0.0025 & 0.75 & 236 \\
81 & 54361.9530 & 0.0003 & 0.0004 & 0.80 & 534 \\
\hline
\end{tabular}
\end{center}
\end{table}

\addtocounter{table}{-1}
\begin{table}
\caption{Superhump maxima of V455 And (2007) (continued).}
\begin{center}
\begin{tabular}{cccccc}
\hline\hline
$E$ & max & error & $O-C$ & phase & $N$ \\
\hline
134 & 54364.9943 & 0.0003 & 0.0086 & 0.81 & 173 \\
135 & 54365.0494 & 0.0002 & 0.0065 & 0.79 & 314 \\
136 & 54365.1085 & 0.0003 & 0.0083 & 0.83 & 275 \\
137 & 54365.1631 & 0.0003 & 0.0057 & 0.80 & 347 \\
138 & 54365.2185 & 0.0003 & 0.0039 & 0.79 & 432 \\
139 & 54365.2757 & 0.0006 & 0.0038 & 0.80 & 325 \\
140 & 54365.3337 & 0.0003 & 0.0046 & 0.83 & 447 \\
141 & 54365.3910 & 0.0003 & 0.0046 & 0.85 & 237 \\
142 & 54365.4472 & 0.0003 & 0.0036 & 0.85 & 207 \\
143 & 54365.5042 & 0.0005 & 0.0034 & 0.86 & 235 \\
144 & 54365.5634 & 0.0007 & 0.0054 & 0.91 & 220 \\
145 & 54365.6164 & 0.0003 & 0.0012 & 0.85 & 316 \\
146 & 54365.6753 & 0.0004 & 0.0029 & 0.90 & 159 \\
147 & 54365.7308 & 0.0003 & 0.0011 & 0.89 & 160 \\
148 & 54365.7882 & 0.0003 & 0.0013 & 0.90 & 119 \\
149 & 54365.8467 & 0.0005 & 0.0025 & 0.94 & 127 \\
150 & 54365.9029 & 0.0007 & 0.0015 & 0.94 & 124 \\
151 & 54365.9589 & 0.0002 & 0.0003 & 0.94 & 748 \\
152 & 54366.0151 & 0.0013 & $-$0.0007 & 0.93 & 480 \\
153 & 54366.0705 & 0.0004 & $-$0.0025 & 0.92 & 529 \\
\hline
\end{tabular}
\end{center}
\end{table}

\begin{table}
\caption{Superhump maxima of V455 And during the Post-Superoutburst Stage (2007).}\label{tab:v455andoc2007post}
\begin{center}
\begin{tabular}{cccccc}
\hline\hline
$E$ & max$^a$ & error & $O-C^b$ & phase$^c$ & $N^d$ \\
\hline
0 & 54367.2376 & 0.0052 & $-$0.0213 & 0.65 & 88 \\
1 & 54367.3004 & 0.0010 & $-$0.0157 & 0.76 & 240 \\
2 & 54367.3639 & 0.0006 & $-$0.0093 & 0.89 & 168 \\
4 & 54367.4756 & 0.0004 & $-$0.0119 & 0.87 & 85 \\
5 & 54367.5306 & 0.0005 & $-$0.0141 & 0.85 & 179 \\
6 & 54367.5895 & 0.0004 & $-$0.0123 & 0.89 & 323 \\
7 & 54367.6474 & 0.0004 & $-$0.0116 & 0.92 & 214 \\
\end{tabular}
\end{center}
\end{table}

\addtocounter{table}{-1}
\begin{table}
\caption{Superhump maxima of V455 And during the Post-Superoutburst Stage (2007) (continued).}
\begin{center}
\begin{tabular}{cccccc}
\hline\hline
$E$ & max & error & $O-C$ & phase & $N$ \\
\hline
108 & 54373.4425 & 0.0009 & 0.0114 & 0.84 & 28 \\
109 & 54373.4987 & 0.0016 & 0.0105 & 0.84 & 74 \\
110 & 54373.5539 & 0.0006 & 0.0085 & 0.82 & 99 \\
123 & 54374.2885 & 0.0011 & 0.0002 & 0.86 & 58 \\
124 & 54374.3461 & 0.0009 & 0.0007 & 0.89 & 83 \\
125 & 54374.4001 & 0.0006 & $-$0.0025 & 0.85 & 86 \\
126 & 54374.4567 & 0.0008 & $-$0.0030 & 0.85 & 77 \\
\hline
\end{tabular}
\end{center}
\end{table}

\addtocounter{table}{-1}
\begin{table}
\caption{Superhump maxima of V455 And during the Post-Superoutburst Stage (2007) (continued).}
\begin{center}
\begin{tabular}{cccccc}
\hline\hline
$E$ & max & error & $O-C$ & phase & $N$ \\
\hline
209 & 54379.1986 & 0.0005 & $-$0.0046 & 0.06 & 435 \\
210 & 54379.2470 & 0.0010 & $-$0.0133 & 0.92 & 364 \\
212 & 54379.3782 & 0.0005 & 0.0036 & 0.25 & 279 \\
213 & 54379.4322 & 0.0006 & 0.0004 & 0.21 & 278 \\
214 & 54379.4949 & 0.0012 & 0.0060 & 0.32 & 163 \\
215 & 54379.5460 & 0.0004 & $-$0.0000 & 0.23 & 399 \\
216 & 54379.6081 & 0.0008 & 0.0049 & 0.33 & 529 \\
\hline
\end{tabular}
\end{center}
\end{table}

\addtocounter{table}{-1}
\begin{table}
\caption{Superhump maxima of V455 And during the Post-Superoutburst Stage (2007) (continued).}
\begin{center}
\begin{tabular}{cccccc}
\hline\hline
$E$ & max & error & $O-C$ & phase & $N$ \\
\hline
309 & 54384.9306 & 0.0005 & 0.0125 & 0.86 & 160 \\
310 & 54384.9850 & 0.0004 & 0.0098 & 0.82 & 355 \\
311 & 54385.0440 & 0.0002 & 0.0117 & 0.87 & 377 \\
312 & 54385.0932 & 0.0004 & 0.0037 & 0.74 & 326 \\
314 & 54385.2160 & 0.0004 & 0.0122 & 0.92 & 376 \\
315 & 54385.2644 & 0.0005 & 0.0035 & 0.79 & 223 \\
321 & 54385.6065 & 0.0009 & 0.0027 & 0.86 & 67 \\
\hline
\end{tabular}
\end{center}
\end{table}

\addtocounter{table}{-1}
\begin{table}
\caption{Superhump maxima of V455 And during the Post-Superoutburst Stage (2007) (continued).}
\begin{center}
\begin{tabular}{cccccc}
\hline\hline
$E$ & max & error & $O-C$ & phase & $N$ \\
\hline
485 & 54394.9930 & 0.0006 & 0.0167 & 0.56 & 232 \\
486 & 54395.0338 & 0.0007 & 0.0003 & 0.28 & 197 \\
487 & 54395.0864 & 0.0016 & $-$0.0042 & 0.21 & 165 \\
489 & 54395.2112 & 0.0018 & 0.0063 & 0.43 & 51 \\
490 & 54395.2625 & 0.0023 & 0.0004 & 0.34 & 79 \\
497 & 54395.6482 & 0.0010 & $-$0.0139 & 0.19 & 68 \\
509 & 54396.3290 & 0.0011 & $-$0.0189 & 0.28 & 122 \\
\hline
\end{tabular}
\end{center}
\end{table}

\addtocounter{table}{-1}
\begin{table}
\caption{Superhump maxima of V455 And during the Post-Superoutburst Stage (2007) (continued).}
\begin{center}
\begin{tabular}{cccccc}
\hline\hline
$E$ & max & error & $O-C$ & phase & $N$ \\
\hline
715 & 54408.0882 & 0.0006 & $-$0.0325 & 0.11 & 91 \\
715 & 54408.1401 & 0.0015 & 0.0194 & 0.04 & 92 \\
725 & 54408.7080 & 0.0011 & 0.0159 & 0.12 & 77 \\
730 & 54408.9503 & 0.0025 & $-$0.0276 & 0.42 & 79 \\
742 & 54409.6787 & 0.0005 & 0.0150 & 0.36 & 77 \\
744 & 54409.7458 & 0.0004 & $-$0.0321 & 0.55 & 76 \\
744 & 54409.7826 & 0.0016 & 0.0047 & 0.21 & 74 \\
\hline
\end{tabular}
\end{center}
\end{table}

\addtocounter{table}{-1}
\begin{table}
\caption{Superhump maxima of V455 And during the Post-Superoutburst Stage (2007) (continued).}
\begin{center}
\begin{tabular}{cccccc}
\hline\hline
$E$ & max & error & $O-C$ & phase & $N$ \\
\hline
1161 & 54433.6227 & 0.0011 & 0.0136 & 0.58 & 78 \\
1201 & 54435.8936 & 0.0061 & $-$0.0016 & 0.91 & 62 \\
1202 & 54435.9601 & 0.0010 & 0.0078 & 0.09 & 81 \\
1204 & 54436.0419 & 0.0009 & $-$0.0246 & 0.55 & 165 \\
1204 & 54436.0933 & 0.0027 & 0.0267 & 0.46 & 111 \\
1239 & 54438.0607 & 0.0026 & $-$0.0061 & 0.40 & 52 \\
1272 & 54439.9187 & 0.0010 & $-$0.0340 & 0.40 & 82 \\
\hline
\end{tabular}
\end{center}
\end{table}

\subsection{V466 Andromedae}\label{obj:v466and}

\begin{figure}
  \begin{center}
    \FigureFile(88mm,110mm){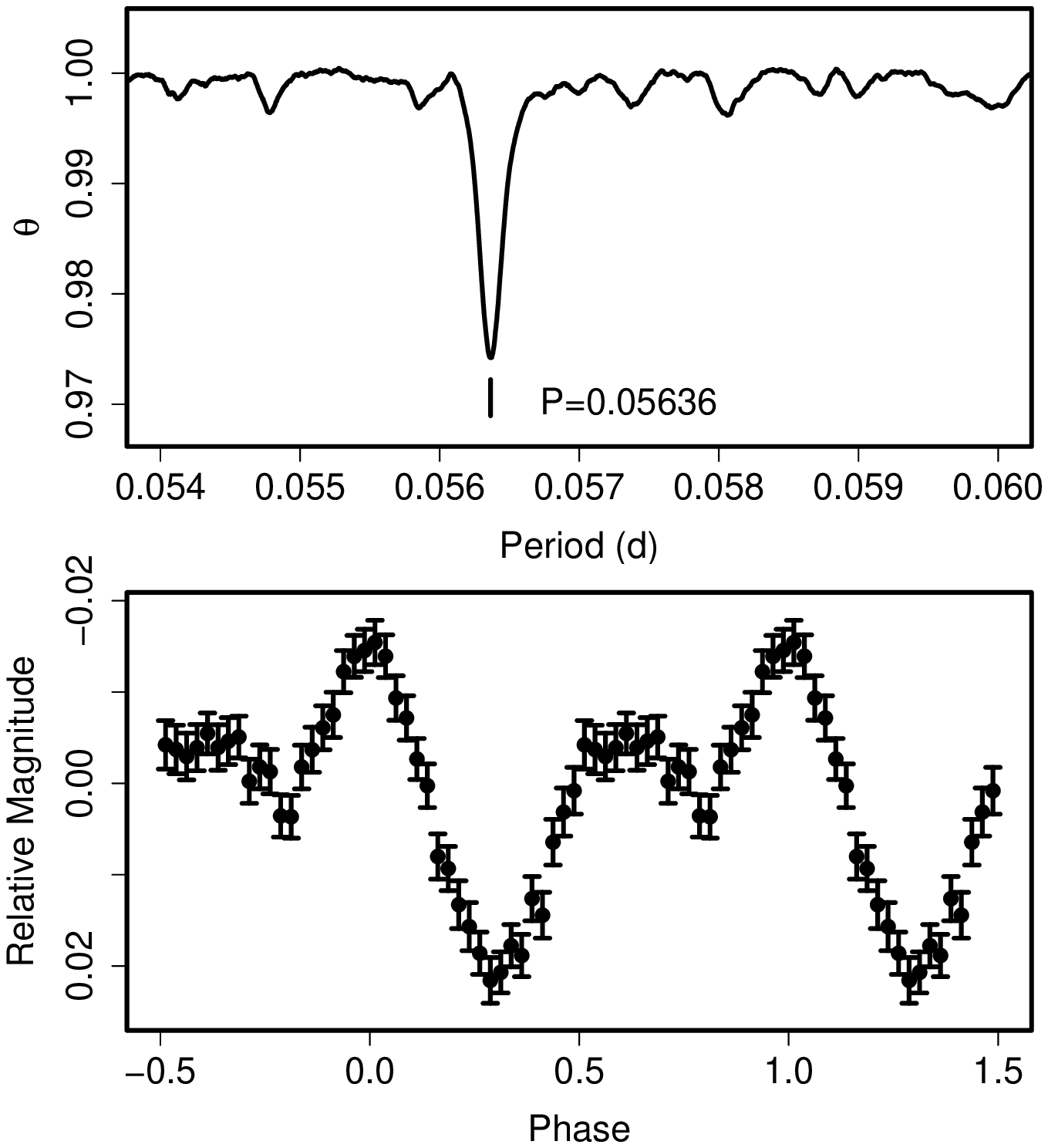}
  \end{center}
  \caption{Early superhumps in V466 And (2008). (Upper): PDM analysis.
     (Lower): Phase-averaged profile.}
  \label{fig:v466eshpdm}
\end{figure}

\begin{figure}
  \begin{center}
    \FigureFile(88mm,110mm){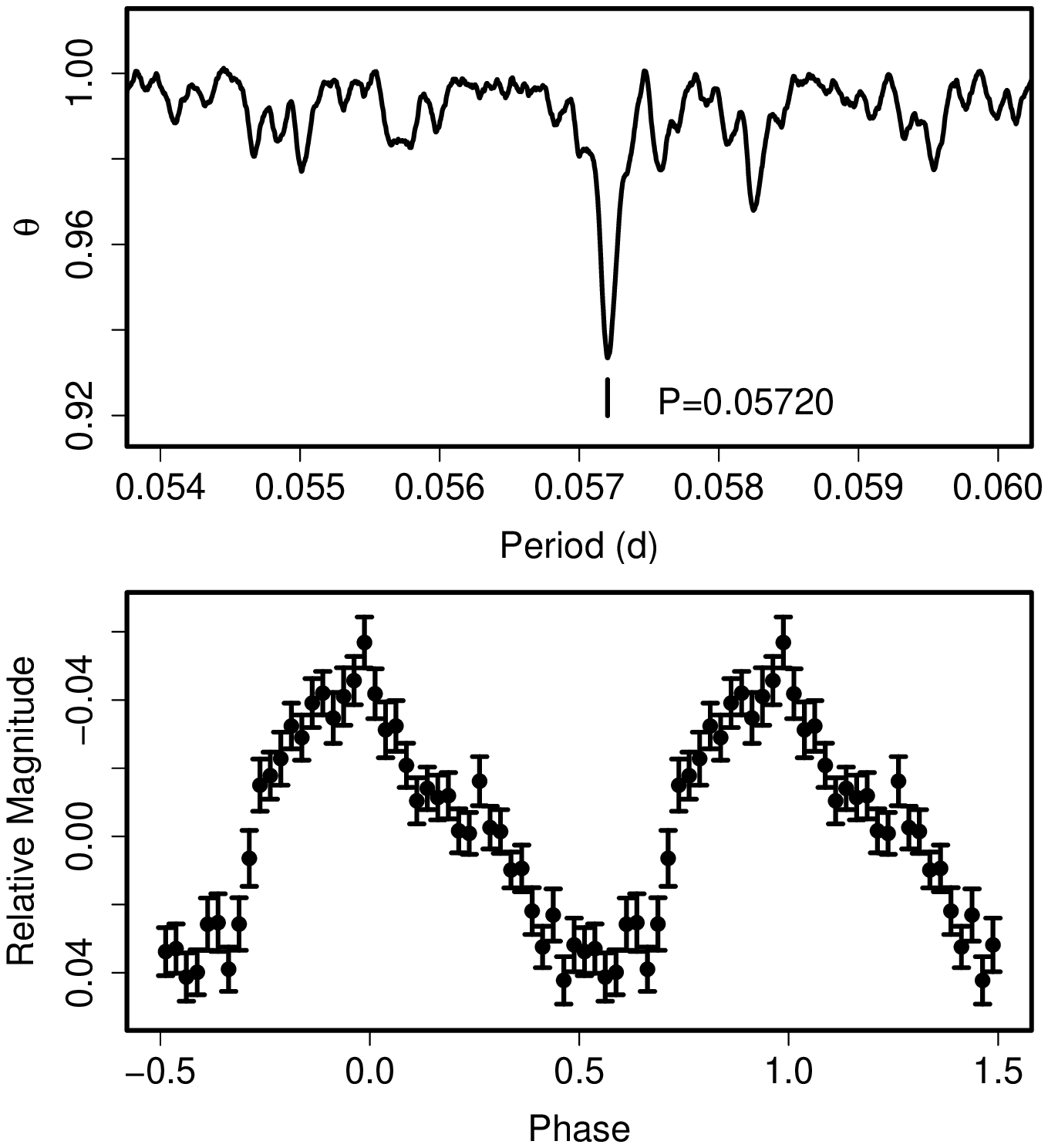}
  \end{center}
  \caption{Ordinary superhumps in V466 And (2008). (Upper): PDM analysis.
     (Lower): Phase-averaged profile.}
  \label{fig:v466shpdm}
\end{figure}

\begin{figure}
  \begin{center}
    \FigureFile(88mm,90mm){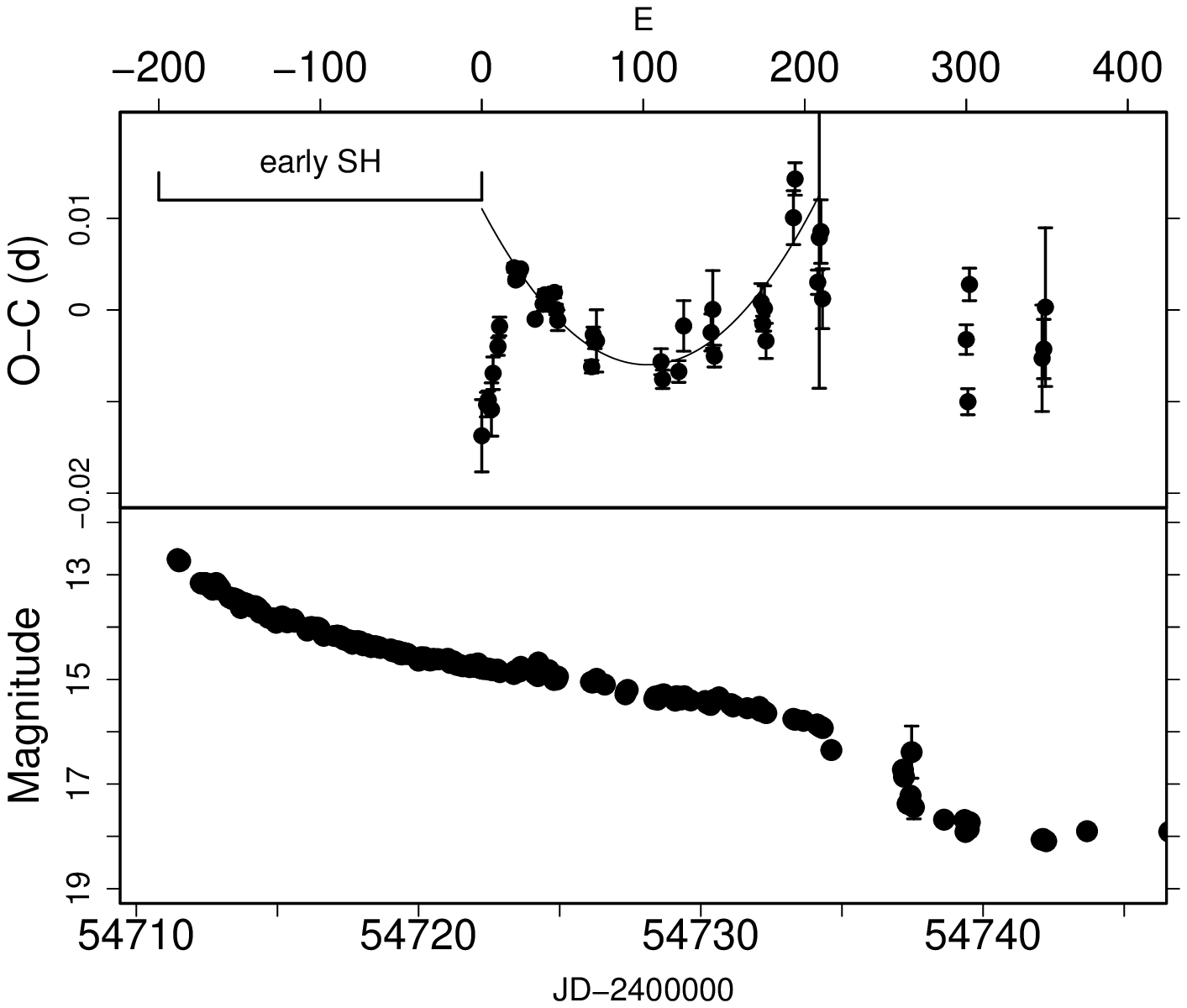}
  \end{center}
  \caption{$O-C$ of superhumps V466 And (2008).
  (Upper): $O-C$ diagram.  The $O-C$ values were against the mean period
  for the stage B ($20 \le E \le 194$, thin curve).
  (Lower): Light curve.
  }
  \label{fig:v466andoc}
\end{figure}

   The object was discovered by K. Itagaki \citep{yam08v466andiauc8971}.
The object was soon recognized as a WZ Sge-type dwarf nova based on
the presence of early superhumps with a period of 0.056365(7) d
(vsnet-alert 10518; period refined in this paper,
figure \ref{fig:v466eshpdm}).
The object later developed ordinary superhumps (mean period
0.057203(10) d with the PDM method; figure \ref{fig:v466shpdm}).
We only deal with ordinary superhumps here (table \ref{tab:v466andoc2008}).
The $O-C$ diagram (figure \ref{fig:v466andoc}) shows the clear presence
of stages A--C.
The $P_{\rm dot}$ during stage B was $+5.7(0.7) \times 10^{-5}$
($20 \le E \le 194$).  More detailed discussion will be presented in
Ohshima et al., in preparation.

\begin{table}
\caption{Superhump maxima of V466 And (2008).}\label{tab:v466andoc2008}
\begin{center}
\begin{tabular}{ccccc}
\hline\hline
$E$ & max$^a$ & error & $O-C^b$ & $N^c$ \\
\hline
0 & 54722.2300 & 0.0039 & $-$0.0115 & 207 \\
3 & 54722.4050 & 0.0013 & $-$0.0081 & 68 \\
4 & 54722.4628 & 0.0010 & $-$0.0076 & 33 \\
6 & 54722.5761 & 0.0029 & $-$0.0087 & 11 \\
7 & 54722.6373 & 0.0018 & $-$0.0048 & 21 \\
10 & 54722.8118 & 0.0010 & $-$0.0019 & 35 \\
11 & 54722.8712 & 0.0010 & 0.0003 & 38 \\
\end{tabular}
\end{center}
\end{table}

\subsection{DH Aquilae}\label{obj:dhaql}

   \citet{nog95dhaql} established the SU UMa-type nature of this object.
We further observed the 2002, 2003 and 2008 superoutbursts
(tables \ref{tab:dhaqloc2002}, \ref{tab:dhaqloc2003}, \ref{tab:dhaqloc2008}).
The global $P_{\rm dot}$ during the 2002 superoutburst was
$-8.4(0.8) \times 10^{-5}$, excluding $E = 0$ taken during the
early evolutionary stage (cf. figure \ref{fig:ocsamp}).
A likely stage B--C transition was recorded during the 2003 superoutburst.
The 2007 and 2008 observations most likely recorded stage C superhumps.
Mean periods 0.07952(4) d and 0.07949(4) d, respectively, determined with
the PDM method were adopted in table \ref{tab:perlist}.

\begin{figure}
  \begin{center}
    \FigureFile(88mm,70mm){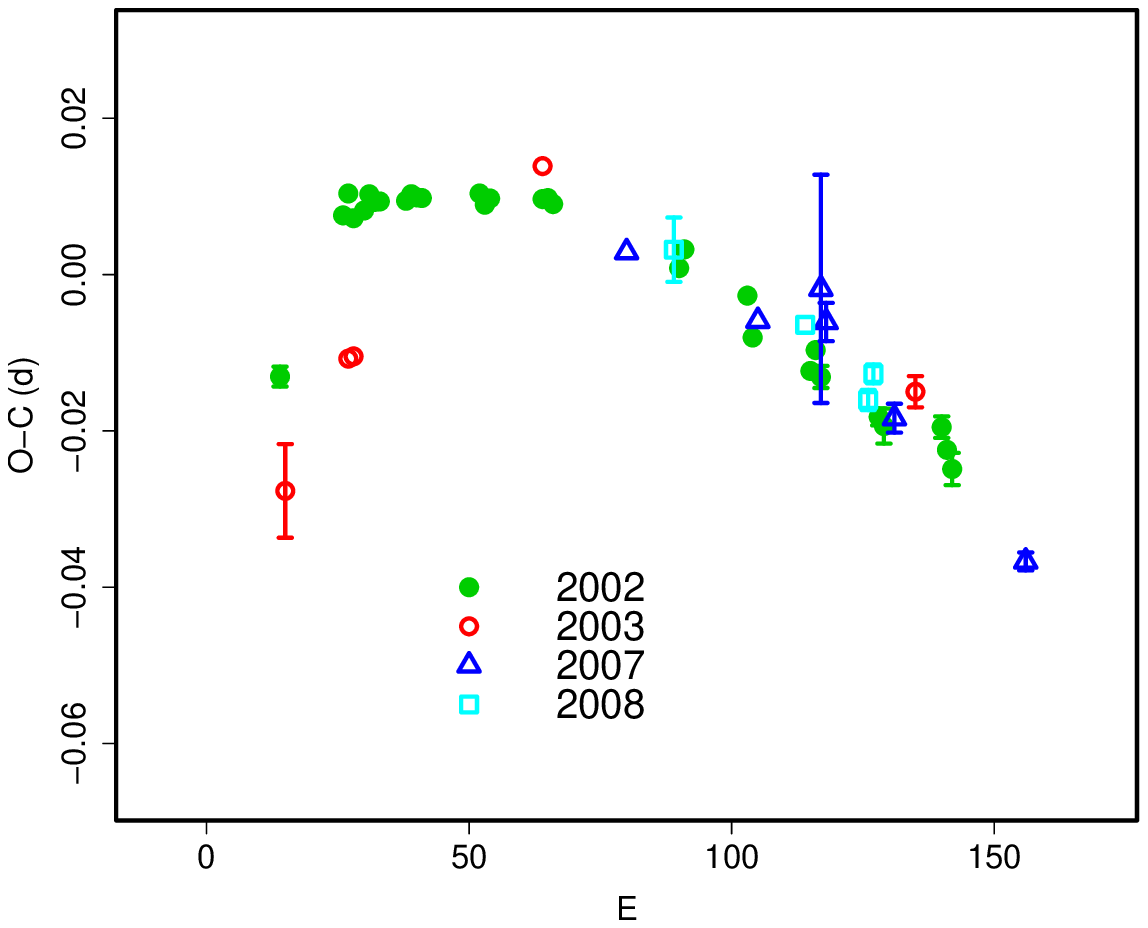}
  \end{center}
  \caption{Comparison of $O-C$ diagrams of DH Aql between different
  superoutbursts.  A period of 0.08000 d was used to draw this figure.
  Approximate cycle counts ($E$) after the start of the
  superoutburst were used.  Since the start of the 2007 superoutburst
  was not well constrained, we shifted the $O-C$ diagrams
  to best fit the others.
  }
  \label{fig:dhaqlcomp}
\end{figure}

   A comparison of $O-C$ diagrams of DH Aql between different
superoutbursts is shown in figure \ref{fig:dhaqlcomp}.

\begin{table}
\caption{Superhumps Maxima of DH Aql (2002).}\label{tab:dhaqloc2002}
\begin{center}
\begin{tabular}{ccccc}
\hline\hline
$E$ & max$^a$ & error & $O-C^b$ & $N^c$ \\
\hline
0 & 52483.9809 & 0.0013 & $-$0.0272 & 93 \\
12 & 52484.9615 & 0.0001 & $-$0.0037 & 267 \\
13 & 52485.0443 & 0.0001 & $-$0.0006 & 256 \\
14 & 52485.1212 & 0.0002 & $-$0.0035 & 250 \\
16 & 52485.2822 & 0.0007 & $-$0.0021 & 135 \\
17 & 52485.3642 & 0.0001 & 0.0003 & 222 \\
18 & 52485.4433 & 0.0001 & $-$0.0005 & 223 \\
19 & 52485.5233 & 0.0002 & $-$0.0002 & 221 \\
24 & 52485.9234 & 0.0001 & 0.0012 & 290 \\
25 & 52486.0043 & 0.0001 & 0.0023 & 472 \\
26 & 52486.0839 & 0.0002 & 0.0021 & 748 \\
27 & 52486.1638 & 0.0003 & 0.0023 & 628 \\
38 & 52487.0443 & 0.0004 & 0.0055 & 306 \\
39 & 52487.1229 & 0.0003 & 0.0044 & 376 \\
40 & 52487.2037 & 0.0002 & 0.0054 & 584 \\
50 & 52488.0036 & 0.0004 & 0.0078 & 228 \\
51 & 52488.0838 & 0.0003 & 0.0082 & 331 \\
52 & 52488.1630 & 0.0003 & 0.0076 & 430 \\
76 & 52490.0748 & 0.0004 & 0.0054 & 221 \\
77 & 52490.1572 & 0.0004 & 0.0080 & 207 \\
89 & 52491.1113 & 0.0009 & 0.0050 & 147 \\
90 & 52491.1859 & 0.0005 & $-$0.0001 & 281 \\
101 & 52492.0616 & 0.0009 & $-$0.0017 & 127 \\
102 & 52492.1443 & 0.0006 & 0.0013 & 152 \\
103 & 52492.2209 & 0.0014 & $-$0.0019 & 120 \\
114 & 52493.0958 & 0.0011 & $-$0.0043 & 151 \\
115 & 52493.1746 & 0.0022 & $-$0.0053 & 136 \\
126 & 52494.0544 & 0.0014 & $-$0.0027 & 117 \\
127 & 52494.1315 & 0.0010 & $-$0.0053 & 325 \\
128 & 52494.2091 & 0.0021 & $-$0.0076 & 128 \\
\hline
  \multicolumn{5}{l}{$^{a}$ BJD$-$2400000.} \\
  \multicolumn{5}{l}{$^{b}$ Against $max = 2452484.0082 + 0.079754 E$.} \\
  \multicolumn{5}{l}{$^{c}$ Number of points used to determine the maximum.} \\
\end{tabular}
\end{center}
\end{table}

\begin{table}
\caption{Superhumps Maxima of DH Aql (2003).}\label{tab:dhaqloc2003}
\begin{center}
\begin{tabular}{ccccc}
\hline\hline
$E$ & max$^a$ & error & $O-C^b$ & $N^c$ \\
\hline
0 & 52886.0802 & 0.0060 & $-$0.0154 & 81 \\
12 & 52887.0571 & 0.0002 & 0.0008 & 124 \\
13 & 52887.1374 & 0.0003 & 0.0010 & 150 \\
49 & 52890.0418 & 0.0007 & 0.0233 & 76 \\
120 & 52895.6929 & 0.0020 & $-$0.0097 & 76 \\
\hline
  \multicolumn{5}{l}{$^{a}$ BJD$-$2400000.} \\
  \multicolumn{5}{l}{$^{b}$ Against $max = 2452886.0957 + 0.080058 E$.} \\
  \multicolumn{5}{l}{$^{c}$ Number of points used to determine the maximum.} \\
\end{tabular}
\end{center}
\end{table}

\begin{table}
\caption{Superhumps Maxima of DH Aql (2007).}\label{tab:dhaqloc2007}
\begin{center}
\begin{tabular}{ccccc}
\hline\hline
$E$ & max$^a$ & error & $O-C^b$ & $N^c$ \\
\hline
0 & 54232.2271 & 0.0005 & $-$0.0058 & 121 \\
25 & 54234.2183 & 0.0008 & $-$0.0016 & 138 \\
37 & 54235.1824 & 0.0146 & 0.0087 & 125 \\
38 & 54235.2582 & 0.0024 & 0.0050 & 88 \\
51 & 54236.2859 & 0.0018 & $-$0.0005 & 80 \\
76 & 54238.2676 & 0.0011 & $-$0.0059 & 84 \\
\hline
  \multicolumn{5}{l}{$^{a}$ BJD$-$2400000.} \\
  \multicolumn{5}{l}{$^{b}$ Against $max = 2454232.2329 + 0.079481 E$.} \\
  \multicolumn{5}{l}{$^{c}$ Number of points used to determine the maximum.} \\
\end{tabular}
\end{center}
\end{table}

\begin{table}
\caption{Superhumps Maxima of DH Aql (2008).}\label{tab:dhaqloc2008}
\begin{center}
\begin{tabular}{ccccc}
\hline\hline
$E$ & max$^a$ & error & $O-C^b$ & $N^c$ \\
\hline
0 & 54628.1832 & 0.0041 & $-$0.0005 & 60 \\
25 & 54630.1736 & 0.0006 & 0.0016 & 142 \\
37 & 54631.1240 & 0.0013 & $-$0.0025 & 138 \\
38 & 54631.2073 & 0.0012 & 0.0014 & 86 \\
\hline
  \multicolumn{5}{l}{$^{a}$ BJD$-$2400000.} \\
  \multicolumn{5}{l}{$^{b}$ Against $max = 2454628.1837 + 0.079534 E$.} \\
  \multicolumn{5}{l}{$^{c}$ Number of points used to determine the maximum.} \\
\end{tabular}
\end{center}
\end{table}

\subsection{V725 Aquilae}\label{sec:v725aql}\label{obj:v725aql}

\begin{figure}
  \begin{center}
    \FigureFile(88mm,70mm){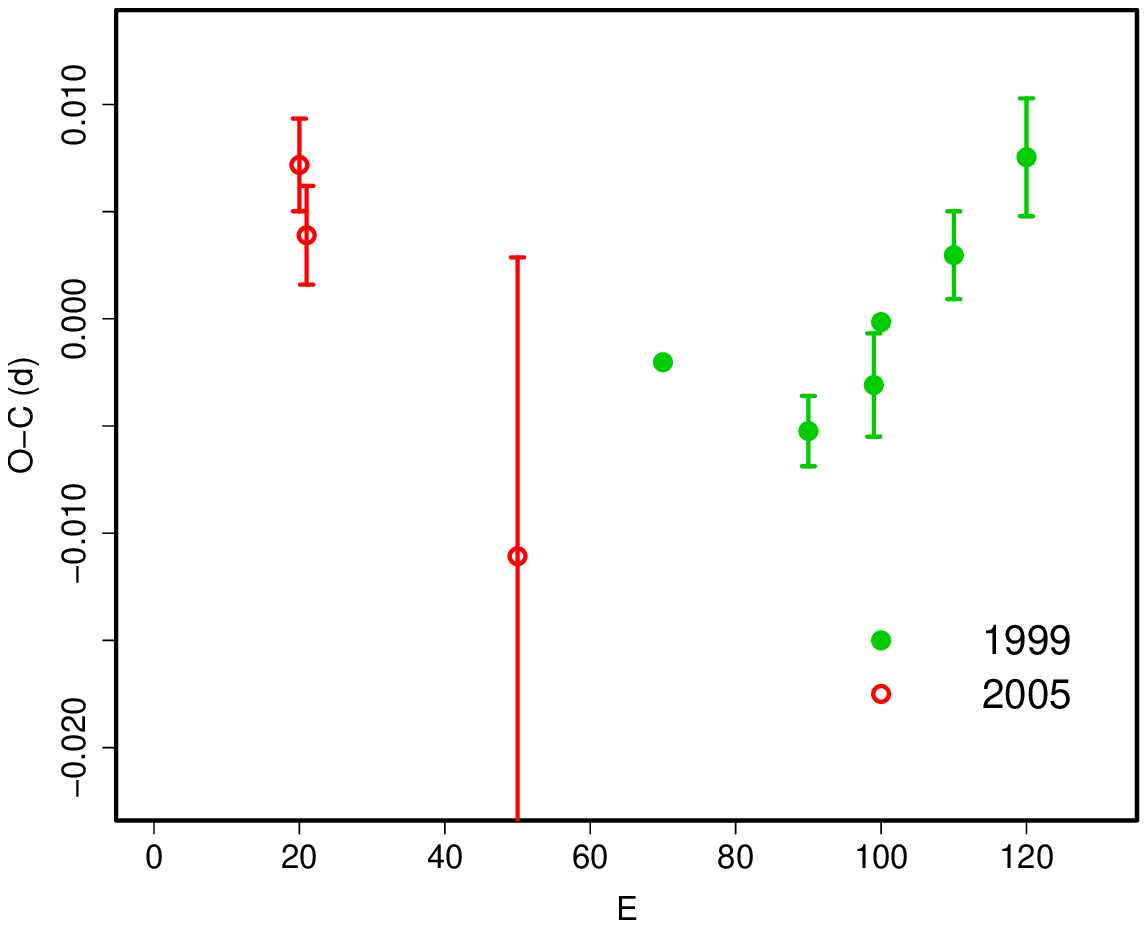}
  \end{center}
  \caption{Comparison of $O-C$ diagrams of V725 Aql between different
  superoutbursts.  A period of 0.06350 d was used to draw this figure.
  Approximate cycle counts ($E$) after the start of the
  superoutburst were used.
  }
  \label{fig:v725aqlcomp}
\end{figure}

   We have reanalyzed the 1999 superoutburst \citep{uem01v725aql}.
The times of superhump maxima are listed in table \ref{tab:v725aqloc1999}.
As shown in \citet{uem01v725aql}, superhumps were only sufficiently
observed mainly after the brightening before termination of the plateau,
presumably corresponding to the stage C.  This would explain the apparently
zero period derivative in \citet{uem01v725aql}.  Although the present
data nominally yielded an overall positive $P_{\rm dot}$ of
$+34.9(15.4) \times 10^{-5}$, the times of maxima for $E \ge 20$ are
well-expressed by a constant period of 0.09977(13) d.
There seems to have been a transition in the period of superhumps
around $E = 20$, associated by a lengthening, rather than shortening
in many SU UMa-type dwarf novae (see also SDSS J1702 for a possible
lengthening of the superhump period in a long-$P_{\rm SH}$ system,
subsection \ref{sec:j1702}).
A better coverage of the early stage of a next superoutburst
is vital to test whether this object indeed has a nearly zero
$P_{\rm dot}$.
The times of superhump maxima during the 2005 superoutburst are
also listed in table \ref{tab:v725aqloc2005}.
A combined $O-C$ diagram (figure \ref{fig:v725aqlcomp}) suggests
a positive $P_{\rm dot}$, which needs to be confirmed by further
observations.

\begin{table}
\caption{Superhump maxima of V725 Aql (1999).}\label{tab:v725aqloc1999}
\begin{center}
\begin{tabular}{ccccc}
\hline\hline
$E$ & max$^a$ & error & $O-C^b$ & $N^c$ \\
\hline
0 & 51447.9596 & 0.0066 & 0.0008 & 116 \\
1 & 51448.0635 & 0.0038 & 0.0055 & 137 \\
2 & 51448.1632 & 0.0072 & 0.0062 & 85 \\
4 & 51448.3580 & 0.0009 & 0.0026 & 56 \\
10 & 51448.9388 & 0.0019 & $-$0.0113 & 157 \\
11 & 51449.0486 & 0.0077 & $-$0.0006 & 185 \\
12 & 51449.1689 & 0.0132 & 0.0205 & 147 \\
20 & 51449.9295 & 0.0046 & $-$0.0120 & 141 \\
21 & 51450.0275 & 0.0091 & $-$0.0131 & 181 \\
24 & 51450.3365 & 0.0016 & $-$0.0015 & 25 \\
30 & 51450.9305 & 0.0040 & $-$0.0023 & 143 \\
31 & 51451.0263 & 0.0199 & $-$0.0056 & 178 \\
32 & 51451.1225 & 0.0087 & $-$0.0086 & 160 \\
33 & 51451.2305 & 0.0024 & 0.0003 & 30 \\
34 & 51451.3325 & 0.0010 & 0.0032 & 53 \\
44 & 51452.3266 & 0.0021 & 0.0059 & 33 \\
54 & 51453.3220 & 0.0027 & 0.0100 & 36 \\
\hline
  \multicolumn{5}{l}{$^{a}$ BJD$-$2400000.} \\
  \multicolumn{5}{l}{$^{b}$ Against $max = 2451447.9588 + 0.099134 E$.} \\
  \multicolumn{5}{l}{$^{c}$ Number of points used to determine the maximum.} \\
\end{tabular}
\end{center}
\end{table}

\begin{table}
\caption{Superhump maxima of V725 Aql (2005).}\label{tab:v725aqloc2005}
\begin{center}
\begin{tabular}{ccccc}
\hline\hline
$E$ & max$^a$ & error & $O-C^b$ & $N^c$ \\
\hline
0 & 53685.5353 & 0.0022 & 0.0013 & 120 \\
1 & 53685.6311 & 0.0023 & $-$0.0014 & 97 \\
30 & 53688.4897 & 0.0139 & 0.0000 & 30 \\
\hline
  \multicolumn{5}{l}{$^{a}$ BJD$-$2400000.} \\
  \multicolumn{5}{l}{$^{b}$ Against $max = 2453685.5339 + 0.098525 E$.} \\
  \multicolumn{5}{l}{$^{c}$ Number of points used to determine the maximum.} \\
\end{tabular}
\end{center}
\end{table}

\subsection{V1141 Aquilae}\label{obj:v1141aql}

   \citet{ole03v1141aql} reported the detection of superhumps during
the 2002 superoutburst.  The reported period was 0.05930(5) d.
Although \citet{ole03v1141aql} attempted to make a comparison with
the system SW UMa, which has similar outburst properties and superhump
period, they failed to detect a positive period derivative.

\begin{figure}
  \begin{center}
    \FigureFile(88mm,110mm){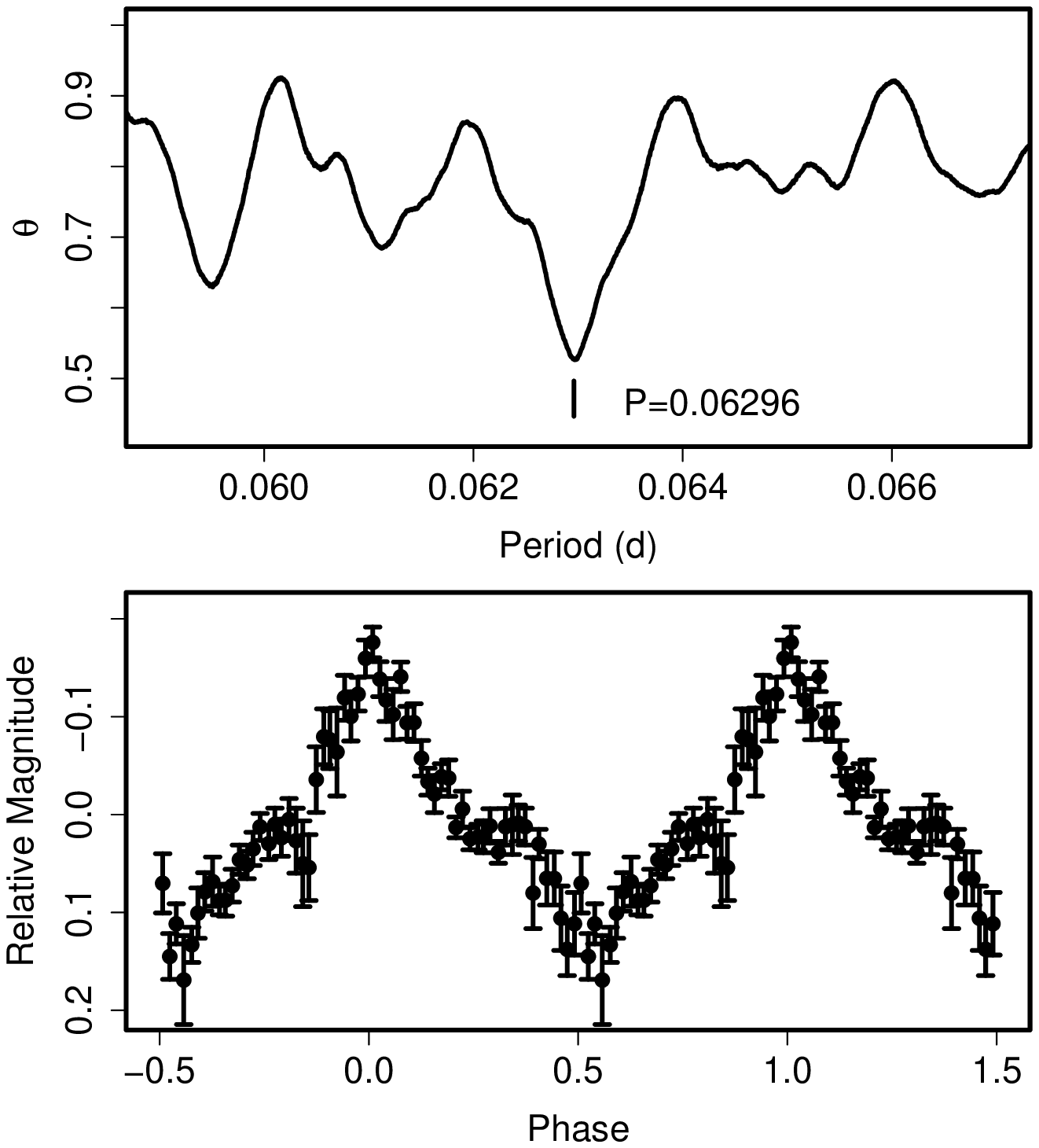}
  \end{center}
  \caption{Superhumps in V1141 Aql (2003). (Upper): PDM analysis.
     (Lower): Phase-averaged profile.}
  \label{fig:v1141aqlshpdm}
\end{figure}

   During the 2003 superoutburst, we performed time-series
photometry on consecutive five nights.  The resultant timings of
superhump maxima are presented in table \ref{tab:v1141aqloc}.
The period by \citet{ole03v1141aql} did not well fit our
observations; instead, a period of 0.06296(2) d well expressed our
observations (figure \ref{fig:v1141aqlshpdm}).
The cycle numbers given in table \ref{tab:v1141aqloc}
refer to this period.  The observed times were well expressed by
a $P_{\rm dot}$ of $+13.4(1.6) \times 10^{-5}$.

\begin{figure}
  \begin{center}
    \FigureFile(88mm,70mm){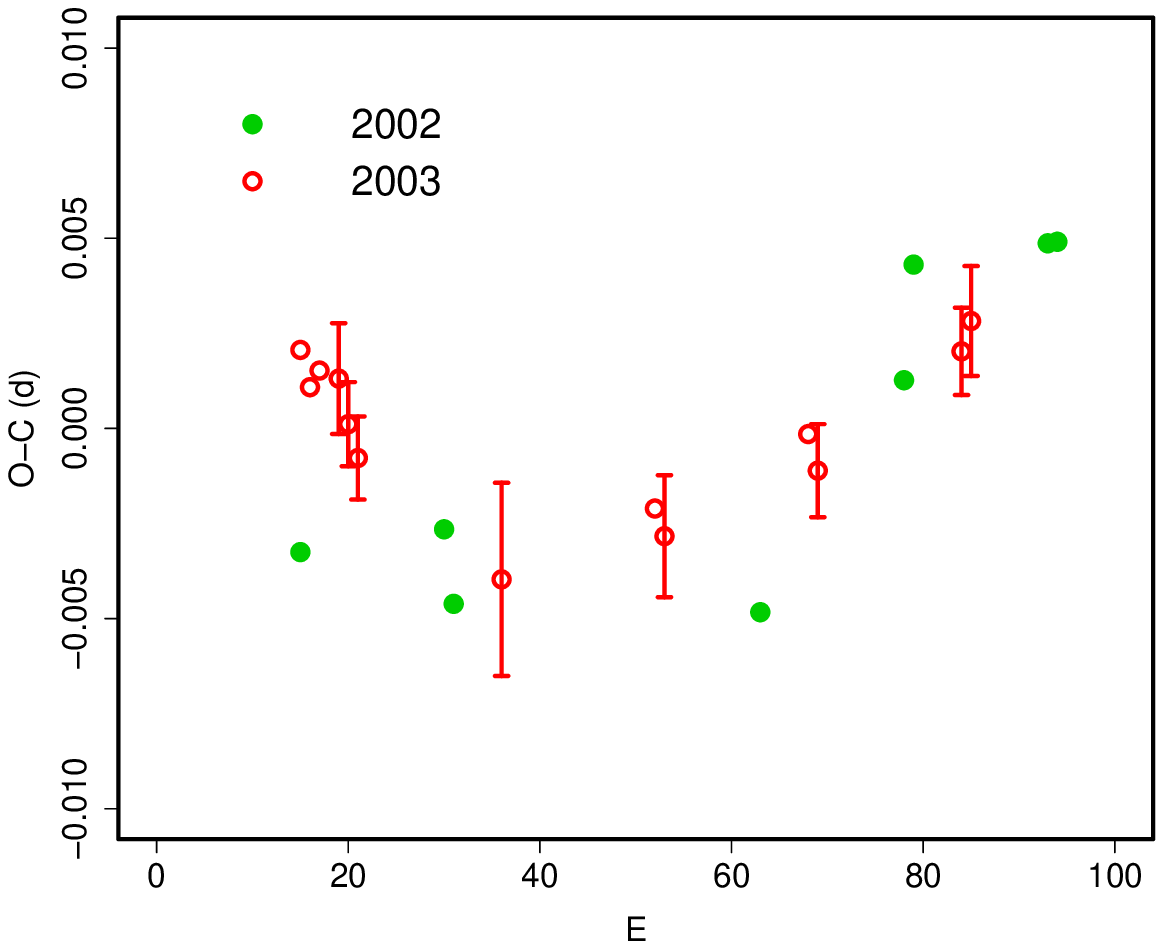}
  \end{center}
  \caption{Comparison of $O-C$ diagrams of V1141 Aql between different
  superoutbursts.  A period of 0.06296 d was used to draw this figure.
  Approximate cycle counts ($E$) after the start of the
  superoutburst were used.
  }
  \label{fig:v1141aqlcomp}
\end{figure}

  A comparison of $O-C$ diagrams of V1141 Aql between different
superoutbursts is shown in figure \ref{fig:v1141aqlcomp}.

\begin{table}
\caption{Superhump maxima of V1141 Aql (2003).}\label{tab:v1141aqloc}
\begin{center}
\begin{tabular}{ccccc}
\hline\hline
$E$ & max$^a$ & error & $O-C$ & $N^b$ \\
\hline
0 & 52823.0212 & 0.0005 & 0.0021 & 137 \\
1 & 52823.0831 & 0.0005 & 0.0011 & 102 \\
2 & 52823.1465 & 0.0007 & 0.0015 & 81 \\
4 & 52823.2723 & 0.0015 & 0.0013 & 17 \\
5 & 52823.3340 & 0.0011 & 0.0001 & 18 \\
6 & 52823.3961 & 0.0011 & $-$0.0008 & 18 \\
21 & 52824.3373 & 0.0025 & $-$0.0040 & 17 \\
37 & 52825.3465 & 0.0010 & $-$0.0021 & 17 \\
38 & 52825.4087 & 0.0016 & $-$0.0028 & 17 \\
53 & 52826.3558 & 0.0008 & $-$0.0002 & 17 \\
54 & 52826.4178 & 0.0012 & $-$0.0011 & 17 \\
69 & 52827.3654 & 0.0011 & 0.0020 & 17 \\
70 & 52827.4291 & 0.0014 & 0.0028 & 17 \\
\hline
  \multicolumn{5}{l}{$^{a}$ BJD$-$2400000.} \\
  \multicolumn{5}{l}{$^{b}$ Number of points used to determine the maximum.} \\
\end{tabular}
\end{center}
\end{table}

   By correctly identifying the cycle numbers based on this period,
the reported times of maxima in
\citet{ole03v1141aql} can also be well fit by a mean period of
0.06308(3) d and $P_{\rm dot}$ of $+9.3(4.3) \times 10^{-5}$.
These period derivatives are indeed similar to that of SW UMa.
The new period is also compatible with the proposed orbital
period of 0.0620 d from single-night quiescent photometry
\citep{hae04v1141aql}.  By literally adopting this proposed orbital
period, we obtain a fractional superhump excess of 1.5\%.
A comparison of $O-C$ diagrams between 2002 and 2003 superoutbursts
is shown in figure \ref{fig:v1141aqlcomp}.

\subsection{VY Aquarii}\label{obj:vyaqr}

   VY Aqr had long been supposed to be a recurrent nova that erupted
in 1907 and 1962 (\cite{str62vyaqr}; \cite{hut62vyaqr}).
While the detection of the 1973 outburst \citep{mcn82vyaqr} suggested
a shorter recurrence time, the detection of additional outbursts
(\cite{ric83vyaqr}; \cite{ric83vyaqraditional}; \cite{lil83vyaqr})
led to a more likely classification as a WZ Sge-type dwarf nova.
Further outbursts were recorded almost yearly (e.g. \cite{mca83vyaqriauc};
\cite{lub86vyaqriauc}; \cite{hur87vyaqriauc}), confirming the
dwarf nova-type classification.  \citet{pat93vyaqr} first established
the SU UMa-type classification based on photoelectric observations
during the 1986 outburst.

\begin{figure}
  \begin{center}
    \FigureFile(88mm,90mm){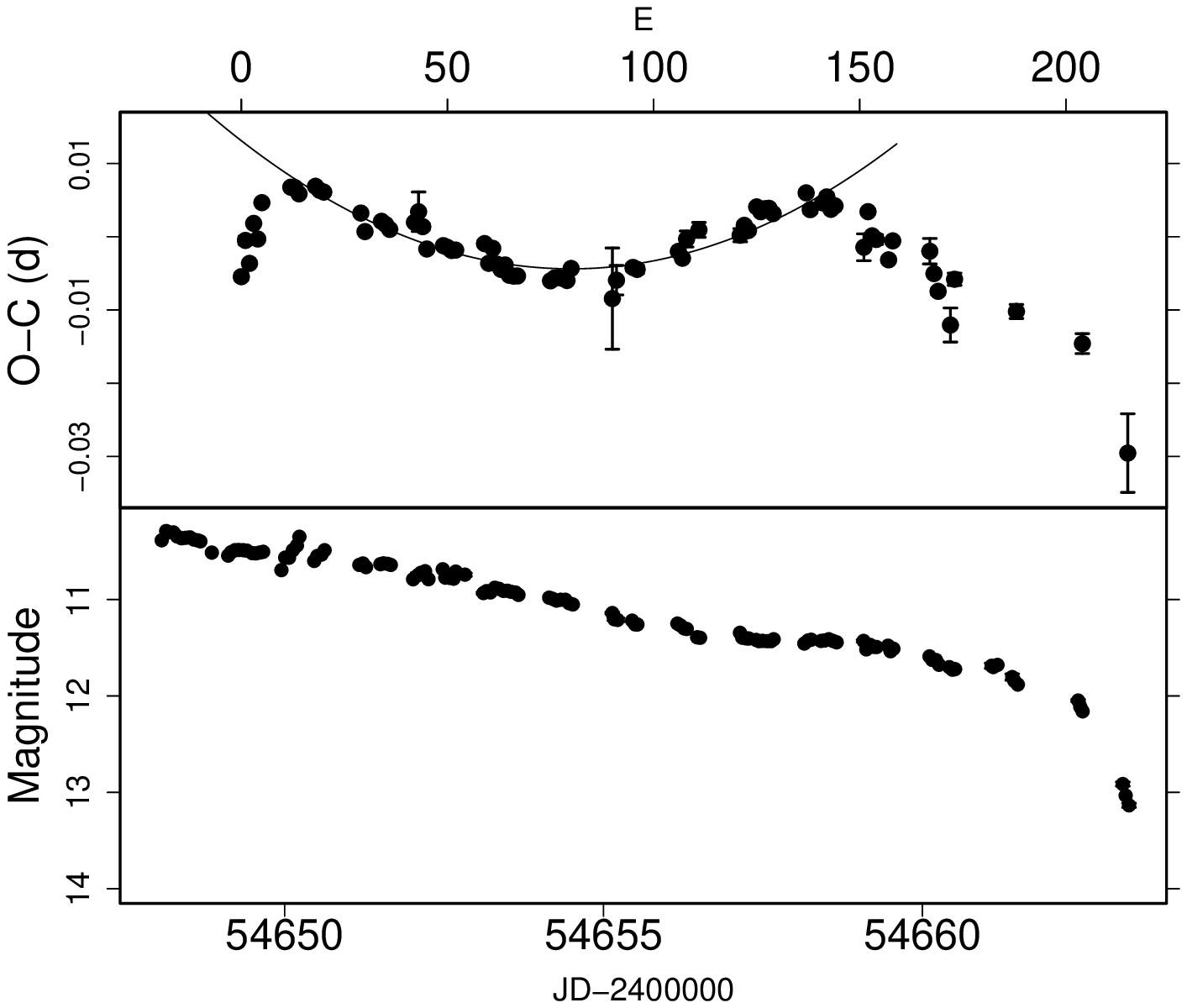}
  \end{center}
  \caption{$O-C$ of superhumps VY Aqr (2008).
  (Upper): $O-C$ diagram.  The $O-C$ values were against the mean period
  for the stage B ($12 \le E \le 144$, thin curve)
  (Lower): Light curve.  A brightening associated with the start of the
  stage C is clearly seen.
  }
  \label{fig:vyaqr2008oc}
\end{figure}

   The only available timing observation of superhumps in the past
\citep{pat93vyaqr} reported a negative $P_{\rm dot}$.
The existence of a negative $P_{\rm dot}$
with this relatively short superhump period had been a mystery.
The fresh outburst in 2008 has enabled us to finally establish
$P_{\rm dot}$ of this object.  The outburst was well-observed
during the entire superoutburst plateau and subsequent decline,
a rebrightening, and final fading.  We only deal with superhumps
during the plateau phase (table \ref{tab:vyaqroc2008}).  The $O-C$ diagram
shows all the distinct stages A--C (figure \ref{fig:vyaqr2008oc}).
The $P_{\rm dot}$ during stage B was $+8.5(0.5) \times 10^{-5}$
($12 \le E \le 144$).

\begin{figure}
  \begin{center}
    \FigureFile(88mm,70mm){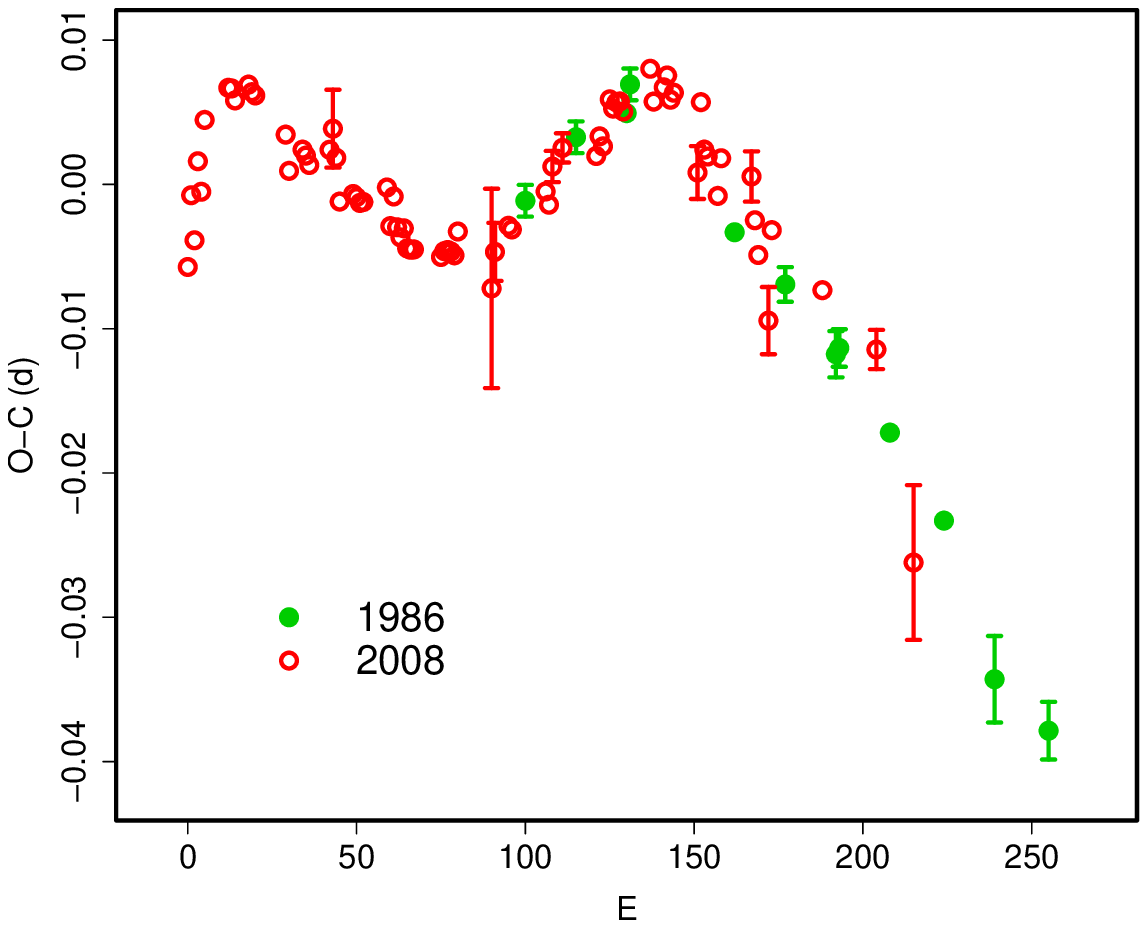}
  \end{center}
  \caption{Comparison of $O-C$ diagrams of VY Aqr between different
  superoutbursts.  A period of 0.06464 d was used to draw this figure.
  Approximate cycle counts ($E$) after the estimated appearance of the
  superhumps were used.
  }
  \label{fig:vyaqrcomp}
\end{figure}

The times of superhump maxima of the 1986 superoutburst
determined from the scanned figure are given in table \ref{tab:vyaqroc1986}.
The negative $P_{\rm dot}$ in \citet{pat93vyaqr} was probably a
result of the stage B--C transition around $E=30-31$
(figure \ref{fig:vyaqrcomp}).
For more details of this outburst, see Ohshima et al., in preparation.

\begin{table}
\caption{Superhump maxima of VY Aqr (2008).}\label{tab:vyaqroc2008}
\begin{center}
\begin{tabular}{ccccc}
\hline\hline
$E$ & max$^a$ & error & $O-C^b$ & $N^c$ \\
\hline
0 & 54649.3119 & 0.0005 & $-$0.0075 & 122 \\
1 & 54649.3815 & 0.0006 & $-$0.0025 & 122 \\
2 & 54649.4430 & 0.0004 & $-$0.0056 & 41 \\
3 & 54649.5132 & 0.0002 & $-$0.0002 & 412 \\
4 & 54649.5757 & 0.0001 & $-$0.0022 & 378 \\
5 & 54649.6453 & 0.0001 & 0.0028 & 286 \\
12 & 54650.1000 & 0.0002 & 0.0051 & 109 \\
\end{tabular}
\end{center}
\end{table}

\addtocounter{table}{-1}
\begin{table}
\caption{Superhump maxima of VY Aqr (2008). (continued)}
\begin{center}
\begin{tabular}{ccccc}
\hline\hline
$E$ & max & error & $O-C$ & $N$ \\
\hline
129 & 54657.6612 & 0.0002 & 0.0059 & 218 \\
137 & 54658.1813 & 0.0004 & 0.0090 & 158 \\
138 & 54658.2437 & 0.0005 & 0.0068 & 157 \\
141 & 54658.4386 & 0.0005 & 0.0078 & 406 \\
142 & 54658.5041 & 0.0005 & 0.0087 & 320 \\
143 & 54658.5670 & 0.0002 & 0.0070 & 278 \\
144 & 54658.6321 & 0.0002 & 0.0075 & 272 \\
151 & 54659.0791 & 0.0018 & 0.0021 & 136 \\
152 & 54659.1486 & 0.0003 & 0.0070 & 506 \\
153 & 54659.2100 & 0.0002 & 0.0037 & 358 \\
154 & 54659.2741 & 0.0007 & 0.0033 & 130 \\
157 & 54659.4653 & 0.0004 & 0.0006 & 145 \\
158 & 54659.5326 & 0.0004 & 0.0032 & 195 \\
167 & 54660.1131 & 0.0017 & 0.0022 & 93 \\
168 & 54660.1747 & 0.0003 & $-$0.0008 & 290 \\
169 & 54660.2369 & 0.0005 & $-$0.0032 & 221 \\
172 & 54660.4263 & 0.0023 & $-$0.0077 & 40 \\
173 & 54660.4972 & 0.0008 & $-$0.0014 & 57 \\
188 & 54661.4626 & 0.0010 & $-$0.0053 & 63 \\
204 & 54662.4928 & 0.0014 & $-$0.0091 & 64 \\
215 & 54663.1890 & 0.0054 & $-$0.0236 & 119 \\
\hline
\end{tabular}
\end{center}
\end{table}

\begin{table}
\caption{Superhump maxima of VY Aqr (1986).}\label{tab:vyaqroc1986}
\begin{center}
\begin{tabular}{cccc}
\hline\hline
$E$ & max$^a$ & error & $O-C^b$ \\
\hline
0 & 46559.9687 & 0.0011 & $-$0.0113 \\
15 & 46560.9427 & 0.0011 & $-$0.0028 \\
30 & 46561.9140 & 0.0010 & 0.0030 \\
31 & 46561.9806 & 0.0011 & 0.0053 \\
62 & 46563.9742 & 0.0008 & 0.0035 \\
77 & 46564.9402 & 0.0012 & 0.0040 \\
92 & 46565.9050 & 0.0016 & 0.0033 \\
93 & 46565.9700 & 0.0013 & 0.0040 \\
108 & 46566.9338 & 0.0009 & 0.0023 \\
124 & 46567.9619 & 0.0009 & 0.0006 \\
139 & 46568.9205 & 0.0030 & $-$0.0063 \\
155 & 46569.9512 & 0.0020 & $-$0.0055 \\
\hline
  \multicolumn{4}{l}{$^{a}$ BJD$-$2400000.} \\
  \multicolumn{4}{l}{$^{b}$ Against $max = 2446559.9800 + 0.064365 E$.} \\
  \multicolumn{4}{l}{$^{c}$ Number of points used to determine the maximum.} \\
\end{tabular}
\end{center}
\end{table}

\subsection{EG Aquarii}\label{obj:egaqr}

   The 2006 superoutburst of this object was extensively studied
by \citet{ima08egaqr}.  We further observed the 2008 superoutburst
(table \ref{tab:egaqroc2008}).
Since the outburst detection was not noticed early enough, the observation
only covered the middle part of the superoutburst.  The resultant
$P_{\rm dot}$ was similar to that obtained during the 2006
superoutburst.  The supercycle of this object is likely $\sim$ 750 d.

\begin{table}
\caption{Superhump maxima of EG Aqr (2008).}\label{tab:egaqroc2008}
\begin{center}
\begin{tabular}{ccccc}
\hline\hline
$E$ & max$^a$ & error & $O-C^b$ & $N^c$ \\
\hline
0 & 54802.9893 & 0.0003 & $-$0.0014 & 216 \\
3 & 54803.2277 & 0.0006 & 0.0007 & 46 \\
12 & 54803.9365 & 0.0004 & 0.0006 & 262 \\
13 & 54804.0154 & 0.0005 & 0.0007 & 173 \\
38 & 54805.9831 & 0.0004 & $-$0.0006 & 242 \\
63 & 54807.9527 & 0.0006 & 0.0000 & 244 \\
\hline
  \multicolumn{5}{l}{$^{a}$ BJD$-$2400000.} \\
  \multicolumn{5}{l}{$^{b}$ Against $max = 2454802.9908 + 0.078760 E$.} \\
  \multicolumn{5}{l}{$^{c}$ Number of points used to determine the maximum.} \\
\end{tabular}
\end{center}
\end{table}

\subsection{BF Arae}\label{obj:bfara}

   \citet{kat03bfara} studied the 2002 superoutburst of this object.
A reanalysis of the same data has yielded improved determination of
superhump maxima than those obtained by eye estimates
(table \ref{tab:bfaraoc}).  The resultant $P_{\rm dot}$ was
$-2.8(1.6) \times 10^{-5}$, giving a slightly smaller value in
\citet{kat03bfara}.  \citet{ole07bfara} obtained photometry in
quiescence and yielded a likely orbital period of 0.084176(21) d,
giving a fractional superhump excess of 4.4 \%.

\begin{table}
\caption{Superhump maxima of BF Ara (2002).}\label{tab:bfaraoc}
\begin{center}
\begin{tabular}{ccccc}
\hline\hline
$E$ & max$^a$ & error & $O-C^b$ & $N^c$ \\
\hline
0 & 52504.9878 & 0.0005 & $-$0.0031 & 143 \\
1 & 52505.0774 & 0.0005 & $-$0.0014 & 120 \\
2 & 52505.1643 & 0.0005 & $-$0.0023 & 114 \\
12 & 52506.0456 & 0.0007 & 0.0001 & 87 \\
13 & 52506.1311 & 0.0005 & $-$0.0023 & 89 \\
14 & 52506.2266 & 0.0024 & 0.0053 & 49 \\
23 & 52507.0148 & 0.0008 & 0.0025 & 82 \\
24 & 52507.1025 & 0.0007 & 0.0024 & 83 \\
25 & 52507.1869 & 0.0008 & $-$0.0012 & 89 \\
35 & 52508.0672 & 0.0006 & 0.0003 & 88 \\
36 & 52508.1565 & 0.0008 & 0.0017 & 88 \\
60 & 52510.2640 & 0.0014 & $-$0.0001 & 102 \\
61 & 52510.3510 & 0.0014 & $-$0.0010 & 100 \\
80 & 52512.0207 & 0.0016 & $-$0.0011 & 30 \\
90 & 52512.9025 & 0.0023 & 0.0018 & 22 \\
91 & 52512.9920 & 0.0017 & 0.0034 & 25 \\
102 & 52513.9504 & 0.0021 & $-$0.0050 & 27 \\
\hline
  \multicolumn{5}{l}{$^{a}$ BJD$-$2400000.} \\
  \multicolumn{5}{l}{$^{b}$ Against $max = 2452504.9909 + 0.087887 E$.} \\
  \multicolumn{5}{l}{$^{c}$ Number of points used to determine the maximum.} \\
\end{tabular}
\end{center}
\end{table}

\subsection{V663 Arae}\label{obj:v663ara}

\begin{figure}
  \begin{center}
    \FigureFile(88mm,110mm){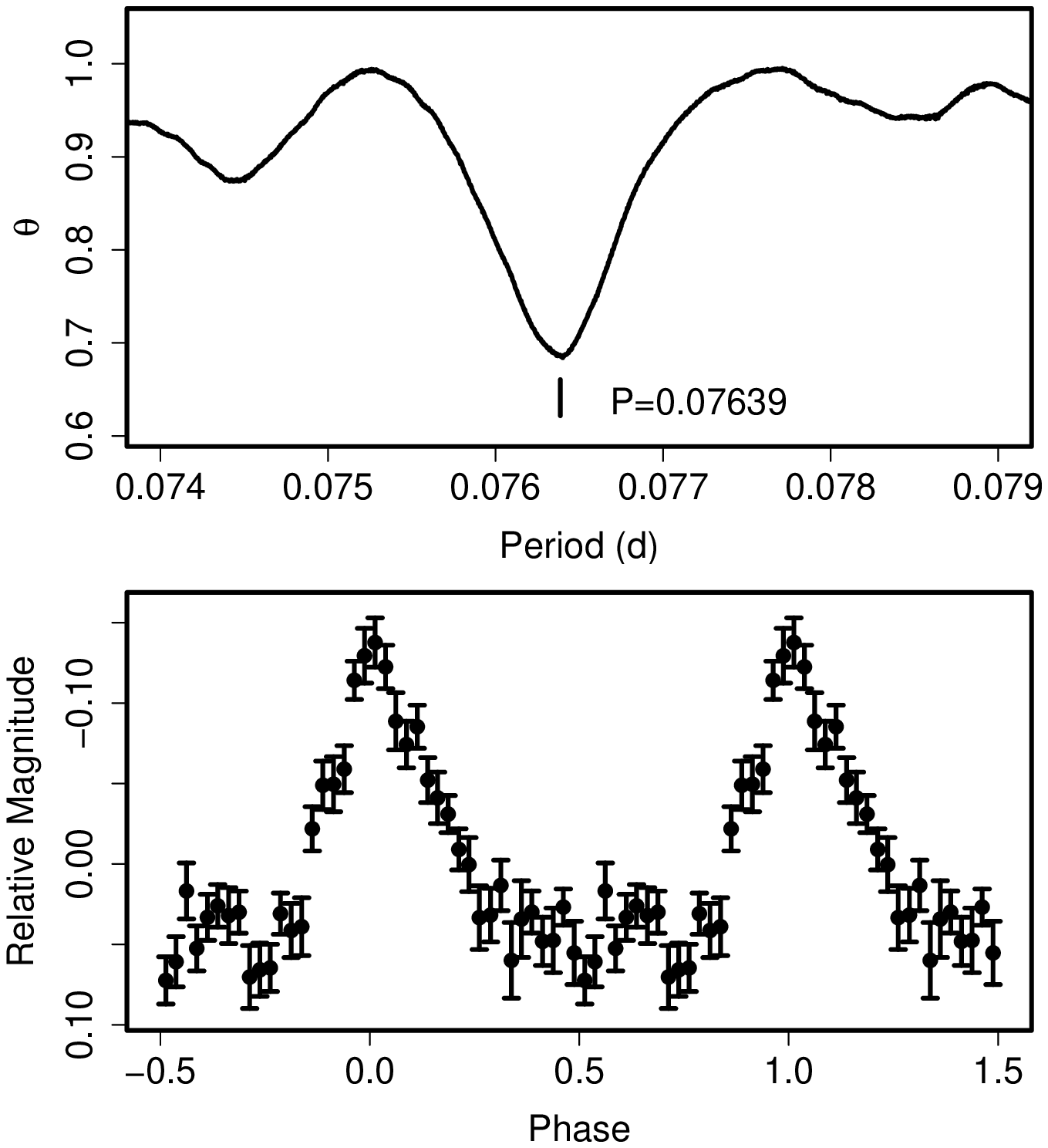}
  \end{center}
  \caption{Superhumps in V663 Ara (2004). (Upper): PDM analysis.
     (Lower): Phase-averaged profile.}
  \label{fig:v663arashpdm}
\end{figure}

   V663 Ara was discovered by \citet{ges74v663ara} as a long-period
variable star.  \citet{DownesCVatlas3} listed this object as a CV.
The SU UMa-type nature of this object was established by B. Monard
(vsnet-alert 8231, 8232).
We obtained a mean superhump period of 0.07639(2) d from observations
on four nights (figure \ref{fig:v663arashpdm}).
The times of superhump maxima are listed in table
\ref{tab:v663araoc2004}.  The resultant $P_{\rm dot}$ was
$-6.2(9.4) \times 10^{-5}$.

\begin{table}
\caption{Superhump maxima of V663 Ara (2004).}\label{tab:v663araoc2004}
\begin{center}
\begin{tabular}{ccccc}
\hline\hline
$E$ & max$^a$ & error & $O-C^b$ & $N^c$ \\
\hline
0 & 53195.4364 & 0.0007 & 0.0001 & 82 \\
11 & 53196.2782 & 0.0013 & 0.0020 & 63 \\
12 & 53196.3546 & 0.0010 & 0.0020 & 83 \\
13 & 53196.4265 & 0.0015 & $-$0.0025 & 81 \\
14 & 53196.5017 & 0.0018 & $-$0.0037 & 81 \\
37 & 53198.2603 & 0.0112 & $-$0.0015 & 47 \\
38 & 53198.3415 & 0.0016 & 0.0033 & 86 \\
39 & 53198.4187 & 0.0020 & 0.0042 & 86 \\
40 & 53198.4907 & 0.0016 & $-$0.0002 & 84 \\
50 & 53199.2517 & 0.0021 & $-$0.0029 & 76 \\
51 & 53199.3304 & 0.0015 & $-$0.0005 & 81 \\
52 & 53199.4068 & 0.0055 & $-$0.0005 & 35 \\
\hline
  \multicolumn{5}{l}{$^{a}$ BJD$-$2400000.} \\
  \multicolumn{5}{l}{$^{b}$ Against $max = 2453195.4362 + 0.076366 E$.} \\
  \multicolumn{5}{l}{$^{c}$ Number of points used to determine the maximum.} \\
\end{tabular}
\end{center}
\end{table}

\subsection{V877 Arae}\label{sec:v877ara}\label{obj:v877ara}

   \citet{kat03v877arakktelpucma} observed the 2002 superoutburst
and reported a strongly negative period derivative.  The variation
of the superhump period occurred during the earliest stage of the
superoutburst, and the originally reported $P_{\rm dot}$ more likely
reflected the early stage of period shift from the stage A to B,
as seen in the similar long-period system DT Oct
(subsection \ref{sec:dtoct}).
The refined superhump maxima are listed in table \ref{tab:v877ara}.
By neglecting the early portion ($E < 24$), we obtained a
$P_{\rm dot}$ of $-5.7(2.9) \times 10^{-5}$, typical for an usual
SU UMa-type dwarf nova.

\begin{table}
\caption{Superhump maxima of V877 Ara (2002).}\label{tab:v877ara}
\begin{center}
\begin{tabular}{ccccc}
\hline\hline
$E$ & max$^a$ & error & $O-C^b$ & $N^c$ \\
\hline
0 & 52434.9582 & 0.0011 & $-$0.0123 & 61 \\
1 & 52435.0470 & 0.0003 & $-$0.0076 & 84 \\
24 & 52436.9922 & 0.0011 & 0.0045 & 64 \\
25 & 52437.0757 & 0.0007 & 0.0039 & 111 \\
26 & 52437.1645 & 0.0005 & 0.0087 & 87 \\
27 & 52437.2473 & 0.0007 & 0.0075 & 86 \\
72 & 52441.0235 & 0.0015 & 0.0013 & 46 \\
73 & 52441.1103 & 0.0008 & 0.0041 & 69 \\
74 & 52441.1923 & 0.0007 & 0.0021 & 65 \\
95 & 52442.9508 & 0.0013 & $-$0.0044 & 23 \\
96 & 52443.0389 & 0.0017 & $-$0.0004 & 24 \\
97 & 52443.1186 & 0.0012 & $-$0.0047 & 51 \\
98 & 52443.2047 & 0.0009 & $-$0.0027 & 61 \\
\hline
  \multicolumn{5}{l}{$^{a}$ BJD$-$2400000.} \\
  \multicolumn{5}{l}{$^{b}$ Against $max = 2452434.9704 + 0.084050 E$.} \\
  \multicolumn{5}{l}{$^{c}$ Number of points used to determine the maximum.} \\
\end{tabular}
\end{center}
\end{table}

\subsection{BB Arietis}\label{sec:bbari}\label{obj:bbari}

   This object was recognized during the identification project of
the New Catalogue of Suspected Variable Stars (NSV, \cite{NSV})
objects against ROSAT X-ray source (Kato, vsnet-chat 3317).
The proximity of a ROSAT source to the position of NSV 907,
a large-amplitude variable star, suggested that the object may be
a dwarf nova, as we have seen in
DT Oct = NSV 10934 \citep{kat02gzcncnsv10934}.

\begin{figure}
  \begin{center}
    \FigureFile(88mm,110mm){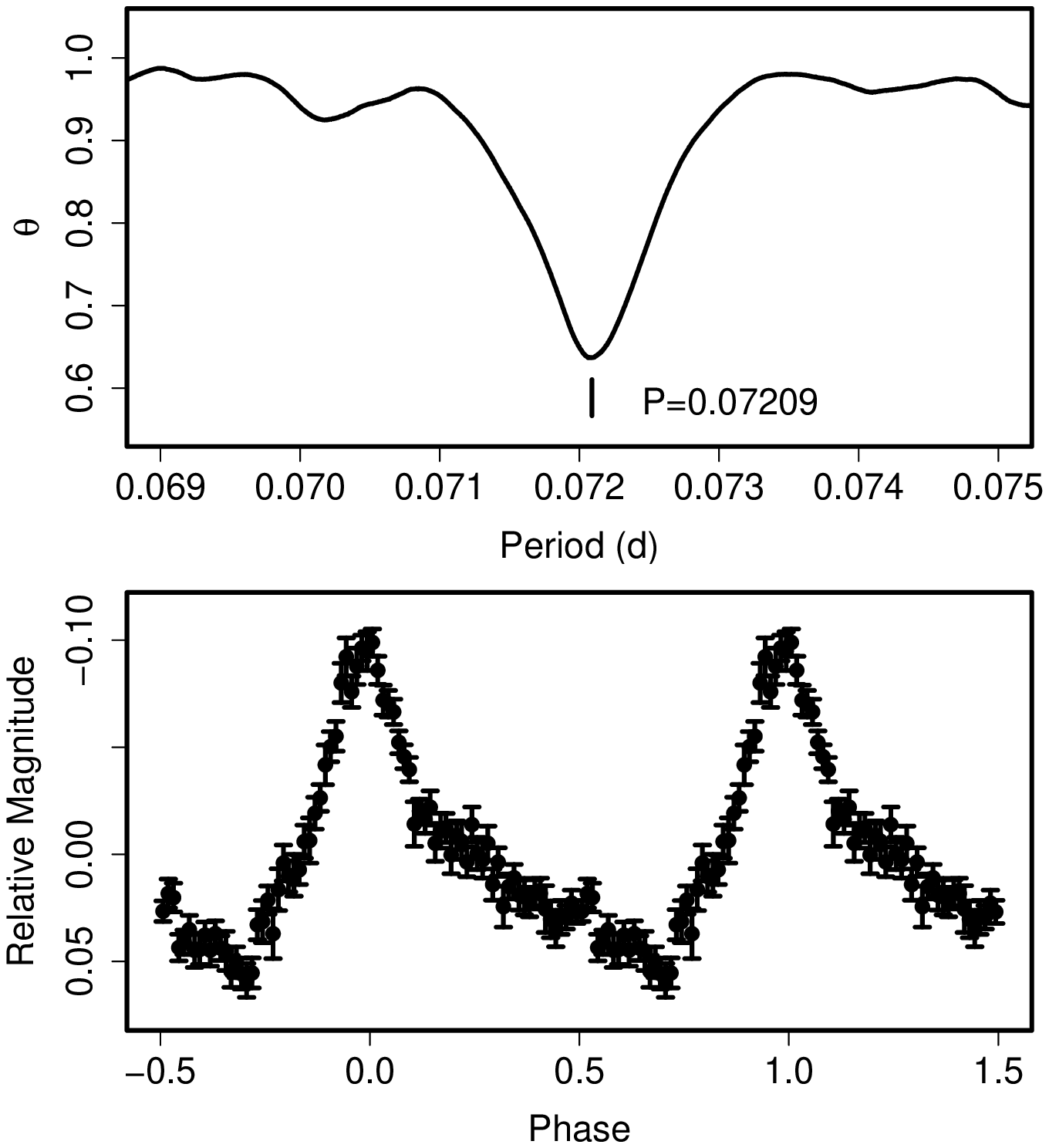}
  \end{center}
  \caption{Superhumps in BB Ari (2004). (Upper): PDM analysis.
     (Lower): Phase-averaged profile.}
  \label{fig:bbarishpdm}
\end{figure}

   The object has been monitored since, and the first outburst was
detected by P. Schmeer on 2004 March 2 at an unfiltered CCD magnitude
of 13.5.  It is unclear how long this outburst lasted.
On 2004 November 1, P. Schmeer detected another outburst at magnitude
13.7.  Following this alert, we started time-resolved photometry and
detected superhumps.  A PDM analysis yielded a mean superhump period
of 0.07209(1) d (figure \ref{fig:bbarishpdm}).
The times of superhump maxima are listed in
table \ref{tab:bbarioc2004}.  We obtained $P_{\rm dot}$ =
$+1.6(3.0) \times 10^{-5}$.  Since the superoutburst was likely detected
during its final stage, the superhump period and period derivative most
likely reflect the behavior after transition to the stage C.
Although the object has been well monitored since, only two normal
outbursts was recorded in 2006 November in 2009 February.
The outburst frequency may be as low as UV Per and VY Aqr,
having similar superhump periods.

\begin{table}
\caption{Superhump maxima of BB Ari (2004).}\label{tab:bbarioc2004}
\begin{center}
\begin{tabular}{ccccc}
\hline\hline
$E$ & max$^a$ & error & $O-C^b$ & $N^c$ \\
\hline
0 & 53311.4608 & 0.0004 & $-$0.0005 & 77 \\
1 & 53311.5325 & 0.0005 & $-$0.0009 & 44 \\
12 & 53312.3280 & 0.0011 & 0.0012 & 113 \\
22 & 53313.0479 & 0.0012 & $-$0.0001 & 166 \\
23 & 53313.1210 & 0.0006 & 0.0009 & 183 \\
24 & 53313.1933 & 0.0008 & 0.0011 & 305 \\
25 & 53313.2648 & 0.0007 & 0.0005 & 366 \\
35 & 53313.9865 & 0.0005 & 0.0009 & 131 \\
36 & 53314.0586 & 0.0005 & 0.0010 & 260 \\
37 & 53314.1318 & 0.0003 & 0.0020 & 320 \\
38 & 53314.2018 & 0.0005 & $-$0.0002 & 226 \\
39 & 53314.2753 & 0.0003 & 0.0013 & 304 \\
42 & 53314.4881 & 0.0007 & $-$0.0023 & 71 \\
43 & 53314.5592 & 0.0008 & $-$0.0034 & 71 \\
60 & 53315.7811 & 0.0020 & $-$0.0075 & 16 \\
61 & 53315.8605 & 0.0017 & $-$0.0002 & 17 \\
73 & 53316.7295 & 0.0028 & 0.0033 & 22 \\
74 & 53316.8010 & 0.0014 & 0.0027 & 23 \\
75 & 53316.8705 & 0.0021 & 0.0000 & 16 \\
\hline
  \multicolumn{5}{l}{$^{a}$ BJD$-$2400000.} \\
  \multicolumn{5}{l}{$^{b}$ Against $max = 2453311.4613 + 0.072122 E$.} \\
  \multicolumn{5}{l}{$^{c}$ Number of points used to determine the maximum.} \\
\end{tabular}
\end{center}
\end{table}

\subsection{HV Aurigae}\label{obj:hvaur}

  \citet{nog95hvaur} reported a superhump period of 0.0855(1) d.
During the 2002 superoutburst, we undertook an observing campaign.
The observation confirmed
the periodicity in \citet{nog95hvaur}.  The measured superhump maxima
are listed in table \ref{tab:hvauroc}.  The data did not show a clear
tendency of period changes.  A high-quality subset ($O-C$'s with errors
less than 0.0015 d) of superhump times gives a virtually zero
($-3.5(5.0) \times 10^{-5}$) period change.  The object looks similar
to BF Ara, another long-period system with a relatively constant
superhump period, although we can not exclude the possibility that we
observed only the stage C superhumps since the start of the outburst
was unknown.

\begin{table}
\caption{Superhump maxima of HV Aur (2002).}\label{tab:hvauroc}
\begin{center}
\begin{tabular}{ccccc}
\hline\hline
$E$ & max$^a$ & error & $O-C^b$ & $N^c$ \\
\hline
0 & 52605.9359 & 0.0038 & $-$0.0044 & 83 \\
1 & 52606.0296 & 0.0049 & 0.0037 & 106 \\
2 & 52606.1087 & 0.0022 & $-$0.0028 & 148 \\
13 & 52607.0547 & 0.0009 & 0.0020 & 477 \\
14 & 52607.1391 & 0.0010 & 0.0009 & 316 \\
15 & 52607.2239 & 0.0008 & 0.0001 & 579 \\
16 & 52607.3117 & 0.0020 & 0.0023 & 437 \\
17 & 52607.3956 & 0.0006 & 0.0007 & 71 \\
18 & 52607.4795 & 0.0007 & $-$0.0009 & 60 \\
20 & 52607.6508 & 0.0006 & $-$0.0008 & 58 \\
24 & 52607.9912 & 0.0044 & $-$0.0026 & 214 \\
25 & 52608.0808 & 0.0010 & 0.0015 & 217 \\
29 & 52608.4213 & 0.0005 & $-$0.0003 & 140 \\
30 & 52608.5063 & 0.0005 & $-$0.0009 & 139 \\
32 & 52608.6785 & 0.0006 & 0.0001 & 44 \\
33 & 52608.7650 & 0.0008 & 0.0011 & 40 \\
41 & 52609.4485 & 0.0003 & 0.0001 & 88 \\
42 & 52609.5330 & 0.0003 & $-$0.0009 & 68 \\
47 & 52609.9713 & 0.0050 & 0.0095 & 134 \\
48 & 52610.0441 & 0.0038 & $-$0.0033 & 166 \\
50 & 52610.2187 & 0.0009 & 0.0002 & 160 \\
51 & 52610.2989 & 0.0041 & $-$0.0052 & 223 \\
59 & 52610.9945 & 0.0045 & 0.0060 & 161 \\
60 & 52611.0745 & 0.0024 & 0.0004 & 119 \\
62 & 52611.2385 & 0.0050 & $-$0.0067 & 224 \\
\hline
  \multicolumn{5}{l}{$^{a}$ BJD$-$2400000.} \\
  \multicolumn{5}{l}{$^{b}$ Against $max = 2452605.9403 + 0.085563 E$.} \\
  \multicolumn{5}{l}{$^{c}$ Number of points used to determine the maximum.} \\
\end{tabular}
\end{center}
\end{table}

\subsection{TT Bootis}\label{obj:ttboo}

   \citet{ole04ttboo} reported on period variation during the 2004
superoutburst.  We observed the same superoutburst and obtained
superhump maxima with higher precision than those in \citet{ole04ttboo}.
A combined list of superhump maxima and the $O-C$ diagram are given
in table \ref{tab:ttboooc2004} and figure \ref{fig:ocsamp}.
We applied a systematic correction of $+$0.0031 d to the times of
\citet{ole04ttboo} and disregarded maxima measured using Cook's
observations, which are included in our own data set and were analyzed
with a higher precision.  Although \citet{ole04ttboo} proposed
a different treatment in dividing the $O-C$ diagram, we derived
$P_{\rm dot}$ = $+8.3(0.7) \times 10^{-5}$ from the segment
$13 \le E \le 120$ (stage B) by analogy with other systems with similar
$O-C$ behavior (subsection \ref{sec:tendency}).
The extreme values in \citet{ole04ttboo} reflected
a transition from the stage A to B
with positive $P_{\rm dot}$, and a transition to the stage C
observed during the late course of the superoutburst.

\begin{table}
\caption{Superhump maxima of TT Boo (2004).}\label{tab:ttboooc2004}
\begin{center}
\begin{tabular}{ccccc}
\hline\hline
$E$ & max$^a$ & error & $O-C^b$ & $N^c$ \\
\hline
0 & 53161.7568 & 0.0004 & $-$0.0150 & 141 \\
1 & 53161.8360 & 0.0003 & $-$0.0138 & 213 \\
2 & 53161.9152 & 0.0003 & $-$0.0126 & 158 \\
9 & 53162.4705 & 0.0004 & $-$0.0029 & 80 \\
13 & 53162.7845 & 0.0002 & $-$0.0007 & 157 \\
14 & 53162.8624 & 0.0002 & $-$0.0008 & 138 \\
15 & 53162.9406 & 0.0002 & $-$0.0006 & 123 \\
\end{tabular}
\end{center}
\end{table}

\subsection{UZ Bootis}\label{obj:uzboo}

   UZ Boo had long been suspected to be a WZ Sge-type dwarf nova
\citep{bai79wzsge}.  Due to the lack of an outburst since 1978,
it was only in 1994 when the SU UMa-type nature of this object was
established (cf. \cite{kat01hvvir}).

\begin{figure}
  \begin{center}
    \FigureFile(88mm,110mm){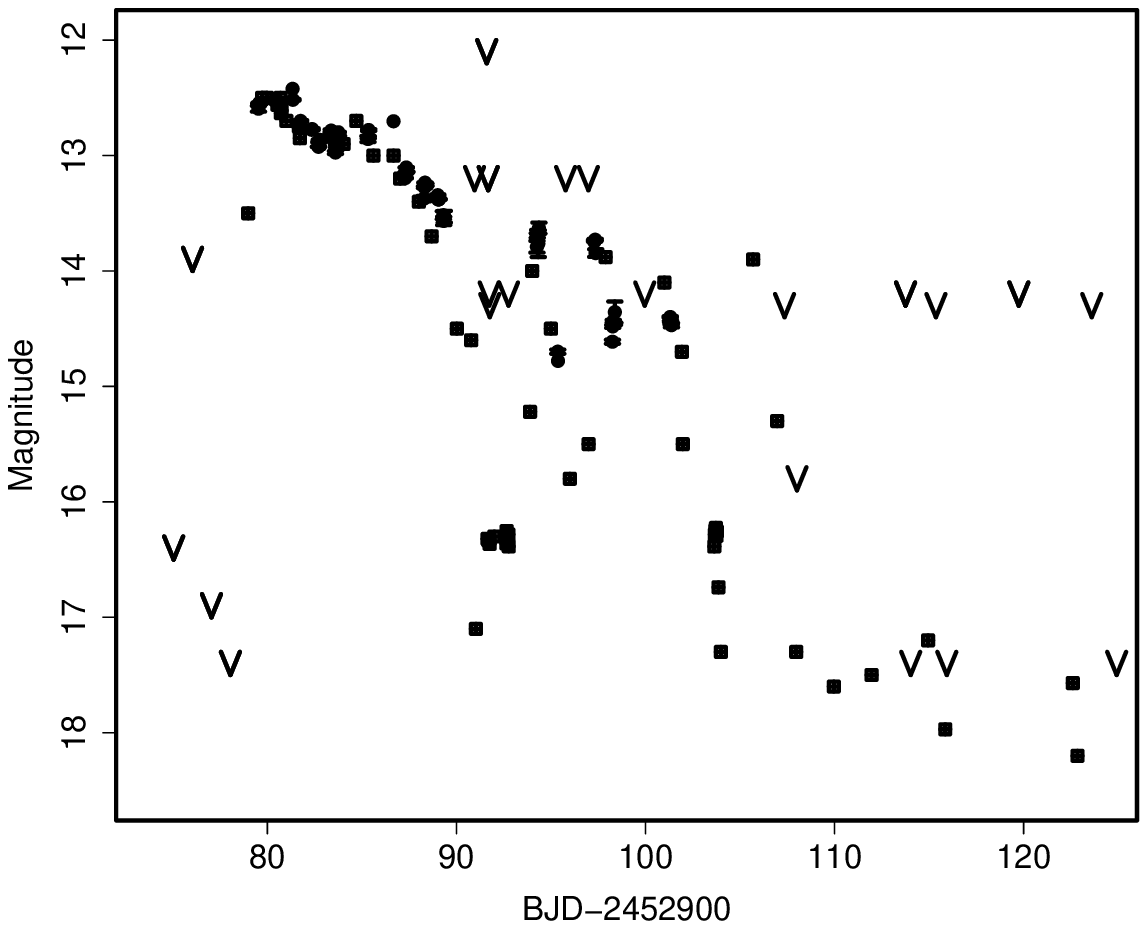}
  \end{center}
  \caption{Superoutburst of UZ Boo in 2003--2004.  The data are a combination
  of our observations (filled squares), and AAVSO and VSNET observations
  (filled squares; the ``V''-marks indicate upper limits).
  Four post-superoutburst rebrightenings were recorded.}
  \label{fig:uzboo2003lc}
\end{figure}

\begin{figure}
  \begin{center}
    \FigureFile(88mm,110mm){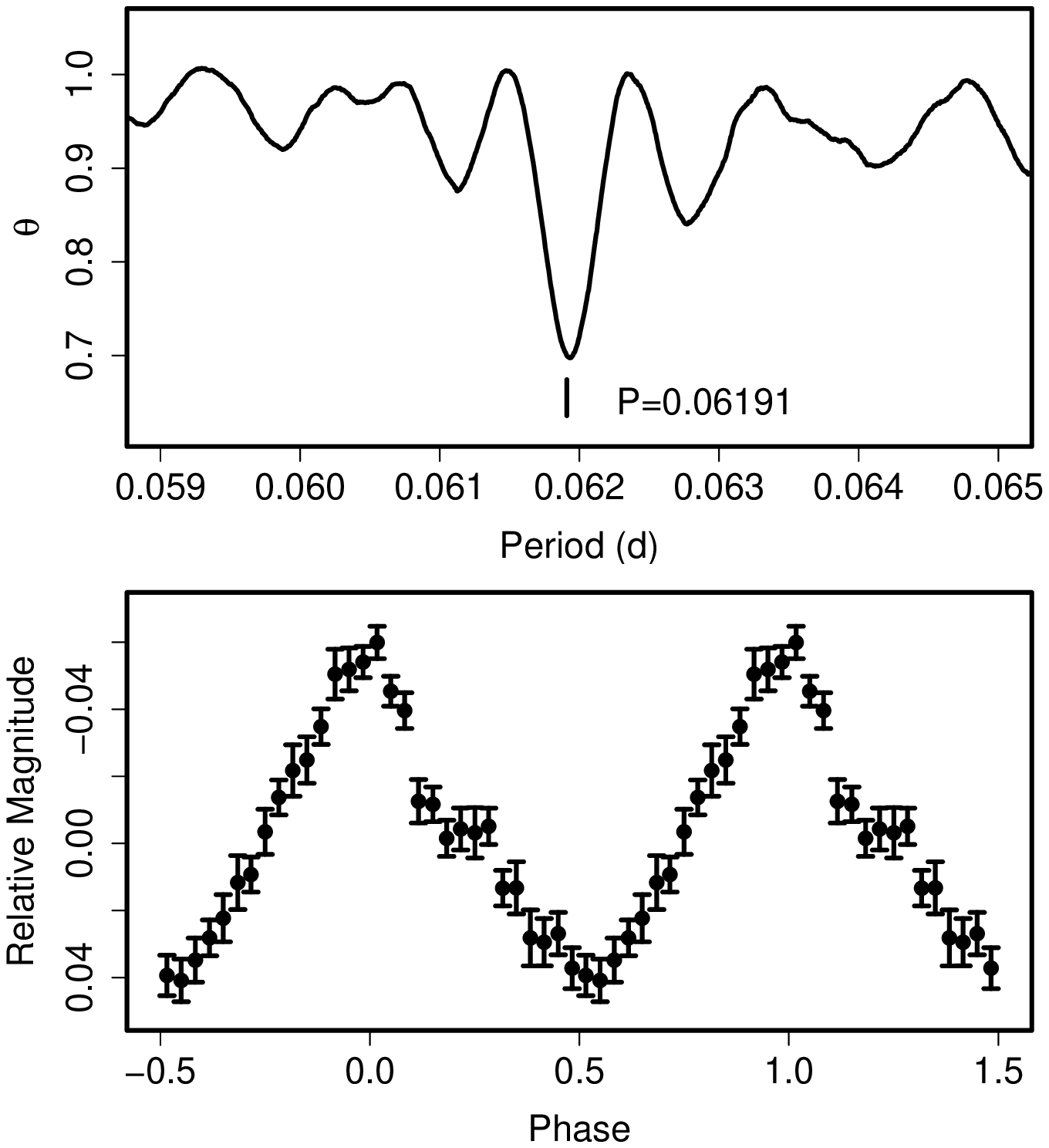}
  \end{center}
  \caption{Superhumps in UZ Boo (2003). (Upper): PDM analysis between
     BJD 2452983 and 2452991.
     (Lower): Phase-averaged profile.}
  \label{fig:uzbooshpdm}
\end{figure}

   The 2003--2004 superoutburst was detected by P. Dubovsky on 2003
December 5 (vsnet-alert 7937).  The true superhump period was finally
identified (vsnet-alert 7952).  The object underwent four rebrightenings
(vsnet-alert 7954, 7960, 7962, 7967) following the main
superoutburst (figure \ref{fig:uzboo2003lc}).\footnote{
  The 1994 superoutburst possibly had two rebrightenings \citep{kuu96TOAD}.
}
Due to the poor seasonal location, the quality of the observation
was not always very good.  We selected the superhump period of
0.06191(2) d with the PDM method (figure \ref{fig:uzbooshpdm})
for the best sampled segment between BJD 2452983 and 2452991.
The times of superhump maxima and cycle counts identified with this
period are listed in table \ref{tab:uzboooc2003}.
Although the superhump period was almost constant for $E \ge 30$
(with mean $P_{\rm SH}$ and $P_{\rm dot}$ of 0.06192(3) d and
$-1.9(6.3) \times 10^{-5}$, respectively), there was clear evidence
of period evolution before $E = 30$.  We identified this segment
to be stage A with a mean $P_{\rm SH}$ of 0.0635(2) d, lasting for
$\sim$ 30 superhump cycles.
The relatively long $P_{\rm SH}$, the lack of period
variation during the stage B and the presence of multiple rebrightenings
make UZ Boo a system analogous to EG Cnc
(\cite{pat98egcnc}; \cite{kat04egcnc}).

   The times of superhump maxima during the 1994 superoutburst were
analyzed using the $P_{\rm SH}$ identified during the 2003 superoutburst
(table \ref{tab:uzboooc1994}).  Although the maximum at $E=0$ was on
a smooth extrapolation of later maxima, this could have been an
early superhump.  The resultant $P_{\rm SH}$ and
$P_{\rm dot}$ were 0.06174(4) d and $-1.5(2.5) \times 10^{-5}$,
respectively.

\begin{table}
\caption{Superhump maxima of UZ Boo (1994).}\label{tab:uzboooc1994}
\begin{center}
\begin{tabular}{ccccc}
\hline\hline
$E$ & max$^a$ & error & $O-C^b$ & $N^c$ \\
\hline
0 & 49582.9956 & 0.0009 & $-$0.0012 & 106 \\
81 & 49587.9933 & 0.0019 & $-$0.0047 & 40 \\
97 & 49588.9930 & 0.0036 & 0.0071 & 41 \\
161 & 49592.9429 & 0.0027 & 0.0055 & 63 \\
177 & 49593.9195 & 0.0077 & $-$0.0058 & 57 \\
178 & 49593.9862 & 0.0019 & $-$0.0009 & 48 \\
\hline
  \multicolumn{5}{l}{$^{a}$ BJD$-$2400000.} \\
  \multicolumn{5}{l}{$^{b}$ Against $max = 2449582.9968 + 0.061743 E$.} \\
  \multicolumn{5}{l}{$^{c}$ Number of points used to determine the maximum.} \\
\end{tabular}
\end{center}
\end{table}

\begin{table}
\caption{Superhump maxima of UZ Boo (2003--2004).}\label{tab:uzboooc2003}
\begin{center}
\begin{tabular}{ccccc}
\hline\hline
$E$ & max$^a$ & error & $O-C^b$ & $N^c$ \\
\hline
0 & 52981.7156 & 0.0009 & $-$0.0263 & 79 \\
10 & 52982.3523 & 0.0029 & $-$0.0109 & 90 \\
16 & 52982.7409 & 0.0007 & 0.0049 & 72 \\
30 & 52983.6210 & 0.0007 & 0.0153 & 34 \\
31 & 52983.6789 & 0.0005 & 0.0111 & 82 \\
32 & 52983.7415 & 0.0005 & 0.0116 & 74 \\
58 & 52985.3526 & 0.0020 & 0.0073 & 35 \\
90 & 52987.3377 & 0.0009 & 0.0044 & 239 \\
106 & 52988.3281 & 0.0016 & 0.0008 & 225 \\
107 & 52988.3793 & 0.0033 & $-$0.0101 & 92 \\
117 & 52989.0067 & 0.0042 & $-$0.0040 & 76 \\
118 & 52989.0687 & 0.0006 & $-$0.0042 & 87 \\
\hline
  \multicolumn{5}{l}{$^{a}$ BJD$-$2400000.} \\
  \multicolumn{5}{l}{$^{b}$ Against $max = 2452981.7419 + 0.062127 E$.} \\
  \multicolumn{5}{l}{$^{c}$ Number of points used to determine the maximum.} \\
\end{tabular}
\end{center}
\end{table}

\subsection{NN Camelopardalis}\label{obj:nncam}

   NN Cam = NSV 1485 is a recently identified SU UMa-type dwarf nova
(\cite{khr05nsv1485}; for more historical information, see
vsnet-alert 9557), whose outburst was detected
on 2007 September 11.  Although this outburst rapidly faded,
a genuine superoutburst followed after eight days (vsnet-alert 9598).

\begin{figure}
  \begin{center}
    \FigureFile(88mm,90mm){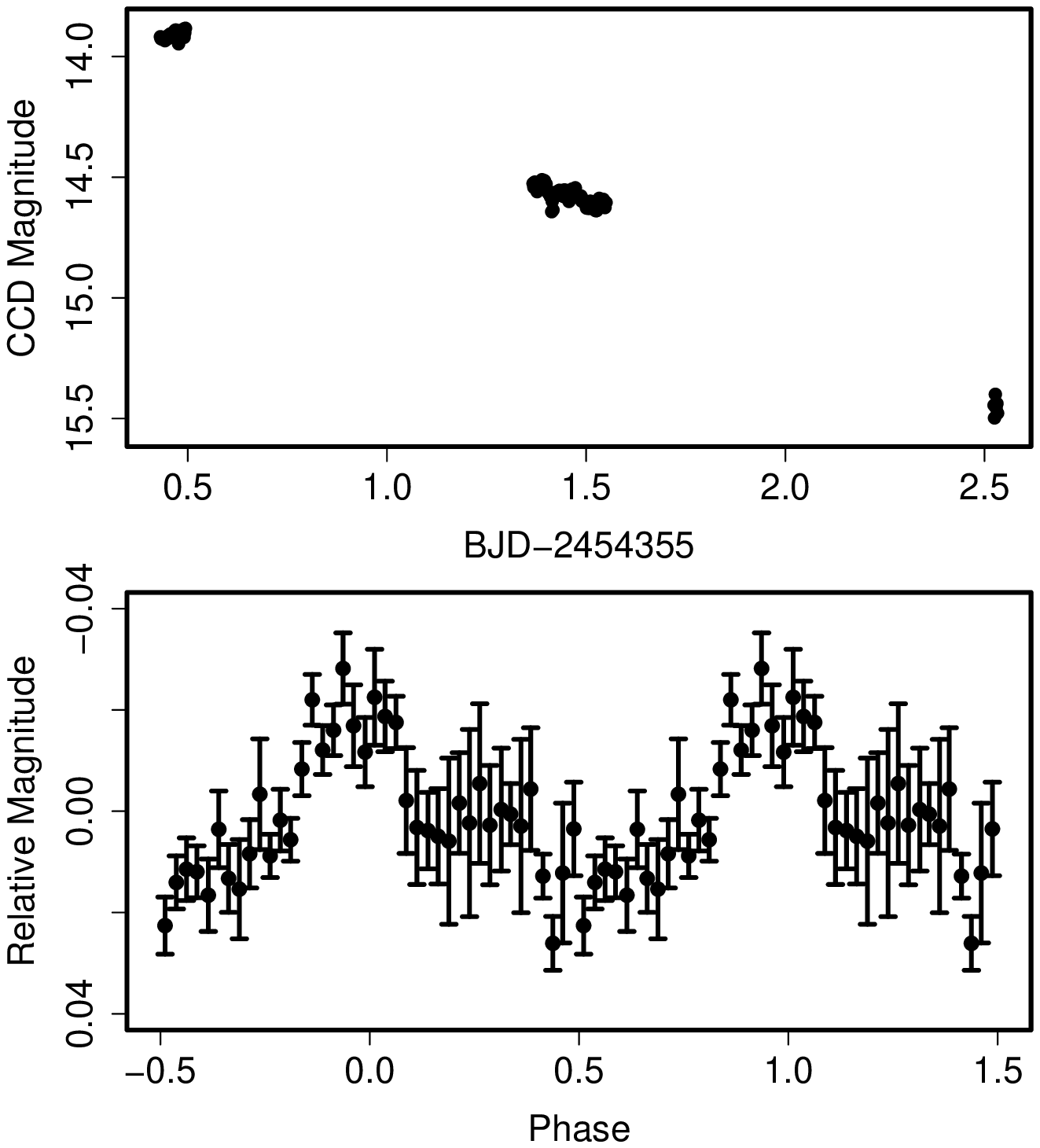}
  \end{center}
  \caption{Precursor outburst of NN Cam in 2007 (Upper): Light curve.
     (Lower): Phase-averaged profile referring to the orbital period.}
  \label{fig:nncamprec}
\end{figure}

   The times of superhump maxima obtained during this superoutburst
are listed in table \ref{tab:nncamoc2007}.  A stage B--C transition was
probably recorded.  Using the orbital period of 0.0717 d determined
photometrically (vsnet-alert 9557), we obtained a fractional superhump
excess for $P_2$ of 3.0 \%.
During the precursor, low-amplitude modulations
were observed (figure \ref{fig:nncamprec}).  Although the duration of
the observation was not long enough, the period of the variations is
consistent with the suggested orbital period.  If this period is
confirmed, the outburst makes the second example of a transition from
the orbital period to the superhump period after the case of
the 1993 superoutburst of QZ Vir \citep{kat97tleo}.

   The object underwent normal outbursts in 2008 March and October
(vsnet-alert 10588).  Photometric observations during the 2008 October
outburst did not record modulations similar to those recorded during
the precursor outburst in 2007 September.  This suggests that some
kind of (immature) superhumps were indeed excited during this
precursor outburst in 2007.

\begin{table}
\caption{Superhump maxima of NN Cam.}\label{tab:nncamoc2007}
\begin{center}
\begin{tabular}{ccccc}
\hline\hline
$E$ & max$^a$ & error & $O-C^b$ & $N^c$ \\
\hline
0 & 54363.5492 & 0.0004 & $-$0.0070 & 83 \\
13 & 54364.5159 & 0.0002 & $-$0.0014 & 260 \\
14 & 54364.5891 & 0.0003 & $-$0.0022 & 301 \\
24 & 54365.3323 & 0.0005 & 0.0017 & 57 \\
25 & 54365.4071 & 0.0006 & 0.0026 & 83 \\
26 & 54365.4816 & 0.0005 & 0.0032 & 173 \\
27 & 54365.5551 & 0.0003 & 0.0027 & 387 \\
28 & 54365.6289 & 0.0006 & 0.0025 & 191 \\
39 & 54366.4393 & 0.0003 & $-$0.0003 & 87 \\
40 & 54366.5162 & 0.0003 & 0.0027 & 324 \\
41 & 54366.5879 & 0.0002 & 0.0005 & 402 \\
54 & 54367.5477 & 0.0005 & $-$0.0009 & 207 \\
55 & 54367.6229 & 0.0003 & 0.0004 & 250 \\
66 & 54368.4375 & 0.0006 & 0.0016 & 87 \\
67 & 54368.5094 & 0.0005 & $-$0.0004 & 86 \\
81 & 54369.5403 & 0.0007 & $-$0.0046 & 267 \\
82 & 54369.6179 & 0.0007 & $-$0.0009 & 218 \\
\hline
  \multicolumn{5}{l}{$^{a}$ BJD$-$2400000.} \\
  \multicolumn{5}{l}{$^{b}$ Against $max = 2454363.5562 + 0.073935 E$.} \\
  \multicolumn{5}{l}{$^{c}$ Number of points used to determine the maximum.} \\
\end{tabular}
\end{center}
\end{table}

\subsection{SY Capriconi}\label{obj:sycap}

\begin{figure}
  \begin{center}
    \FigureFile(88mm,110mm){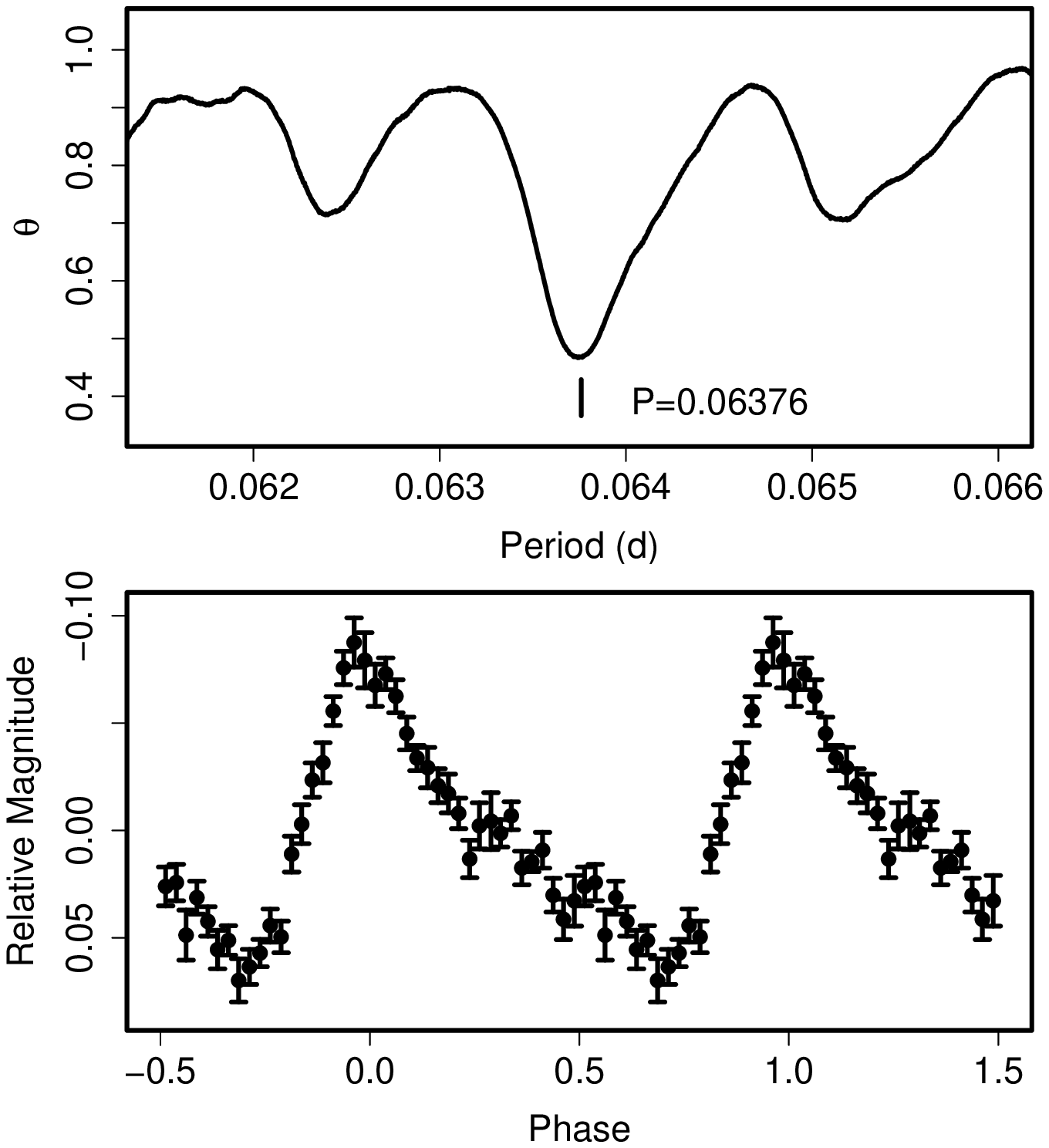}
  \end{center}
  \caption{Superhumps in SY Cap (2008). (Upper): PDM analysis.
     (Lower): Phase-averaged profile.}
  \label{fig:sycapshpdm}
\end{figure}

   SY Cap was originally classified as a long-period variable \citep{GCVS}.
The dwarf nova-type nature was pointed out by one of the authors
(T. Kato, vsnet-alert 10025).  Observations during the 2008 outburst
established the SU UMa-type nature of this object (vsnet-alert 10453,
figure \ref{fig:sycapshpdm}).
The times of superhump maxima are listed in table \ref{tab:sycapoc2008}.
The mean superhump period and global $P_{\rm dot}$ were 0.06376(2) d and
$-11.4(9.0) \times 10^{-5}$, respectively.  The negative value of
$P_{\rm dot}$ is probably a result of transition between stages B and C.
The object resembles CI UMa in its short supercycles, combined with
relatively few normal outbursts and the short duration of superoutbursts
(cf. \cite{nog97ciuma}).

\begin{table}
\caption{Superhump maxima of SY Cap (2008).}\label{tab:sycapoc2008}
\begin{center}
\begin{tabular}{ccccc}
\hline\hline
$E$ & max$^a$ & error & $O-C^b$ & $N^c$ \\
\hline
0 & 54700.0560 & 0.0006 & $-$0.0009 & 121 \\
1 & 54700.1206 & 0.0006 & 0.0000 & 120 \\
2 & 54700.1844 & 0.0009 & 0.0000 & 81 \\
14 & 54700.9505 & 0.0005 & 0.0011 & 78 \\
47 & 54703.0539 & 0.0018 & 0.0004 & 70 \\
48 & 54703.1186 & 0.0006 & 0.0013 & 92 \\
49 & 54703.1790 & 0.0010 & $-$0.0020 & 99 \\
\hline
  \multicolumn{5}{l}{$^{a}$ BJD$-$2400000.} \\
  \multicolumn{5}{l}{$^{b}$ Against $max = 2454700.0568 + 0.063759 E$.} \\
  \multicolumn{5}{l}{$^{c}$ Number of points used to determine the maximum.} \\
\end{tabular}
\end{center}
\end{table}

\subsection{AX Capriconi}\label{obj:axcap}

   AX Cap was serendipitously discovered as a dwarf nova during a
search for asteroids \citep{how94CVspec3}.  \citet{how94CVspec3}
reported a spectrum during a faint outburst.  An exceptionally
bright (15.4 mag) outburst was reported on 2004 July 17 (R. Stubbings,
vsnet-obs 50216).  The confirmation of superhumps (vsnet-campaign-dn 4337)
led to a classification as a long-$P_{\rm SH}$ SU UMa-type dwarf nova.
Table \ref{tab:axcapoc2004} lists the observed superhump maxima.
During $E \le 2$, the superhumps were still evolving.  The period smoothly
decreased with a large negative $P_{\rm dot}$ until $E=34$, then
it apparently shifted to a shorter one (figure \ref{fig:lp2}).
The $P_{\rm dot}$ for the former interval ($8 \le E \le 34$) was
$P_{\rm dot} = -87(65) \times 10^{-5}$.

   Among SU UMa-type dwarf novae, AX Cap has the second longest
$P_{\rm SH}$ next to TU Men.  Together with the large period variation
similar to MN Dra, this object certainly deserves a further detailed
study.

\begin{table}
\caption{Superhump maxima of AX Cap (2004).}\label{tab:axcapoc2004}
\begin{center}
\begin{tabular}{ccccc}
\hline\hline
$E$ & max$^a$ & error & $O-C^b$ & $N^c$ \\
\hline
0 & 53204.3758 & 0.0069 & $-$0.0824 & 259 \\
1 & 53204.4990 & 0.0101 & $-$0.0722 & 265 \\
2 & 53204.6508 & 0.0042 & $-$0.0336 & 194 \\
8 & 53205.3625 & 0.0044 & $-$0.0007 & 176 \\
9 & 53205.4690 & 0.0026 & $-$0.0073 & 256 \\
10 & 53205.5742 & 0.0053 & $-$0.0153 & 258 \\
15 & 53206.1685 & 0.0033 & 0.0134 & 62 \\
17 & 53206.4050 & 0.0014 & 0.0237 & 260 \\
18 & 53206.5227 & 0.0031 & 0.0282 & 373 \\
19 & 53206.6365 & 0.0016 & 0.0290 & 239 \\
34 & 53208.3657 & 0.0007 & 0.0612 & 257 \\
35 & 53208.4751 & 0.0007 & 0.0575 & 258 \\
36 & 53208.5862 & 0.0011 & 0.0554 & 213 \\
50 & 53210.1405 & 0.0024 & 0.0259 & 200 \\
51 & 53210.2576 & 0.0058 & 0.0299 & 119 \\
85 & 53214.0435 & 0.0034 & $-$0.0305 & 116 \\
86 & 53214.1470 & 0.0061 & $-$0.0401 & 317 \\
99 & 53215.6157 & 0.0066 & $-$0.0421 & 155 \\
\hline
  \multicolumn{5}{l}{$^{a}$ BJD$-$2400000.} \\
  \multicolumn{5}{l}{$^{b}$ Against $max = 2453204.4582 + 0.113128 E$.} \\
  \multicolumn{5}{l}{$^{c}$ Number of points used to determine the maximum.} \\
\end{tabular}
\end{center}
\end{table}

\subsection{GX Cassiopeiae}\label{obj:gxcas}

   The object has one of the longest superhump periods among known
SU UMa-type dwarf novae.  \citet{nog98gxcasv419lyr} reported the detection
of superhumps during the 1994 superoutburst.

   We further observed the 1999 and 2006 superoutbursts from the start of
the appearance of superhumps.  We also analyzed the AAVSO data of the
1996 superoutburst.
The determined times of superhump maxima are listed in
tables \ref{tab:gxcasoc1994}, \ref{tab:gxcasoc1996}, \ref{tab:gxcasoc1999} and
\ref{tab:gxcasoc2006}.

   The early and late stages were observed during the 1996 superoutburst.
Based on the identification of the $P_{\rm SH}$ during other superoutbursts,
we can unambiguously determine $E$ for each superhumps.  The results
demonstrate the clear presence of stages A and C.  The parameters are
listed in table \ref{tab:perlist}.  It might be worth noting that a PDM
analysis gave a false period (0.0862 d) due to the strong period variation,
similar to the case in CTCV J0549$-$4921 \citep{ima08fltractcv0549}.

   During the 1999 superoutburst, the object showed
a significantly longer period ($P > 0.0964$ d) for $E < 21$,
probably reflecting the stage A as in the 1996 superoutburst.
The rest of the superoutburst showed a relatively regular decrease of
the superhump period.
The mean $P_{\rm dot}$ was $-7.6(2.5) \times 10^{-5}$.
The present analysis confirmed the period identification
in \citet{nog98gxcasv419lyr}.

   The 2006 superoutburst showed a similar tendency of a
large period change during the early stage.  Such large variations
of superhump periods appear to be common in long-period SU UMa-type
dwarf novae (cf. subsection \ref{sec:longp}; \cite{rut07v419lyr}).

\begin{figure}
  \begin{center}
    \FigureFile(88mm,70mm){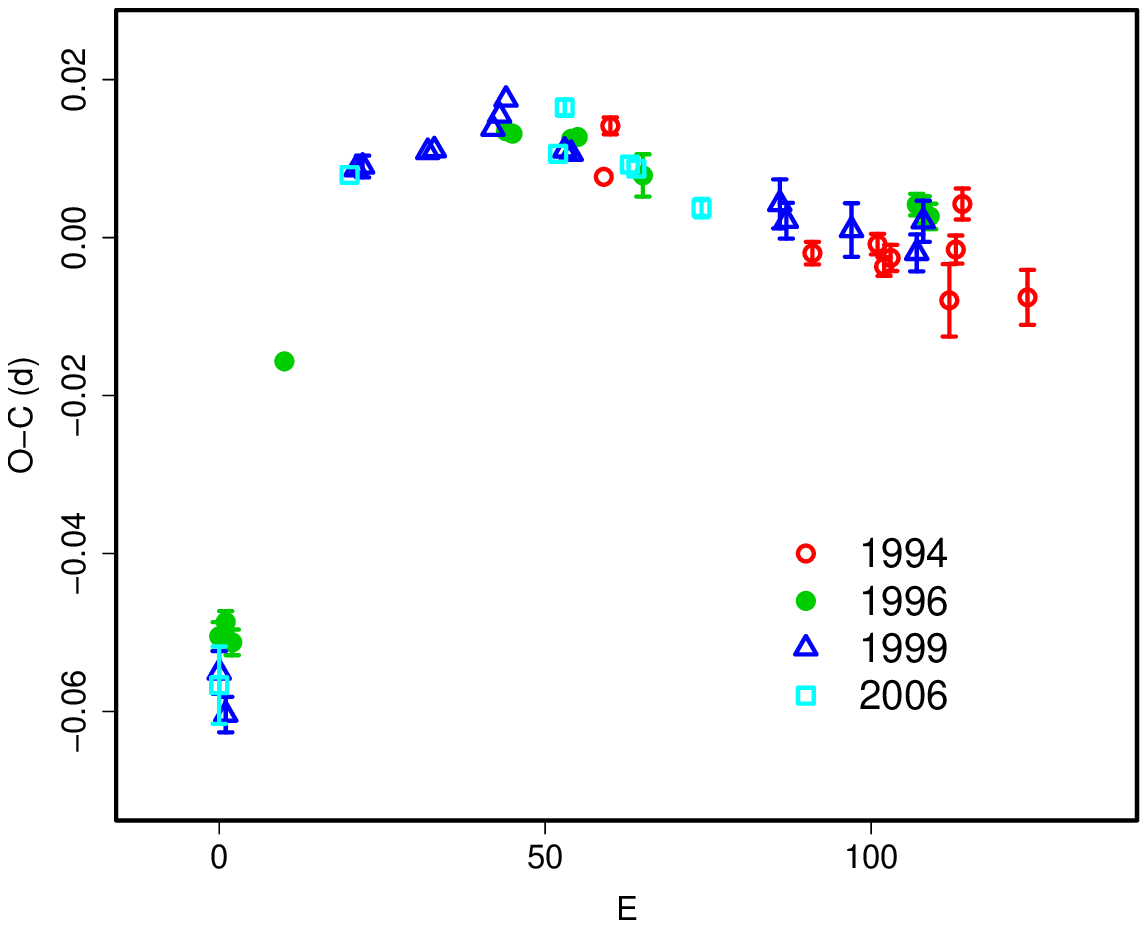}
  \end{center}
  \caption{Comparison of $O-C$ diagrams of GX Cas between different
  superoutbursts.  A period of 0.09320 d was used to draw this figure.
  Approximate cycle counts ($E$) after the start of the superoutburst
  were used.
  }
  \label{fig:gxcascomp}
\end{figure}

   A combined $O-C$ diagram (figure \ref{fig:gxcascomp}) now clearly
illustrates the period variation of superhumps in this system.
Having observed $\sim$ 5 d after the outburst detection, the 1994
observation recorded the stage C superhumps.

\begin{table}
\caption{Superhump maxima of GX Cas (1994).}\label{tab:gxcasoc1994}
\begin{center}
\begin{tabular}{ccccc}
\hline\hline
$E$ & max$^a$ & error & $O-C^b$ & $N^c$ \\
\hline
0 & 49585.1684 & 0.0003 & $-$0.0021 & 48 \\
1 & 49585.2681 & 0.0011 & 0.0046 & 20 \\
32 & 49588.1412 & 0.0014 & $-$0.0037 & 126 \\
42 & 49589.0743 & 0.0013 & $-$0.0000 & 49 \\
43 & 49589.1647 & 0.0012 & $-$0.0026 & 65 \\
44 & 49589.2590 & 0.0017 & $-$0.0013 & 67 \\
53 & 49590.0924 & 0.0046 & $-$0.0044 & 33 \\
54 & 49590.1920 & 0.0018 & 0.0023 & 32 \\
55 & 49590.2910 & 0.0020 & 0.0083 & 49 \\
65 & 49591.2112 & 0.0035 & $-$0.0010 & 39 \\
\hline
  \multicolumn{5}{l}{$^{a}$ BJD$-$2400000.} \\
  \multicolumn{5}{l}{$^{b}$ Against $max = 2449585.1706 + 0.092947 E$.} \\
  \multicolumn{5}{l}{$^{c}$ Number of points used to determine the maximum.} \\
\end{tabular}
\end{center}
\end{table}

\begin{table}
\caption{Superhump maxima of GX Cas (1996).}\label{tab:gxcasoc1996}
\begin{center}
\begin{tabular}{ccccc}
\hline\hline
$E$ & max$^a$ & error & $O-C^b$ & $N^c$ \\
\hline
0 & 50358.5124 & 0.0018 & $-$0.0199 & 28 \\
1 & 50358.6074 & 0.0014 & $-$0.0185 & 27 \\
2 & 50358.6980 & 0.0016 & $-$0.0216 & 11 \\
10 & 50359.4792 & 0.0009 & 0.0104 & 38 \\
44 & 50362.6771 & 0.0006 & 0.0242 & 26 \\
45 & 50362.7700 & 0.0007 & 0.0234 & 26 \\
54 & 50363.6082 & 0.0007 & 0.0187 & 22 \\
55 & 50363.7016 & 0.0007 & 0.0185 & 21 \\
65 & 50364.6287 & 0.0027 & 0.0091 & 11 \\
107 & 50368.5394 & 0.0014 & $-$0.0136 & 22 \\
108 & 50368.6320 & 0.0017 & $-$0.0147 & 20 \\
109 & 50368.7243 & 0.0016 & $-$0.0160 & 22 \\
\hline
  \multicolumn{5}{l}{$^{a}$ BJD$-$2400000.} \\
  \multicolumn{5}{l}{$^{b}$ Against $max = 2450358.5323 + 0.093652 E$.} \\
  \multicolumn{5}{l}{$^{c}$ Number of points used to determine the maximum.} \\
\end{tabular}
\end{center}
\end{table}

\begin{table}
\caption{Superhump maxima of GX Cas (1999).}\label{tab:gxcasoc1999}
\begin{center}
\begin{tabular}{ccccc}
\hline\hline
$E$ & max$^a$ & error & $O-C^b$ & $N^c$ \\
\hline
0 & 51472.1436 & 0.0027 & $-$0.0416 & 140 \\
1 & 51472.2314 & 0.0022 & $-$0.0472 & 188 \\
21 & 51474.1645 & 0.0005 & 0.0167 & 187 \\
22 & 51474.2580 & 0.0014 & 0.0168 & 142 \\
32 & 51475.1919 & 0.0006 & 0.0160 & 187 \\
33 & 51475.2853 & 0.0008 & 0.0160 & 165 \\
42 & 51476.1268 & 0.0009 & 0.0164 & 121 \\
43 & 51476.2217 & 0.0005 & 0.0178 & 185 \\
44 & 51476.3169 & 0.0008 & 0.0195 & 121 \\
53 & 51477.1493 & 0.0005 & 0.0107 & 187 \\
54 & 51477.2421 & 0.0005 & 0.0101 & 187 \\
86 & 51480.2181 & 0.0031 & $-$0.0046 & 181 \\
87 & 51480.3092 & 0.0023 & $-$0.0070 & 138 \\
97 & 51481.2400 & 0.0034 & $-$0.0108 & 143 \\
107 & 51482.1691 & 0.0023 & $-$0.0162 & 184 \\
108 & 51482.2663 & 0.0026 & $-$0.0125 & 188 \\
\hline
  \multicolumn{5}{l}{$^{a}$ BJD$-$2400000.} \\
  \multicolumn{5}{l}{$^{b}$ Against $max = 2451472.1852 + 0.093459 E$.} \\
  \multicolumn{5}{l}{$^{c}$ Number of points used to determine the maximum.} \\
\end{tabular}
\end{center}
\end{table}

\begin{table}
\caption{Superhump maxima of GX Cas (2006).}\label{tab:gxcasoc2006}
\begin{center}
\begin{tabular}{ccccc}
\hline\hline
$E$ & max$^a$ & error & $O-C^b$ & $N^c$ \\
\hline
0 & 54069.0790 & 0.0049 & $-$0.0240 & 66 \\
20 & 54071.0076 & 0.0006 & 0.0266 & 78 \\
52 & 54073.9927 & 0.0004 & 0.0068 & 99 \\
53 & 54074.0917 & 0.0010 & 0.0119 & 72 \\
63 & 54075.0165 & 0.0009 & $-$0.0023 & 130 \\
64 & 54075.1092 & 0.0009 & $-$0.0035 & 136 \\
74 & 54076.0362 & 0.0012 & $-$0.0155 & 129 \\
\hline
  \multicolumn{5}{l}{$^{a}$ BJD$-$2400000.} \\
  \multicolumn{5}{l}{$^{b}$ Against $max = 54069.1030 + 0.093902 E$.} \\
  \multicolumn{5}{l}{$^{c}$ Number of points used to determine the maximum.} \\
\end{tabular}
\end{center}
\end{table}

\subsection{HT Cassiopeiae}\label{obj:htcas}

   The only superoutburst observed for superhumps was in 1985
\citep{zha86htcas}, who reported $P_{\rm SH}$ of 0.076077 d without
giving details.  Although this observations were based on only two
nights, we extracted the observations from the published light curves
by referring to published times of eclipses and obtained times of
superhump maxima (table \ref{tab:htcasoc1985}).  Since the determination
of the maximum at $E=0$ was affected by the lack of observations before
the maximum, we calculated the period by using two remaining maxima.
The nominal $P_{\rm SH}$ was 0.07592(2) d, giving a slightly smaller
$\epsilon$ of 3.0 \% than in \citet{zha86htcas}.

\begin{table}
\caption{Superhump maxima of HT Cas (1985).}\label{tab:htcasoc1985}
\begin{center}
\begin{tabular}{ccccc}
\hline\hline
$E$ & max$^a$ & error & $O-C^b$ \\
\hline
0 & 46084.5704 & 0.0004 & $-$0.0069 \\
1 & 46084.6612 & 0.0001 & 0.0074 \\
14 & 46085.6482 & 0.0001 & $-$0.0005 \\
\hline
  \multicolumn{4}{l}{$^{a}$ BJD$-$2400000.} \\
  \multicolumn{4}{l}{$^{b}$ Against $max = 2446084.5773 + 0.076529 E$.} \\
\end{tabular}
\end{center}
\end{table}

\subsection{KP Cassiopeiae}\label{obj:kpcas}

\begin{figure}
  \begin{center}
    \FigureFile(88mm,110mm){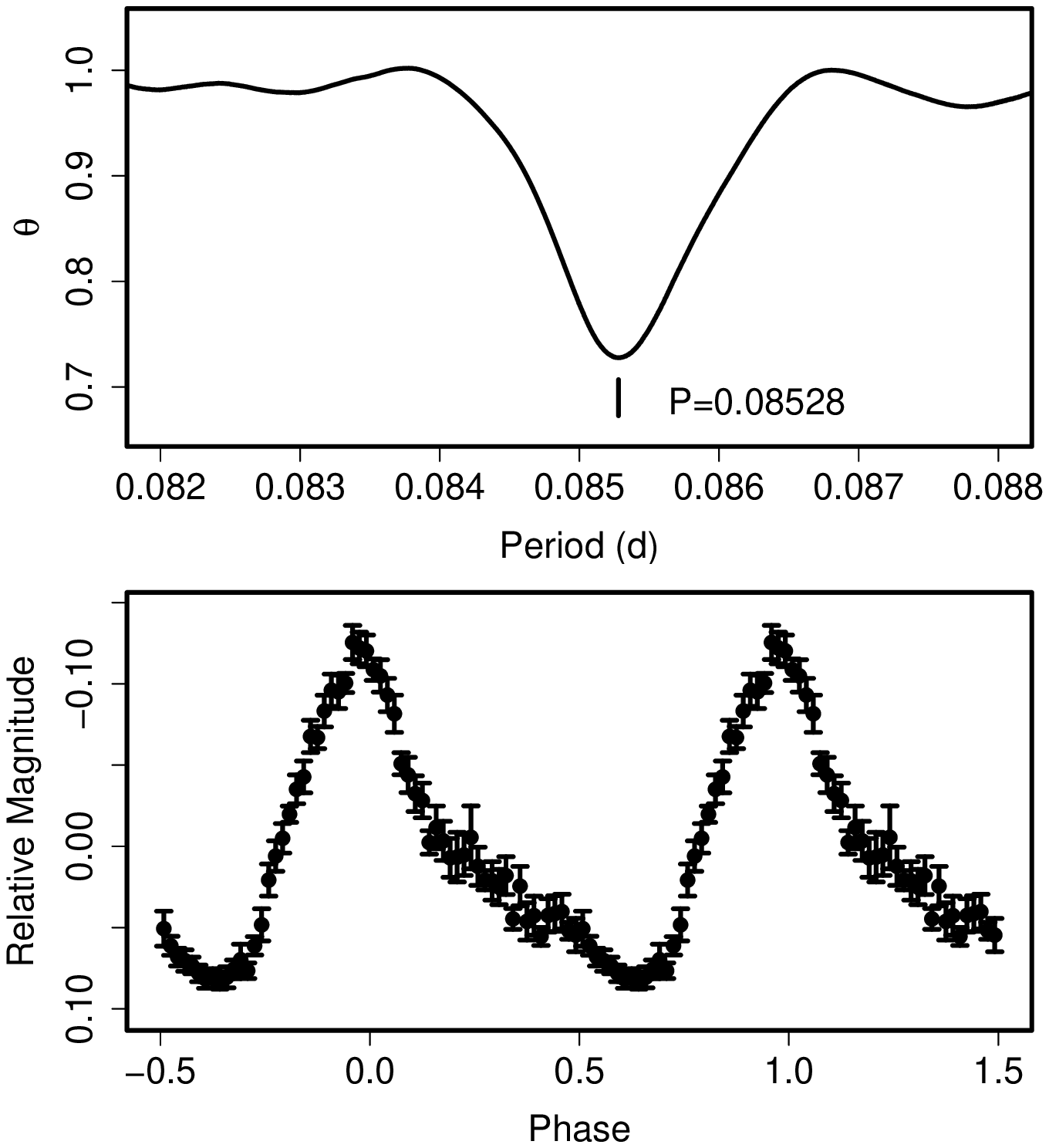}
  \end{center}
  \caption{Superhumps in KP Cas (2008). (Upper): PDM analysis.
     (Lower): Phase-averaged profile.}
  \label{fig:kpcasshpdm}
\end{figure}

   Little had been known about KP Cas before the detection of a bright
outburst by Y. Sano (vsnet-alert 10629).  The outburst soon turned
out to be a superoutburst.  The mean superhump period with the PDM
method was 0.085283(12) d (figure \ref{fig:kpcasshpdm}).
The times of superhump maxima are listed
in table \ref{tab:kpcasoc2008}.  The outburst was apparently detected
during the middle-to-late stage, and a clear transition of the superhump
period (stage B to C) was detected around $E = 15$.

\begin{table}
\caption{Superhump maxima of KP Cas (2008).}\label{tab:kpcasoc2008}
\begin{center}
\begin{tabular}{ccccc}
\hline\hline
$E$ & max$^a$ & error & $O-C^b$ & $N^c$ \\
\hline
0 & 54767.0256 & 0.0003 & $-$0.0020 & 130 \\
1 & 54767.1100 & 0.0004 & $-$0.0029 & 133 \\
3 & 54767.2827 & 0.0005 & $-$0.0008 & 71 \\
4 & 54767.3600 & 0.0014 & $-$0.0087 & 139 \\
5 & 54767.4551 & 0.0014 & 0.0011 & 183 \\
6 & 54767.5440 & 0.0012 & 0.0047 & 136 \\
8 & 54767.7096 & 0.0005 & $-$0.0002 & 37 \\
\end{tabular}
\end{center}
\end{table}

\subsection{V452 Cassiopeiae}\label{obj:v452cas}

\begin{figure}
  \begin{center}
    \FigureFile(88mm,70mm){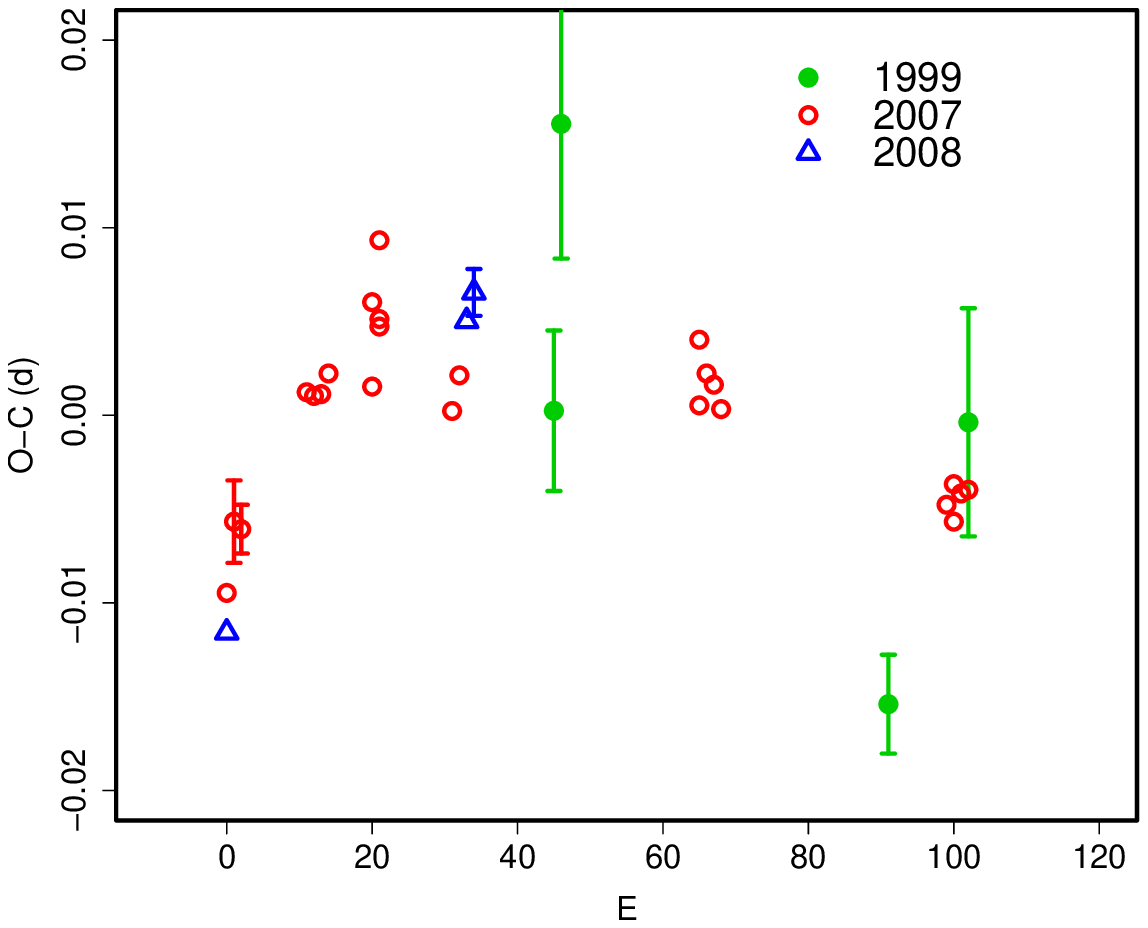}
  \end{center}
  \caption{Comparison of $O-C$ diagrams of V452 Cas between different
  superoutbursts.  A period of 0.08880 d was used to draw this figure.
  Approximate cycle counts ($E$) after the start of the superoutburst
  were used.
  }
  \label{fig:v452cascomp}
\end{figure}

   In addition to \citet{she08v452cas}, we analyzed the 1999 superoutburst
and the AAVSO data during the 2008 December superoutburst (tables
\ref{tab:v452casoc1999}, \ref{tab:v452casoc2008}).
The 1999 observation covered the middle-to-late stage of the superoutburst.
A PDM analysis yielded a mean $P_{\rm SH}$ of 0.08856(6) d, which
probably corresponds to the stage C superhumps.
The 2008 observations recorded the early part of this superoutburst
and yielded a slightly shorter $P_{\rm SH}$ of 0.08932(3) d than
in \citet{she08v452cas}.  A combined $O-C$ diagram is shown in
figure \ref{fig:v452cascomp}.

\begin{table}
\caption{Superhump maxima of V452 Cas (1999).}\label{tab:v452casoc1999}
\begin{center}
\begin{tabular}{ccccc}
\hline\hline
$E$ & max$^a$ & error & $O-C^b$ & $N^c$ \\
\hline
0 & 51496.2273 & 0.0043 & $-$0.0067 & 153 \\
1 & 51496.3314 & 0.0072 & 0.0088 & 121 \\
46 & 51500.2964 & 0.0026 & $-$0.0100 & 17 \\
57 & 51501.2882 & 0.0061 & 0.0079 & 54 \\
\hline
  \multicolumn{5}{l}{$^{a}$ BJD$-$2400000.} \\
  \multicolumn{5}{l}{$^{b}$ Against $max = 2451496.2340 + 0.088532 E$.} \\
  \multicolumn{5}{l}{$^{c}$ Number of points used to determine the maximum.} \\
\end{tabular}
\end{center}
\end{table}

\begin{table}
\caption{Superhump maxima of V452 Cas (2008).}\label{tab:v452casoc2008}
\begin{center}
\begin{tabular}{ccccc}
\hline\hline
$E$ & max$^a$ & error & $O-C^b$ & $N^c$ \\
\hline
0 & 54805.3812 & 0.0005 & 0.0000 & 76 \\
33 & 54808.3282 & 0.0009 & $-$0.0005 & 100 \\
34 & 54808.4186 & 0.0012 & 0.0005 & 74 \\
\hline
  \multicolumn{5}{l}{$^{a}$ BJD$-$2400000.} \\
  \multicolumn{5}{l}{$^{b}$ Against $max = 2454805.3812 + 0.089319 E$.} \\
  \multicolumn{5}{l}{$^{c}$ Number of points used to determine the maximum.} \\
\end{tabular}
\end{center}
\end{table}

\subsection{V359 Centauri}\label{obj:v359cen}

   We reanalyzed the data of the 2002 superoutburst
\citep{kat02v359cen}.  The result (table \ref{tab:v359cenoc2002})
generally confirmed the conclusion
in \citet{kat02v359cen}: the global $P_{\rm dot}$ was
$-16.3(1.7) \times 10^{-5}$ while $P_{\rm dot}$ for $E > 22$
was $-9.4(3.0) \times 10^{-5}$ (see discussion in \cite{kat02v359cen}
for a selection of the interval).  We adopted the latter as being
the representative $P_{\rm dot}$ for this object.

\begin{table}
\caption{Superhump maxima of V359 Cen (2002).}\label{tab:v359cenoc2002}
\begin{center}
\begin{tabular}{ccccc}
\hline\hline
$E$ & max$^a$ & error & $O-C^b$ & $N^c$ \\
\hline
0 & 52423.9416 & 0.0004 & $-$0.0137 & 106 \\
1 & 52424.0235 & 0.0003 & $-$0.0129 & 130 \\
23 & 52425.8215 & 0.0004 & 0.0025 & 109 \\
35 & 52426.7972 & 0.0007 & 0.0058 & 94 \\
36 & 52426.8804 & 0.0006 & 0.0080 & 195 \\
37 & 52426.9619 & 0.0009 & 0.0084 & 71 \\
49 & 52427.9337 & 0.0004 & 0.0078 & 166 \\
50 & 52428.0149 & 0.0009 & 0.0080 & 107 \\
62 & 52428.9865 & 0.0004 & 0.0072 & 86 \\
84 & 52430.7602 & 0.0007 & $-$0.0018 & 47 \\
85 & 52430.8391 & 0.0018 & $-$0.0040 & 11 \\
102 & 52432.2163 & 0.0008 & $-$0.0043 & 91 \\
103 & 52432.2971 & 0.0009 & $-$0.0046 & 92 \\
104 & 52432.3762 & 0.0012 & $-$0.0065 & 92 \\
\hline
  \multicolumn{5}{l}{$^{a}$ BJD$-$2400000.} \\
  \multicolumn{5}{l}{$^{b}$ Against $max = 2452423.9553 + 0.081033 E$.} \\
  \multicolumn{5}{l}{$^{c}$ Number of points used to determine the maximum.} \\
\end{tabular}
\end{center}
\end{table}

\subsection{V485 Centauri}\label{obj:v485cen}

   The period evolution of this ultrashort-$P_{\rm orb}$ SU UMa-type
dwarf nova was studied by \citet{ole97v485cen}, yielding a positive
$P_{\rm dot}$ (the value has been corrected in this paper, see
subsection \ref{sec:pdotb}).

We observed the 2001 superoutburst (table \ref{tab:v485cenoc2001}).
Although the data were rather sparse, there was again no indication of
an exceptionally large $P_{\rm dot}$.

We also examined the 2004 superoutburst using the AAVSO data
(table \ref{tab:v485cenoc2004}).  The data clearly showed a stage B--C
transition around $E=166$.  The $P_{\rm dot}$ during the stage B was
$+3.1(0.9) \times 10^{-5}$, strengthening our interpretation that
this object has an usual $P_{\rm dot}$.  The existence of the stage C
has been demonstrated for this class of objects first time in this
superoutburst.

\begin{figure}
  \begin{center}
    \FigureFile(88mm,70mm){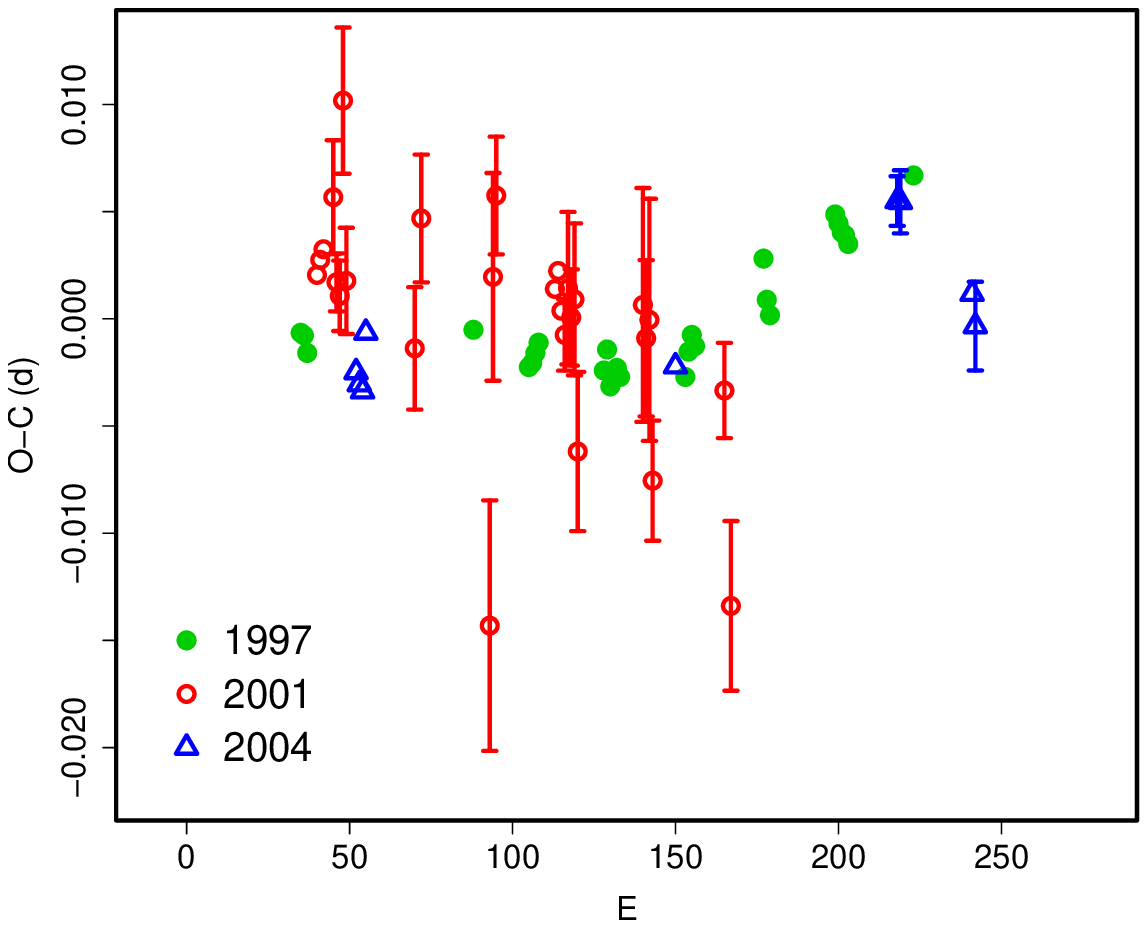}
  \end{center}
  \caption{Comparison of $O-C$ diagrams of V485 Cen between different
  superoutbursts.  A period of 0.04212 d was used to draw this figure.
  Approximate cycle counts ($E$) after the start of the superoutburst
  were used.
  }
  \label{fig:v485cencomp}
\end{figure}

A comparison of $O-C$ diagrams between different superoutbursts
is presented in figure \ref{fig:v485cencomp}.

\begin{table}
\caption{Superhump maxima of V485 Cen (2001).}\label{tab:v485cenoc2001}
\begin{center}
\begin{tabular}{ccccc}
\hline\hline
$E$ & max$^a$ & error & $O-C^b$ & $N^c$ \\
\hline
0 & 51999.8526 & 0.0004 & $-$0.0019 & 37 \\
1 & 51999.8954 & 0.0004 & $-$0.0012 & 39 \\
2 & 51999.9380 & 0.0004 & $-$0.0006 & 40 \\
5 & 52000.0668 & 0.0027 & 0.0020 & 79 \\
6 & 52000.1049 & 0.0014 & $-$0.0019 & 82 \\
7 & 52000.1464 & 0.0016 & $-$0.0024 & 82 \\
8 & 52000.1976 & 0.0034 & 0.0068 & 65 \\
9 & 52000.2314 & 0.0025 & $-$0.0016 & 53 \\
30 & 52001.1127 & 0.0029 & $-$0.0033 & 75 \\
32 & 52001.2030 & 0.0030 & 0.0029 & 68 \\
53 & 52002.0686 & 0.0058 & $-$0.0146 & 60 \\
54 & 52002.1269 & 0.0048 & 0.0018 & 76 \\
55 & 52002.1729 & 0.0027 & 0.0056 & 71 \\
73 & 52002.9267 & 0.0007 & 0.0025 & 38 \\
74 & 52002.9696 & 0.0007 & 0.0034 & 39 \\
75 & 52003.0099 & 0.0005 & 0.0017 & 40 \\
76 & 52003.0509 & 0.0017 & 0.0006 & 21 \\
77 & 52003.0952 & 0.0036 & 0.0029 & 76 \\
78 & 52003.1359 & 0.0022 & 0.0016 & 71 \\
79 & 52003.1789 & 0.0035 & 0.0025 & 44 \\
80 & 52003.2139 & 0.0037 & $-$0.0046 & 82 \\
100 & 52004.0632 & 0.0055 & 0.0037 & 81 \\
101 & 52004.1037 & 0.0036 & 0.0022 & 82 \\
102 & 52004.1467 & 0.0056 & 0.0031 & 81 \\
103 & 52004.1813 & 0.0028 & $-$0.0043 & 81 \\
125 & 52005.1122 & 0.0022 & 0.0014 & 78 \\
127 & 52005.1864 & 0.0040 & $-$0.0085 & 74 \\
\hline
  \multicolumn{5}{l}{$^{a}$ BJD$-$2400000.} \\
  \multicolumn{5}{l}{$^{b}$ Against $max = 2451999.8544 + 0.042050 E$.} \\
  \multicolumn{5}{l}{$^{c}$ Number of points used to determine the maximum.} \\
\end{tabular}
\end{center}
\end{table}

\begin{table}
\caption{Superhump maxima of V485 Cen (2004).}\label{tab:v485cenoc2004}
\begin{center}
\begin{tabular}{ccccc}
\hline\hline
$E$ & max$^a$ & error & $O-C^b$ & $N^c$ \\
\hline
0 & 53140.0437 & 0.0003 & 0.0000 & 74 \\
1 & 53140.0853 & 0.0002 & $-$0.0006 & 66 \\
2 & 53140.1271 & 0.0003 & $-$0.0009 & 74 \\
3 & 53140.1719 & 0.0004 & 0.0018 & 39 \\
98 & 53144.1718 & 0.0007 & $-$0.0024 & 42 \\
166 & 53147.0436 & 0.0012 & 0.0034 & 37 \\
167 & 53147.0857 & 0.0015 & 0.0033 & 38 \\
189 & 53148.0081 & 0.0008 & $-$0.0016 & 33 \\
190 & 53148.0487 & 0.0021 & $-$0.0031 & 41 \\
\hline
  \multicolumn{5}{l}{$^{a}$ BJD$-$2400000.} \\
  \multicolumn{5}{l}{$^{b}$ Against $max = 2453140.0437 + 0.042148 E$.} \\
  \multicolumn{5}{l}{$^{c}$ Number of points used to determine the maximum.} \\
\end{tabular}
\end{center}
\end{table}

\subsection{V1040 Centauri}\label{obj:v1040cen}

   V1040 Cen (=RX J1155.4$-$5641) is an ROSAT-selected CV
\citep{mot98ROSATCV}.  \citet{pat03suumas} reported a $P_{\rm SH}$
of 0.06215(10) d for the 2002 superoutburst.  We analyzed the same
superoutburst using the available data.  The times of superhump maxima
are listed in table \ref{tab:v1040cenoc2002}.  Except $50 \le E \le 54$,
the overall $O-C$ diagram showed typical stage A--C transitions.
The epochs of $50 \le E \le 54$ were affected by strong variation
in the superhump profile (broad maxima), which may be due to overlapping
orbital signals.  Disregarding these epochs, we obtained a strongly
positive $P_{\rm dot}$ of $+27.1(2.2) \times 10^{-5}$
($17 \le E \le 86$).  Other parameters are listed in table
\ref{tab:perlist}.  The behavior is somewhat reminiscent to ER UMa
stars (subsection \ref{sec:erumastars}).  A further analysis and
observations might shed light to further understanding period variations
and the evolution of stage C superhumps in this object and ER UMa stars.

   We used BJD 2452383--2452402 (post-outburst
rebrightening and subsequent phase) and obtained a refined photometric
period of 0.060296(8) d, which has been attributed to $P_{\rm orb}$
\citep{pat03suumas}.  No strong superhump signals were evident during
this stage.  This period, however, was not dominant during the quiescence
in 2008 (BJD 2454547--2454574).  The exact identification of $P_{\rm orb}$
should await a spectroscopic study.

\begin{table}
\caption{Superhump maxima of V1040 Cen (2002).}\label{tab:v1040cenoc2002}
\begin{center}
\begin{tabular}{ccccc}
\hline\hline
$E$ & max$^a$ & error & $O-C^b$ & $N^c$ \\
\hline
0 & 52366.2446 & 0.0002 & 0.0006 & 93 \\
1 & 52366.3075 & 0.0001 & 0.0013 & 92 \\
2 & 52366.3688 & 0.0003 & 0.0004 & 93 \\
3 & 52366.4303 & 0.0003 & $-$0.0002 & 93 \\
4 & 52366.4925 & 0.0003 & $-$0.0002 & 93 \\
5 & 52366.5560 & 0.0003 & 0.0012 & 93 \\
6 & 52366.6173 & 0.0003 & 0.0003 & 93 \\
\end{tabular}
\end{center}
\end{table}

\subsection{WX Ceti}\label{sec:wxcet}\label{obj:wxcet}

   We reanalyzed the 1998 data in \citet{kat01wxcet} combined with the AAVSO
data and obtained refined times of maxima (table \ref{tab:wxcetoc1998}).
Several newly determined maxima are also included.
The new $O-C$ diagram basically confirms the finding in \citet{kat01wxcet},
but now clearly shows three stages of A--C.
The timings of ``late superhumps'' in \citet{kat01wxcet} were
somewhat contaminated by the incorrect phase identification in the
stage C.  We obtained $P_{\rm dot}$ = $+6.4(1.0) \times 10^{-5}$ for the
stage B ($15 \le E \le 157$).

   We analyzed the 2001 superoutburst after combining our data and
those in \citet{ste07wxcet}.  The resultant times of maxima are
listed in table \ref{tab:wxcetoc2001}.  The observation well covered
the middle part of the superoutburst and yielded
$P_{\rm dot}$ = $+7.5(1.1) \times 10^{-5}$.

   The 2004 observation (table \ref{tab:wxcetoc2004}) also covered
the stages A--C.  The $P_{\rm dot}$ of the stage B was
$+5.5(1.8) \times 10^{-5}$ ($E \le 137$).

   We also analyzed the data for the 1989 superoutburst
\citep{odo91wzsge} after extracting the data from the scanned figure.
Although systematic errors may be significantly larger than the errors
given in the table, we could extract times of superhump maxima
(table \ref{tab:wxcetoc1989}).  The $O-C$ diagram clearly exhibits
stages A--C.  The $P_{\rm dot}$ of the stage B was
$+10.3(1.4) \times 10^{-5}$ ($33 \le E \le 185$).
The difficulty in determining the period in \citet{odo91wzsge} was
probably a result from this strong period variation.

\begin{figure}
  \begin{center}
    \FigureFile(88mm,70mm){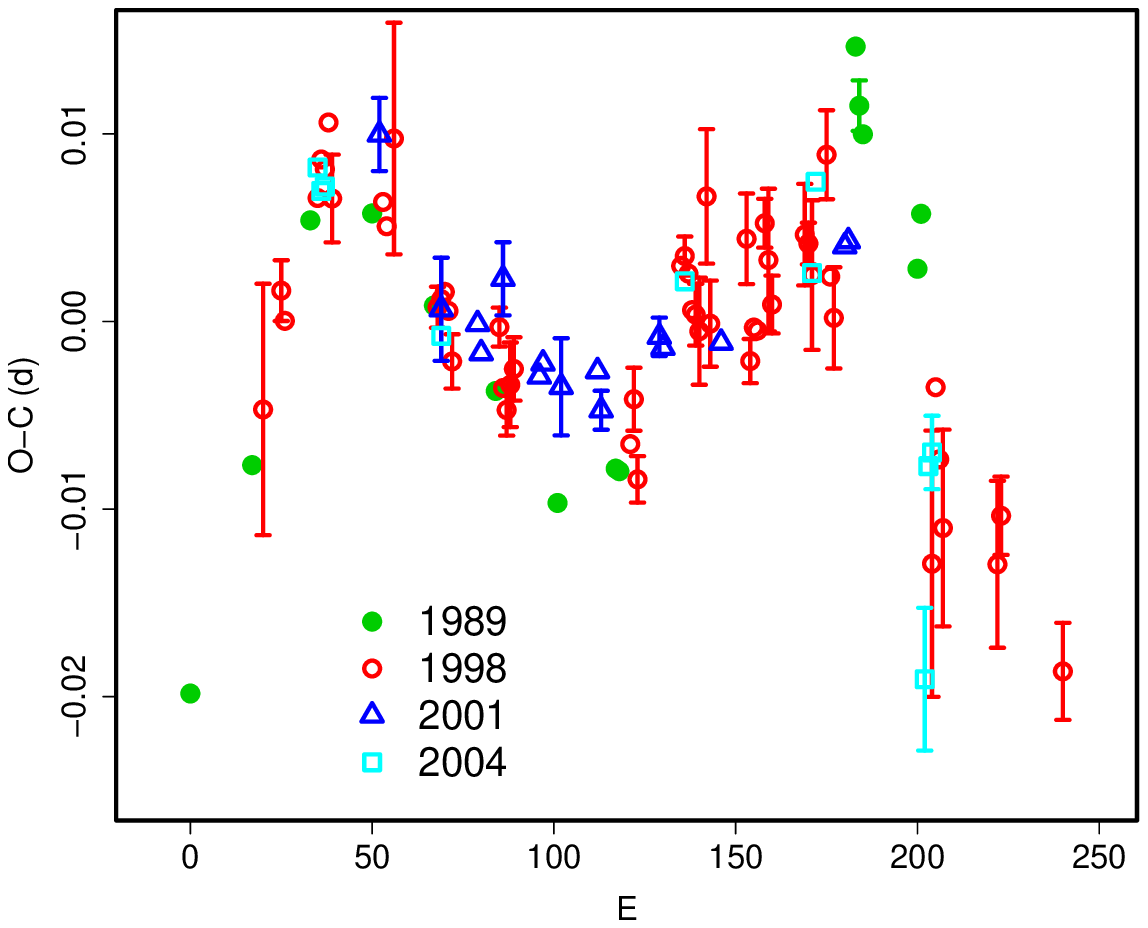}
  \end{center}
  \caption{Comparison of $O-C$ diagrams of WX Cet between different
  superoutbursts.  A period of 0.05955 d was used to draw this figure.
  Estimated cycle counts ($E$) after the appearance of the
  superhumps were used.
  }
  \label{fig:wxcetcomp}
\end{figure}

   In summary, all observed superoutbursts of WX Cet showed a similar
pattern of $O-C$ and $P_{\rm dot}$ was always positive in the
middle of the plateau phase (figure \ref{fig:wxcetcomp}).

\begin{table}
\caption{Superhump maxima of WX Cet (1998).}\label{tab:wxcetoc1998}
\begin{center}
\begin{tabular}{ccccc}
\hline\hline
$E$ & max$^a$ & error & $O-C^b$ & $N^c$ \\
\hline
0 & 51129.0492 & 0.0067 & $-$0.0100 & 114 \\
5 & 51129.3533 & 0.0016 & $-$0.0034 & 35 \\
6 & 51129.4112 & 0.0009 & $-$0.0050 & 34 \\
15 & 51129.9537 & 0.0005 & 0.0020 & 138 \\
16 & 51130.0153 & 0.0005 & 0.0041 & 128 \\
17 & 51130.0743 & 0.0006 & 0.0037 & 85 \\
18 & 51130.1364 & 0.0006 & 0.0062 & 118 \\
\end{tabular}
\end{center}
\end{table}

\begin{table}
\caption{Superhump maxima of WX Cet (2001).}\label{tab:wxcetoc2001}
\begin{center}
\begin{tabular}{ccccc}
\hline\hline
$E$ & max$^a$ & error & $O-C^b$ & $N^c$ \\
\hline
0 & 52092.2687 & 0.0019 & 0.0099 & 41 \\
17 & 52093.2717 & 0.0027 & 0.0006 & 50 \\
27 & 52093.8664 & 0.0002 & $-$0.0002 & 73 \\
28 & 52093.9244 & 0.0002 & $-$0.0017 & 67 \\
34 & 52094.2857 & 0.0019 & 0.0022 & 72 \\
44 & 52094.8760 & 0.0002 & $-$0.0030 & 72 \\
45 & 52094.9362 & 0.0003 & $-$0.0023 & 58 \\
50 & 52095.2327 & 0.0026 & $-$0.0035 & 54 \\
60 & 52095.8291 & 0.0005 & $-$0.0026 & 12 \\
61 & 52095.8865 & 0.0010 & $-$0.0047 & 15 \\
77 & 52096.8433 & 0.0010 & $-$0.0008 & 15 \\
78 & 52096.9022 & 0.0004 & $-$0.0014 & 67 \\
94 & 52097.8553 & 0.0006 & $-$0.0011 & 58 \\
128 & 52099.8851 & 0.0004 & 0.0041 & 62 \\
129 & 52099.9450 & 0.0004 & 0.0044 & 40 \\
\hline
  \multicolumn{5}{l}{$^{a}$ BJD$-$2400000.} \\
  \multicolumn{5}{l}{$^{b}$ Against $max = 2452092.2588 + 0.059549 E$.} \\
  \multicolumn{5}{l}{$^{c}$ Number of points used to determine the maximum.} \\
\end{tabular}
\end{center}
\end{table}

\begin{table}
\caption{Superhump maxima of WX Cet (2004).}\label{tab:wxcetoc2004}
\begin{center}
\begin{tabular}{ccccc}
\hline\hline
$E$ & max$^a$ & error & $O-C^b$ & $N^c$ \\
\hline
0 & 53347.9434 & 0.0001 & 0.0007 & 282 \\
1 & 53348.0017 & 0.0002 & $-$0.0005 & 279 \\
2 & 53348.0614 & 0.0002 & $-$0.0002 & 331 \\
34 & 53349.9591 & 0.0002 & $-$0.0055 & 322 \\
101 & 53353.9518 & 0.0007 & 0.0029 & 213 \\
136 & 53356.0365 & 0.0004 & 0.0062 & 213 \\
137 & 53356.1009 & 0.0006 & 0.0112 & 108 \\
167 & 53357.8609 & 0.0038 & $-$0.0129 & 196 \\
168 & 53357.9318 & 0.0008 & $-$0.0014 & 304 \\
169 & 53357.9921 & 0.0020 & $-$0.0006 & 262 \\
\hline
  \multicolumn{5}{l}{$^{a}$ BJD$-$2400000.} \\
  \multicolumn{5}{l}{$^{b}$ Against $max = 2453347.9427 + 0.0594678 E$.} \\
  \multicolumn{5}{l}{$^{c}$ Number of points used to determine the maximum.} \\
\end{tabular}
\end{center}
\end{table}

\begin{table}
\caption{Superhump maxima of WX Cet (1989).}\label{tab:wxcetoc1989}
\begin{center}
\begin{tabular}{cccc}
\hline\hline
$E$ & max$^a$ & error & $O-C^b$ \\
\hline
0 & 47683.6324 & 0.0009 & $-$0.0105 \\
17 & 47684.6569 & 0.0024 & 0.0003 \\
33 & 47685.6227 & 0.0006 & 0.0120 \\
50 & 47686.6355 & 0.0006 & 0.0109 \\
67 & 47687.6429 & 0.0003 & 0.0045 \\
84 & 47688.6507 & 0.0003 & $-$0.0015 \\
101 & 47689.6571 & 0.0006 & $-$0.0089 \\
117 & 47690.6117 & 0.0003 & $-$0.0084 \\
118 & 47690.6711 & 0.0012 & $-$0.0087 \\
183 & 47694.5645 & 0.0015 & 0.0084 \\
184 & 47694.6209 & 0.0039 & 0.0052 \\
185 & 47694.6789 & 0.0009 & 0.0036 \\
200 & 47695.5650 & 0.0021 & $-$0.0049 \\
201 & 47695.6275 & 0.0012 & $-$0.0020 \\
\hline
  \multicolumn{4}{l}{$^{a}$ BJD$-$2400000.} \\
  \multicolumn{4}{l}{$^{b}$ Against $max = 2447683.6428 + 0.059635 E$.} \\
  \multicolumn{4}{l}{$^{c}$ Number of points used to determine the maximum.} \\
\end{tabular}
\end{center}
\end{table}

\subsection{RX Chameleontis}\label{obj:rxcha}

\begin{figure}
  \begin{center}
    \FigureFile(88mm,110mm){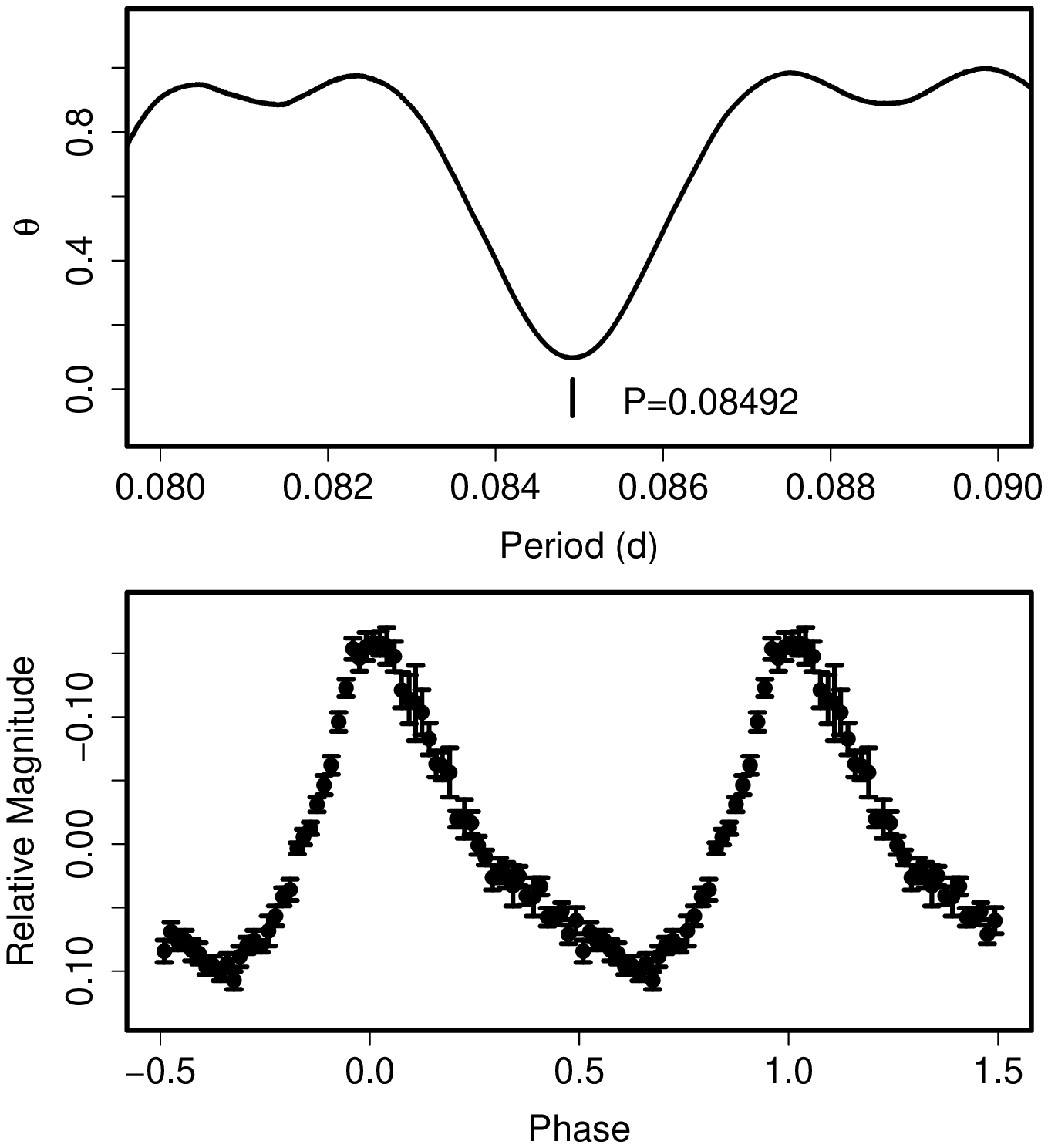}
  \end{center}
  \caption{Superhumps in RX Cha (2009). (Upper): PDM analysis excluding
     the early evolutionary stage before BJD 2454857.5).
     (Lower): Phase-averaged profile.}
  \label{fig:rxchashpdm}
\end{figure}

   \citet{kat01rxcha} analyzed the 1998 outburst and reported a superhump
period of 0.084 d.  We observed the 2009 superoutburst during the early
stage (table \ref{tab:rxchaoc2009}).  The $O-C$ diagram showed a typical
stage A--B transition.  The mean superhump period during the stage B
with the PDM method was 0.08492(2) d (figure \ref{fig:rxchashpdm}),
confirming the long-$P_{\rm SH}$ nature claimed in \citet{kat01rxcha}.

\begin{table}
\caption{Superhump maxima of RX Cha (2009).}\label{tab:rxchaoc2009}
\begin{center}
\begin{tabular}{ccccc}
\hline\hline
$E$ & max$^a$ & error & $O-C^b$ & $N^c$ \\
\hline
0 & 54857.1087 & 0.0007 & $-$0.0077 & 106 \\
10 & 54857.9797 & 0.0007 & 0.0087 & 76 \\
22 & 54859.0007 & 0.0003 & 0.0043 & 140 \\
34 & 54860.0165 & 0.0004 & $-$0.0054 & 110 \\
\hline
  \multicolumn{5}{l}{$^{a}$ BJD$-$2400000.} \\
  \multicolumn{5}{l}{$^{b}$ Against $max = 2454857.1164 + 0.085455 E$.} \\
  \multicolumn{5}{l}{$^{c}$ Number of points used to determine the maximum.} \\
\end{tabular}
\end{center}
\end{table}

\subsection{BZ Circini}\label{obj:bzcir}

\begin{figure}
  \begin{center}
    \FigureFile(88mm,110mm){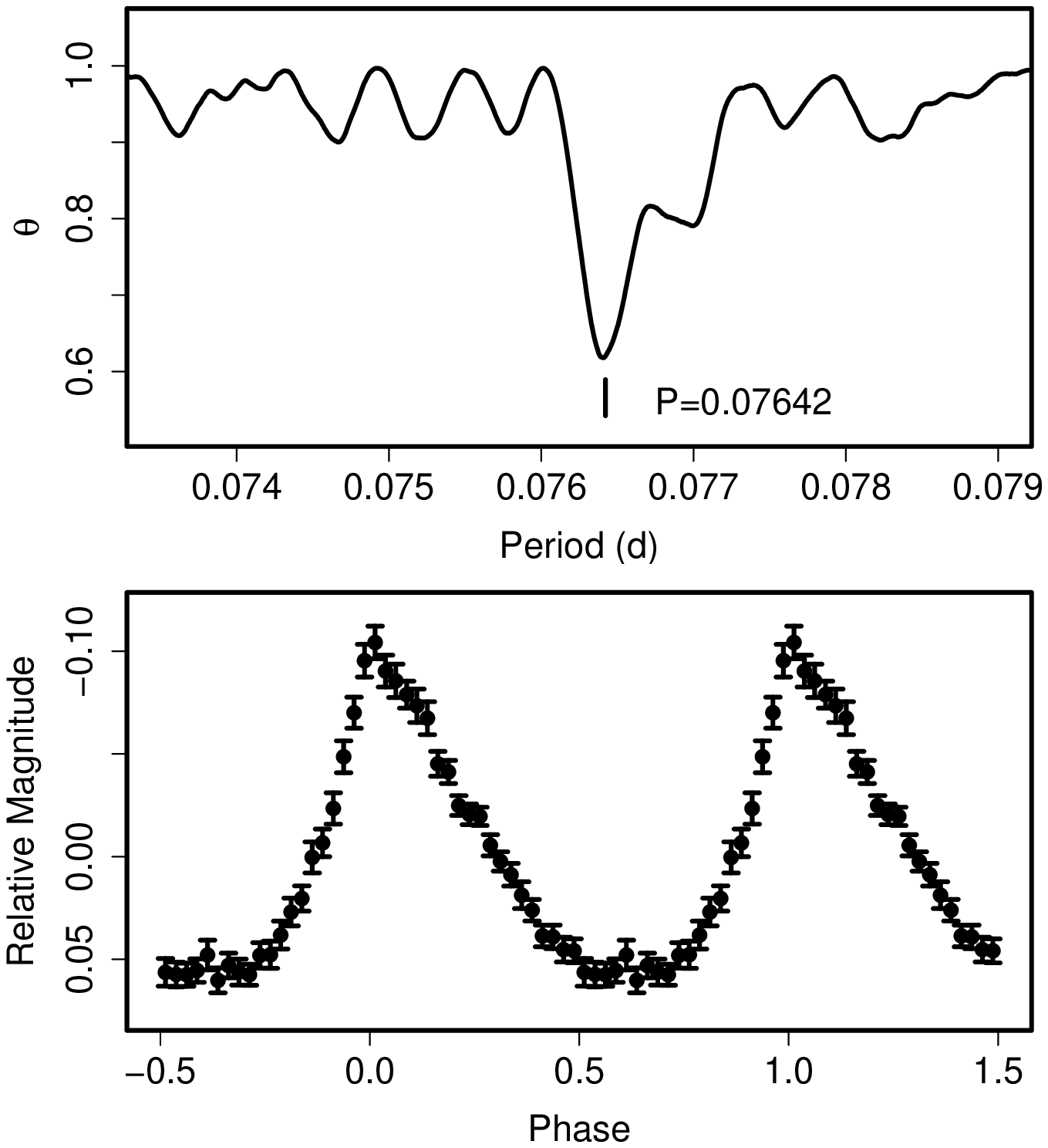}
  \end{center}
  \caption{Superhumps in BZ Cir (2004). (Upper): PDM analysis.
     (Lower): Phase-averaged profile.}
  \label{fig:bzcirshpdm}
\end{figure}

   BZ Cir is an X-ray selected CV (=1E 1449.7$-$6804:
\cite{gri87bzciriauc}; \cite{her90XrayCVs}).
The first recorded outburst was detected by B. Monard in 2004 June
(vsnet-alert 8194).  The outburst soon turned out to be a superoutburst
(vsnet-alert 8201).  We analyzed this superoutburst.
The mean superhump period with the PDM method was 0.076422(5) d
(figure \ref{fig:bzcirshpdm}).
The times of superhump maxima are listed in table \ref{tab:bzciroc2004}.
While the global $P_{\rm dot}$ corresponds to
$-6.9(0.5) \times 10^{-5}$, there was an apparent transition
of periods around $E = 68$.  The $P_{\rm dot}$ of the
middle segment ($13 \le E \le 68$) was $-0.5(3.8) \times 10^{-5}$
(cf. figure \ref{fig:ocsamp}).

\begin{table}
\caption{Superhump maxima of BZ Cir (2004).}\label{tab:bzciroc2004}
\begin{center}
\begin{tabular}{ccccc}
\hline\hline
$E$ & max$^a$ & error & $O-C^b$ & $N^c$ \\
\hline
0 & 53183.2842 & 0.0004 & $-$0.0102 & 86 \\
13 & 53184.2835 & 0.0002 & $-$0.0044 & 142 \\
14 & 53184.3612 & 0.0002 & $-$0.0031 & 172 \\
15 & 53184.4375 & 0.0002 & $-$0.0033 & 163 \\
16 & 53184.5156 & 0.0004 & $-$0.0016 & 114 \\
26 & 53185.2797 & 0.0003 & $-$0.0017 & 173 \\
27 & 53185.3597 & 0.0006 & 0.0019 & 83 \\
42 & 53186.5081 & 0.0009 & 0.0040 & 92 \\
43 & 53186.5809 & 0.0019 & 0.0003 & 75 \\
53 & 53187.3473 & 0.0004 & 0.0025 & 155 \\
54 & 53187.4272 & 0.0007 & 0.0060 & 137 \\
66 & 53188.3453 & 0.0005 & 0.0070 & 155 \\
67 & 53188.4219 & 0.0004 & 0.0072 & 171 \\
68 & 53188.4986 & 0.0007 & 0.0075 & 108 \\
94 & 53190.4820 & 0.0005 & 0.0039 & 172 \\
132 & 53193.3791 & 0.0011 & $-$0.0030 & 145 \\
145 & 53194.3683 & 0.0006 & $-$0.0074 & 166 \\
146 & 53194.4465 & 0.0004 & $-$0.0056 & 167 \\
\hline
  \multicolumn{5}{l}{$^{a}$ BJD$-$2400000.} \\
  \multicolumn{5}{l}{$^{b}$ Against $max = 2453183.2944 + 0.076422 E$.} \\
  \multicolumn{5}{l}{$^{c}$ Number of points used to determine the maximum.} \\
\end{tabular}
\end{center}
\end{table}

\subsection{CG Canis Majoris}\label{obj:cgcma}

   CG CMa was originally classified as a classical nova in 1934
\citep{due87novaatlas}.  A new outburst in 1999 finally led to
a classification as a WZ Sge-type dwarf nova (\cite{due99cgcma};
\cite{kat99cgcma}).  We reanalyzed photometric data reported in
\citet{kat99cgcma}.  The period around $\sim$0.063 d reported in
\citet{kat99cgcma} appears viable, although the faintness of
the object and the existence of a close companion made the uncertainty
large.  We determined $O-C$'s based on this period selection
(table \ref{tab:cgcmaoc1999}).  If this period is the true
period, the $P_{\rm dot}$ is almost zero at $+0.5(1.6) \times 10^{-5}$.
Since this variation was detected during the early stage of the
outburst, this period likely refers to that of early superhumps,
rather than superhumps suggested in \citet{kat99cgcma}.
Other candidate periods could not express observations well.

\begin{table}
\caption{Maxima of (Early) Superhumps in CG CMa (1999).}\label{tab:cgcmaoc1999}
\begin{center}
\begin{tabular}{ccccc}
\hline\hline
$E$ & max$^a$ & error & $O-C^b$ & $N^c$ \\
\hline
0 & 51232.1013 & 0.0077 & 0.0005 & 101 \\
45 & 51234.9526 & 0.0036 & 0.0045 & 127 \\
47 & 51235.0719 & 0.0042 & $-$0.0027 & 121 \\
78 & 51237.0438 & 0.0048 & 0.0076 & 56 \\
79 & 51237.0974 & 0.0036 & $-$0.0020 & 62 \\
92 & 51237.9213 & 0.0130 & $-$0.0007 & 59 \\
94 & 51238.0369 & 0.0066 & $-$0.0117 & 127 \\
95 & 51238.1084 & 0.0058 & $-$0.0034 & 78 \\
108 & 51238.9387 & 0.0054 & 0.0044 & 127 \\
110 & 51239.0654 & 0.0036 & 0.0044 & 102 \\
141 & 51241.0137 & 0.0039 & $-$0.0088 & 118 \\
171 & 51242.9313 & 0.0123 & 0.0107 & 62 \\
173 & 51243.0503 & 0.0036 & 0.0031 & 36 \\
187 & 51243.9274 & 0.0049 & $-$0.0057 & 81 \\
189 & 51244.0595 & 0.0053 & $-$0.0002 & 85 \\
\hline
  \multicolumn{5}{l}{$^{a}$ BJD$-$2400000.} \\
  \multicolumn{5}{l}{$^{b}$ Against $max = 2451232.1007 + 0.063275 E$.} \\
  \multicolumn{5}{l}{$^{c}$ Number of points used to determine the maximum.} \\
\end{tabular}
\end{center}
\end{table}

\subsection{PU Canis Majoris}\label{tab:pucma}\label{obj:pucma}

   The SU UMa-type nature of PU CMa was pointed out by
\citet{kat03v877arakktelpucma}, but they were unable to uniquely
determine the superhump period.  Thanks to three superoutbursts
in 2003, 2005 and 2008, we have been able to firmly establish the
superhump period.  The times of superhump maxima are summarized
in tables \ref{tab:pucmaoc2003}, \ref{tab:pucmaoc2005} and
\ref{tab:pucmaoc2008}.

The 2003 superoutburst was observed during its later course
and a clear transition from the stage B to C was observed.
We also included some of post-outburst hump maxima having the
same phase as in stage C superhumps.  No clear phase shift,
expected for traditional ``late superhumps'',
was observed during and soon after the rapidly fading stage.

\begin{figure}
  \begin{center}
    \FigureFile(88mm,90mm){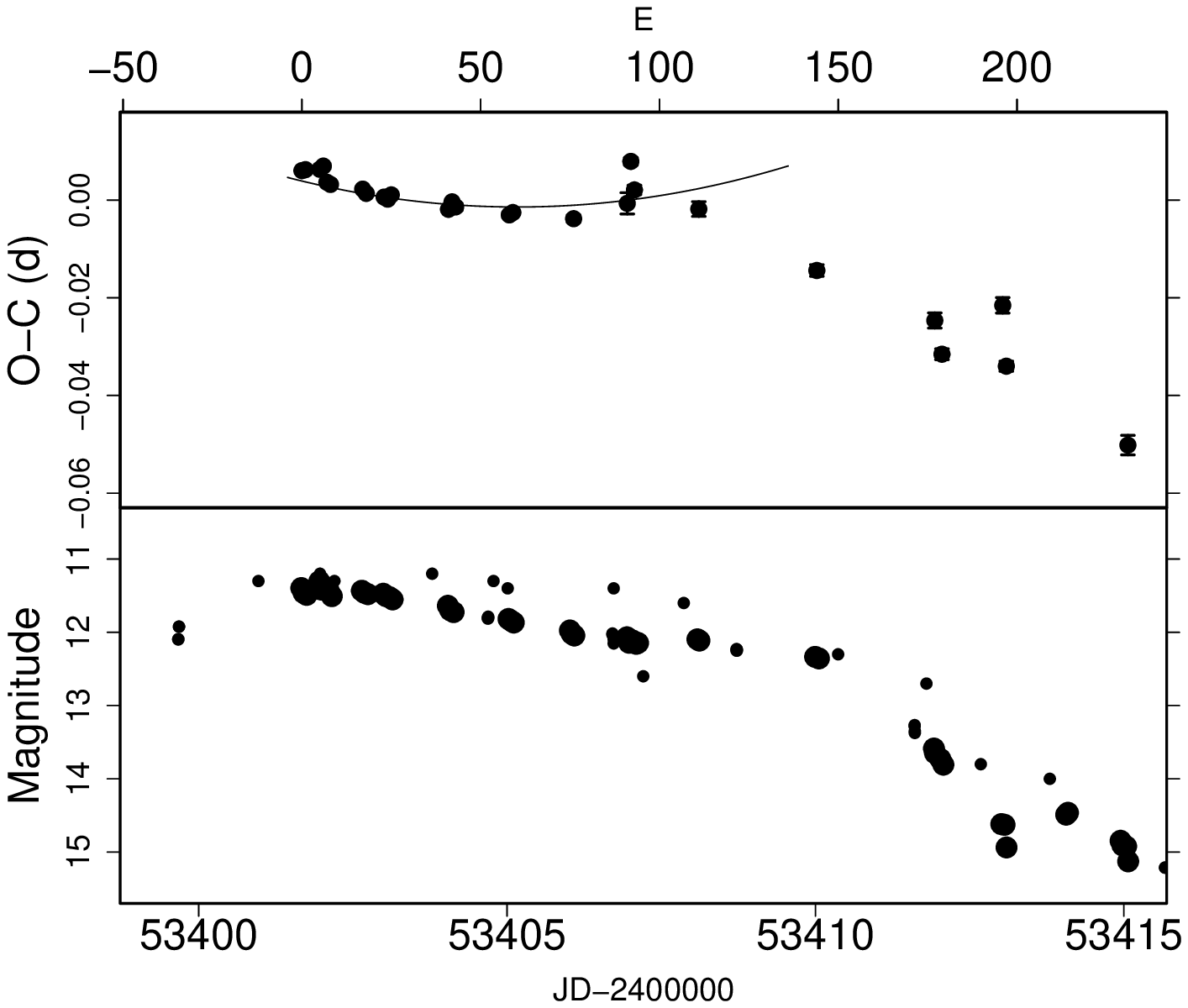}
  \end{center}
  \caption{$O-C$ of superhumps PU CMa (2005).
  (Upper): $O-C$ diagram.  The $O-C$ values were against the mean period
  for the stage B ($E \le 93$, thin curve)
  (Lower): Light curve.  Large dots are our CCD observations and small
  dots are visual observation from the VSNET database.}
  \label{fig:pucma2005oc}
\end{figure}

The 2005 and 2008 superoutbursts were observed during their earlier
stages and the superhump period showed an increase during
the superoutburst plateau.  The $P_{\rm dot}$'s were
$+11.4(1.8) \times 10^{-5}$ and $+4.4(3.1) \times 10^{-5}$,
respectively.  The 2005 superoutburst showed a clear transition
to the stage C (figure \ref{fig:pucma2005oc};
$P_2 =$ 0.05768(2) d, disregarding $E=196$ and $E=215$).

\begin{figure}
  \begin{center}
    \FigureFile(88mm,90mm){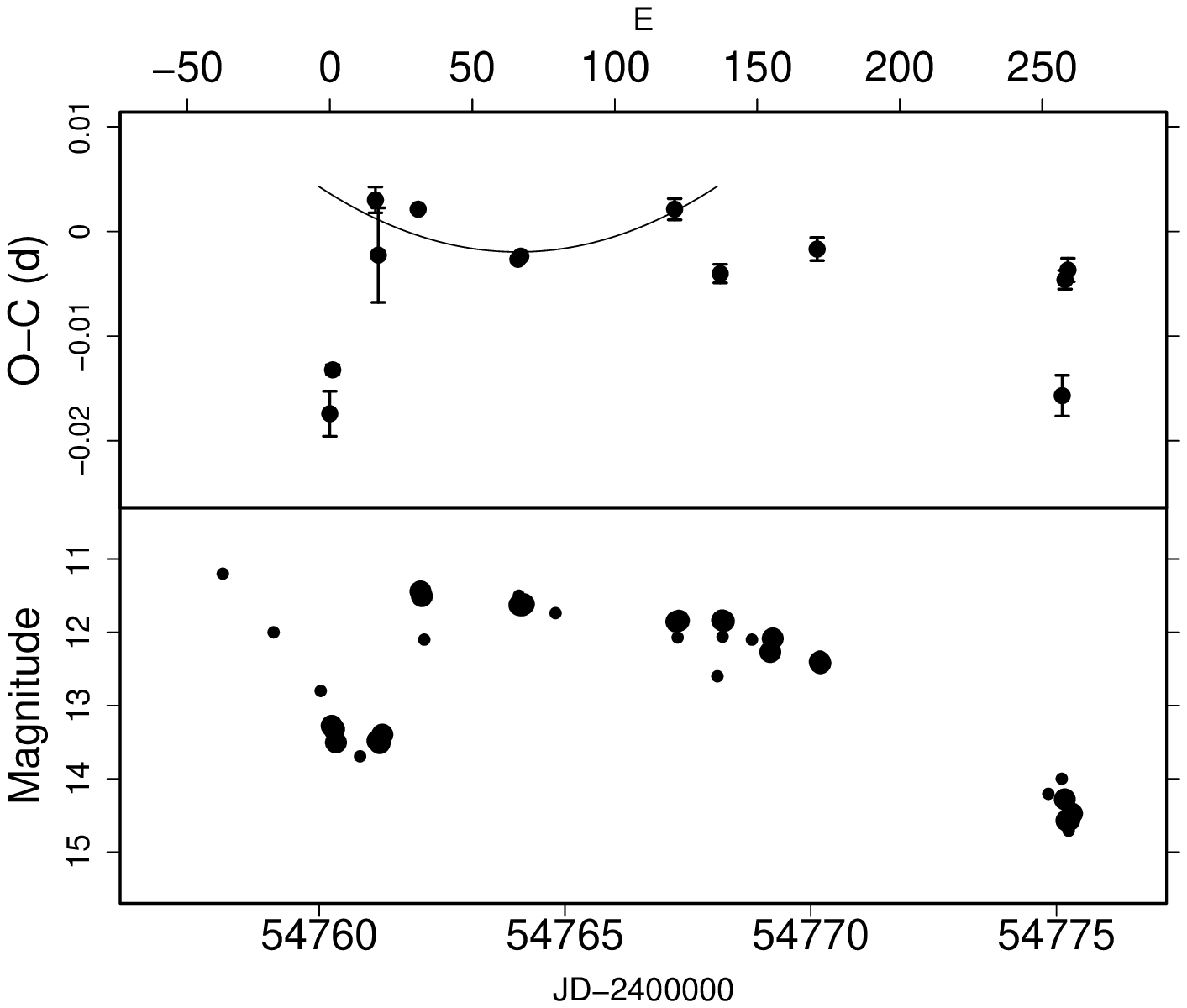}
  \end{center}
  \caption{$O-C$ of superhumps PU CMa (2008).
  (Upper): $O-C$ diagram.  The $O-C$ values were against the mean period
  for the stage B ($16 \le E \le 121$, thin curve)
  (Lower): Light curve.  Large dots are our CCD observations and small
  dots are visual observation from the VSNET database.}
  \label{fig:pucma2008oc}
\end{figure}

The 2008 superoutburst was preceded by a distinct
precursor (corresponding to $E \le 17$), during which a longer
$P_{\rm SH}$ was observed (figure \ref{fig:pucma2008oc}).
The fractional superhump excess
was 2.3 \% (mean period) against the orbital period by
\citet{tho03kxaqlftcampucmav660herdmlyr}.

\begin{table}
\caption{Superhump maxima of PU CMa (2003).}\label{tab:pucmaoc2003}
\begin{center}
\begin{tabular}{ccccc}
\hline\hline
$E$ & max$^a$ & error & $O-C^b$ & $N^c$ \\
\hline
0 & 52784.8969 & 0.0007 & $-$0.0104 & 248 \\
23 & 52786.2337 & 0.0007 & $-$0.0005 & 187 \\
40 & 52787.2190 & 0.0003 & 0.0041 & 116 \\
49 & 52787.7365 & 0.0005 & 0.0024 & 127 \\
51 & 52787.8539 & 0.0006 & 0.0045 & 215 \\
57 & 52788.1995 & 0.0005 & 0.0040 & 131 \\
69 & 52788.8885 & 0.0004 & 0.0007 & 356 \\
74 & 52789.1770 & 0.0009 & 0.0007 & 92 \\
86 & 52789.8705 & 0.0004 & 0.0020 & 232 \\
92 & 52790.2129 & 0.0005 & $-$0.0017 & 132 \\
109 & 52791.1955 & 0.0021 & 0.0002 & 59 \\
144 & 52793.2083 & 0.0008 & $-$0.0060 & 34 \\
\hline
  \multicolumn{5}{l}{$^{a}$ BJD$-$2400000.} \\
  \multicolumn{5}{l}{$^{b}$ Against $max = 2452784.9074 + 0.057688 E$.} \\
  \multicolumn{5}{l}{$^{c}$ Number of points used to determine the maximum.} \\
\end{tabular}
\end{center}
\end{table}

\begin{table}
\caption{Superhump maxima of PU CMa (2005).}\label{tab:pucmaoc2005}
\begin{center}
\begin{tabular}{ccccc}
\hline\hline
$E$ & max$^a$ & error & $O-C^b$ & $N^c$ \\
\hline
0 & 53401.6702 & 0.0002 & $-$0.0033 & 45 \\
1 & 53401.7285 & 0.0002 & $-$0.0029 & 57 \\
5 & 53401.9608 & 0.0004 & $-$0.0019 & 60 \\
6 & 53402.0195 & 0.0006 & $-$0.0011 & 37 \\
7 & 53402.0742 & 0.0004 & $-$0.0042 & 90 \\
8 & 53402.1319 & 0.0003 & $-$0.0044 & 106 \\
17 & 53402.6534 & 0.0003 & $-$0.0034 & 56 \\
18 & 53402.7106 & 0.0003 & $-$0.0041 & 56 \\
23 & 53403.0002 & 0.0004 & $-$0.0037 & 167 \\
24 & 53403.0578 & 0.0004 & $-$0.0040 & 261 \\
25 & 53403.1167 & 0.0004 & $-$0.0029 & 226 \\
41 & 53404.0427 & 0.0003 & $-$0.0023 & 161 \\
42 & 53404.1023 & 0.0003 & $-$0.0006 & 251 \\
43 & 53404.1593 & 0.0008 & $-$0.0014 & 122 \\
58 & 53405.0286 & 0.0006 & 0.0002 & 201 \\
59 & 53405.0871 & 0.0004 & 0.0009 & 266 \\
76 & 53406.0729 & 0.0008 & 0.0033 & 180 \\
91 & 53406.9469 & 0.0022 & 0.0097 & 58 \\
92 & 53407.0135 & 0.0008 & 0.0185 & 85 \\
93 & 53407.0657 & 0.0010 & 0.0129 & 84 \\
111 & 53408.1069 & 0.0015 & 0.0129 & 82 \\
144 & 53410.0103 & 0.0012 & 0.0074 & 57 \\
177 & 53411.9160 & 0.0016 & 0.0043 & 59 \\
179 & 53412.0252 & 0.0011 & $-$0.0021 & 80 \\
196 & 53413.0222 & 0.0016 & 0.0116 & 74 \\
197 & 53413.0678 & 0.0010 & $-$0.0007 & 80 \\
215 & 53414.0804 & 0.0012 & $-$0.0292 & 85 \\
231 & 53415.0256 & 0.0020 & $-$0.0095 & 81 \\
\hline
  \multicolumn{5}{l}{$^{a}$ BJD$-$2400000.} \\
  \multicolumn{5}{l}{$^{b}$ Against $max = 2453401.6735 + 0.057842 E$.} \\
  \multicolumn{5}{l}{$^{c}$ Number of points used to determine the maximum.} \\
\end{tabular}
\end{center}
\end{table}

\begin{table}
\caption{Superhump maxima of PU CMa (2008).}\label{tab:pucmaoc2008}
\begin{center}
\begin{tabular}{ccccc}
\hline\hline
$E$ & max$^a$ & error & $O-C^b$ & $N^c$ \\
\hline
0 & 54760.2505 & 0.0022 & $-$0.0132 & 154 \\
1 & 54760.3128 & 0.0005 & $-$0.0090 & 255 \\
16 & 54761.1995 & 0.0012 & 0.0073 & 86 \\
17 & 54761.2523 & 0.0045 & 0.0020 & 58 \\
31 & 54762.0691 & 0.0002 & 0.0064 & 214 \\
66 & 54764.0955 & 0.0003 & 0.0018 & 177 \\
67 & 54764.1538 & 0.0002 & 0.0021 & 166 \\
121 & 54767.2920 & 0.0010 & 0.0068 & 110 \\
137 & 54768.2144 & 0.0009 & 0.0007 & 56 \\
171 & 54770.1899 & 0.0011 & 0.0032 & 37 \\
257 & 54775.1667 & 0.0020 & $-$0.0104 & 38 \\
258 & 54775.2358 & 0.0009 & 0.0007 & 59 \\
259 & 54775.2948 & 0.0011 & 0.0016 & 59 \\
\hline
  \multicolumn{5}{l}{$^{a}$ BJD$-$2400000.} \\
  \multicolumn{5}{l}{$^{b}$ Against $max = 2454760.2158 + 0.057977 E$.} \\
  \multicolumn{5}{l}{$^{c}$ Number of points used to determine the maximum.} \\
\end{tabular}
\end{center}
\end{table}

\subsection{YZ Cancri}\label{obj:yzcnc}

   YZ Cnc is one of the oldest known SU UMa-type dwarf nova.
The superhump period of 0.09204 d \citep{pat79SH} has long been
widely used.  We, however, noticed that this period was incorrect.
We obtained the times of superhump maxima from the observations of
the 2007 February superoutburst (table \ref{tab:yzcncoc}).
The period 0.09204 d could not fit the observation.
A PDM analysis and superhump timing analysis yielded mean periods
of 0.09042(4) d and 0.09031(5) d, respectively.
The corresponding fractional superhump excess was 4.0 \%.
The $P_{\rm dot}$ was $-5.1(4.7) \times 10^{-5}$.

\begin{table}
\caption{Superhump maxima of YZ Cnc (2007).}\label{tab:yzcncoc}
\begin{center}
\begin{tabular}{ccccc}
\hline\hline
$E$ & max$^a$ & error & $O-C^b$ & $N^c$ \\
\hline
0 & 54144.0639 & 0.0006 & $-$0.0017 & 112 \\
1 & 54144.1553 & 0.0008 & $-$0.0006 & 110 \\
22 & 54146.0535 & 0.0010 & 0.0012 & 104 \\
23 & 54146.1421 & 0.0007 & $-$0.0005 & 110 \\
34 & 54147.1341 & 0.0022 & $-$0.0019 & 81 \\
35 & 54147.2321 & 0.0025 & 0.0058 & 49 \\
65 & 54149.9319 & 0.0031 & $-$0.0036 & 77 \\
66 & 54150.0270 & 0.0014 & 0.0012 & 113 \\
\hline
  \multicolumn{5}{l}{$^{a}$ BJD$-$2400000.} \\
  \multicolumn{5}{l}{$^{b}$ Against $max = 2454144.0656 + 0.090307 E$.} \\
  \multicolumn{5}{l}{$^{c}$ Number of points used to determine the maximum.} \\
\end{tabular}
\end{center}
\end{table}

\subsection{AK Cancri}\label{obj:akcnc}

   \citet{kat94akcnc} first detected superhumps in this object, and
reported a period of 0.06735(5) d.  We measured times of superhump maxima
from these observations (table \ref{tab:akcncoc1992}).  The first
two nights of the observation were likely taken during stage B,
while the last night was likely during stage C.
\citet{men96akcnc} further observed the 1995 superoutburst and yielded
a mean period of 0.06749(1) d.

   We analyzed the 1999 superoutburst using the AAVSO data and
the 2003 superoutburst using the data by VSNET Collaboration.
The superhump maxima are given in table \ref{tab:akcncoc1999} and
\ref{tab:akcncoc2003}.  The 1999 superoutburst was preceded by
a precursor outburst 9 d before.
The $O-C$ diagram during the 2003 superoutburst (figure \ref{fig:octrans})
showed a feature characteristic to a short-$P_{\rm orb}$ SU UMa-type
dwarf nova:
following the stage B with a  positive $P_{\rm dot}$, the period
switched to a shorter one (stage C) before the termination of
the plateau phase.
The $P_{\rm dot}$ for the first interval ($E < 101$) was
$+4.8(3.2) \times 10^{-5}$.

\begin{figure}
  \begin{center}
    \FigureFile(88mm,70mm){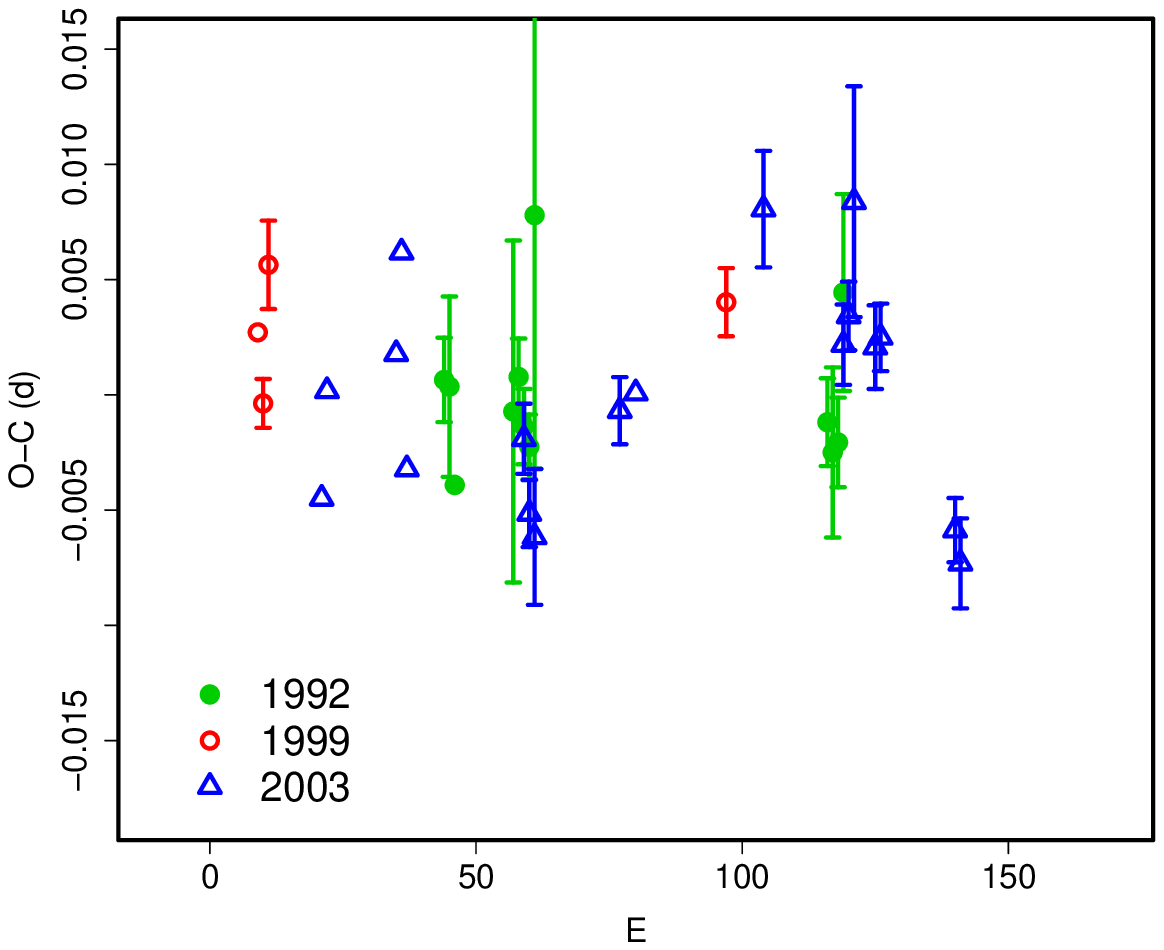}
  \end{center}
  \caption{Comparison of $O-C$ diagrams of AK Cnc between different
  superoutbursts.  A period of 0.06736 d was used to draw this figure.
  Approximate cycle counts ($E$) after the start of the superoutburst
  were used.
  }
  \label{fig:akcnccomp}
\end{figure}

\begin{table}
\caption{Superhump maxima of AK Cnc (1992).}\label{tab:akcncoc1992}
\begin{center}
\begin{tabular}{ccccc}
\hline\hline
$E$ & max$^a$ & error & $O-C^b$ & $N^c$ \\
\hline
0 & 48639.1631 & 0.0018 & 0.0006 & 34 \\
1 & 48639.2302 & 0.0039 & 0.0003 & 29 \\
2 & 48639.2933 & 0.0010 & $-$0.0040 & 45 \\
13 & 48640.0374 & 0.0074 & $-$0.0008 & 59 \\
14 & 48640.1063 & 0.0017 & 0.0007 & 101 \\
15 & 48640.1715 & 0.0016 & $-$0.0014 & 89 \\
16 & 48640.2380 & 0.0014 & $-$0.0023 & 86 \\
17 & 48640.3154 & 0.0110 & 0.0078 & 89 \\
72 & 48644.0112 & 0.0019 & $-$0.0011 & 54 \\
73 & 48644.0772 & 0.0037 & $-$0.0024 & 58 \\
74 & 48644.1450 & 0.0019 & $-$0.0020 & 64 \\
75 & 48644.2189 & 0.0043 & 0.0045 & 56 \\
\hline
  \multicolumn{5}{l}{$^{a}$ BJD$-$2400000.} \\
  \multicolumn{5}{l}{$^{b}$ Against $max = 2448639.1625 + 0.067358 E$.} \\
  \multicolumn{5}{l}{$^{c}$ Number of points used to determine the maximum.} \\
\end{tabular}
\end{center}
\end{table}

\begin{table}
\caption{Superhump maxima of AK Cnc (1999).}\label{tab:akcncoc1999}
\begin{center}
\begin{tabular}{ccccc}
\hline\hline
$E$ & max$^a$ & error & $O-C^b$ & $N^c$ \\
\hline
0 & 51261.4195 & 0.0009 & 0.0001 & 30 \\
1 & 51261.4838 & 0.0011 & $-$0.0030 & 36 \\
2 & 51261.5572 & 0.0019 & 0.0030 & 21 \\
88 & 51267.3485 & 0.0015 & $-$0.0000 & 36 \\
\hline
  \multicolumn{5}{l}{$^{a}$ BJD$-$2400000.} \\
  \multicolumn{5}{l}{$^{b}$ Against $max = 2451261.4195 + 0.067376 E$.} \\
  \multicolumn{5}{l}{$^{c}$ Number of points used to determine the maximum.} \\
\end{tabular}
\end{center}
\end{table}

\begin{table}
\caption{Superhump maxima of AK Cnc (2003).}\label{tab:akcncoc2003}
\begin{center}
\begin{tabular}{ccccc}
\hline\hline
$E$ & max$^a$ & error & $O-C^b$ & $N^c$ \\
\hline
0 & 52722.3924 & 0.0004 & $-$0.0035 & 65 \\
1 & 52722.4644 & 0.0005 & 0.0011 & 52 \\
14 & 52723.3417 & 0.0008 & 0.0025 & 31 \\
15 & 52723.4135 & 0.0008 & 0.0069 & 66 \\
16 & 52723.4714 & 0.0007 & $-$0.0025 & 70 \\
38 & 52724.9547 & 0.0015 & $-$0.0015 & 118 \\
39 & 52725.0188 & 0.0015 & $-$0.0048 & 108 \\
40 & 52725.0851 & 0.0029 & $-$0.0058 & 115 \\
56 & 52726.1683 & 0.0015 & $-$0.0006 & 144 \\
59 & 52726.3712 & 0.0004 & 0.0001 & 67 \\
83 & 52727.9958 & 0.0025 & 0.0077 & 70 \\
98 & 52729.0003 & 0.0017 & 0.0016 & 71 \\
99 & 52729.0689 & 0.0015 & 0.0028 & 90 \\
100 & 52729.1413 & 0.0050 & 0.0078 & 74 \\
104 & 52729.4044 & 0.0018 & 0.0014 & 32 \\
105 & 52729.4722 & 0.0015 & 0.0018 & 57 \\
119 & 52730.4068 & 0.0014 & $-$0.0068 & 45 \\
120 & 52730.4728 & 0.0020 & $-$0.0082 & 34 \\
\hline
  \multicolumn{5}{l}{$^{a}$ BJD$-$2400000.} \\
  \multicolumn{5}{l}{$^{b}$ Against $max = 2452722.3959 + 0.067375 E$.} \\
  \multicolumn{5}{l}{$^{c}$ Number of points used to determine the maximum.} \\
\end{tabular}
\end{center}
\end{table}

\subsection{CC Cancri}\label{obj:cccnc}

   \citet{kat97cccnc} first reported the detection of superhumps in this
object.  \citet{kat02cccnc} further reported the result of a more extensive
campaign in 2001, yielding a strongly negative
$P_{\rm dot}$ = $-10.2(1.3) \times 10^{-5}$.
Based on our new knowledge, this period derivative can be better understood
to represent the rapid period decrease (stage A to B) during
the early stage of a superoutburst.  We thereby reexamined the 2001 data
and obtained the times of maxima (table \ref{tab:cccncoc}).
The $O-C$ diagram can be interpreted
as a combination of stage A evolution with a long superhump period ($E < 20$),
and the stage B with a more stabilized superhump period.
The $P_{\rm dot}$ of the latter interval was $-7.3(2.5) \times 10^{-5}$.

\begin{table}
\caption{Superhump maxima of CC Cnc (2001).}\label{tab:cccncoc}
\begin{center}
\begin{tabular}{ccccc}
\hline\hline
$E$ & max$^a$ & error & $O-C^b$ & $N^c$ \\
\hline
0 & 52226.3217 & 0.0039 & $-$0.0107 & 70 \\
11 & 52227.1401 & 0.0029 & $-$0.0232 & 79 \\
12 & 52227.2344 & 0.0015 & $-$0.0044 & 93 \\
13 & 52227.3099 & 0.0030 & $-$0.0044 & 75 \\
24 & 52228.1482 & 0.0008 & 0.0031 & 145 \\
25 & 52228.2212 & 0.0011 & 0.0006 & 147 \\
26 & 52228.3014 & 0.0013 & 0.0053 & 147 \\
27 & 52228.3780 & 0.0012 & 0.0064 & 77 \\
37 & 52229.1273 & 0.0030 & 0.0003 & 55 \\
38 & 52229.2072 & 0.0007 & 0.0047 & 147 \\
39 & 52229.2823 & 0.0009 & 0.0043 & 146 \\
40 & 52229.3592 & 0.0020 & 0.0057 & 19 \\
50 & 52230.1126 & 0.0034 & 0.0038 & 53 \\
51 & 52230.1918 & 0.0010 & 0.0074 & 100 \\
52 & 52230.2637 & 0.0013 & 0.0038 & 103 \\
53 & 52230.3439 & 0.0020 & 0.0085 & 82 \\
64 & 52231.1685 & 0.0017 & 0.0023 & 89 \\
65 & 52231.2473 & 0.0028 & 0.0055 & 20 \\
79 & 52232.3036 & 0.0014 & 0.0044 & 120 \\
80 & 52232.3756 & 0.0013 & 0.0009 & 89 \\
90 & 52233.1340 & 0.0015 & 0.0041 & 145 \\
91 & 52233.2064 & 0.0017 & 0.0009 & 146 \\
92 & 52233.2818 & 0.0020 & 0.0008 & 146 \\
93 & 52233.3532 & 0.0010 & $-$0.0033 & 82 \\
104 & 52234.1860 & 0.0025 & $-$0.0013 & 147 \\
105 & 52234.2532 & 0.0013 & $-$0.0097 & 147 \\
106 & 52234.3374 & 0.0016 & $-$0.0010 & 124 \\
116 & 52235.0789 & 0.0069 & $-$0.0147 & 115 \\
119 & 52235.3201 & 0.0017 & $-$0.0001 & 146 \\
\hline
  \multicolumn{5}{l}{$^{a}$ BJD$-$2400000.} \\
  \multicolumn{5}{l}{$^{b}$ Against $max = 2452226.3324 + 0.075528 E$.} \\
  \multicolumn{5}{l}{$^{c}$ Number of points used to determine the maximum.} \\
\end{tabular}
\end{center}
\end{table}

\subsection{AL Comae Berenices}\label{obj:alcom}

   We have reanalyzed the data in \citet{nog97alcom} of this well-known
WZ Sge-type dwarf nova.  The combined list of superhump maxima from
\citet{how96alcom}, \citet{pyc95alcom} and \citet{pat96alcom} is presented
in table \ref{tab:alcomoc1995}.
The $O-C$ diagram clearly showed the same characteristics to that of another
WZ Sge-type dwarf nova, HV Vir.  The $P_{\rm dot}$ of the middle segment
(stage B) was $+1.9(0.5) \times 10^{-5}$ ($24 \le E \le 229$).  This value
supersedes the published period derivative in \citet{nog97alcom}.

\begin{figure}
  \begin{center}
    \FigureFile(88mm,70mm){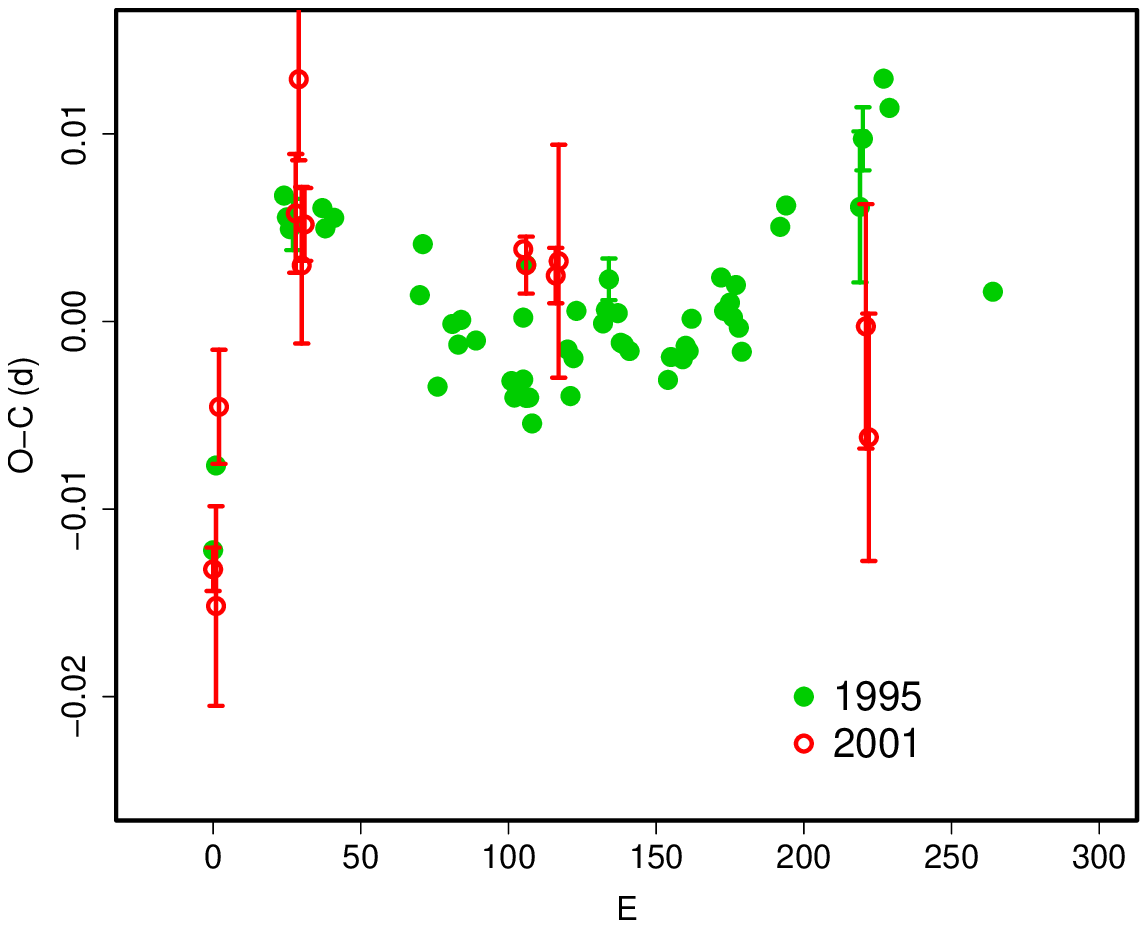}
  \end{center}
  \caption{Comparison of $O-C$ diagrams of AL Com Cnc between different
  superoutbursts.  A period of 0.05728 d was used to draw this figure.
  Approximate cycle counts ($E$) after the appearance of the ordinary
  superhumps were used.
  }
  \label{fig:alcomcomp}
\end{figure}

   The refined times of superhump maxima during the 2001 superoutburst
\citep{ish02wzsgeletter} are listed in table \ref{tab:alcomoc2001}.
The $P_{\rm dot}$ during the stage B was
$-0.2(0.8) \times 10^{-5}$ ($28 \le E \le 222$).
A comparison of the $O-C$ diagrams is shown in figure \ref{fig:alcomcomp}.

   The object underwent another superoutburst in 2007 \citep{uem08alcom}.
This behavior of this superoutburst was different from those in
1995 and 2001 in that the object showed separate rebrightenings
(type-B).  Although the observations was incomplete due to the poor
seasonal location, a weak periodicity of 0.05717(1) d was detected
during this rebrightening stage.  Since the object showed $P_{\rm SH}$
during the type-A superoutburst in 1995, we adopted this period
as the $P_{\rm SH}$ of the 2007 superoutburst in table \ref{tab:perlist}.

\begin{table}
\caption{Superhump maxima of AL Com (1995).}\label{tab:alcomoc1995}
\begin{center}
\begin{tabular}{ccccc}
\hline\hline
$E$ & max$^a$ & error & $O-C^b$ & $N^c$ \\
\hline
0 & 49823.5693 & -- & $-$0.0139 & R3 \\
1 & 49823.6311 & -- & $-$0.0094 & R5 \\
24 & 49824.9629 & 0.0007 & 0.0053 & 25 \\
25 & 49825.0190 & 0.0007 & 0.0042 & 29 \\
26 & 49825.0757 & 0.0008 & 0.0035 & 29 \\
27 & 49825.1332 & 0.0014 & 0.0038 & 29 \\
37 & 49825.7069 & -- & 0.0048 & R5 \\
\end{tabular}
\end{center}
\end{table}

\addtocounter{table}{-1}
\begin{table}
\caption{Superhump maxima of AL Com (1995) (continued).}
\begin{center}
\begin{tabular}{ccccc}
\hline\hline
$E$ & max & error & $O-C$ & $N$ \\
\hline
179 & 49833.8330 & -- & $-$0.0010 & R5 \\
192 & 49834.5843 & -- & 0.0058 & R3 \\
194 & 49834.7000 & -- & 0.0070 & R5 \\
219 & 49836.1319 & 0.0040 & 0.0073 & 24 \\
220 & 49836.1928 & 0.0017 & 0.0109 & 22 \\
227 & 49836.5970 & -- & 0.0142 & R3 \\
229 & 49836.7100 & -- & 0.0127 & R5 \\
264 & 49838.7050 & -- & 0.0033 & R5 \\
349 & 49843.5732 & -- & 0.0038 & R3 \\
360 & 49844.1662 & 0.0089 & $-$0.0331 & 21 \\
\hline
\end{tabular}
\end{center}
\end{table}

\begin{table}
\caption{Superhump maxima of AL Com (2001).}\label{tab:alcomoc2001}
\begin{center}
\begin{tabular}{ccccc}
\hline\hline
$E$ & max$^a$ & error & $O-C^b$ & $N^c$ \\
\hline
0 & 52056.3598 & 0.0012 & $-$0.0124 & 142 \\
1 & 52056.4151 & 0.0053 & $-$0.0144 & 142 \\
2 & 52056.4830 & 0.0030 & $-$0.0038 & 85 \\
28 & 52057.9826 & 0.0032 & 0.0062 & 67 \\
29 & 52058.0470 & 0.0043 & 0.0134 & 89 \\
30 & 52058.0944 & 0.0042 & 0.0035 & 78 \\
31 & 52058.1538 & 0.0019 & 0.0056 & 77 \\
105 & 52062.3912 & 0.0007 & 0.0036 & 51 \\
106 & 52062.4477 & 0.0015 & 0.0027 & 42 \\
116 & 52063.0199 & 0.0015 & 0.0021 & 104 \\
117 & 52063.0779 & 0.0062 & 0.0028 & 101 \\
221 & 52069.0316 & 0.0065 & $-$0.0017 & 55 \\
222 & 52069.0830 & 0.0066 & $-$0.0076 & 69 \\
\hline
  \multicolumn{5}{l}{$^{a}$ BJD$-$2400000.} \\
  \multicolumn{5}{l}{$^{b}$ Against $max = 2452056.3722 + 0.057290 E$.} \\
  \multicolumn{5}{l}{$^{c}$ Number of points used to determine the maximum.} \\
\end{tabular}
\end{center}
\end{table}

\subsection{GO Comae Berenices}\label{obj:gocom}

\begin{figure}
  \begin{center}
    \FigureFile(88mm,70mm){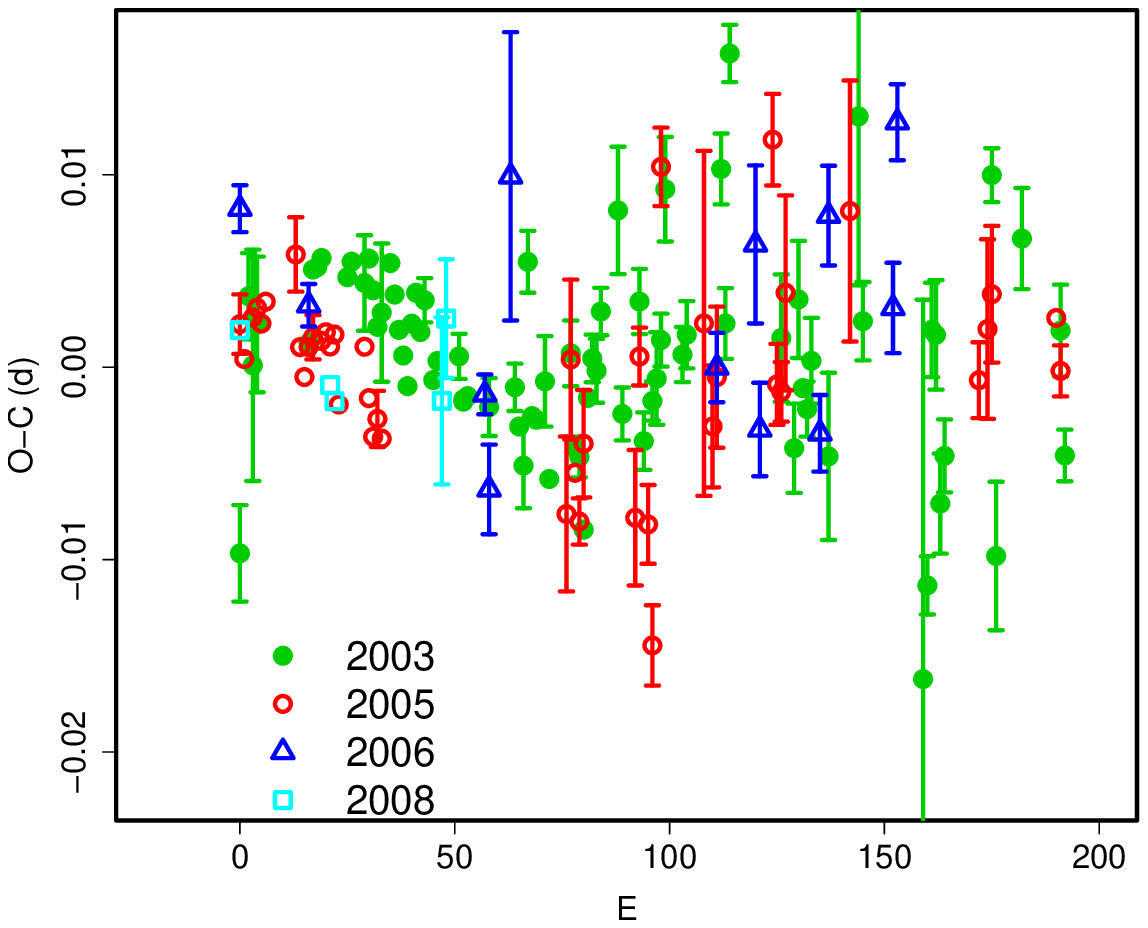}
  \end{center}
  \caption{Comparison of $O-C$ diagrams of GO Com between different
  superoutbursts.  A period of 0.063059 d was used to draw this figure.
  Approximate cycle counts ($E$) after the start of the
  superoutburst were used.
  }
  \label{fig:gocomcomp}
\end{figure}

   We reanalyzed the data used in \citet{ima05gocom}, combined with
Crimea (Pav) data, and new data for the 2005 and 2006 superoutbursts
(tables \ref{tab:gocomoc2003}, \ref{tab:gocomoc2005}, \ref{tab:gocomoc2006}).
The values of $P_{\rm dot}$ were
$+15.5(2.3) \times 10^{-5}$ (2003, $16 \le E \le 115$),
$+6.9(1.5) \times 10^{-5}$ (2005, $13 \le E \le 142$).
$+4.6(3.4) \times 10^{-5}$ (2006, excluding $E=64$ and $E=136$).
The 2008 superoutburst was also observed (table \ref{tab:gocomoc2008}).
A marginally significant $P_{\rm dot}$ = $+16(11) \times 10^{-5}$
was recorded.  The new observations in 2003 indicated that the stage C
superhumps persisted even during the post-superoutburst stage ($E \ge 230$).
The $O-C$ diagrams did not drastically vary between different
superoutbursts (figure \ref{fig:gocomcomp}).

\begin{table}
\caption{Superhump maxima of GO Com (2003).}\label{tab:gocomoc2003}
\begin{center}
\begin{tabular}{ccccc}
\hline\hline
$E$ & max$^a$ & error & $O-C^b$ & $N^c$ \\
\hline
0 & 52794.1340 & 0.0025 & $-$0.0141 & 225 \\
2 & 52794.2734 & 0.0022 & $-$0.0007 & 65 \\
3 & 52794.3329 & 0.0060 & $-$0.0042 & 72 \\
4 & 52794.3981 & 0.0035 & $-$0.0020 & 39 \\
16 & 52795.1538 & 0.0008 & $-$0.0024 & 109 \\
17 & 52795.2207 & 0.0003 & 0.0015 & 36 \\
18 & 52795.2839 & 0.0003 & 0.0016 & 37 \\
\end{tabular}
\end{center}
\end{table}

\addtocounter{table}{-1}
\begin{table}
\caption{Superhump maxima of GO Com (2003) (continued).}
\begin{center}
\begin{tabular}{ccccc}
\hline\hline
$E$ & max & error & $O-C$ & $N$ \\
\hline
96 & 52800.1955 & 0.0010 & $-$0.0015 & 35 \\
97 & 52800.2598 & 0.0024 & $-$0.0002 & 35 \\
98 & 52800.3248 & 0.0014 & 0.0018 & 60 \\
99 & 52800.3957 & 0.0027 & 0.0097 & 67 \\
103 & 52800.6394 & 0.0014 & 0.0013 & 41 \\
104 & 52800.7035 & 0.0017 & 0.0024 & 28 \\
112 & 52801.2166 & 0.0018 & 0.0114 & 34 \\
113 & 52801.2716 & 0.0018 & 0.0034 & 24 \\
114 & 52801.3487 & 0.0015 & 0.0175 & 36 \\
115 & 52801.4228 & 0.0035 & 0.0286 & 33 \\
126 & 52802.0906 & 0.0033 & 0.0033 & 71 \\
129 & 52802.2740 & 0.0023 & $-$0.0023 & 16 \\
130 & 52802.3448 & 0.0030 & 0.0055 & 35 \\
131 & 52802.4033 & 0.0007 & 0.0009 & 132 \\
132 & 52802.4652 & 0.0014 & $-$0.0001 & 120 \\
133 & 52802.5308 & 0.0022 & 0.0025 & 40 \\
137 & 52802.7781 & 0.0043 & $-$0.0023 & 29 \\
144 & 52803.2372 & 0.0088 & 0.0157 & 34 \\
145 & 52803.2896 & 0.0020 & 0.0051 & 35 \\
159 & 52804.1538 & 0.0197 & $-$0.0128 & 24 \\
160 & 52804.2217 & 0.0015 & $-$0.0079 & 34 \\
161 & 52804.2981 & 0.0025 & 0.0055 & 59 \\
162 & 52804.3609 & 0.0028 & 0.0053 & 38 \\
163 & 52804.4152 & 0.0026 & $-$0.0035 & 77 \\
164 & 52804.4807 & 0.0019 & $-$0.0009 & 70 \\
175 & 52805.1889 & 0.0014 & 0.0142 & 34 \\
176 & 52805.2322 & 0.0039 & $-$0.0055 & 33 \\
182 & 52805.6271 & 0.0026 & 0.0112 & 41 \\
191 & 52806.1898 & 0.0024 & 0.0069 & 24 \\
192 & 52806.2464 & 0.0013 & 0.0005 & 34 \\
230 & 52808.6277 & 0.0072 & $-$0.0126 & 26 \\
231 & 52808.6935 & 0.0036 & $-$0.0098 & 40 \\
241 & 52809.3289 & 0.0016 & $-$0.0044 & 13 \\
262 & 52810.6366 & 0.0017 & $-$0.0200 & 24 \\
\hline
\end{tabular}
\end{center}
\end{table}

\begin{table}
\caption{Superhump maxima of GO Com (2005).}\label{tab:gocomoc2005}
\begin{center}
%
\end{center}
\end{table}

\begin{table}
\caption{Superhump maxima of GO Com (2006).}\label{tab:gocomoc2006}
\begin{center}
%
\end{center}
\end{table}

\begin{table}
\caption{Superhump maxima of GO Com (2008).}\label{tab:gocomoc2008}
\begin{center}
%
\end{center}
\end{table}

\subsection{V728 Coronae Australis}\label{obj:v728cra}

\begin{figure}
  \begin{center}
    \FigureFile(88mm,110mm){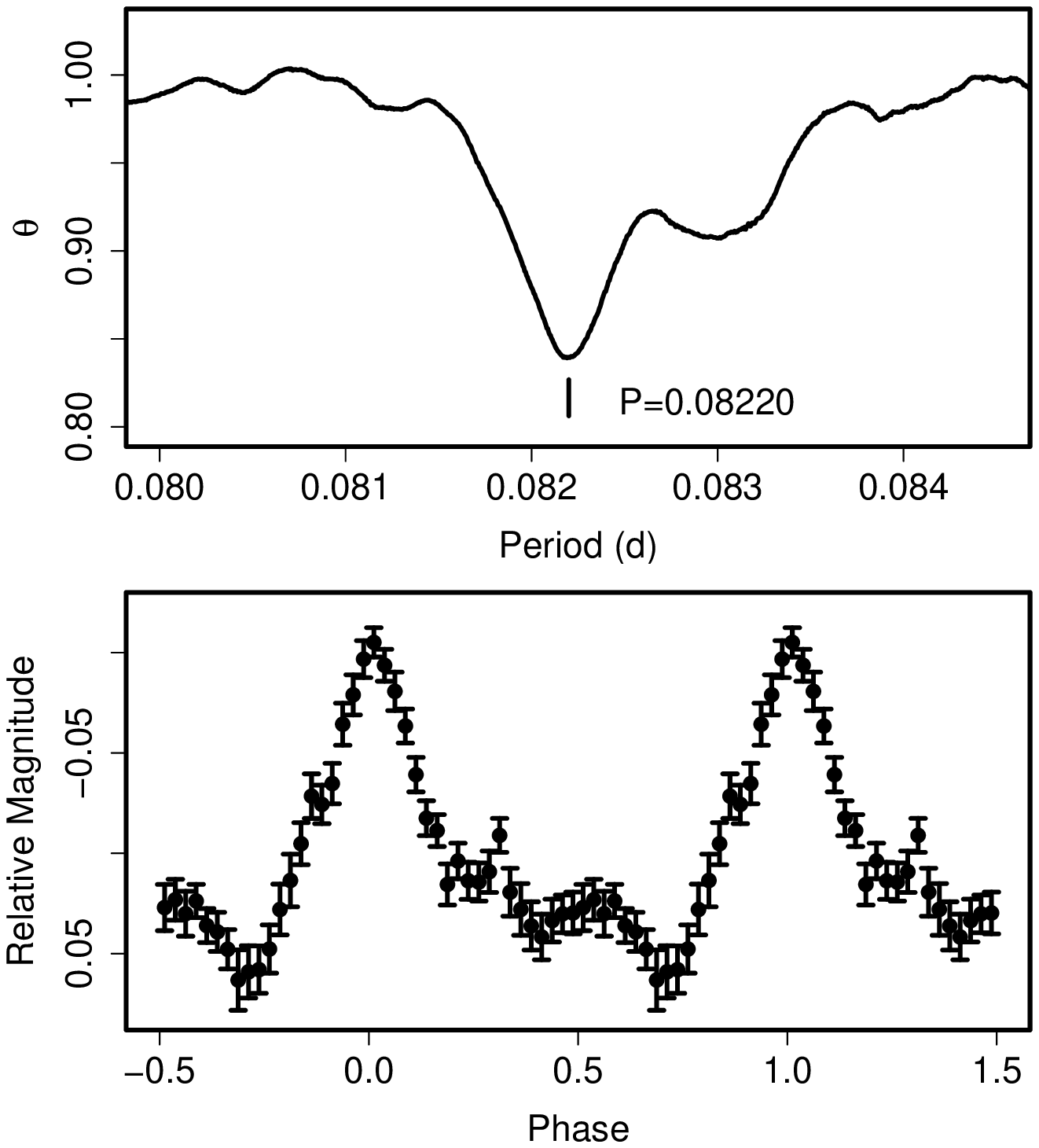}
  \end{center}
  \caption{Superhumps in V728 CrA (2003). (Upper): PDM analysis.
     (Lower): Phase-averaged profile.}
  \label{fig:v728crashpdm}
\end{figure}

   This object was selected during the identification project of
NSV objects against ROSAT X-ray source (Kato, vsnet-id-rosat 11).
The proximity of the ROSAT position to the position of NSV 9923
suggested that the object may be a dwarf nova, as we have seen in
BB Ari (subsection \ref{sec:bbari}) and DT Oct \citep{kat02gzcncnsv10934}.
Following this suggestion, the object was monitored for outbursts.
An outburst detection was announced on 2003 June 28
(R. Stubbings, vsnet-alert 7787).
The mean superhump period with the PDM method was 0.082200(13) d
(figure \ref{fig:v728crashpdm}).
The times of superhump maxima are listed in table \ref{tab:v728craoc}.
Although the original observations
included later stage at $E > 50$, the superhump signal became weaker
and irregular, sometimes with multiple peaks.  We therefore restricted
our $O-C$ analysis to $E \leq 50$.  The situation may be similar to
another long-period system SS UMi \citep{ole06ssumi}.
The resultant $P_{\rm dot}$ was $-2.3(3.4) \times 10^{-5}$.
Upon announcement of this observation, the variable has been given
a General Catalogue of Variable Stars (GCVS) designation V728 CrA
\citep{NameList78}.

\begin{table}
\caption{Superhump maxima of V728 CrA (2003).}\label{tab:v728craoc}
\begin{center}
\begin{tabular}{ccccc}
\hline\hline
$E$ & max$^a$ & error & $O-C^b$ & $N^c$ \\
\hline
0 & 52820.3295 & 0.0006 & 0.0006 & 36 \\
1 & 52820.4098 & 0.0003 & $-$0.0014 & 47 \\
2 & 52820.4925 & 0.0003 & $-$0.0011 & 47 \\
12 & 52821.3170 & 0.0010 & $-$0.0004 & 24 \\
13 & 52821.4015 & 0.0010 & 0.0017 & 41 \\
14 & 52821.4828 & 0.0006 & 0.0007 & 37 \\
15 & 52821.5651 & 0.0004 & 0.0005 & 43 \\
23 & 52822.2260 & 0.0005 & 0.0024 & 48 \\
24 & 52822.3031 & 0.0005 & $-$0.0028 & 41 \\
25 & 52822.3891 & 0.0007 & 0.0008 & 47 \\
26 & 52822.4706 & 0.0004 & $-$0.0001 & 46 \\
27 & 52822.5525 & 0.0005 & $-$0.0005 & 43 \\
35 & 52823.2133 & 0.0010 & 0.0012 & 41 \\
36 & 52823.2926 & 0.0005 & $-$0.0018 & 44 \\
37 & 52823.3784 & 0.0005 & 0.0016 & 48 \\
38 & 52823.4586 & 0.0006 & $-$0.0006 & 48 \\
39 & 52823.5405 & 0.0014 & $-$0.0012 & 44 \\
44 & 52823.9541 & 0.0005 & 0.0006 & 87 \\
45 & 52824.0362 & 0.0005 & 0.0003 & 93 \\
46 & 52824.1181 & 0.0007 & $-$0.0001 & 60 \\
48 & 52824.2860 & 0.0009 & 0.0030 & 46 \\
49 & 52824.3637 & 0.0007 & $-$0.0017 & 45 \\
50 & 52824.4460 & 0.0008 & $-$0.0017 & 45 \\
\hline
  \multicolumn{5}{l}{$^{a}$ BJD$-$2400000.} \\
  \multicolumn{5}{l}{$^{b}$ Against $max = 2452820.3288 + 0.082378 E$.} \\
  \multicolumn{5}{l}{$^{c}$ Number of points used to determine the maximum.} \\
\end{tabular}
\end{center}
\end{table}

\subsection{VW Coronae Borealis}\label{obj:vwcrb}

  \citet{nog04vwcrb} presented an analysis of 2003 superoutburst and
other recorded superoutbursts.  We reanalyzed the 2003 data and yielded
refined times of superhump maxima (table \ref{tab:vwcrboc2003}).
The resultant $O-C$ diagram basically confirmed the finding
in \citet{nog04vwcrb}, giving
$P_{\rm dot}$ = $+7.7(0.8) \times 10^{-5}$ for $E \le 142$.

   As discussed in \citet{nog04vwcrb}, positive period derivatives are
rare in systems with long superhump periods ($P_{\rm SH} > 0.07$ d).
This phenomenon may be analogous to the one observed in
TT Boo \citep{ole04ttboo}, another SU UMa-type dwarf nova with
a relatively long superhump period and long superoutbursts
(see also FQ Mon, subsection \ref{sec:fqmon}).
For objects with positive $P_{\rm dot}$, also see subsections RU Hor
(\ref{sec:ruhor}) and QY Per (\ref{sec:qyper}).

   We also included times of superhump maxima during the 2001 and 2006
superoutbursts (tables \ref{tab:vwcrboc2001}, \ref{tab:vwcrboc2006}).
Although the superhump signal was present, we did not use the 2001
superoutburst to determine $P_{\rm dot}$ because of the lower quality of
the data.  This outburst was only observed for its late stage, and the
observed superhumps were likely stage C superhumps.
The 2006 superoutburst was observed for its early part.
The derived $P_{\rm SH}$ = 0.07268(6) d, shorter than the mean $P_1$
for the 2003 superoutburst, also supports that the $P_{\rm SH}$ was
shorter (i.e. with a probable positive $P_{\rm dot}$) near the start of
this superoutburst.

\begin{figure}
  \begin{center}
    \FigureFile(88mm,70mm){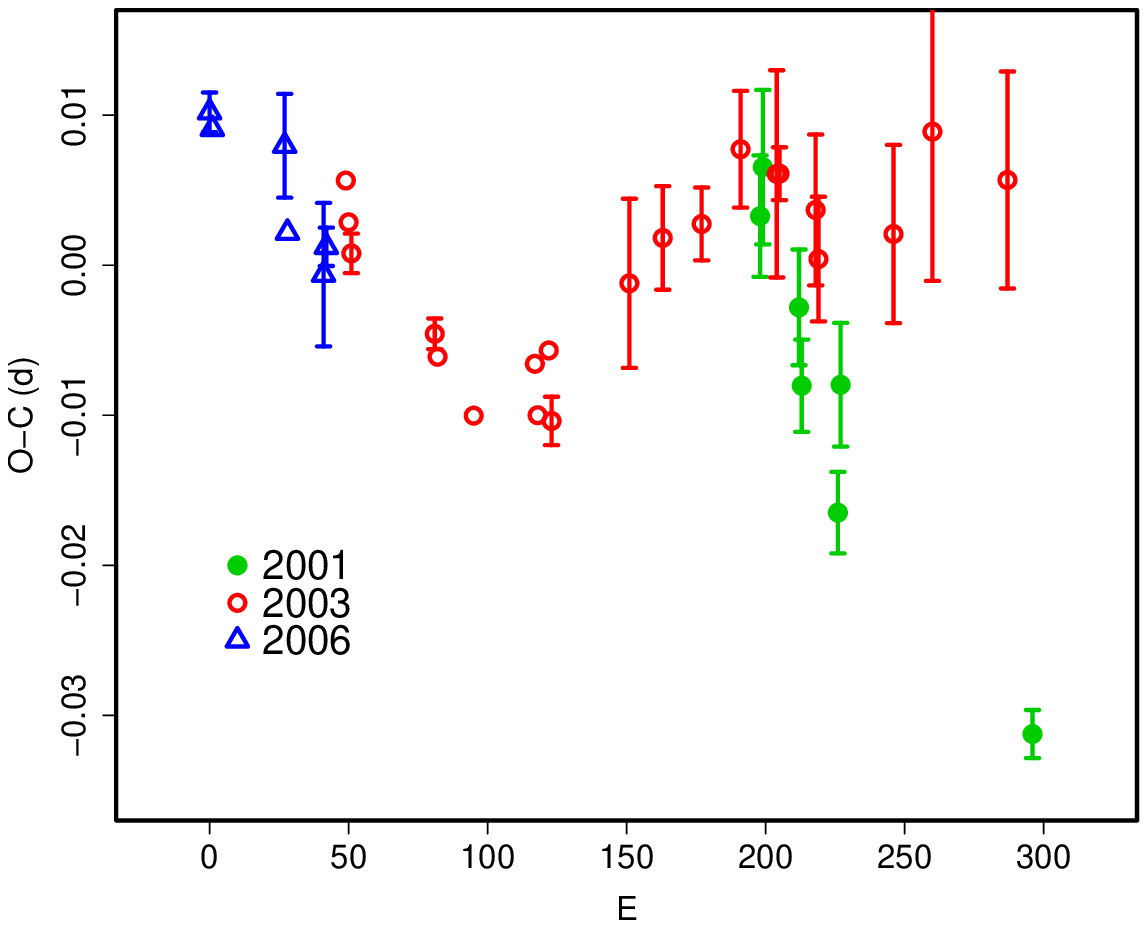}
  \end{center}
  \caption{Comparison of $O-C$ diagrams of VW CrB between different
  superoutbursts.  A period of 0.07290 d was used to draw this figure.
  Approximate cycle counts ($E$) after the start of the
  superoutburst were used.
  }
  \label{fig:vwcrbcomp}
\end{figure}

   A combined $O-C$ diagram is presented in figure \ref{fig:vwcrbcomp}.
The stage C behavior may have been different between the 2001 and 2003
superoutbursts.

\begin{table}
\caption{Superhump maxima of VW CrB (2003).}\label{tab:vwcrboc2003}
\begin{center}

\end{center}
\end{table}

\begin{table}
\caption{Superhump maxima of VW CrB (2001).}\label{tab:vwcrboc2001}
\begin{center}
%
\end{center}
\end{table}

\begin{table}
\caption{Superhump maxima of VW CrB (2006).}\label{tab:vwcrboc2006}
\begin{center}
%
\end{center}
\end{table}

\subsection{TU Crateris}\label{obj:tucrt}

   TU Crt had long been suspected to be an SU UMa-type candidate since
the discovery (cf. \cite{maz92tucrt}; \cite{haz93tucrt}; \cite{wen93tucrt}).
It was only in 1998 when its SU UMa-type nature was confirmed
\citep{men98tucrt}.  \citet{men98tucrt} reported an superhump period
of 0.08535(5) d and $P_{\rm dot}$ of $-7.2(0.9) \times 10^{-5}$
(the reference apparently had an error in conversion from coefficients
to $P_{\rm dot}$).

   We observed the 2001 and 2009 superoutbursts.  The times of superhump
maxima are listed in tables \ref{tab:tucrtoc2001} and \ref{tab:tucrtoc2009}.
The mean superhump period of the 2001 superoutburst determined with
PDM method was 0.08532(8) d.
The $P_{\rm dot}$ was $-12.3(9.2) \times 10^{-5}$.

\begin{figure}
  \begin{center}
    \FigureFile(88mm,70mm){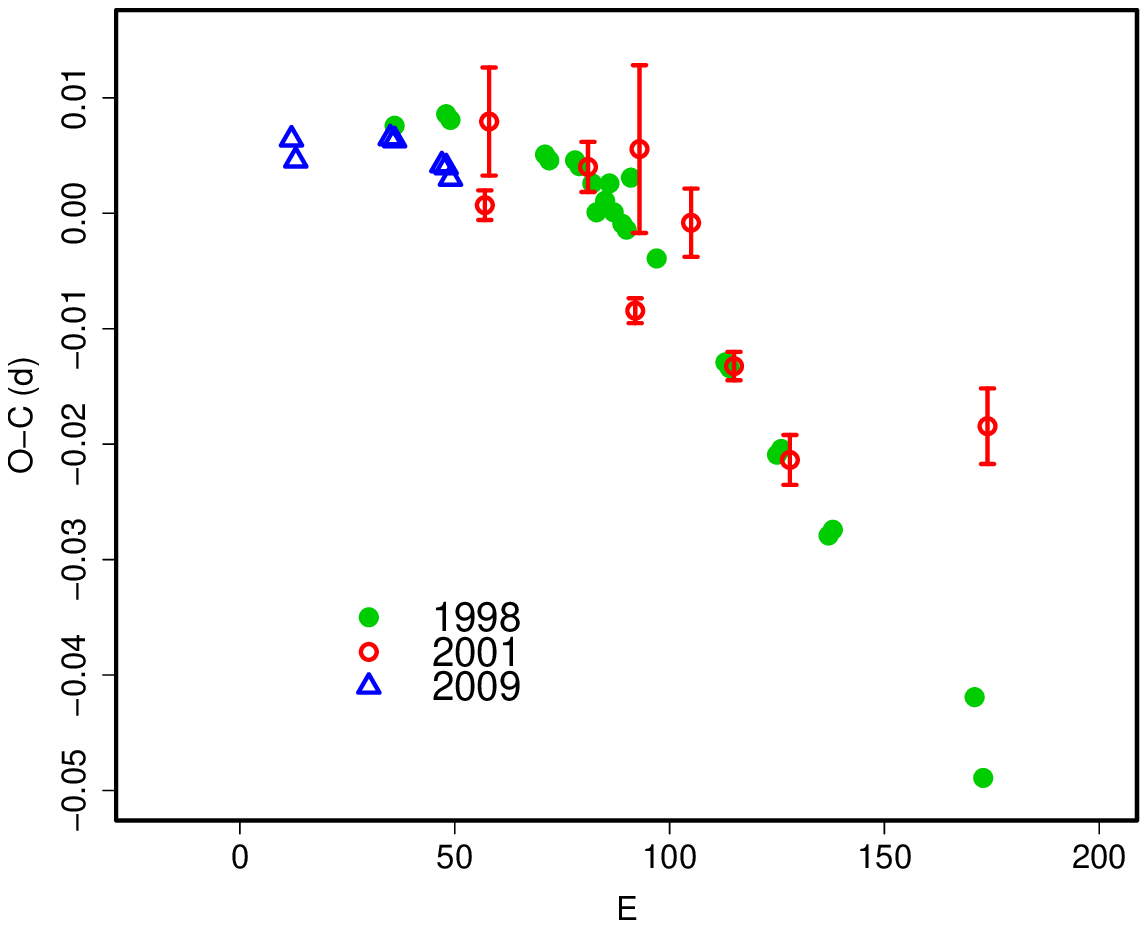}
  \end{center}
  \caption{Comparison of $O-C$ diagrams of TU Crt between different
  superoutbursts.  A period of 0.08550 d was used to draw this figure.
  Approximate cycle counts ($E$) after the start of the
  superoutburst were used.
  }
  \label{fig:tucrtcomp}
\end{figure}

   A combined $O-C$ diagram is presented in figure \ref{fig:tucrtcomp}.
The early part of the 2001 superoutburst was likely missed and we shifted
the $O-C$ diagram to best fit the 1998 superoutburst.  None of
observations yet recorded the epoch of stage A evolution.

\begin{table}
\caption{Superhump maxima of TU Crt (2001).}\label{tab:tucrtoc2001}
\begin{center}
\begin{tabular}{ccccc}
\hline\hline
$E$ & max$^a$ & error & $O-C^b$ & $N^c$ \\
\hline
0 & 52010.0402 & 0.0013 & $-$0.0070 & 101 \\
1 & 52010.1329 & 0.0047 & 0.0005 & 69 \\
24 & 52012.0955 & 0.0022 & 0.0040 & 109 \\
35 & 52013.0235 & 0.0011 & $-$0.0050 & 57 \\
36 & 52013.1230 & 0.0073 & 0.0094 & 59 \\
48 & 52014.1427 & 0.0030 & 0.0068 & 18 \\
58 & 52014.9852 & 0.0012 & $-$0.0024 & 158 \\
71 & 52016.0886 & 0.0022 & $-$0.0063 & 141 \\
\hline
  \multicolumn{5}{l}{$^{a}$ BJD$-$2400000.} \\
  \multicolumn{5}{l}{$^{b}$ Against $max = 2452010.0460 + 0.085175 E$.} \\
  \multicolumn{5}{l}{$^{c}$ Number of points used to determine the maximum.} \\
\end{tabular}
\end{center}
\end{table}

\begin{table}
\caption{Superhump maxima of TU Crt (2009).}\label{tab:tucrtoc2009}
\begin{center}
\begin{tabular}{ccccc}
\hline\hline
$E$ & max$^a$ & error & $O-C^b$ & $N^c$ \\
\hline
0 & 54881.1051 & 0.0001 & 0.0003 & 287 \\
1 & 54881.1888 & 0.0002 & $-$0.0014 & 173 \\
23 & 54883.0717 & 0.0005 & 0.0015 & 205 \\
24 & 54883.1570 & 0.0003 & 0.0014 & 261 \\
35 & 54884.0954 & 0.0004 & $-$0.0002 & 80 \\
36 & 54884.1807 & 0.0005 & $-$0.0003 & 86 \\
37 & 54884.2652 & 0.0008 & $-$0.0013 & 88 \\
\hline
  \multicolumn{5}{l}{$^{a}$ BJD$-$2400000.} \\
  \multicolumn{5}{l}{$^{b}$ Against $max = 2454881.1048 + 0.085452 E$.} \\
  \multicolumn{5}{l}{$^{c}$ Number of points used to determine the maximum.} \\
\end{tabular}
\end{center}
\end{table}

\subsection{TV Corvi}\label{sec:tvcrv}\label{obj:tvcrv}

   We reanalyzed the 2001, 2003 and 2004 data published in \citet{uem05tvcrv}
(tables \ref{tab:tvcrvoc2001}, \ref{tab:tvcrvoc2003} and
\ref{tab:tvcrvoc2004}).
Regarding the 2001 superoutburst, we obtained a result similar to that
in \citet{uem05tvcrv}.  The $P_{\rm dot}$ was $+6.2(1.5) \times 10^{-5}$
($1 \le E \le 109$).
We, however, obtained a different result for the 2004 superoutburst
The $O-C$ diagram was similar to that of 2001 one, contrary to the analysis
in \citet{uem05tvcrv}
(subsection \ref{sec:different}; figure \ref{fig:tvcrvcomp}).
We obtained $P_{\rm dot}$ = $+9.5(3.1) \times 10^{-5}$ ($16 \le E \le 103$),
excluding the initial stage of early evolution (stage A) and last segment
(stage C) after a period decrease.

\begin{table}
\caption{Superhump maxima of TV Crv (2001).}\label{tab:tvcrvoc2001}
\begin{center}

\end{center}
\end{table}

\begin{table}
\caption{Superhump maxima of TV Crv (2003).}\label{tab:tvcrvoc2003}
\begin{center}
%
\end{center}
\end{table}

\begin{table}
\caption{Superhump maxima of TV Crv (2004).}\label{tab:tvcrvoc2004}
\begin{center}
%
\end{center}
\end{table}

\subsection{V337 Cygni}\label{obj:v337cyg}

   Although V337 Cyg had long been registered as a dwarf nova,
the identification of the true object was made only recently
by J. Manek based on archival Sonneberg plate (vsnet 775, 780\footnote{
  $<$http://www.kusastro.kyoto-u.ac.jp/vsnet/DNe/v337cyg.html$>$
}; see also \cite{boy07v337cyg}).

\begin{figure}
  \begin{center}
    \FigureFile(88mm,110mm){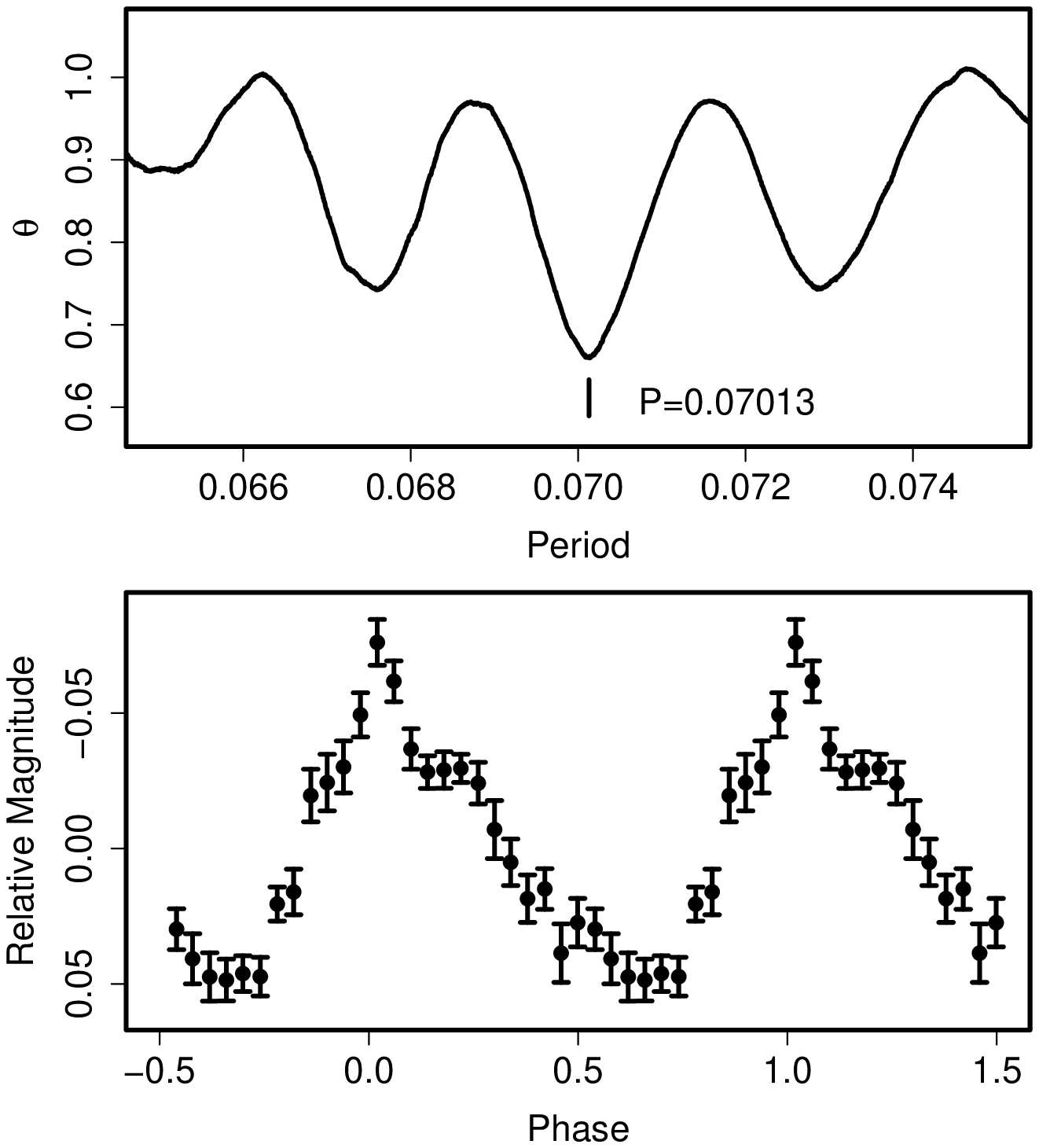}
  \end{center}
  \caption{Superhumps in V337 Cyg (2006). (Upper): PDM analysis.
     (Lower): Phase-averaged profile.}
  \label{fig:v337cygshpdm}
\end{figure}

   We analyzed the AAVSO data of the 2006 superoutburst, the same
outburst reported in \citet{boy07v337cyg}.  These observations
were performed during the late stage of the superoutburst, and the
superhumps were most likely stage C superhumps.
The times of maxima are given in table \ref{tab:v337cygoc2006}.
The mean $P_{\rm SH}$ was determined with the PDM method
to be 0.07013(3) d (figure \ref{fig:v337cygshpdm}).
This outburst was followed by a rebrightening according to the
AAVSO data.  We might expect a positive $P_{\rm dot}$ if observations
covered the earlier stage of this superoutburst.

\begin{table}
\caption{Superhump maxima of V337 Cyg (2006).}\label{tab:v337cygoc2006}
\begin{center}
\begin{tabular}{ccccc}
\hline\hline
$E$ & max$^a$ & error & $O-C^b$ & $N^c$ \\
\hline
0 & 53886.4495 & 0.0010 & $-$0.0021 & 69 \\
1 & 53886.5209 & 0.0014 & $-$0.0008 & 61 \\
4 & 53886.7305 & 0.0009 & $-$0.0011 & 72 \\
5 & 53886.8052 & 0.0010 & 0.0035 & 71 \\
6 & 53886.8741 & 0.0009 & 0.0024 & 71 \\
7 & 53886.9404 & 0.0012 & $-$0.0012 & 72 \\
28 & 53888.4064 & 0.0081 & $-$0.0053 & 40 \\
29 & 53888.4885 & 0.0014 & 0.0068 & 117 \\
30 & 53888.5495 & 0.0009 & $-$0.0023 & 119 \\
\hline
  \multicolumn{5}{l}{$^{a}$ BJD$-$2400000.} \\
  \multicolumn{5}{l}{$^{b}$ Against $max = 2453886.4516 + 0.070003 E$.} \\
  \multicolumn{5}{l}{$^{c}$ Number of points used to determine the maximum.} \\
\end{tabular}
\end{center}
\end{table}

\subsection{V503 Cygni}\label{obj:v503cyg}

   \citet{har95v503cyg} established the SU UMa-type nature of this
object and reported a mean $P_{\rm SH}$ of 0.08101(4) d.

   We observed the 2002 July superoutburst.  The times of superhump
maxima are listed in table \ref{tab:v503cygoc2002}.
Although the coverage of the observation was not sufficient,
a likely stage B--C transition was recorded.  The parameters are
listed in table \ref{tab:perlist}.
The observation of the 2008 December superoutburst is given in
table \ref{tab:v503cygoc2008}.  There was an apparent break in the
$O-C$ around $E=49$.  Due to the limited phase coverage,
we determined superhump periods for the first (before BJD 2454824)
and the second (after BJD 2454823) intervals with the PDM method.
The periods were 0.081767(45) d and 0.081022(18) d, respectively.
These periods were adopted in table \ref{tab:perlist}.

   This object is of particular interest since its
supercycle is one of the next shortest to ER UMa stars and MN Dra
(\cite{har95v503cyg}; \cite{kat02v503cyg}) and there appears to be
a hint of superhump evolution similar to ER UMa stars
(\cite{har95v503cyg}, figure 7).
It would be worth studying whether a phase reversal, or early
emergence of stage C superhumps (cf. subsection \ref{sec:erumastars}),
also takes place in this system.

\begin{table}
\caption{Superhump maxima of V503 Cyg (2002).}\label{tab:v503cygoc2002}
\begin{center}
\begin{tabular}{ccccc}
\hline\hline
$E$ & max$^a$ & error & $O-C^b$ & $N^c$ \\
\hline
0 & 52478.2155 & 0.0004 & $-$0.0110 & 312 \\
13 & 52479.2861 & 0.0014 & 0.0047 & 196 \\
17 & 52479.5975 & 0.0047 & $-$0.0085 & 28 \\
18 & 52479.7013 & 0.0036 & 0.0142 & 28 \\
25 & 52480.2501 & 0.0008 & $-$0.0051 & 309 \\
30 & 52480.6656 & 0.0008 & 0.0047 & 44 \\
31 & 52480.7429 & 0.0006 & 0.0009 & 55 \\
37 & 52481.2303 & 0.0009 & 0.0014 & 324 \\
38 & 52481.3145 & 0.0008 & 0.0044 & 180 \\
49 & 52482.2026 & 0.0007 & $-$0.0000 & 238 \\
76 & 52484.3913 & 0.0008 & $-$0.0023 & 50 \\
77 & 52484.4713 & 0.0014 & $-$0.0034 & 65 \\
\hline
  \multicolumn{5}{l}{$^{a}$ BJD$-$2400000.} \\
  \multicolumn{5}{l}{$^{b}$ Against $max = 2452478.2265 + 0.081145 E$.} \\
  \multicolumn{5}{l}{$^{c}$ Number of points used to determine the maximum.} \\
\end{tabular}
\end{center}
\end{table}

\begin{table}
\caption{Superhump maxima of V503 Cyg (2008).}\label{tab:v503cygoc2008}
\begin{center}
\begin{tabular}{ccccc}
\hline\hline
$E$ & max$^a$ & error & $O-C^b$ & $N^c$ \\
\hline
0 & 54819.9455 & 0.0012 & $-$0.0035 & 81 \\
36 & 54822.8735 & 0.0017 & 0.0029 & 78 \\
49 & 54823.9288 & 0.0010 & 0.0033 & 100 \\
98 & 54827.8995 & 0.0017 & $-$0.0027 & 64 \\
\hline
  \multicolumn{5}{l}{$^{a}$ BJD$-$2400000.} \\
  \multicolumn{5}{l}{$^{b}$ Against $max = 2454819.9490 + 0.081155 E$.} \\
  \multicolumn{5}{l}{$^{c}$ Number of points used to determine the maximum.} \\
\end{tabular}
\end{center}
\end{table}

\subsection{V550 Cygni}\label{obj:v550cyg}

   Although V550 Cyg had long been known as a dwarf nova, the supposed
identification became available only in 1999 \citep{ski99VSID}.
Two outbursts were detected in 2000 (vsnet-alert 3993, 5191).
Superhumps were detected during the August outburst (vsnet-alert 5196).
H. Yamaoka provided astrometry from outburst images (vsnet-alert 5210),
which slightly differed from the position in \citet{ski99VSID}, making
the full amplitude of outbursts larger than five magnitudes.

\begin{figure}
  \begin{center}
    \FigureFile(88mm,110mm){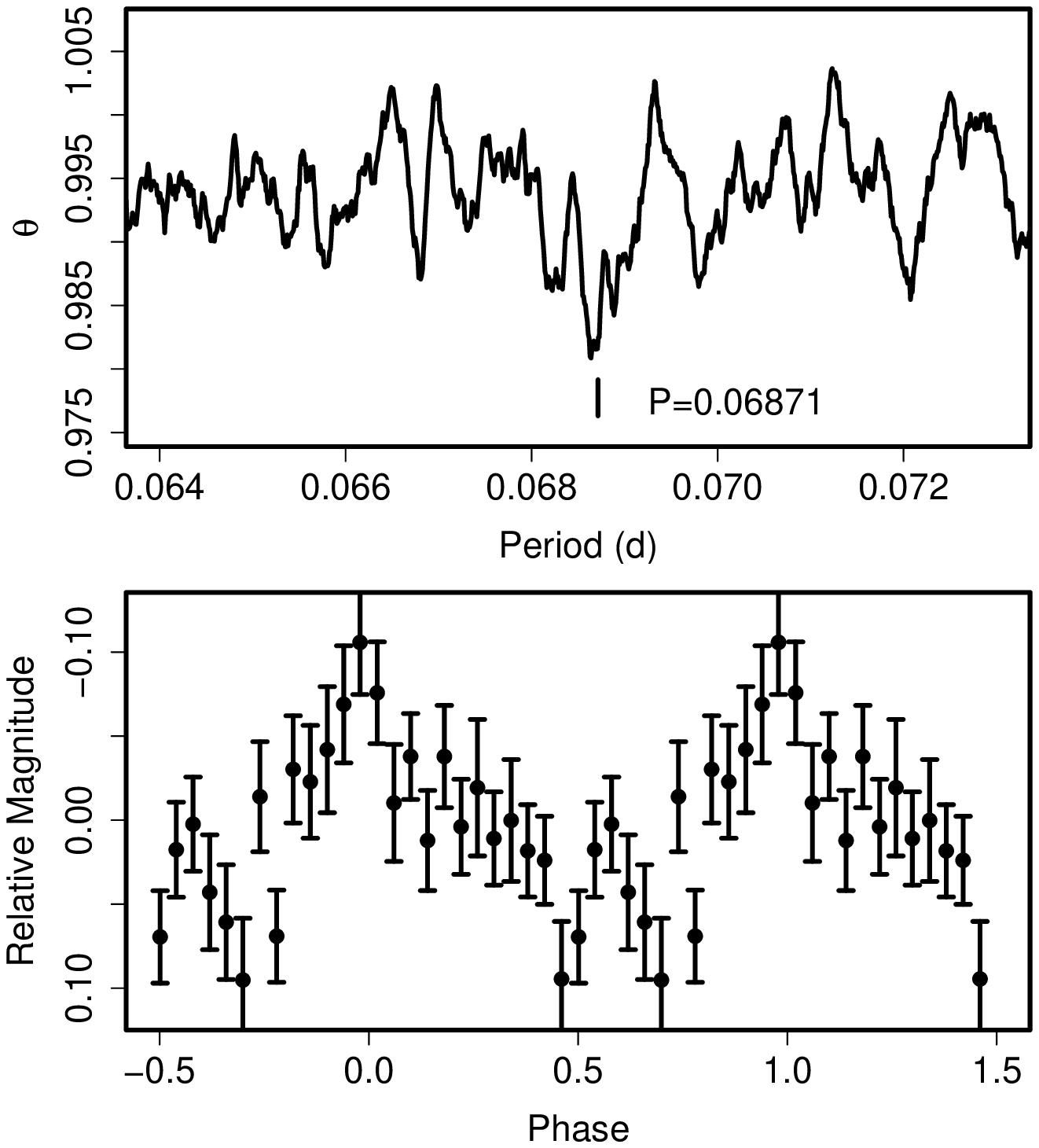}
  \end{center}
  \caption{Superhumps in V550 Cyg (2000). (Upper): PDM analysis.
     (Lower): Phase-averaged profile.}
  \label{fig:v550cygshpdm}
\end{figure}

   The mean superhump period with the PDM method was 0.06871(6) d
(figure \ref{fig:v550cygshpdm}).
The times of superhump maxima are listed in table \ref{tab:v550cygoc2000}.
The outburst was apparently observed during its middle-to-late course,
and a stage B--C transition was recorded.
The mean $P_{\rm SH}$ for stages B and C were 0.06917(26) d and
0.06848(6) d, respectively.

\begin{table}
\caption{Superhump maxima of V550 Cyg (2000).}\label{tab:v550cygoc2000}
\begin{center}
\begin{tabular}{ccccc}
\hline\hline
$E$ & max$^a$ & error & $O-C^b$ & $N^c$ \\
\hline
0 & 51777.0086 & 0.0018 & $-$0.0003 & 129 \\
14 & 51777.9680 & 0.0011 & $-$0.0031 & 148 \\
15 & 51778.0417 & 0.0016 & 0.0018 & 149 \\
16 & 51778.0926 & 0.0011 & $-$0.0161 & 147 \\
17 & 51778.1784 & 0.0040 & 0.0010 & 147 \\
18 & 51778.2396 & 0.0018 & $-$0.0066 & 265 \\
32 & 51779.2210 & 0.0088 & 0.0126 & 87 \\
33 & 51779.2910 & 0.0094 & 0.0138 & 116 \\
35 & 51779.4164 & 0.0011 & 0.0017 & 34 \\
50 & 51780.4497 & 0.0009 & 0.0040 & 41 \\
61 & 51781.2028 & 0.0015 & 0.0009 & 104 \\
62 & 51781.2697 & 0.0044 & $-$0.0009 & 124 \\
64 & 51781.4099 & 0.0016 & 0.0018 & 11 \\
65 & 51781.4741 & 0.0014 & $-$0.0027 & 26 \\
76 & 51782.2311 & 0.0088 & $-$0.0019 & 106 \\
79 & 51782.4381 & 0.0012 & $-$0.0010 & 35 \\
91 & 51783.2589 & 0.0143 & $-$0.0051 & 130 \\
\hline
  \multicolumn{5}{l}{$^{a}$ BJD$-$2400000.} \\
  \multicolumn{5}{l}{$^{b}$ Against $max = 2451777.0088 + 0.068738 E$.} \\
  \multicolumn{5}{l}{$^{c}$ Number of points used to determine the maximum.} \\
\end{tabular}
\end{center}
\end{table}

\subsection{V630 Cygni}\label{obj:v630cyg}

   The SU UMa-type nature of this dwarf nova was established by
\citet{nog01v630cyg}.  The times of superhump maxima during the 1996
superoutburst measured from these data are listed in
table \ref{tab:v630cygoc1996}.

   We further observed the 2008 superoutburst (table \ref{tab:v630cygoc2008}).
The $O-C$'s apparently showed a stage B--C transition.
The mean $P_{\rm SH}$ and $P_{\rm dot}$ for the stage B were
0.07918(7) d and $+27.4(7.7) \times 10^{-5}$, respectively.
Since the value was derived from a limited sample, the large positive
$P_{\rm dot}$ needs to be confirmed by further observations.

\begin{table}
\caption{Superhump maxima of V630 Cyg (1996).}\label{tab:v630cygoc1996}
\begin{center}
\begin{tabular}{ccccc}
\hline\hline
$E$ & max$^a$ & error & $O-C^b$ & $N^c$ \\
\hline
0 & 50313.9866 & 0.0012 & 0.0007 & 52 \\
1 & 50314.0667 & 0.0012 & 0.0014 & 45 \\
16 & 50315.2506 & 0.0014 & $-$0.0045 & 55 \\
29 & 50316.2887 & 0.0063 & 0.0024 & 34 \\
\hline
  \multicolumn{5}{l}{$^{a}$ BJD$-$2400000.} \\
  \multicolumn{5}{l}{$^{b}$ Against $max = 2450313.9860 + 0.079320 E$.} \\
  \multicolumn{5}{l}{$^{c}$ Number of points used to determine the maximum.} \\
\end{tabular}
\end{center}
\end{table}

\begin{table}
\caption{Superhump maxima of V630 Cyg (2008).}\label{tab:v630cygoc2008}
\begin{center}
\begin{tabular}{ccccc}
\hline\hline
$E$ & max$^a$ & error & $O-C^b$ & $N^c$ \\
\hline
0 & 54690.0683 & 0.0007 & $-$0.0053 & 129 \\
12 & 54691.0151 & 0.0006 & $-$0.0040 & 81 \\
25 & 54692.0444 & 0.0007 & 0.0010 & 112 \\
26 & 54692.1213 & 0.0004 & $-$0.0009 & 167 \\
39 & 54693.1564 & 0.0261 & 0.0099 & 10 \\
40 & 54693.2342 & 0.0052 & 0.0088 & 70 \\
51 & 54694.0912 & 0.0055 & $-$0.0009 & 51 \\
76 & 54696.0586 & 0.0007 & $-$0.0034 & 165 \\
77 & 54696.1350 & 0.0009 & $-$0.0057 & 118 \\
103 & 54698.1900 & 0.0113 & 0.0005 & 115 \\
\hline
  \multicolumn{5}{l}{$^{a}$ BJD$-$2400000.} \\
  \multicolumn{5}{l}{$^{b}$ Against $max = 2454690.0736 + 0.078795 E$.} \\
  \multicolumn{5}{l}{$^{c}$ Number of points used to determine the maximum.} \\
\end{tabular}
\end{center}
\end{table}

\subsection{V632 Cygni}\label{obj:v632cyg}

   The SU UMa-type nature of this dwarf nova had long been suggested
(cf. \cite{wen89v632cygv630cyg} for a historical record of bright
outbursts).  \citet{she07CVspec} determined its orbital period
to be 0.06377(8) d.  The SU UMa-type nature was finally established
during the 2008 superoutburst.

\begin{figure}
  \begin{center}
    \FigureFile(88mm,110mm){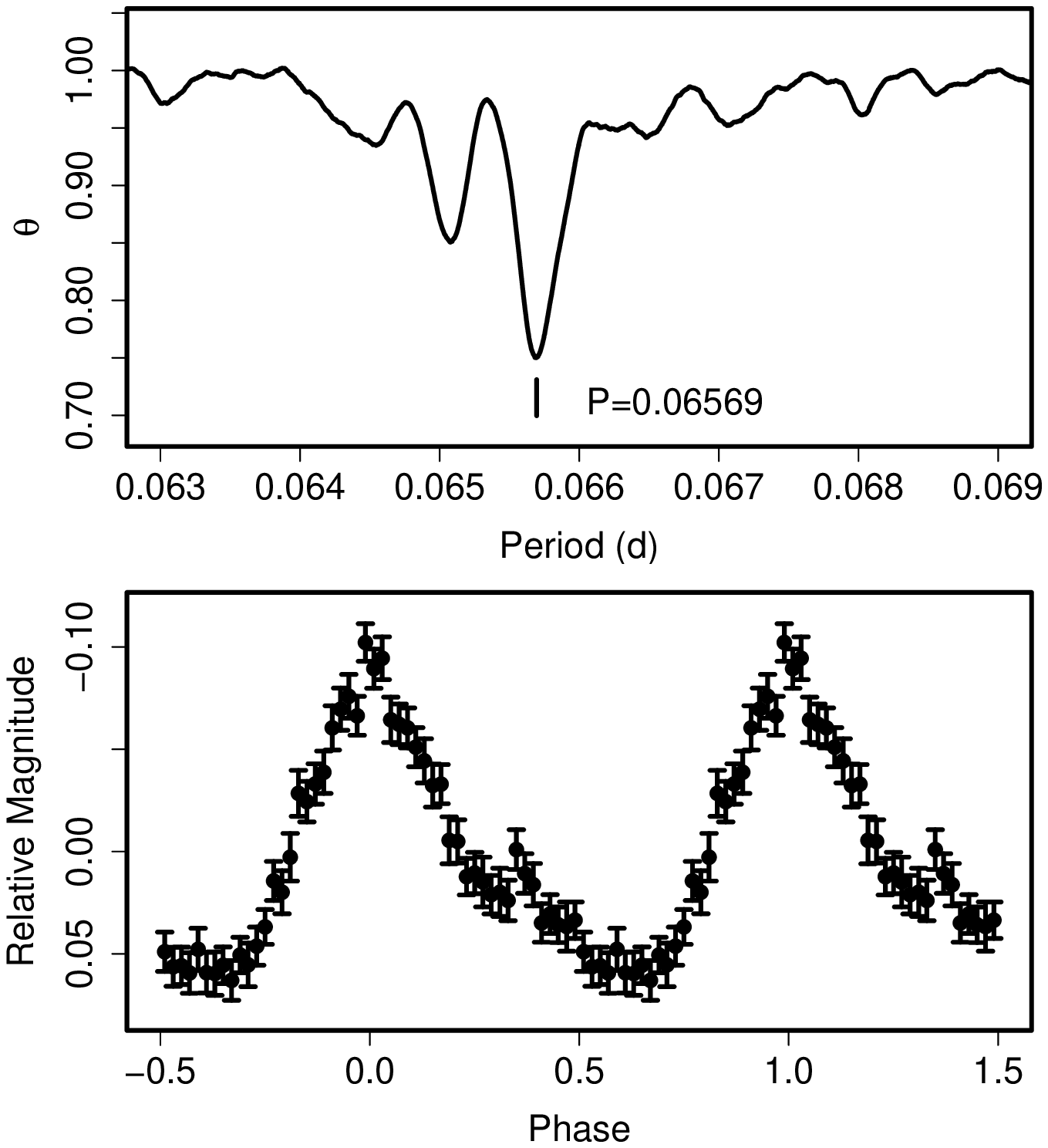}
  \end{center}
  \caption{Superhumps in V632 Cyg (2008). (Upper): PDM analysis.
     (Lower): Phase-averaged profile.}
  \label{fig:v632cygshpdm}
\end{figure}

   The global mean superhump period during the 2008 superoutburst was
0.065695(6) d (PDM method, figure \ref{fig:v632cygshpdm}).
The times of superhump maxima are listed in table \ref{tab:v632cygoc2008}.
Although the stage A--B and B--C transitions were observed,
a gap in the middle of the stage B makes determination of $P_{\rm dot}$
rather uncertain.
The value for $16 \le E \le 82$ was $+17.4(3.0) \times 10^{-5}$.

\begin{table}
\caption{Superhump maxima of V632 Cyg (2008).}\label{tab:v632cygoc2008}
\begin{center}
\begin{tabular}{ccccc}
\hline\hline
$E$ & max$^a$ & error & $O-C^b$ & $N^c$ \\
\hline
0 & 54782.3640 & 0.0079 & $-$0.0094 & 112 \\
1 & 54782.4310 & 0.0048 & $-$0.0081 & 69 \\
9 & 54782.9596 & 0.0014 & $-$0.0052 & 35 \\
13 & 54783.2258 & 0.0003 & $-$0.0018 & 125 \\
14 & 54783.2922 & 0.0005 & $-$0.0010 & 102 \\
15 & 54783.3567 & 0.0007 & $-$0.0023 & 96 \\
16 & 54783.4264 & 0.0018 & 0.0017 & 39 \\
23 & 54783.8856 & 0.0003 & 0.0010 & 106 \\
24 & 54783.9505 & 0.0003 & 0.0002 & 129 \\
30 & 54784.3441 & 0.0005 & $-$0.0004 & 63 \\
31 & 54784.4103 & 0.0005 & 0.0001 & 155 \\
32 & 54784.4755 & 0.0008 & $-$0.0004 & 81 \\
65 & 54786.6457 & 0.0008 & 0.0016 & 46 \\
66 & 54786.7131 & 0.0008 & 0.0033 & 64 \\
67 & 54786.7801 & 0.0007 & 0.0047 & 55 \\
80 & 54787.6374 & 0.0005 & 0.0078 & 55 \\
81 & 54787.7034 & 0.0006 & 0.0081 & 66 \\
82 & 54787.7729 & 0.0016 & 0.0119 & 31 \\
104 & 54789.2107 & 0.0004 & 0.0042 & 105 \\
105 & 54789.2759 & 0.0008 & 0.0037 & 115 \\
106 & 54789.3450 & 0.0020 & 0.0071 & 46 \\
110 & 54789.6027 & 0.0006 & 0.0020 & 45 \\
111 & 54789.6694 & 0.0006 & 0.0031 & 68 \\
112 & 54789.7352 & 0.0007 & 0.0031 & 69 \\
115 & 54789.9278 & 0.0011 & $-$0.0014 & 36 \\
116 & 54789.9920 & 0.0009 & $-$0.0029 & 49 \\
130 & 54790.9178 & 0.0013 & 0.0031 & 123 \\
131 & 54790.9802 & 0.0013 & $-$0.0002 & 128 \\
145 & 54791.8925 & 0.0013 & $-$0.0077 & 63 \\
156 & 54792.6065 & 0.0014 & $-$0.0165 & 48 \\
157 & 54792.6792 & 0.0020 & $-$0.0095 & 32 \\
\hline
  \multicolumn{5}{l}{$^{a}$ BJD$-$2400000.} \\
  \multicolumn{5}{l}{$^{b}$ Against $max = 2454782.3734 + 0.065702 E$.} \\
  \multicolumn{5}{l}{$^{c}$ Number of points used to determine the maximum.} \\
\end{tabular}
\end{center}
\end{table}

\subsection{V1028 Cygni}\label{sec:v1028cyg}\label{obj:v1028cyg}

   \citet{bab00v1028cyg} reported the detection of positive period
derivative during the 1995 superoutburst.  This outburst was indeed
one of the earliest with significantly positive $P_{\rm dot}$'s.
We reanalyzed the data, combined with the AAVSO observations,
for an improvement of the parameters.
The results generally confirmed the conclusion by \citet{bab00v1028cyg}
(table \ref{tab:v1028cygoc1995}).
The $P_{\rm dot}$ for the interval $15 \le E \le 148$ (stage B) was
$+8.2(1.2) \times 10^{-5}$.

   We further analyzed the 1996, 1999, 2001, 2002, 2004 and 2008
superoutbursts
(tables \ref{tab:v1028cygoc1996}, \ref{tab:v1028cygoc1999},
\ref{tab:v1028cygoc2001}, \ref{tab:v1028cygoc2002},
\ref{tab:v1028cygoc2004}, \ref{tab:v1028cygoc2008}).
The observation in 2001 and 2002 covered
the middle-to-late portion of the superoutburst, and the $O-C$ diagram
commonly showed a transition to a shorter period (stage C).
For the 1999 and 2002 superoutbursts, we obtained $P_{\rm dot}$
before this transition as follows:
$P_{\rm dot}$ = $+12.2(3.1) \times 10^{-5}$ (1999, $E \le 148$) and
$P_{\rm dot}$ = $+14.7(5.5) \times 10^{-5}$ (2002, $E \le 55$).
Although the 1996 and 2004 superoutbursts were preceded
by a distinct precursor, only the late stage of the superoutburst was
meaningfully observed.

\begin{figure}
  \begin{center}
    \FigureFile(88mm,70mm){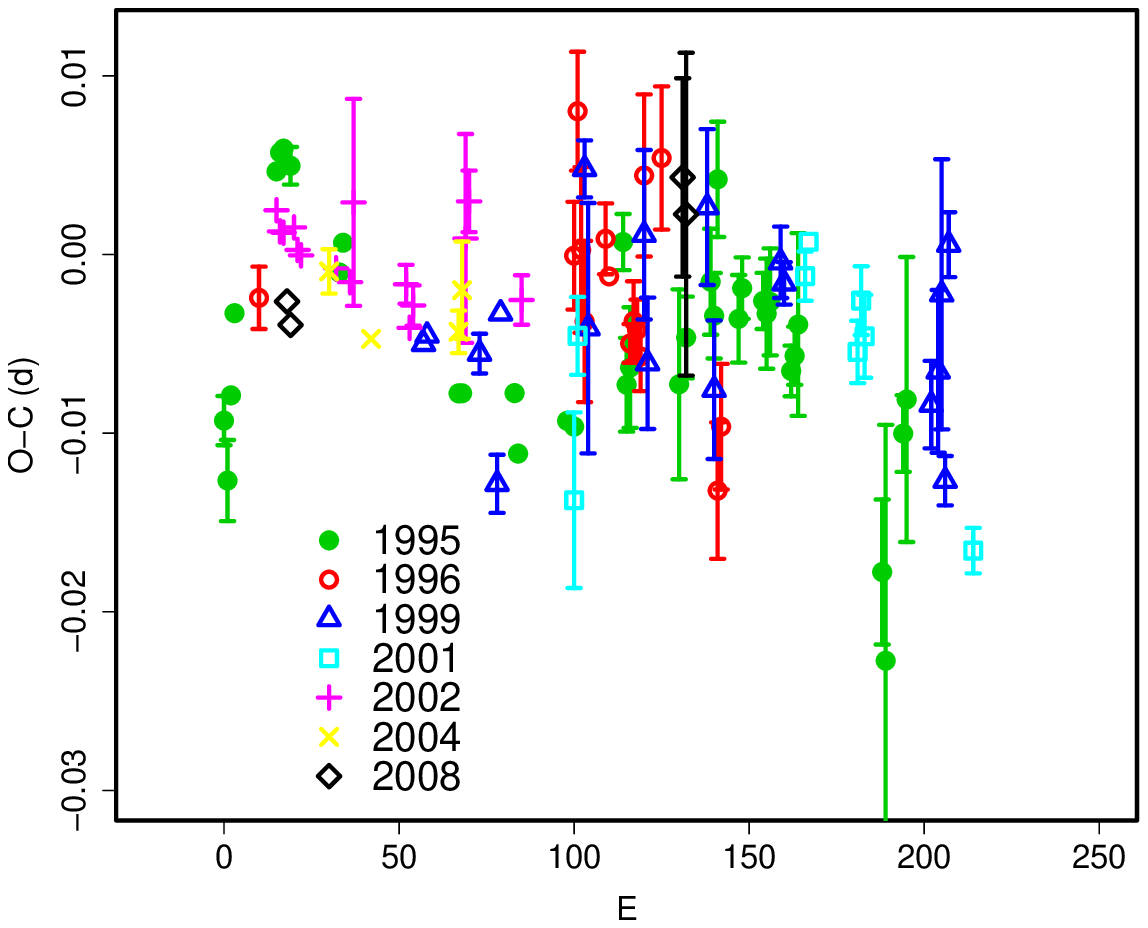}
  \end{center}
  \caption{Comparison of $O-C$ diagrams of V1028 Cyg between different
  superoutbursts.  A period of 0.06180 d was used to draw this figure.
  Approximate cycle counts ($E$) after the start of the
  superoutburst (the start of the main superoutburst when preceded by
  a precursor) were used.  The $E$ for the 2008 superoutburst was
  somewhat uncertain due to the lack of observations at the early stage.
  }
  \label{fig:v1028cygcomp}
\end{figure}

   A comparison of $O-C$ diagrams is shown in figure \ref{fig:v1028cygcomp}.
There appears to be a slight variation in the $O-C$ behavior during the
late stage (stage B--C).  This may have been caused by the difference
in the extent between superoutbursts.

\begin{table}
\caption{Superhump maxima of V1028 Cyg (1995).}\label{tab:v1028cygoc1995}
\begin{center}

\end{center}
\end{table}

\begin{table}
\caption{Superhump maxima of V1028 Cyg (1996).}\label{tab:v1028cygoc1996}
\begin{center}
%
\end{center}
\end{table}

\begin{table}
\caption{Superhump maxima of V1028 Cyg (1999).}\label{tab:v1028cygoc1999}
\begin{center}
%
\end{center}
\end{table}

\begin{table}
\caption{Superhump maxima of V1028 Cyg (2001).}\label{tab:v1028cygoc2001}
\begin{center}
%
\end{center}
\end{table}

\subsection{V1113 Cygni}\label{obj:v1113cyg}

   We have reanalyzed the observation in \citet{kat96v1113cyg}
and obtained new observations during the 2008 superoutburst.
Both observations covered the relatively early stages of the
superoutbursts.
The times of superhump maxima are listed in tables \ref{tab:v1113cygoc1994}
and \ref{tab:v1113cygoc2008}, respectively.
The resultant global $P_{\rm dot}$'s were $-19.2(6.8) \times 10^{-5}$
and $-5.2(4.7) \times 10^{-5}$, respectively.
The former strongly negative value
can be interpreted as a result of a possible stage A--B transition.

\begin{table}
\caption{Superhump maxima of V1113 Cyg (1994).}\label{tab:v1113cygoc1994}
\begin{center}
\begin{tabular}{ccccc}
\hline\hline
$E$ & max$^a$ & error & $O-C^b$ & $N^c$ \\
\hline
0 & 49598.0086 & 0.0022 & $-$0.0046 & 24 \\
14 & 49599.1249 & 0.0007 & 0.0022 & 29 \\
26 & 49600.0790 & 0.0005 & 0.0052 & 47 \\
51 & 49602.0541 & 0.0005 & $-$0.0009 & 35 \\
64 & 49603.0835 & 0.0005 & $-$0.0018 & 25 \\
\hline
  \multicolumn{5}{l}{$^{a}$ BJD$-$2400000.} \\
  \multicolumn{5}{l}{$^{b}$ Against $max = 2449598.0132 + 0.079253 E$.} \\
  \multicolumn{5}{l}{$^{c}$ Number of points used to determine the maximum.} \\
\end{tabular}
\end{center}
\end{table}

\begin{table}
\caption{Superhump maxima of V1113 Cyg (2008).}\label{tab:v1113cygoc2008}
\begin{center}
\begin{tabular}{ccccc}
\hline\hline
$E$ & max$^a$ & error & $O-C^b$ & $N^c$ \\
\hline
0 & 54757.2763 & 0.0003 & $-$0.0012 & 109 \\
1 & 54757.3568 & 0.0004 & 0.0002 & 139 \\
2 & 54757.4356 & 0.0004 & $-$0.0001 & 135 \\
8 & 54757.9108 & 0.0005 & 0.0009 & 143 \\
9 & 54757.9866 & 0.0008 & $-$0.0024 & 81 \\
13 & 54758.3069 & 0.0003 & 0.0017 & 156 \\
14 & 54758.3849 & 0.0004 & 0.0007 & 159 \\
21 & 54758.9386 & 0.0009 & 0.0010 & 118 \\
34 & 54759.9650 & 0.0009 & $-$0.0002 & 155 \\
46 & 54760.9125 & 0.0013 & $-$0.0014 & 73 \\
47 & 54760.9936 & 0.0009 & 0.0007 & 68 \\
\hline
  \multicolumn{5}{l}{$^{a}$ BJD$-$2400000.} \\
  \multicolumn{5}{l}{$^{b}$ Against $max = 2454757.2775 + 0.079051 E$.} \\
  \multicolumn{5}{l}{$^{c}$ Number of points used to determine the maximum.} \\
\end{tabular}
\end{center}
\end{table}

\subsection{V1251 Cygni}\label{sec:v1251cyg}\label{obj:v1251cyg}

   The history of V1251 Cyg was summarized in \citet{kat95v1251cyg}.
Only five outbursts (1963, 1991, 1994--1995, 1997 and 2008) have been
recorded.  All of these outbursts were superoutbursts, and were associated
with a rebrightening (1997, 2008).  Despite the long $P_{\rm SH}$,
\citet{kat01hvvir} included this object as a candidate WZ Sge-type
dwarf nova based on the long recurrence time, the large outburst amplitude
and the lack of normal outbursts.

   We observed the 1991 \citep{kat95v1251cyg}, 1994--1995, and 2008
superoutbursts.  The 1995 observation was performed on single night,
only confirming the presence of a superhump.  The times of superhump maxima
(refined times for the 1991 superoutburst) are listed in tables
\ref{tab:v1251cygoc1991} and \ref{tab:v1251cygoc2008}.

\begin{figure}
  \begin{center}
    \FigureFile(88mm,110mm){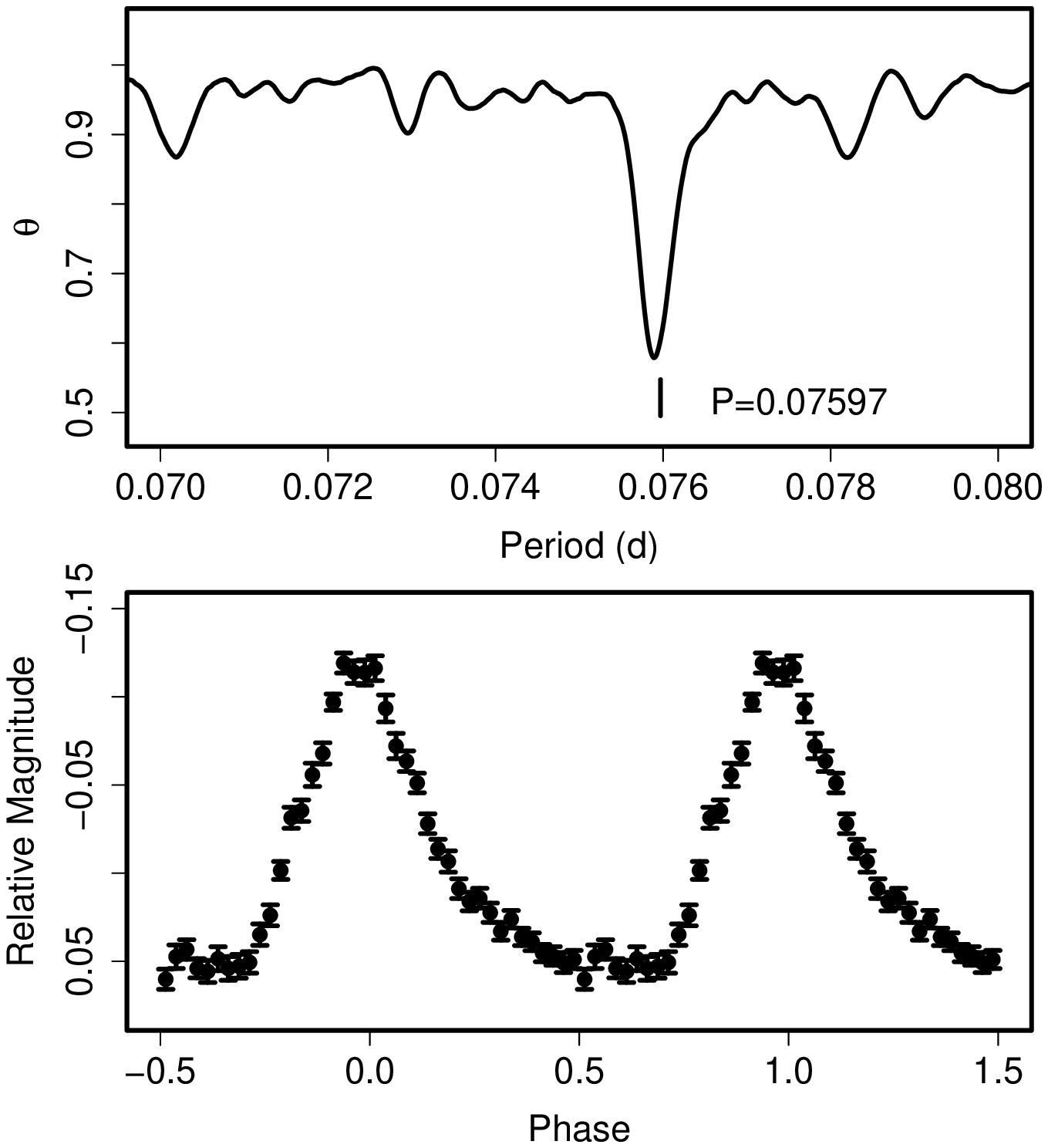}
  \end{center}
  \caption{Ordinary superhumps in V1251 Cyg (2008). (Upper): PDM analysis.
     (Lower): Phase-averaged profile.}
  \label{fig:v1251shpdm}
\end{figure}

\begin{figure}
  \begin{center}
    \FigureFile(88mm,110mm){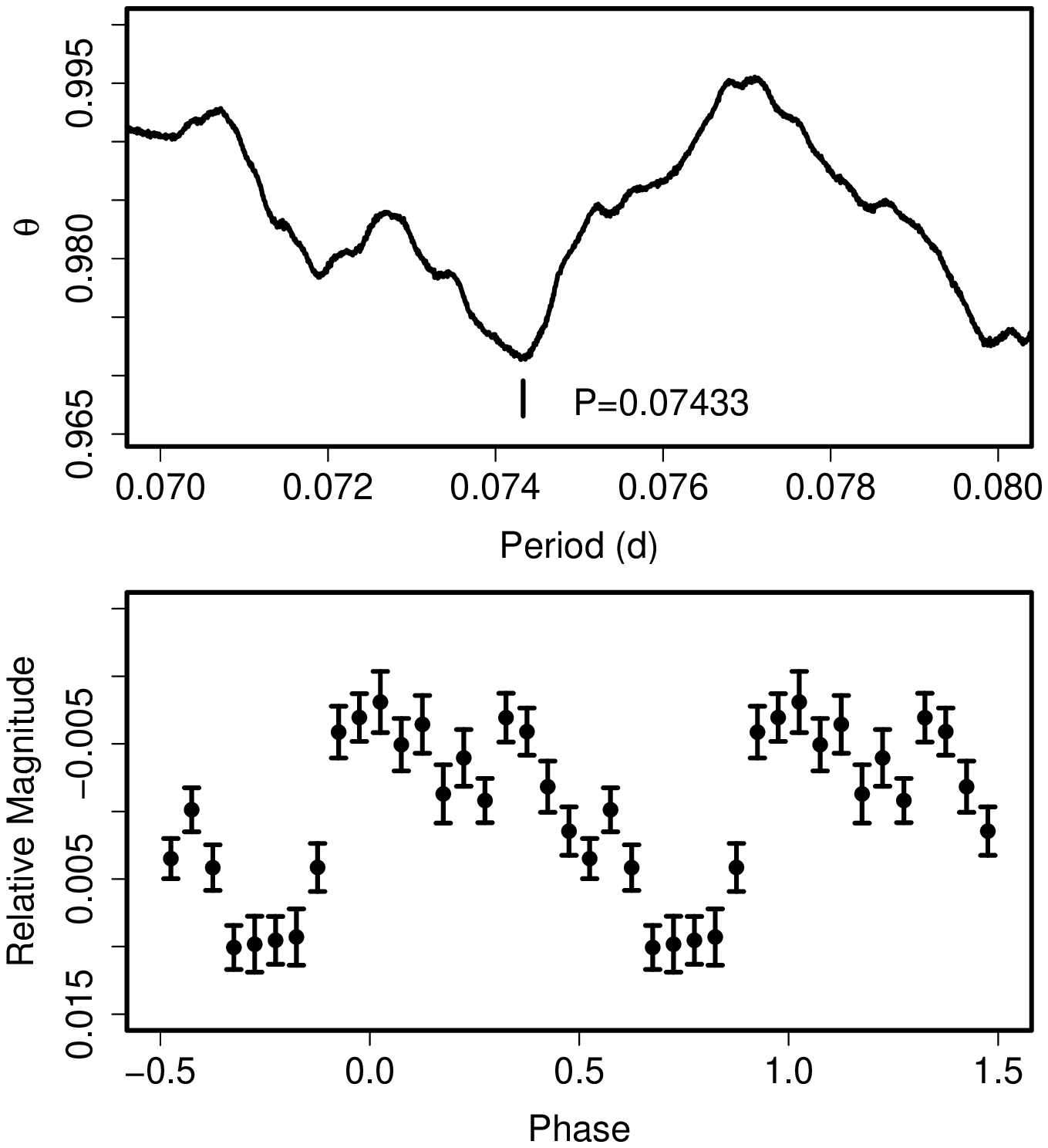}
  \end{center}
  \caption{Early superhumps in V1251 Cyg (2008). (Upper): PDM analysis.
     (Lower): Phase-averaged profile.}
  \label{fig:v1251eshpdm}
\end{figure}

\begin{figure}
  \begin{center}
    \FigureFile(88mm,70mm){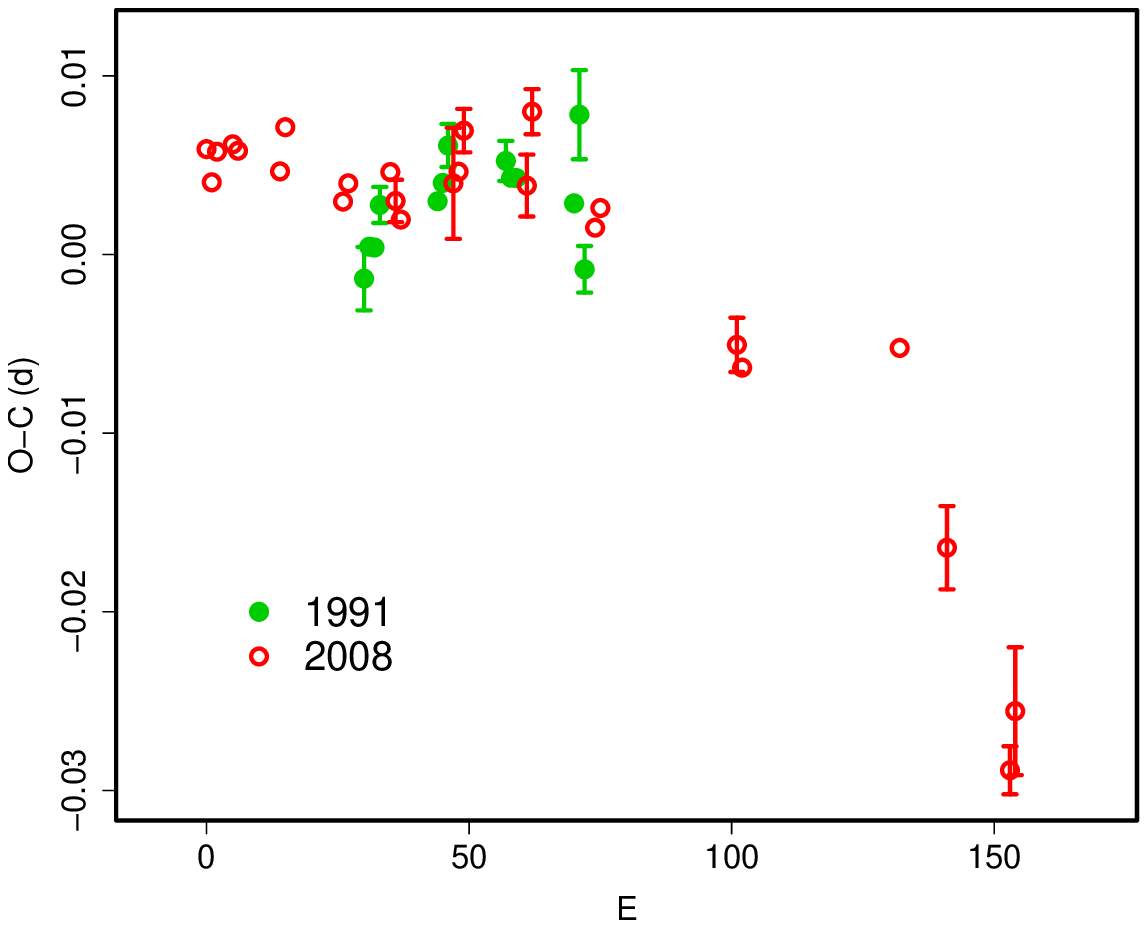}
  \end{center}
  \caption{Comparison of $O-C$ diagrams of V1251 Cyg between different
  superoutbursts.  A period of 0.06180 d was used to draw this figure.
  Approximate cycle counts ($E$, estimated ones for the 1991 superoutbursts)
  after the appearance of superhumps were used.
  }
  \label{fig:v1251cygcomp}
\end{figure}

   The 2008 superoutburst was clearly composed of stages B and C.
The mean $P_{\rm SH}$ and $P_{\rm dot}$ for the stage B were
0.07597(2) d and $+6.0(2.7) \times 10^{-5}$, respectively.
($0 \le E \le 62$).  The last part of the stage C includes superhumps
during the rapid fading stage ($E=141$) and the post-superoutburst stage
($E=153, 154$).  A phase shift expected for traditional late superhumps
was not recorded.
It took five days ordinary superhumps (figure \ref{fig:v1251shpdm}) to appear
after the onset of the outburst, which is unusually long for an SU UMa-type
dwarf nova with this $P_{\rm SH}$ (this anomaly was already addressed
in \cite{kat91v1251cygiauc}).  During this stage, double-wave modulations
similar to early superhumps in WZ Sge-type dwarf novae were observed
(figure \ref{fig:v1251eshpdm}).
The period (0.07433(6) d, vsnet-alert 10612; refined in this paper)
is 2.2 \% shorter than the above $P_{\rm SH}$ and can be good candidate
for $P_{\rm orb}$.
Despite its long $P_{\rm SH}$, V1251 Cyg is extremely analogous to
WZ Sge-type dwarf novae.  The implication for the presence of such
a long-$P_{\rm SH}$ WZ Sge-like objects was discussed in \citet{ish01rzleo},
\citet{kat02wzsgeESH}.  Compared to the 2008 superoutburst, only later
half of the stage B was likely recorded during the 1991 superoutburst
(figure \ref{fig:v1251cygcomp}).

\begin{table}
\caption{Superhump maxima of V1251 Cyg (1991).}\label{tab:v1251cygoc1991}
\begin{center}

\end{center}
\end{table}

\begin{table}
\caption{Superhump maxima of V1251 Cyg (2008).}\label{tab:v1251cygoc2008}
\begin{center}
%
\end{center}
\end{table}

\subsection{V1316 Cygni}\label{obj:v1316cyg}

   Although V1316 Cyg was listed as an SU UMa-type dwarf nova in
the GCVS \citep{GCVS}, the misidentification on the original discovery
paper \citep{rom69v1316cyg} led to a long-lasting confusion.
\citet{hen97sequence} suggested a nearby faint blue star to be
the genuine V1316 Cyg, whose variability in quiescence was confirmed
in 2000 (B. Sumner, AAVSO discussion message).
This suggestion was confirmed by the later
detection of an outburst in 2002 (M. Moriyama, vsnet-campaign-dn 2910).
Subsequent observations starting in 2003 recorded a number of outbursts.
It has now been established that the object a short cycle length
of outbursts \citep{she06v1316cyg} as originally reported
by \citet{rom69v1316cyg}.

   \citet{boy08v1316cyg} observed the 2006 superoutburst of this object
and reported a phase shift around $E=90$, which they interpreted
as the appearance of (traditional) late superhumps.  The phase shift
was so large that it is difficult to attribute it to the stage B period
increase.  The relatively early appearance of late superhumps, or
the occurrence of a phase reversal, is somewhat reminiscent to
ER UMa (section \ref{sec:erumastars}).  Judging from the short
($\sim$ 10 d) outburst \citep{she06v1316cyg}, this object appears to have
a high mass-transfer rate that could enable ER UMa-like evolution of
superhumps.  The other parameters, such as the duration of the superoutburst
and $P_{\rm SH}$ are, however, unlike those of ER UMa and resemble
those of a long $P_{\rm SH}$-system BF Ara \citep{kat03bfara}.
Since \citet{boy08v1316cyg} used a different method in extracting
maxima times, a reanalysis of their data and tracking maxima of the
original superhumps as in ER UMa (section \ref{sec:erumastars}) might be
helpful in better understanding this system and its relation to ER UMa.
We identified the period for $E \ge 94$ as the stage C superhumps
and listed in table \ref{tab:perlist}.

\subsection{V1454 Cygni}\label{obj:v1454cyg}

   V1454 Cyg is a poorly-known dwarf nova.  Although discovery observations
suggested the existence of long and short outbursts resembling an SU UMa-type
dwarf nova (\cite{pin72v1454cyg}; \cite{los79v1454cyg}),
spectroscopic observation could not confirm the CV nature of the suggested
quiescent counterpart (\cite{liu99CVspec2}; this later turned out to be
a false identification).

\begin{figure}
  \begin{center}
    \FigureFile(88mm,110mm){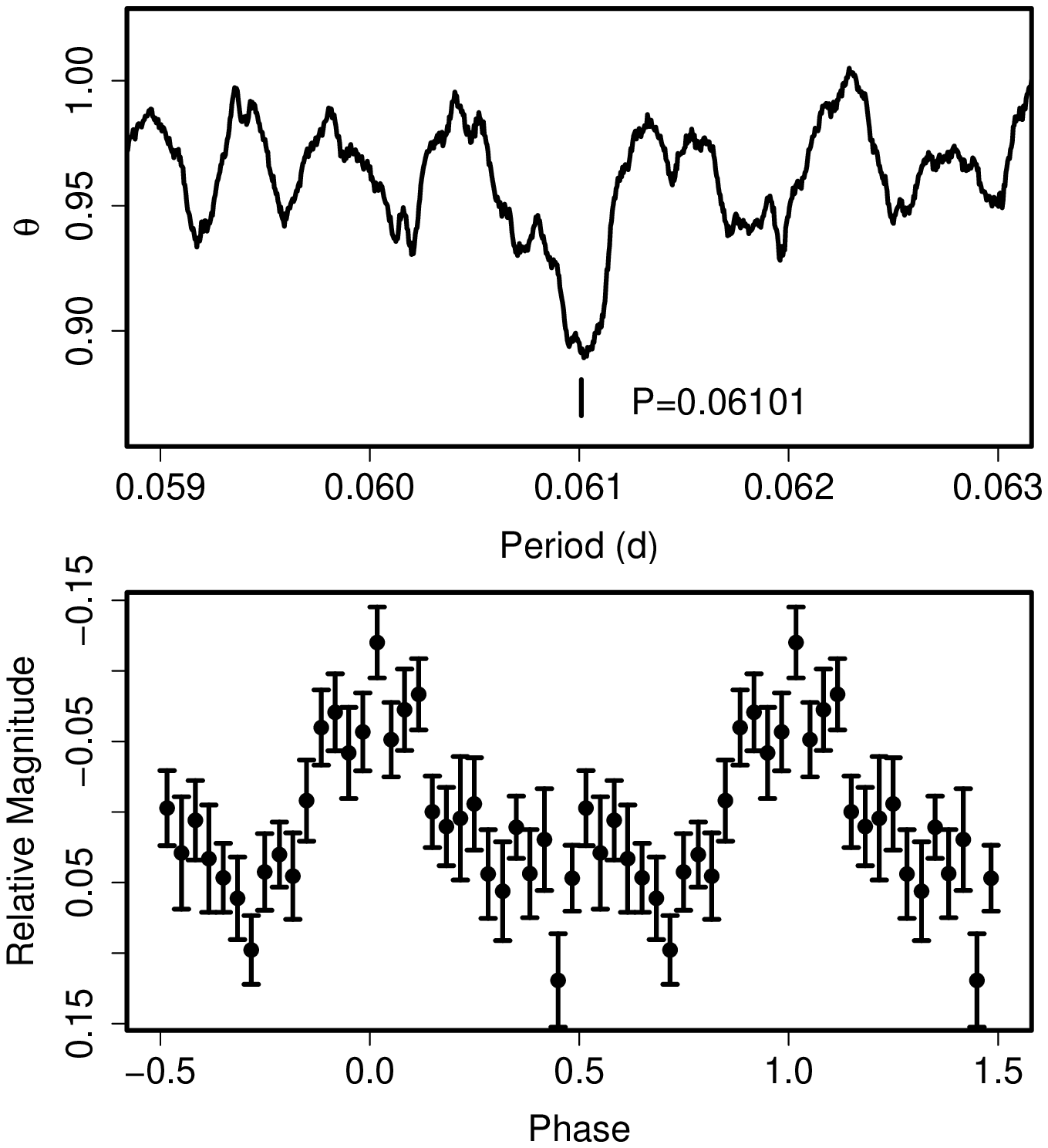}
  \end{center}
  \caption{Superhumps in V1454 Cyg (2006) for BJD 2454070.5--2454076.5.
     (Upper): PDM analysis.
     (Lower): Phase-averaged profile.}
  \label{fig:v1454cygshpdm}
\end{figure}

   The object underwent a long, bright outburst in 1996 (vsnet-obs 4039).
During the 2006 outburst, announced by J. Shears (November 23),
one of the authors (Njh) undertook
time-resolved CCD photometry, and detected superhumps.
During the first seven days, the superhump signal was very weak.
The superhumps showed a remarkable growth on December 1 and were
followed until December 6.  On December 11, the object showed a
trend of rebrightening around the termination of the plateau
stage (cf. \cite{kat03hodel}).  We used the data for December
1--6 to determine the superhump period and its variation.
A PDM analysis yielded a mean period of 0.06101(2) d
(figure \ref{fig:v1454cygshpdm}).
One-day aliases appear to be excluded from the December 1 data.
The times of maxima identified with this $P_{\rm SH}$
are listed in table \ref{tab:v1454cygoc2006}, likely composed of
a stage B--C transition and a possible stage A observation at $E=0$.
The $P_{\rm dot}$ for $113 \le E \le 196$ (stage B)
was $+15.0(4.3) \times 10^{-5}$.

\begin{table}
\caption{Superhump maxima of V1454 Cyg (2006).}\label{tab:v1454cygoc2006}
\begin{center}
\begin{tabular}{ccccc}
\hline\hline
$E$ & max$^a$ & error & $O-C^b$ & $N^c$ \\
\hline
0 & 54063.9562 & 0.0019 & $-$0.0274 & 83 \\
113 & 54070.8827 & 0.0008 & 0.0113 & 88 \\
114 & 54070.9395 & 0.0012 & 0.0071 & 87 \\
135 & 54072.2185 & 0.0007 & 0.0061 & 23 \\
163 & 54073.9221 & 0.0048 & 0.0029 & 84 \\
179 & 54074.9054 & 0.0017 & 0.0110 & 88 \\
195 & 54075.8850 & 0.0020 & 0.0153 & 66 \\
196 & 54075.9445 & 0.0031 & 0.0138 & 66 \\
234 & 54078.2569 & 0.0024 & 0.0100 & 34 \\
261 & 54079.8787 & 0.0221 & $-$0.0140 & 49 \\
262 & 54079.9336 & 0.0047 & $-$0.0201 & 73 \\
278 & 54080.9128 & 0.0027 & $-$0.0161 & 89 \\
\hline
  \multicolumn{5}{l}{$^{a}$ BJD$-$2400000.} \\
  \multicolumn{5}{l}{$^{b}$ Against $max = 2454063.9836 + 0.060955 E$.} \\
  \multicolumn{5}{l}{$^{c}$ Number of points used to determine the maximum.} \\
\end{tabular}
\end{center}
\end{table}

\subsection{V1504 Cygni}\label{obj:v1504cyg}

   \citet{raj87v1504cyg} suggested that this object is an SU UMa-type
dwarf nova based on the presence of two types of outbursts.
\citet{nog97v1504cyg} indeed confirmed the presence of superhumps
during the 1994 outburst.  \citet{tho97uvpervyaqrv1504cyg} reported
spectroscopic orbital period.  Since the alias selection was incorrect
in \citet{nog97v1504cyg}, we (re)analyzed the 1994, 2008 and 2009
superoutbursts (tables \ref{tab:v1504cygoc2008}, \ref{tab:v1504cygoc2009})
to determine the superhump period.  The results are summarized in
table \ref{tab:perlist}.  The 1994 and 2008 superoutbursts were probably
observed during the stage B, and the 2009 was probably observed during
the stage C.  \citet{pav02v1504cygproc} also reported correct period
identification.

\begin{table}
\caption{Superhump maxima of V1504 Cyg (2008).}\label{tab:v1504cygoc2008}
\begin{center}
\begin{tabular}{ccccc}
\hline\hline
$E$ & max$^a$ & error & $O-C^b$ & $N^c$ \\
\hline
0 & 54710.1735 & 0.0004 & $-$0.0001 & 143 \\
13 & 54711.1116 & 0.0009 & 0.0016 & 189 \\
14 & 54711.1805 & 0.0011 & $-$0.0015 & 134 \\
\hline
  \multicolumn{5}{l}{$^{a}$ BJD$-$2400000.} \\
  \multicolumn{5}{l}{$^{b}$ Against $max = 2454710.1736 + 0.072028 E$.} \\
  \multicolumn{5}{l}{$^{c}$ Number of points used to determine the maximum.} \\
\end{tabular}
\end{center}
\end{table}

\begin{table}
\caption{Superhump maxima of V1504 Cyg (2009).}\label{tab:v1504cygoc2009}
\begin{center}
\begin{tabular}{ccccc}
\hline\hline
$E$ & max$^a$ & error & $O-C^b$ & $N^c$ \\
\hline
0 & 54950.2615 & 0.0011 & 0.0000 & 142 \\
41 & 54953.2040 & 0.0019 & $-$0.0005 & 97 \\
42 & 54953.2768 & 0.0015 & 0.0005 & 132 \\
\hline
  \multicolumn{5}{l}{$^{a}$ BJD$-$2400000.} \\
  \multicolumn{5}{l}{$^{b}$ Against $max = 2454950.2615 + 0.071782 E$.} \\
  \multicolumn{5}{l}{$^{c}$ Number of points used to determine the maximum.} \\
\end{tabular}
\end{center}
\end{table}

\subsection{V2176 Cygni}\label{obj:v2176cyg}

   V2176 Cyg was discovered by \citet{hu97v2176cygiauc}.
\citet{van97v2176cygiauc} reported the detection of superhumps
with a period of 0.0561(4) d.  The object soon entered a 2-mag
``dip'' characteristic to a WZ Sge-type outburst (type-A outburst)
and exhibited a long-lasting second plateau stage following
a short precursor-like maximum \citep{nov01v2176cyg}.
Since only insufficient data were available before the dip,
we analyzed the second plateau stage.  The data used for analysis
were from AAVSO database and ones extracted from electronic figures
in \citet{nov01v2176cyg} (the data for their figure 3 were not
included for analysis).  A PDM analysis after removing the overall
trend yielded a strong periodicity of 0.056239(12) d.
The period agrees with that by \citet{van97v2176cygiauc} within
their errors, and we regard it as a refined value of $P_{\rm SH}$.
The times of superhump maxima are listed in table \ref{tab:v2176cygoc1997b}.

   We also determined times of maxima during the initial superoutburst
plateau using the data in \citet{kwa98v2176cyg}
(table \ref{tab:v2176cygoc1997a}).  These maxima could not be directly
linked by the above period.  By assuming phase continuity,
we obtained a mean period of 0.05607(5) d between BJD 2450696 and
2450703.  Since early observations by \citet{van97v2176cygiauc}
are unavailable, the possibility remains open whether the $P_{\rm SH}$
increased after the dip, or whether there was a phase discontinuity.
By allowing a 0.5 phase shift, the period from the combined data
is 0.05630(4) d.

\begin{figure}
  \begin{center}
    \FigureFile(88mm,110mm){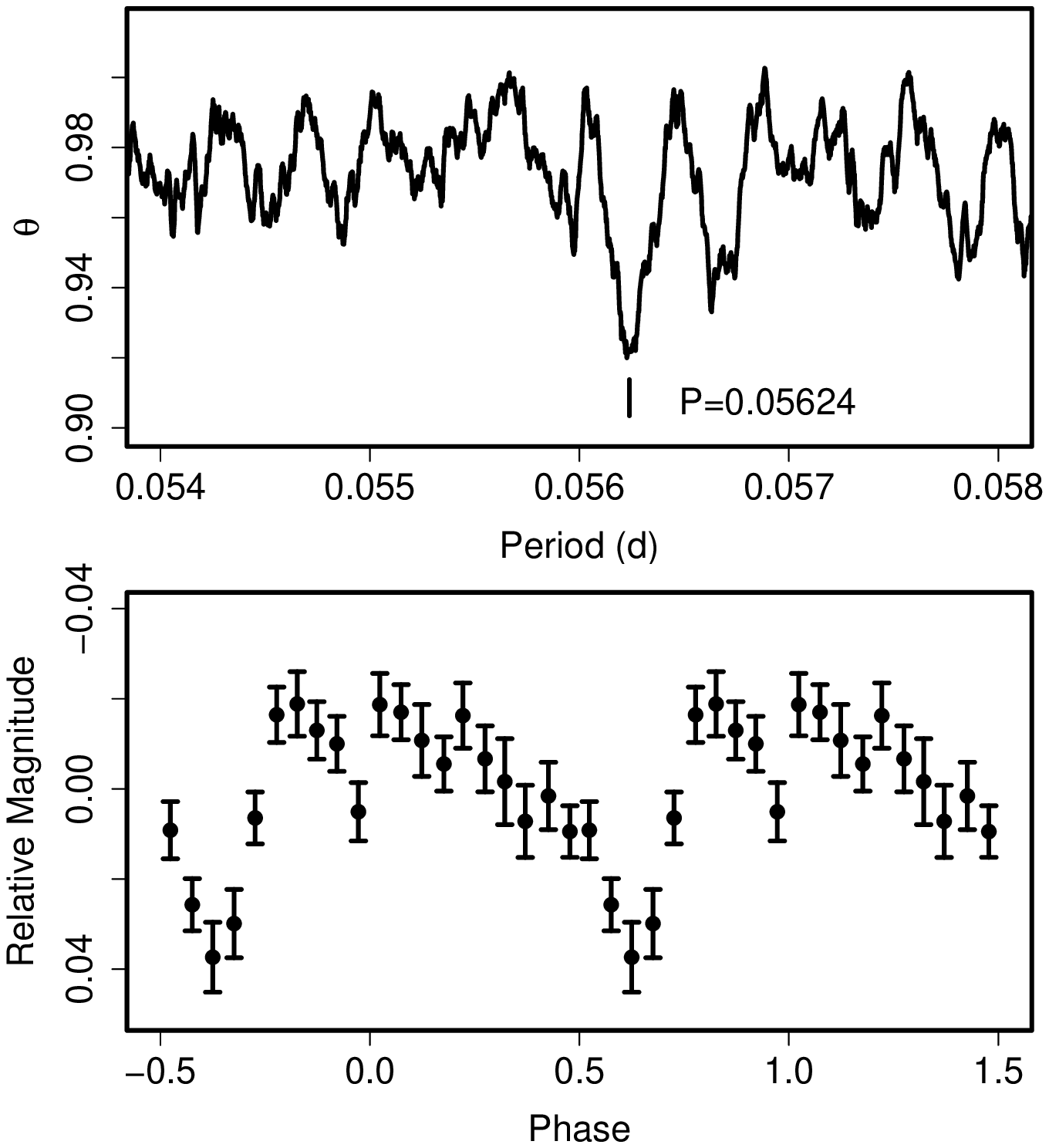}
  \end{center}
  \caption{Superhumps in V2176 Cyg after the dip (1997). (Upper): PDM analysis.
     (Lower): Phase-averaged profile.}
  \label{fig:v2176shpdm}
\end{figure}

\begin{table}
\caption{Superhump maxima of V2176 Cyg after the dip (1997).}\label{tab:v2176cygoc1997b}
\begin{center}
\begin{tabular}{ccccc}
\hline\hline
$E$ & max$^a$ & error & $O-C^b$ & $N^c$ \\
\hline
0 & 50702.4013 & 0.0063 & 0.0121 & -- \\
1 & 50702.4365 & 0.0018 & $-$0.0089 & -- \\
3 & 50702.5537 & 0.0013 & $-$0.0043 & -- \\
4 & 50702.5991 & 0.0020 & $-$0.0151 & -- \\
19 & 50703.4775 & 0.0046 & 0.0197 & -- \\
20 & 50703.5277 & 0.0031 & 0.0137 & -- \\
52 & 50705.3081 & 0.0020 & $-$0.0055 & 21 \\
53 & 50705.3589 & 0.0022 & $-$0.0110 & 22 \\
123 & 50709.3007 & 0.0255 & $-$0.0058 & 34 \\
124 & 50709.3531 & 0.0028 & $-$0.0097 & 34 \\
125 & 50709.4246 & 0.0055 & 0.0056 & 30 \\
141 & 50710.3297 & 0.0201 & 0.0109 & 29 \\
142 & 50710.3605 & 0.0046 & $-$0.0145 & 28 \\
143 & 50710.4327 & 0.0035 & 0.0015 & 22 \\
152 & 50710.9360 & 0.0015 & $-$0.0014 & 53 \\
153 & 50710.9964 & 0.0016 & 0.0028 & 54 \\
154 & 50711.0464 & 0.0200 & $-$0.0035 & 33 \\
159 & 50711.3335 & 0.0033 & 0.0025 & 30 \\
160 & 50711.3982 & 0.0042 & 0.0109 & 25 \\
\hline
  \multicolumn{5}{l}{$^{a}$ BJD$-$2400000.} \\
  \multicolumn{5}{l}{$^{b}$ Against $max = 2450702.3893 + 0.056238 E$.} \\
  \multicolumn{5}{l}{$^{c}$ Number of points used to determine the maximum.} \\
\end{tabular}
\end{center}
\end{table}

\begin{table}
\caption{Superhump maxima of V2176 Cyg before the dip (1997).}\label{tab:v2176cygoc1997a}
\begin{center}
\begin{tabular}{ccccc}
\hline\hline
$E$ & max$^a$ & error & $O-C^b$ & $N^c$ \\
\hline
0 & 50696.3318 & 0.0008 & 0.0005 & 26 \\
1 & 50696.3893 & 0.0011 & 0.0008 & 26 \\
2 & 50696.4425 & 0.0013 & $-$0.0030 & 27 \\
3 & 50696.5041 & 0.0012 & 0.0015 & 27 \\
4 & 50696.5597 & 0.0017 & 0.0001 & 19 \\
\hline
  \multicolumn{5}{l}{$^{a}$ BJD$-$2400000.} \\
  \multicolumn{5}{l}{$^{b}$ Against $max = 2450696.3314 + 0.057048 E$.} \\
  \multicolumn{5}{l}{$^{c}$ Number of points used to determine the maximum.} \\
\end{tabular}
\end{center}
\end{table}

\subsection{HO Delphini}\label{obj:hodel}

   \citet{kat03hodel} reported on three superoutbursts in 1994, 1996
and 2001.  \citet{kat03hodel} did not attempt to determine $P_{\rm dot}$
because of the decaying signal of the superhumps.  We present
times of superhump maxima for the 1994 and 2001 superoutbursts
(tables \ref{tab:hodeloc1994}, \ref{tab:hodeloc2001}).

   The 2008 superoutburst was well-observed.  This outburst was preceded
by a precursor outburst and followed by a rebrightening.
The times of superhump maxima are listed in table \ref{tab:hodeloc2008}.
The $O-C$ diagram (figure \ref{fig:octrans}) was clearly composed of
the stage A ($E \le 2$), the stage B with a positive $P_{\rm dot}$,
and a transition to the stage C with a shorter period,
associated with the brightening near the termination of the superoutburst
(cf. \cite{kat03hodel}).
The $P_{\rm dot}$ for the stage B was $+6.4(1.5) \times 10^{-5}$.

\begin{figure}
  \begin{center}
    \FigureFile(88mm,70mm){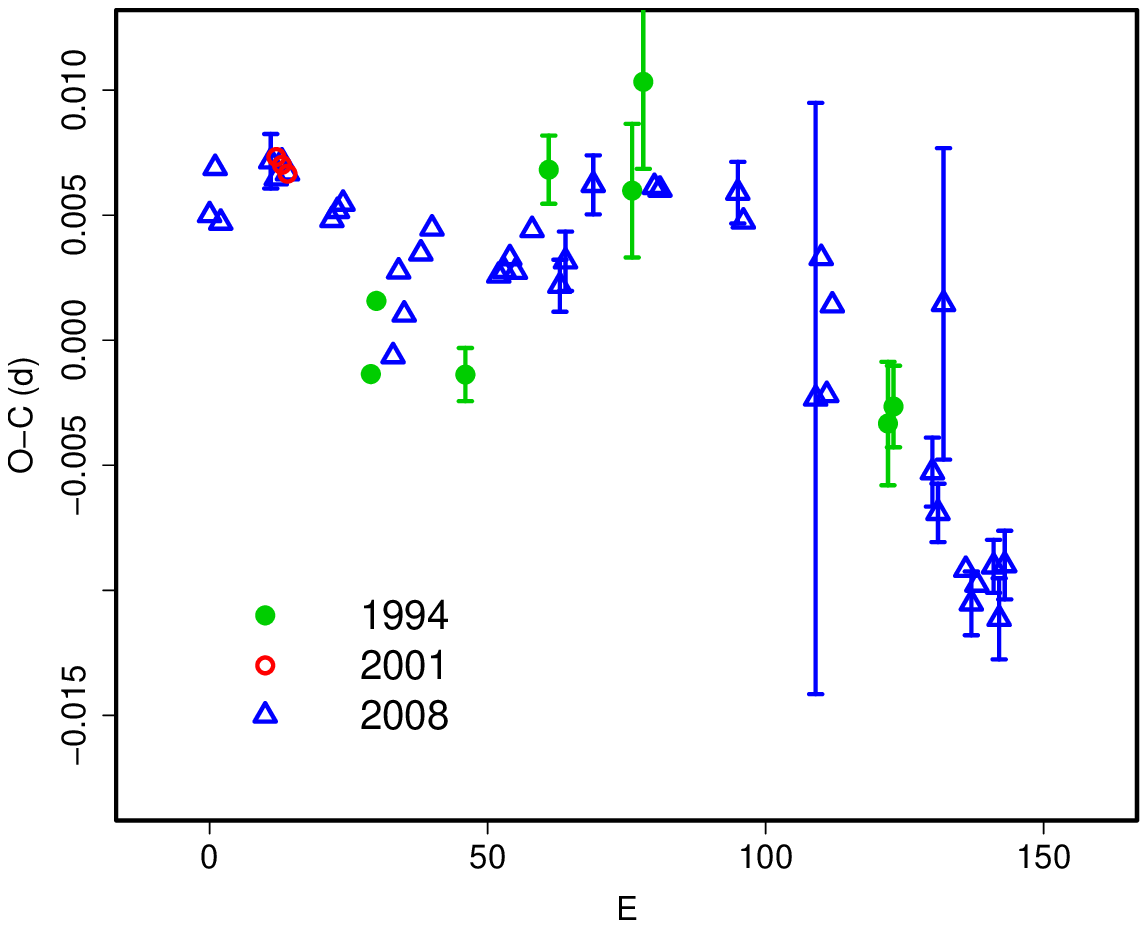}
  \end{center}
  \caption{Comparison of $O-C$ diagrams of HO Del between different
  superoutbursts.  A period of 0.06437 d was used to draw this figure.
  Approximate cycle counts ($E$) after the start of the
  superoutburst (the start of the main superoutburst when preceded by
  a precursor) were used.
  }
  \label{fig:hodelcomp}
\end{figure}

   A comparison of $O-C$ diagrams between different superoutbursts
(figure \ref{fig:hodelcomp}) now clearly indicate that the 1994 observation
recorded the stage B--C transition, in good agreement with the presence
of a terminal brightening, and the short $P_{\rm SH}$ during the
2001 superoutburst reflects the short $P_{\rm SH}$ at the start of
the stage B.

\begin{table}
\caption{Superhump maxima of HO Del (1994).}\label{tab:hodeloc1994}
\begin{center}

\end{center}
\end{table}

\begin{table}
\caption{Superhump maxima of HO Del (2001).}\label{tab:hodeloc2001}
\begin{center}
%
\end{center}
\end{table}

\begin{table}
\caption{Superhump maxima of HO Del (2008).}\label{tab:hodeloc2008}
\begin{center}
%
\end{center}
\end{table}

\subsection{BC Doradus}\label{obj:bcdor}

   \citet{kat04nsv10934mmscoabnorcal86} suggested the SU UMa-type
classification of BC Dor = CAL 86.  This suggestion was confirmed
by the detection of superhumps during the 2003 November superoutburst.
The times of superhump maxima are listed in table \ref{tab:bcdoroc2003}.
Since the superhumps were still growing on the first night and since
the object was already fading on the last night, we used the middle
two nights and determined the mean superhump period of 0.06850(12) d.
The $O-C$'s were strongly negative on the first and last night,
suggesting that early (stage A to B) and late (stage B to C) evolution
took place.  Although the global $P_{\rm dot}$ of
$-8.9(0.5) \times 10^{-5}$ was obtained, this value should be treated
with caution since it was determined from presumably segments of different
types of behavior.

\begin{table}
\caption{Superhump maxima of BC Dor (2003).}\label{tab:bcdoroc2003}
\begin{center}
\begin{tabular}{ccccc}
\hline\hline
$E$ & max$^a$ & error & $O-C^b$ & $N^c$ \\
\hline
0 & 52958.0575 & 0.0005 & $-$0.0118 & 226 \\
45 & 52961.1395 & 0.0005 & 0.0021 & 34 \\
59 & 52962.0985 & 0.0006 & 0.0066 & 35 \\
60 & 52962.1658 & 0.0005 & 0.0057 & 32 \\
61 & 52962.2336 & 0.0006 & 0.0053 & 24 \\
146 & 52968.0157 & 0.0012 & $-$0.0078 & 100 \\
\hline
  \multicolumn{5}{l}{$^{a}$ BJD$-$2400000.} \\
  \multicolumn{5}{l}{$^{b}$ Against $max = 2452958.0693 + 0.068180 E$.} \\
  \multicolumn{5}{l}{$^{c}$ Number of points used to determine the maximum.} \\
\end{tabular}
\end{center}
\end{table}

\subsection{CP Draconis}\label{obj:cpdra}

   CP Dra was initially discovered as a suspect supernova in NGC 3147.
Subsequent observations established the dwarf nova-type
nature of the object (\cite{kho72cpdra}; \cite{kol79cpdraciuma}).
The object has been regularly monitored by visual observers.
During the 2001 outburst, T. Vanmunster detected superhumps
with a period of 0.0687(7) d (vsnet-alert 5709).  The period, however,
did not agree with later observations.

   During the 2003 superoutburst, we succeeded in identifying the
superhump period from the high-quality observations on first two nights.
The best period determined from the entire outburst was 0.08348(10) d.
The times of superhump maxima are shown in table \ref{tab:cpdraoc2003}.
The period decreased at $P_{\rm dot}$ = $-22.6(4.6) \times 10^{-5}$,
probably reflecting the stage B--C transition.

   The 2009 superoutburst was well-observed during its middle-to-late
stage (table \ref{tab:cpdraoc2009}).  A clear stage B--C transition
was recorded.  The mean $P_{\rm SH}$ during the stage C was
0.083323(11) d (PDM method).  The other parameters are listed in
table \ref{tab:perlist}.

\begin{table}
\caption{Superhump maxima of CP Dra (2003).}\label{tab:cpdraoc2003}
\begin{center}

\end{center}
\end{table}

\begin{table}
\caption{Superhump maxima of CP Dra (2009).}\label{tab:cpdraoc2009}
\begin{center}
%
\end{center}
\end{table}

\subsection{DM Draconis}\label{obj:dmdra}

   DM Dra was discovered as a dwarf nova by \citet{ste82dmdra}.
\citet{kat02dmdra} studied the 2001 outburst and reported superhumps
with a period of 0.07561(3) d.  The coverage of this outburst was
insufficient to determine $P_{\rm dot}$.  We undertook a more extensive
campaign during the 2003 superoutburst.  The times of superhump maxima
are listed in table \ref{tab:dmdraoc2003}.  We obtained a global
$P_{\rm dot}$ = $-15.3(1.8) \times 10^{-5}$.  Excluding the first two
maxima, which may have been recorded during the stage A,
we obtained $P_{\rm dot}$ = $-13.6(2.3) \times 10^{-5}$
(cf. figure \ref{fig:octrans}).

\begin{table}
\caption{Superhump maxima of DM Dra (2003).}\label{tab:dmdraoc2003}
\begin{center}
\begin{tabular}{ccccc}
\hline\hline
$E$ & max$^a$ & error & $O-C^b$ & $N^c$ \\
\hline
0 & 52706.2096 & 0.0005 & $-$0.0086 & 182 \\
1 & 52706.2893 & 0.0005 & $-$0.0045 & 245 \\
12 & 52707.1236 & 0.0045 & $-$0.0008 & 80 \\
13 & 52707.1984 & 0.0014 & $-$0.0015 & 80 \\
14 & 52707.2737 & 0.0021 & $-$0.0017 & 79 \\
15 & 52707.3546 & 0.0039 & 0.0037 & 49 \\
25 & 52708.1070 & 0.0011 & 0.0010 & 176 \\
26 & 52708.1834 & 0.0011 & 0.0018 & 185 \\
27 & 52708.2588 & 0.0007 & 0.0017 & 185 \\
28 & 52708.3360 & 0.0014 & 0.0035 & 163 \\
38 & 52709.0884 & 0.0013 & 0.0007 & 81 \\
39 & 52709.1661 & 0.0006 & 0.0029 & 176 \\
40 & 52709.2405 & 0.0006 & 0.0018 & 236 \\
41 & 52709.3206 & 0.0006 & 0.0064 & 239 \\
51 & 52710.0709 & 0.0011 & 0.0016 & 79 \\
52 & 52710.1467 & 0.0012 & 0.0019 & 81 \\
53 & 52710.2205 & 0.0011 & 0.0001 & 81 \\
54 & 52710.2987 & 0.0073 & 0.0028 & 77 \\
80 & 52712.2539 & 0.0006 & $-$0.0052 & 96 \\
81 & 52712.3270 & 0.0016 & $-$0.0076 & 73 \\
\hline
  \multicolumn{5}{l}{$^{a}$ BJD$-$2400000.} \\
  \multicolumn{5}{l}{$^{b}$ Against $max = 2452706.2183 + 0.075510 E$.} \\
  \multicolumn{5}{l}{$^{c}$ Number of points used to determine the maximum.} \\
\end{tabular}
\end{center}
\end{table}

\subsection{DV Draconis}\label{obj:dvdra}

\begin{figure}
  \begin{center}
    \FigureFile(88mm,110mm){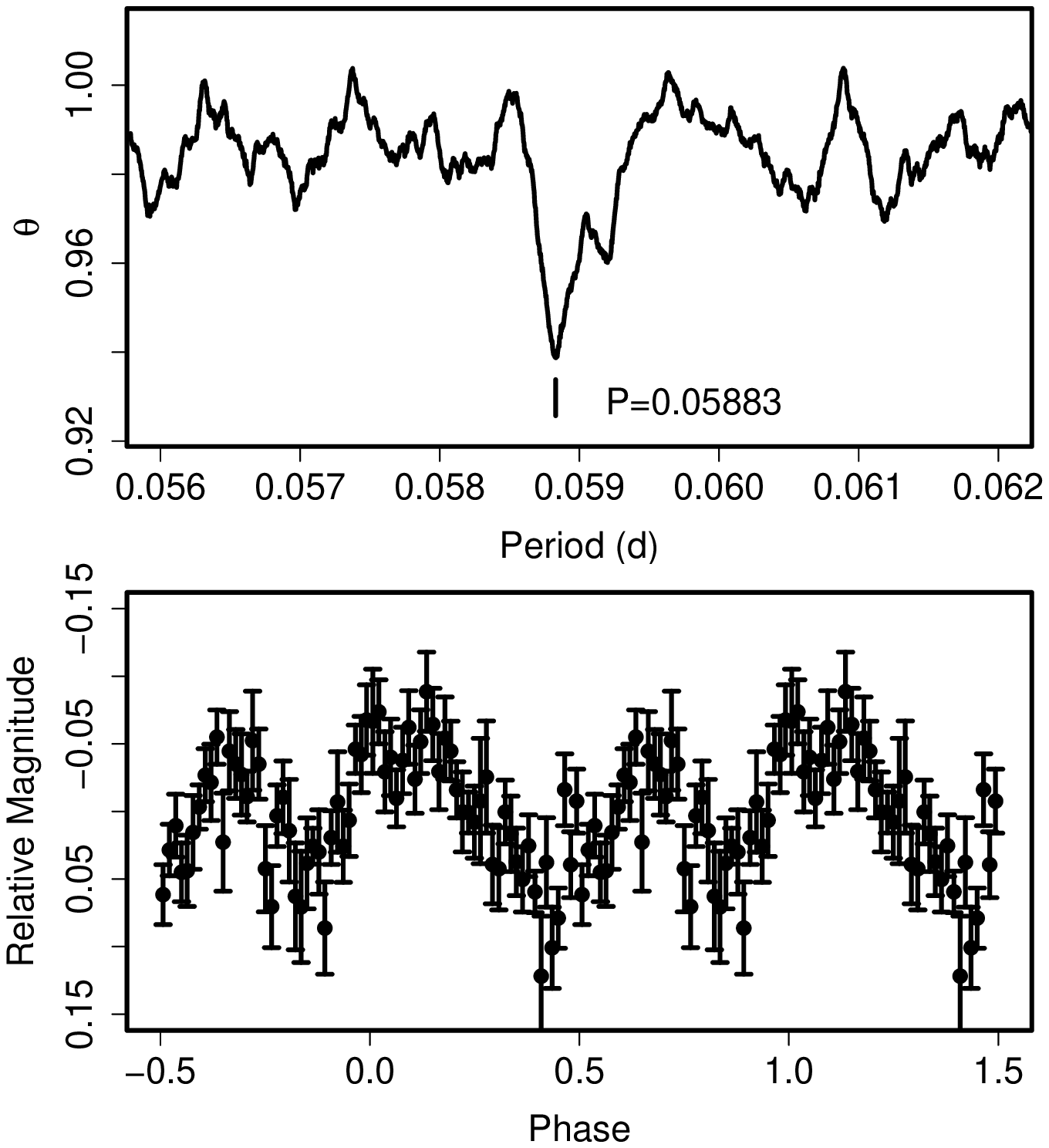}
  \end{center}
  \caption{Early superhumps in DV Dra (2005). (Upper): PDM analysis.
     (Lower): Phase-averaged profile.}
  \label{fig:dvdraeshpdm}
\end{figure}

   DV Dra is a dwarf nova discovered by \citet{pav85dvdra}.  The object
had long been suspected to be a WZ Sge-type dwarf nova \citep{wen91dvdra}.
\citet{iid95dvdra} claimed a detection of a new outburst,
but was later confirmed to be a false recognition of a field star
(vsnet-id 182, 183).
In 2005 November, P. Schmeer detected an outburst at an unfiltered
CCD magnitude of 15.0 (vsnet-alert 8749).  T. Vanmunster reported
the detection of double-wave early superhumps (cvnet-outburst 790).
We observed the outburst between November 22 (just preceding Vanmunster's
observation) and December 6.  Early superhumps with a mean period of
0.05883(2) d were detected at least until November 27
(figure \ref{fig:dvdraeshpdm}).
Due to the short visibility, we could not convincingly detect the appearance
of ordinary superhumps.
We include this object for improving the statistics of WZ Sge-type
dwarf novae.

\subsection{KV Draconis}\label{obj:kvdra}

   The 2000 superoutburst was observed by two teams, \citet{nog00kvdra}
and \citet{van00kvdra}.  Although \citet{van00kvdra} reported a slight
increase of the superhump period from 0.0601 d to 0.0603 d, we could not
calculate $P_{\rm dot}$ because they did not publish the times of maxima.
\citet{nog00kvdra} reported a candidate period of 0.06019(2) d based on
observations separated by seven days.  The period by \citet{nog00kvdra}
was severely suffered from an aliasing problem, particularly when
the period was changing, due to the large gap in observation.
Although the SU UMa-type nature was well-established
upon this superoutburst, we still needed a better coverage to determine
the superhump period and its derivative.

   The 2002 superoutburst was relatively well-observed during the
most of the course of the outburst (table \ref{tab:kvdraoc2002}).
Although we obtained $P_{\rm dot}$ = $+11.4(3.9) \times 10^{-5}$
for $E \le 108$, the period variation appeared rather abrupt,
giving a relatively constant period of 0.06002(3) d for $E \le 59$.
The similar pattern of period variation was also observed during
the 2008 superoutburst of AQ Eri (subsection \ref{sec:aqeri}).
A stage B--C transition was also recorded.

   The 2004 superoutburst was well-observed for the early stage
(table \ref{tab:kvdraoc2004}).
The $P_{\rm dot}$ = $+43.4(8.5) \times 10^{-5}$ for $E \le 96$ appears
too large.  There might have been a phase shift between $E=70$ and
$E=79$.  The period for $E \le 24$ was relatively constant at
0.06001(8) d, a period very close to the 2002 one.
The rather anomalous $O-C$ behavior during the late course of
the superoutburst in this system requires further investigation.

   We also observed the 2005 superoutburst, covering the middle-to-later
portion of the plateau phase.  The estimated times of superhump
maxima are listed in table \ref{tab:kvdraoc2005}.  The data gave
a significantly longer mean period of 0.06034(3) d, which is in
better agreement with the longer value in \citet{van00kvdra}.
The $P_{\rm dot}$ from these data was $+11.2(4.2) \times 10^{-5}$.

   The 2009 superoutburst was observed during the stage A--B transition
and a later stage (table \ref{tab:kvdraoc2009}).  The $P_1$ in
table \ref{tab:perlist} refers to the mean period of the early part
of the stage B, shorter than $P_1$'s of other superoutbursts.
The maximum of $E=100$ was not included in calculating the $P_2$.
This maximum may have been a final part of the stage B.

\begin{figure}
  \begin{center}
    \FigureFile(88mm,70mm){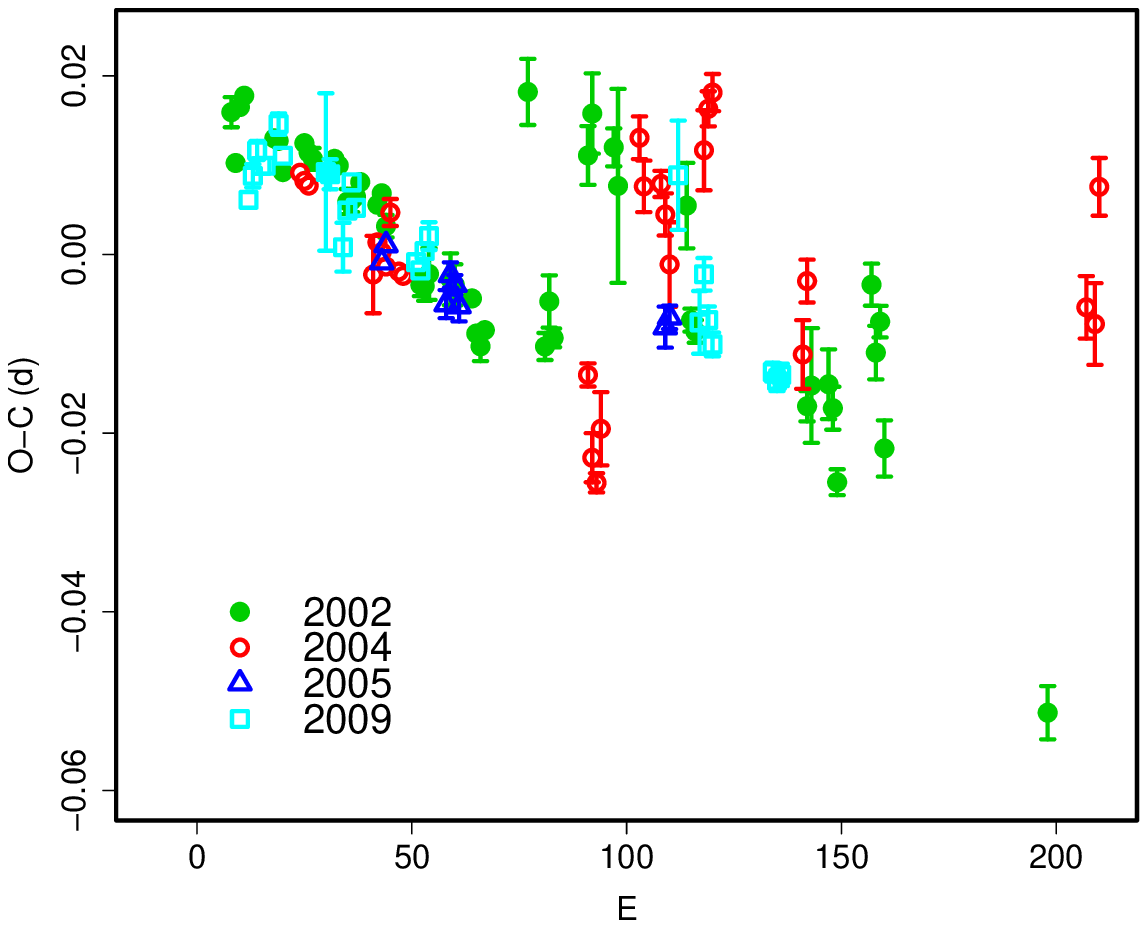}
  \end{center}
  \caption{Comparison of $O-C$ diagrams of KV Dra between different
  superoutbursts.  A period of 0.06044 d was used to draw this figure.
  Approximate cycle counts ($E$) after the start of the
  superoutburst were used.
  }
  \label{fig:kvdracomp}
\end{figure}

   A comparison of $O-C$ diagrams between different superoutbursts
is given in figure \ref{fig:kvdracomp}.  The stage B in this system
appears to be composed of two linear segments rather than a continuous
period change.  The behavior of the late stage B was different
between 2002 and 2004 superoutbursts.  The difference may be
a result of early appearance of stage C superhumps during the
2002 superoutburst.

\begin{table}
\caption{Superhump maxima of KV Dra (2002).}\label{tab:kvdraoc2002}
\begin{center}

\end{center}
\end{table}

\begin{table}
\caption{Superhump maxima of KV Dra (2004).}\label{tab:kvdraoc2004}
\begin{center}
%
\end{center}
\end{table}

\begin{table}
\caption{Superhump maxima of KV Dra (2005).}\label{tab:kvdraoc2005}
\begin{center}
%
\end{center}
\end{table}

\begin{table}
\caption{Superhump maxima of KV Dra (2009).}\label{tab:kvdraoc2009}
\begin{center}
%
\end{center}
\end{table}

\subsection{MN Draconis}\label{obj:mndra}

   This object was discovered by \citet{ant02var73dra}.  \citet{nog03var73dra}
presented an extensive study of this object and established its unusual
properties: long $P_{\rm SH}$ of 0.104--0.106 d and unusually short 
($\sim$ 60 d) supercycle length.  The difference of periods between
two superoutbursts can be attributed to different stages observed:
stage C in 2002 October and stage B--C transition in 2002 December
(figure \ref{fig:lp1}).
We present periods based on this interpretation in table \ref{tab:perlist}.

   Since the photometric orbital period (0.10424 d) mentioned in
\citep{nog03var73dra} is extremely close to the $P_2$ in the present
identification, we analyzed the corresponding segment in our data
and obtained a period a periodicity around 0.1042--0.1047 d.
We suspect that this period was not the true
orbital period, but persisting (or permanent) superhumps having
a period close to $P_2$.  The presence of permanent superhumps,
if confirmed, would strengthen the resemblance of MN Dra to
ER UMa stars (e.g. \cite{gao99erumaSH}; \cite{ole08rzlmi}).
If the true orbital period is shorter, the problem of an exceptionally
small fractional superhump excess \citet{nog03var73dra} will be solved.

   We further point out that the 2003 April outburst was a superoutburst
(table \ref{tab:mndraoc2003apr}).  The derived superhump period
of 0.10480(5) d with the PDM method is in good agreement with the mean
period of the 2002 October superoutburst.  This superoutburst occurred
$\sim$ 65 d after the 2003 February superoutburst mentioned in
\citet{nog03var73dra}, confirming the relatively stable, short
supercycle.  A PDM analysis of the 2008 July superoutburst yielded
a mean period of 0.10514(14) d (table \ref{tab:mndraoc2008jul}).

\begin{table}
\caption{Superhump maxima of MN Dra (2003 April).}\label{tab:mndraoc2003apr}
\begin{center}
\begin{tabular}{ccccc}
\hline\hline
$E$ & max$^a$ & error & $O-C^b$ & $N^c$ \\
\hline
0 & 52750.4325 & 0.0008 & $-$0.0017 & 91 \\
9 & 52751.3812 & 0.0013 & 0.0039 & 51 \\
10 & 52751.4814 & 0.0011 & $-$0.0007 & 43 \\
19 & 52752.4238 & 0.0021 & $-$0.0015 & 43 \\
\hline
  \multicolumn{5}{l}{$^{a}$ BJD$-$2400000.} \\
  \multicolumn{5}{l}{$^{b}$ Against $max = 2452750.4342 + 0.10479 E$.} \\
  \multicolumn{5}{l}{$^{c}$ Number of points used to determine the maximum.} \\
\end{tabular}
\end{center}
\end{table}

\begin{table}
\caption{Superhump maxima of MN Dra (2008 July).}\label{tab:mndraoc2008jul}
\begin{center}
\begin{tabular}{ccccc}
\hline\hline
$E$ & max$^a$ & error & $O-C^b$ & $N^c$ \\
\hline
0 & 54677.5447 & 0.0014 & $-$0.0001 & 60 \\
9 & 54678.4928 & 0.0015 & 0.0009 & 60 \\
10 & 54678.5963 & 0.0016 & $-$0.0008 & 41 \\
\hline
  \multicolumn{5}{l}{$^{a}$ BJD$-$2400000.} \\
  \multicolumn{5}{l}{$^{b}$ Against $max = 2454677.5448 + 0.10524 E$.} \\
  \multicolumn{5}{l}{$^{c}$ Number of points used to determine the maximum.} \\
\end{tabular}
\end{center}
\end{table}

\subsection{XZ Eridani}\label{obj:xzeri}

   XZ Eri is an eclipsing SU UMa-type dwarf nova with a short
orbital period (\cite{uem04xzeri}; \cite{wou01v359cenxzeriyytel}).
We reanalyzed the observations presented in \citet{uem04xzeri}
and determined the times of superhump maxima (table \ref{tab:xzerioc2003a}).
Although the scatter was rather large, we can see an earlier segment
with a positive $P_{\rm dot}$ (stage B) followed by a transition to
a shorter period (stage C).  The $P_{\rm dot}$ for the stage B
($E \le 77$) was $+15.3(5.6) \times 10^{-5}$, strengthening the
suggestion in \citet{uem04xzeri}.

   We also observed two superoutbursts in 2003 December (table
\ref{tab:xzerioc2003b}), in 2007 (table \ref{tab:xzerioc2007})
and in 2008 (table \ref{tab:xzerioc2008}, combined data with the AAVSO
observations).
We only recorded the transition to a shorter period during the first
superoutburst, while we managed to mainly record the stages of early
evolution (stage A to B) and a positive $P_{\rm dot}$.
The $P_{\rm dot}$ for the 2007
superoutburst was $+7.6(1.0) \times 10^{-5}$ ($15 \le E \le 138$).
The 2008 superoutburst showed all stages of A--C.
The $P_{\rm dot}$ during the stage B was $+22.5(4.7) \times 10^{-5}$
($23 \le E \le 92$).

\begin{figure}
  \begin{center}
    \FigureFile(88mm,70mm){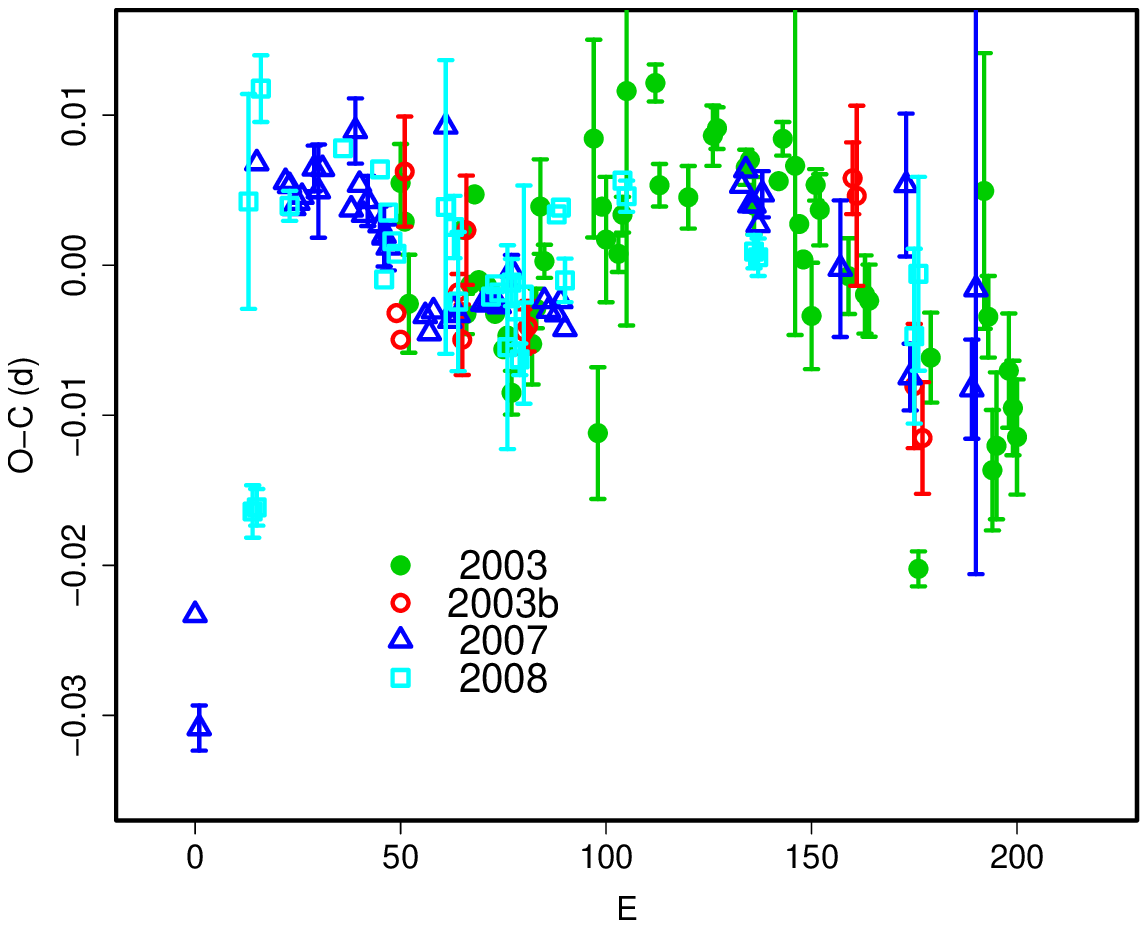}
  \end{center}
  \caption{Comparison of $O-C$ diagrams of XZ Eri between different
  superoutbursts.  A period of 0.06283 d was used to draw this figure.
  Approximate cycle counts ($E$) after the start of the
  superoutburst were used.
  }
  \label{fig:xzericomp}
\end{figure}

   A comparison of $O-C$ diagrams between different superoutbursts
is presented in figure \ref{fig:xzericomp}.

\begin{table}
\caption{Superhump maxima of XZ Eri (2003a).}\label{tab:xzerioc2003a}
\begin{center}

\end{center}
\end{table}

\begin{table}
\caption{Superhump maxima of XZ Eri (2003b).}\label{tab:xzerioc2003b}
\begin{center}
%
\end{center}
\end{table}

\begin{table}
\caption{Superhump maxima of XZ Eri (2007).}\label{tab:xzerioc2007}
\begin{center}
%
\end{center}
\end{table}

\begin{table}
\caption{Superhump maxima of XZ Eri (2008).}\label{tab:xzerioc2008}
\begin{center}
%
\end{center}
\end{table}

\subsection{AQ Eridani}\label{sec:aqeri}\label{obj:aqeri}

   \citet{kat91aqeri} observed the 1991 superoutburst and reported
a period of 0.06225 d.  \citet{kat01aqeri} reported a single-night
observation of the 1992 superoutburst, and found an anomalously
long (0.0642(4) d) superhump period.

   Although the original data for the 1991 superoutburst is already
unavailable, we reanalyzed the 1992 data together with unpublished
observations (table \ref{tab:aqerioc1992}).  The anomalously long
$P_{\rm SH}$ has been confirmed (0.0638(7) d for $0 \le E \le 3$).
The period, however, of the entire observation is 0.0616(2).
The observation likely caught the transition from the stage B to C.

   We further observed the 2006 superoutburst during its late plateau
stage (table \ref{tab:aqerioc2006}).  Because the observation was
performed when the superhumps had small amplitudes and relatively
irregular profiles, the quality of the $O-C$ analysis was not
satisfactory.  The mean period (likely $P_2$) was 0.0617(1) d.

\begin{figure}
  \begin{center}
    \FigureFile(88mm,90mm){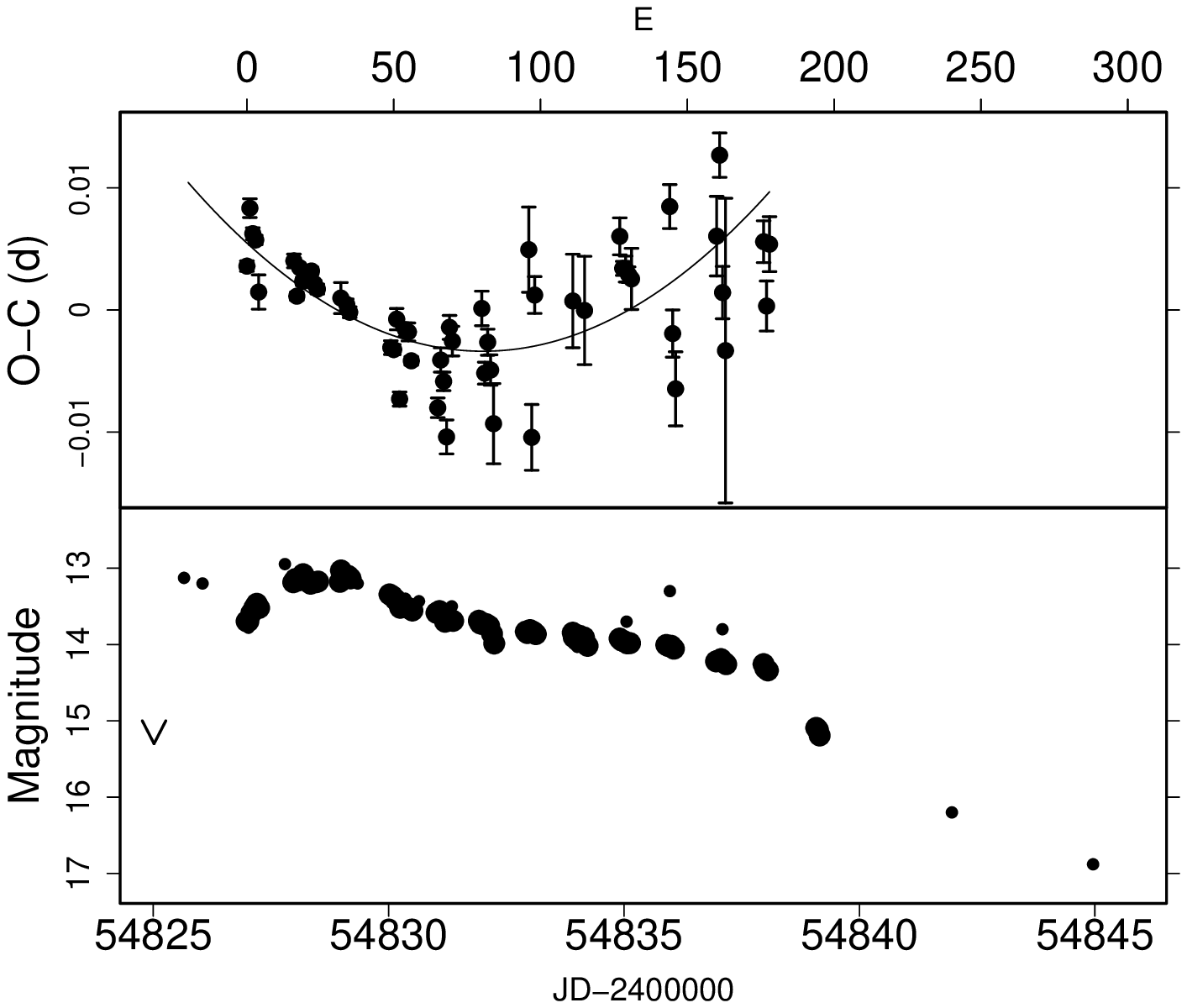}
  \end{center}
  \caption{$O-C$ of superhumps AQ Eri (2008).
  (Upper): $O-C$ diagram.  The $O-C$ values were against the mean period
  for the stage B ($E \le 163$, thin curve)
  (Lower): Light curve.  Large dots are our CCD observations and small
  dots are visual and $V$ observation from the VSOLJ database and ASAS-3
  observations.}
  \label{fig:aqeri2008oc}
\end{figure}

   The 2008 superoutburst was well observed (table \ref{tab:aqerioc2008}),
first clearly establishing the positive $P_{\rm dot}$ of
$+4.4(0.8) \times 10^{-5}$ (figure \ref{fig:aqeri2008oc}).
This superoutburst was preceded by a distinct precursor,
strengthening that the overall behavior of period derivatives are
not strongly affected by the presence of a precursor outburst.

\begin{table}
\caption{Superhump maxima of AQ Eri (1992).}\label{tab:aqerioc1992}
\begin{center}

\end{center}
\end{table}

\begin{table}
\caption{Superhump maxima of AQ Eri (2006).}\label{tab:aqerioc2006}
\begin{center}
%
\end{center}
\end{table}

\begin{table}
\caption{Superhump maxima of AQ Eri (2008).}\label{tab:aqerioc2008}
\begin{center}
%
\end{center}
\end{table}

\subsection{UV Geminorum}\label{sec:uvgem}\label{obj:uvgem}

   UV Gem has long been known as a dwarf nova \citep{GCVS}.
\citet{kat01uvgemfsandaspsc} suggested the SU UMa-type classification
based on the long-term light curve consisting of a likely superoutburst
and short outbursts with a short cycle length.
T. Vanmunster (vsnet-alert 3821) first reported the detection of
superhumps with a period of 0.0902(6) d.
During the 2003 superoutburst, we conducted an extensive campaign
and obtained a high-quality set of superhump times
(table \ref{tab:uvgemoc2003}).  The large variation in the $O-C$
diagram indicates a strong period decrease (figure \ref{fig:lp1}).
Using all the data, the $P_{\rm dot}$ was $-53.4(3.6) \times 10^{-5}$.
Even if we exclude the early part ($E \le 5$), the $P_{\rm dot}$ was
$-33.5(2.0) \times 10^{-5}$, still extreme.
The situation is particularly similar to a long-$P_{\rm orb}$ system
MN Dra \citep{nog03var73dra}, who reported a global $P_{\rm dot}$
of $-170(20) \times 10^{-5}$.  The present data of UV Gem
neither has cycle ambiguity nor a large gap in observation,
thereby firmly demonstrating the existence of an exceptionally
strong decrease in the superhump period.
Long-$P_{\rm orb}$ systems appear to share this tendency of period
variation.

   The times of superhump maxima for the 2008 superoutburst are also
given (table \ref{tab:uvgemoc2008}).  This superoutburst was probably
observed during its late stage.

\begin{table}
\caption{Superhump maxima of UV Gem (2003).}\label{tab:uvgemoc2003}
\begin{center}

\end{center}
\end{table}

\begin{table}
\caption{Superhump maxima of UV Gem (2008).}\label{tab:uvgemoc2008}
\begin{center}
%
\end{center}
\end{table}

\subsection{AW Geminorum}\label{obj:awgem}

   \citet{kat96awgem} observed the 1995 superoutburst during its
early stage.  The refined times of superhump maxima are listed in table
\ref{tab:awgemoc1995}.  The $O-C$ diagram shows a similar trend to
those of V877 Ara and DT Oct (a period shift from a longer period
during the earliest stage).  Excluding the early part (stage A, $E \le 1$),
we obtained the mean $P_{\rm SH}$ = 0.07935(9) d and
$P_{\rm dot}$ = $-3.2(1.5) \times 10^{-5}$.
We also observed the 2008 superoutburst (table \ref{tab:awgemoc2008})
during its early stage and the 2009 superoutburst
(table \ref{tab:awgemoc2009}).
A strong period variation, as recorded in the 1995 superoutburst,
was recorded during the latter superoutburst.

\begin{table}
\caption{Superhump maxima of AW Gem (1995).}\label{tab:awgemoc1995}
\begin{center}

\end{center}
\end{table}

\begin{table}
\caption{Superhump maxima of AW Gem (2008).}\label{tab:awgemoc2008}
\begin{center}
%
\end{center}
\end{table}

\begin{table}
\caption{Superhump maxima of AW Gem (2009).}\label{tab:awgemoc2009}
\begin{center}
%
\end{center}
\end{table}

\subsection{CI Geminorum}\label{obj:cigem}

\begin{figure}
  \begin{center}
    \FigureFile(88mm,110mm){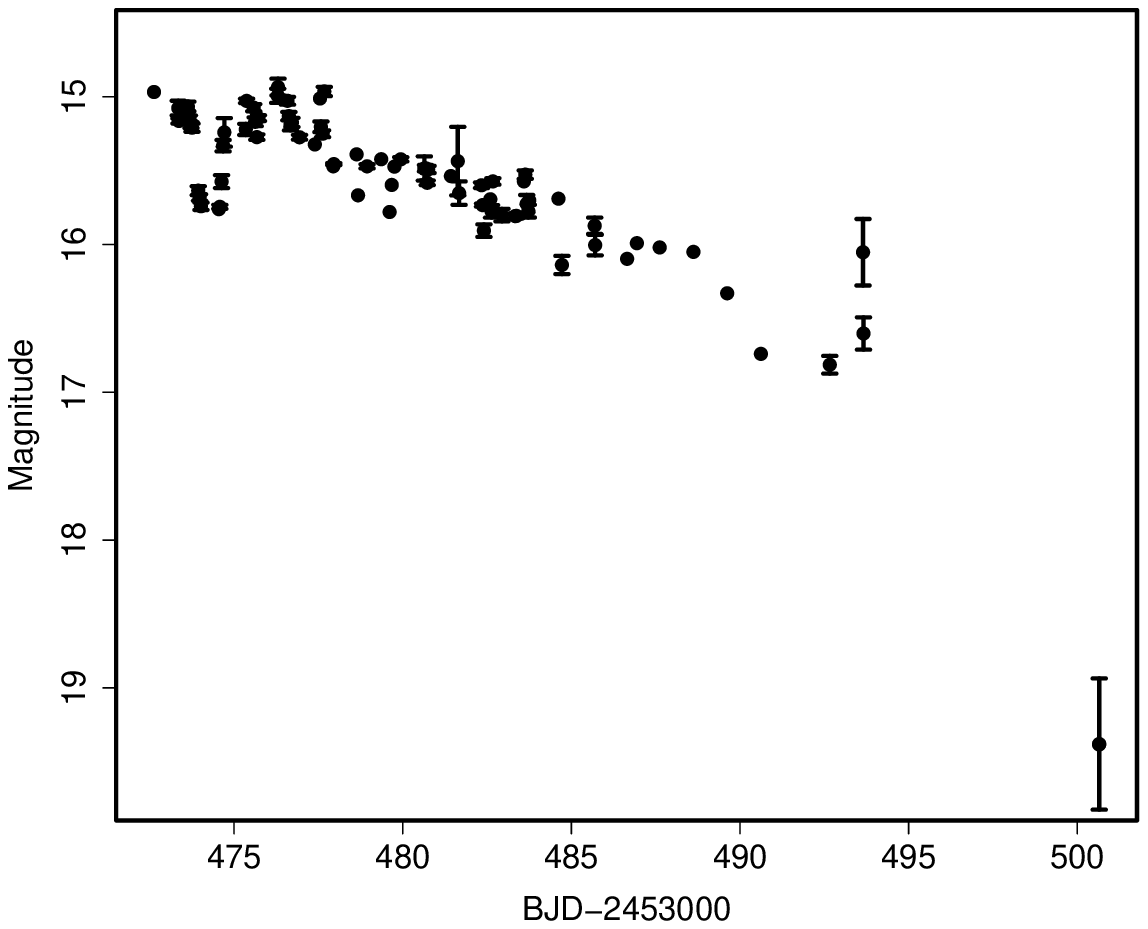}
  \end{center}
  \caption{Superoutburst of CI Gem in 2005.  The data are a combination
  of the AAVSO data and our observations.  The fading around BJD
  2453473--2453474 is a precursor outburst.}
  \label{fig:cigemlc}
\end{figure}

   \citet{wen90cigem} suggested the SU UMa-type classification of this
object based on the existence of long and short outbursts.
The large rate of decline of a short outburst in 1999 was consistent
with that of a normal outburst of an SU UMa-type dwarf nova
\citep{kat99cigem}, although \citet{sch99cigem} favored the SS Cyg-type
classification.
The object underwent a long outburst in 2005 April, consisting of
a precursor and a long plateau (figure \ref{fig:cigemlc}).

\begin{figure}
  \begin{center}
    \FigureFile(88mm,110mm){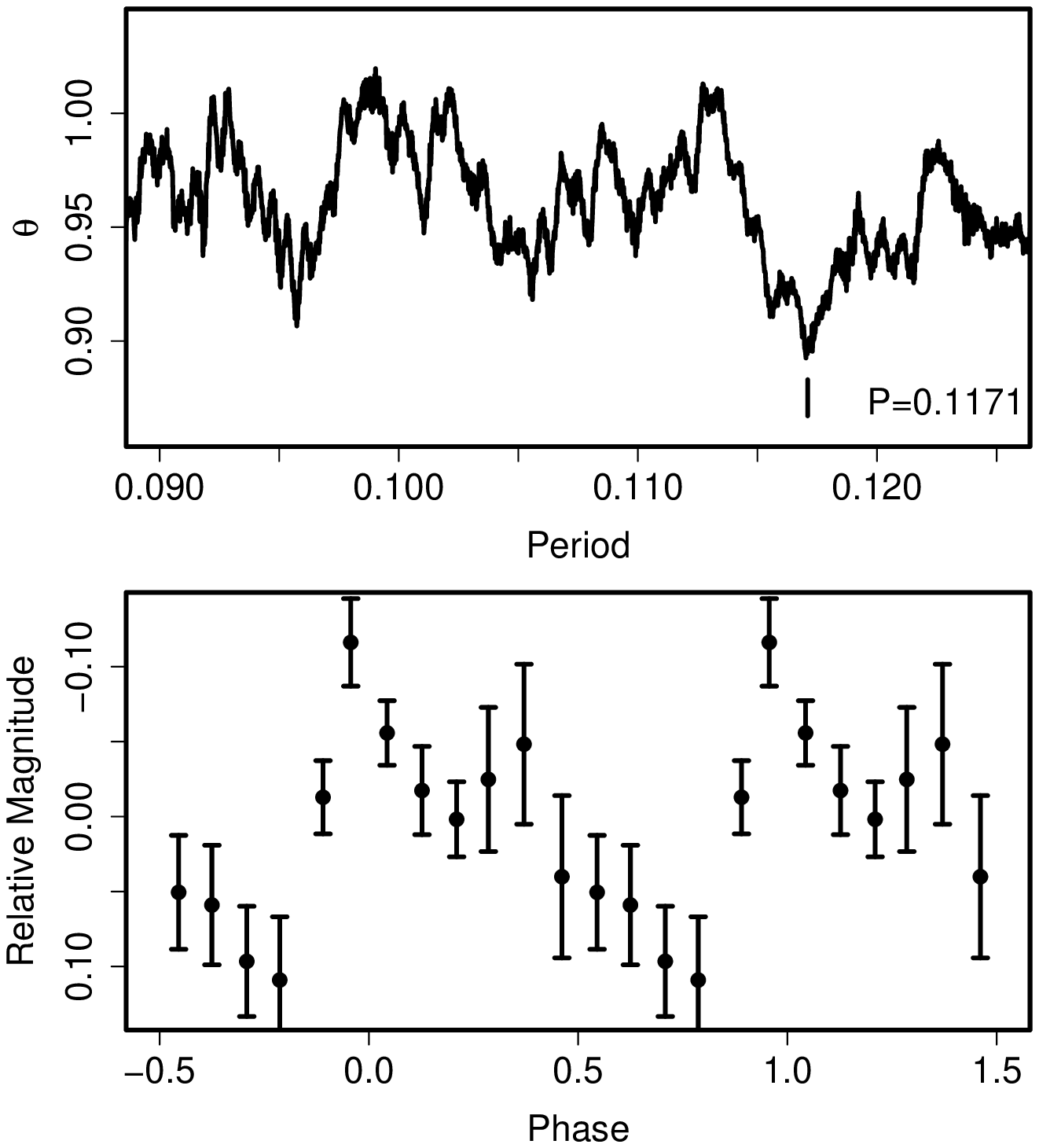}
  \end{center}
  \caption{Superhumps in CI Gem for the early stage the plateau
     (BJD 2453474.5 -- 2453476). (Upper): PDM analysis.
     (Lower): Phase-averaged profile.}
  \label{fig:cigemshpdm}
\end{figure}

   Although the presence of superhumps with a period about $\sim$0.1 d
is apparent in the sparse raw data, a PDM analysis of the entire set of
data did not yield a significant period.  The situation appears similar to 
CTCV J0549 with a long $P_{\rm SH}$ and large period variation.
We therefore analyzed the data in separate segments, measured superhump
maxima (table \ref{tab:cigemoc2005}) and searched for a likely period.
The period of $\sim$0.117 d with a significant period decrease only can
naturally express the available observations (figure \ref{fig:cigemshpdm}).
Although the exact identification of the period should await further
observations, the present analysis suggests that CI Gem is an excellent
candidate for a dwarf nova in the period gap.

\begin{table}
\caption{Superhump maxima of CI Gem (2005).}\label{tab:cigemoc2005}
\begin{center}
\begin{tabular}{ccccc}
\hline\hline
$E$ & max$^a$ & error & $O-C^b$ & $N^c$ \\
\hline
0 & 53474.6646 & 0.0077 & $-$0.0206 & 55 \\
6 & 53475.3707 & 0.0017 & $-$0.0072 & 16 \\
9 & 53475.7219 & 0.0048 & $-$0.0023 & 91 \\
16 & 53476.5655 & 0.0144 & 0.0333 & 21 \\
17 & 53476.6938 & 0.0011 & 0.0461 & 25 \\
25 & 53477.5421 & 0.0059 & $-$0.0290 & 7 \\
26 & 53477.6663 & 0.0023 & $-$0.0203 & 19 \\
\hline
  \multicolumn{5}{l}{$^{a}$ BJD$-$2400000.} \\
  \multicolumn{5}{l}{$^{b}$ Against $max = 2453474.6853 + 0.115433 E$.} \\
  \multicolumn{5}{l}{$^{c}$ Number of points used to determine the maximum.} \\
\end{tabular}
\end{center}
\end{table}

\subsection{IR Geminorum}\label{obj:irgem}

   We measured times of superhump maxima (table \ref{tab:irgemoc1991})
from observations reported in \citet{kat01irgem}.
We also observed the 2009 superoutburst (table \ref{tab:irgemoc2009}).
Although the data were limited, we can see a likely stage B--C
transition (the presence of a phase shift between $E=27$ and $E=86$
is not completely excluded).
Because the profile of the superhumps was rather irregular, we determined
the mean period for stage B with the PDM method as 0.07093(3) d.

\begin{table}
\caption{Superhump maxima of IR Gem (1991).}\label{tab:irgemoc1991}
\begin{center}
\begin{tabular}{ccccc}
\hline\hline
$E$ & max$^a$ & error & $O-C^b$ & $N^c$ \\
\hline
0 & 48333.9835 & 0.0006 & 0.0018 & 268 \\
1 & 48334.0507 & 0.0006 & $-$0.0019 & 260 \\
14 & 48334.9742 & 0.0009 & 0.0009 & 269 \\
15 & 48335.0434 & 0.0010 & $-$0.0007 & 261 \\
\hline
  \multicolumn{5}{l}{$^{a}$ BJD$-$2400000.} \\
  \multicolumn{5}{l}{$^{b}$ Against $max = 2448333.9818 + 0.070821 E$.} \\
  \multicolumn{5}{l}{$^{c}$ Number of points used to determine the maximum.} \\
\end{tabular}
\end{center}
\end{table}

\begin{table}
\caption{Superhump maxima of IR Gem (2009).}\label{tab:irgemoc2009}
\begin{center}
\begin{tabular}{ccccc}
\hline\hline
$E$ & max$^a$ & error & $O-C^b$ & $N^c$ \\
\hline
0 & 54838.0172 & 0.0003 & $-$0.0048 & 196 \\
2 & 54838.1611 & 0.0007 & $-$0.0018 & 73 \\
3 & 54838.2338 & 0.0006 & 0.0005 & 74 \\
4 & 54838.3022 & 0.0008 & $-$0.0016 & 63 \\
27 & 54839.9357 & 0.0007 & 0.0117 & 63 \\
86 & 54844.0724 & 0.0019 & $-$0.0080 & 61 \\
87 & 54844.1521 & 0.0036 & 0.0013 & 24 \\
100 & 54845.0674 & 0.0010 & 0.0007 & 193 \\
101 & 54845.1414 & 0.0009 & 0.0043 & 185 \\
102 & 54845.2089 & 0.0015 & 0.0014 & 145 \\
103 & 54845.2744 & 0.0019 & $-$0.0036 & 53 \\
\hline
  \multicolumn{5}{l}{$^{a}$ BJD$-$2400000.} \\
  \multicolumn{5}{l}{$^{b}$ Against $max = 2454838.0220 + 0.070447 E$.} \\
  \multicolumn{5}{l}{$^{c}$ Number of points used to determine the maximum.} \\
\end{tabular}
\end{center}
\end{table}

\subsection{CI Gruis}\label{obj:cigru}

\begin{figure}
  \begin{center}
    \FigureFile(88mm,110mm){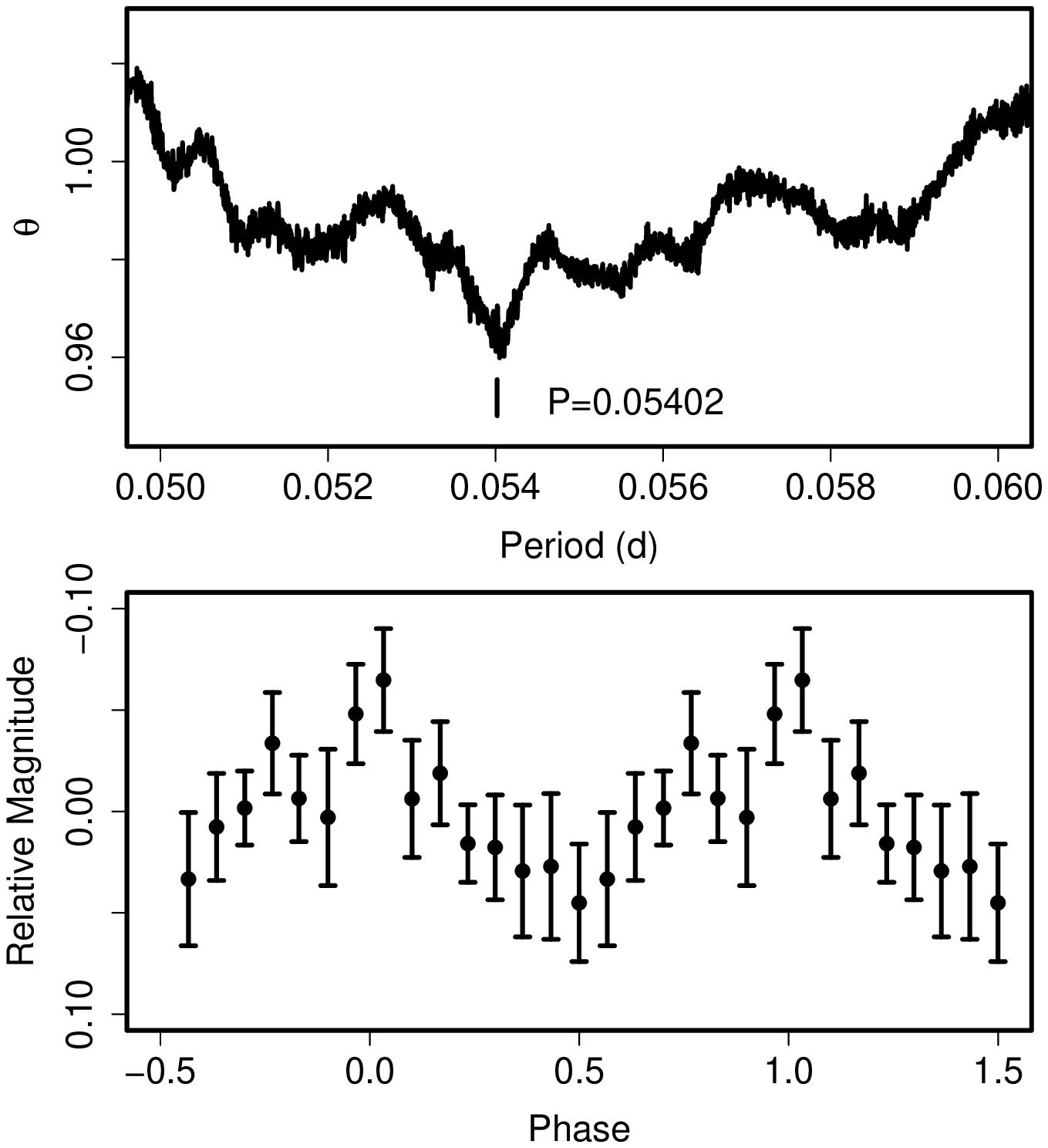}
  \end{center}
  \caption{Superhumps in CI Gru (2004). (Upper): PDM analysis.
     (Lower): Phase-average profile.}
  \label{fig:cigrushpdm}
\end{figure}

   CI Gru was discovered as an outbursting CV \citep{haw83cegruchgrucigru}.
\citet{hae95cfgrucigru} reported semi-periodic variations with a period
of 0.056 d during the possible fading stage of an outburst.
B. Monard detected an outburst on 2004 June 4 at a CCD magnitude of 16.2.
The outburst lasted at least for five days, accompanied by a rapid fading.
The overall behavior suggests that the object underwent a superoutburst.
Based on a single-night observation covering for 7.7 hours, likely
superhumps were detected
(figure \ref{fig:cigrushpdm}, table \ref{tab:cigruoc2004}).
The best period was 0.05402(14) d.  Although this value needs to be
confirmed by future observations, this object would be a candidate for
a very short-$P_{\rm orb}$ SU UMa-type dwarf nova.
The object underwent another outburst (possibly a superoutburst)
in 2006 September at a visual magnitude of 15.4 (Stubbings,
vsnet-alert 9023).

\begin{table}
\caption{Superhump maxima of CI Gru (2004).}\label{tab:cigruoc2004}
\begin{center}
\begin{tabular}{ccccc}
\hline\hline
$E$ & max$^a$ & error & $O-C^b$ & $N^c$ \\
\hline
0 & 53162.3545 & 0.0036 & 0.0008 & 35 \\
1 & 53162.3994 & 0.0050 & $-$0.0078 & 62 \\
2 & 53162.4689 & 0.0199 & 0.0081 & 62 \\
3 & 53162.5175 & 0.0091 & 0.0032 & 62 \\
4 & 53162.5642 & 0.0083 & $-$0.0037 & 62 \\
5 & 53162.6208 & 0.0030 & $-$0.0007 & 59 \\
\hline
  \multicolumn{5}{l}{$^{a}$ BJD$-$2400000.} \\
  \multicolumn{5}{l}{$^{b}$ Against $max = 2453162.3537 + 0.053553 E$.} \\
  \multicolumn{5}{l}{$^{c}$ Number of points used to determine the maximum.} \\
\end{tabular}
\end{center}
\end{table}

\subsection{V844 Herculis}\label{sec:v844her}\label{obj:v844her}

   \citet{oiz07v844her} summarized the analysis of past outbursts.
We present observation of the 2008 superoutburst, an analysis the
AAVSO data for the 1997 superoutburst and a reanalysis of
the 1999 superoutburst \citep{kat00v844her}.
The times of superhumps maxima are listed in
tables \ref{tab:v844heroc1997}, \ref{tab:v844heroc1999},
\ref{tab:v844heroc2008}.

   During the 1999 superoutburst, we obtained
$P_{\rm dot}$ = $+4.5(2.8) \times 10^{-5}$.
No significant period variation was recorded during the 1997 superoutburst.
This was probably due to the limited sampling near the end of the
stage B.

\begin{figure}
  \begin{center}
    \FigureFile(88mm,70mm){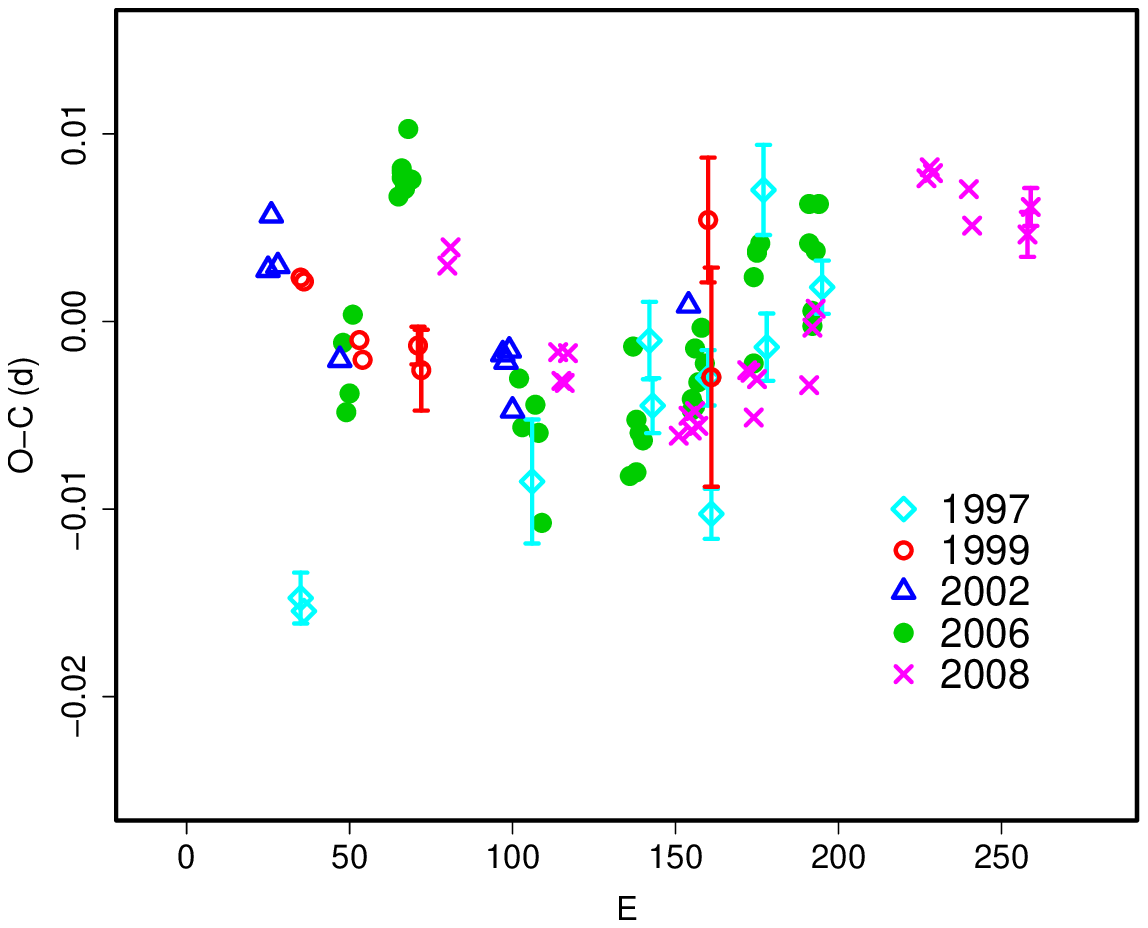}
  \end{center}
  \caption{Comparison of $O-C$ diagrams of V844 Her between different
  superoutbursts.  A period of 0.05590 d was used to draw this figure.
  Approximate cycle counts ($E$) after the start of the
  superoutburst were used.
  The evolution of superhumps apparently started earlier during
  a faint (maximum 12.4 mag) superoutburst in 2002 and later
  during a bright (12.1 mag) superoutburst in 2006.
  }
  \label{fig:v844hercomp}
\end{figure}

   A comparison of $O-C$ diagrams between different superoutburst
is given in figure \ref{fig:v844hercomp}.  While $P_{\rm dot}$'s were
relatively similar, the start of the stage B was different between
different superoutbursts:
the stage B started earlier during a faint (maximum 12.4 mag)
superoutburst in 2002 and later during a bright (12.1 mag)
superoutburst in 2006.
This result further supports the earlier claim \citep{kat08wzsgelateSH}
that the duration before the start of the stage B (or the appearance of
superhumps) depends on the extent of the superoutburst
(see also \cite{soe09swuma}).

   During the 2008 superoutburst we obtained
$P_{\rm dot}$ = $+7.1(0.4) \times 10^{-5}$ for $E \le 149$ (stage B).
There was, however, a phase reversal (associated with
secondary maxima) on BJD 2454584.  These maxima were omitted for
calculating the $P_{\rm dot}$.  This phenomenon may have been similar
to the one observed in OT J055718$+$683226 \citep{uem09j0557}.

   A full description of the outburst will be discussed in Ohshima et al.,
in preparation.

\begin{table}
\caption{Superhump maxima of V844 Her (1997).}\label{tab:v844heroc1997}
\begin{center}

\end{center}
\end{table}

\begin{table}
\caption{Superhump maxima of V844 Her (1999).}\label{tab:v844heroc1999}
\begin{center}
%
\end{center}
\end{table}

\begin{table}
\caption{Superhump maxima of V844 Her (2008).}\label{tab:v844heroc2008}
\begin{center}
%
\end{center}
\end{table}

\subsection{V1108 Herculis}\label{obj:v1108her}

   V1108 Her was discovered by Y. Nakamura on 2004 June 16
\citep{nak04v1108her}.  The earliest positive detection of the
outburst was on 2004 Jun 12 (unfiltered CCD magnitude of 12.0)
by A. Takao (vsnet-alert 8190).
Due to the delayed announcement of the discovery, only the late part
of the superoutburst (11 d after the initial detection) was observed.
We used a combined data set of ours and from the AAVSO data,
which were used in \citet{pri04v1108her}.
The times of superhump maxima are listed in table \ref{tab:v1108heroc2004}.
As in WZ Sge, a strong hump feature appeared and surpassed in amplitude
in the late stage of the outburst.  For the interval $E \ge 79$,
we used a fit to a smaller width $\pm$0.1 $P_{\rm SH}$ around the
peaks whose phases can be smoothly linked to earlier peaks,
as in V455 And and WZ Sge.  The resultant data clearly showed
a transition from a longer period to a shorter one around $E = 29$.
The mean period for $E \le 29$ was 0.05880(18) d, while the period
for $E \ge 29$ was 0.05748(3) d.  
The maxima of secondary (but stronger in the final
fading stage) peaks are listed in table \ref{tab:v1108heroc2004sec}.
For the interval $81 \le E \le 108$, they had a relatively stable
periodicity of 0.05703(8) d.  By analogy with WZ Sge, this periodicity
might be considered to be the orbital period.\footnote{
   This period, though, might refer to a variety of superhumps.
   \citet{pri04v1108her} reported another candidate periodicity
   of 0.05686(7) d marginally detected in post-superoutburst stage.
   The exact identification of the periodicities should await
   future observations.
}
Using this period, we
obtained the fractional superhump excesses for the two segments
($E \le 29$ and $E \ge 29$) of 3.1(3) \% and 0.8(1) \%, respectively.
These period excesses might be attributed to stage B and C superhumps.
The unusually large fractional superhump excess (3.1 \%) might be
a result of lengthening in the $P_{\rm SH}$ during the stage B
(see an example of AQ Eri, subsection \ref{sec:aqeri}).  This value
might not be used to derive system parameters such as $q$.
This object, with relatively frequent historical outbursts
\citep{pri04v1108her}, appears more analogous to positive-$P_{\rm dot}$
systems such as HV Vir and AL Com rather than extreme WZ Sge-type
dwarf novae with little variation in the superhump period (e.g. WZ Sge
and V455 And).

\begin{table}
\caption{Superhump maxima of V1108 Her (2004).}\label{tab:v1108heroc2004}
\begin{center}

\end{center}
\end{table}

\begin{table}
\caption{Secondary Superhump of V1108 Her (2004).}\label{tab:v1108heroc2004sec}
\begin{center}
%
\end{center}
\end{table}

\subsection{RU Horologii}\label{sec:ruhor}\label{obj:ruhor}

   The times of superhump maxima obtained during the 2003 superoutburst are
listed in table \ref{tab:ruhoroc2003}.  The object clearly showed
brightening near the termination of a superoutburst (cf. \cite{kat03hodel}
and discussion in subsection \ref{sec:lland}), after which ($E > 80$)
the superhump period remarkably decreased
(cf. figure \ref{fig:octrans}).
Using the timings for the interval $0 \le E \le 76$, we obtained
$P_{\rm dot}$ = $+7.5(1.1) \times 10^{-5}$ and a mean superhump
period of 0.07095(2) d.

   The 2008 superoutburst (table \ref{tab:ruhoroc2008}) was observed
during the middle-to-late stage of the plateau phase.  There is
a clear signature of a transition to a shorter period (stage B to C).
The $P_{\rm dot}$ before this transition, disregarding the slightly
discrepant point $E = 0$, was $+6.5(3.2) \times 10^{-5}$
($1 \le E \le 44$).

\begin{figure}
  \begin{center}
    \FigureFile(88mm,70mm){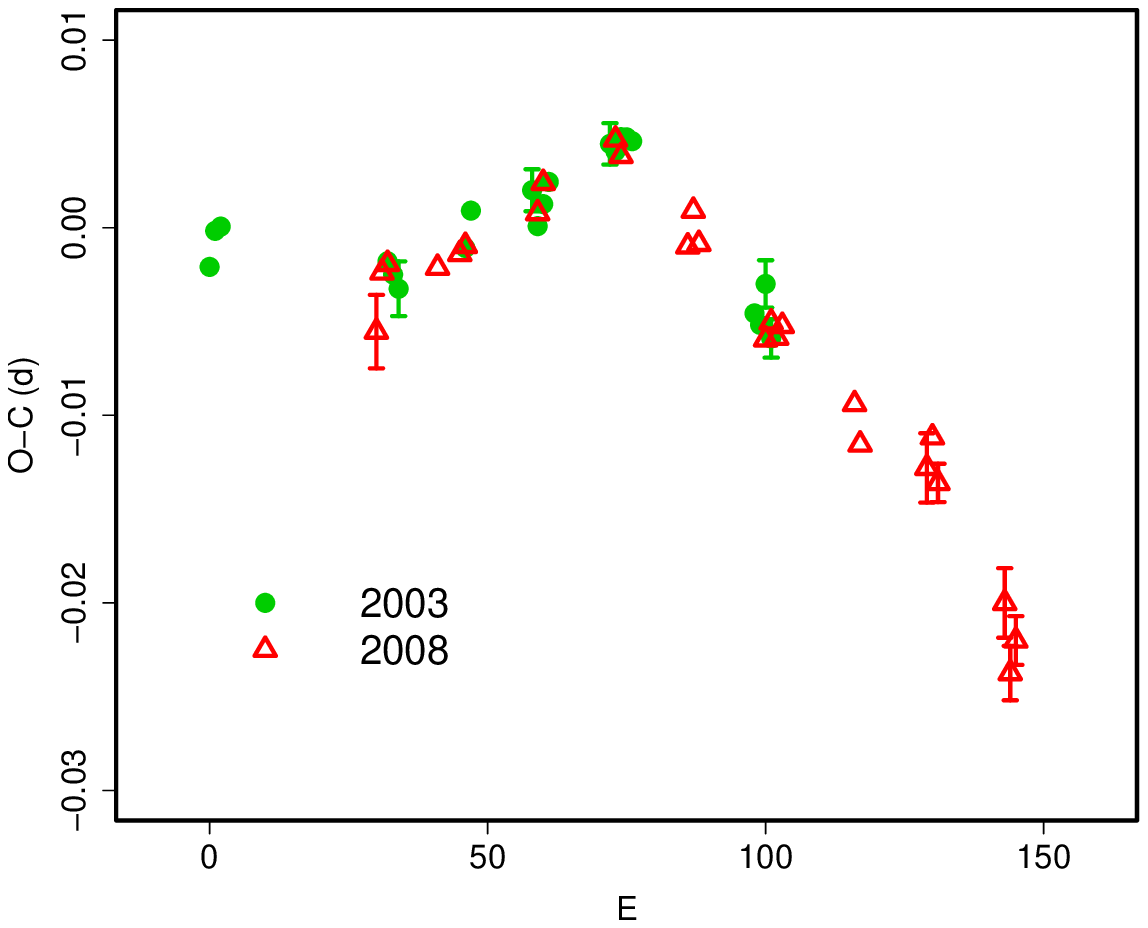}
  \end{center}
  \caption{Comparison of $O-C$ diagrams of RU Hor between different
  superoutbursts.  A period of 0.07087 d was used to draw this figure.
  Although the actual start of the outburst was not well constrained,
  the $O-C$ diagram of the 2008 superoutburst almost perfectly fits
  the 2003 one by assuming a 50-cycle difference in $E$.
  }
  \label{fig:ruhorcomp}
\end{figure}

   A comparison of $O-C$ diagrams of RU Hor between different
superoutbursts is given in figure \ref{fig:ruhorcomp}.
Although the actual start of the outburst was not well constrained,
the $O-C$ diagram of the 2008 superoutburst almost perfectly fits
the 2003 one by assuming a 50-cycle difference in $E$.

\begin{table}
\caption{Superhump maxima of RU Hor (2003).}\label{tab:ruhoroc2003}
\begin{center}

\end{center}
\end{table}

\begin{table}
\caption{Superhump maxima of RU Hor (2008).}\label{tab:ruhoroc2008}
\begin{center}
%
\end{center}
\end{table}

\subsection{CT Hydrae}\label{obj:cthya}

   Superhumps during two superoutbursts (1995 and 1999) were reported
in the past literature (\cite{nog96cthya}; \cite{kat99cthya}).
We reanalyzed the 1999 observations in view of
the modern knowledge.  The times of superhump maxima are listed in table
\ref{tab:cthyaoc1999}.  The Brno data were removed before the
analysis because of the yet unsolved phase problem (cf. \cite{kat99cthya}).
Although \citet{kat99cthya} stated that the change in the superhump
period was negligible, the present analysis seems to show a tendency
of a period increase.  The negative $O-C$ of the last ($E = 105$)
being likely a result of the period decrease associated with
a stage B--C transition, we excluded this point and obtained
$P_{\rm dot}$ = $+7.0(4.3) \times 10^{-5}$.
If we include this point, the resultant $P_{\rm dot}$ is almost zero
($-1.0(8.7) \times 10^{-5}$), confirming the analysis in
\citet{kat99cthya}.  We further present the superoutbursts in 2000,
2002 February, 2002 November and 2009 January.
(tables \ref{tab:cthyaoc2000}, \ref{tab:cthyaoc2002a},
\ref{tab:cthyaoc2002b}, \ref{tab:cthyaoc2009}).
The resultant values of $P_{\rm dot}$ for the stage B
were $+9.6(5.2) \times 10^{-5}$ (2000, $E \le 78$),
$+11.6(3.8) \times 10^{-5}$ (2002 February, $E \ge 14$)
and $+13.2(3.1) \times 10^{-5}$ (2002 November, $E \le 90$), respectively.

\begin{figure}
  \begin{center}
    \FigureFile(88mm,70mm){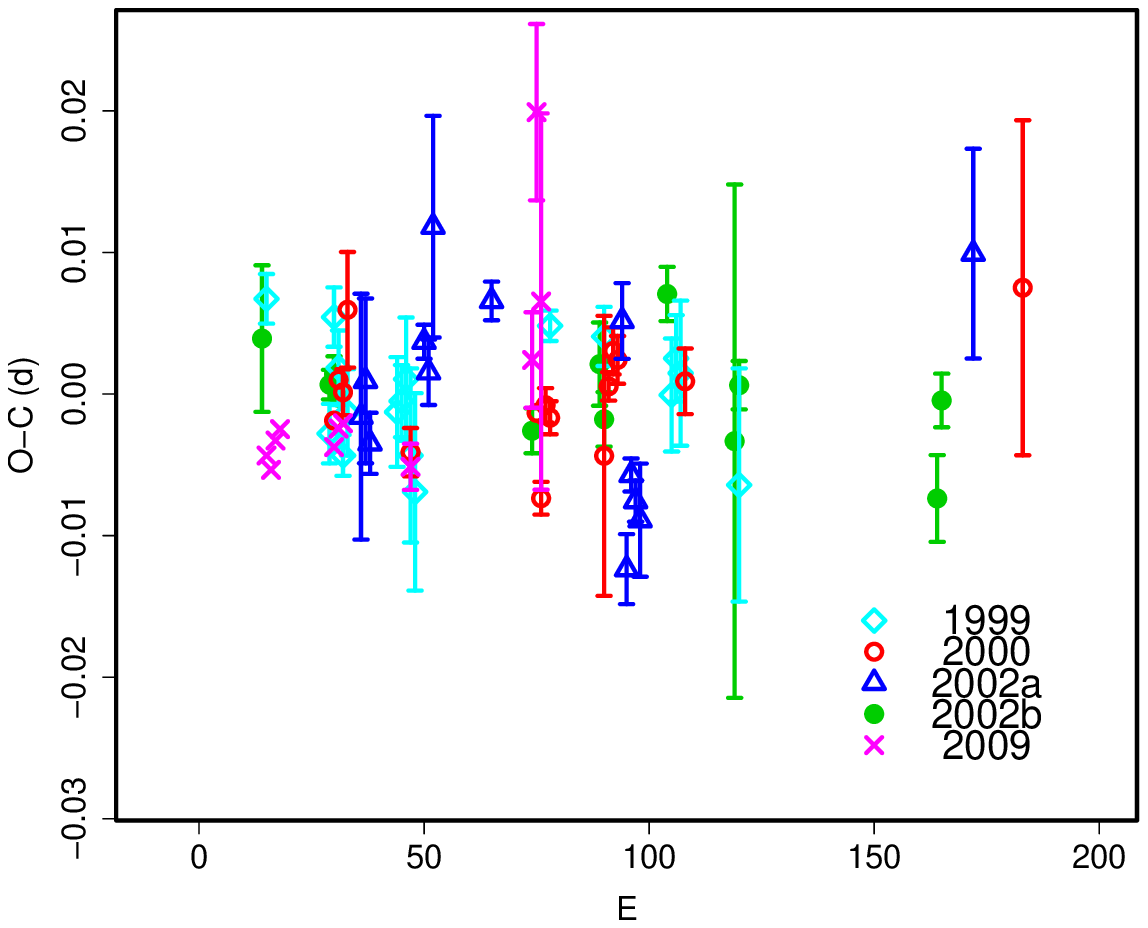}
  \end{center}
  \caption{Comparison of $O-C$ diagrams of CT Hya between different
  superoutbursts.  A period of 0.06640 d was used to draw this figure.
  Approximate cycle counts ($E$) after the start of the
  superoutburst were used.
  }
  \label{fig:cthyacomp}
\end{figure}

   A comparison of $O-C$ diagrams between different superoutbursts
is given in figure \ref{fig:cthyacomp}.  The relatively large error in
$O-C$'s in this system makes a comparison rather difficult.
The behavior (and diversity) of the late stage B is somewhat reminiscent
to KV Dra.

\begin{table}
\caption{Superhump maxima of CT Hya (1999).}\label{tab:cthyaoc1999}
\begin{center}

\end{center}
\end{table}

\begin{table}
\caption{Superhump maxima of CT Hya (2000).}\label{tab:cthyaoc2000}
\begin{center}
%
\end{center}
\end{table}

\begin{table}
\caption{Superhump maxima of CT Hya (2002a).}\label{tab:cthyaoc2002a}
\begin{center}
%
\end{center}
\end{table}

\begin{table}
\caption{Superhump maxima of CT Hya (2002b).}\label{tab:cthyaoc2002b}
\begin{center}
%
\end{center}
\end{table}

\begin{table}
\caption{Superhump maxima of CT Hya (2009).}\label{tab:cthyaoc2009}
\begin{center}
%
\end{center}
\end{table}

\subsection{MM Hydrae}\label{obj:mmhya}

   MM Hya, selected by the Palomer-Green survey \citep{gre82PGsurveyCV},
had long been suspected to be a WZ Sge-like object based on the
short orbital period \citep{mis95PGCV}.  The object was soon confirmed
to undergo long outbursts approximately once per year, indicating
a more usual SU UMa-type dwarf nova rather than a WZ Wge-like object.
\citet{pat03suumas} reported a mean $P_{\rm SH}$ of 0.05868 d during the
1998 superoutburst without giving the details.
We analyzed the AAVSO observations of the 1998 superoutburst and obtained
times of superhump maxima (table \ref{tab:mmhyaoc1998}).
The mean $P_{\rm SH}$ determined with the PDM method was 0.05894(3) d.
This period is significantly longer than that by \citet{pat03suumas}.
The present observation was probably obtained near the end of the stage B.
A possible decrease in $O-C$, although the uncertainty is large, in
$E=65-66$ may be a result of a transition to the stage C.

\begin{figure}
  \begin{center}
    \FigureFile(88mm,110mm){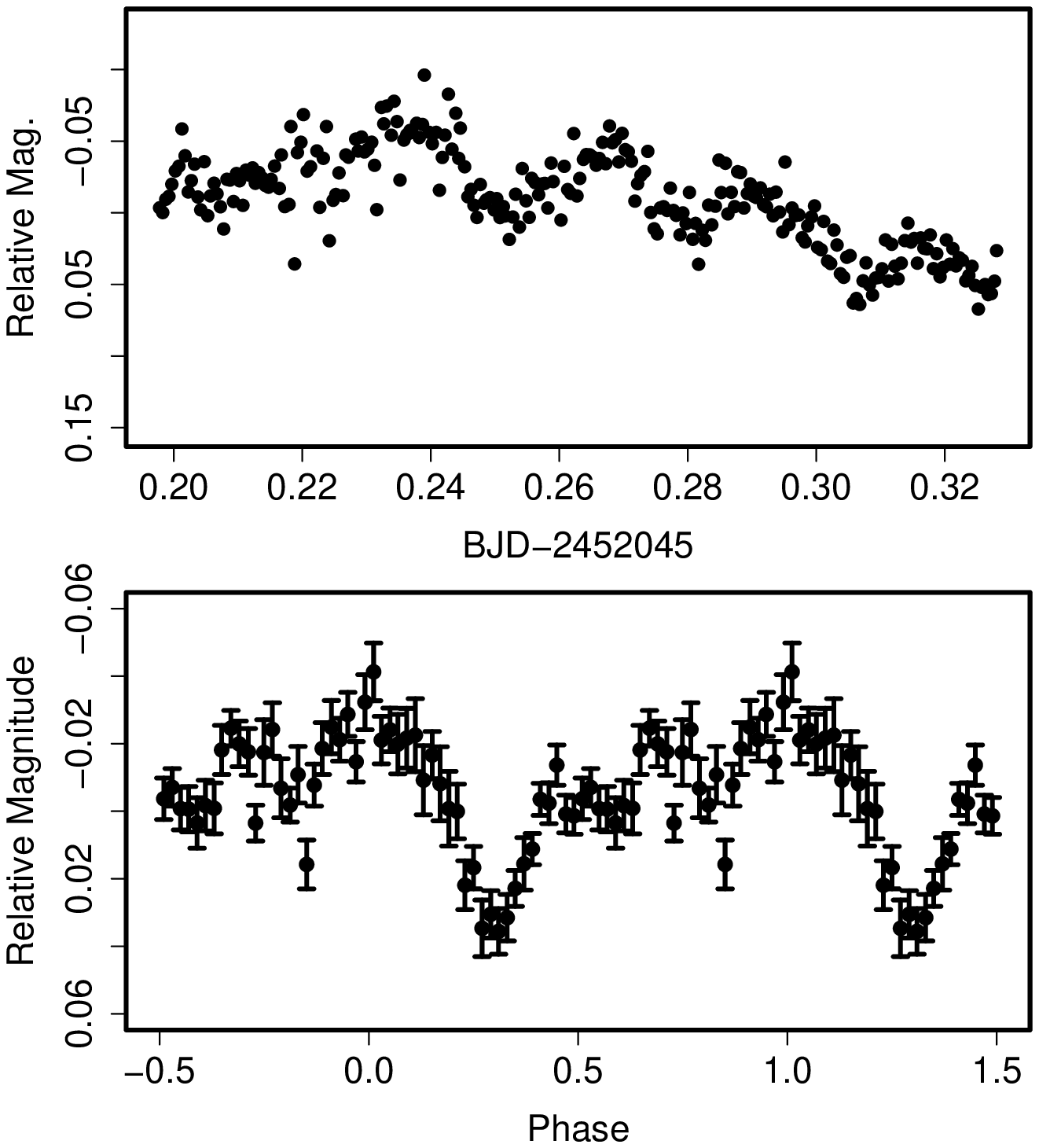}
  \end{center}
  \caption{Double-wave humps in MM Hya (2001) (Upper): Light curve.
     (Lower): Phase-averaged profile referring to the orbital period.}
  \label{fig:mmhyaearly}
\end{figure}

   We also observed the 2001 superoutburst during its earliest stage
(table \ref{tab:mmhyaoc2001}).  The observations corresponded to
the stage A--B transition.  The mean periods during the stage A was
0.0603(3) d.  The observed length of the stage B was too short to
determine the period.  On the first night of the observation
(2001 May 15), double-wave modulations similar to early superhumps
in WZ Sge-type dwarf novae were observed (figure \ref{fig:mmhyaearly}).
Although the length of observations was insufficient to discriminate
between $P_{\rm SH}$ and $P_{\rm orb}$, the profile strongly suggests
the presence of early superhumps.  The object is similar to
BC UMa (\cite{pat03suumas}; \cite{mae07bcuma}) and RZ Leo
(\cite{ish01rzleo}; \cite{pat03suumas}) that showed early superhumps
during the earliest stage of their superoutbursts.

\begin{table}
\caption{Superhump maxima of MM Hya (1998).}\label{tab:mmhyaoc1998}
\begin{center}
\begin{tabular}{ccccc}
\hline\hline
$E$ & max$^a$ & error & $O-C^b$ & $N^c$ \\
\hline
0 & 50882.2853 & 0.0076 & $-$0.0050 & 17 \\
1 & 50882.3503 & 0.0023 & 0.0011 & 33 \\
2 & 50882.4123 & 0.0021 & 0.0042 & 34 \\
3 & 50882.4671 & 0.0012 & 0.0001 & 29 \\
15 & 50883.1732 & 0.0026 & $-$0.0011 & 58 \\
51 & 50885.3016 & 0.0030 & 0.0058 & 23 \\
52 & 50885.3517 & 0.0016 & $-$0.0031 & 34 \\
65 & 50886.1191 & 0.0058 & $-$0.0018 & 48 \\
66 & 50886.1796 & 0.0029 & $-$0.0002 & 48 \\
\hline
  \multicolumn{5}{l}{$^{a}$ BJD$-$2400000.} \\
  \multicolumn{5}{l}{$^{b}$ Against $max = 2450882.2903 + 0.058931 E$.} \\
  \multicolumn{5}{l}{$^{c}$ Number of points used to determine the maximum.} \\
\end{tabular}
\end{center}
\end{table}

\begin{table}
\caption{Superhump maxima of MM Hya (2001).}\label{tab:mmhyaoc2001}
\begin{center}
\begin{tabular}{ccccc}
\hline\hline
$E$ & max$^a$ & error & $O-C^b$ & $N^c$ \\
\hline
0 & 52045.9826 & 0.0025 & $-$0.0000 & 89 \\
4 & 52046.2195 & 0.0007 & $-$0.0022 & 615 \\
17 & 52047.0068 & 0.0021 & 0.0082 & 71 \\
21 & 52047.2359 & 0.0003 & $-$0.0017 & 707 \\
22 & 52047.2931 & 0.0001 & $-$0.0043 & 703 \\
\hline
  \multicolumn{5}{l}{$^{a}$ BJD$-$2400000.} \\
  \multicolumn{5}{l}{$^{b}$ Against $max = 2452045.9826 + 0.059762 E$.} \\
  \multicolumn{5}{l}{$^{c}$ Number of points used to determine the maximum.} \\
\end{tabular}
\end{center}
\end{table}

\subsection{VW Hydri}\label{obj:vwhyi}

   We analyzed the 2000 May superoutburst (table \ref{tab:vwhyioc2002}).
The observation covered the early stage of the superoutburst and
we obtained a mean $P_{\rm SH}$ of 0.07699(6) d.
The observations were slightly insufficient
to estimate a $P_{\rm dot}$.
The $O-C$ variation in \citet{vog74vwhyi} confirmed the presence
of the stage B--C transition.
\citet{lil96vwhyi} reported little evidence for period variation
of superhumps during the 1995 November superoutburst.  We did not,
however, include this observation because the periods were
not based on an $O-C$ analysis nor times of superhumps were given.
The reported period of 0.076646(3) d with the Fourier analysis
was between $P_1$ and $P_2$ of the 2000 superoutburst.

\begin{table}
\caption{Superhump maxima of VW Hyi (2000).}\label{tab:vwhyioc2002}
\begin{center}
\begin{tabular}{ccccc}
\hline\hline
$E$ & max$^a$ & error & $O-C^b$ & $N^c$ \\
\hline
0 & 51680.9054 & 0.0022 & $-$0.0014 & 6 \\
8 & 51681.5219 & 0.0020 & $-$0.0007 & 7 \\
13 & 51681.9050 & 0.0074 & $-$0.0026 & 7 \\
33 & 51683.4515 & 0.0032 & 0.0042 & 6 \\
34 & 51683.5298 & 0.0059 & 0.0056 & 7 \\
46 & 51684.4467 & 0.0045 & $-$0.0014 & 5 \\
47 & 51684.5265 & 0.0024 & 0.0014 & 7 \\
52 & 51684.9081 & 0.0099 & $-$0.0019 & 7 \\
60 & 51685.5228 & 0.0022 & $-$0.0031 & 9 \\
\hline
  \multicolumn{5}{l}{$^{a}$ BJD$-$2400000.} \\
  \multicolumn{5}{l}{$^{b}$ Against $max = 2451680.9067 + 0.076986 E$.} \\
  \multicolumn{5}{l}{$^{c}$ Number of points used to determine the maximum.} \\
\end{tabular}
\end{center}
\end{table}

\subsection{RZ Leonis}\label{obj:rzleo}

   We reanalyzed a combination of \citet{ish01rzleo} and the AAVSO data.
The times of superhump maxima are given in table \ref{tab:rzleooc2000}.
Although \citet{ish01rzleo}
identified earlier maxima ($E \le 6$) as being early superhumps, we examined
whether these maxima can be tracked back as in the stage A
of other SU UMa-type dwarf novae.  Although we could track back
the maxima with a slightly longer period for $\sim$ 1 d, as in the stage A
of other SU UMa-type dwarf novae, this attempt failed to express earlier
($E < 0$) epochs.
This analysis also supports the identification of these humps as being
early superhumps, rather than a smooth extension of ordinary superhumps.
The $O-C$ diagram showed a transition to the stage C after $E = 100$.
For the interval $13 \le E \le 100$ (stage B), we obtained
$P_{\rm dot}$ = $+4.9(1.7) \times 10^{-5}$.
The value is in good agreement with \citet{ish01rzleo}.

\begin{figure}
  \begin{center}
    \FigureFile(88mm,70mm){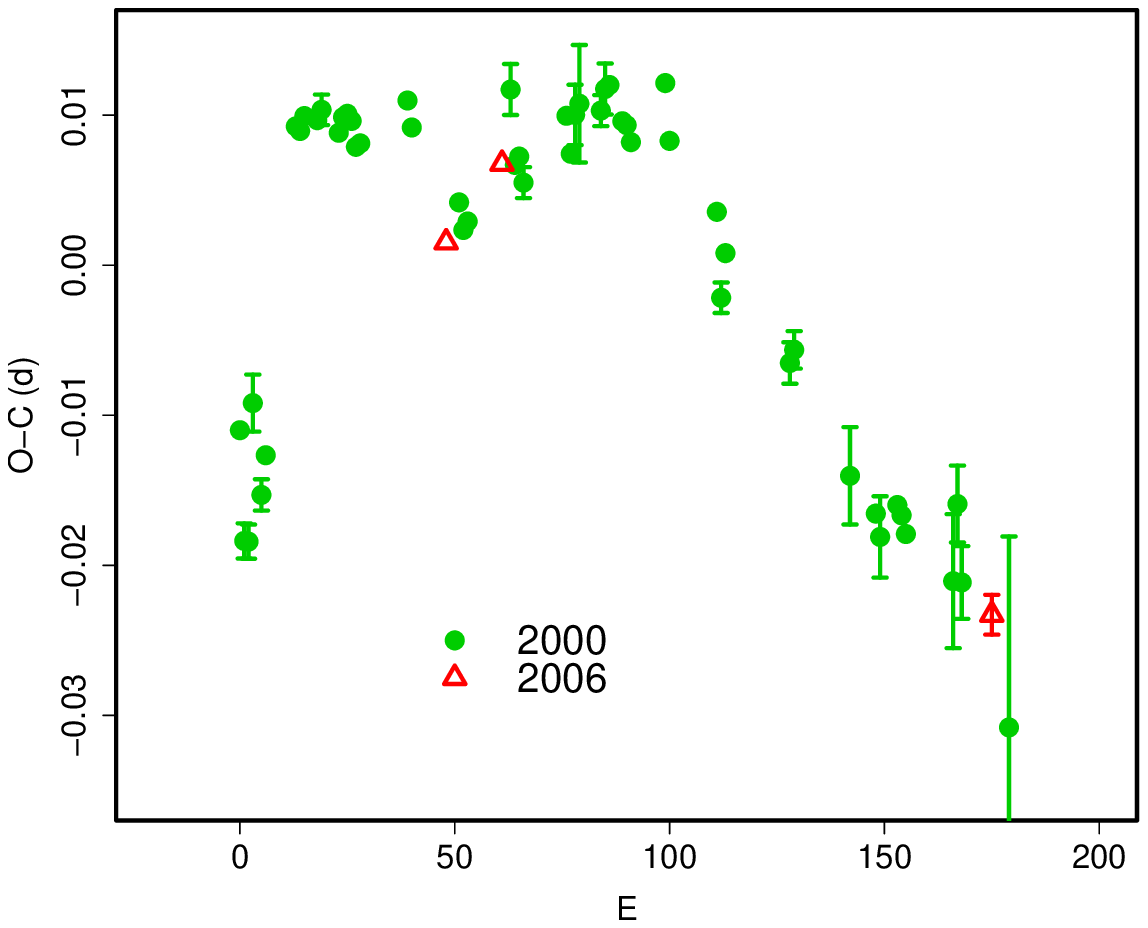}
  \end{center}
  \caption{Comparison of $O-C$ diagrams of RZ Leo between different
  superoutbursts.  A period of 0.07865 d was used to draw this figure.
  Approximate cycle counts ($E$) after the start of the
  superoutburst were used.
  }
  \label{fig:rzleocomp}
\end{figure}

   The object underwent another superoutburst in 2006.
Although the seasonal condition was poor, we obtained several superhump
maxima (table \ref{tab:rzleooc2006}).  The $O-C$'s against the mean period
of 2000 suggest that the observation caught the increasing period
during the first two nights, and last observation with a strongly negative
$O-C$ should have caught the late transition from the stage B to C
(see figure \ref{fig:rzleocomp}).
We did not attempt to derive a global $P_{\rm dot}$.

\begin{table}
\caption{Superhump maxima of RZ Leo (2000--2001).}\label{tab:rzleooc2000}
\begin{center}

\end{center}
\end{table}

\begin{table}
\caption{Superhump maxima of RZ Leo (2006).}\label{tab:rzleooc2006}
\begin{center}
%
\end{center}
\end{table}

\subsection{GW Librae}\label{sec:gwlib}\label{obj:gwlib}

   GW Lib, originally reported as a nova in 1983 \citep{maz83gwlibiauc},
and long suspected to be a WZ Sge-type dwarf nova, underwent a spectacular
outburst in 2007 (R. Stubbings, vsnet-alert 9279; \cite{waa07gwlibiauc}).
The object initially showed only very low-amplitude modulations similar to
early superhumps, whose period was not well determined.  After $\sim$ 11
days, ordinary superhumps emerged (vsnet-alert 9315, 9316).

\begin{figure}
  \begin{center}
    \FigureFile(88mm,110mm){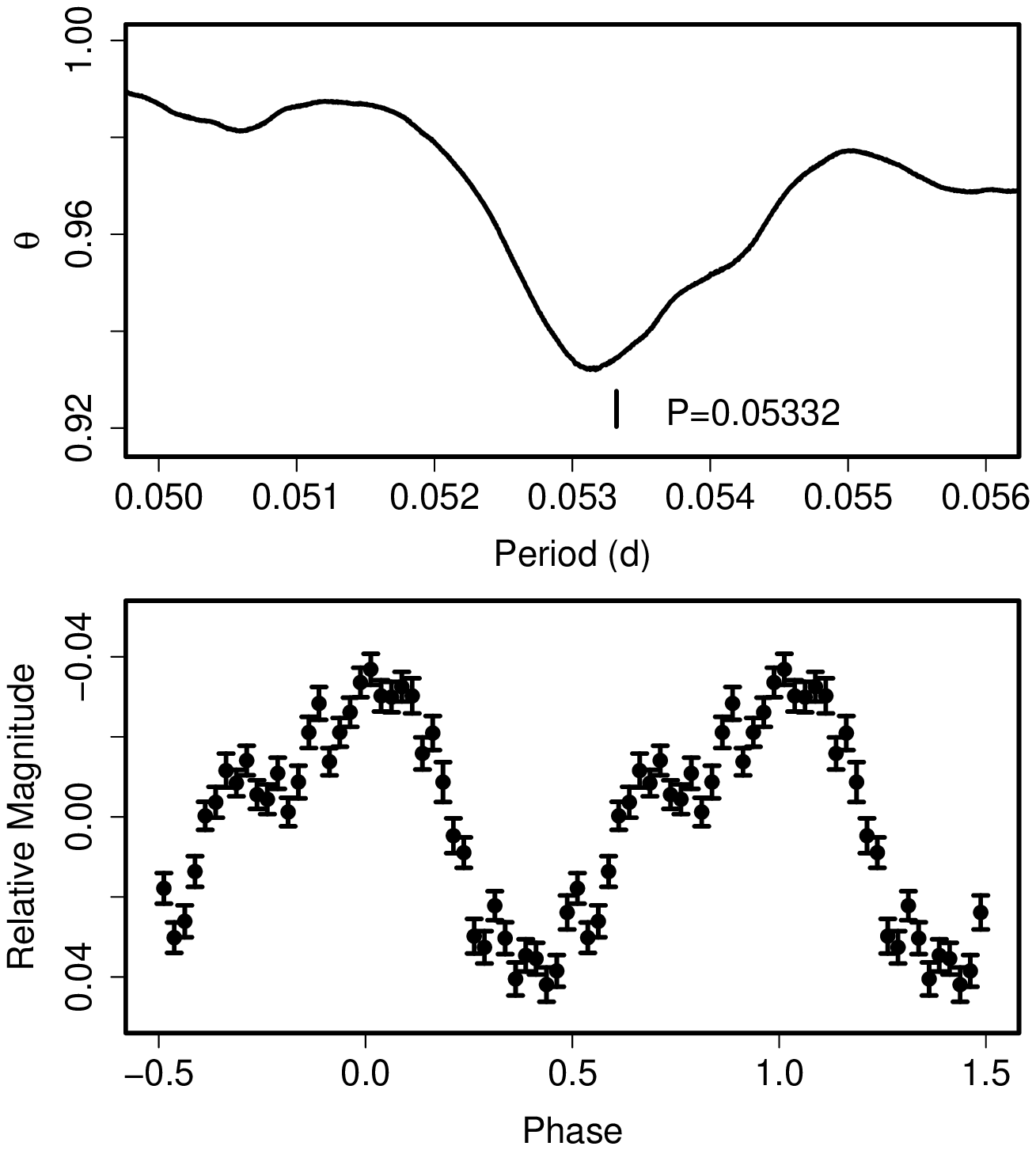}
  \end{center}
  \caption{Orbital humps in GW Lib (2007) during the late stage
     (BJD 2454227--2454230) of the superoutburst plateau.
     (Upper): PDM analysis.  The tick mark is given at the
     orbital period.
     (Lower): Phase-averaged profile.}
  \label{fig:gwporb}
\end{figure}

   The maxima times of ordinary superhumps are listed
in table \ref{tab:gwliboc2007}.
The $O-C$ diagram (figure \ref{fig:gwlibhumpall}) very clearly
consisted of the stage A with a long superhump period
($E \le 39$), the stage B with a definitely positive $P_{\rm dot}$,
and later stage ($E \ge 289$) with noticeably negative $O-C$'s.
For the stage B ($51 \le E \le 278$), we obtained $P_{\rm dot}$ =
$+4.0(0.1) \times 10^{-5}$.  It seems that the phase was discontinuous
between the middle and the last segments.  This may be attributed
to the appearance of the orbital humps (figure \ref{fig:gwporb}).
At an estimated orbital inclination of 11$^{\circ}$
\citep{tho02gwlibv844herdiuma}, the appearance of orbital humps
is surprising.  The orbital inclination is either higher or
there is a special mechanism for manifesting orbital humps during
the late stage of the plateau phase of WZ Sge-type dwarf novae
(see also V455 And and WZ Sge, subsections \ref{sec:v455and} amd
\ref{sec:wzsge}.  The double-wave profile in GW Lib might suggest
that a sort of the 2:1 resonance, as in early superhumps, is somehow
excited, or persists, during the last stage of the superoutburst
plateau of WZ Sge-type dwarf novae.

   More detailed analysis of the outburst will be presented by
Imada et al., in preparation.

\begin{table}
\caption{Superhump maxima of GW Lib (2007).}\label{tab:gwliboc2007}
\begin{center}

\end{center}
\end{table}

\addtocounter{table}{-1}
\begin{table}
\caption{Superhump maxima of GW Lib (2007) (continued).}
\begin{center}
%
\end{center}
\end{table}

\addtocounter{table}{-1}
\begin{table}
\caption{Superhump maxima of GW Lib (2007) (continued).}
\begin{center}
%
\end{center}
\end{table}

\addtocounter{table}{-1}
\begin{table}
\caption{Superhump maxima of GW Lib (2007) (continued).}
\begin{center}
%
\end{center}
\end{table}

\subsection{RZ Leonis Minoris}\label{sec:rzlmi}\label{obj:rzlmi}

   We analyzed the 2005 April superoutburst of RZ LMi
(table \ref{tab:rzlmioc2005}).  This superoutburst had a marginally
positive $P_{\rm dot}$ of $+2.3(1.1) \times 10^{-5}$, as in the 2004
superoutburst \citet{ole08rzlmi}.  The maxima for $E \ge 118$
(during the rapid fading stage) were either phase 0.5 offset
(traditional late superhumps), stage C superhumps with a period
of 0.05875(8) d ($E \ge 84$), or even a candidate for orbital humps.
Since none of these kinds of phenomena have not yet been reported
in RZ LMi \citep{ole08rzlmi}, further detailed observations during
the rapid fading stage might provide crucial information.

\begin{table}
\caption{Superhump maxima of RZ LMi (2005).}\label{tab:rzlmioc2005}
\begin{center}
\begin{tabular}{ccccc}
\hline\hline
$E$ & max$^a$ & error & $O-C^b$ & $N^c$ \\
\hline
0 & 53473.7101 & 0.0004 & $-$0.0054 & 40 \\
1 & 53473.7688 & 0.0004 & $-$0.0058 & 36 \\
2 & 53473.8282 & 0.0003 & $-$0.0057 & 36 \\
33 & 53475.6684 & 0.0005 & $-$0.0003 & 45 \\
34 & 53475.7284 & 0.0006 & 0.0005 & 36 \\
35 & 53475.7862 & 0.0006 & $-$0.0009 & 35 \\
36 & 53475.8467 & 0.0009 & 0.0004 & 30 \\
50 & 53476.6803 & 0.0008 & 0.0053 & 37 \\
51 & 53476.7379 & 0.0006 & 0.0037 & 40 \\
52 & 53476.7962 & 0.0005 & 0.0028 & 38 \\
53 & 53476.8563 & 0.0006 & 0.0037 & 37 \\
84 & 53478.7001 & 0.0008 & 0.0126 & 40 \\
86 & 53478.8169 & 0.0017 & 0.0111 & 41 \\
118 & 53480.7007 & 0.0079 & 0.0007 & 20 \\
119 & 53480.7497 & 0.0025 & $-$0.0094 & 20 \\
120 & 53480.8145 & 0.0017 & $-$0.0038 & 20 \\
136 & 53481.7560 & 0.0069 & $-$0.0094 & 20 \\
\hline
  \multicolumn{5}{l}{$^{a}$ BJD$-$2400000.} \\
  \multicolumn{5}{l}{$^{b}$ Against $max = 2453473.7154 + 0.059191 E$.} \\
  \multicolumn{5}{l}{$^{c}$ Number of points used to determine the maximum.} \\
\end{tabular}
\end{center}
\end{table}

\subsection{SS Leonis Minoris}\label{obj:sslmi}

   SS LMi was discovered as an extragalactic nova or an unusual
dwarf nova \citep{alk80sslmi}.  Although \citet{har91sslmiiauc} reported
a ``red'' outburst in 1991, the nature of this outburst remained
unclear.\footnote{
   See also \citet{how91sslmiiauc}; the unusual color in these
   observations could have been a combination with a nearby field star
   \citet{she08sslmi}.
}
   \citet{she08sslmi} reported the 2006 superoutburst.
Based on their times of superhump maxima, we obtained
$P_{\rm dot}$ = $+0.3(0.4) \times 10^{-5}$ ($E \le 128$).
Since these superhumps were detected during the initial stage of
a likely WZ Sge-type outburst, they can be interpreted as
early superhumps rather than ordinary superhumps.  The lack of
period variation and a hint of double-wave modulations
\citep{she08sslmi} may support this interpretation.
We listed the period in table \ref{tab:perlist} based on this
identification.

\subsection{SX Leonis Minoris}\label{obj:sxlmi}

   \citet{nog97sxlmi} reported on the 1994 superoutburst.  We reanalyzed
the data during this superoutburst.  The resultant times of superhump maxima
are listed in table \ref{tab:sxlmioc1994}.
The overall $P_{\rm dot}$ was $-8.2(1.1) \times 10^{-5}$, in good
agreement with \citet{nog97sxlmi}.

   We also observed the 2001 and 2002 superoutbursts
(tables \ref{tab:sxlmioc2001}, \ref{tab:sxlmioc2002}).
The resultant values of $P_{\rm dot}$ were $-3.3(3.0) \times 10^{-5}$
and $-4.1(1.5) \times 10^{-5}$ (excluding $E = 0$), respectively.
The 2002 result might be interpreted as a sudden shift to a shorter
superhump period (stage B to C) between $E = 116$ and $E = 130$.
Using the interval of $14 \le E \le 115$, the resultant period
change was almost zero, $P_{\rm dot}$ = $-0.7(0.5) \times 10^{-5}$.

\begin{table}
\caption{Superhump maxima of SX LMi (1994).}\label{tab:sxlmioc1994}
\begin{center}

\end{center}
\end{table}

\begin{table}
\caption{Superhump maxima of SX LMi (2001).}\label{tab:sxlmioc2001}
\begin{center}
%
\end{center}
\end{table}

\begin{table}
\caption{Superhump maxima of SX LMi (2002).}\label{tab:sxlmioc2002}
\begin{center}
%
\end{center}
\end{table}

\subsection{BR Lupi}\label{obj:brlup}

\begin{figure}
  \begin{center}
    \FigureFile(88mm,70mm){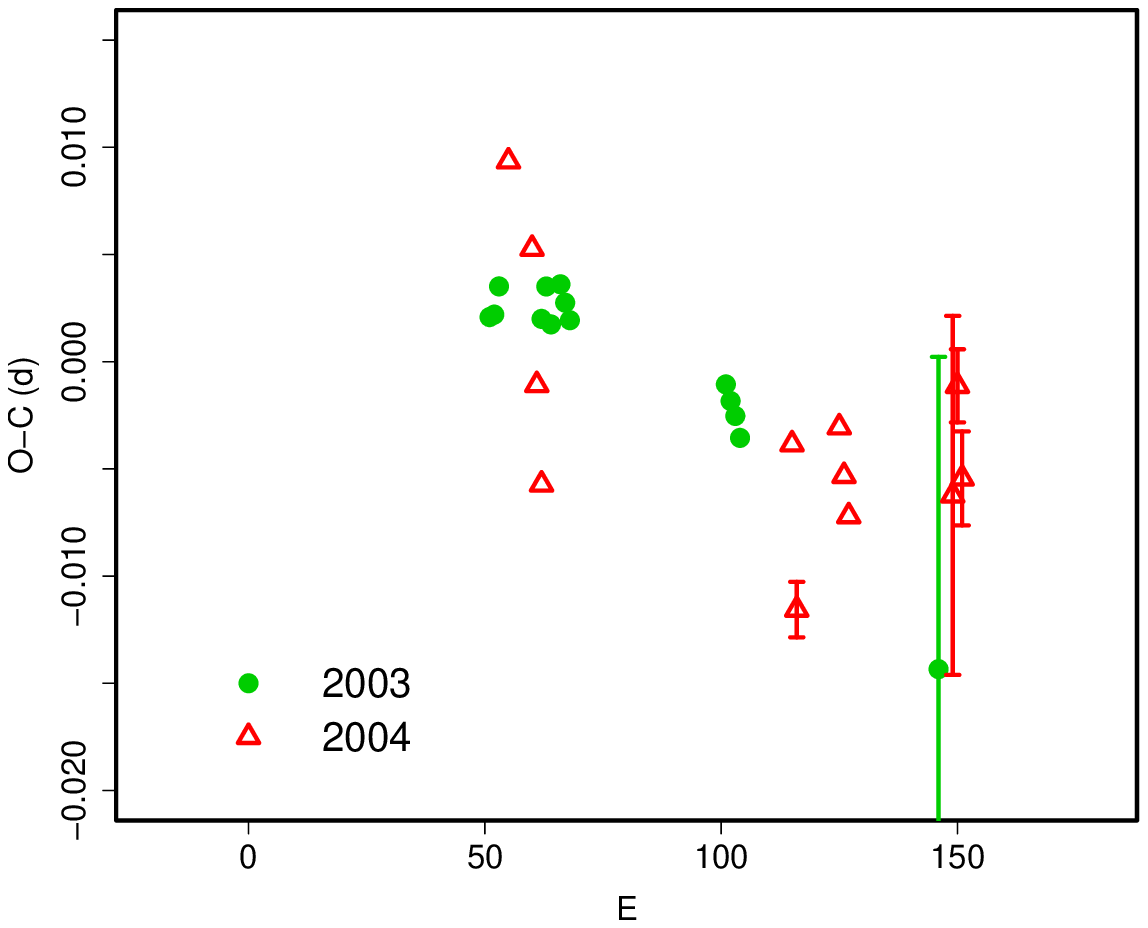}
  \end{center}
  \caption{Comparison of $O-C$ diagrams of BR Lup between different
  superoutbursts.  A period of 0.08228 d was used to draw this figure.
  Approximate cycle counts ($E$) after the start of the
  superoutburst were used.
  }
  \label{fig:brlupcomp}
\end{figure}

   We observed the 2003 and 2004 superoutbursts.  The times of
superhump maxima are listed in tables \ref{tab:brlupoc2003} and
\ref{tab:brlupoc2004}.
The both observations covered the relatively late stages of the
superoutbursts (figure \ref{fig:brlupcomp}).
A stage B--C transition was probably caught during the 2003 superoutburst
and the only the stage C was likely recorded during the 2004 superoutburst.
We give parameters in table \ref{tab:perlist} based on this interpretation.

\begin{table}
\caption{Superhump maxima of BR Lup (2003).}\label{tab:brlupoc2003}
\begin{center}
\begin{tabular}{ccccc}
\hline\hline
$E$ & max$^a$ & error & $O-C^b$ & $N^c$ \\
\hline
0 & 52737.2349 & 0.0005 & $-$0.0023 & 83 \\
1 & 52737.3172 & 0.0004 & $-$0.0020 & 83 \\
2 & 52737.4008 & 0.0006 & $-$0.0006 & 59 \\
11 & 52738.1398 & 0.0005 & $-$0.0007 & 66 \\
12 & 52738.2236 & 0.0004 & 0.0010 & 83 \\
13 & 52738.3041 & 0.0005 & $-$0.0006 & 82 \\
15 & 52738.4706 & 0.0006 & 0.0016 & 89 \\
16 & 52738.5520 & 0.0005 & 0.0009 & 94 \\
17 & 52738.6335 & 0.0006 & 0.0002 & 78 \\
50 & 52741.3457 & 0.0008 & 0.0025 & 69 \\
51 & 52741.4272 & 0.0006 & 0.0019 & 88 \\
52 & 52741.5088 & 0.0008 & 0.0013 & 79 \\
53 & 52741.5901 & 0.0009 & 0.0005 & 58 \\
95 & 52745.0350 & 0.0146 & $-$0.0037 & 18 \\
\hline
  \multicolumn{5}{l}{$^{a}$ BJD$-$2400000.} \\
  \multicolumn{5}{l}{$^{b}$ Against $max = 2452737.2372 + 0.082121 E$.} \\
  \multicolumn{5}{l}{$^{c}$ Number of points used to determine the maximum.} \\
\end{tabular}
\end{center}
\end{table}

\begin{table}
\caption{Superhump maxima of BR Lup (2004).}\label{tab:brlupoc2004}
\begin{center}
\begin{tabular}{ccccc}
\hline\hline
$E$ & max$^a$ & error & $O-C^b$ & $N^c$ \\
\hline
0 & 53139.6291 & 0.0003 & 0.0077 & 176 \\
5 & 53140.0364 & 0.0006 & 0.0041 & 44 \\
6 & 53140.1124 & 0.0008 & $-$0.0022 & 40 \\
7 & 53140.1900 & 0.0006 & $-$0.0067 & 38 \\
60 & 53144.5527 & 0.0005 & $-$0.0002 & 186 \\
61 & 53144.6273 & 0.0013 & $-$0.0079 & 186 \\
70 & 53145.3763 & 0.0006 & 0.0014 & 186 \\
71 & 53145.4563 & 0.0005 & $-$0.0008 & 186 \\
72 & 53145.5367 & 0.0006 & $-$0.0026 & 186 \\
94 & 53147.3479 & 0.0084 & 0.0003 & 167 \\
95 & 53147.4352 & 0.0017 & 0.0055 & 181 \\
96 & 53147.5132 & 0.0022 & 0.0013 & 141 \\
\hline
  \multicolumn{5}{l}{$^{a}$ BJD$-$2400000.} \\
  \multicolumn{5}{l}{$^{b}$ Against $max = 2453139.6214 + 0.082193 E$.} \\
  \multicolumn{5}{l}{$^{c}$ Number of points used to determine the maximum.} \\
\end{tabular}
\end{center}
\end{table}

\subsection{AY Lyrae}\label{obj:aylyr}

\begin{figure}
  \begin{center}
    \FigureFile(88mm,70mm){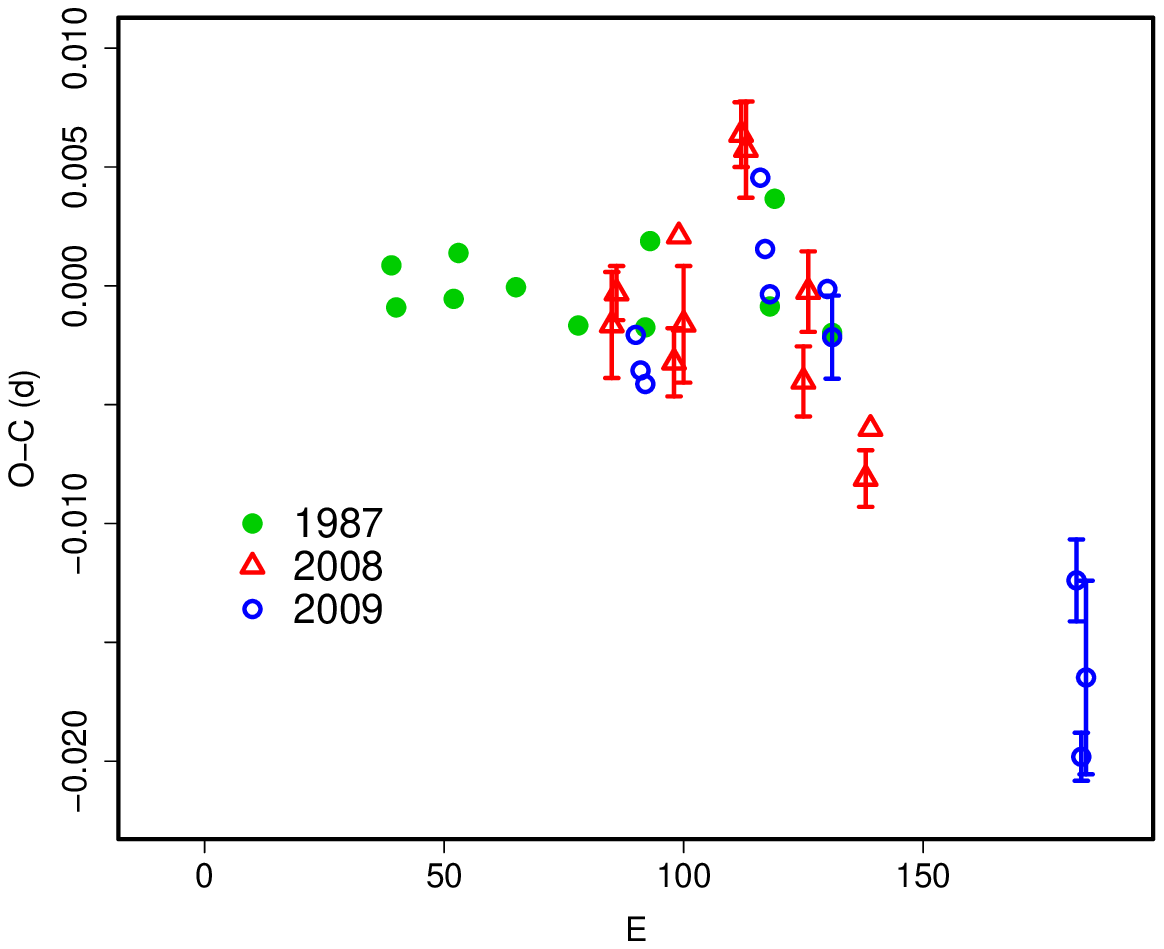}
  \end{center}
  \caption{Comparison of $O-C$ diagrams of AY Lyr between different
  superoutbursts.  A period of 0.07597 d was used to draw this figure.
  Approximate cycle counts ($E$) after the start of the
  superoutburst were used.
  Since the start of the 2009 superoutburst was not well constrained,
  we shifted the $O-C$ diagrams to best fit the others.
  }
  \label{fig:aylyrcomp}
\end{figure}

   Although AY Lyr has long been known a representative SU UMa-type
dwarf nova, little is known about the variation of the superhump period
except for the classical study by \citet{uda88aylyr}.
We observed the 2008 and 2009 superoutbursts
(tables \ref{tab:aylyroc2008}, \ref{tab:aylyroc2009}).
Although we only observed five consecutive nights during the 2008
superoutburst, a transition from stage B to C was apparently recorded.
The early stage of the 2009 superoutburst was likely missed.
The period variation probably reflects a stage B--C transition.
A comparison of $O-C$ diagrams of between different superoutbursts
is given in figure \ref{fig:aylyrcomp}.

\begin{table}
\caption{Superhump maxima of AY Lyr (2008).}\label{tab:aylyroc2008}
\begin{center}
\begin{tabular}{ccccc}
\hline\hline
$E$ & max$^a$ & error & $O-C^b$ & $N^c$ \\
\hline
0 & 54754.9197 & 0.0022 & $-$0.0031 & 87 \\
1 & 54754.9970 & 0.0011 & $-$0.0016 & 140 \\
13 & 54755.9057 & 0.0014 & $-$0.0034 & 77 \\
14 & 54755.9870 & 0.0009 & 0.0020 & 243 \\
15 & 54756.0593 & 0.0025 & $-$0.0017 & 119 \\
27 & 54756.9789 & 0.0014 & 0.0075 & 82 \\
28 & 54757.0542 & 0.0020 & 0.0069 & 67 \\
40 & 54757.9561 & 0.0015 & $-$0.0017 & 58 \\
41 & 54758.0359 & 0.0017 & 0.0022 & 75 \\
53 & 54758.9396 & 0.0012 & $-$0.0046 & 141 \\
54 & 54759.0177 & 0.0008 & $-$0.0024 & 105 \\
\hline
  \multicolumn{5}{l}{$^{a}$ BJD$-$2400000.} \\
  \multicolumn{5}{l}{$^{b}$ Against $max = 2454754.9228 + 0.075876 E$.} \\
  \multicolumn{5}{l}{$^{c}$ Number of points used to determine the maximum.} \\
\end{tabular}
\end{center}
\end{table}

\begin{table}
\caption{Superhump maxima of AY Lyr (2009).}\label{tab:aylyroc2009}
\begin{center}
\begin{tabular}{ccccc}
\hline\hline
$E$ & max$^a$ & error & $O-C^b$ & $N^c$ \\
\hline
0 & 54963.1200 & 0.0006 & $-$0.0038 & 151 \\
1 & 54963.1944 & 0.0005 & $-$0.0052 & 269 \\
2 & 54963.2698 & 0.0008 & $-$0.0056 & 139 \\
26 & 54965.1018 & 0.0005 & 0.0071 & 190 \\
27 & 54965.1748 & 0.0004 & 0.0043 & 256 \\
28 & 54965.2488 & 0.0004 & 0.0026 & 266 \\
40 & 54966.1607 & 0.0004 & 0.0048 & 125 \\
41 & 54966.2346 & 0.0017 & 0.0029 & 75 \\
92 & 54970.0989 & 0.0017 & 0.0013 & 99 \\
93 & 54970.1674 & 0.0010 & $-$0.0060 & 114 \\
94 & 54970.2467 & 0.0041 & $-$0.0025 & 99 \\
\hline
  \multicolumn{5}{l}{$^{a}$ BJD$-$2400000.} \\
  \multicolumn{5}{l}{$^{b}$ Against $max = 2454963.1238 + 0.075802 E$.} \\
  \multicolumn{5}{l}{$^{c}$ Number of points used to determine the maximum.} \\
\end{tabular}
\end{center}
\end{table}

\subsection{DM Lyrae}\label{obj:dmlyr}

   \citet{nog03dmlyr} studied the 1996 and 1997 outbursts and confirmed
the SU UMa-type nature of this object (the times of superhump maxima
measured from the 1997 data are listed in table \ref{tab:dmlyroc1997}).
We further observed the 2002 superoutburst (table \ref{tab:dmlyroc2002}).
As in 1996 and 1997 ones, the 2002 superoutburst was observed during its
later stage.  Although we could not determine $P_{\rm dot}$ for the stage B,
other parameters are given in table \ref{tab:perlist}.

\begin{table}
\caption{Superhump maxima of DM Lyr (1997).}\label{tab:dmlyroc1997}
\begin{center}
\begin{tabular}{ccccc}
\hline\hline
$E$ & max$^a$ & error & $O-C^b$ & $N^c$ \\
\hline
0 & 50509.2862 & 0.0015 & $-$0.0001 & 61 \\
45 & 50512.3171 & 0.0011 & 0.0066 & 63 \\
46 & 50512.3713 & 0.0014 & $-$0.0065 & 33 \\
\hline
  \multicolumn{5}{l}{$^{a}$ BJD$-$2400000.} \\
  \multicolumn{5}{l}{$^{b}$ Against $max = 2450509.2863 + 0.067205 E$.} \\
  \multicolumn{5}{l}{$^{c}$ Number of points used to determine the maximum.} \\
\end{tabular}
\end{center}
\end{table}

\begin{table}
\caption{Superhump maxima of DM Lyr (2002).}\label{tab:dmlyroc2002}
\begin{center}
\begin{tabular}{ccccc}
\hline\hline
$E$ & max$^a$ & error & $O-C^b$ & $N^c$ \\
\hline
0 & 52580.0178 & 0.0073 & $-$0.0019 & 100 \\
58 & 52583.9153 & 0.0027 & $-$0.0001 & 101 \\
59 & 52583.9862 & 0.0010 & 0.0037 & 103 \\
104 & 52587.0065 & 0.0097 & 0.0015 & 41 \\
119 & 52588.0084 & 0.0108 & $-$0.0041 & 58 \\
134 & 52589.0209 & 0.0068 & 0.0009 & 60 \\
\hline
  \multicolumn{5}{l}{$^{a}$ BJD$-$2400000.} \\
  \multicolumn{5}{l}{$^{b}$ Against $max = 2452580.0197 + 0.067166 E$.} \\
  \multicolumn{5}{l}{$^{c}$ Number of points used to determine the maximum.} \\
\end{tabular}
\end{center}
\end{table}

\subsection{V344 Lyrae}\label{obj:v344lyr}

   \citet{kat93v344lyr} reported on the 1993 superoutburst.
We determined superhump maxima from these observations
(table \ref{tab:v344lyroc1993}).  The resultant $P_{\rm dot}$ was
$-7.1(4.3) \times 10^{-5}$.  Since this object has one of the longest
$P_{\rm orb}$, more complicated period variation may be expected
as in MN Dra and UV Gem.  Future better observations are needed
to test this possibility.

\begin{table}
\caption{Superhump maxima of V344 Lyr (1993).}\label{tab:v344lyroc1993}
\begin{center}
\begin{tabular}{ccccc}
\hline\hline
$E$ & max$^a$ & error & $O-C^b$ & $N^c$ \\
\hline
0 & 49133.1065 & 0.0014 & $-$0.0042 & 49 \\
1 & 49133.2004 & 0.0011 & $-$0.0015 & 48 \\
2 & 49133.2949 & 0.0019 & 0.0016 & 29 \\
12 & 49134.2104 & 0.0024 & 0.0035 & 34 \\
34 & 49136.2187 & 0.0016 & 0.0020 & 47 \\
78 & 49140.2348 & 0.0046 & $-$0.0014 & 23 \\
\hline
  \multicolumn{5}{l}{$^{a}$ BJD$-$2400000.} \\
  \multicolumn{5}{l}{$^{b}$ Against $max = 2449133.1106 + 0.091354 E$.} \\
  \multicolumn{5}{l}{$^{c}$ Number of points used to determine the maximum.} \\
\end{tabular}
\end{center}
\end{table}

\subsection{V358 Lyrae}\label{obj:v358lyr}

   Although the object was originally discovered as a nova
\citep{hof67an289205}, \citet{ric86v358lyr} suggested that it is
a WZ Sge-type dwarf nova based on its faintness and the similarity
in the light curve with that of WZ Sge.  \citet{ant04v358lyr}
pointed out that the reported maximum in \citet{ric86v358lyr} referred
to a plate defect and presented the correct identification.
The maximum recorded photographic magnitude was 16.42.

   J. Shears detected a new outburst on 2008 November 22 at an unfiltered
CCD magnitude of 16.26 (vsnet-outburst 9714).  The object experienced
a ``dip''-like fading characteristic to (type-A) WZ Sge-type superoutbursts
and exhibited a long-lasting second plateau stage.

\begin{figure}
  \begin{center}
    \FigureFile(88mm,130mm){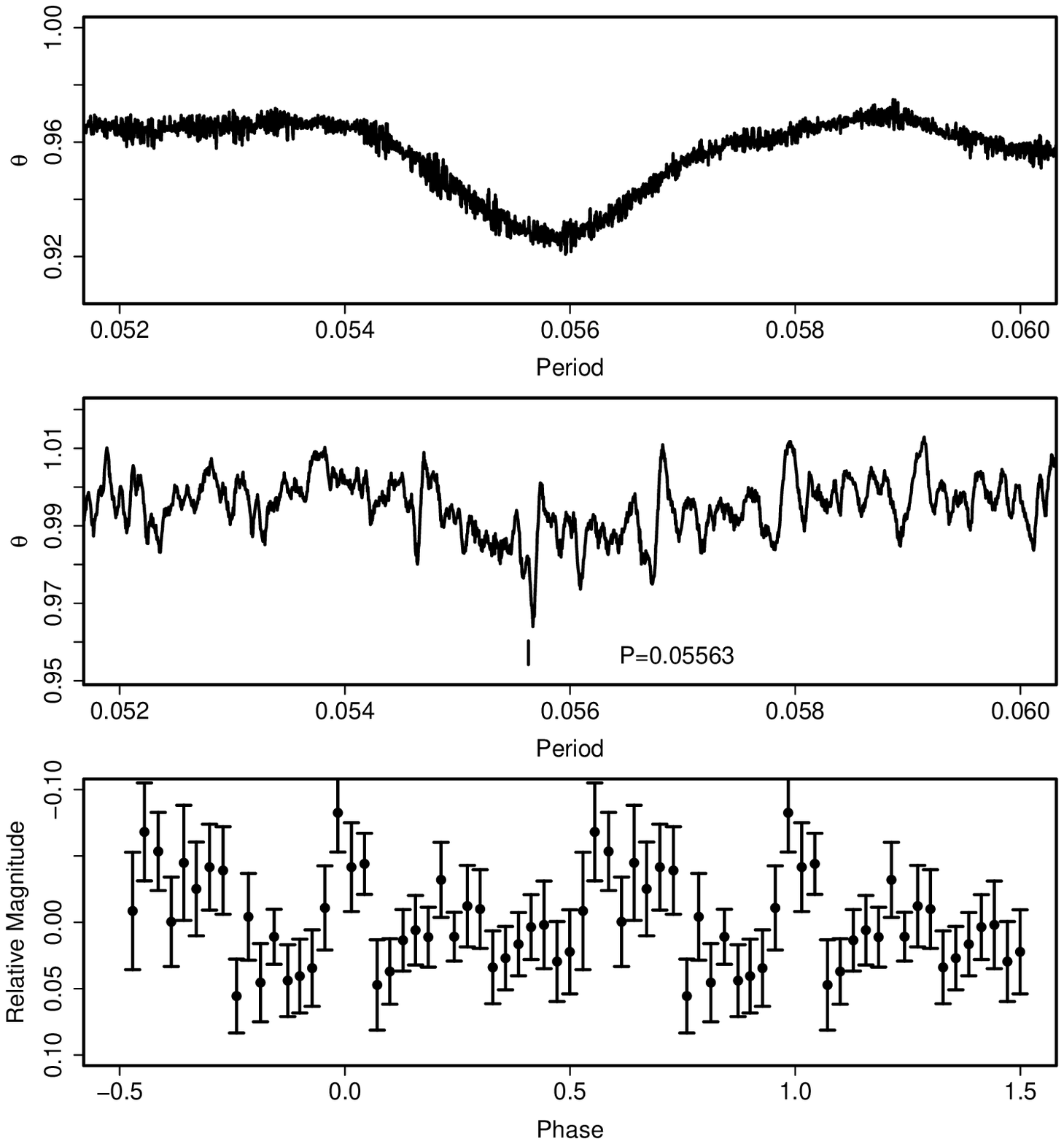}
  \end{center}
  \caption{Superhumps in V358 Lyr (2008). (Upper): PDM analysis of the
  interval BJD 2454793.8--2454794.3.  (Middle): PDM analysis of the
  interval BJD 2454793.8--2454797.0.  (Lower): Phase-averaged profile.}
  \label{fig:v358lyrshpdm}
\end{figure}

   Due to the low amplitudes of variations and faintness of the object,
we mainly focus on the variation before the dip.  Using the best segments
of observations, we obtained a $P_{\rm SH}$ of 0.05563(3) with the
PDM method (figure \ref{fig:v358lyrshpdm}).  The profile of variation
appeared doubly humped.  Although the profile resembles those of
early superhumps, we identified these variations as ordinary superhumps
because these variations were observed $\sim$ 10 d before the dip,
at an epoch when all well-observed WZ Sge-type dwarf novae exhibited
ordinary superhumps.  The low-amplitude, double-wave modulations may
have been a result of temporary reduction of amplitudes of superhumps
frequently seen in many systems in the middle-to-late stage of
a superoutburst plateau (see e.g. \cite{kat03hodel}).
Although we measured times of superhump maxima 
(table \ref{tab:v358lyroc2008}), the quality was not sufficient
because of this complexity in the profile.
\citet{she09v358lyr} reported possible detection of small-scale periodic
signals including a candidate period of 0.05556(32) d.

\begin{figure}
  \begin{center}
    \FigureFile(88mm,110mm){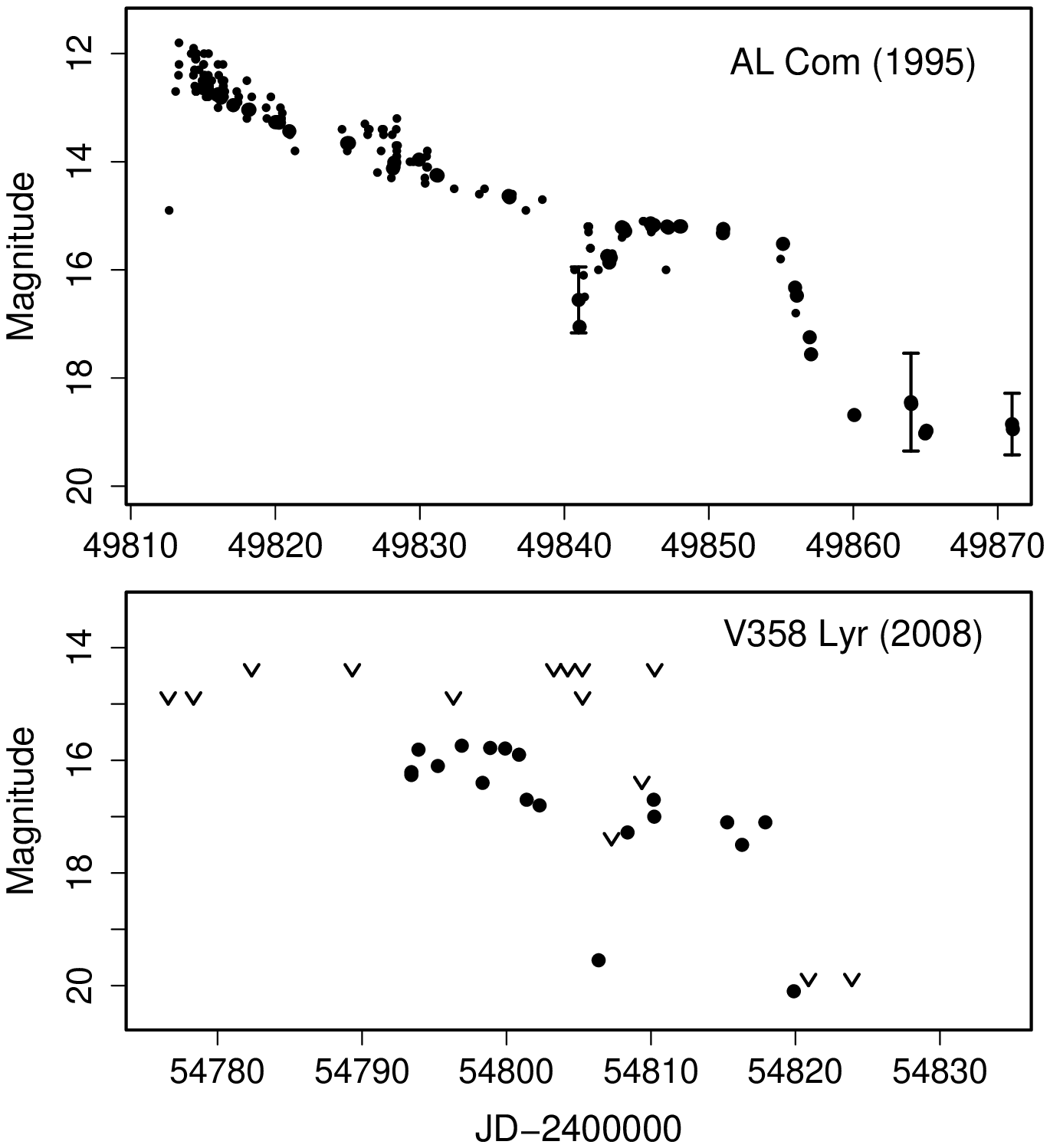}
  \end{center}
  \caption{Comparison of light curves of AL Com and V358 Lyr.
  (Upper) AL Com in 1995.  The data are from \citet{kat96alcom}.
  (Lower) V358 Lyr.  The ``v'' marks indicate upper limits.}
  \label{fig:v358lyrlc}
\end{figure}

   The overall light curve of the superoutburst bears strong similarity
to that of AL Com in 1995 (figure \ref{fig:v358lyrlc}).  The earlier
stage of the outburst, potentially with early superhumps, may have
been unfortunately missed below the detection limit of visual observations.

\begin{table}
\caption{Superhump maxima of V358 Lyr (2008).}\label{tab:v358lyroc2008}
\begin{center}
\begin{tabular}{ccccc}
\hline\hline
$E$ & max$^a$ & error & $O-C^b$ & $N^c$ \\
\hline
0 & 54793.8615 & 0.0049 & 0.0039 & 55 \\
1 & 54793.9053 & 0.0226 & $-$0.0080 & 130 \\
13 & 54794.5960 & 0.0030 & 0.0133 & 13 \\
26 & 54795.3027 & 0.0036 & $-$0.0051 & 48 \\
27 & 54795.3578 & 0.0030 & $-$0.0057 & 47 \\
48 & 54796.5339 & 0.0014 & $-$0.0010 & 14 \\
55 & 54796.9275 & 0.0043 & 0.0022 & 44 \\
109 & 54799.9377 & 0.0043 & 0.0004 & 30 \\
\hline
  \multicolumn{5}{l}{$^{a}$ BJD$-$2400000.} \\
  \multicolumn{5}{l}{$^{b}$ Against $max = 2454793.8576 + 0.055777 E$.} \\
  \multicolumn{5}{l}{$^{c}$ Number of points used to determine the maximum.} \\
\end{tabular}
\end{center}
\end{table}

\subsection{V419 Lyrae}\label{obj:v419lyr}

   \citet{nog98gxcasv419lyr} reported the detection of superhumps in
this object and proposed candidate periods.  Although their observations
were not long enough to discriminate the possibilities, the long superhump
period already made V419 Lyr an outstanding object.
We observed the 1999 superoutburst, and obtained the following superhump
maxima and first identified the correct superhump period (table
\ref{tab:v419lyroc1999}).  The superhump period apparently largely
varied between $E = 0$ and $E = 3$.  Excluding the point of $E = 11$
(observation of the maximum somewhat affected by thin clouds),
and $E \le 3$ epochs, we obtained $P_{\rm dot}$ of
$-32.4(2.4) \times 10^{-5}$.

  \citet{rut07v419lyr} obtained
$P_{\rm dot}$ = $-24.8(2.2) \times 10^{-5}$ during the 2006 superoutburst.
We analyzed the available data (from the AAVSO database and Dubovsky's data)
and combined with \citet{rut07v419lyr} after adding a systematic correction
of 0.0026 d to \citet{rut07v419lyr} and removing maxima of Boyd's
observations, which were included in our own analysis
(table \ref{tab:v419lyroc2006})

\begin{figure}
  \begin{center}
    \FigureFile(88mm,70mm){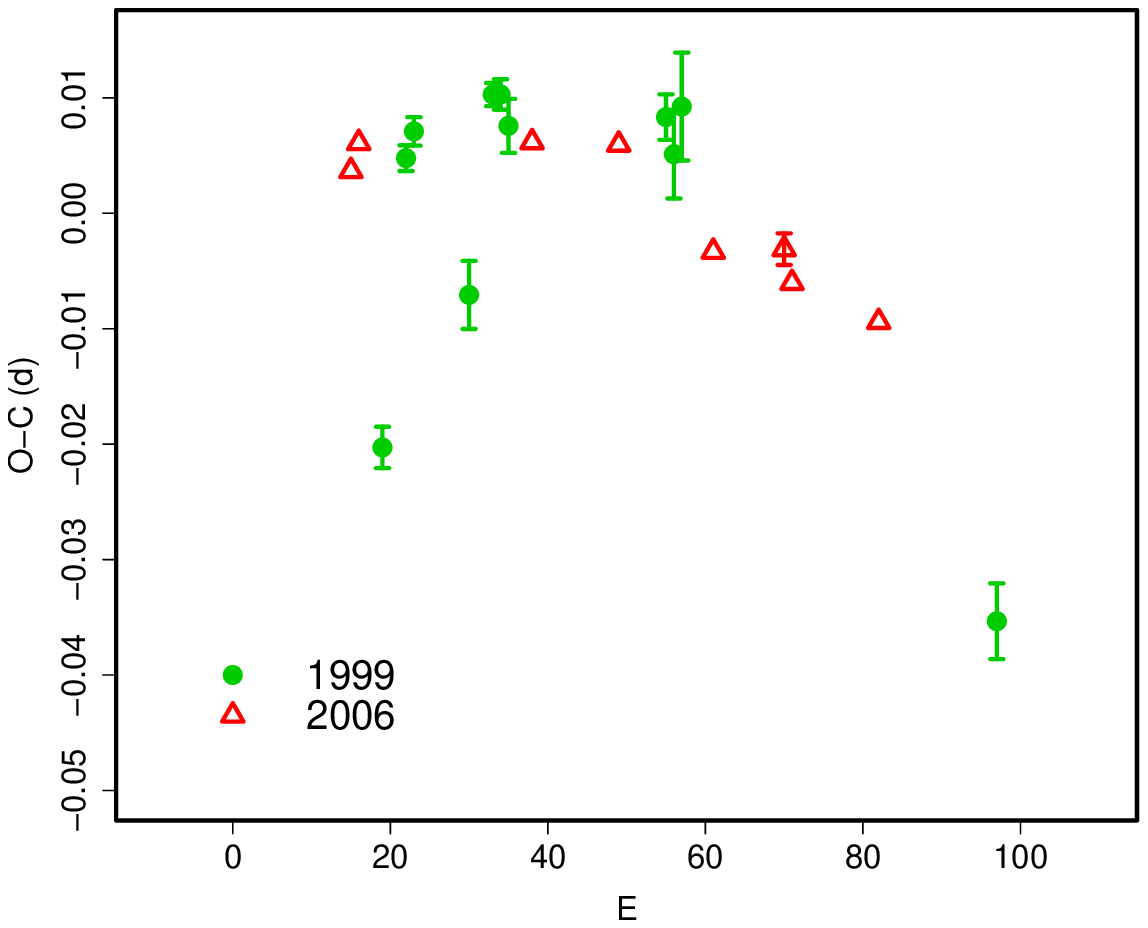}
  \end{center}
  \caption{Comparison of $O-C$ diagrams of V419 Lyr between different
  superoutbursts.  A period of 0.09005 d was used to draw this figure.
  Approximate cycle counts ($E$) after the start of the
  superoutburst were used.
  }
  \label{fig:v419lyrcomp}
\end{figure}

   A comparison of $O-C$ diagrams between different superoutbursts
is given in figure \ref{fig:v419lyrcomp}.

   This long-period system resembles UV Gem, MN Dra and NY Ser in its strongly
negative superhump derivative.  It would be notable that V419 Lyr
shows frequent normal outbursts (intervals 9--12 d), which is also
reminiscent of the behavior in UV Gem \citep{kat01uvgemfsandaspsc}
and NY Ser \citep{iid95nyser}.  Very long-$P_{\rm SH}$ systems with
frequent normal outbursts may be associated with strongly negative
$P_{\rm dot}$ (see subsection \ref{sec:longp}).

\begin{table}
\caption{Superhump maxima of V419 Lyr (1999).}\label{tab:v419lyroc1999}
\begin{center}

\end{center}
\end{table}

\begin{table}
\caption{Superhump maxima of V419 Lyr (2006).}\label{tab:v419lyroc2006}
\begin{center}
%
\end{center}
\end{table}

\subsection{V585 Lyrae}\label{obj:v585lyr}

   V585 Lyr was discovered by \citet{kry01v585lyrv587lyr}.
An extensive photometric campaign was undertaken during the 2003
superoutburst.  The times of superhump maxima during this superoutburst
are listed in table \ref{tab:v585lyroc2003}.
The interval $32 \le E \le 150$ (stage B) showed
a positive $P_{\rm dot}$ of $+10.7(1.2) \times 10^{-5}$, then followed
by the emergence of a shorter period (stage C) and a regrowth
of superhumps, typical behavior for a short-period system
(cf. figure \ref{fig:ocsamp}).

\begin{table}
\caption{Superhump maxima of V585 Lyr (2003).}\label{tab:v585lyroc2003}
\begin{center}

\end{center}
\end{table}

\addtocounter{table}{-1}
\begin{table}
\caption{Superhump maxima of V585 Lyr (2003) (continued).}
\begin{center}
%
\end{center}
\end{table}

\subsection{AD Mensae}\label{obj:admen}

   AD Men was discovered as a variable star in the region of the
Large Magellanic Cloud.  The GCVS \citep{GCVS} listed the object
as an SS Cyg-type dwarf nova with an outburst cycle length of $\sim$ 30 d.

   The object underwent a bright outburst in 2003 March at a visual
magnitude of 14.0.  The existence of superhumps was inconclusive
during this outburst.\footnote{
 $<$http://vsnet.kusastro.kyoto-u.ac.jp/vsnet/DNe/admen.html$>$.
}

\begin{figure}
  \begin{center}
    \FigureFile(88mm,110mm){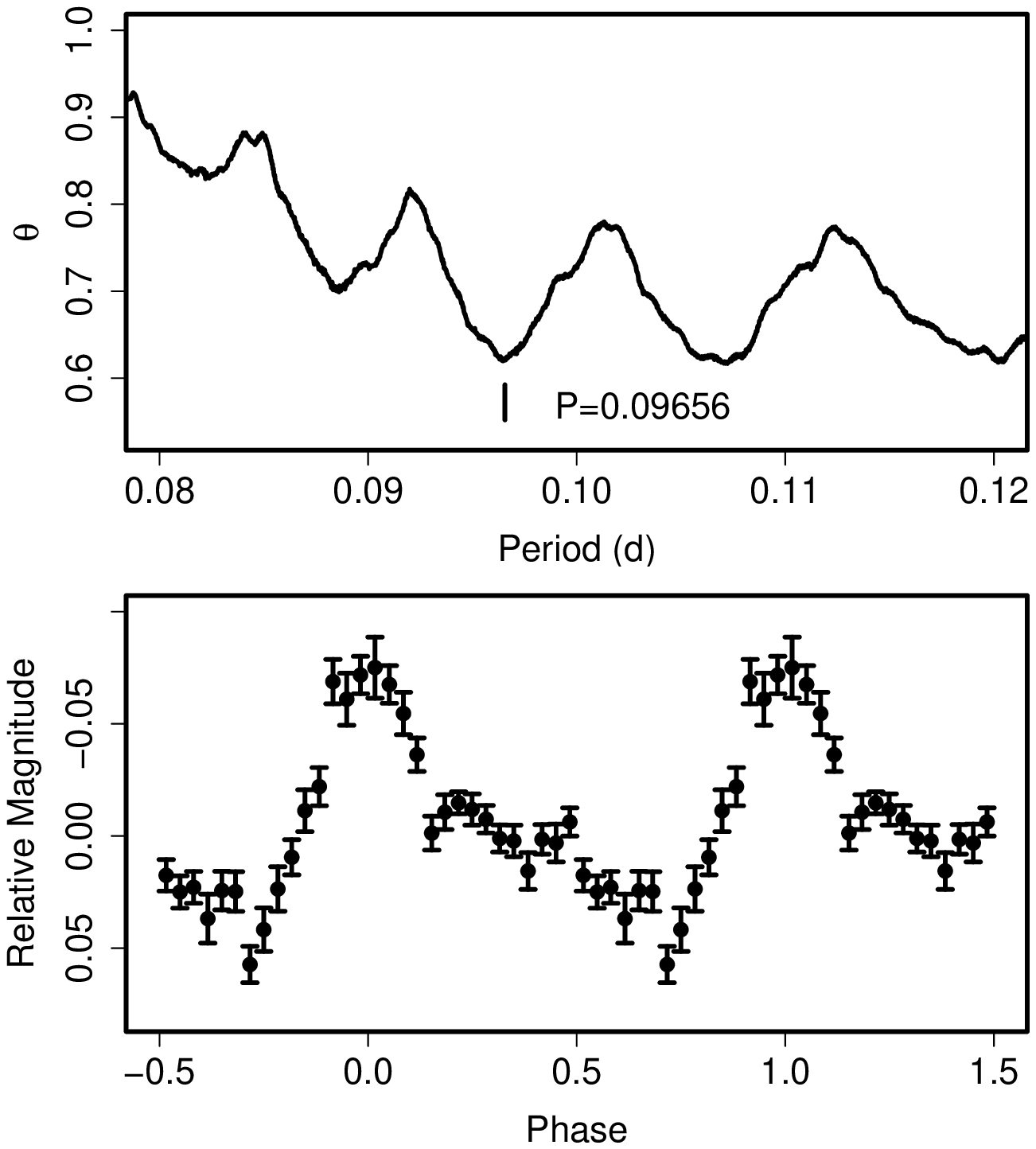}
  \end{center}
  \caption{Superhumps in AD Men (2004). (Upper): PDM analysis.
     (Lower): Phase-averaged profile.}
  \label{fig:admenshpdm}
\end{figure}

   The object underwent another bright outburst in 2004 March.
The existence of superhumps was confirmed during this outburst,
establishing the SU UMa-type nature of this object.
Although a single superhump maximum of BJD 2453090.3137(7) was
obtained, a PDM analysis and the examination of the single-night
observation yielded the most likely period of 0.0966(2) d
(figure \ref{fig:admenshpdm}).  The object is an SU UMa-type dwarf nova
likely in the period gap.  The present $P_{\rm SH}$ is consistent
with a photometric measurement of $P_{\rm orb}$ = 0.0917(10) d
\citep{sch06admen}.  The fractional superhump excess is $\sim$ 5 \%.

\subsection{FQ Monocerotis}\label{sec:fqmon}\label{obj:fqmon}

   FQ Mon, originally classified as a possible Mira-type variable
\citep{GCVS}, was suspected to be a CV (vsnet-chat 3063,3066).
The first known outburst since the discovery was recorded in 2004
(vsnet-alert 8048).
We observed the 2004, 2006, 2007--2008 superoutbursts.

   The 2004 superoutburst was relatively well observed
(table \ref{tab:fqmonoc2004}).  The $O-C$ diagram was composed of
a typical stage B--C transition.  The $P_{\rm dot}$ for the
stage B was $+9.2(2.4) \times 10^{-5}$ ($E \le 111$).

The later part of the 2006 superoutburst was observed
(table \ref{tab:fqmonoc2006}).  Compared to other superoutbursts,
the fairly constant period after $E=51$ likely corresponds to $P_2$.

\begin{figure}
  \begin{center}
    \FigureFile(88mm,70mm){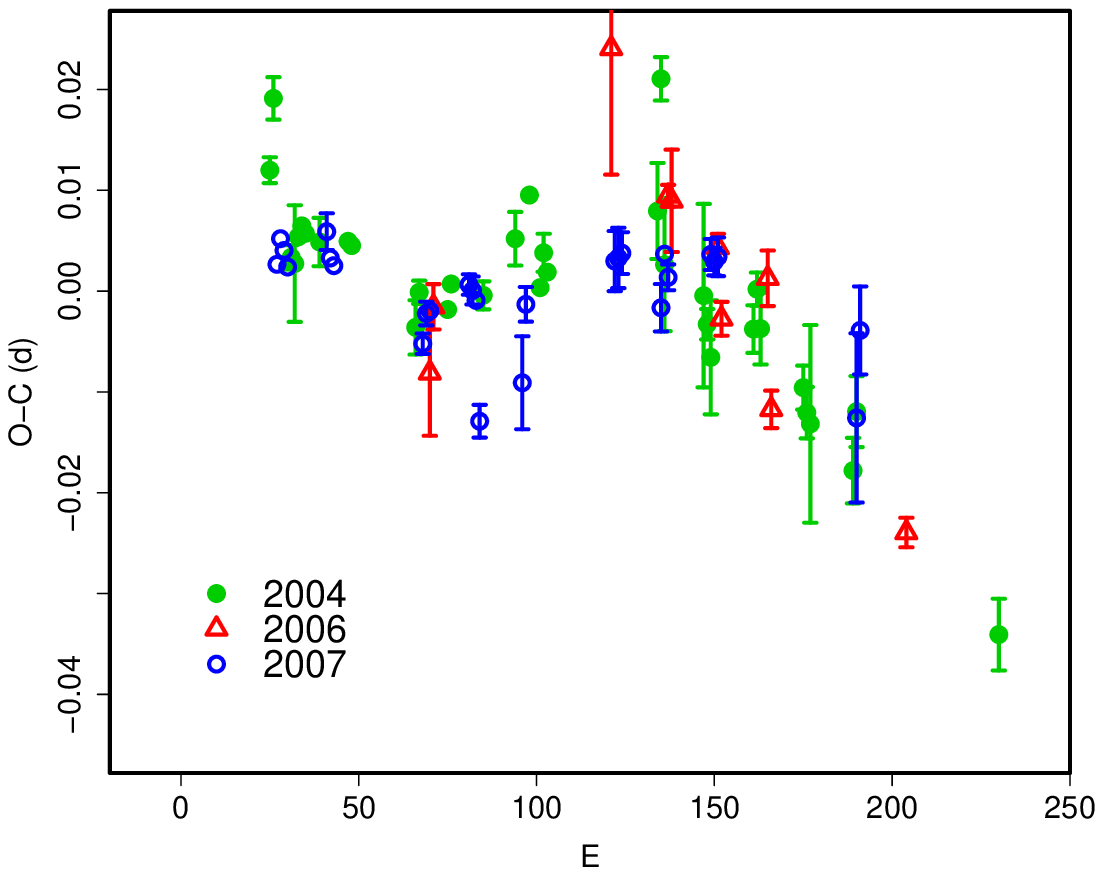}
  \end{center}
  \caption{Comparison of $O-C$ diagrams of FQ Mon between different
  superoutbursts.  A period of 0.07335 d was used to draw this figure.
  Approximate cycle counts ($E$) after the start of the
  superoutburst were used.
  }
  \label{fig:fqmoncomp}
\end{figure}

   A photometric campaign was undertaken during the 2007--2008
superoutburst.  The times of superhump maxima are listed in table
\ref{tab:fqmonoc2007}.  The object reached the maximum
light around $E = 40$.  Although superhumps were still prominent
before this epoch, the period was significantly shorter than in the
later epoch.  The combined $O-C$ diagram (figure \ref{fig:fqmoncomp})
suggests that stages B and C were recorded during this superoutburst.
The $P_{\rm dot}$ for the stage B was $+5.4(1.3) \times 10^{-5}$
($E \le 124$).

   The overall behavior resembles the $O-C$ variation in TT Boo
\citep{ole04ttboo}.  These two objects have common
properties of a relatively long superhump period (0.07--0.08 d),
unusually long superoutburst ($\ge$ 15 d) and relatively few normal
outbursts.

\begin{table}
\caption{Superhump maxima of FQ Mon (2004).}\label{tab:fqmonoc2004}
\begin{center}

\end{center}
\end{table}

\begin{table}
\caption{Superhump maxima of FQ Mon (2006).}\label{tab:fqmonoc2006}
\begin{center}
%
\end{center}
\end{table}

\begin{table}
\caption{Superhump maxima of FQ Mon (2007--2008).}\label{tab:fqmonoc2007}
\begin{center}
%
\end{center}
\end{table}

\subsection{AB Normae}\label{obj:abnor}

   \citet{kat04nsv10934mmscoabnorcal86} reported the detection of
superhumps in AB Nor during its 2002 superoutburst.   Due to the
observational gap and apparent period variation, the identification
of the correct $P_{\rm SH}$ was rather ambiguous.  Based on the
improved knowledge of period variations in long-$P_{\rm SH}$ systems,
we succeeded in identifying a more likely $P_{\rm SH}$
(table \ref{tab:abnoroc2002}).  For $E \le 16$, the system showed
the stage A period evolution associated with the growth of
superhumps.  The mean period and $P_{\rm dot}$'s were
0.07962(3) d and $-8.1(2.7) \times 10^{-5}$, respectively
($15 \le E \le 142$) or 0.07955(3) d and $-6.1(5.2) \times 10^{-5}$,
respectively ($37 \le E \le 142$).

\begin{table}
\caption{Superhump maxima of AB Nor (2002).}\label{tab:abnoroc2002}
\begin{center}
\begin{tabular}{ccccc}
\hline\hline
$E$ & max$^a$ & error & $O-C^b$ & $N^c$ \\
\hline
0 & 52518.9807 & 0.0013 & $-$0.0189 & 36 \\
4 & 52519.3045 & 0.0020 & $-$0.0141 & 86 \\
15 & 52520.2092 & 0.0007 & 0.0134 & 41 \\
16 & 52520.2844 & 0.0015 & 0.0089 & 31 \\
37 & 52521.9670 & 0.0003 & 0.0167 & 40 \\
124 & 52528.8911 & 0.0022 & 0.0027 & 18 \\
141 & 52530.2420 & 0.0019 & $-$0.0022 & 87 \\
142 & 52530.3175 & 0.0061 & $-$0.0065 & 49 \\
\hline
  \multicolumn{5}{l}{$^{a}$ BJD$-$2400000.} \\
  \multicolumn{5}{l}{$^{b}$ Against $max = 2452518.9996 + 0.079749 E$.} \\
  \multicolumn{5}{l}{$^{c}$ Number of points used to determine the maximum.} \\
\end{tabular}
\end{center}
\end{table}

\subsection{DT Octantis}\label{sec:dtoct}\label{obj:dtoct}

\begin{figure}
  \begin{center}
    \FigureFile(88mm,70mm){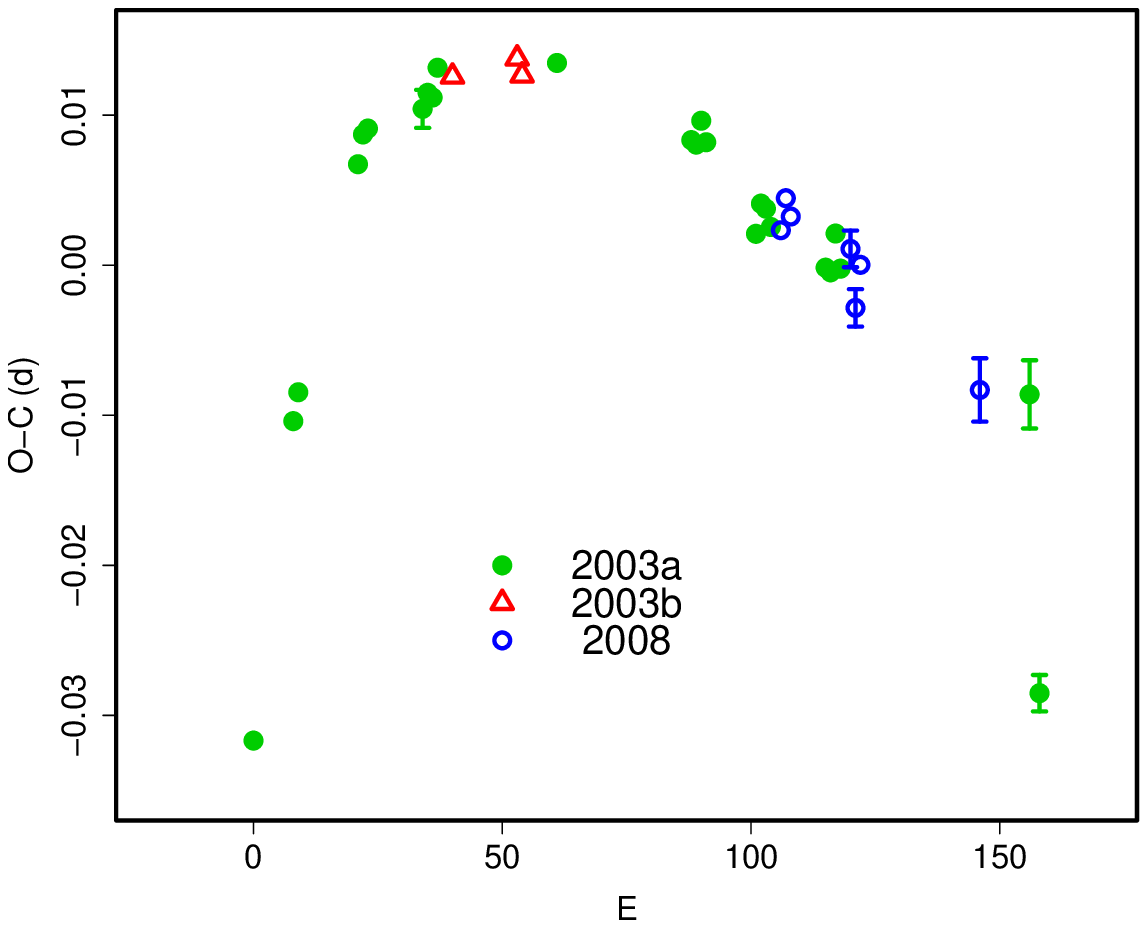}
  \end{center}
  \caption{Comparison of $O-C$ diagrams of DT Oct between different
  superoutbursts.  A period of 0.07485 d was used to draw this figure.
  Approximate cycle counts ($E$) after the start of the
  superoutburst were used.
  }
  \label{fig:dtoctcomp}
\end{figure}

   \citet{kat04nsv10934mmscoabnorcal86} reported the detection of
superhumps in DT Oct = NSV 10934 during its 2003 January superoutburst.
Table \ref{tab:dtoctoc2003a} gives an upgraded list of superhump maxima.
The epochs $156 \le E \le 158$ correspond to the post-superoutburst
stage.  There was a hint of double-wave modulation at this stage
and was possibly from classical ``late superhumps''.
Disregarding this stage and the stage A ($E \le 9$), the global
$P_{\rm dot}$ corresponds to $-9.0(1.1) \times 10^{-5}$.
The times of superhump maxima during the 2003 November superoutburst
and the 2008 superoutburst are also given for a supplementary purpose
(tables \ref{tab:dtoctoc2003b}, \ref{tab:dtoctoc2008}).  The latter
superoutburst probably recorded the stage C superhumps
(see figure \ref{fig:dtoctcomp}).

\begin{table}
\caption{Superhump maxima of DT Oct (2003a).}\label{tab:dtoctoc2003a}
\begin{center}

\end{center}
\end{table}

\begin{table}
\caption{Superhump maxima of DT Oct (2003b).}\label{tab:dtoctoc2003b}
\begin{center}
%
\end{center}
\end{table}

\begin{table}
\caption{Superhump maxima of DT Oct (2008).}\label{tab:dtoctoc2008}
\begin{center}
%
\end{center}
\end{table}

\subsection{V699 Ophiuchi}\label{sec:v699oph}\label{obj:v699oph}

   Until very recently, the nature of V699 Oph remained controversial.
The object was originally discovered as a possible dwarf nova.
\citet{wal58CVchart} presented a finding chart, but later spectroscopic
studies have shown that the marked object is a normal star
(\cite{zwi96CVspec}; \cite{liu99CVspec2}; Kato et al., unpublished).

   On 1999 April 16, A. Pearce discovered an outburst of this object
(vsnet-alert 2877).  Astrometry and photometry of the outbursting object
indicated that the true V699 Oph is an unresolved companion to a 16-th
magnitude star (vsnet-alert 2878, vsnet-chat 1810, 1868).

\begin{figure}
  \begin{center}
    \FigureFile(88mm,110mm){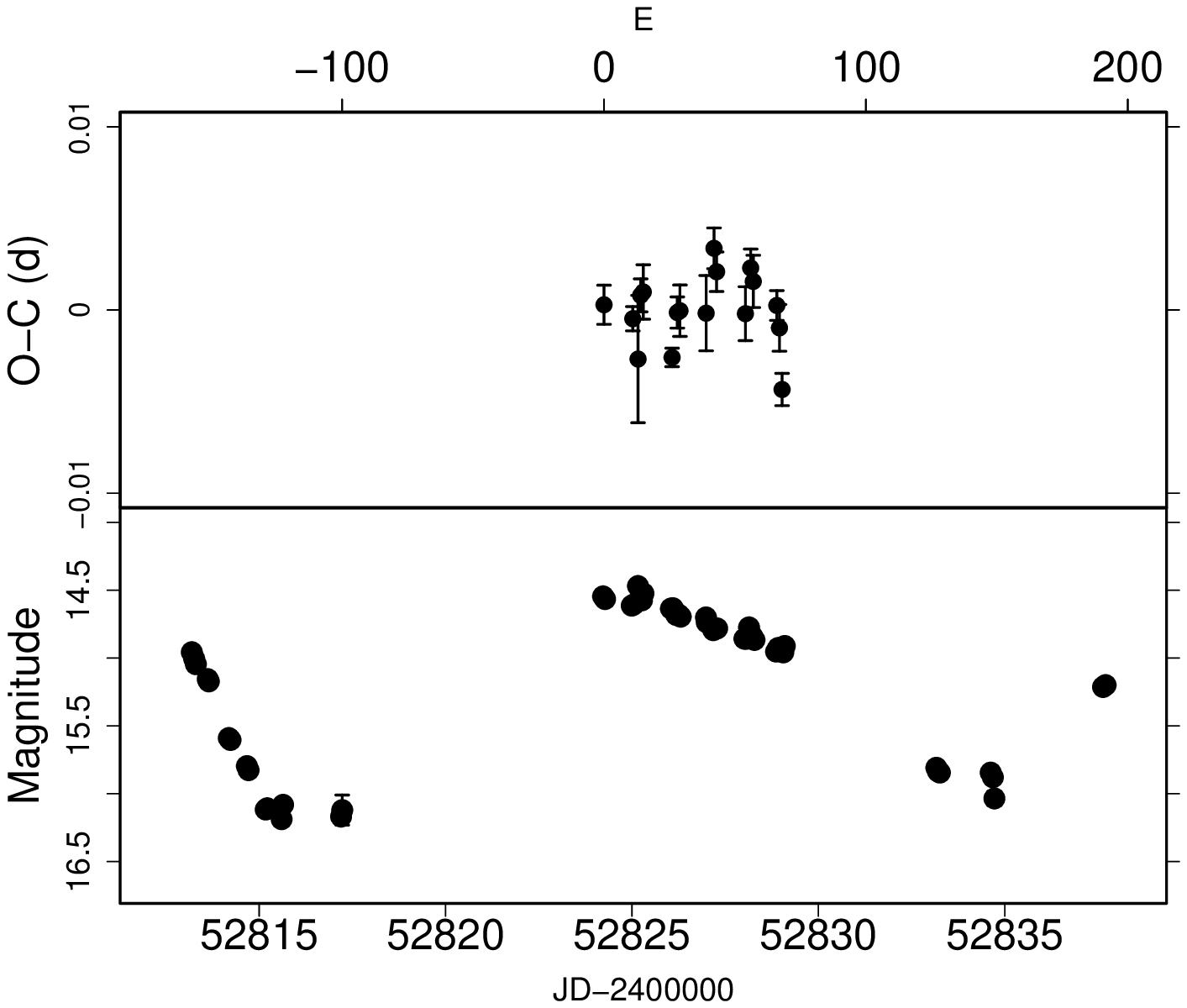}
  \end{center}
  \caption{$O-C$ of superhumps V699 Oph (2003).
  (Upper): $O-C$ diagram.
  (Lower): Light curve.  The superoutburst was preceded by a precursor
  and followed by a rebrightening.  The flat bottom at magnitude
  $\sim$ 16.1 was a result of an unresolved companion.
  }
  \label{fig:v699oph2003oc}
\end{figure}

\begin{figure}
  \begin{center}
    \FigureFile(88mm,110mm){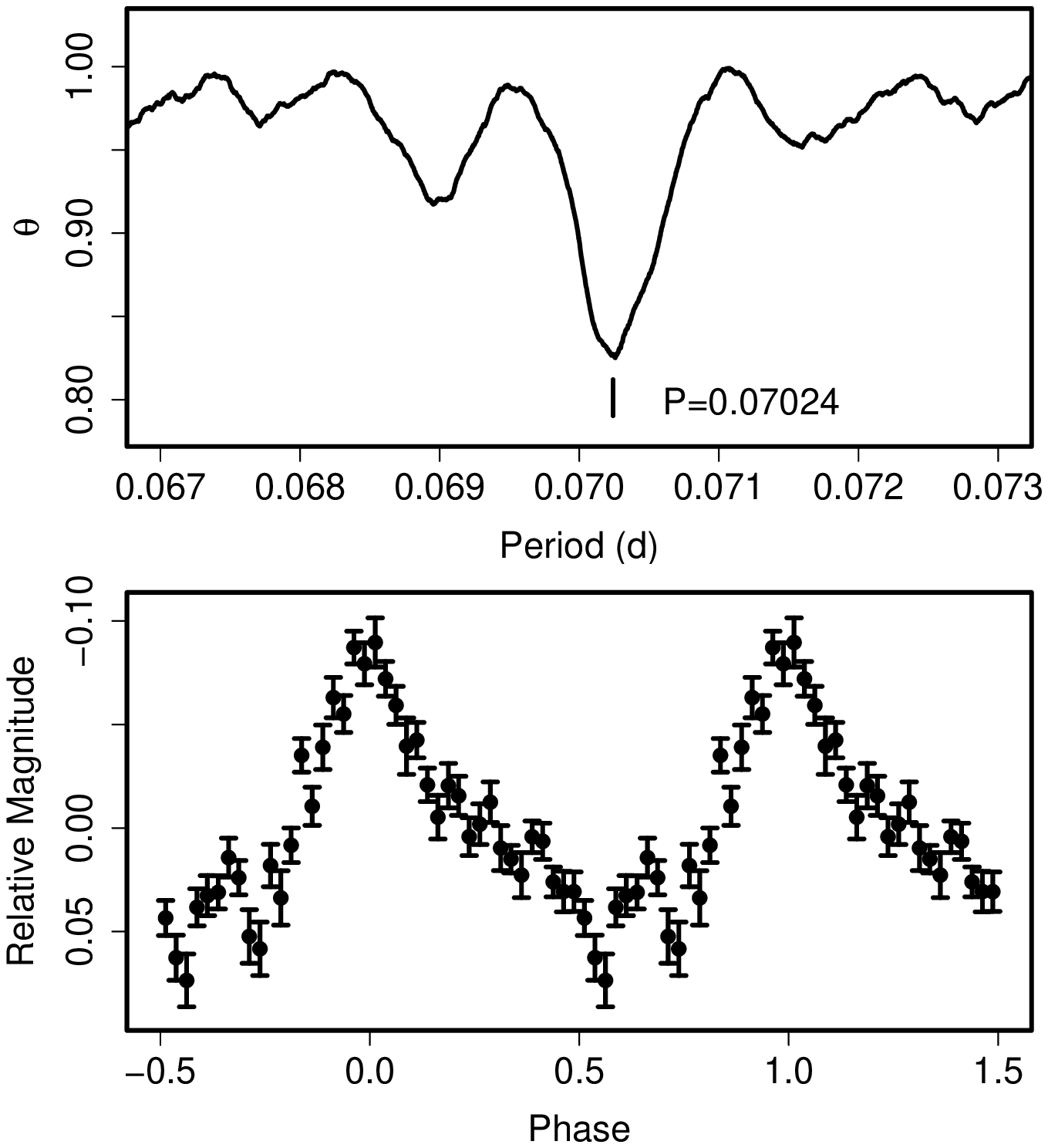}
  \end{center}
  \caption{Superhumps in V699 Oph (2003). (Upper): PDM analysis.
     (Lower): Phase-averaged profile.}
  \label{fig:v699ophshpdm}
\end{figure}

   The 2003 superoutburst was noteworthy in that it was preceded by
a precursor outburst (vsnet-alert 7768, 7795) 11 d before the onset
of the superoutburst and followed by a rebrightening
(figure \ref{fig:v699oph2003oc}).  The mean superhump period
with the PDM method was 0.070242(12) d (figure \ref{fig:v699ophshpdm}).
The superhump maxima during the plateau stage are listed in table
\ref{tab:v699ophoc2003}.  There was likely a stage B--C transition around
$E=43$.  The $P_{\rm dot}$ during the stage B was $+14.2(7.7) \times 10^{-5}$.
There was marginal evidence for $\sim$ 0.02 mag modulation with
a period of 0.0689(2) d during the first two days of the precursor,
which might be related to orbital modulations.

   The 2008 superoutburst (table \ref{tab:v699ophoc2008}) lacked good
coverage in the middle of the superoutburst.  The maxima with $E \ge 87$
were obtained during the late-stage decline of the superoutburst and
most likely correspond to the stage C.
Using all the superhump maxima, we obtained a global
$P_{\rm dot}$ of $-6.9(1.4) \times 10^{-5}$.  The $P_{\rm dot}$ before
the supposed stage B--C transition should have been closer to
zero than this global value.

\begin{table}
\caption{Superhump maxima of V699 Oph (2003).}\label{tab:v699ophoc2003}
\begin{center}
\begin{tabular}{ccccc}
\hline\hline
$E$ & max$^a$ & error & $O-C^b$ & $N^c$ \\
\hline
0 & 52824.2510 & 0.0011 & 0.0003 & 38 \\
11 & 52825.0233 & 0.0007 & $-$0.0005 & 389 \\
13 & 52825.1616 & 0.0035 & $-$0.0027 & 20 \\
14 & 52825.2354 & 0.0009 & 0.0008 & 39 \\
15 & 52825.3058 & 0.0015 & 0.0010 & 27 \\
26 & 52826.0753 & 0.0005 & $-$0.0026 & 121 \\
28 & 52826.2183 & 0.0009 & $-$0.0001 & 39 \\
29 & 52826.2887 & 0.0014 & $-$0.0000 & 39 \\
39 & 52826.9913 & 0.0021 & $-$0.0002 & 93 \\
42 & 52827.2056 & 0.0011 & 0.0034 & 39 \\
43 & 52827.2746 & 0.0011 & 0.0021 & 38 \\
54 & 52828.0453 & 0.0015 & $-$0.0002 & 220 \\
56 & 52828.1884 & 0.0010 & 0.0023 & 37 \\
57 & 52828.2579 & 0.0014 & 0.0016 & 38 \\
66 & 52828.8891 & 0.0008 & 0.0002 & 150 \\
67 & 52828.9581 & 0.0013 & $-$0.0010 & 192 \\
68 & 52829.0250 & 0.0009 & $-$0.0043 & 291 \\
\hline
  \multicolumn{5}{l}{$^{a}$ BJD$-$2400000.} \\
  \multicolumn{5}{l}{$^{b}$ Against $max = 2452824.2508 + 0.070274 E$.} \\
  \multicolumn{5}{l}{$^{c}$ Number of points used to determine the maximum.} \\
\end{tabular}
\end{center}
\end{table}

\begin{table}
\caption{Superhump maxima of V699 Oph (2008).}\label{tab:v699ophoc2008}
\begin{center}
\begin{tabular}{ccccc}
\hline\hline
$E$ & max$^a$ & error & $O-C^b$ & $N^c$ \\
\hline
0 & 54618.1103 & 0.0026 & $-$0.0063 & 72 \\
1 & 54618.1851 & 0.0006 & $-$0.0017 & 141 \\
14 & 54619.1009 & 0.0007 & 0.0030 & 131 \\
15 & 54619.1697 & 0.0008 & 0.0017 & 121 \\
16 & 54619.2392 & 0.0007 & 0.0011 & 115 \\
17 & 54619.3097 & 0.0010 & 0.0015 & 128 \\
87 & 54624.2193 & 0.0019 & 0.0047 & 100 \\
128 & 54627.0849 & 0.0039 & $-$0.0034 & 144 \\
129 & 54627.1577 & 0.0063 & $-$0.0007 & 61 \\
\hline
  \multicolumn{5}{l}{$^{a}$ BJD$-$2400000.} \\
  \multicolumn{5}{l}{$^{b}$ Against $max = 2454618.1166 + 0.070091 E$.} \\
  \multicolumn{5}{l}{$^{c}$ Number of points used to determine the maximum.} \\
\end{tabular}
\end{center}
\end{table}

\subsection{V2051 Ophiuchi}\label{obj:v2051oph}

   V2051 Oph is an eclipsing dwarf nova whose SU UMa-type nature was
established by \citet{kiy1998v2051oph}.  \citet{pat03suumas}
observed the 1999 superoutburst and reported a representative superhump
period.

   We observed the 1999, 2003 and 2009 superoutburst.
The times of superhump maxima
(tables \ref{tab:v2051ophoc1999}, \ref{tab:v2051ophoc2003},
\ref{tab:v2051ophoc2009}) were obtained after removing observations
within 0.07 $P_{\rm orb}$ of eclipses.
The 1999 observation covered the later part of a superoutburst
and 2003 mostly covered the earlier part.
We could not reliably determine superhump maxima during the
later course of the 2003 superoutburst because of the complex superhump
profile and the presence of eclipses and the orbital signature.
The 1999 $O-C$ diagram clearly showed a shift to a shorter superhump
period (stage B to C) associated with a regrowth of superhumps.
Using the $0 \le E \le 113$ segment,
we obtained $P_{\rm dot}$ = $+2.9(2.9) \times 10^{-5}$.
Using the entire data ($0 \le E \le 48$) of the 2003 superoutburst,
we obtained $P_{\rm dot}$ = $-44.8(15.1) \times 10^{-5}$.
Such a large decrease in the period
was most likely due to the early development of the superhump period
from a longer period (stage A to B).
Using the interval of $E \le 16$, we obtained a mean superhump period
of 0.06380(8) d and $P_{\rm dot}$ of $+14.0(26.8) \times 10^{-5}$.
\citet{pat03suumas} reported a possible period decrease from 0.0641 d
to 0.0637 d during the 1998 superoutburst, which may have been a similar
phenomenon as seen in the 2003 superoutburst.
Combining the 1999 and 2003 results, the behavior of the period change
was not dramatically different from those of other SU UMa-type dwarf
novae with similar superhump periods.
More comprehensive observations covering the entire superoutburst are
needed to clearly identify the superhump period and its evolution.

\begin{figure}
  \begin{center}
    \FigureFile(88mm,70mm){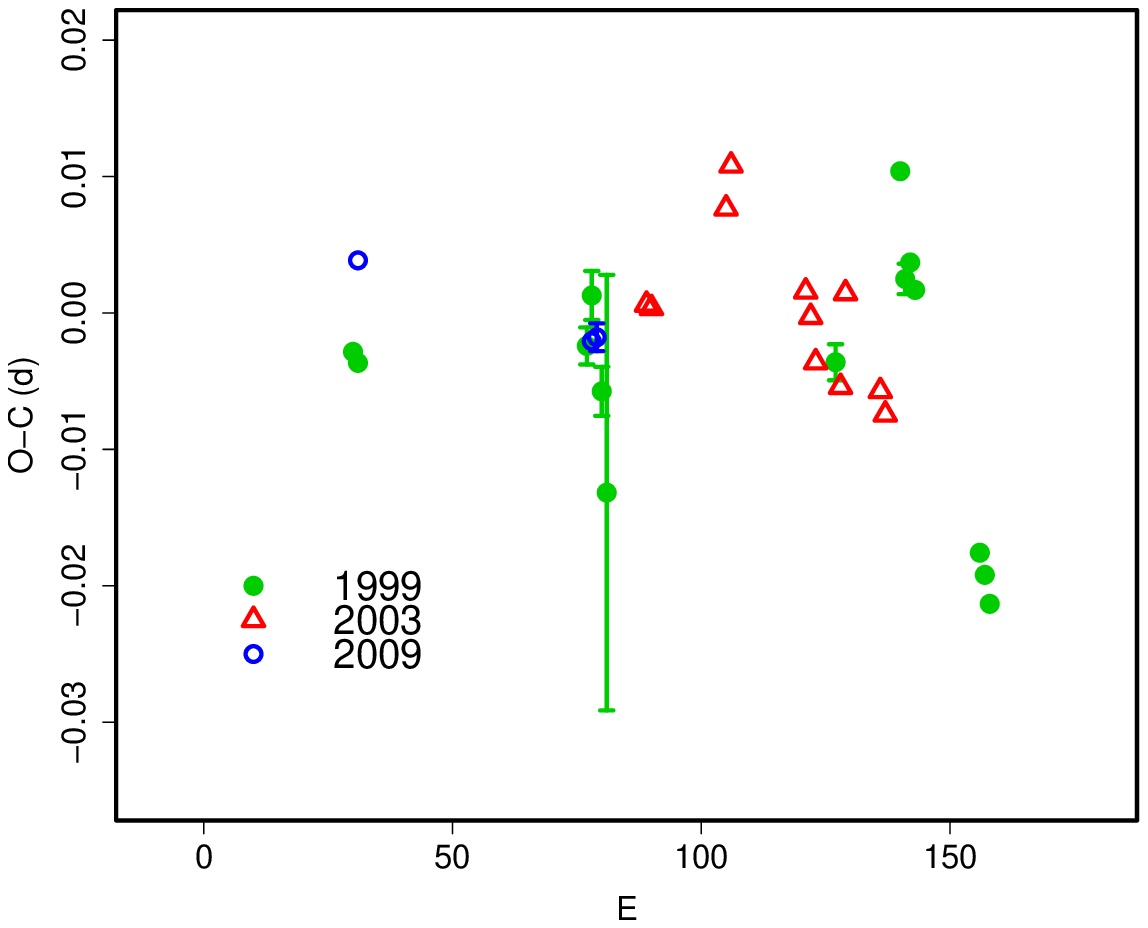}
  \end{center}
  \caption{Comparison of $O-C$ diagrams of V2051 Oph between different
  superoutbursts.  A period of 0.06430 d was used to draw this figure.
  Approximate cycle counts ($E$) after the start of the
  superoutburst were used.
  }
  \label{fig:v2051ophcomp}
\end{figure}

   A comparison of $O-C$ diagrams of V2051 Oph between different
superoutbursts is shown in figure \ref{fig:v2051ophcomp}.

\begin{table}
\caption{Superhump maxima of V2051 Oph (1999).}\label{tab:v2051ophoc1999}
\begin{center}

\end{center}
\end{table}

\begin{table}
\caption{Superhump maxima of V2051 Oph (2003).}\label{tab:v2051ophoc2003}
\begin{center}
%
\end{center}
\end{table}

\begin{table}
\caption{Superhump maxima of V2051 Oph (2009).}\label{tab:v2051ophoc2009}
\begin{center}
%
\end{center}
\end{table}

\subsection{V2527 Ophiuchi}\label{obj:v2527oph}

   V2527 Oph was an X-ray selected CV, 1E1719.1$-$1946
\citep{her90XrayCVs}.  The low absolute magnitude in quiescence inferred
from spectroscopy was already suggestive of a short-period SU UMa-type
dwarf nova.
The first detection of an outburst was reported in 1999 October
(P. Schmeer).

\begin{figure}
  \begin{center}
    \FigureFile(88mm,110mm){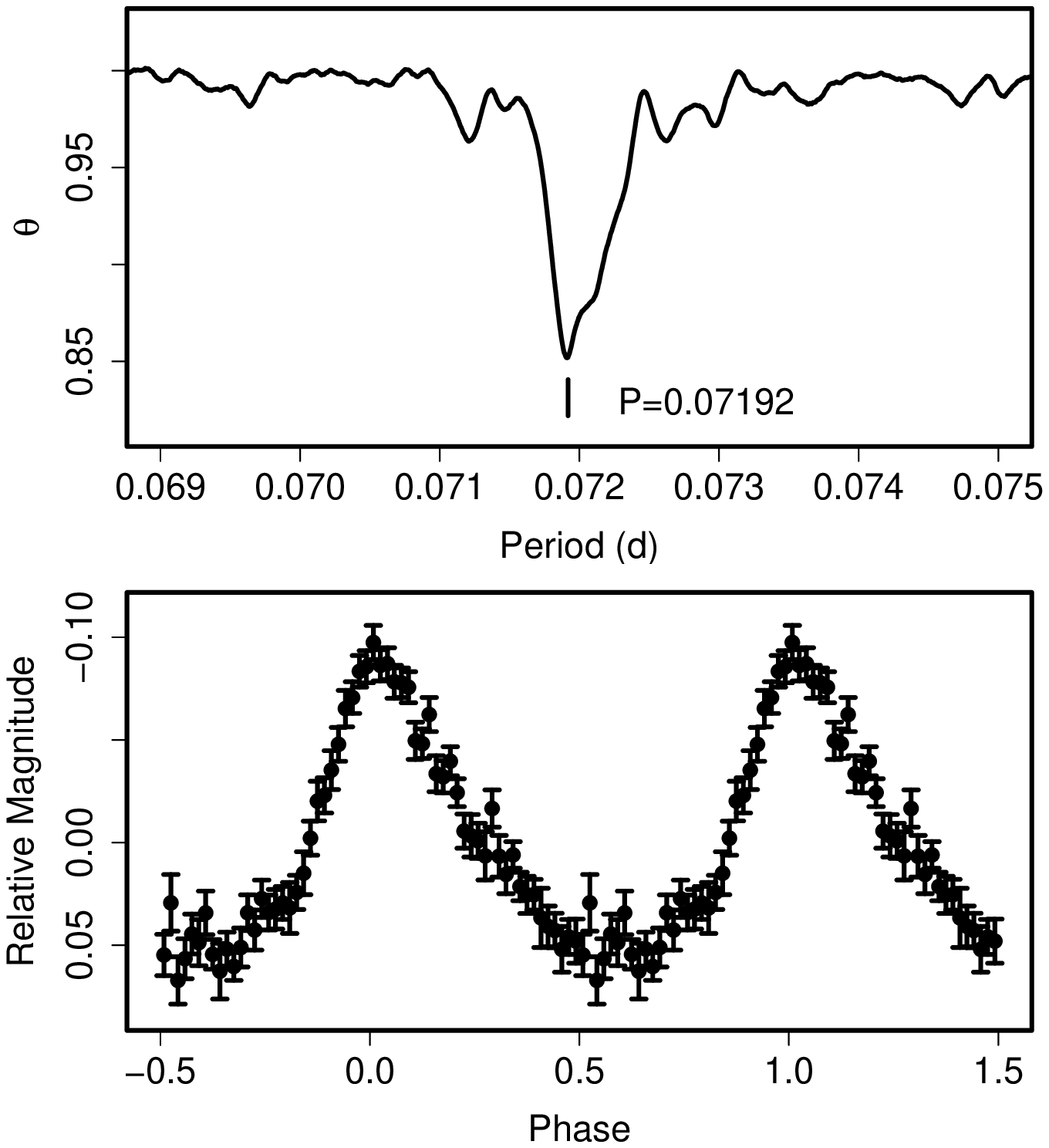}
  \end{center}
  \caption{Superhumps in V2527 Oph (2004). (Upper): PDM analysis.
     (Lower): Phase-averaged profile.}
  \label{fig:v2527ophshpdm}
\end{figure}

   The 2004 superoutburst was very well observed.
The mean superhump period during the entire outburst was 0.071919(5) d
(PDM method, figure \ref{fig:v2527ophshpdm}).
This superoutburst had a distinct precursor outburst,
during which superhumps already started emerging.
The times of superhump maxima are listed in
table \ref{tab:v2527ophoc2004}.  The portion $E \le 7$ corresponds to
the precursor, and $20 \le E \le 22$ rising stage from the minimum
following the precursor.
The superhump period showed stage A ($E \le 29$),
stage B with a positive period derivative,
and a transition to the stage C with a shorter period ($E \ge 103$).
Using the stage B, we obtained $P_{\rm dot}$ = $+6.0(1.7) \times 10^{-5}$
($29 \le E \le 103$).
The times of superhump maxima and period analyses of two other
superoutbursts in 2006 and 2008 are also given
(tables \ref{tab:v2527ophoc2006} and
\ref{tab:v2527ophoc2008}).  A comparison of $O-C$ diagrams between
different superoutbursts is given in figure \ref{fig:v2527ophcomp}.

\begin{figure}
  \begin{center}
    \FigureFile(88mm,110mm){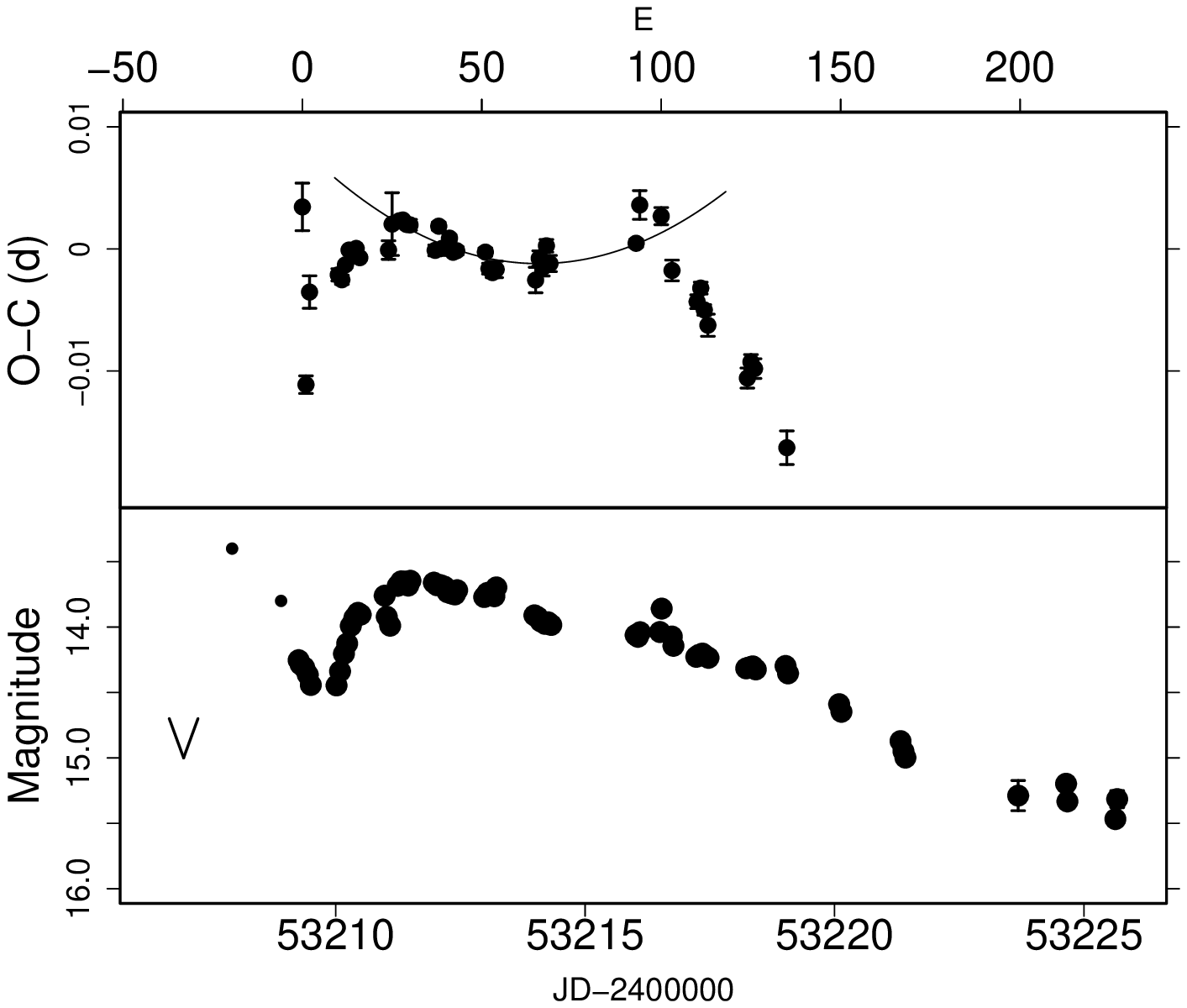}
  \end{center}
  \caption{$O-C$ of superhumps V2527 Oph (2004).
  (Upper): $O-C$ diagram.  The curve represents a quadratic fit to
  $29 \le E \le 103$.
  (Lower): Light curve.  The superoutburst was preceded by a precursor.
  Large dots represent CCD observations.  Small dots and a ``V'' mark
  represent visual observations and a upper limit, respectively.
  }
  \label{fig:v2527oph2004oc}
\end{figure}

\begin{figure}
  \begin{center}
    \FigureFile(88mm,70mm){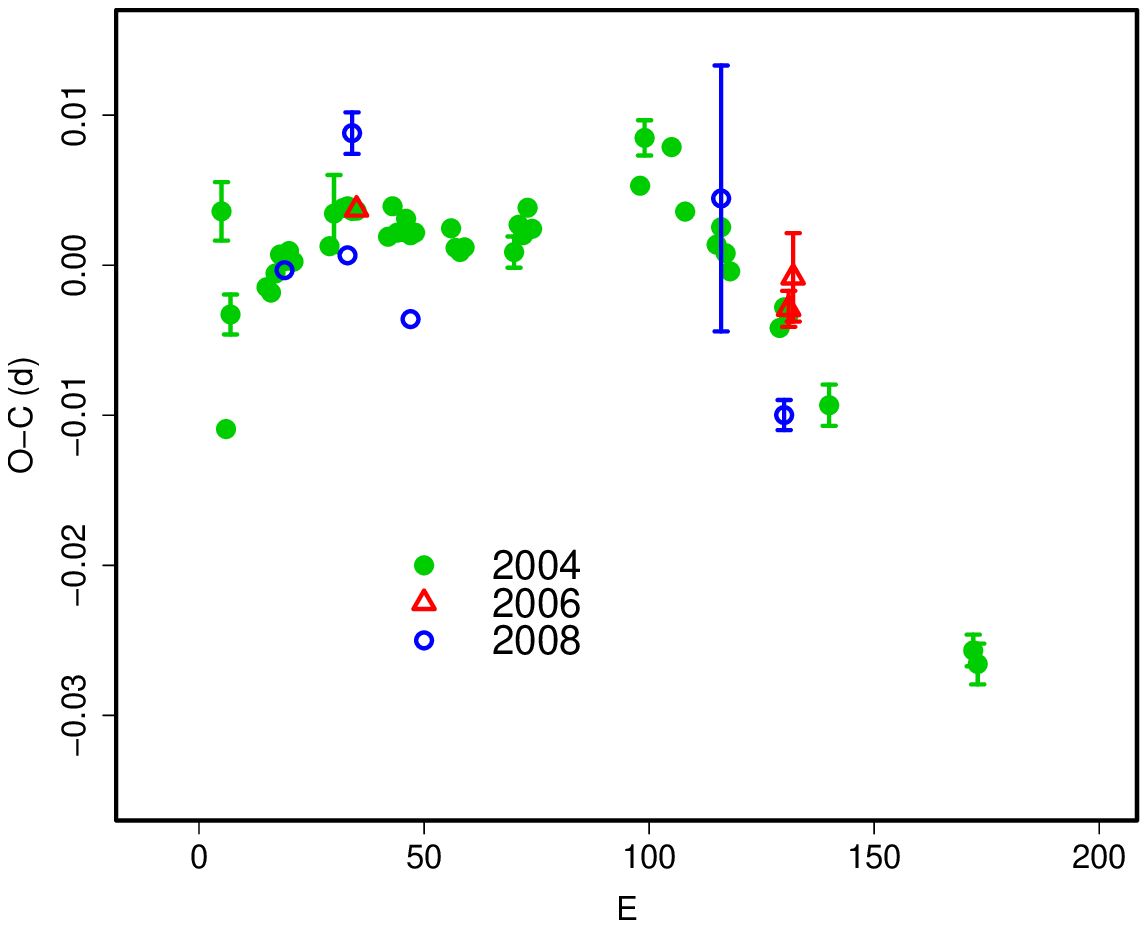}
  \end{center}
  \caption{Comparison of $O-C$ diagrams of V2527 Oph between different
  superoutbursts.  A period of 0.07200 d was used to draw this figure.
  Approximate cycle counts ($E$) after the start of the
  superoutburst were used.
  }
  \label{fig:v2527ophcomp}
\end{figure}

\begin{table}
\caption{Superhump maxima of V2527 Oph (2004).}\label{tab:v2527ophoc2004}
\begin{center}

\end{center}
\end{table}

\begin{table}
\caption{Superhump maxima of V2527 Oph (2006).}\label{tab:v2527ophoc2006}
\begin{center}
%
\end{center}
\end{table}

\begin{table}
\caption{Superhump maxima of V2527 Oph (2008).}\label{tab:v2527ophoc2008}
\begin{center}
%
\end{center}
\end{table}

\subsection{V1159 Orionis}\label{obj:v1159ori}

   V1159 Ori is a member of ER UMa stars (\cite{nog95v1159ori};
\cite{rob95eruma}; \cite{pat95v1159ori}), having outburst characteristics
similar to those of the prototype ER UMa itself.
We analyzed the 2002 November--December superoutburst
(table \ref{tab:v1159orioc2002}).  Since the waveform of superhumps
in ER UMa stars are relatively complex and sometimes show double peaks
(cf. \cite{kat03erumaSH}; \cite{pat95v1159ori}), we only deal with
prominent maxima and do not discuss on secondary maxima.\footnote{
  In table \ref{tab:v1159orioc2002}, two maxima are given for  $E = 110$.
  These maxima, with nearly equal amplitudes, were probably a result
  of manifestation of the secondary maximum.  We only used the latter
  maximum, which fits the trend of the rest of superhump maxima,
  in the analysis.
}
There appears to be a $\sim$0.5 phase shift before $E = 93$ as reported
in ER UMa \citep{kat03erumaSH}.  The nominal $P_{\rm dot}$ for the
segment $E \le 63$ was $+14.9(5.4) \times 10^{-5}$.
After $E = 93$, the object showed a fairly constant $P_{\rm SH}$
of 0.06409(5) d, which likely corresponds to $P_2$ in ordinary
SU UMa-type dwarf novae.  The overall feature is similar to
that reported by \citet{pat95v1159ori}.  The times of superhump
minima listed in \citet{pat95v1159ori} can be expressed
by a segment with a positive $P_{\rm dot}$, followed by a transition
(without a phase shift) to a shorter period which was very close to ours.
Note, however, the difference may have been caused by different methods
(\cite{pat95v1159ori} used superhump minima rather than maxima)
in determining period variation.

\begin{table}
\caption{Superhump maxima of V1159 Ori (2002).}\label{tab:v1159orioc2002}
\begin{center}

\end{center}
\end{table}

\subsection{V344 Pavonis}\label{obj:v344pav}

   We analyzed the data in \citet{uem04v344pav} and obtained the times
of superhump maxima (table \ref{tab:v344pavoc2004}).
Since the observation started during the late stage of the superoutburst,
we did not attempt to determine a $P_{\rm dot}$.
It would be noteworthy that no phase reversal, expected
for traditional late superhumps, was recorded even after the rapid
fading.

\begin{table}
\caption{Superhump maxima of V344 Pav (2004).}\label{tab:v344pavoc2004}
\begin{center}
%
\end{center}
\end{table}

\subsection{EF Pegasi}\label{obj:efpeg}

\begin{figure}
  \begin{center}
    \FigureFile(88mm,70mm){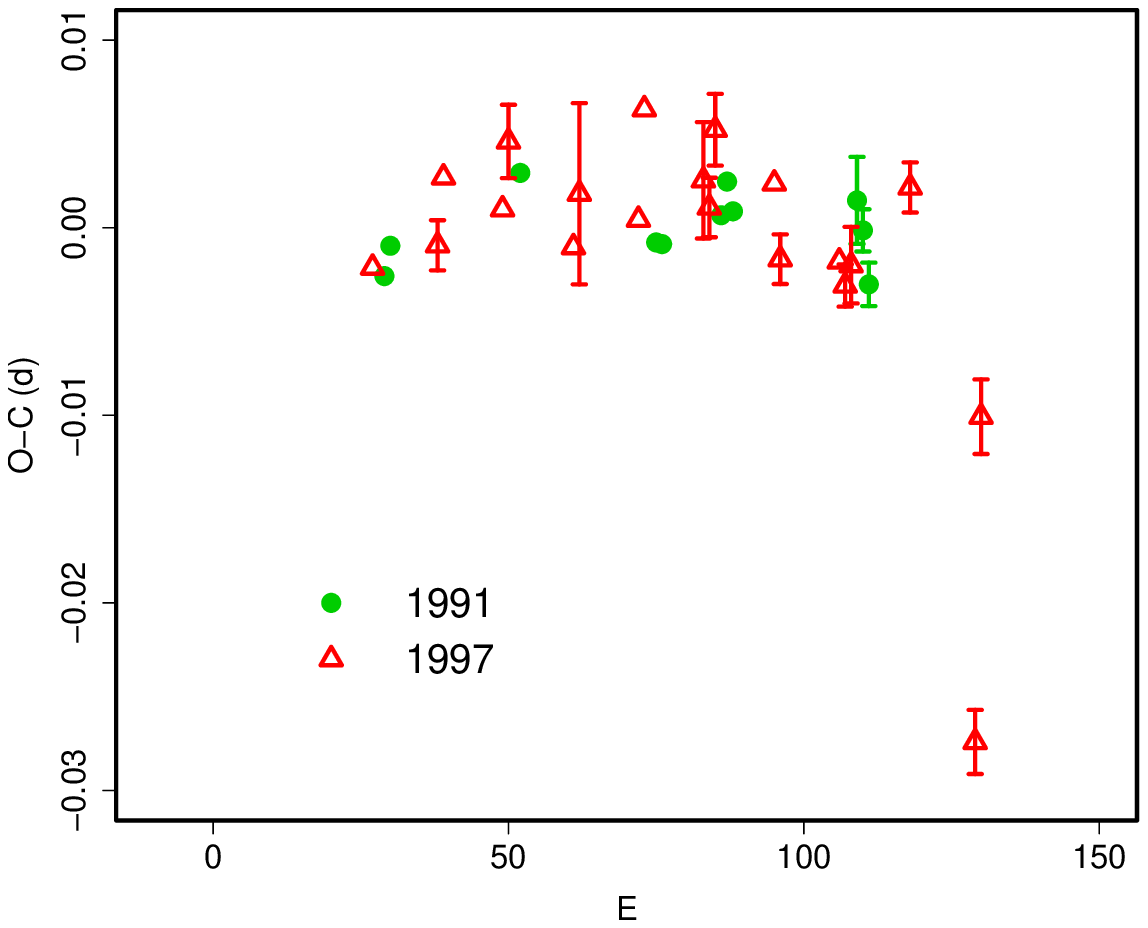}
  \end{center}
  \caption{Comparison of $O-C$ diagrams of EF Peg between different
  superoutbursts.  A period of 0.08705 d was used to draw this figure.
  Approximate cycle counts ($E$) after the start of the
  superoutburst were used.
  }
  \label{fig:efpegcomp}
\end{figure}

   \citet{how93efpeg} and \citet{kat02efpeg} reported on the 1991
superoutburst.  \citet{kat02efpeg} reported a period decrease
at $P_{\rm dot}$ = $-5.1(0.7) \times 10^{-5}$ after combination
with the times of maxima in \citet{how93efpeg}.
We further observed this object during the 1997
superoutburst.  The times of superhump maxima are listed in table
\ref{tab:efpegoc1997}.  The $P_{\rm dot}$ determined from these data,
excluding the last two maxima, corresponds to
$-4.2(2.1) \times 10^{-5}$, similar to the one in 1991.
There was an indication that the earliest superhump maxima of
the 1991 were obtained during the evolutionary stage of superhumps.
An exclusion of these maxima has only yielded an insignificant $P_{\rm dot}$
due to the fragmentary observational coverage.
We thus regard the 1997 result more reliable based on
homogeneous set of observations.
This result supersedes the preliminary argument on period changes
in \citet{kat02efpeg}.  A comparison of 1991 and 1997 $O-C$ variations
is presented in figure \ref{fig:efpegcomp}.

\begin{table}
\caption{Superhump maxima of EF Peg (1997).}\label{tab:efpegoc1997}
\begin{center}
\begin{tabular}{ccccc}
\hline\hline
$E$ & max$^a$ & error & $O-C^b$ & $N^c$ \\
\hline
0 & 50757.0034 & 0.0005 & $-$0.0073 & 152 \\
11 & 50757.9621 & 0.0013 & $-$0.0049 & 117 \\
12 & 50758.0528 & 0.0006 & $-$0.0011 & 126 \\
22 & 50758.9216 & 0.0003 & $-$0.0017 & 193 \\
23 & 50759.0123 & 0.0020 & 0.0021 & 145 \\
34 & 50759.9642 & 0.0008 & $-$0.0023 & 162 \\
35 & 50760.0541 & 0.0048 & 0.0007 & 56 \\
45 & 50760.9232 & 0.0004 & 0.0004 & 165 \\
46 & 50761.0162 & 0.0007 & 0.0064 & 126 \\
56 & 50761.8829 & 0.0031 & 0.0038 & 100 \\
57 & 50761.9685 & 0.0016 & 0.0025 & 160 \\
58 & 50762.0597 & 0.0019 & 0.0067 & 143 \\
68 & 50762.9273 & 0.0009 & 0.0050 & 186 \\
69 & 50763.0103 & 0.0013 & 0.0011 & 163 \\
79 & 50763.8807 & 0.0008 & 0.0021 & 115 \\
80 & 50763.9665 & 0.0011 & 0.0010 & 151 \\
81 & 50764.0546 & 0.0020 & 0.0022 & 62 \\
91 & 50764.9292 & 0.0013 & 0.0075 & 23 \\
102 & 50765.8572 & 0.0017 & $-$0.0208 & 44 \\
103 & 50765.9616 & 0.0020 & $-$0.0034 & 132 \\
\hline
  \multicolumn{5}{l}{$^{a}$ BJD$-$2400000.} \\
  \multicolumn{5}{l}{$^{b}$ Against $max = 2450757.0107 + 0.086934 E$.} \\
  \multicolumn{5}{l}{$^{c}$ Number of points used to determine the maximum.} \\
\end{tabular}
\end{center}
\end{table}

\subsection{V364 Pegasi}\label{obj:v364peg}

\begin{figure}
  \begin{center}
    \FigureFile(88mm,110mm){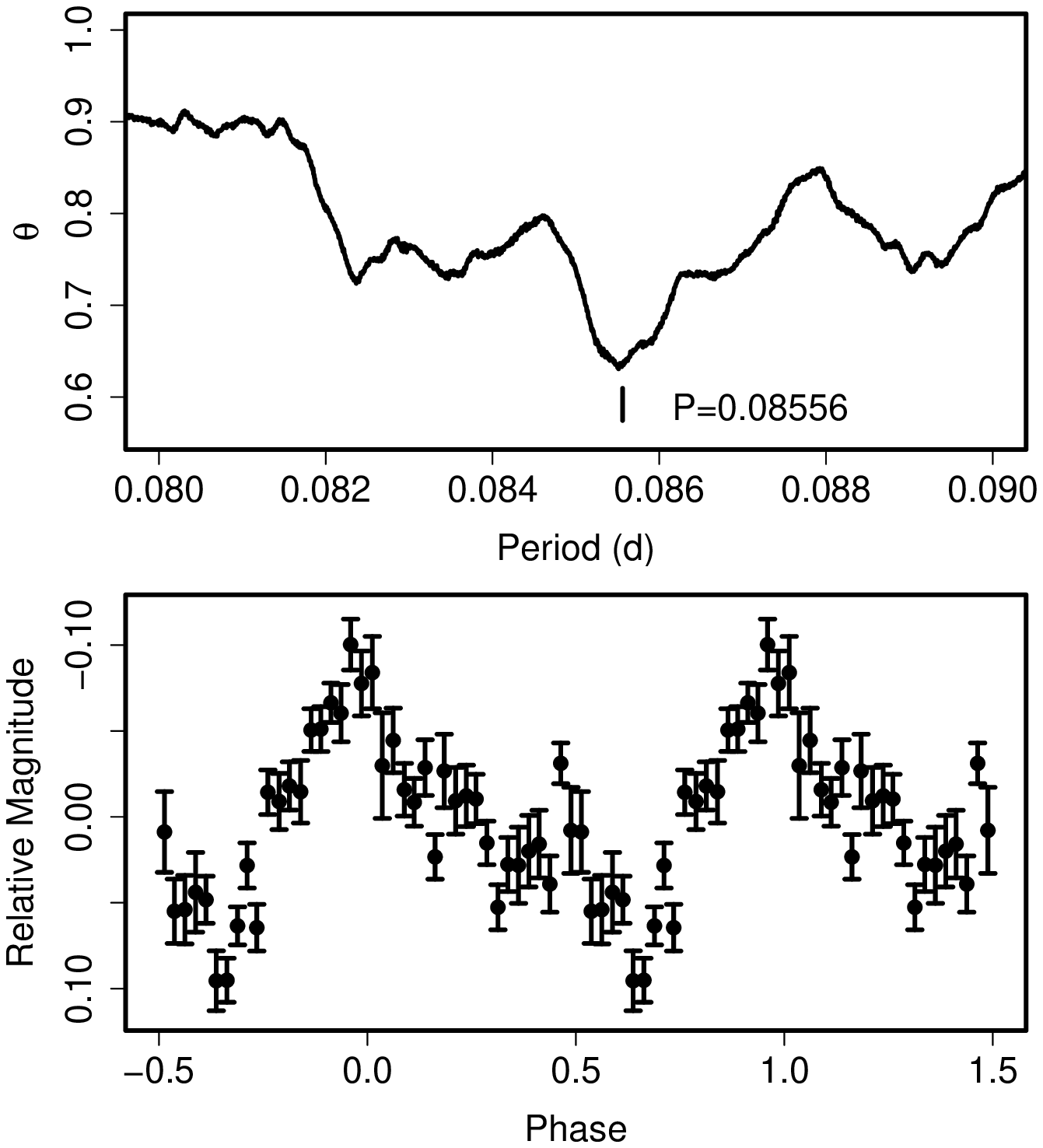}
  \end{center}
  \caption{Superhumps in V364 Peg (2004). (Upper): PDM analysis.
     (Lower): Phase-averaged profile.}
  \label{fig:v364pegshpdm}
\end{figure}

   V364 Peg is a dwarf nova discovered during the supernova survey
\citep{qiu97v364pegiauc6746}.  \citet{kat99v364peg} reported, based on
time-resolved photometry during the 1997 November outburst, that this
object is a likely SU UMa-type dwarf nova with a long superhump period.
This suggestion has been confirmed during the 2004 outburst
(T. Vanmunster, aavso-photometry message), reporting a superhump
period of 0.0882(70) d.
We have refined the period to 0.08556(5) d with the PDM method
(figure \ref{fig:v364pegshpdm}).  The times of superhump maxima
are listed in table \ref{tab:v364pegoc2004}.
If there was a stage B--C transition as in many SU UMa-type dwarf novae,
this period likely represents $P_2$.
The inferred orbital period lies close to the lower edge of the
period gap.  The object appears to show rather frequent outbursts
(cf. \cite{qiu97v364pegiauc6772}).

\begin{table}
\caption{Superhump maxima of V364 Peg (2004).}\label{tab:v364pegoc2004}
\begin{center}
\begin{tabular}{ccccc}
\hline\hline
$E$ & max$^a$ & error & $O-C^b$ & $N^c$ \\
\hline
0 & 53329.2263 & 0.0007 & 0.0008 & 57 \\
1 & 53329.3101 & 0.0007 & $-$0.0007 & 70 \\
2 & 53329.3967 & 0.0024 & 0.0006 & 30 \\
4 & 53329.5654 & 0.0024 & $-$0.0015 & 48 \\
5 & 53329.6529 & 0.0012 & 0.0008 & 60 \\
27 & 53331.5302 & 0.0364 & 0.0006 & 29 \\
28 & 53331.6144 & 0.0032 & $-$0.0005 & 46 \\
\hline
  \multicolumn{5}{l}{$^{a}$ BJD$-$2400000.} \\
  \multicolumn{5}{l}{$^{b}$ Against $max = 2453329.2255 + 0.085338 E$.} \\
  \multicolumn{5}{l}{$^{c}$ Number of points used to determine the maximum.} \\
\end{tabular}
\end{center}
\end{table}

\subsection{V368 Pegasi}\label{obj:v368peg}

\begin{figure}
  \begin{center}
    \FigureFile(88mm,110mm){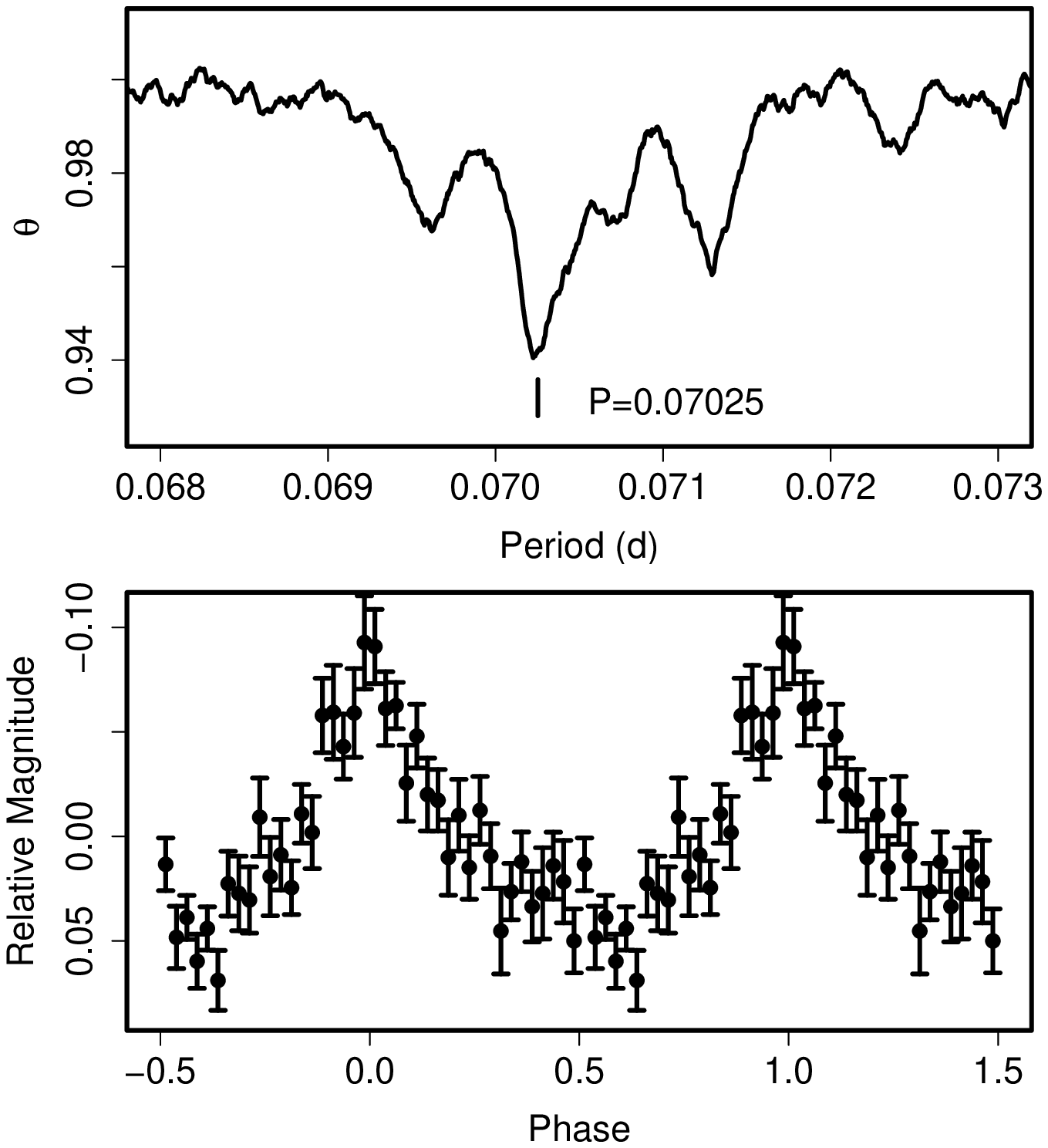}
  \end{center}
  \caption{Superhumps in V368 Peg (2000). (Upper): PDM analysis.
     (Lower): Phase-averaged profile.}
  \label{fig:v368pegshpdm}
\end{figure}

\begin{figure}
  \begin{center}
    \FigureFile(88mm,70mm){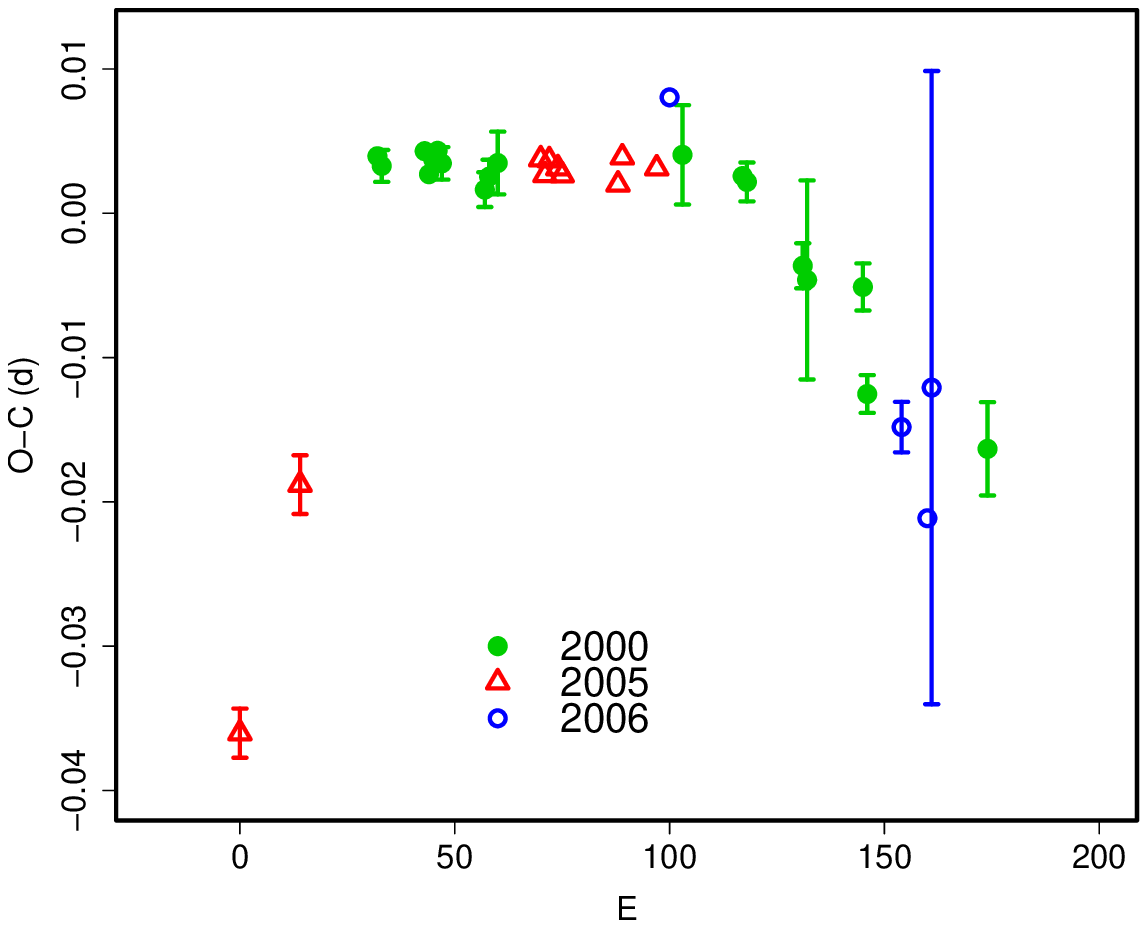}
  \end{center}
  \caption{Comparison of $O-C$ diagrams of V368 Peg between different
  superoutbursts.  A period of 0.07039 d was used to draw this figure.
  Approximate cycle counts ($E$) after the start of the
  superoutburst were used.
  }
  \label{fig:v368pegcomp}
\end{figure}

   V368 Peg is a dwarf nova discovered by
\citet{ant99v368pegftcamv367pegv2209cyg}.
The SU UMa-type nature of this object was established by J. Pietz
during the 1999 superoutburst (vsnet-alert 3317).
We observed the 2000 superoutburst.
The mean superhump period with the PDM method was 0.070253(17) d
(figure \ref{fig:v368pegshpdm}).
The times of superhump maxima are listed in table \ref{tab:v368pegoc2000}.
There was a clear transition in the superhump period around $E = 86$.
The mean $P_{\rm SH}$ and $P_{\rm dot}$ for $E \le 86$ were
0.070380(8) d and $+0.5(1.2) \times 10^{-5}$, respectively.
We also observed the 2005 superoutburst (table \ref{tab:v368pegoc2005})
during the growing stage of superhumps.  A likely stage A--B transition
was recorded ($E \le 14$).  Combined with the AAVSO
observations, we obtained $P_{\rm SH}$ = 0.07038(3) d for $70 \le E \le 97$.
The 2006 superoutburst was observed during its late stage
(table \ref{tab:v368pegoc2006}), yielding $P_2$ = 0.069945(18) d with
the PDM method.
A comparison of $O-C$ diagrams between different superoutbursts
is shown in figure \ref{fig:v368pegcomp}.

\begin{table}
\caption{Superhump maxima of V368 Peg (2000).}\label{tab:v368pegoc2000}
\begin{center}

\end{center}
\end{table}

\begin{table}
\caption{Superhump maxima of V368 Peg (2005).}\label{tab:v368pegoc2005}
\begin{center}
%
\end{center}
\end{table}

\begin{table}
\caption{Superhump maxima of V368 Peg (2006).}\label{tab:v368pegoc2006}
\begin{center}
%
\end{center}
\end{table}

\subsection{V369 Pegasi}\label{obj:v369peg}

   The SU UMa-type nature of V369 Peg (=KUV 23012$+$1702) was established
during the 1999 superoutburst \citep{kat01v369peg}.
We reanalyzed the data in \citet{kat01v369peg} and AAVSO observations.
The times of superhump maxima are listed in table \ref{tab:v369pegoc1999}.
The $O-C$ diagram shows a stage B--C transition.

\begin{table}
\caption{Superhump maxima of V369 Peg (1999).}\label{tab:v369pegoc1999}
\begin{center}
%
\end{center}
\end{table}

\subsection{UV Persei}\label{sec:uvper}\label{obj:uvper}

   UV Per is a well-known SU UMa-type dwarf nova with a relatively long
recurrence time and a large outburst amplitude.
\citet{uda92uvper} detected superhumps during the 1989 superoutburst.
\citet{uda92uvper} reported that they did not detect a significant
quadratic term ($P_{\rm dot}$), probably due to the short ($\sim 3$ d)
coverage.

   We observed four superoutbursts in 1991--1992, 2000, 2003 and 2007
(tables \ref{tab:uvperoc1991}, \ref{tab:uvperoc2000}, \ref{tab:uvperoc2003},
\ref{tab:uvperoc2007}).
The 2000 observation covered the entire superoutburst, including the
growing stage of superhumps and the rapid fading stage, but with
lower signal statistics.  The 2003 observation covered the
superoutburst with higher statistics.  The $O-C$ diagrams of these
outbursts can be interpreted as a well-demonstrated sequence of
stages A--C.
The $P_{\rm dot}$'s of the stage B corresponded to
$+9.5(6.0) \times 10^{-5}$ ($14 \le E \le 62$)
for the 2000 superoutburst, and $+5.1(1.0) \times 10^{-5}$
($20 \le E \le 109$) for the 2003 superoutburst, respectively.
The 1991--1992 and 2007 superoutbursts were observed during the
(middle-to-)final stage
of the plateau and clearly showed a transition to a shorter period
(stage B to C).
Although the $P_{\rm dot}$ of the entire 2007 data was
$-7.0(0.9) \times 10^{-5}$, this value should be used carefully since
the measured segment of the $O-C$ diagram was different from those
in the 2000 and 2003 superoutburst.

\begin{table}
\caption{Superhump maxima of UV Per (1991--1992).}\label{tab:uvperoc1991}
\begin{center}

\end{center}
\end{table}

\begin{table}
\caption{Superhump maxima of UV Per (2000).}\label{tab:uvperoc2000}
\begin{center}
%
\end{center}
\end{table}

\begin{table}
\caption{Superhump maxima of UV Per (2003).}\label{tab:uvperoc2003}
\begin{center}
%
\end{center}
\end{table}

\addtocounter{table}{-1}
\begin{table}
\caption{Superhump maxima of UV Per (2003). (continued)}
\begin{center}
%
\end{center}
\end{table}

\addtocounter{table}{-1}
\begin{table}
\caption{Superhump maxima of UV Per (2003). (continued)}
\begin{center}
%
\end{center}
\end{table}

\begin{table}
\caption{Superhump maxima of UV Per (2007).}\label{tab:uvperoc2007}
\begin{center}
%
\end{center}
\end{table}

\subsection{PU Persei}\label{obj:puper}

\begin{figure}
  \begin{center}
    \FigureFile(88mm,110mm){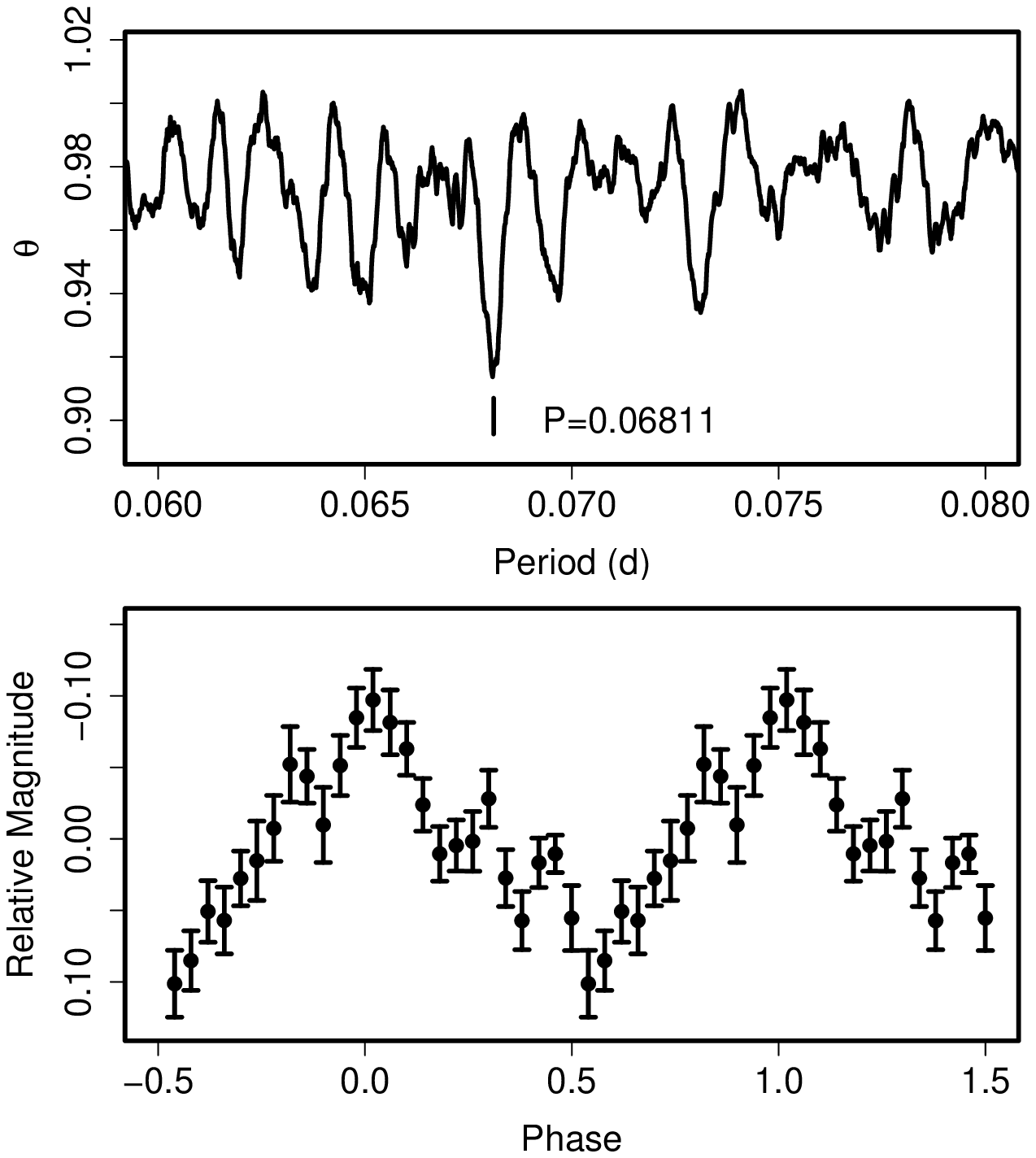}
  \end{center}
  \caption{Superhumps in PU Per (2009). (Upper): PDM analysis.
     (Lower): Phase-averaged profile.}
  \label{fig:pupershpdm}
\end{figure}

   PU Per was discovered as a dwarf nova by \citet{hof67an29043}.
The object has a relatively long outburst recurrence time and
a large outburst amplitude (cf. \cite{rom76DN}; \cite{bus79VS17};
\cite{kat95puper}).
Although the detection of superhumps in this object was reported in
\citet{kat99puper}, the identification of the period had awaited further
observation.  We observed the object during the 2009 superoutburst
and identified the superhump period as 0.06811(3) d
(figure \ref{fig:pupershpdm}), a one-day alias to \citet{kat99puper}.

   The times of superhump maxima are listed in table \ref{tab:puperoc2009}.
Since the object faded shortly after the last observation, it was most
likely that we observed the later part of the superoutburst, corresponding
to the stage B--C transition.  We presented the measured periods based on
this interpretation in table \ref{tab:perlist}.

\begin{table}
\caption{Superhump maxima of PU Per (2009).}\label{tab:puperoc2009}
\begin{center}
\begin{tabular}{ccccc}
\hline\hline
$E$ & max$^a$ & error & $O-C^b$ & $N^c$ \\
\hline
0 & 54837.8888 & 0.0021 & 0.0026 & 42 \\
1 & 54837.9477 & 0.0025 & $-$0.0067 & 70 \\
2 & 54838.0191 & 0.0038 & $-$0.0034 & 69 \\
3 & 54838.0902 & 0.0050 & $-$0.0004 & 38 \\
18 & 54839.1206 & 0.0031 & 0.0082 & 139 \\
60 & 54841.9793 & 0.0014 & 0.0060 & 132 \\
61 & 54842.0340 & 0.0021 & $-$0.0073 & 133 \\
75 & 54843.0002 & 0.0039 & 0.0053 & 58 \\
76 & 54843.0678 & 0.0086 & 0.0048 & 72 \\
89 & 54843.9468 & 0.0013 & $-$0.0018 & 142 \\
90 & 54844.0095 & 0.0018 & $-$0.0072 & 125 \\
\hline
  \multicolumn{5}{l}{$^{a}$ BJD$-$2400000.} \\
  \multicolumn{5}{l}{$^{b}$ Against $max = 2454837.8863 + 0.068116 E$.} \\
  \multicolumn{5}{l}{$^{c}$ Number of points used to determine the maximum.} \\
\end{tabular}
\end{center}
\end{table}

\subsection{PV Persei}\label{obj:pvper}

\begin{figure}
  \begin{center}
    \FigureFile(88mm,110mm){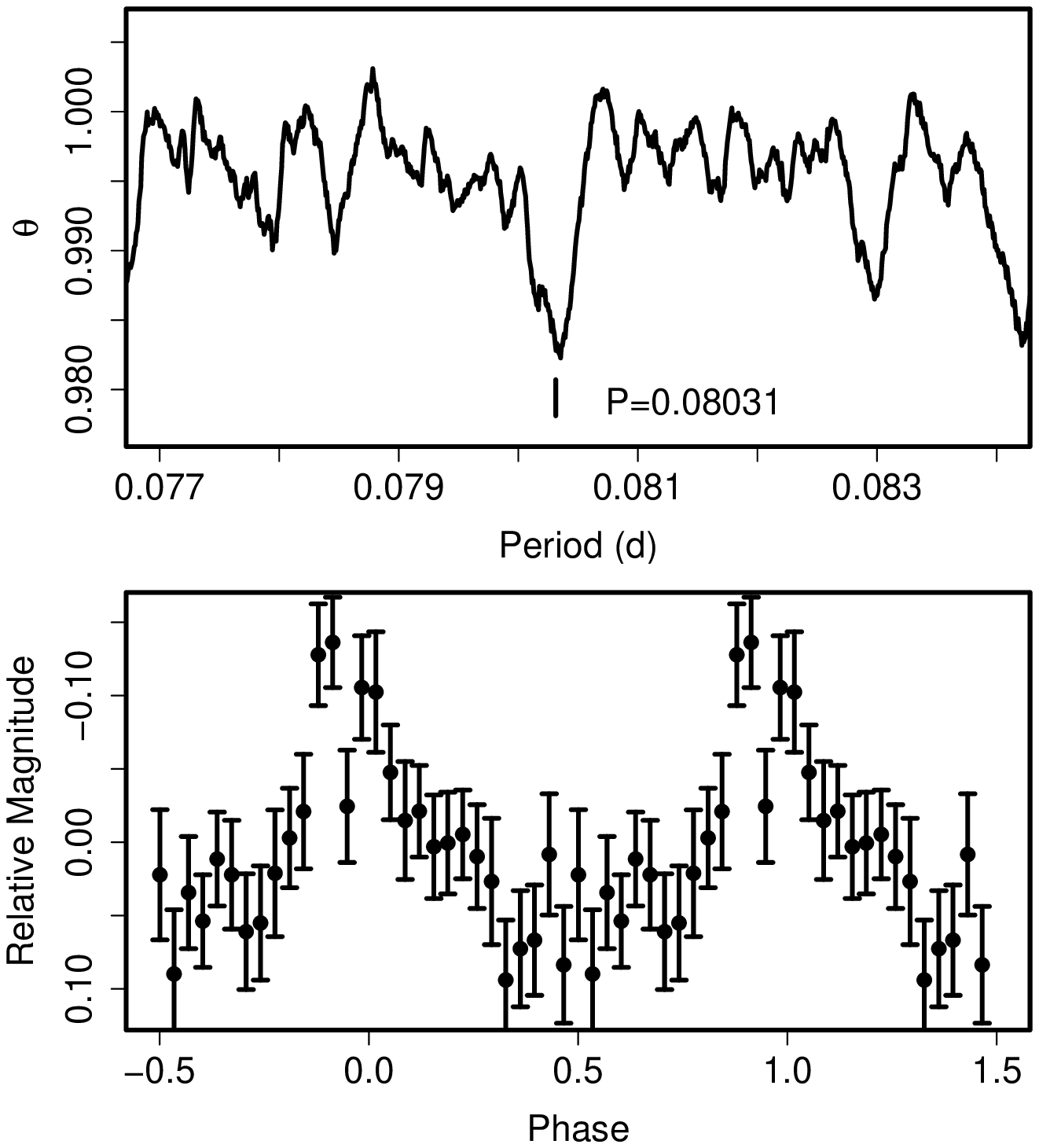}
  \end{center}
  \caption{Superhumps in PV Per (2008). (Upper): PDM analysis.
     (Lower): Phase-averaged profile.}
  \label{fig:pvpershpdm}
\end{figure}

   In contrast to PU Per, discovered at the same time \citep{hof67an29043},
PV Per shows frequent outbursts (\cite{rom76DN}; \cite{bus79VS17}).
The SU UMa-type nature of PV Per was established by \citet{van97pvper},
who reported a period of 0.0805(1) d.
We further observed the 2008 superoutburst.
The mean superhump period with the PDM method was 0.08031(4) d
(figure \ref{fig:pvpershpdm}).
The times of superhump maxima during the 2008 superoutburst
are listed in table \ref{tab:pvperoc2008}.  The observation mainly
covered the later stage of the superoutburst.  Although the global
$P_{\rm dot}$ was $-4.4(2.1) \times 10^{-5}$, this change can be
interpreted as a result of a stage B--C transition
(see also table \ref{tab:perlist}).

\begin{table}
\caption{Superhump maxima of PV Per (2008).}\label{tab:pvperoc2008}
\begin{center}
\begin{tabular}{ccccc}
\hline\hline
$E$ & max$^a$ & error & $O-C^b$ & $N^c$ \\
\hline
0 & 54745.3856 & 0.0006 & $-$0.0074 & 60 \\
1 & 54745.4668 & 0.0008 & $-$0.0066 & 86 \\
35 & 54748.2132 & 0.0007 & 0.0057 & 150 \\
36 & 54748.2952 & 0.0011 & 0.0073 & 136 \\
73 & 54751.2686 & 0.0005 & 0.0054 & 170 \\
74 & 54751.3372 & 0.0036 & $-$0.0064 & 89 \\
121 & 54755.1253 & 0.0011 & 0.0022 & 151 \\
122 & 54755.2073 & 0.0033 & 0.0038 & 132 \\
123 & 54755.2909 & 0.0048 & 0.0070 & 127 \\
135 & 54756.2568 & 0.0068 & 0.0079 & 153 \\
146 & 54757.1335 & 0.0027 & 0.0000 & 149 \\
147 & 54757.2170 & 0.0046 & 0.0032 & 151 \\
159 & 54758.1612 & 0.0033 & $-$0.0177 & 168 \\
160 & 54758.2475 & 0.0055 & $-$0.0118 & 149 \\
161 & 54758.3471 & 0.0075 & 0.0074 & 78 \\
\hline
  \multicolumn{5}{l}{$^{a}$ BJD$-$2400000.} \\
  \multicolumn{5}{l}{$^{b}$ Against $max = 2454745.3930 + 0.080414 E$.} \\
  \multicolumn{5}{l}{$^{c}$ Number of points used to determine the maximum.} \\
\end{tabular}
\end{center}
\end{table}

\subsection{QY Persei}\label{sec:qyper}\label{obj:qyper}

   QY Per is a dwarf nova discovered by \citet{hof66an289139}.
The object had long been suspected to be an excellent candidate
for a WZ Sge-like object based on the large outburst amplitude
and long recurrence time.

\begin{figure}
  \begin{center}
    \FigureFile(88mm,70mm){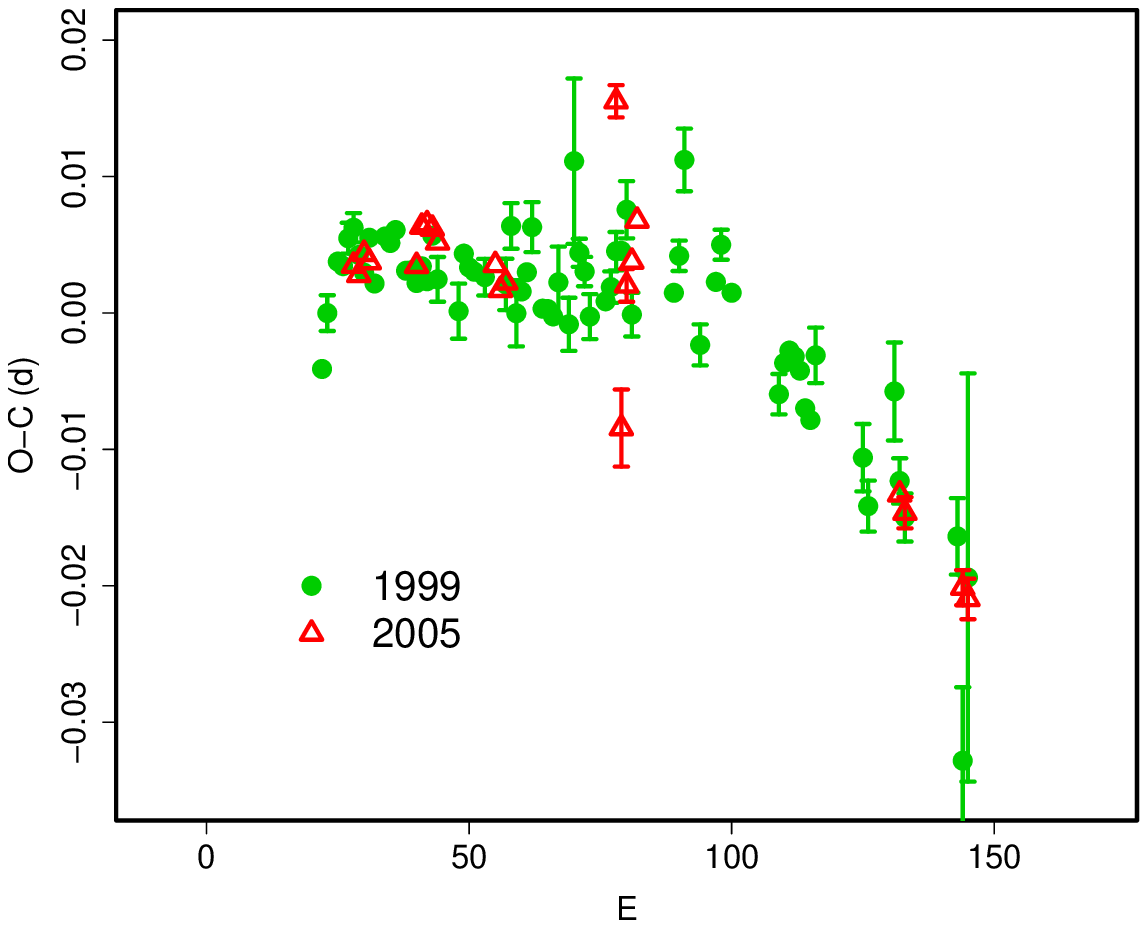}
  \end{center}
  \caption{Comparison of $O-C$ diagrams of QY Per between different
  superoutbursts.  A period of 0.07862 d was used to draw this figure.
  Approximate cycle counts ($E$) after the start of the
  superoutburst were used.
  }
  \label{fig:qypercomp}
\end{figure}

   We observed two superoutbursts in 1999 (\cite{mat99qyperiauc};
\cite{kat00qyperiauc}) and 2005.
The 1999 superoutburst (table \ref{tab:qyperoc1999}) was
one of the the best sampled superoutbursts among all SU UMa-type
dwarf novae.  The $O-C$ diagram consisted of all stages A--C
(cf. figure \ref{fig:ocsamp}).
The $P_{\rm dot}$ during the stage B corresponds to
$+7.8(3.1) \times 10^{-5}$ ($5 \le E \le 69$).
This example demonstrates that a positive $P_{\rm dot}$ system
is present among systems with longer superhump periods.
A stage B--C transition was recorded
during the 2005 superoutburst (table \ref{tab:qyperoc2005}).
A comparison of $O-C$ diagrams between different superoutbursts
is shown in figure \ref{fig:qypercomp}.

\begin{table}
\caption{Superhump maxima of QY Per (1999).}\label{tab:qyperoc1999}
\begin{center}

\end{center}
\end{table}

\addtocounter{table}{-1}
\begin{table}
\caption{Superhump maxima of QY Per (1999) (continued).}
\begin{center}
%
\end{center}
\end{table}

\begin{table}
\caption{Superhump maxima of QY Per (2005).}\label{tab:qyperoc2005}
\begin{center}
%
\end{center}
\end{table}

\subsection{V518 Persei}\label{sec:v518per}\label{obj:v518per}

   This object (=GRO J0422+32) is a BHXT (see subsection \ref{sec:BHXT}).
We present reanalysis of observations in \citet{kat95v518per}.
A new analysis has yielded a slightly longer superhump period of 0.2159(3) d.
(table \ref{tab:v518peroc1992}).
The fractional superhump excess is 1.8(1) \%.  Using the relation in
subsection \ref{tab:v518peroc1992}), we can expect $q =$ 0.096(7),
reasonably consistent with the determination from radial-velocity studies
($q =$ 0.116(8), \cite{har99v518per}).

\begin{figure}
  \begin{center}
    \FigureFile(88mm,110mm){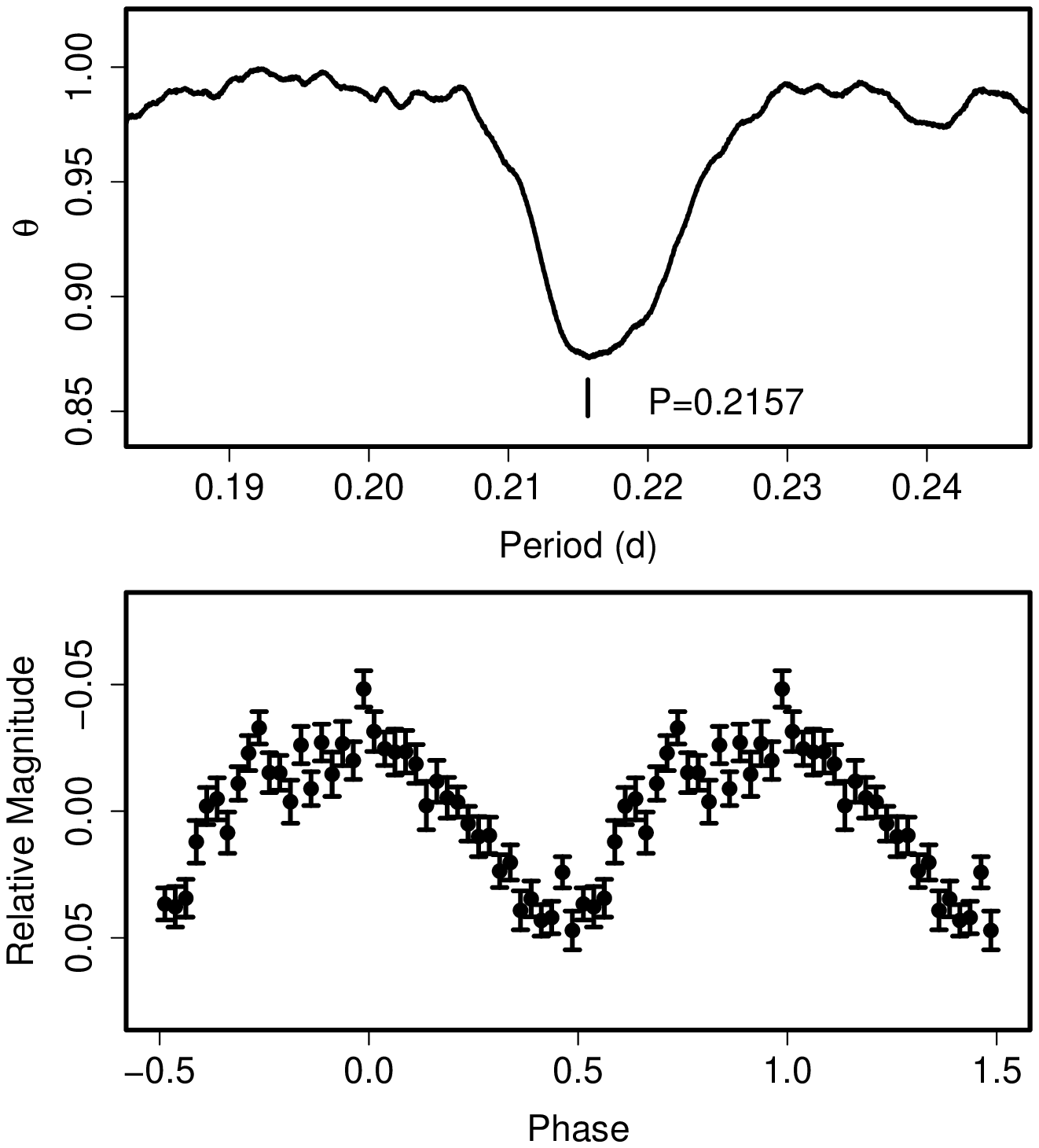}
  \end{center}
  \caption{Superhumps in V518 Per (1992). (Upper): PDM analysis.
     (Lower): Phase-averaged profile.}
  \label{fig:v518pershpdm}
\end{figure}

\begin{table}
\caption{Superhump maxima of V518 Per (1992).}\label{tab:v518peroc1992}
\begin{center}
\begin{tabular}{ccccc}
\hline\hline
$E$ & max$^a$ & error & $O-C^b$ & $N^c$ \\
\hline
0 & 48948.0905 & 0.0061 & $-$0.0064 & 249 \\
1 & 48948.3060 & 0.0051 & $-$0.0099 & 228 \\
4 & 48948.9895 & 0.0052 & 0.0168 & 247 \\
5 & 48949.1934 & 0.0033 & 0.0017 & 254 \\
9 & 48950.0745 & 0.0027 & 0.0069 & 265 \\
13 & 48950.9336 & 0.0195 & $-$0.0099 & 174 \\
14 & 48951.1657 & 0.0029 & 0.0033 & 293 \\
18 & 48952.0357 & 0.0023 & $-$0.0026 & 260 \\
\hline
  \multicolumn{5}{l}{$^{a}$ BJD$-$2400000.} \\
  \multicolumn{5}{l}{$^{b}$ Against $max = 2448948.0969 + 0.21897 E$.} \\
  \multicolumn{5}{l}{$^{c}$ Number of points used to determine the maximum.} \\
\end{tabular}
\end{center}
\end{table}

\subsection{TY Piscis Austrini}\label{obj:typsa}

   Although TY PsA is among the SU UMa-type dwarf novae earliest
discovered \citep{bar82typsa}, the only published $P_{\rm SH}$ was
0.08765 d, determined from the relatively limited data taken during
the 1984 superoutburst \citep{war89typsa}.

   We observed the 2008 superoutburst starting 2 d after the initial
detection of the outburst.
The times of superhump maxima are listed in table \ref{tab:typsaoc2008}.
Although a stage B--C transition was likely present around $E = 40$,
the times of maxima were not very well determined because the durations
of each observations were comparable to the superhump period and the maxima
often fell close to start or end of the observation.  We therefore
determined periods for the stage B ($E \le 34$) and the stage C ($E \ge 46$)
using the PDM method.  The values were 0.087990(17) d and 0.087730(30) d,
respectively, and these values are adopted in table \ref{tab:perlist}.
The latter period is close to one reported by \citet{war89typsa}, suggesting
that \citet{war89typsa} recorded the stage C superhumps.

\begin{table}
\caption{Superhump maxima of TY PsA (2008).}\label{tab:typsaoc2008}
\begin{center}
\begin{tabular}{ccccc}
\hline\hline
$E$ & max$^a$ & error & $O-C^b$ & $N^c$ \\
\hline
0 & 54798.9021 & 0.0003 & $-$0.0043 & 148 \\
11 & 54799.8674 & 0.0019 & $-$0.0044 & 89 \\
12 & 54799.9677 & 0.0005 & 0.0081 & 99 \\
23 & 54800.9228 & 0.0004 & $-$0.0021 & 156 \\
34 & 54801.8976 & 0.0007 & 0.0074 & 74 \\
x46 & 54802.9482 & 0.0010 & 0.0048 & 86 \\
57 & 54803.9093 & 0.0006 & 0.0006 & 163 \\
68 & 54804.8764 & 0.0012 & 0.0024 & 128 \\
69 & 54804.9378 & 0.0028 & $-$0.0240 & 98 \\
80 & 54805.9325 & 0.0020 & 0.0053 & 75 \\
91 & 54806.8987 & 0.0009 & 0.0062 & 197 \\
\hline
  \multicolumn{5}{l}{$^{a}$ BJD$-$2400000.} \\
  \multicolumn{5}{l}{$^{b}$ Against $max = 2454798.9064 + 0.087759 E$.} \\
  \multicolumn{5}{l}{$^{c}$ Number of points used to determine the maximum.} \\
\end{tabular}
\end{center}
\end{table}

\subsection{TY Piscium}\label{obj:typsc}

   TY Psc has long been known as an SU UMa-type dwarf nova
(cf. \cite{szk88DNnovaIR}), though accurate determination of
the superhump period has not yet been published.  Although
\citet{kun01typsc} reported observations of the 2000 superoutburst,
the resultant period had a large uncertainty.

\begin{figure}
  \begin{center}
    \FigureFile(88mm,70mm){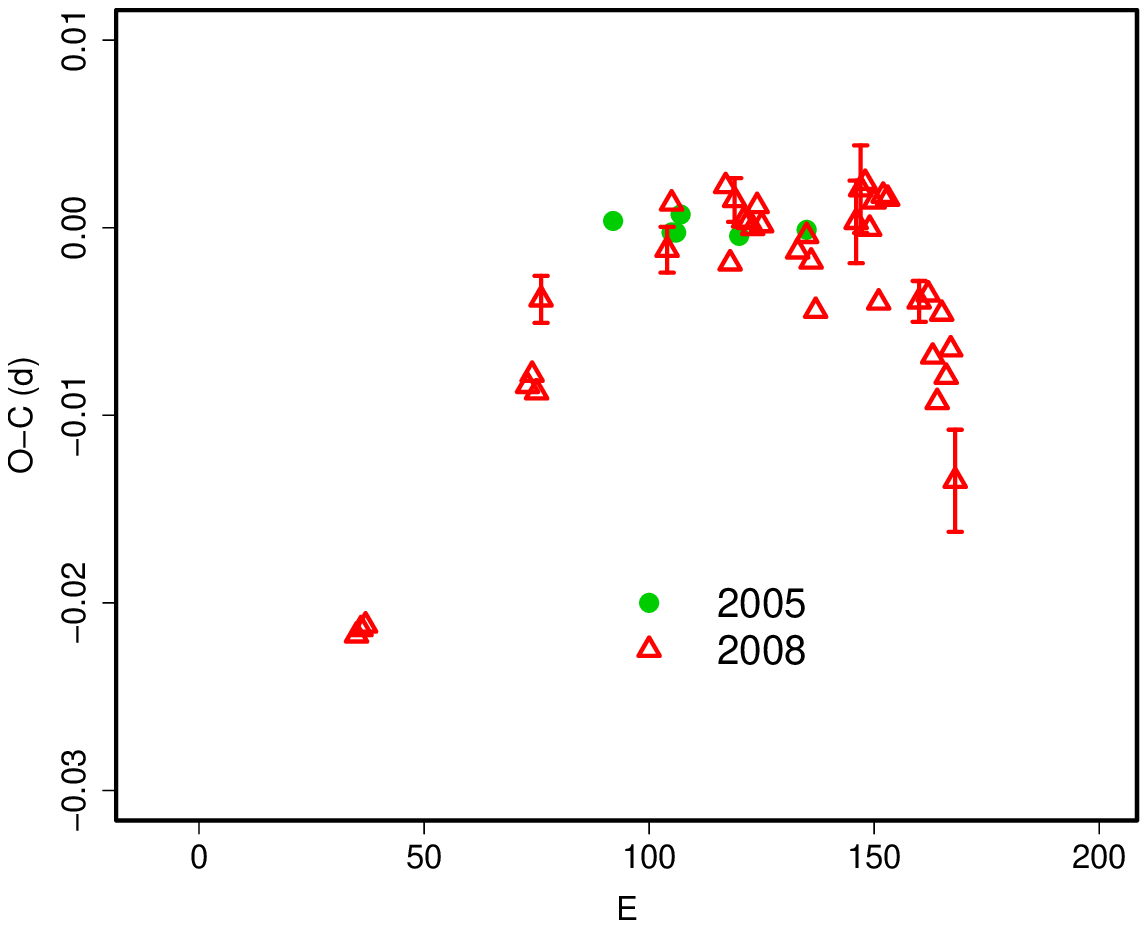}
  \end{center}
  \caption{Comparison of $O-C$ diagrams of TY Psc between different
  superoutbursts.  A period of 0.07035 d was used to draw this figure.
  Approximate cycle counts ($E$) after the start of the
  superoutburst were used.
  }
  \label{fig:typsccomp}
\end{figure}

   We observed the 2005 and 2008 superoutbursts
(tables \ref{tab:typscoc2005}, \ref{tab:typscoc2008}).
The global $P_{\rm dot}$ during the 2005 superoutburst was
$+1.5(3.0) \times 10^{-5}$.  A stage B--C transition
was observed during the 2008 superoutburst, although this outburst
may have had a prolonged state with stage A superhumps
(figure \ref{fig:typsccomp}).
The nominal global superhump period and derivative were 0.07045(2) d and
$P_{\rm dot}$ = $-9.2(0.8) \times 10^{-5}$, respectively.

\begin{table}
\caption{Superhump maxima of TY Psc (2005).}\label{tab:typscoc2005}
\begin{center}

\end{center}
\end{table}

\begin{table}
\caption{Superhump maxima of TY Psc (2008).}\label{tab:typscoc2008}
\begin{center}
%
\end{center}
\end{table}

\subsection{EI Piscium}\label{obj:eipsc}

\begin{figure}
  \begin{center}
    \FigureFile(88mm,110mm){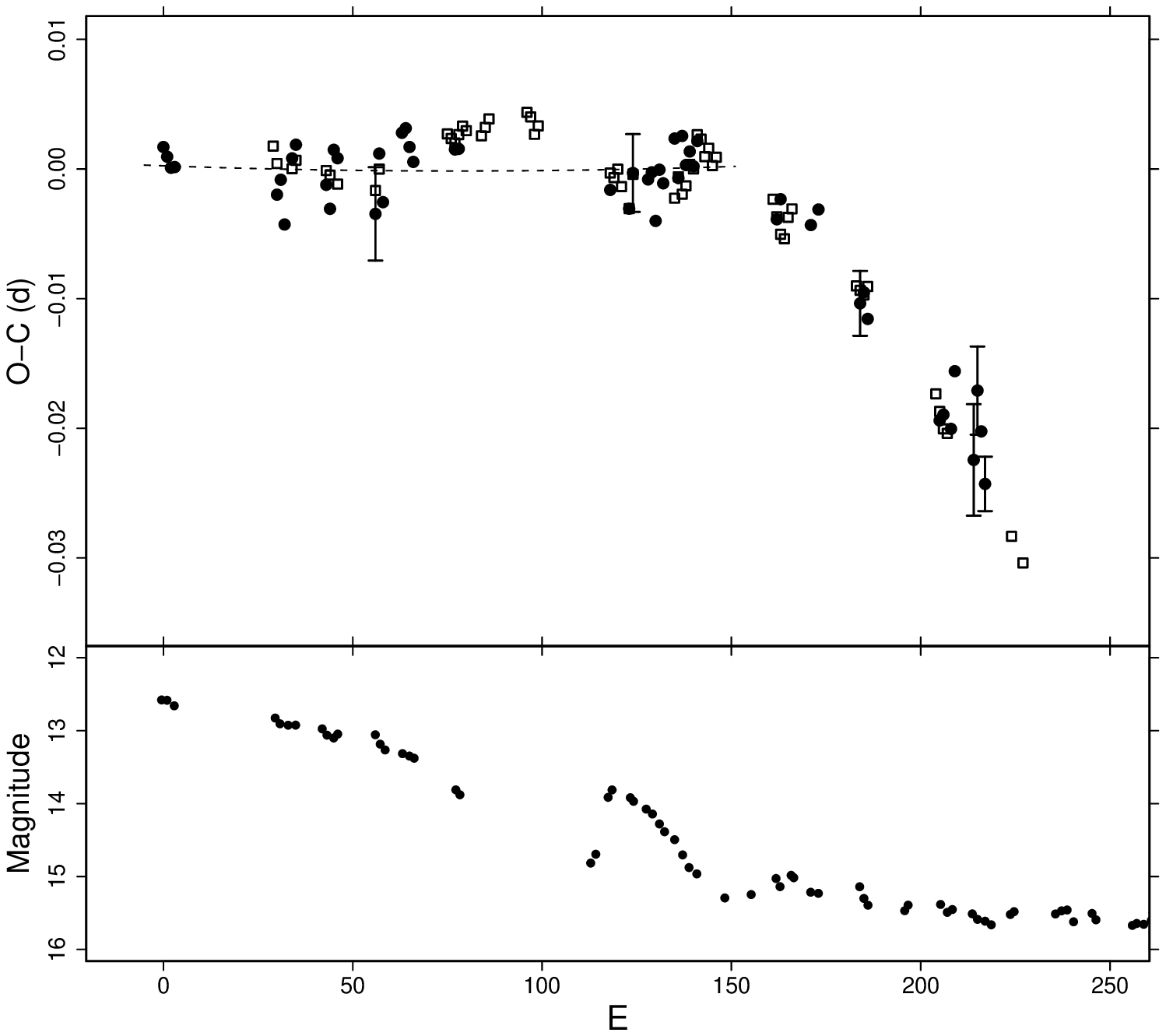}
  \end{center}
  \caption{$O-C$ diagram of EI Psc during the superoutburst in 2001.
  (Upper) $O-C$ diagram.  The filled circles and open squares represent
  maxima presented here and maxima reported in \citet{ski02j2329}.
  We used only the former set of maxima in order to avoid a systematic
  error potentially caused by superimposed orbital modulations.
  The dashed curve corresponds to $P_{\rm dot}$ = $+0.3 \times 10^{-5}$.
  (Lower) light curve.
  }
  \label{fig:eipscoc2001}
\end{figure}

   EI Psc (=1RXS J232953.9$+$062814) is one of two (the other being
V485 Cen) unusually short-$P_{\rm SH}$ SU UMa-type dwarf novae with
evolved secondaries (\cite{uem02j2329letter}; \cite{ski02j2329}).
Since the orbital variation, arising from the ellipsoidal variation of
the secondary star, is strong, we subtracted the mean orbital variation
from the raw data in \citet{uem02j2329}.  The resultant times of superhump
maxima are listed in table \ref{tab:eipscoc2001}.
A combination of the times of reported superhumps in \citet{ski02j2329}
yielded a slightly discontinuous $O-C$ variation, although the transition
to a shorter period was recorded in both sets of observations
(figure \ref{fig:eipscoc2001}).
The discrepancy between these analyses was largest between the fading stage
of the main superoutburst and the rebrightening, suggesting that the
times of maxima in \citet{ski02j2329} were more affected by orbital variations.
We therefore used times in \citet{uem02j2329}, updated here, and obtained
$P_{\rm dot}$ = $+0.3(0.8) \times 10^{-5}$ ($E \le 141$).
The period then experienced a transition to a shorter one 0.046090(12) d.
We regard this transition as a stage B--C transition based on the
$O-C$ characteristics.  Since this transition is usually observed during
the superoutburst plateau in most SU UMa-type dwarf novae, the existence
of a transition around the rebrightening looks peculiar to EI Psc.

\begin{figure}
  \begin{center}
    \FigureFile(88mm,70mm){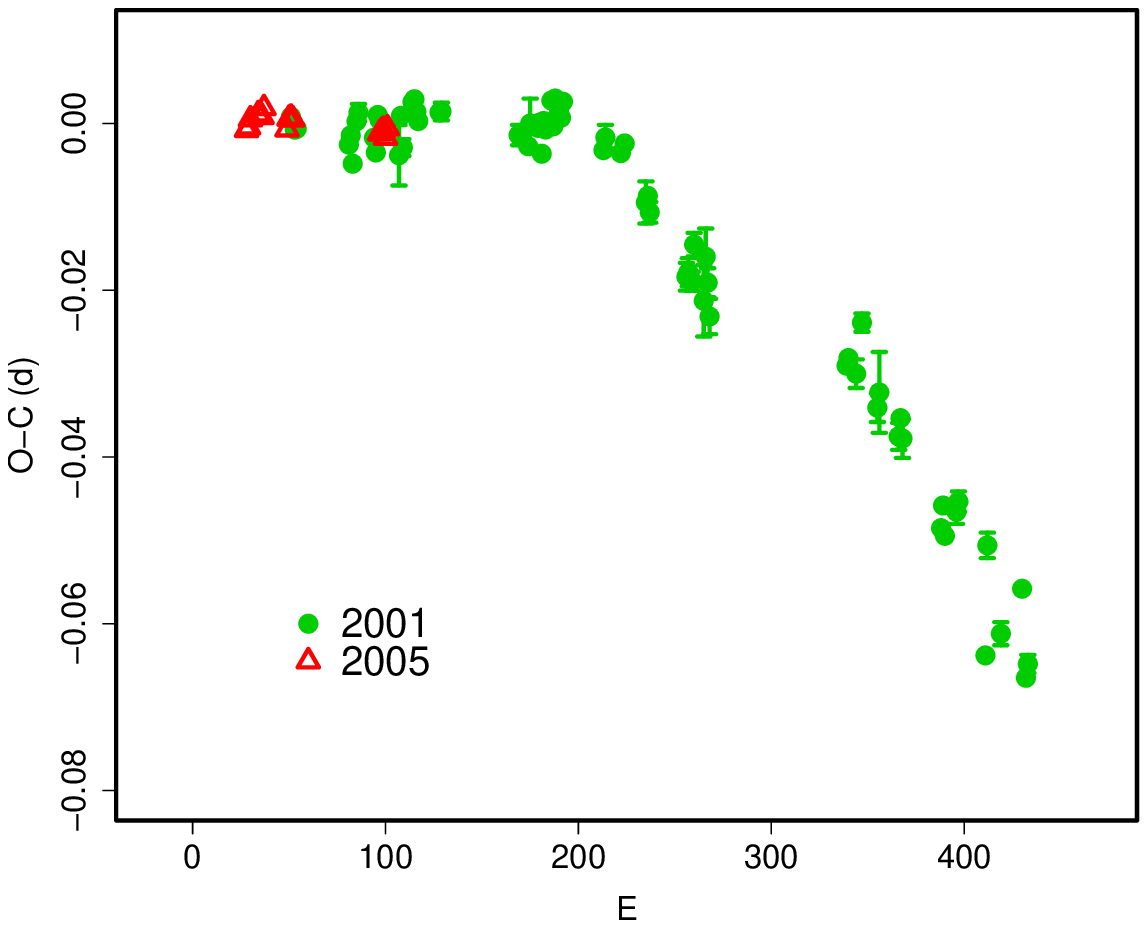}
  \end{center}
  \caption{Comparison of $O-C$ diagrams of EI Psc between different
  superoutbursts.  A period of 0.04634 d was used to draw this figure.
  Approximate cycle counts ($E$) after the start of the
  superoutburst were used.
  Since the start of the 2001 superoutburst was unknown, the $E$ was
  shifted assuming that the two superoutbursts have the same duration
  of the plateau phase.
  }
  \label{fig:eipsccomp}
\end{figure}

   We also analyzed the 2005 superoutburst and obtained the times of
superhump maxima (table \ref{tab:eipscoc2005}).  The global $P_{\rm dot}$
was $-2.8(2.0) \times 10^{-5}$, although there may have been a break
in the $O-C$ diagram around $E=9$.  This possible break may be a stage
A--B transition (cf. figure \ref{fig:eipsccomp}).
This superoutburst exhibited a rebrightening in a similar way as in
the 2001 one.

\begin{table}
\caption{Superhump maxima of EI Psc (2001).}\label{tab:eipscoc2001}
\begin{center}

\end{center}
\end{table}

\addtocounter{table}{-1}
\begin{table}
\caption{Superhump maxima of EI Psc (2001). (continued)}
\begin{center}
%
\end{center}
\end{table}

\begin{table}
\caption{Superhump maxima of EI Psc (2005).}\label{tab:eipscoc2005}
\begin{center}
%
\end{center}
\end{table}

\subsection{VZ Pyxidis}\label{obj:vzpyx}

   VZ Pyx was identified as an SU UMa-type dwarf nova by \citet{kat97vzpyx}.
We observed the 2008 superoutburst (table \ref{tab:vzpyxoc2008}).
Since multiple maxima apparently appeared around and after the rapid
fading stage ($E \ge 120$), we restricted our analysis to $E < 120$.
Although a global $P_{\rm dot}$ = $-16.3(1.3) \times 10^{-5}$
($E \le 80$) was obtained, there was apparently a break in the period
between $E=27$ and $E=54$.  In table \ref{tab:perlist}, we presented
the periods based in this interpretation.
We also included a reanalysis of \citet{kat97vzpyx}
(table \ref{tab:vzpyxoc1996})
and the times of superhump maxima during the 2002 and 2004 superoutbursts
(tables \ref{tab:vzpyxoc2000}, \ref{tab:vzpyxoc2004}).
The 2000 superoutburst was observed during the terminal stage and
the 2004 superoutburst was observed between 5 and 9 d from the onset of
the outburst.  The period for the 2000 superoutburst could be considered
as a typical period for stage C superhumps in this object.

\begin{table}
\caption{Superhump maxima of VZ Pyx (1996).}\label{tab:vzpyxoc1996}
\begin{center}

\end{center}
\end{table}

\begin{table}
\caption{Superhump maxima of VZ Pyx (2000).}\label{tab:vzpyxoc2000}
\begin{center}
%
\end{center}
\end{table}

\begin{table}
\caption{Superhump maxima of VZ Pyx (2004).}\label{tab:vzpyxoc2004}
\begin{center}
%
\end{center}
\end{table}

\begin{table}
\caption{Superhump maxima of VZ Pyx (2008).}\label{tab:vzpyxoc2008}
\begin{center}
%
\end{center}
\end{table}

\subsection{DV Scorpii}\label{obj:dvsco}

\begin{figure}
  \begin{center}
    \FigureFile(88mm,110mm){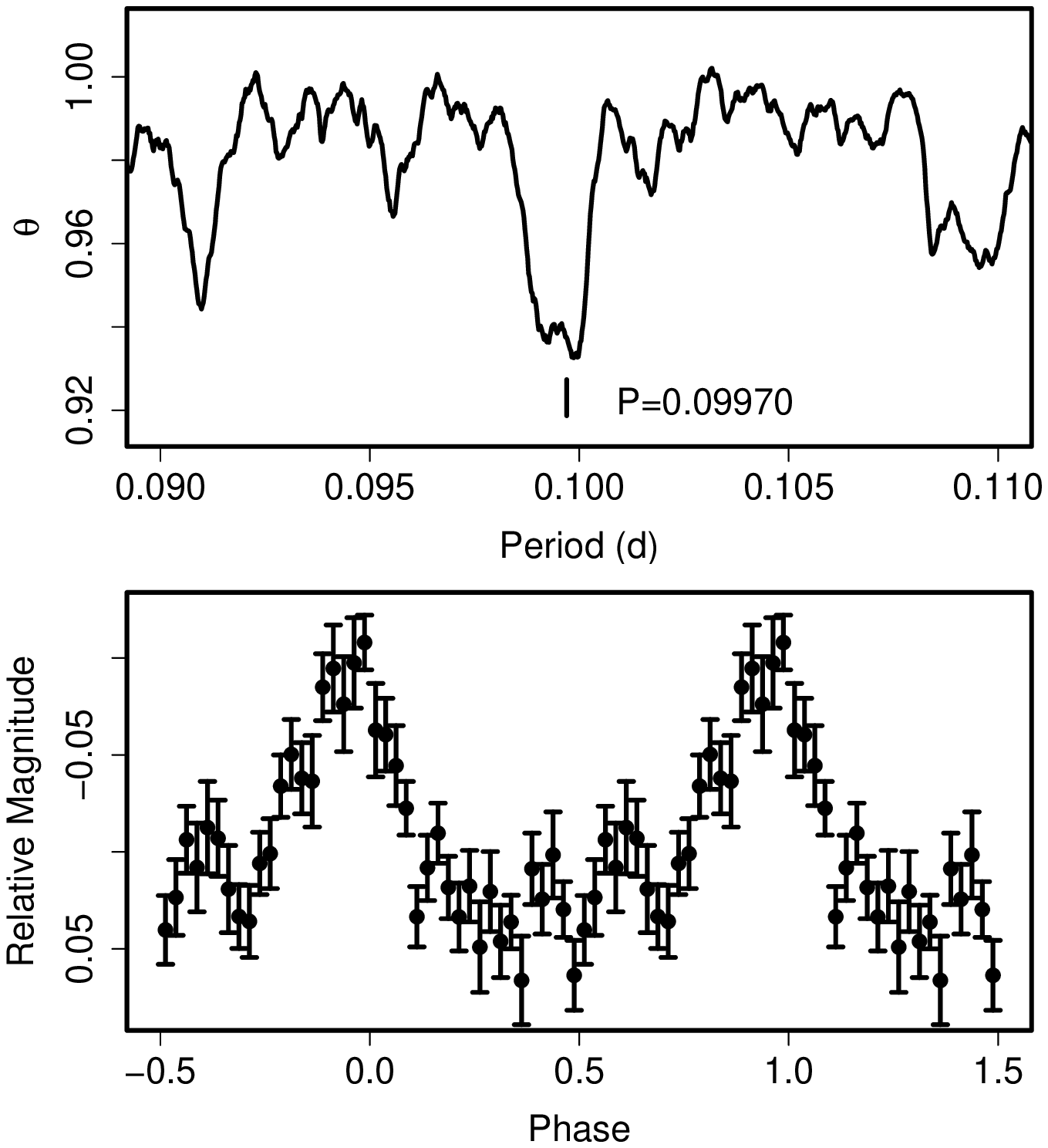}
  \end{center}
  \caption{Superhumps in DV Sco (2004). (Upper): PDM analysis.
     (Lower): Phase-averaged profile.}
  \label{fig:dvscoshpdm}
\end{figure}

   DV Sco was recently reclassified as a likely dwarf nova \citep{pas03dvsco}.
The SU UMa-type nature of this dwarf nova was established by B. Monard
(cf. vsnet-alert 8321, 8322) during its 2004 outburst.
This object is a dwarf nova in the period gap (vsnet-alert 8325).
We analyzed this superoutburst (table \ref{tab:dvscooc2004})
and another in 2008 (table \ref{tab:dvscooc2008}).
The mean superhump period with the PDM method was 0.09970(7) d
for the 2004 superoutburst (figure \ref{fig:dvscoshpdm}).
The resultant global $P_{\rm dot}$ for the 2004 superoutburst
was $-15.1(5.5) \times 10^{-5}$.  The 2008 superoutburst was observed
during its late course to its final decline.  Due to relatively large
error in maxima times and the short coverage, we did not attempt to
determine $P_{\rm dot}$.

\begin{table}
\caption{Superhump maxima of DV Sco (2004).}\label{tab:dvscooc2004}
\begin{center}

\end{center}
\end{table}

\begin{table}
\caption{Superhump maxima of DV Sco (2008).}\label{tab:dvscooc2008}
\begin{center}
%
\end{center}
\end{table}

\subsection{MM Scorpii}\label{obj:mmsco}

   The updated times of superhump maxima from the 2002 data
\citep{kat04nsv10934mmscoabnorcal86} are listed in table \ref{tab:mmscooc2002}.
The observation was likely performed in the middle of the stage B,
after 5 d of the visual maximum \citep{kat04nsv10934mmscoabnorcal86}.

\begin{table}
\caption{Superhump maxima of MM Sco (2002).}\label{tab:mmscooc2002}
\begin{center}
%
\end{center}
\end{table}

\subsection{NY Serpentis}\label{obj:nyser}

   We used the data in \citet{nog98nyser} to determine the refined times
of superhump maxima (table \ref{tab:nyseroc1996}).
Although the initial maximum was recorded during the developmental stage
of superhumps, we adopted $P_{\rm dot}$ = $-144(8) \times 10^{-5}$
using all maxima times because the effect of the evolutionary stage is
relatively small in systems with strongly negative $P_{\rm dot}$'s
(cf. UV Gem: subsection \ref{sec:uvgem}).
Excluding the initial maximum, the $P_{\rm dot}$ amounted to
$-117(27) \times 10^{-5}$.
More observations are needed to see if such an extreme period variation
is indeed present during the entire superoutburst.

\begin{table}
\caption{Superhump maxima of NY Ser (1996).}\label{tab:nyseroc1996}
\begin{center}
\begin{tabular}{ccccc}
\hline\hline
$E$ & max$^a$ & error & $O-C^b$ & $N^c$ \\
\hline
0 & 50195.2477 & 0.0029 & $-$0.0103 & 69 \\
18 & 50197.2027 & 0.0006 & 0.0145 & 71 \\
27 & 50198.1604 & 0.0007 & 0.0071 & 75 \\
28 & 50198.2642 & 0.0008 & 0.0037 & 59 \\
37 & 50199.2107 & 0.0014 & $-$0.0150 & 31 \\
\hline
  \multicolumn{5}{l}{$^{a}$ BJD$-$2400000.} \\
  \multicolumn{5}{l}{$^{b}$ Against $max = 2450195.2580 + 0.107235 E$.} \\
  \multicolumn{5}{l}{$^{c}$ Number of points used to determine the maximum.} \\
\end{tabular}
\end{center}
\end{table}

\subsection{RZ Sagittae}\label{obj:rzsge}

\begin{figure}
  \begin{center}
    \FigureFile(88mm,70mm){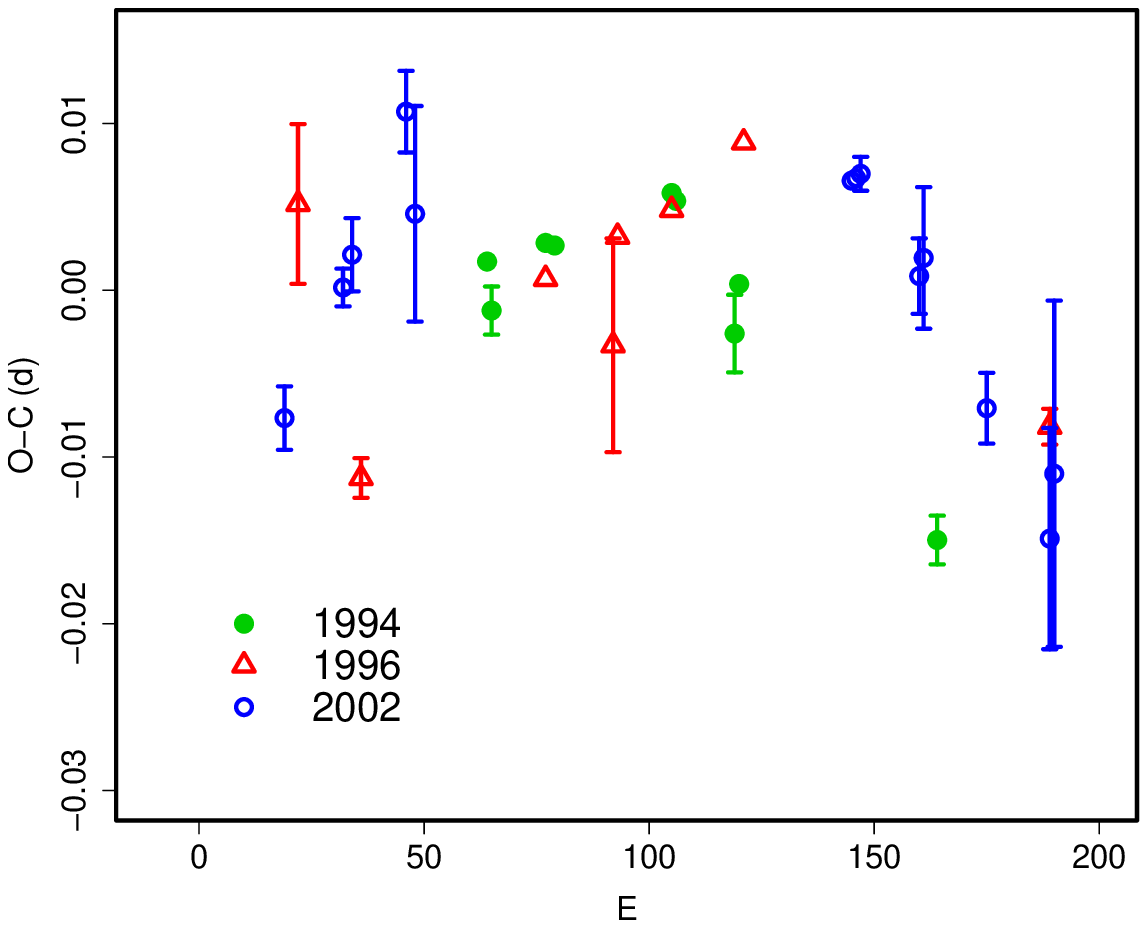}
  \end{center}
  \caption{Comparison of $O-C$ diagrams of RZ Sge between different
  superoutbursts.  A period of 0.07045 d was used to draw this figure.
  Approximate cycle counts ($E$) after the start of the
  superoutburst were used.
  The 1994 superoutburst probably had a shorter stage B.
  }
  \label{fig:rzsgecomp}
\end{figure}

   \citet{kat96rzsge} reported on the 1994 superoutburst, giving
$P_{\rm dot}$ = $-10(2) \times 10^{-5}$.  Table \ref{tab:rzsgeoc1994}
gives refined and newly measured times of superhump maxima from the
data used in \citet{kat96rzsge}.
The refined global $P_{\rm dot}$ corresponds to $-11.0(2.2) \times 10^{-5}$.
The 1996 superoutburst was observed by us and by \citet{sem97rzsge}.
A combined list of superhump maxima is given in table \ref{tab:rzsgeoc1996}.
The global $P_{\rm dot}$ corresponds to $-6.9(1.6) \times 10^{-5}$.
The difference in $P_{\rm dot}$ from \citet{sem97rzsge} was probably
because they only observed the late stage of the superoutburst.
There is an indication of a transition from a longer to a shorter
period (already somewhat evident on the figure 4 in \cite{sem97rzsge}),
corresponding to a stage B--C transition.
If we restrict the fit to $E < 100$, we obtain $P_{\rm dot}$ =
$+0.6(5.1) \times 10^{-5}$ indicating a relatively constant superhump
period.  This phenomenon may be analogous to the one observed in
TT Boo \citep{ole04ttboo}, another SU UMa-type dwarf nova with
a relatively long superhump period and long superoutbursts
(see also FQ Mon, subsection \ref{sec:fqmon}).
We also observed the 2002 superoutburst (table \ref{tab:rzsgeoc2002}).
Although the coverage was not sufficient (our observation covered the
early to middle stage of the superoutburst), we obtained the global
$P_{\rm dot}$ = $-4.9(3.0) \times 10^{-5}$.
A comparison of $O-C$ diagrams between different superoutbursts
is shown in figure \ref{fig:rzsgecomp}.  The 1994 superoutburst may have
had a shorter stage B than in other superoutbursts.

\begin{table}
\caption{Superhump maxima of RZ Sge (1994).}\label{tab:rzsgeoc1994}
\begin{center}

\end{center}
\end{table}

\begin{table}
\caption{Superhump maxima of RZ Sge (1996).}\label{tab:rzsgeoc1996}
\begin{center}
%
\end{center}
\end{table}

\begin{table}
\caption{Superhump maxima of RZ Sge (2002).}\label{tab:rzsgeoc2002}
\begin{center}
%
\end{center}
\end{table}

\subsection{WZ Sagittae}\label{sec:wzsge}\label{obj:wzsge}

\begin{figure}
  \begin{center}
    \FigureFile(88mm,110mm){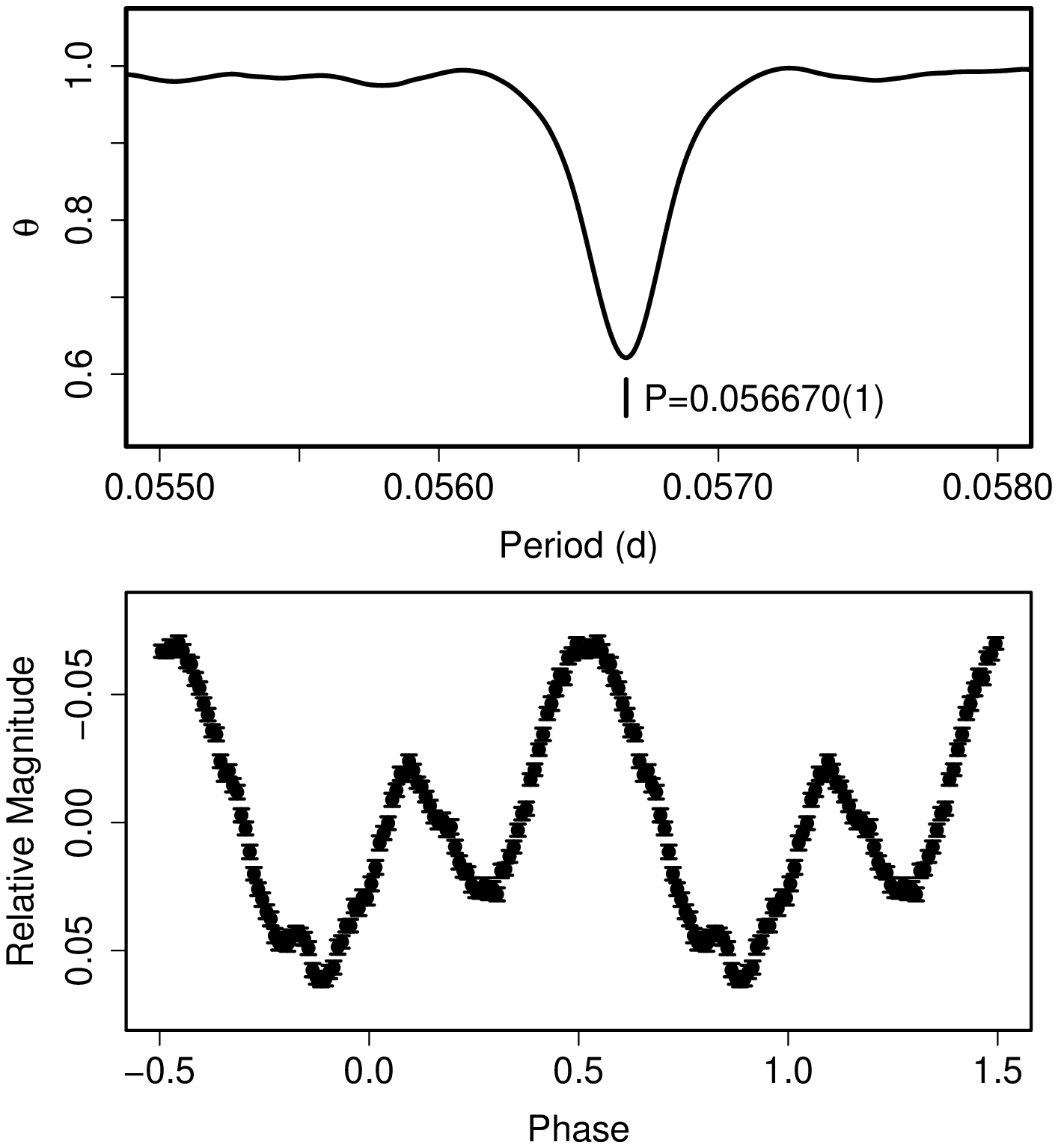}
  \end{center}
  \caption{Early superhumps in WZ Sge (2001). (Upper): PDM analysis.
  (Lower): Phase-averaged profile.  The phase zero corresponds
  to eclipses in quiescence.}
  \label{fig:wzsgeeshpdm}
\end{figure}

   Several authors reported on the 2001 superoutburst of WZ Sge
(\cite{pat02wzsge}; \cite{ish02wzsgeletter}).  We used the data
used in \citet{ish02wzsgeletter} to determine superhump maxima.
We deal with ordinary superhumps and give only a representative
figure of early superhumps (figure \ref{fig:wzsgeeshpdm}).

   We extracted times of superhump maxima after subtracting the
general trend of the outburst and subtracting averaged orbital
variation as in V455 And.  The interval for averaging the orbital
variation was 4--6 d during the superoutburst plateau and 1 d
for the final stage of early superhumps and the final stage of
the superoutburst plateau.

   The tables of maxima are separately given for the earlier half
before double humps became apparent (table \ref{tab:wzsgeoc2001}) and
the final stage when newly arising humps became apparent
(table \ref{tab:wzsgeoc2001b}) because different base periods were
used for calculating the $O-C$'s.  The humps having orbital phases
$0.6 < {\rm phase} < 1.0$ in the latter table can be attributed to
orbital humps.  The situation can be clearly seen on the combined
$O-C$ diagram during this stage and the early part the subsequent
rebrightening phase (figure \ref{fig:wzsgemainoc}).
It is evident from the $O-C$ diagram that our method is less affected
by the orbital (eclipse) feature than in \citet{pat02wzsge}, enabling
a firmer estimate of the period variation.
The interval $E \le 27$ showed an early-stage transition with a
longer period (stage A).  Since the orbital phases of these humps
do not coincide either of two maxima of early superhumps, we regard
them as genuine superhumps.  The mean period was 0.05839(6) d.

   The mean $P_{\rm SH}$ and $P_{\rm dot}$ for $27 \le E \le 177$\footnote{
   The epochs $E > 165$ in this paragraph denotes maxima in table
   \ref{tab:wzsgeoc2001b}.  The epoch $E=0$ in table \ref{tab:wzsgeoc2001b}
   corresponds to $E=169$.
}
(stage B) was 0.057204(5) d and $P_{\rm dot}$ = $+2.0(0.4) \times 10^{-5}$.
During the last stage of the superoutburst plateau, rapid fading
and the dip, the orbital humps dominated (see figure \ref{fig:wzsgemainoc}).
A new series of superhumps with a longer period emerged
(filled circles in figure \ref{fig:wzsgemainoc} for $E>200$)
during the rapid fading and smoothly evolved into superhumps
during the rebrightening phase.  The mean period and period derivative
of these superhumps for $200 \le E \le 400$ were 0.057488(14) d
and $P_{\rm dot}$ = $+5.0(0.7) \times 10^{-5}$.

   We also analyzed the rebrightening phase.  The analysis follows
the similar manner as in SDSS J0804 \citep{kat09j0804}.
The phase-averaged light curve (figure \ref{fig:wzsgerebph}) closely
resembles that of SDSS J0804 and is in good agreement with
the analysis by \citet{pat02wzsge}.  After subtracting orbital light
curves averaged over three days, we extracted the times of measured
maxima (table \ref{tab:wzsgeoc2001reb}).
The $P_{\rm SH}$ was 0.057501(12) d for $E \le 199$ and was
0.057305(11) d for $E \ge 200$ (see figure \ref{fig:wzsgereboc}).
These periods are 0.52(2) \% and 0.18(2) \% longer than the
mean $P_{\rm SH}$ during the main superoutburst, respectively.
These long-period superhumps correspond to long-period
late(-stage) superhumps reported in \citet{kat08wzsgelateSH}.

During the post-superoutburst stage, although eclipses and orbital humps
were prominent (figure \ref{fig:wzsgelateph}), overlapping superhumps
persisted at least for $\sim$ 600 cycles ($\sim$ 30 d).
The times of maxima, determined after subtracting the orbital modulations,
during the post-superoutburst stage are listed in
table \ref{tab:wzsgeoc2001late}.
The interval for averaging the orbital variation was 10 d.
For $E \le 598$, the mean $P_{\rm SH}$ and $P_{\rm dot}$ were
0.057351(3) d and $+0.5(0.1) \times 10^{-5}$.
This period is 0.25(1) \% longer than the mean $P_{\rm SH}$ during
the main superoutburst.
There was some indication of the persisting superhumps after $E = 848$
with a different period before $E=598$.

The overall $O-C$ behavior during the entire course of the superoutburst
is shown in figure \ref{fig:wzsgehumpall}.  The behavior is remarkably
similar to GW Lib (subsection \ref{sec:latestage}).
In WZ Sge, a disturbance in the $O-C$ diagram also appeared during the rapid
fading stage and subsequent ``dip'' phase.  During the rebrightening
and post-superoutburst stages, the superhump period lengthened in
a similar way to GW Lib.  The $O-C$ diagram showed a slightly positive
deviation from this overall trend during the rebrightening phase.
The $O-C$ behavior after the rebrightening phase appears to be a natural
extension of the stage B superhumps, as in GW Lib.

\begin{figure*}
  \begin{center}
    \FigureFile(150mm,30mm){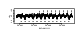}
  \end{center}
  \caption{Transition from early superhumps to ordinary superhumps
  in WZ Sge (2001).  The open circles represent minima of early
  superhumps.  The stage A superhumps (ticks) smoothly developed
  from one of two peaks of early superhumps.}
  \label{fig:wzsgeeshtrans}
\end{figure*}

\begin{figure*}
  \begin{center}
    \FigureFile(140mm,140mm){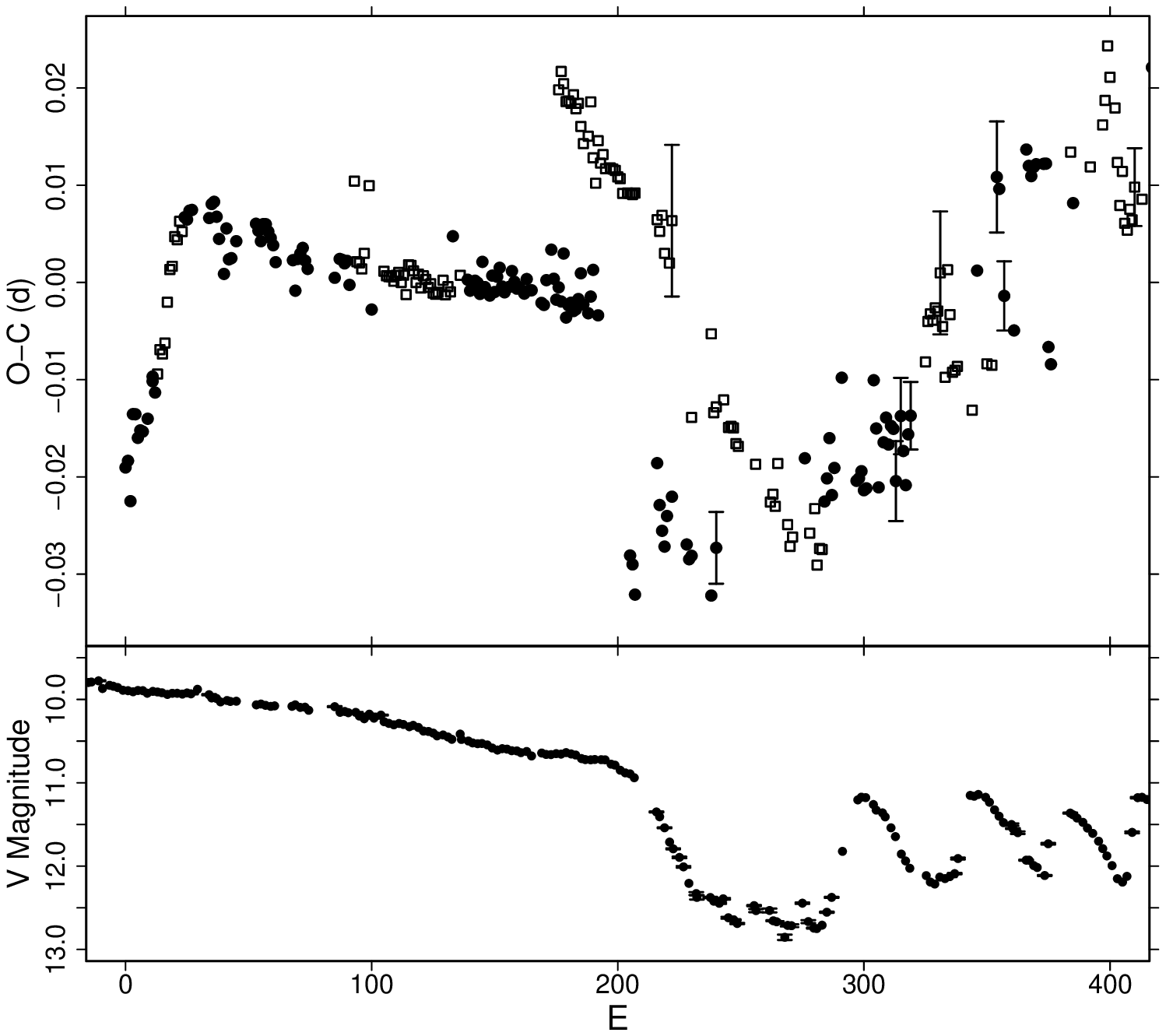}
  \end{center}
  \caption{$O-C$ variation in WZ Sge (2001).  (Upper) $O-C$.
  Open squares indicate humps coinciding with the phase of orbital humps.
  Filled circles are humps outside the phase of orbital humps.
  We used a period of 0.057244 d for calculating the $O-C$'s.
  The evolution of the $O-C$ diagram is remarkably similar
  to that of GW Lib (figure \ref{fig:gwlibhumpall}).
  (Lower) Light curve.
  }
  \label{fig:wzsgemainoc}
\end{figure*}

\begin{figure}
  \begin{center}
    \FigureFile(88mm,70mm){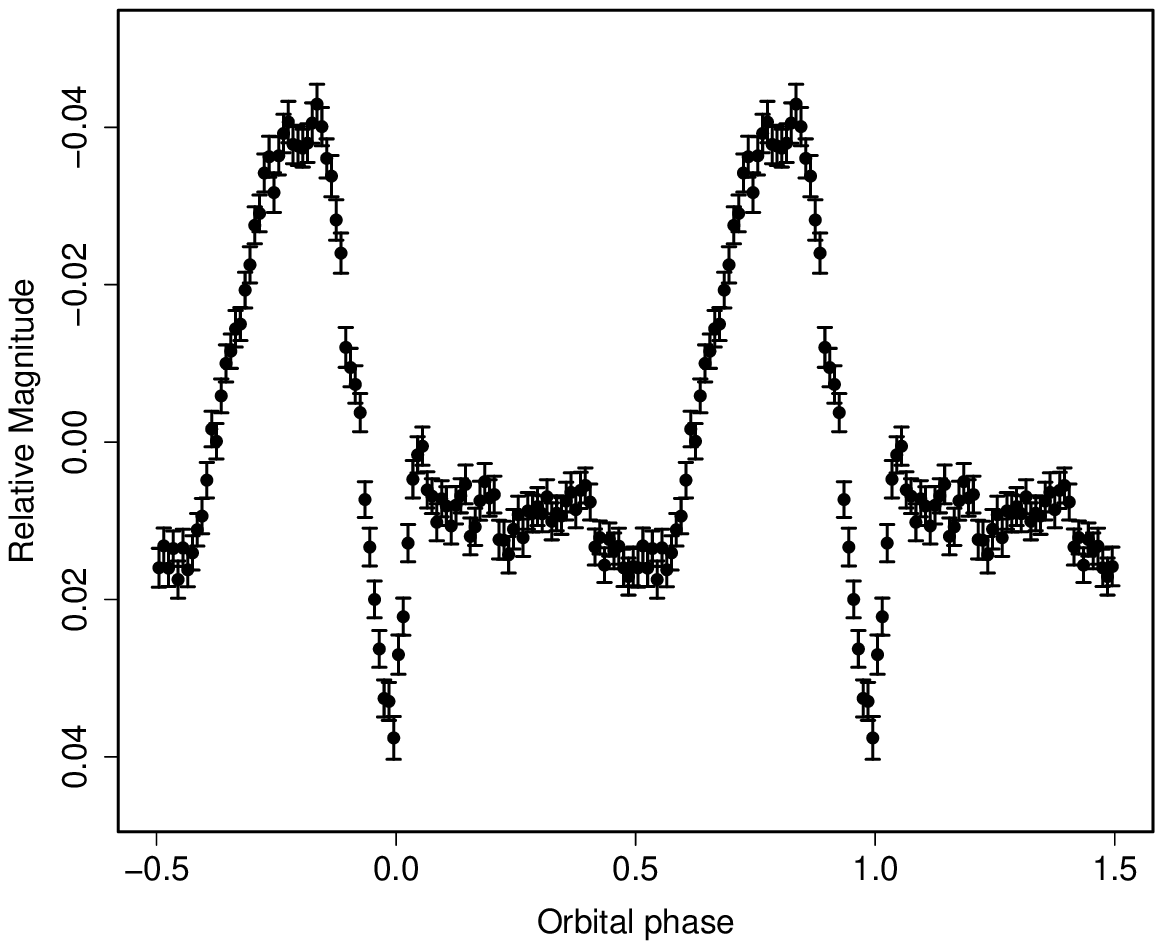}
  \end{center}
  \caption{Orbital light curve of WZ Sge during the rebrightening
  phase of the 2001 superoutburst (BJD 2452141--2452167)}
  \label{fig:wzsgerebph}
\end{figure}

\begin{figure}
  \begin{center}
    \FigureFile(88mm,110mm){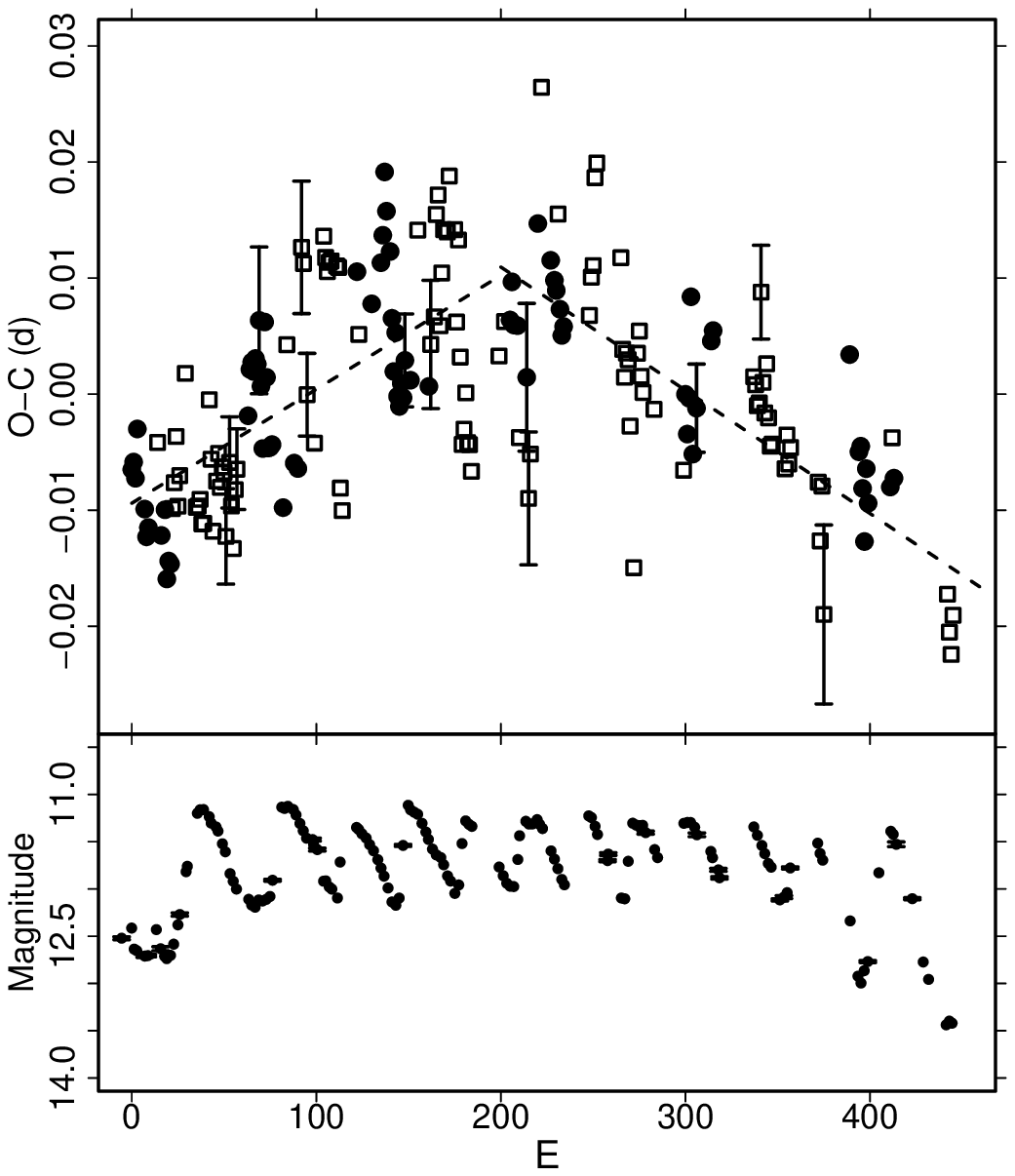}
  \end{center}
  \caption{$O-C$ of humps during the rebrightening phase of WZ Sge (2001).
  (Upper): $O-C$ diagram.  Filled squares and open squares represent
  superhumps and humps coinciding with orbital humps, respectively.
  Two dashes represent the superhump periods of 0.057501(12) d and
  0.057305(11) d.  (Lower): Light curve.}
  \label{fig:wzsgereboc}
\end{figure}

\begin{figure}
  \begin{center}
    \FigureFile(88mm,70mm){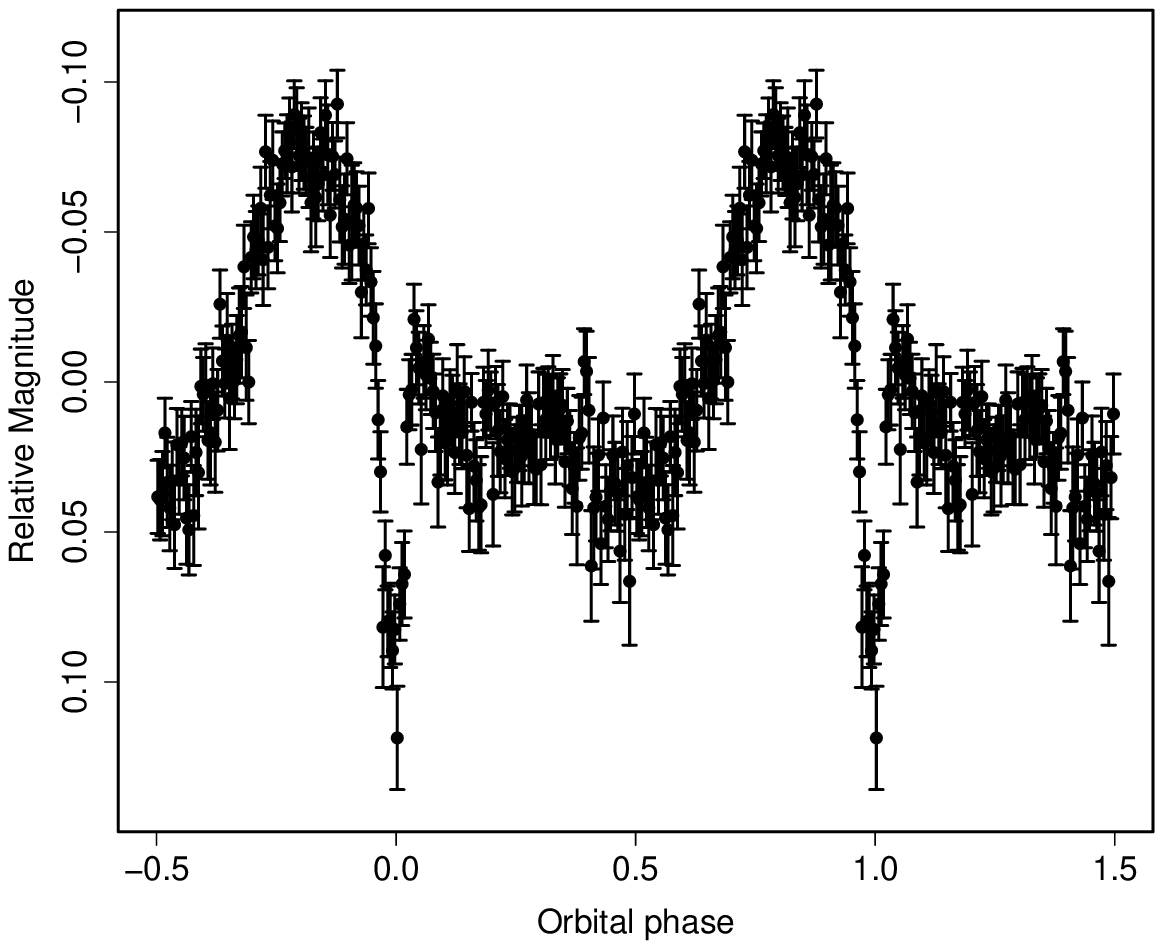}
  \end{center}
  \caption{Orbital light curve of WZ Sge during the post-superoutburst
  stage (BJD 2452167--2452267)}
  \label{fig:wzsgelateph}
\end{figure}

\begin{figure*}
  \begin{center}
    \FigureFile(160mm,160mm){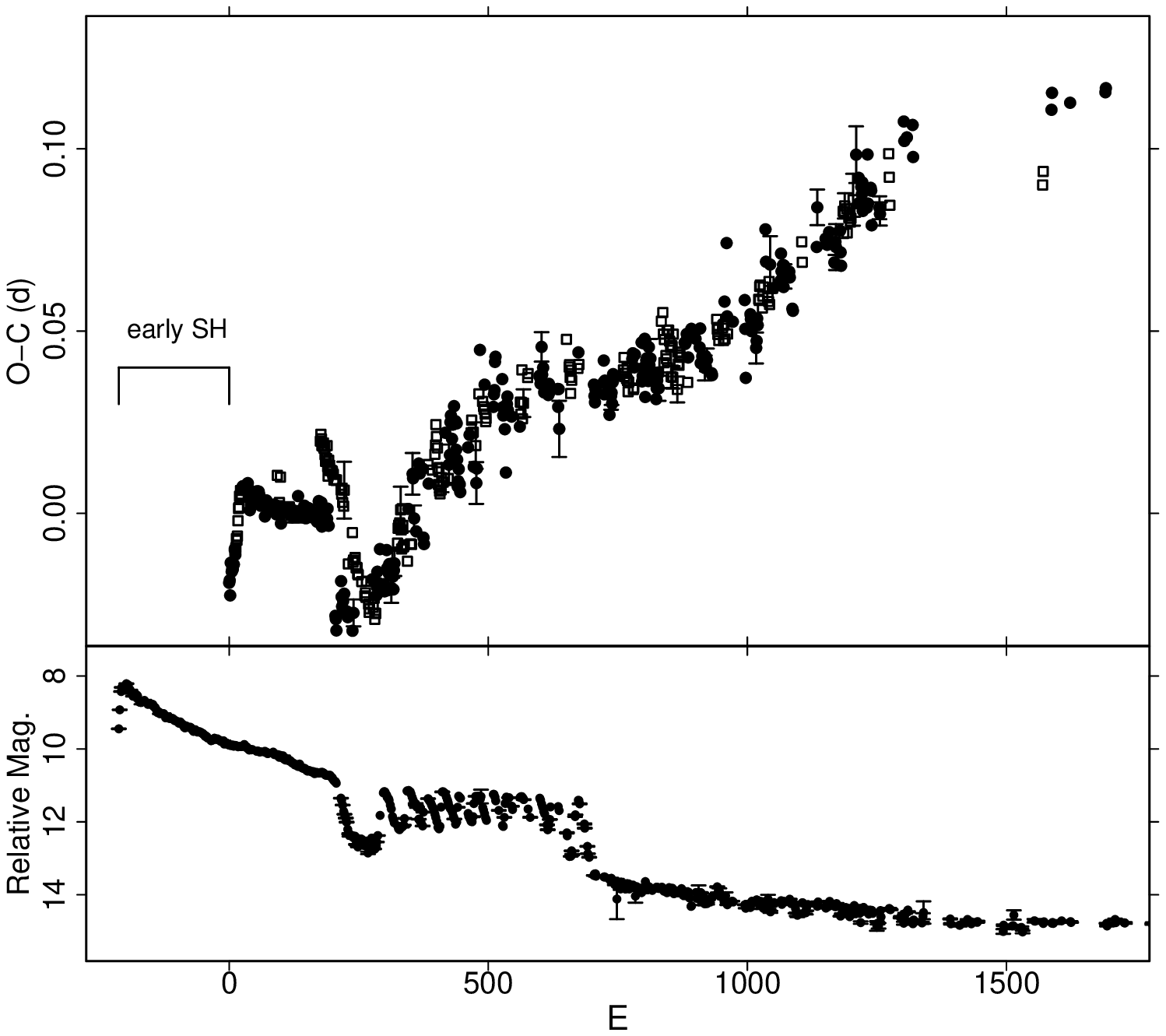}
  \end{center}
  \caption{$O-C$ variation in WZ Sge (2001).  (Upper) $O-C$.
  Open squares and filled circles represent superhumps
  and humps coinciding with orbital humps, respectively.
  We used a period of 0.057244 d for calculating the $O-C$'s.
  The global evolution of the $O-C$ diagram is remarkably similar
  to that of GW Lib (figure \ref{fig:gwlibhumpall}).
  (Lower) Light curve.
  }
  \label{fig:wzsgehumpall}
\end{figure*}

The times of superhump maxima during the 1978 superoutburst are
listed in table \ref{tab:wzsgeoc1978}.  The times were taken from
literature except for \citet{hei79wzsge}, for which we measured
the maxima from individual observations.  We obtained
$P_{\rm dot}$ = $+0.4(0.8) \times 10^{-5}$.

\begin{table}
\caption{Superhump maxima of WZ Sge (2001).}\label{tab:wzsgeoc2001}
\begin{center}

\end{center}
\end{table}

\addtocounter{table}{-1}
\begin{table}
\caption{Superhump maxima of WZ Sge (2001) (continued).}
\begin{center}
%
\end{center}
\end{table}

\addtocounter{table}{-1}
\begin{table}
\caption{Superhump maxima of WZ Sge (2001) (continued).}
\begin{center}
%
\end{center}
\end{table}

\begin{table}
\caption{Hump Maxima of WZ Sge during the end stage of the superoutburst plateau (2001).}\label{tab:wzsgeoc2001b}
\begin{center}
%
\end{center}
\end{table}

\begin{table}
\caption{Hump Maxima of WZ Sge during the end stage of the superoutburst plateau (2001).}
\begin{center}
%
\end{center}
\end{table}

\begin{table}
\caption{Superhump maxima of WZ Sge during the rebrightening phase (2001).}\label{tab:wzsgeoc2001reb}
\begin{center}
%
\end{center}
\end{table}

\addtocounter{table}{-1}
\begin{table}
\caption{Superhump maxima of WZ Sge during the rebrightening phase (2001) (continued).}
\begin{center}
%
\end{center}
\end{table}

\addtocounter{table}{-1}
\begin{table}
\caption{Superhump maxima of WZ Sge during the rebrightening phase (2001) (continued).}
\begin{center}
%
\end{center}
\end{table}

\addtocounter{table}{-1}
\begin{table}
\caption{Superhump maxima of WZ Sge during the rebrightening phase (2001) (continued).}
\begin{center}
%
\end{center}
\end{table}

\begin{table}
\caption{Superhump maxima of WZ Sge during the post-superoutburst stage (2001).}\label{tab:wzsgeoc2001late}
\begin{center}
%
\end{center}
\end{table}

\addtocounter{table}{-1}
\begin{table}
\caption{Superhump maxima of WZ Sge during the post-superoutburst stage (2001) (continued).}
\begin{center}
%
\end{center}
\end{table}

\addtocounter{table}{-1}
\begin{table}
\caption{Superhump maxima of WZ Sge during the post-superoutburst stage (2001) (continued).}
\begin{center}
%
\end{center}
\end{table}

\addtocounter{table}{-1}
\begin{table}
\caption{Superhump maxima of WZ Sge during the post-superoutburst stage (2001) (continued).}
\begin{center}
%
\end{center}
\end{table}

\begin{table}
\caption{Superhump maxima of WZ Sge (1978).}\label{tab:wzsgeoc1978}
\begin{center}
%
\end{center}
\end{table}

\subsection{AW Sagittae}\label{obj:awsge}

\begin{figure}
  \begin{center}
    \FigureFile(88mm,110mm){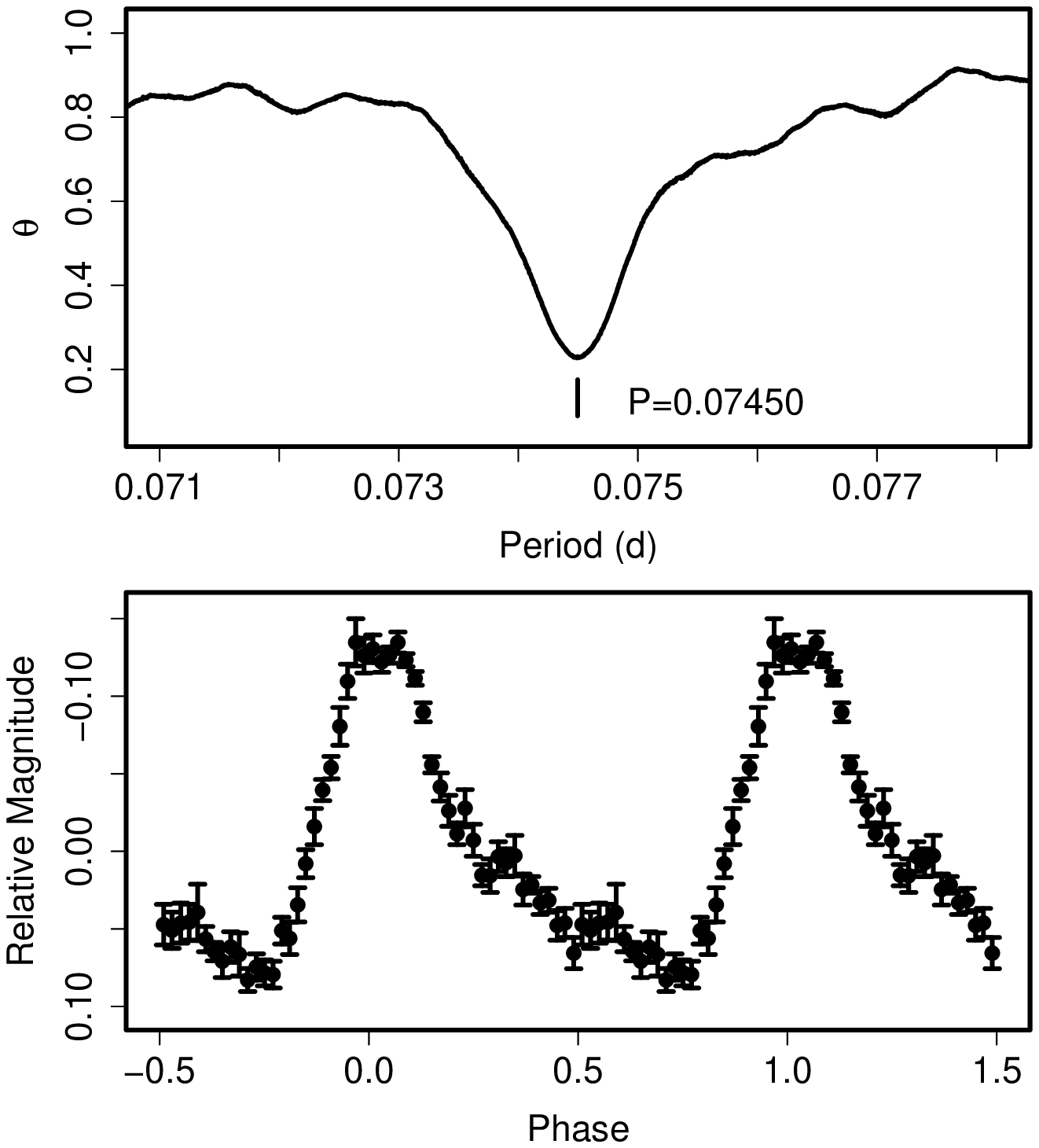}
  \end{center}
  \caption{Superhumps in AW Sge (2006). (Upper): PDM analysis.
     (Lower): Phase-averaged profile.}
  \label{fig:awsgeshpdm}
\end{figure}

   This dwarf nova has long been known since its early discovery
\citep{wol06awsge}.  The SU UMa-type nature was established during
the 2000 superoutburst (vsnet-alert 5111, 5112, 5114).
\citet{llo07awsge} summarized the history of outbursts of this object
and \citet{llo08awsge} presented observations during the 2007 normal outburst.
We analyzed the available AAVSO observation of the 2006 superoutburst,
a part of the data reported in \citet{she08awsge}.
The observation apparently covered the middle-to-late stage
of the superoutburst.
The times of superhump maxima are listed in table \ref{tab:awsgeoc2006}.
The mean $P_{\rm SH}$ with the PDM method was 0.07449(2) d
(figure \ref{fig:awsgeshpdm}) and
$P_{\rm dot}$ = $-7.9(6.4) \times 10^{-5}$, which may be a result
of combination of stage B and C superhumps.
We also give times of superhump maxima during the 2000 superoutburst
(table \ref{tab:awsgeoc2000}).  The mean mean $P_{\rm SH}$ with
the PDM method was 0.07473(8) d.

\begin{table}
\caption{Superhump maxima of AW Sge (2000).}\label{tab:awsgeoc2000}
\begin{center}
\begin{tabular}{ccccc}
\hline\hline
$E$ & max$^a$ & error & $O-C^b$ & $N^c$ \\
\hline
0 & 51741.5174 & 0.0010 & $-$0.0001 & 61 \\
12 & 51742.4131 & 0.0025 & 0.0014 & 40 \\
13 & 51742.4849 & 0.0012 & $-$0.0013 & 52 \\
\hline
  \multicolumn{5}{l}{$^{a}$ BJD$-$2400000.} \\
  \multicolumn{5}{l}{$^{b}$ Against $max = 2451741.5175 + 0.074519 E$.} \\
  \multicolumn{5}{l}{$^{c}$ Number of points used to determine the maximum.} \\
\end{tabular}
\end{center}
\end{table}

\begin{table}
\caption{Superhump maxima of AW Sge (2006).}\label{tab:awsgeoc2006}
\begin{center}
\begin{tabular}{ccccc}
\hline\hline
$E$ & max$^a$ & error & $O-C^b$ & $N^c$ \\
\hline
0 & 54056.3222 & 0.0007 & $-$0.0004 & 53 \\
30 & 54058.5593 & 0.0004 & 0.0009 & 139 \\
31 & 54058.6336 & 0.0003 & 0.0006 & 139 \\
43 & 54059.5258 & 0.0009 & $-$0.0015 & 47 \\
44 & 54059.6022 & 0.0004 & 0.0004 & 115 \\
\hline
  \multicolumn{5}{l}{$^{a}$ BJD$-$2400000.} \\
  \multicolumn{5}{l}{$^{b}$ Against $max = 2454056.3226 + 0.074528 E$.} \\
  \multicolumn{5}{l}{$^{c}$ Number of points used to determine the maximum.} \\
\end{tabular}
\end{center}
\end{table}

\subsection{V551 Sagittarii}\label{obj:v551sgr}

\begin{figure}
  \begin{center}
    \FigureFile(88mm,110mm){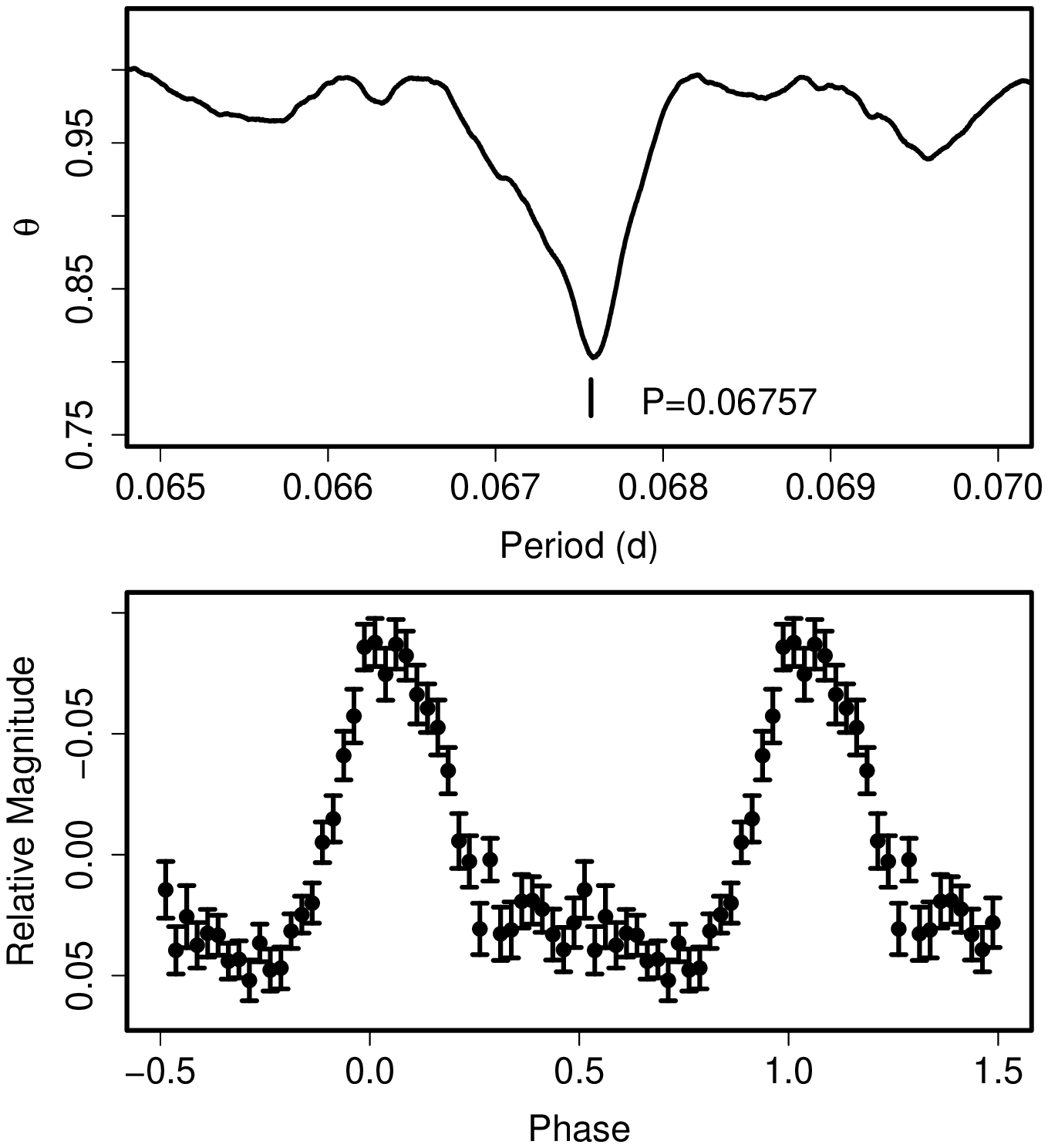}
  \end{center}
  \caption{Superhumps in V551 Sgr (2003). (Upper): PDM analysis.
     (Lower): Phase-averaged profile.}
  \label{fig:v551sgrshpdm}
\end{figure}

\begin{figure}
  \begin{center}
    \FigureFile(88mm,110mm){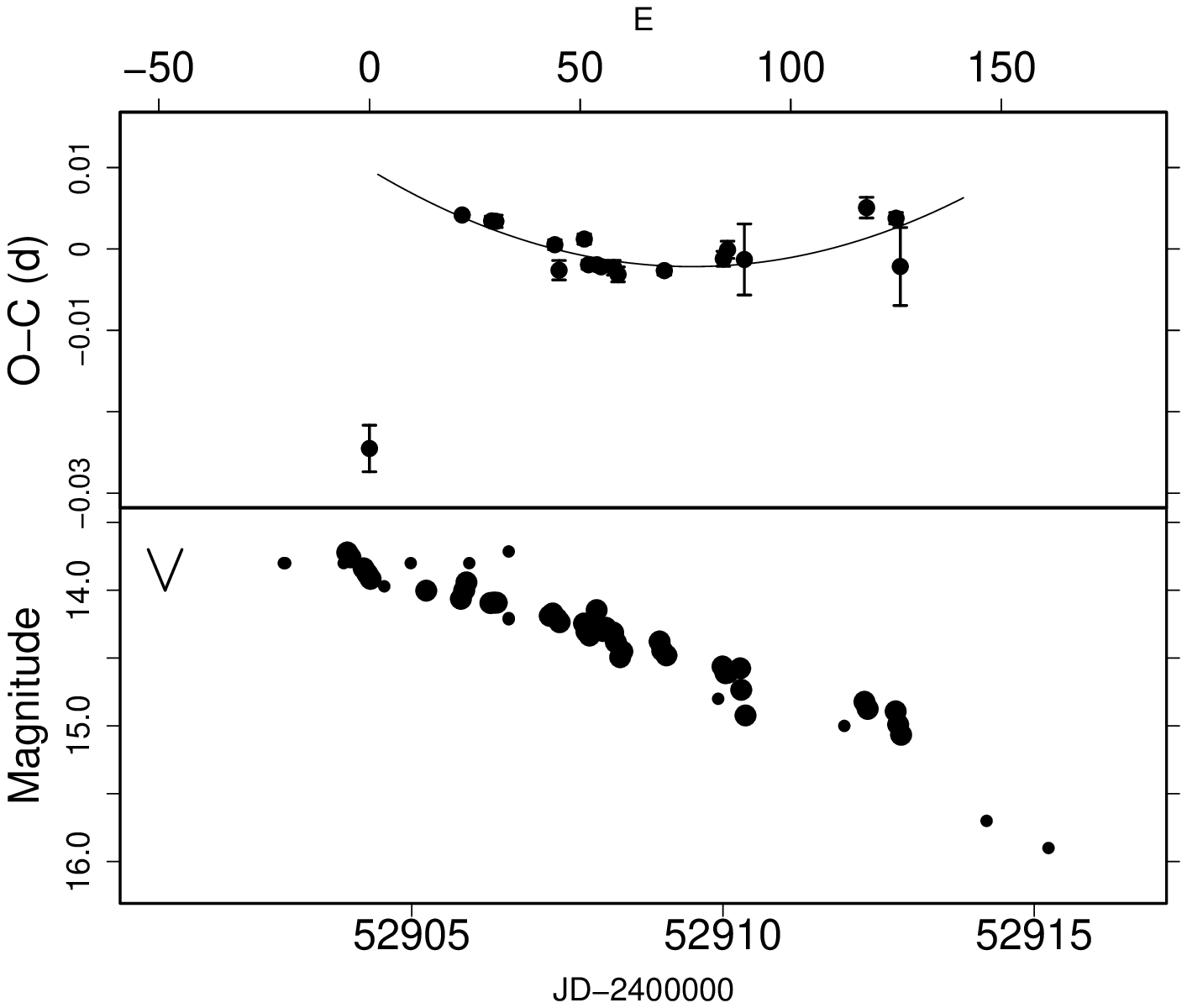}
  \end{center}
  \caption{$O-C$ of superhumps V551 Sgr (2003).
  (Upper): $O-C$ diagram.  The curve represents a quadratic fit to
  $E \ge 22$.
  (Lower): Light curve.
  Large dots represent CCD observations.  Small dots and a ``V'' mark
  represent visual observations and a upper limit, respectively.
  }
  \label{fig:v551sgr2003oc}
\end{figure}

   V551 Sgr has long been suspected to be a candidate WZ Sge-type
dwarf nova (cf. \cite{dow90wxcet}).
During the 2003 superoutburst, we managed to obtain excellent time-series
photometry.  A PDM analysis has yielded a mean period of 0.06757(1) d
(figure \ref{fig:v551sgrshpdm}).  The times of superhump maxima are
listed in table \ref{tab:v551sgroc2003}.
The $O-C$ diagram clearly shows a positive period derivative except
for the earliest part (figure \ref{fig:v551sgr2003oc}).
Excluding $E = 0$ (stage A), we obtained
$P_{\rm dot}$ = $+6.0(1.5) \times 10^{-5}$.
There were no indication of early superhumps.  Together with the
relatively long superhump period, and a likely supercycle of $\sim$ 1 yr,
the object is likely an SU UMa-type dwarf nova similar to UV Per
(subsection \ref{sec:uvper}) and QY Per (subsection \ref{sec:qyper}),
rather than a genuine WZ Sge-type object.

\begin{figure}
  \begin{center}
    \FigureFile(88mm,70mm){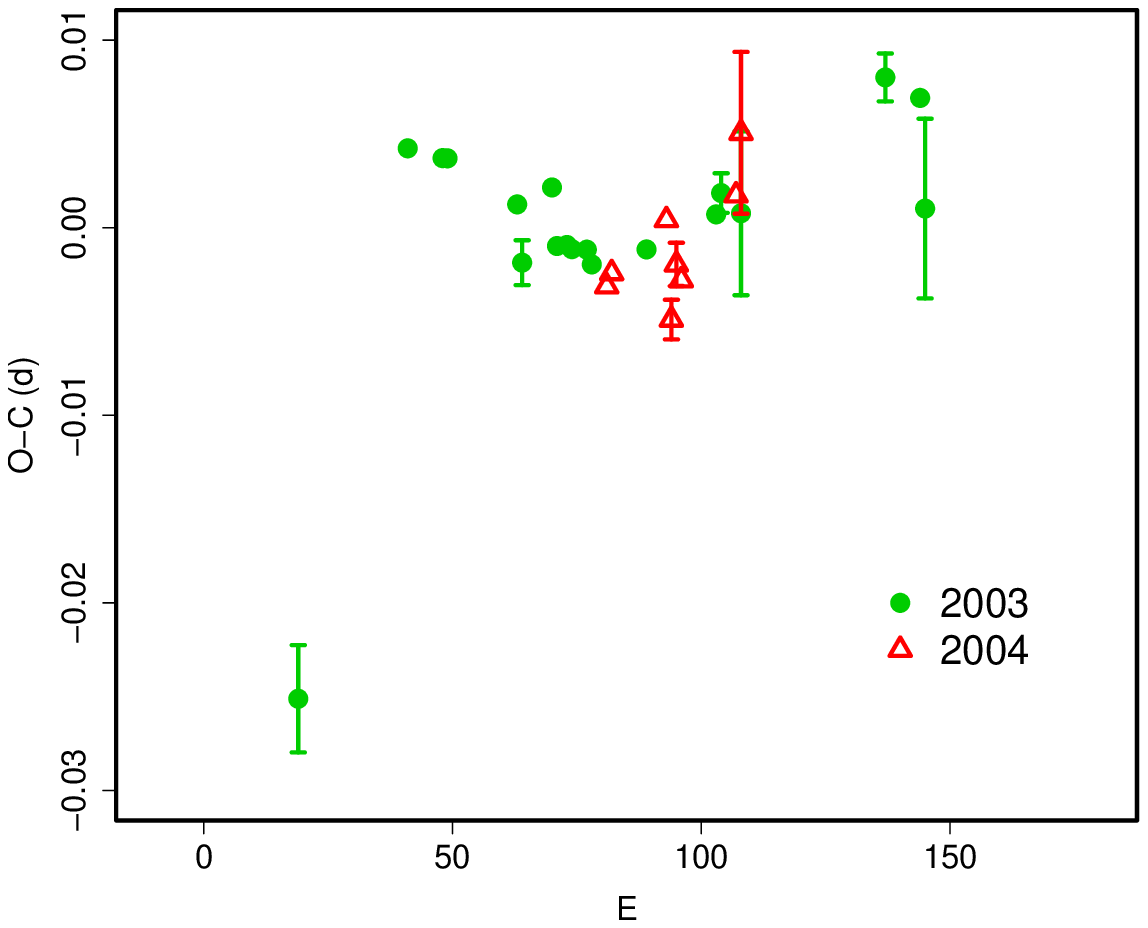}
  \end{center}
  \caption{Comparison of $O-C$ diagrams of V551 Sgr between different
  superoutbursts.  A period of 0.06757 d was used to draw this figure.
  Approximate cycle counts ($E$) after the start of the
  superoutburst were used.
  }
  \label{fig:v551sgrcomp}
\end{figure}

   The 2004 superoutburst was less sufficiently observed
(table \ref{tab:v551sgroc2004}).
The $P_{\rm dot}$ was likely positive and apparently recorded during
the stage B (figure \ref{fig:v551sgrcomp}), but we did not attempt
to measure the $P_{\rm dot}$ because of the short baseline.

\begin{table}
\caption{Superhump maxima of V551 Sgr (2003).}\label{tab:v551sgroc2003}
\begin{center}
\begin{tabular}{ccccc}
\hline\hline
$E$ & max$^a$ & error & $O-C^b$ & $N^c$ \\
\hline
0 & 52904.3030 & 0.0029 & $-$0.0187 & 68 \\
22 & 52905.8189 & 0.0003 & 0.0084 & 165 \\
29 & 52906.2914 & 0.0006 & 0.0072 & 68 \\
30 & 52906.3589 & 0.0008 & 0.0070 & 49 \\
44 & 52907.3025 & 0.0006 & 0.0032 & 74 \\
45 & 52907.3669 & 0.0012 & $-$0.0000 & 62 \\
51 & 52907.7764 & 0.0006 & 0.0034 & 123 \\
52 & 52907.8408 & 0.0006 & 0.0001 & 167 \\
54 & 52907.9760 & 0.0006 & $-$0.0000 & 32 \\
55 & 52908.0433 & 0.0004 & $-$0.0003 & 46 \\
58 & 52908.2460 & 0.0009 & $-$0.0007 & 61 \\
59 & 52908.3128 & 0.0009 & $-$0.0016 & 61 \\
70 & 52909.0569 & 0.0006 & $-$0.0019 & 45 \\
84 & 52910.0047 & 0.0009 & $-$0.0014 & 153 \\
85 & 52910.0734 & 0.0011 & $-$0.0004 & 177 \\
89 & 52910.3426 & 0.0044 & $-$0.0019 & 37 \\
118 & 52912.3094 & 0.0013 & 0.0024 & 63 \\
125 & 52912.7813 & 0.0007 & 0.0006 & 145 \\
126 & 52912.8430 & 0.0048 & $-$0.0054 & 154 \\
\hline
  \multicolumn{5}{l}{$^{a}$ BJD$-$2400000.} \\
  \multicolumn{5}{l}{$^{b}$ Against $max = 2452904.3217 + 0.067672 E$.} \\
  \multicolumn{5}{l}{$^{c}$ Number of points used to determine the maximum.} \\
\end{tabular}
\end{center}
\end{table}

\begin{table}
\caption{Superhump maxima of V551 Sgr (2004).}\label{tab:v551sgroc2004}
\begin{center}
\begin{tabular}{ccccc}
\hline\hline
$E$ & max$^a$ & error & $O-C^b$ & $N^c$ \\
\hline
0 & 53153.5658 & 0.0008 & 0.0010 & 150 \\
1 & 53153.6341 & 0.0007 & 0.0015 & 153 \\
12 & 53154.3802 & 0.0008 & 0.0018 & 151 \\
13 & 53154.4424 & 0.0011 & $-$0.0038 & 153 \\
14 & 53154.5129 & 0.0012 & $-$0.0011 & 153 \\
15 & 53154.5797 & 0.0008 & $-$0.0021 & 153 \\
26 & 53155.3275 & 0.0008 & $-$0.0002 & 131 \\
27 & 53155.3984 & 0.0043 & 0.0029 & 16 \\
\hline
  \multicolumn{5}{l}{$^{a}$ BJD$-$2400000.} \\
  \multicolumn{5}{l}{$^{b}$ Against $max = 2453153.5648 + 0.067803 E$.} \\
  \multicolumn{5}{l}{$^{c}$ Number of points used to determine the maximum.} \\
\end{tabular}
\end{center}
\end{table}

\subsection{V4140 Sagittarii}\label{obj:v4140sgr}

   V4140 Sgr has long been known as an eclipsing CV below the period gap
\citep{jab87v4140sgr}.  The dwarf nova-type nature was confirmed only
very recently \citep{bor05v4140sgr}, who interpreted short outbursts
of this object as being normal outbursts of an SU UMa-type dwarf nova.
In 2004, B. Monard detected a long outburst and reported the existence
of superhumps (vsnet-alert 8313).  We analyzed the data obtained during
this superoutburst.  We used out-of-eclipse observations as was done for
V2051 Oph, using the ephemeris by \citet{bap03v4140sgr}.
The times of superhump maxima are listed in table \ref{tab:v4140sgroc2004}
(the identification of maxima was slightly uncertain for $E \ge 149$
due to the faintness of the object and shortness of the observing runs).
Disregarding the first night, when superhumps were likely still evolving,
the $O-C$ diagram seems to be composed of the stage B with a positive
$P_{\rm dot}$ ($16 \le E \le 70$), followed by a transition to the stage C
with a shorter period.  The $P_{\rm dot}$ for the stage B was
$+25.3(12.3) \times 10^{-5}$.  The mean superhump period from the
first five nights (with better statistics) was 0.06324(3) d,
yielding a fractional superhump excess of 2.9(1) \%.

\begin{table}
\caption{Superhump maxima of V4140 Sgr (2004).}\label{tab:v4140sgroc2004}
\begin{center}
\begin{tabular}{ccccc}
\hline\hline
$E$ & max$^a$ & error & $O-C^b$ & $N^c$ \\
\hline
0 & 53269.2561 & 0.0025 & 0.0047 & 176 \\
1 & 53269.3193 & 0.0013 & 0.0045 & 238 \\
2 & 53269.3858 & 0.0013 & 0.0079 & 240 \\
16 & 53270.2604 & 0.0021 & $-$0.0035 & 118 \\
17 & 53270.3227 & 0.0021 & $-$0.0045 & 116 \\
18 & 53270.3850 & 0.0019 & $-$0.0054 & 107 \\
19 & 53270.4474 & 0.0032 & $-$0.0063 & 113 \\
53 & 53272.6051 & 0.0043 & $-$0.0002 & 97 \\
69 & 53273.6226 & 0.0021 & 0.0049 & 113 \\
70 & 53273.6919 & 0.0026 & 0.0109 & 111 \\
133 & 53277.6735 & 0.0023 & 0.0059 & 93 \\
140 & 53278.1180 & 0.0030 & 0.0075 & 26 \\
155 & 53279.0477 & 0.0023 & $-$0.0120 & 26 \\
156 & 53279.1107 & 0.0050 & $-$0.0123 & 27 \\
165 & 53279.6901 & 0.0098 & $-$0.0024 & 106 \\
181 & 53280.6891 & 0.0131 & $-$0.0159 & 87 \\
275 & 53286.6694 & 0.0025 & 0.0162 & 121 \\
\hline
  \multicolumn{5}{l}{$^{a}$ BJD$-$2400000.} \\
  \multicolumn{5}{l}{$^{b}$ Against $max = 2453269.2514 + 0.063279 E$.} \\
  \multicolumn{5}{l}{$^{c}$ Number of points used to determine the maximum.} \\
\end{tabular}
\end{center}
\end{table}

\subsection{V701 Tauri}\label{obj:v701tau}

   V701 Tau was discovered by \citet{era73v701tau} as an eruptive object.
The SU UMa-type nature was first reported by us during the 1995--1996
outburst (vsnet-alert 303).  \citet{she07v701tau} further reported
the 2005 superoutburst and obtained a superhump period of 0.0690(2) d,
or its one-day alias, 0.0663(2) d.
Based on our 1995--1996 observations, we obtained a mean period
of 0.06898(3) d.  The times of superhump maxima are listed in table
\ref{tab:v701tauoc1995}.
During the interval $0 \le E \le 3$, the superhumps were still in
the growing stage (stage A) and the mean period (0.073(1) d) significantly
differed from the later observations.  The $P_{\rm dot}$ estimated from
the segment of $31 \le E \le 159$ was $-2.6(0.8) \times 10^{-5}$.

\begin{figure}
  \begin{center}
    \FigureFile(88mm,110mm){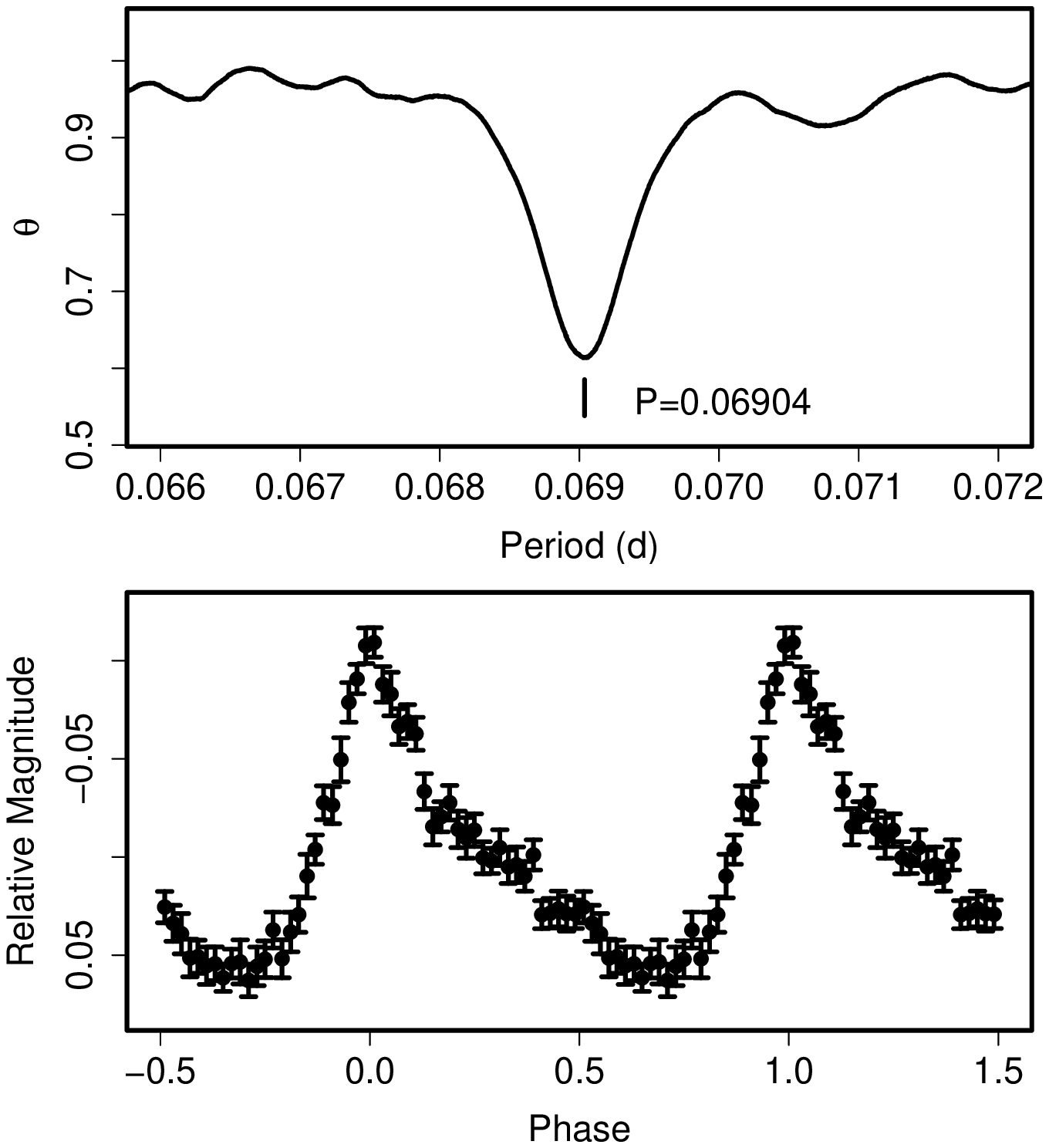}
  \end{center}
  \caption{Superhumps in V701 Tau (2005). (Upper): PDM analysis.
     (Lower): Phase-averaged profile.}
  \label{fig:v701taushpdm}
\end{figure}

   We also analyzed the 2005 superoutburst (table \ref{tab:v701tauoc2005}).
The mean superhump period with the PDM method was 0.069037(12) d
(figure \ref{fig:v701taushpdm}).
The $P_{\rm SH}$ showed a clear increase (stage B) at
$P_{\rm dot}$ = $+11.0(3.5) \times 10^{-5}$.

\begin{table}
\caption{Superhump maxima of V701 Tau (1995--1996).}\label{tab:v701tauoc1995}
\begin{center}
\begin{tabular}{ccccc}
\hline\hline
$E$ & max$^a$ & error & $O-C^b$ & $N^c$ \\
\hline
0 & 50078.9780 & 0.0016 & $-$0.0077 & 61 \\
1 & 50079.0485 & 0.0020 & $-$0.0061 & 65 \\
2 & 50079.1260 & 0.0035 & 0.0024 & 64 \\
3 & 50079.1946 & 0.0013 & 0.0021 & 55 \\
31 & 50081.1266 & 0.0005 & 0.0029 & 65 \\
58 & 50082.9901 & 0.0016 & 0.0042 & 46 \\
59 & 50083.0577 & 0.0116 & 0.0028 & 46 \\
60 & 50083.1280 & 0.0021 & 0.0042 & 44 \\
159 & 50089.9471 & 0.0059 & $-$0.0047 & 37 \\
\hline
  \multicolumn{5}{l}{$^{a}$ BJD$-$2400000.} \\
  \multicolumn{5}{l}{$^{b}$ Against $max = 2450078.9857 + 0.068970 E$.} \\
  \multicolumn{5}{l}{$^{c}$ Number of points used to determine the maximum.} \\
\end{tabular}
\end{center}
\end{table}

\begin{table}
\caption{Superhump maxima of V701 Tau (2005).}\label{tab:v701tauoc2005}
\begin{center}
\begin{tabular}{ccccc}
\hline\hline
$E$ & max$^a$ & error & $O-C^b$ & $N^c$ \\
\hline
0 & 53711.4936 & 0.0007 & 0.0033 & 67 \\
1 & 53711.5650 & 0.0013 & 0.0056 & 34 \\
14 & 53712.4576 & 0.0005 & 0.0008 & 57 \\
23 & 53713.0767 & 0.0007 & $-$0.0015 & 137 \\
24 & 53713.1447 & 0.0007 & $-$0.0025 & 146 \\
25 & 53713.2144 & 0.0009 & $-$0.0019 & 127 \\
27 & 53713.3536 & 0.0007 & $-$0.0007 & 68 \\
28 & 53713.4233 & 0.0020 & $-$0.0000 & 42 \\
36 & 53713.9687 & 0.0095 & $-$0.0070 & 86 \\
37 & 53714.0437 & 0.0009 & $-$0.0010 & 145 \\
38 & 53714.1135 & 0.0009 & $-$0.0002 & 145 \\
39 & 53714.1764 & 0.0047 & $-$0.0063 & 83 \\
40 & 53714.2561 & 0.0012 & 0.0043 & 71 \\
41 & 53714.3195 & 0.0008 & $-$0.0013 & 60 \\
54 & 53715.2204 & 0.0046 & 0.0021 & 121 \\
69 & 53716.2549 & 0.0021 & 0.0010 & 34 \\
70 & 53716.3218 & 0.0016 & $-$0.0011 & 36 \\
71 & 53716.3959 & 0.0007 & 0.0040 & 33 \\
73 & 53716.5323 & 0.0009 & 0.0023 & 88 \\
\hline
  \multicolumn{5}{l}{$^{a}$ BJD$-$2400000.} \\
  \multicolumn{5}{l}{$^{b}$ Against $max = 2453711.4903 + 0.069036 E$.} \\
  \multicolumn{5}{l}{$^{c}$ Number of points used to determine the maximum.} \\
\end{tabular}
\end{center}
\end{table}

\subsection{V1208 Tauri}\label{obj:v1208tau}

   V1208 Tau was originally identified as a CV
during the course of identification of ROSAT sources \citep{mot96CVROSAT}.
P. Schmeer detected the first-ever recorded outburst in 2000
(vsnet-alert 4118).  Time-resolved photometry during this superoutburst
established the SU UMa-type dwarf novae (vsnet-alert 4122).

\begin{figure}
  \begin{center}
    \FigureFile(88mm,110mm){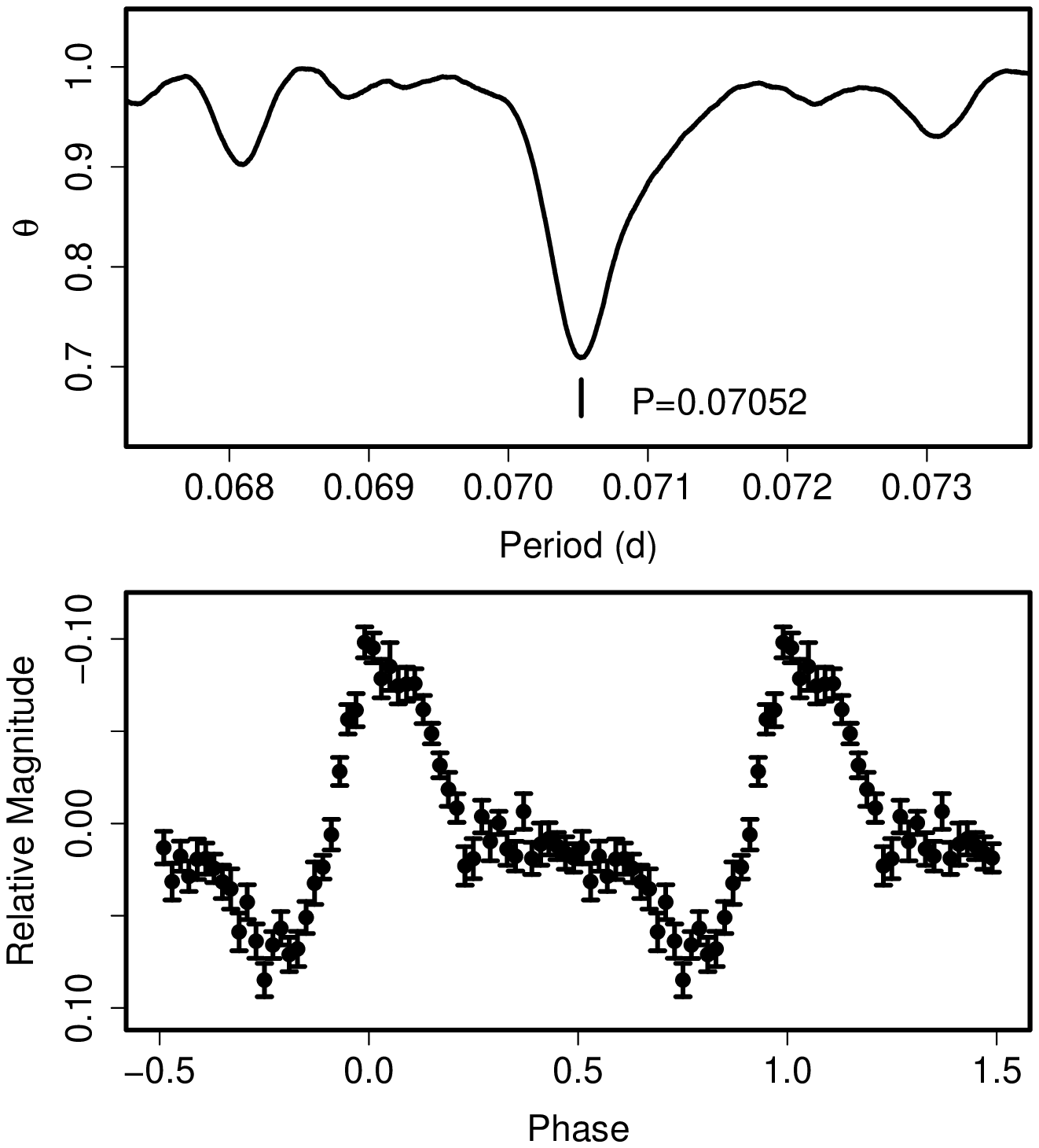}
  \end{center}
  \caption{Superhumps in V1208 Tau (2002--2003). (Upper): PDM analysis.
     (Lower): Phase-averaged profile.}
  \label{fig:v1208taushpdm}
\end{figure}

\begin{figure}
  \begin{center}
    \FigureFile(88mm,70mm){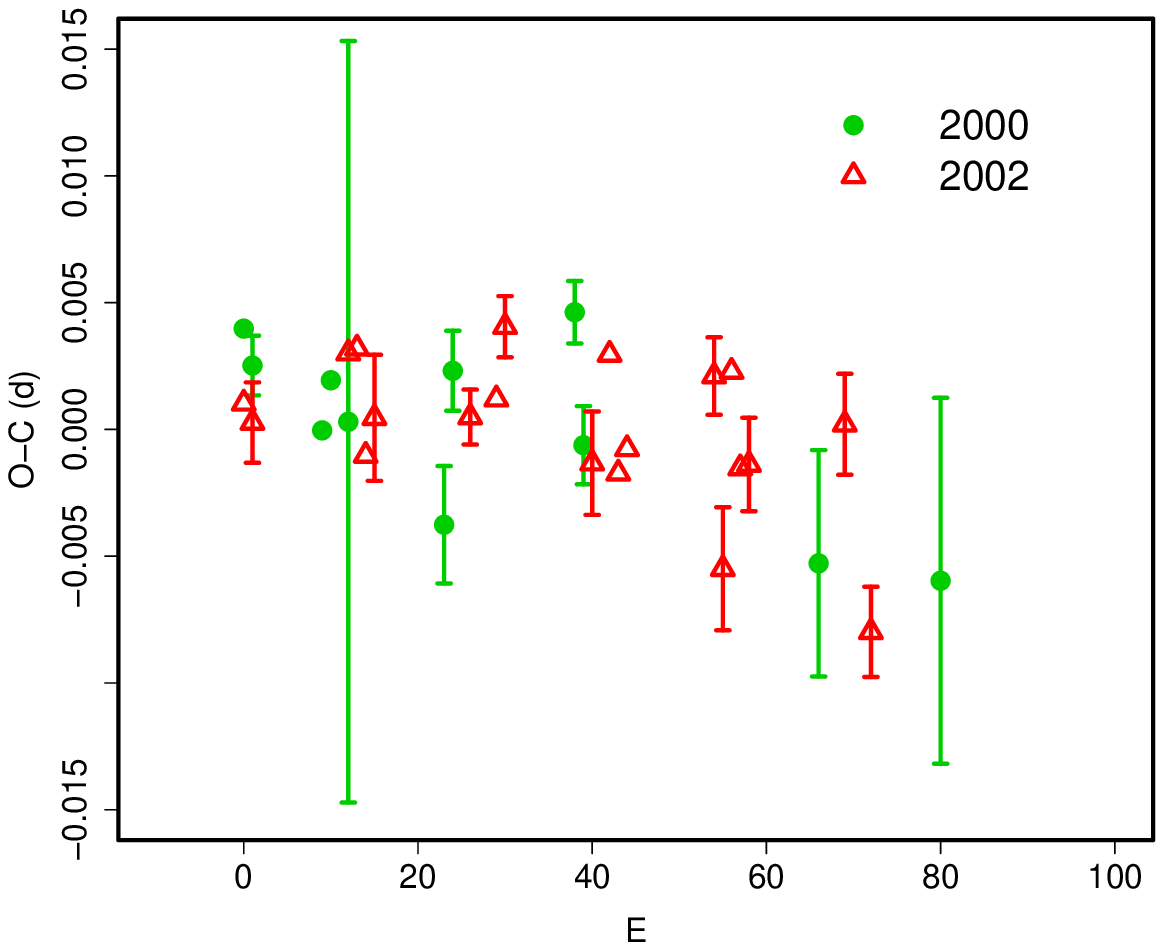}
  \end{center}
  \caption{Comparison of $O-C$ diagrams of V1208 Tau between different
  superoutbursts.  A period of 0.07060 d was used to draw this figure.
  Since the start of the outburst was unknown, the start of time-resolved
  photometry was chosen as $E=0$.
  }
  \label{fig:v1208taucomp}
\end{figure}

   We observed two superoutbursts in 2000 and 2002--2003.
The superhump profile for the 2002--2003 superoutburst is shown
in figure \ref{fig:v1208taushpdm}.
The times of superhump maxima are listed in tables \ref{tab:v1208tauoc2000}
and \ref{tab:v1208tauoc2002}, respectively.
The values of $P_{\rm dot}$ were $-2.8(4.0) \times 10^{-5}$ and
$-6.3(3.8) \times 10^{-5}$, respectively.  These negative values appear
to have resulted from stage B--C transitions (figure \ref{fig:v1208taucomp}).

\begin{table}
\caption{Superhump maxima of V1208 Tau (2000).}\label{tab:v1208tauoc2000}
\begin{center}

\end{center}
\end{table}

\begin{table}
\caption{Superhump maxima of V1208 Tau (2002--2003).}\label{tab:v1208tauoc2002}
\begin{center}
%
\end{center}
\end{table}

\subsection{KK Telescopii}\label{obj:kktel}

\begin{figure}
  \begin{center}
    \FigureFile(88mm,70mm){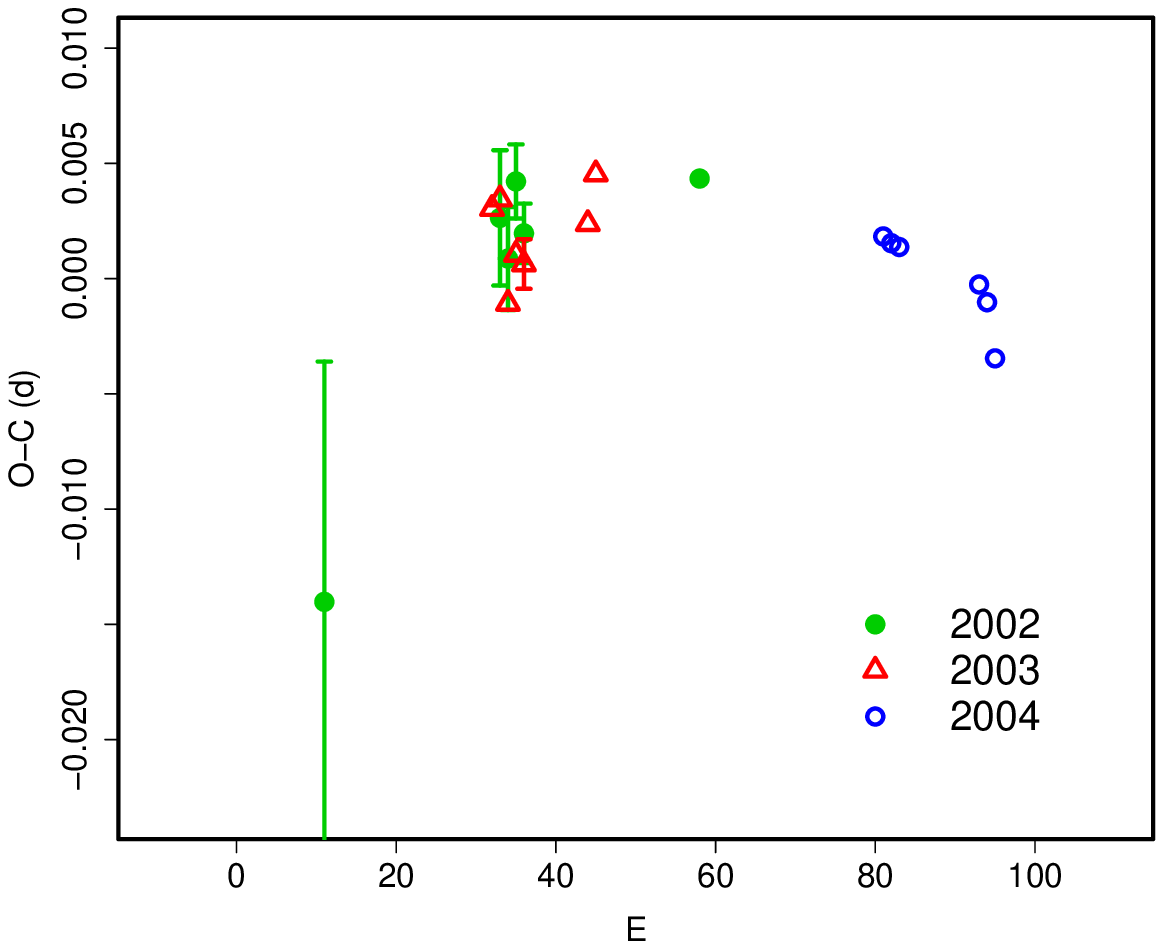}
  \end{center}
  \caption{Comparison of $O-C$ diagrams of KK Tel between different
  superoutbursts.  A period of 0.08761 d was used to draw this figure.
  Approximate cycle counts ($E$) after the start of the
  superoutburst were used.
  }
  \label{fig:kktelcomp}
\end{figure}

   \citet{kat03v877arakktelpucma} reported the detection of superhumps
and derived an exceptionally large rate of period decrease.
We also observed the 2003 superoutburst, and identified an unambiguous
superhump period of 0.08753(5) d, which is in good agreement with
\citet{pat03suumas}, who observed the 2000 superoutburst.
Based on this identification of the period, we give refined
$O-C$'s for the 2002 superoutburst (table \ref{tab:kkteloc2002}).
It is now evident the times of superhumps for $22 \le E \le 47$ are
well expressed by this improved superhump period.
The maximum at $E = 0$ has a strongly negative $O-C$, indicating that
this maximum was observed during the stage A evolution.
The period derivative shown in \citet{kat03v877arakktelpucma} was
thus a result of a stage A--B transition, and should not be used as
a global $P_{\rm dot}$.
The times of superhump maxima during the 2003 superoutburst are
listed in table \ref{tab:kkteloc2003}.
The mean period of 0.08734(6) d determined from the late stage of the
2004 superoutburst (table \ref{tab:kkteloc2004}) suggests that
a shortening of the period (stage C) near the termination of
the superoutburst also occurred in this system
(see also the combined $O-C$ diagram in figure \ref{fig:kktelcomp}).

\begin{table}
\caption{Superhump maxima of KK Tel (2002).}\label{tab:kkteloc2002}
\begin{center}
\begin{tabular}{ccccc}
\hline\hline
$E$ & max$^a$ & error & $O-C^b$ & $N^c$ \\
\hline
0 & 52444.0061 & 0.0104 & $-$0.0139 & 185 \\
22 & 52445.9501 & 0.0029 & 0.0028 & 20 \\
23 & 52446.0360 & 0.0022 & 0.0010 & 30 \\
24 & 52446.1269 & 0.0016 & 0.0043 & 48 \\
25 & 52446.2123 & 0.0013 & 0.0021 & 37 \\
47 & 52448.1421 & 0.0006 & 0.0045 & 90 \\
\hline
  \multicolumn{5}{l}{$^{a}$ BJD$-$2400000.} \\
  \multicolumn{5}{l}{$^{b}$ Against $max = 2452444.2000 + 0.08761 E$.} \\
  \multicolumn{5}{l}{$^{c}$ Number of points used to determine the maximum.} \\
\end{tabular}
\end{center}
\end{table}

\begin{table}
\caption{Superhump maxima of KK Tel (2003).}\label{tab:kkteloc2003}
\begin{center}
\begin{tabular}{ccccc}
\hline\hline
$E$ & max$^a$ & error & $O-C^b$ & $N^c$ \\
\hline
0 & 52816.0225 & 0.0004 & 0.0018 & 91 \\
1 & 52816.1105 & 0.0003 & 0.0020 & 90 \\
2 & 52816.1936 & 0.0005 & $-$0.0026 & 83 \\
3 & 52816.2833 & 0.0004 & $-$0.0007 & 91 \\
4 & 52816.3705 & 0.0011 & $-$0.0012 & 52 \\
12 & 52817.0732 & 0.0008 & $-$0.0006 & 64 \\
13 & 52817.1629 & 0.0009 & 0.0014 & 25 \\
\hline
  \multicolumn{5}{l}{$^{a}$ BJD$-$2400000.} \\
  \multicolumn{5}{l}{$^{b}$ Against $max = 2452816.0207 + 0.087756 E$.} \\
  \multicolumn{5}{l}{$^{c}$ Number of points used to determine the maximum.} \\
\end{tabular}
\end{center}
\end{table}

\begin{table}
\caption{Superhump maxima of KK Tel (2004).}\label{tab:kkteloc2004}
\begin{center}
\begin{tabular}{ccccc}
\hline\hline
$E$ & max$^a$ & error & $O-C^b$ & $N^c$ \\
\hline
0 & 53151.4369 & 0.0003 & $-$0.0001 & 157 \\
1 & 53151.5242 & 0.0003 & $-$0.0001 & 199 \\
2 & 53151.6116 & 0.0003 & $-$0.0000 & 198 \\
12 & 53152.4861 & 0.0003 & 0.0011 & 198 \\
13 & 53152.5730 & 0.0003 & 0.0006 & 199 \\
14 & 53152.6581 & 0.0006 & $-$0.0015 & 120 \\
\hline
  \multicolumn{5}{l}{$^{a}$ BJD$-$2400000.} \\
  \multicolumn{5}{l}{$^{b}$ Against $max = 2453151.4370 + 0.087335 E$.} \\
  \multicolumn{5}{l}{$^{c}$ Number of points used to determine the maximum.} \\
\end{tabular}
\end{center}
\end{table}

\subsection{EK Trianguli Australis}\label{obj:ektra}

\begin{figure}
  \begin{center}
    \FigureFile(88mm,110mm){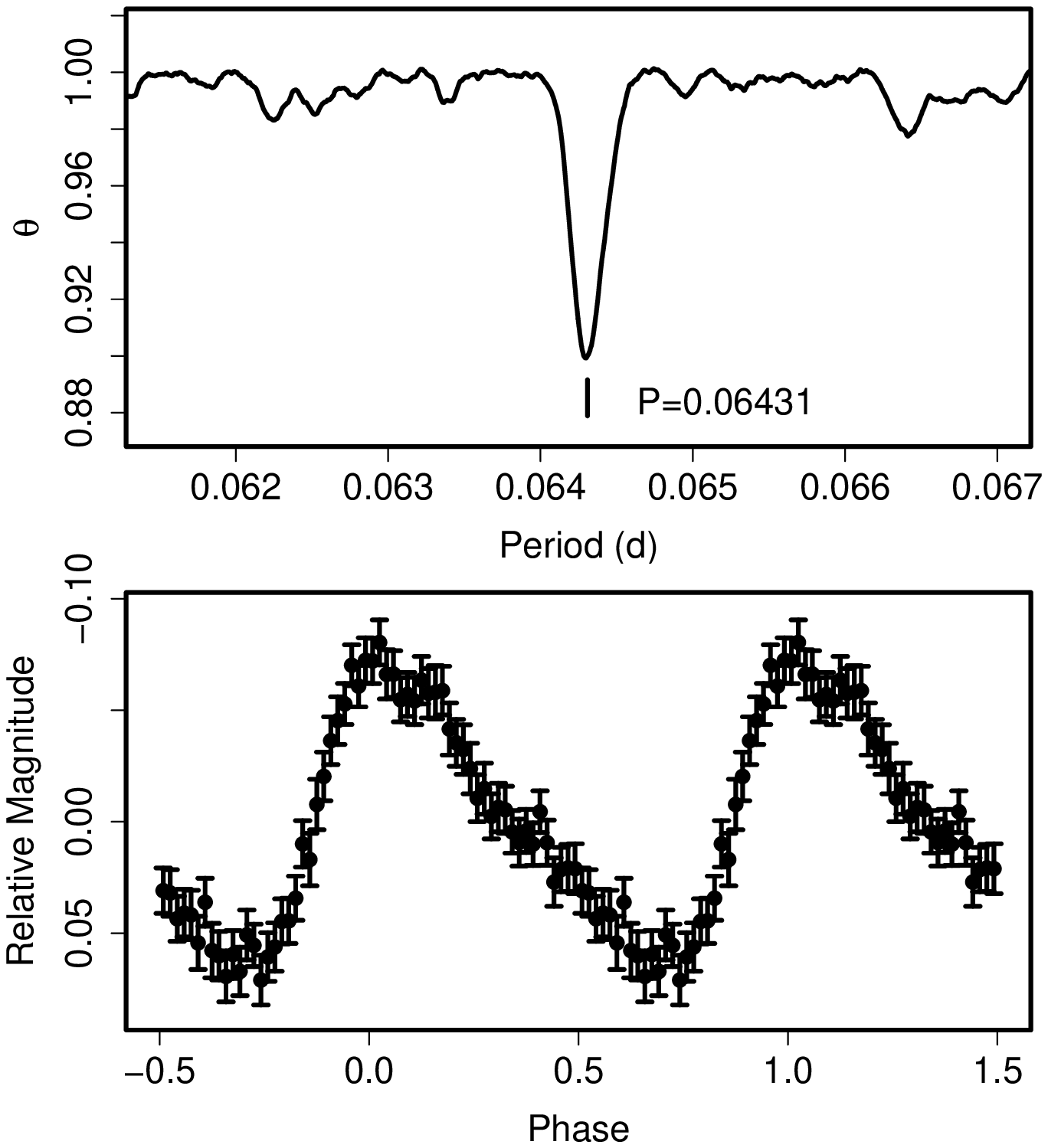}
  \end{center}
  \caption{Superhumps in EK TrA (2007). (Upper): PDM analysis.
     (Lower): Phase-averaged profile.}
  \label{fig:ektrashpdm}
\end{figure}

   Although EK TrA had long been known as an SU UMa-type dwarf nova
\citep{vog80ektra}, the precise superhump period was not reported.
We observed the 2007 superoutburst and obtained a mean superhump
period of 0.064309(6) with the PDM method (figure \ref{fig:ektrashpdm}).
The times of superhump maxima are listed in table \ref{tab:ektraoc2007}.
The $O-C$'s were almost zero, and the $P_{\rm dot}$ for the entire observation
was $-0.5(0.5) \times 10^{-5}$.  There was no noticeable structure in
the $O-C$ diagram.  Individual superhumps, however, showed strongly
variable profiles: double-humped around
$0 \le E \le 1$ (maxima matching the ephemeris were given in the table),
and around $E = 63$, a complex, double wave-like profile emerged with
reduced superhump amplitudes (maxima not determined).
The latter feature somewhat resembled the behavior observed in
OT J055718$+$683226 \citep{uem09j0557}.

\begin{table}
\caption{Superhump maxima of EK TrA (2007).}\label{tab:ektraoc2007}
\begin{center}
\begin{tabular}{ccccc}
\hline\hline
$E$ & max$^a$ & error & $O-C^b$ & $N^c$ \\
\hline
0 & 54294.2815 & 0.0015 & $-$0.0060 & 148 \\
1 & 54294.3559 & 0.0016 & 0.0041 & 91 \\
16 & 54295.3112 & 0.0008 & $-$0.0056 & 148 \\
31 & 54296.2840 & 0.0008 & 0.0021 & 148 \\
32 & 54296.3473 & 0.0011 & 0.0011 & 148 \\
33 & 54296.4101 & 0.0007 & $-$0.0004 & 149 \\
46 & 54297.2488 & 0.0005 & 0.0020 & 148 \\
47 & 54297.3105 & 0.0015 & $-$0.0007 & 148 \\
48 & 54297.3760 & 0.0004 & 0.0005 & 148 \\
49 & 54297.4317 & 0.0016 & $-$0.0081 & 148 \\
77 & 54299.2400 & 0.0008 & $-$0.0012 & 149 \\
78 & 54299.3099 & 0.0011 & 0.0043 & 148 \\
79 & 54299.3763 & 0.0009 & 0.0064 & 148 \\
80 & 54299.4360 & 0.0008 & 0.0017 & 148 \\
93 & 54300.2697 & 0.0010 & $-$0.0009 & 149 \\
94 & 54300.3375 & 0.0012 & 0.0026 & 148 \\
95 & 54300.4075 & 0.0013 & 0.0082 & 149 \\
108 & 54301.2350 & 0.0014 & $-$0.0005 & 148 \\
109 & 54301.2989 & 0.0010 & $-$0.0011 & 148 \\
110 & 54301.3618 & 0.0007 & $-$0.0024 & 149 \\
124 & 54302.2590 & 0.0007 & $-$0.0059 & 148 \\
125 & 54302.3328 & 0.0007 & 0.0035 & 148 \\
126 & 54302.3889 & 0.0007 & $-$0.0047 & 148 \\
127 & 54302.4575 & 0.0007 & $-$0.0005 & 141 \\
155 & 54304.2554 & 0.0009 & $-$0.0039 & 149 \\
156 & 54304.3223 & 0.0012 & $-$0.0013 & 148 \\
157 & 54304.3915 & 0.0010 & 0.0035 & 149 \\
158 & 54304.4523 & 0.0010 & $-$0.0001 & 140 \\
172 & 54305.3570 & 0.0010 & 0.0040 & 149 \\
173 & 54305.4195 & 0.0014 & 0.0021 & 141 \\
186 & 54306.2550 & 0.0018 & 0.0013 & 121 \\
187 & 54306.3107 & 0.0018 & $-$0.0073 & 149 \\
188 & 54306.3916 & 0.0029 & 0.0093 & 150 \\
249 & 54310.2996 & 0.0007 & $-$0.0072 & 142 \\
250 & 54310.3722 & 0.0014 & 0.0011 & 122 \\
\hline
  \multicolumn{5}{l}{$^{a}$ BJD$-$2400000.} \\
  \multicolumn{5}{l}{$^{b}$ Against $max = 2454294.2875 + 0.064335 E$.} \\
  \multicolumn{5}{l}{$^{c}$ Number of points used to determine the maximum.} \\
\end{tabular}
\end{center}
\end{table}

\subsection{UW Trianguli}\label{obj:uwtri}

   This object was originally reported as a nova
(\cite{kur84newCV}; \cite{arg83uwtriiauc}).  The detection of a second
outburst in 1995 by T. Vanmunster led to an identification as
a large-amplitude dwarf nova \citep{kat01uwtri}.
\citet{kat01uwtri} reported a candidate superhump period 0.0569 d,
whose selection was based on the period distribution of known CVs.
Other one-day aliases were not excluded due to the shortness of
observations.

\begin{figure}
  \begin{center}
    \FigureFile(88mm,110mm){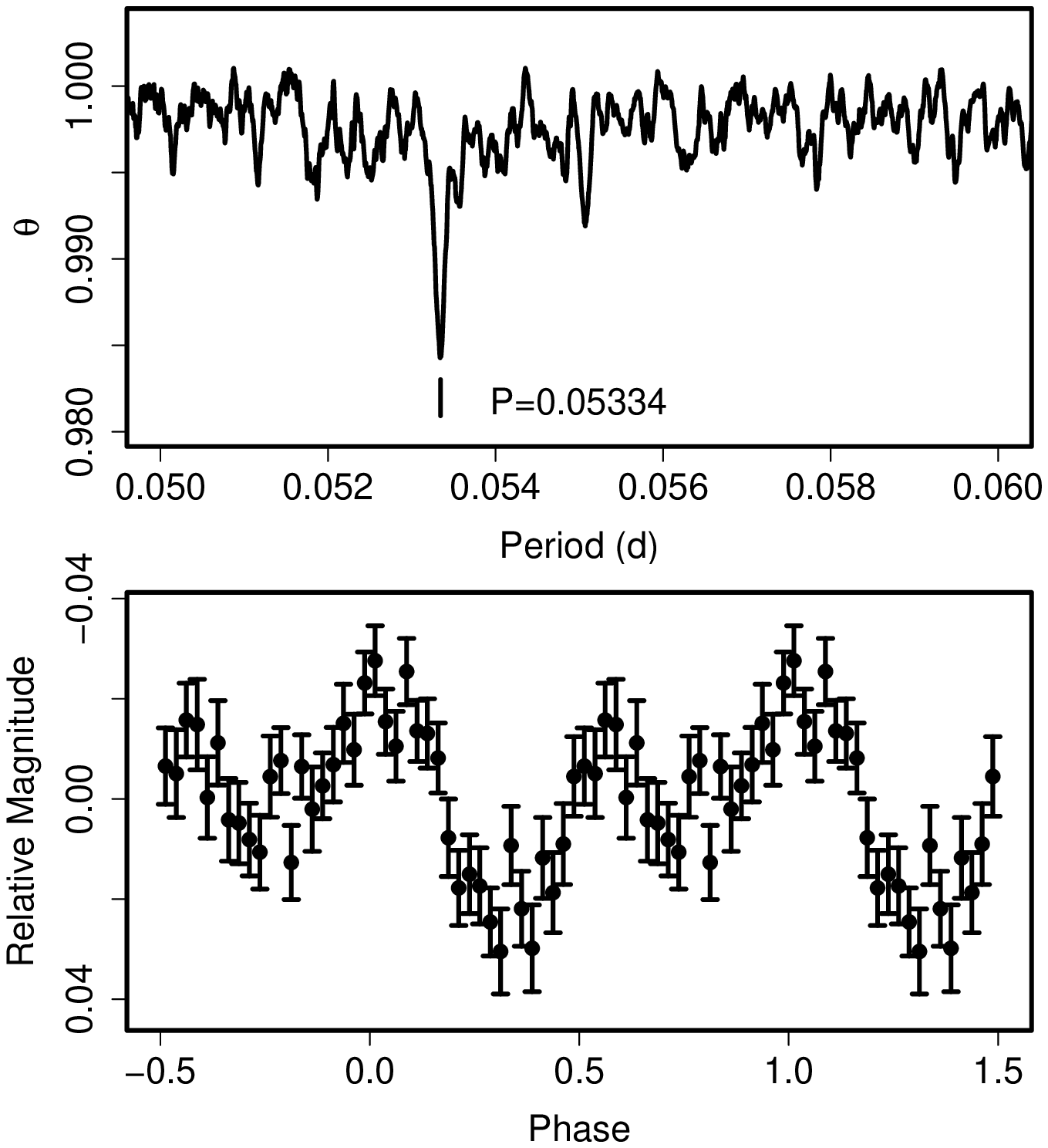}
  \end{center}
  \caption{Early superhumps in UW Tri (2008). (Upper): PDM analysis.
     (Lower): Phase-averaged profile.}
  \label{fig:uwtrieshpdm}
\end{figure}

\begin{figure}
  \begin{center}
    \FigureFile(88mm,110mm){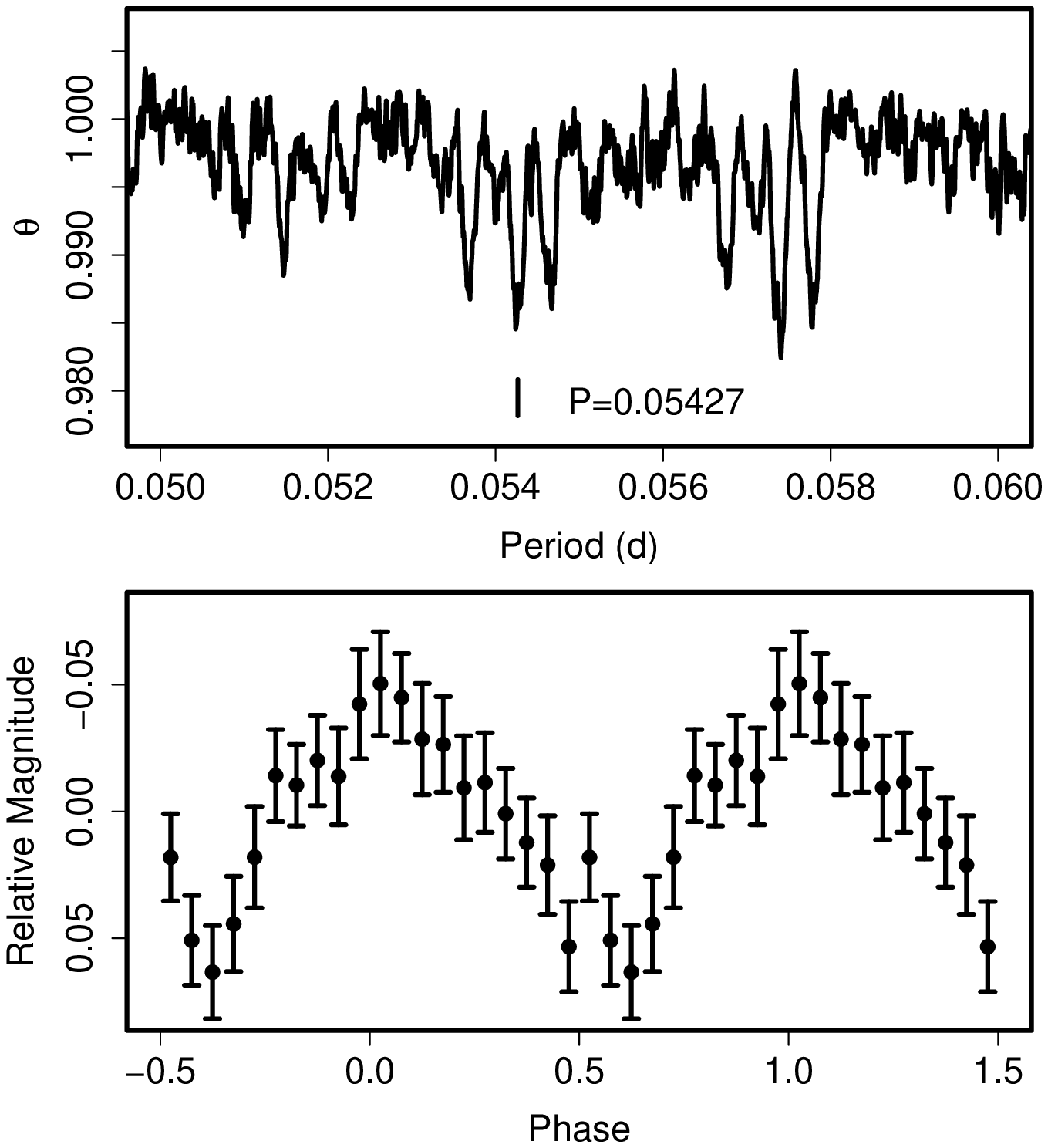}
  \end{center}
  \caption{Ordinary superhumps in UW Tri (2008). (Upper): PDM analysis.
     (Lower): Phase-averaged profile.}
  \label{fig:uwtrishpdm}
\end{figure}

\begin{figure}
  \begin{center}
    \FigureFile(88mm,110mm){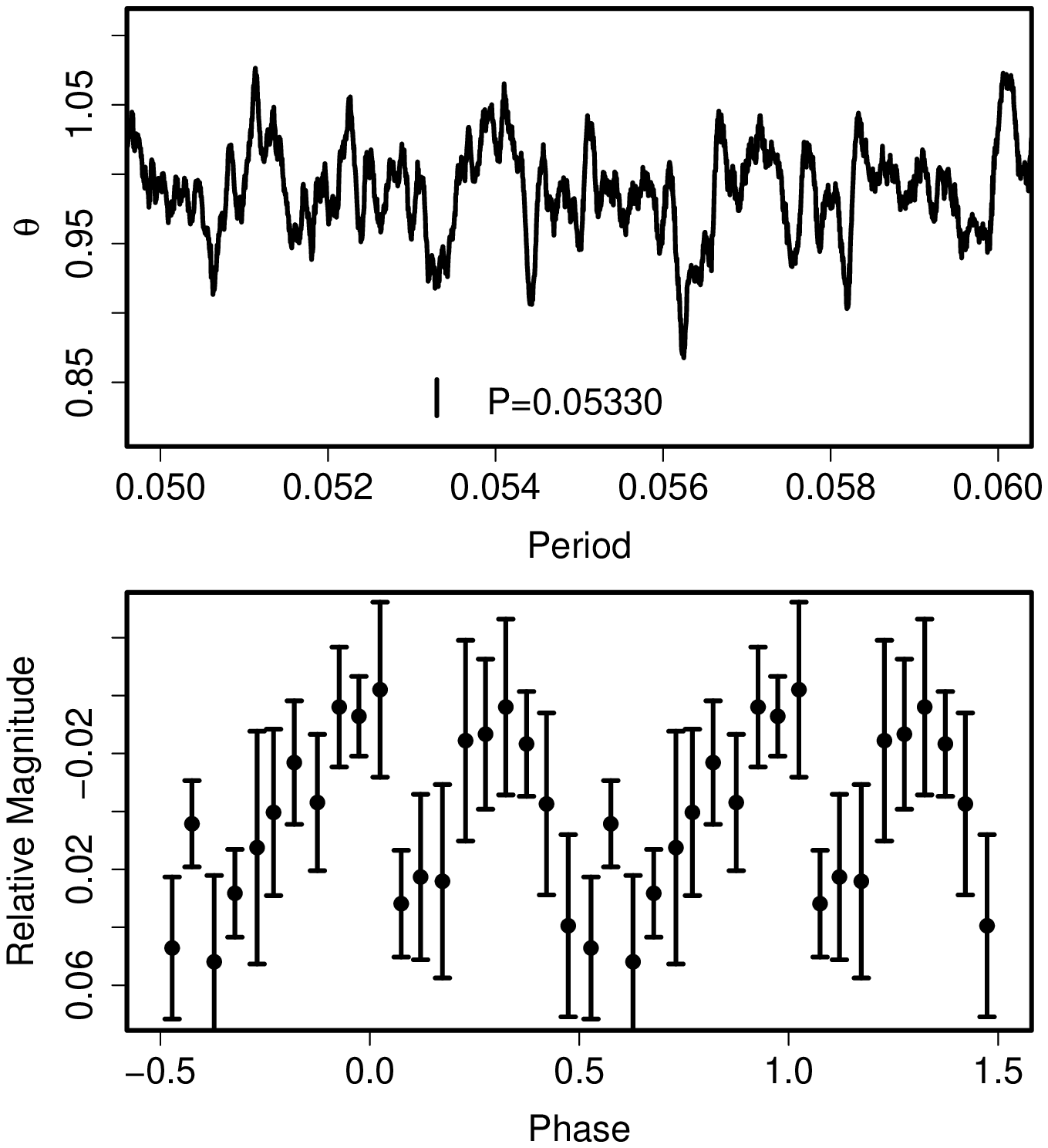}
  \end{center}
  \caption{Early superhumps in UW Tri (1995). (Upper): PDM analysis.
     (Lower): Phase-averaged profile.}
  \label{fig:uwtrieshpdm1995}
\end{figure}

   The object underwent a new outburst in 2008 (vsnet-alert 10635).
The data taken during this superoutburst now strongly favor
a shorter period of 0.05334(2) d for early superhumps and
0.05427(2) for ordinary superhumps
(figures \ref{fig:uwtrieshpdm}, \ref{fig:uwtrishpdm}).
We adopted these values as the basic periods for the following analysis.
We also reanalyzed the data in \citet{kat01uwtri} and yielded
a period of 0.05330(2) d based on the present alias selection.
The light curve of the 1995 observation averaged with this period now
exhibits double-wave modulations characteristic to early superhumps
(figure \ref{fig:uwtrieshpdm1995}).

   The maxima of ordinary superhump in 2008 are listed in table
\ref{tab:uwtrioc2008}.  The resultant $P_{\rm dot}$ was
$+3.7(0.6) \times 10^{-5}$, although there remained some uncertainty
in the constancy of the $P_{\rm dot}$ due to long gaps between observations.
If the $P_{\rm dot}$ is confirmed, the parameters of superhumps
and outbursts resemble those of another short $P_{\rm SH}$ WZ Sge-type
dwarf nova OT J0238 (subsection \ref{sec:j0238}).

   The details will be presented in Ohshima et al., in preparation.

\begin{table}
\caption{Superhump maxima of UW Tri (2008).}\label{tab:uwtrioc2008}
\begin{center}
\begin{tabular}{ccccc}
\hline\hline
$E$ & max$^a$ & error & $O-C^b$ & $N^c$ \\
\hline
0 & 54777.4487 & 0.0007 & 0.0144 & 40 \\
1 & 54777.5041 & 0.0007 & 0.0156 & 40 \\
101 & 54782.9032 & 0.0023 & $-$0.0047 & 86 \\
102 & 54782.9616 & 0.0024 & $-$0.0005 & 110 \\
104 & 54783.0654 & 0.0017 & $-$0.0051 & 109 \\
105 & 54783.1148 & 0.0009 & $-$0.0100 & 113 \\
106 & 54783.1829 & 0.0017 & 0.0040 & 111 \\
107 & 54783.2300 & 0.0014 & $-$0.0031 & 107 \\
108 & 54783.2810 & 0.0024 & $-$0.0063 & 87 \\
123 & 54784.0994 & 0.0015 & $-$0.0008 & 104 \\
124 & 54784.1527 & 0.0052 & $-$0.0017 & 85 \\
125 & 54784.1944 & 0.0017 & $-$0.0142 & 95 \\
126 & 54784.2621 & 0.0054 & $-$0.0007 & 110 \\
217 & 54789.1923 & 0.0023 & $-$0.0022 & 100 \\
233 & 54790.0511 & 0.0146 & $-$0.0105 & 65 \\
234 & 54790.1127 & 0.0012 & $-$0.0031 & 117 \\
235 & 54790.1698 & 0.0012 & $-$0.0002 & 115 \\
236 & 54790.2187 & 0.0039 & $-$0.0055 & 96 \\
271 & 54792.1292 & 0.0059 & 0.0082 & 30 \\
272 & 54792.1780 & 0.0070 & 0.0028 & 44 \\
288 & 54793.0662 & 0.0133 & 0.0239 & 44 \\
\hline
  \multicolumn{5}{l}{$^{a}$ BJD$-$2400000.} \\
  \multicolumn{5}{l}{$^{b}$ Against $max = 2454777.4343 + 0.054194 E$.} \\
  \multicolumn{5}{l}{$^{c}$ Number of points used to determine the maximum.} \\
\end{tabular}
\end{center}
\end{table}

\subsection{WY Trianguli}\label{obj:wytri}

   WY Tri is a dwarf nova discovered by \citet{mei86uztri}.
The SU UMa-type nature was established during its 2000 superoutburst
\citep{van01wytri}.
Since the original data in \citet{van01wytri} were not available,
we extracted the data from a scanned figure.  The quality of the
extracted data were sufficient for the following analysis.
The times of maxima determined from the combined data set with
\citet{van01wytri} are listed in table \ref{tab:wytrioc2000}.
Although the global $P_{\rm dot}$ was $-18.3(5.9) \times 10^{-5}$,
this variation can be attributed to a stage B--C transition.
The parameters are given in table \ref{tab:perlist}.
A PDM analysis of the stage B superhumps yielded a period of
0.07838(5) d.

\begin{table}
\caption{Superhump maxima of WY Tri(2000).}\label{tab:wytrioc2000}
\begin{center}
\begin{tabular}{ccccc}
\hline\hline
$E$ & max$^a$ & error & $O-C^b$ & $N^c$ \\
\hline
0 & 51899.3738 & 0.0015 & $-$0.0038 & -- \\
1 & 51899.4552 & 0.0027 & $-$0.0007 & -- \\
2 & 51899.5306 & 0.0027 & $-$0.0036 & -- \\
12 & 51900.3197 & 0.0015 & 0.0026 & -- \\
13 & 51900.3951 & 0.0024 & $-$0.0002 & -- \\
14 & 51900.4758 & 0.0033 & 0.0022 & -- \\
15 & 51900.5547 & 0.0030 & 0.0028 & -- \\
20 & 51900.9382 & 0.0047 & $-$0.0052 & 82 \\
21 & 51901.0216 & 0.0019 & $-$0.0000 & 113 \\
22 & 51901.0999 & 0.0038 & $-$0.0001 & 62 \\
24 & 51901.2571 & 0.0024 & 0.0006 & -- \\
25 & 51901.3355 & 0.0021 & 0.0007 & -- \\
26 & 51901.4140 & 0.0015 & 0.0009 & -- \\
27 & 51901.4923 & 0.0021 & 0.0009 & -- \\
37 & 51902.2808 & 0.0030 & 0.0066 & -- \\
38 & 51902.3551 & 0.0015 & 0.0025 & -- \\
39 & 51902.4357 & 0.0024 & 0.0049 & -- \\
40 & 51902.5101 & 0.0033 & 0.0009 & -- \\
46 & 51902.9736 & 0.0046 & $-$0.0053 & 82 \\
47 & 51903.0572 & 0.0035 & 0.0001 & 83 \\
58 & 51903.9114 & 0.0129 & $-$0.0069 & 113 \\
\hline
  \multicolumn{5}{l}{$^{a}$ BJD$-$2400000.} \\
  \multicolumn{5}{l}{$^{b}$ Against $max = 2451899.3776 + 0.078287 E$.} \\
  \multicolumn{5}{l}{$^{c}$ Number of points used to determine the maximum.} \\
\end{tabular}
\end{center}
\end{table}

\subsection{SU Ursae Majoris}\label{obj:suuma}

\begin{figure}
  \begin{center}
    \FigureFile(88mm,70mm){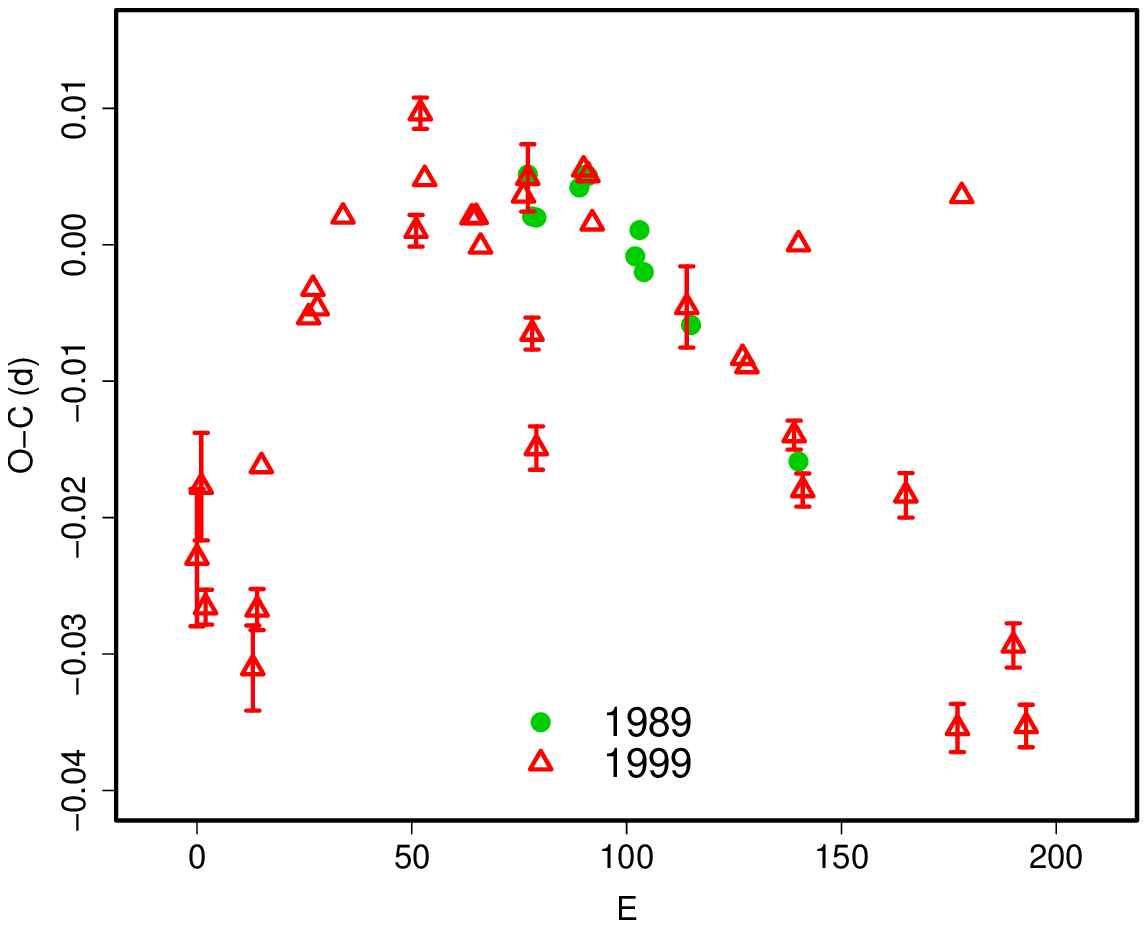}
  \end{center}
  \caption{Comparison of $O-C$ diagrams of SU UMa between different
  superoutbursts.  A period of 0.07908 d was used to draw this figure.
  Approximate cycle counts ($E$) after the start of the
  superoutburst were used.
  }
  \label{fig:suumacomp}
\end{figure}

   We observed the 1999 January superoutburst.  This outburst had
a precursor outburst, and the observation covered the precursor phase.
The times of superhump maxima are listed in table \ref{tab:suumaoc1999}.
The segment of $0 \le E \le 2$ corresponds to the precursor phase,
when the superhump period rapidly evolved.  Since there were multiple
hump maxima within one cycle after $E > 170$ (post-superoutburst stage),
we restricted our analysis to $E \le 165$.
Although the global $P_{\rm dot}$ was $-10.2(1.9) \times 10^{-5}$
($13 \le E \le $165), the $O-C$ diagram can be better interpreted as
a combination of A--C stages.  The $P_{\rm dot}$ for the stage B was
$-0.2(3.9) \times 10^{-5}$ ($34 \le E \le 92$,
disregarding $E = 78$ and $E = 79$).  Other parameters are presented
in table \ref{tab:perlist}.
A comparison between the 1989 and 1999 superoutbursts is shown in
figure \ref{fig:suumacomp}.

\begin{table}
\caption{Superhump maxima of SU UMa (1999).}\label{tab:suumaoc1999}
\begin{center}
\begin{tabular}{ccccc}
\hline\hline
$E$ & max$^a$ & error & $O-C^b$ & $N^c$ \\
\hline
0 & 51185.9020 & 0.0050 & $-$0.0152 & 80 \\
1 & 51185.9863 & 0.0039 & $-$0.0100 & 98 \\
2 & 51186.0566 & 0.0013 & $-$0.0188 & 87 \\
13 & 51186.9220 & 0.0031 & $-$0.0233 & 120 \\
14 & 51187.0053 & 0.0015 & $-$0.0190 & 124 \\
15 & 51187.0949 & 0.0008 & $-$0.0085 & 122 \\
26 & 51187.9757 & 0.0004 & 0.0025 & 134 \\
27 & 51188.0569 & 0.0003 & 0.0046 & 146 \\
28 & 51188.1345 & 0.0002 & 0.0031 & 152 \\
34 & 51188.6157 & 0.0002 & 0.0099 & 51 \\
51 & 51189.9591 & 0.0012 & 0.0089 & 77 \\
52 & 51190.0468 & 0.0011 & 0.0175 & 44 \\
53 & 51190.1210 & 0.0005 & 0.0127 & 50 \\
64 & 51190.9881 & 0.0009 & 0.0099 & 96 \\
65 & 51191.0672 & 0.0010 & 0.0100 & 34 \\
66 & 51191.1441 & 0.0010 & 0.0078 & 68 \\
76 & 51191.9386 & 0.0004 & 0.0116 & 115 \\
77 & 51192.0190 & 0.0025 & 0.0129 & 40 \\
78 & 51192.0867 & 0.0012 & 0.0015 & 67 \\
79 & 51192.1574 & 0.0016 & $-$0.0069 & 21 \\
90 & 51193.0477 & 0.0006 & 0.0135 & 152 \\
91 & 51193.1263 & 0.0007 & 0.0131 & 68 \\
92 & 51193.2019 & 0.0007 & 0.0096 & 91 \\
114 & 51194.9355 & 0.0030 & 0.0035 & 41 \\
127 & 51195.9598 & 0.0009 & $-$0.0001 & 117 \\
128 & 51196.0383 & 0.0004 & $-$0.0008 & 118 \\
139 & 51196.9031 & 0.0011 & $-$0.0058 & 138 \\
140 & 51196.9962 & 0.0009 & 0.0082 & 156 \\
141 & 51197.0573 & 0.0012 & $-$0.0098 & 98 \\
165 & 51198.9548 & 0.0016 & $-$0.0101 & 107 \\
177 & 51199.8867 & 0.0018 & $-$0.0271 & 62 \\
178 & 51200.0048 & 0.0007 & 0.0119 & 143 \\
190 & 51200.9208 & 0.0016 & $-$0.0210 & 109 \\
191 & 51201.0514 & 0.0005 & 0.0306 & 126 \\
193 & 51201.1521 & 0.0016 & $-$0.0269 & 91 \\
\hline
  \multicolumn{5}{l}{$^{a}$ BJD$-$2400000.} \\
  \multicolumn{5}{l}{$^{b}$ Against $max = 2451185.9172 + 0.079077 E$.} \\
  \multicolumn{5}{l}{$^{c}$ Number of points used to determine the maximum.} \\
\end{tabular}
\end{center}
\end{table}

\subsection{SW Ursae Majoris}\label{sec:swuma}\label{obj:swuma}

   We present observations of the 1991, 1997, 2000, 2002,
and 2006 superoutbursts
(tables \ref{tab:swumaoc1991}, \ref{tab:swumaoc1997}, \ref{tab:swumaoc2000},
\ref{tab:swumaoc2002}, \ref{tab:swumaoc2006}), a part of which are
a reanalysis of the data in \citet{soe09swuma}.

\begin{figure}
  \begin{center}
    \FigureFile(70mm,60mm){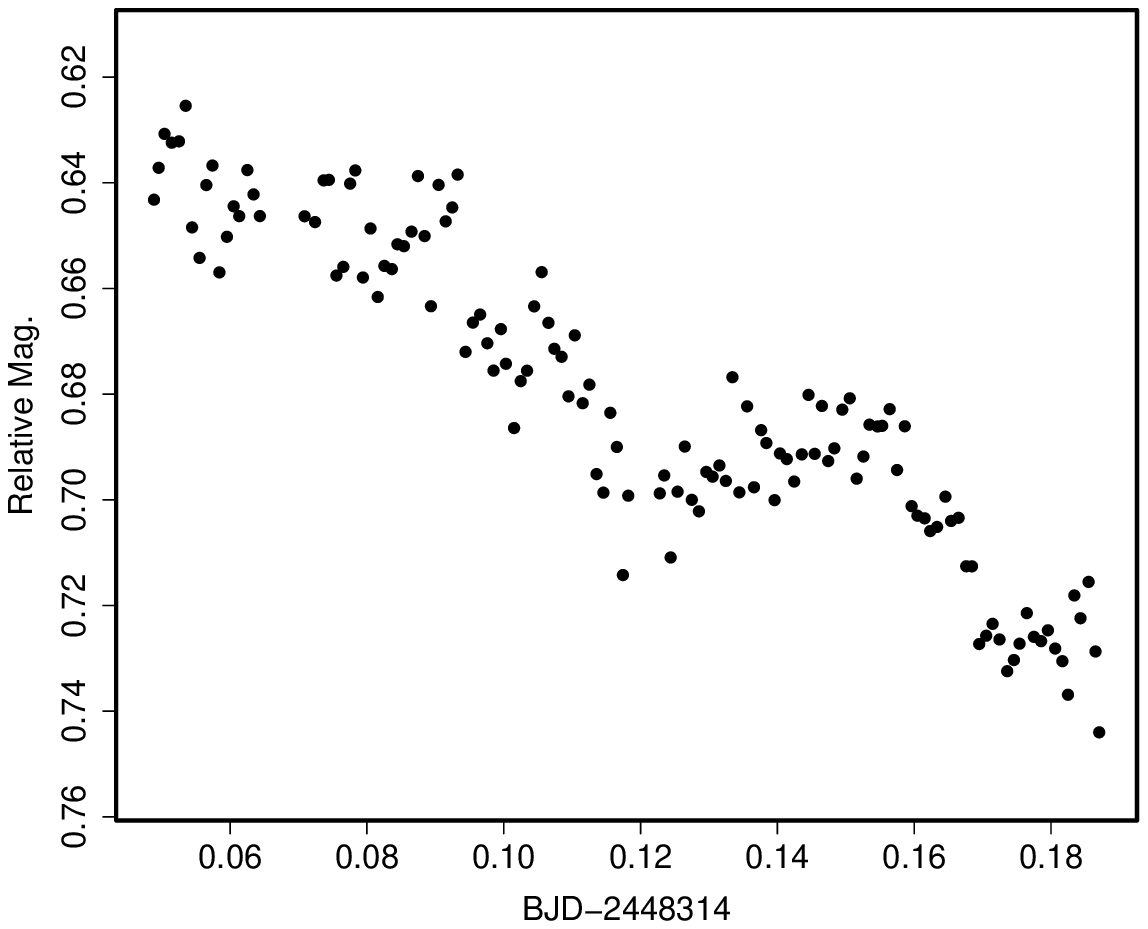}
  \end{center}
  \caption{Precursor outburst of SW UMa on 1991 February 26.}
  \label{fig:swuma1991prec}
\end{figure}

   The 1991 superoutburst was a faint superoutburst reaching a visual
magnitude of $\sim$ 11.0.
The superoutburst was associated with a precursor outburst
(figure \ref{fig:swuma1991prec}).
The identification of $E$ in table \ref{tab:swumaoc1991}\footnote{
  Since original data have become unavailable,
  we extracted observations from printed light curves.  The errors of
  maxima times may be larger than the listed values.
}
is based on the present knowledge, assuming that the object experienced
stage A during the precursor phase and the presence of the stage C
during the post-superoutburst stage.  The segment $52 \le E \le 88$ seems
to be the early phase of the stage B.  The shortness of the mean
$P_{\rm SH}$ = 0.05825(2) d probably reflects a short $P_{\rm SH}$
at the beginning of the stage B.

   The 1997 superoutburst showed $P_{\rm dot}$ = $+8.6(0.5) \times 10^{-5}$.

   The $O-C$ diagram of the 2000 superoutburst was clearly composed of
the three distinct stages A--C.  We obtained
$P_{\rm dot}$ = $+5.1(0.5) \times 10^{-5}$ (stage B, $27 \le E \le 217$).

   During the 2002 superoutburst, we obtained $P_{\rm dot}$ =
$+9.9(0.9) \times 10^{-5}$ for the interval $E \le 142$ (stage B).
After $E = 142$, the $O-C$ diagram showed a clear transition to a shorter
superhump period (stage C).

   During the 2006 superoutburst, we obtained $P_{\rm dot}$ =
$+9.5(0.6) \times 10^{-5}$ during the stage B ($33 \le E \le 189$).
Although this superoutburst was one of the brightest
(reaching a visual magnitude of 10.2) in the last decade, the behavior
in the $O-C$ diagram during the stage B was similar to the ones in other
superoutbursts.
The start of the stage B was $\sim$8.5 d after the initial detection
of the outburst.  The corresponding delay time for the 2000 superoutburst
was $\sim$ 7 d, and the delay time for the 1991 superoutburst was
less than 3 d.  The duration before the start of the stage B (or
the appearance of superhumps) depends on the extent of the superoutburst,
as pointed out by \citet{kat08wzsgelateSH}.  A comparison of the
$O-C$ diagrams further indicates that the stage B evolution was also
different in this superoutburst (figure \ref{fig:swumacomp}).

   During the rapid fading stage of this superoutburst, large-amplitude
quasi-periodic oscillations (QPOs) were recorded (figure \ref{fig:swumaqpo}).
The appearance of large-amplitude QPOs during the rapid fading stage
was recorded during the 2000 and 2002 superoutbursts \citep{soe09swuma}.
The present period of the QPOs is close to theirs (i.e. about the double
of ``super-QPOs'' observed during the 1992 superoutburst,
\cite{kat92swumasuperQPO}).  There must be a common mechanism to excite
these QPOs during the terminal stage of superoutbursts.

   In summary, although the behavior of period variation is generally
similar between different superoutbursts of SW UMa, there was a subtle
dependence on the extent of superoutbursts.

\begin{figure}
  \begin{center}
    \FigureFile(88mm,90mm){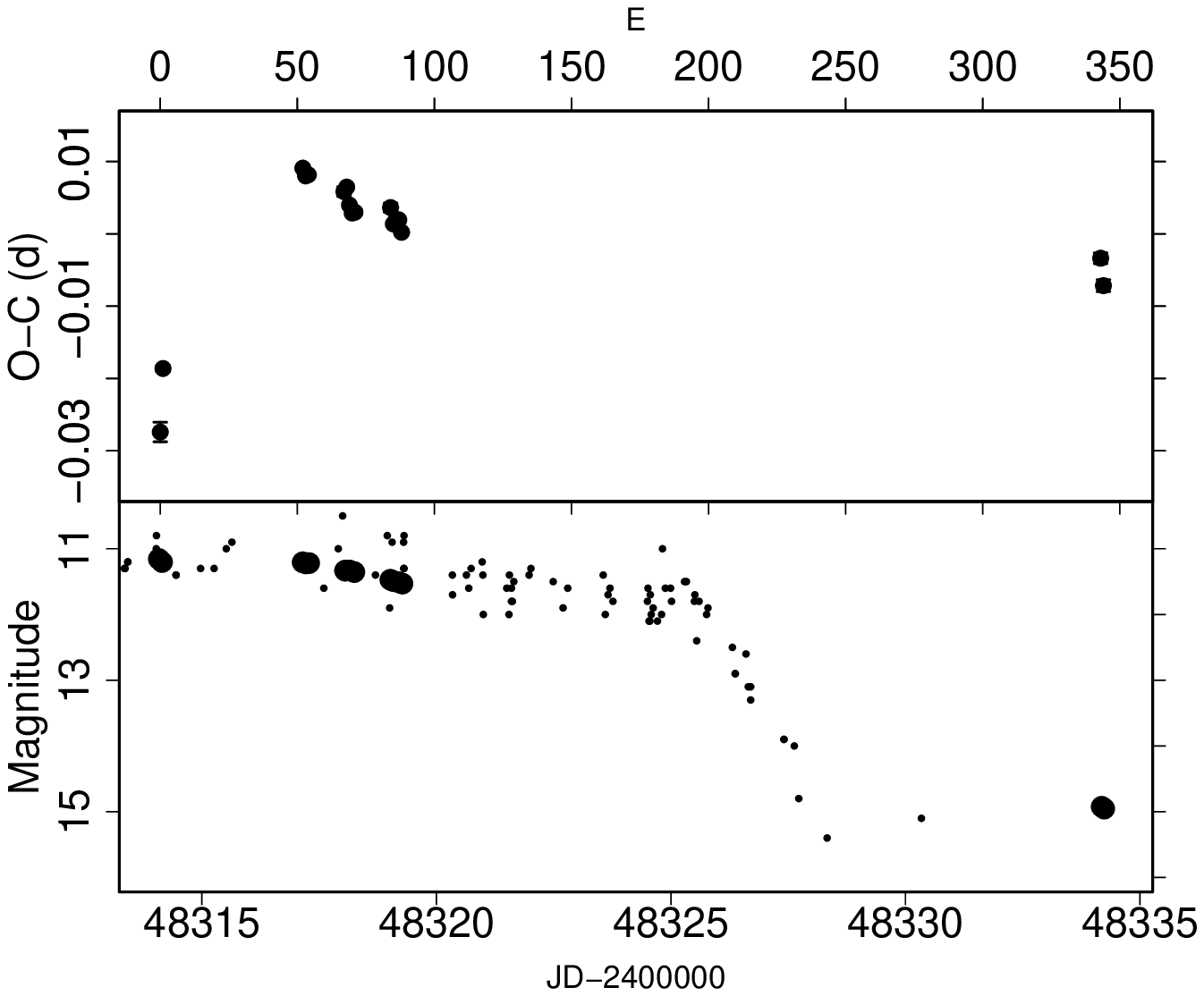}
  \end{center}
  \caption{$O-C$ of superhumps SW UMa (1991).
  (Upper): $O-C$ diagram.
  (Lower): Light curve.  Large dots are our CCD observations and small
  dots are visual observation from the VSOLJ and AAVSO databases.}
  \label{fig:swuma1991oc}
\end{figure}

\begin{figure}
  \begin{center}
    \FigureFile(88mm,70mm){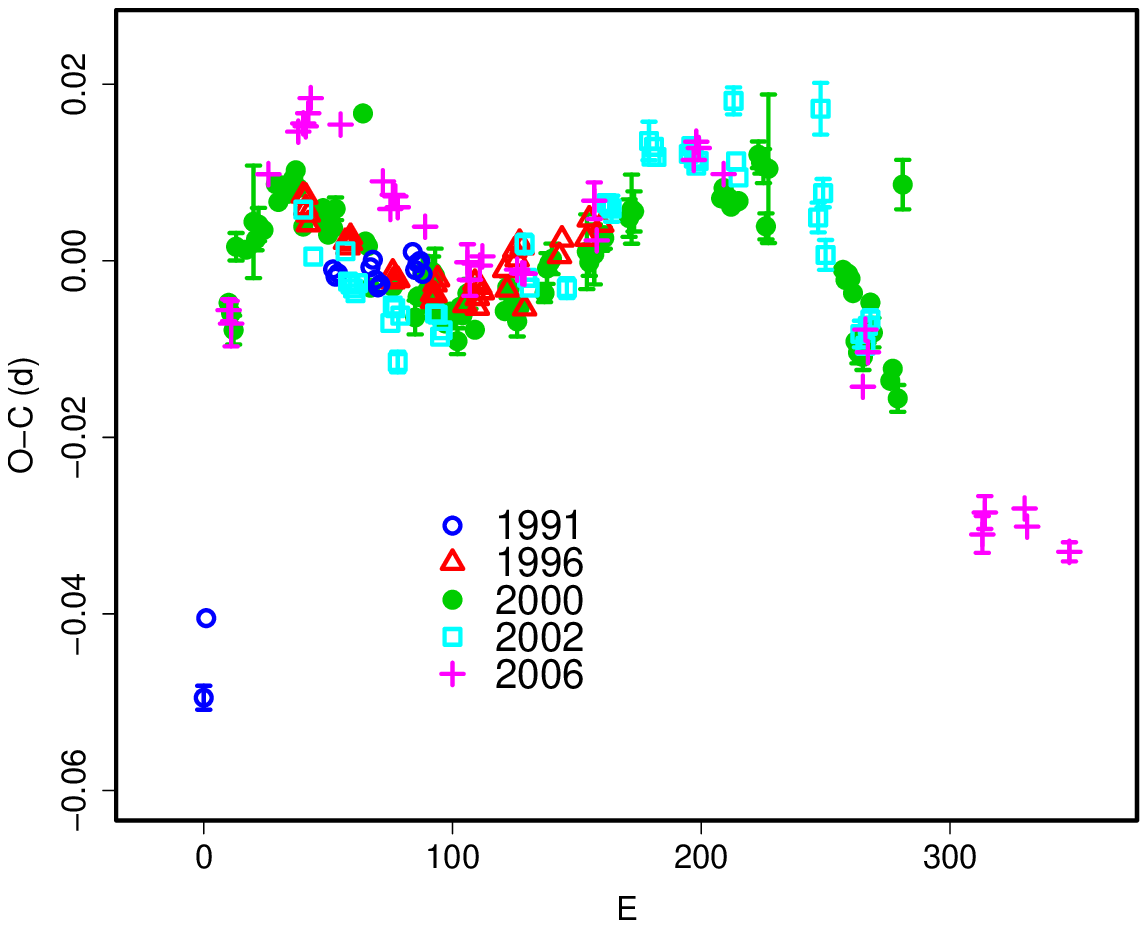}
  \end{center}
  \caption{Comparison of $O-C$ diagrams of SW UMa between different
  superoutbursts.  A period of 0.05822 d was used to draw this figure.
  Since the delay in the appearance of superhumps is known to vary
  in SW UMa, we shifted individual $O-C$ diagrams to get a best
  match (approximately corresponds to a definition of the appearance of
  superhumps to be $E=0$).  The evolution of the bright 2008 superoutburst
  was apparently different from the other superoutbursts.
  }
  \label{fig:swumacomp}
\end{figure}

\begin{figure}
  \begin{center}
    \FigureFile(88mm,110mm){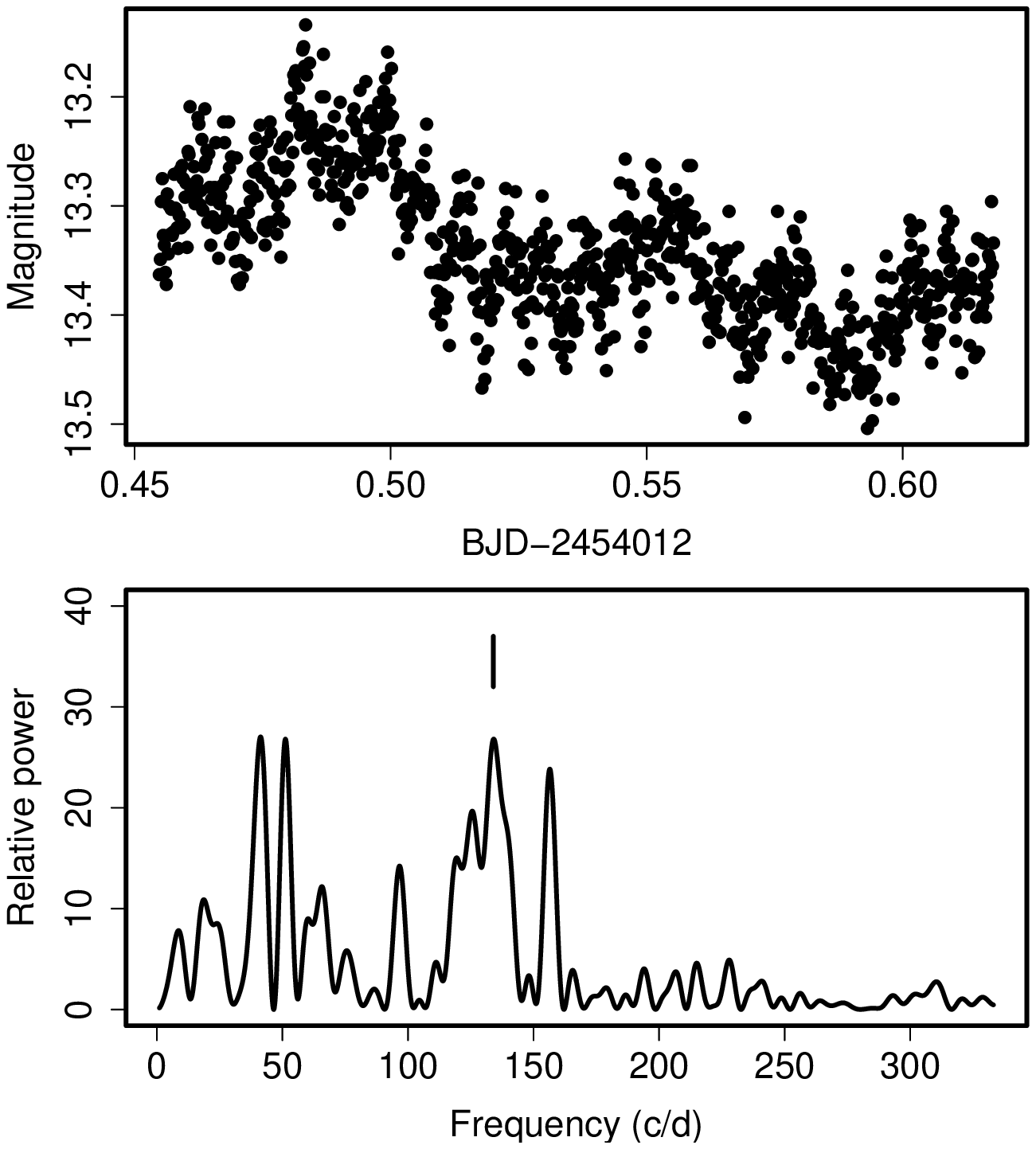}
  \end{center}
  \caption{Quasi-periodic oscillations (QPOs) on 2006 October 4.
  (Upper): Light curve.
  (Lower): Power spectrum after subtracting superhumps.  The signal
  around a frequency 134 cycle/d (11 m) corresponds to the QPOs.}
  \label{fig:swumaqpo}
\end{figure}

\begin{table}
\caption{Superhump maxima of SW UMa (1991).}\label{tab:swumaoc1991}
\begin{center}

\end{center}
\end{table}

\begin{table}
\caption{Superhump maxima of SW UMa (1997).}\label{tab:swumaoc1997}
\begin{center}
%
\end{center}
\end{table}

\begin{table}
\caption{Superhump maxima of SW UMa (2000).}\label{tab:swumaoc2000}
\begin{center}
%
\end{center}
\end{table}

\addtocounter{table}{-1}
\begin{table}
\caption{Superhump maxima of SW UMa (2000) (continued).}
\begin{center}
%
\end{center}
\end{table}

\begin{table}
\caption{Superhump maxima of SW UMa (2002).}\label{tab:swumaoc2002}
\begin{center}
%
\end{center}
\end{table}

\subsection{BC Ursae Majoris}\label{sec:bcuma}\label{obj:bcuma}

   BC UMa was one of the classically known objects displaying
a diversity in the extent of (super)outbursts \citep{rom64bcuma}.
Although the SU UMa-type nature had long been suspected, the definite
detection of superhumps awaited the 1994 detection by M. Iida and
confirmation by C. Kunjaya (unpublished; vsnet-alert 154).\footnote{
$<$http://www.kusastro.kyoto-u.ac.jp/vsnet/DNe/bcuma.html$>$.
}

   \citet{mae07bcuma} observed the 2003 superoutburst and obtained
$P_{\rm dot}$ = $+3.2(0.8) \times 10^{-5}$.  \citet{mae07bcuma} also
detected double-wave ``early superhumps'' before ordinary superhumps
appeared.

   We observed the 2000 and 2003 superoutbursts, the latter
also including the data used in \citet{mae07bcuma}.
The times of superhump maxima are listed in tables
\ref{tab:bcumaoc2000} and \ref{tab:bcumaoc2003}.
These epochs do not include maxima of early superhumps.
The times of maxima for the 2003 superoutburst systematically differ
from those in \citet{mae07bcuma}, probably reflecting the difference
in the template superhump light curve.  This difference was almost
constant during the outburst and did not affect the determination
of the $P_{\rm dot}$.
The both sets of $O-C$'s showed all stages A--C.
We measured $P_{\rm dot}$ for the stage B:
$+4.0(1.4) \times 10^{-5}$ (2000, $16 \le E \le 99$) and
$+4.2(0.8) \times 10^{-5}$ (2003, $15 \le E \le 114$).
The 2003 data also include the times of superhump maxima during the rapidly
fading stage.  The maxima times for $123 \le E \le 189$ were very
well expressed by a constant period of 0.06418(2) d, 0.5 \% shorter
than the mean superhump period.  No apparent phase shift, corresponding
to traditional late superhumps, was detected during
the rapid fading.

\begin{figure}
  \begin{center}
    \FigureFile(88mm,70mm){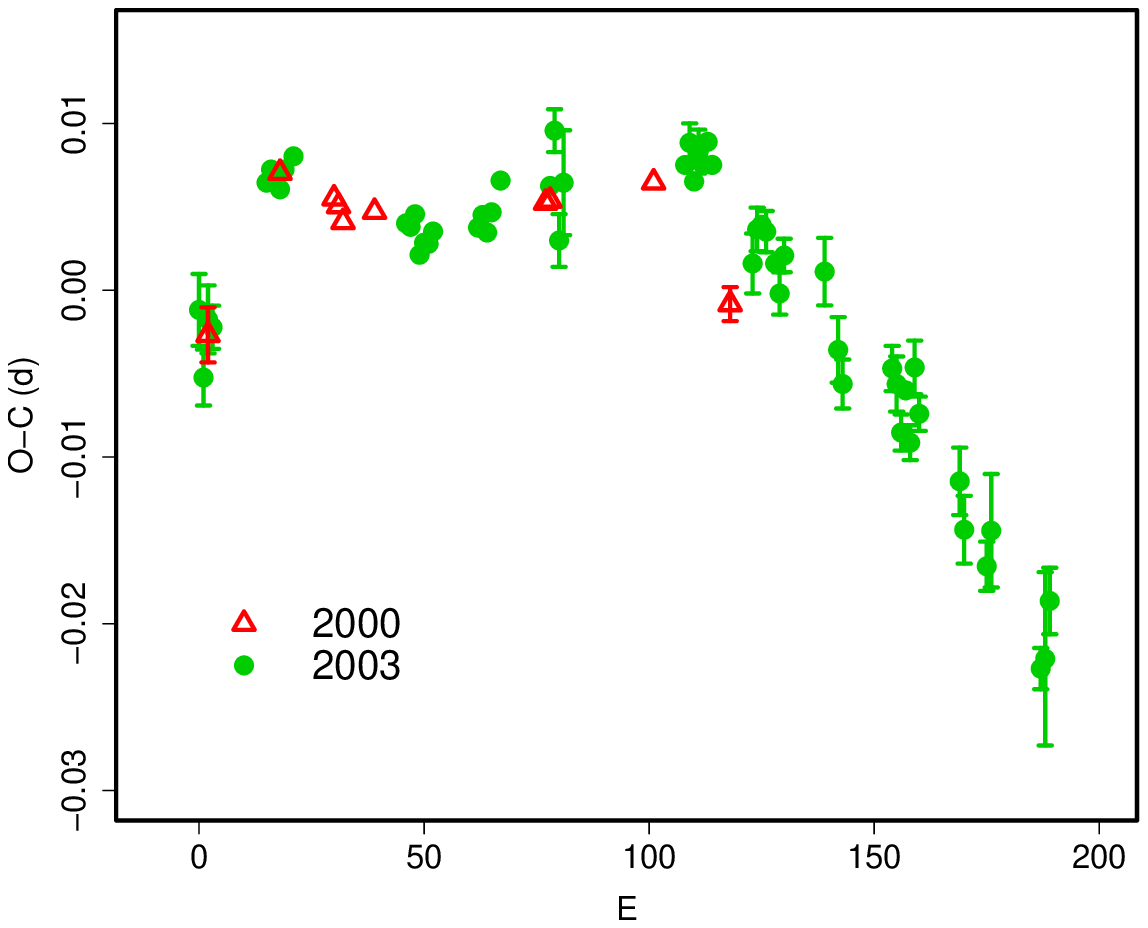}
  \end{center}
  \caption{Comparison of $O-C$ diagrams of BC UMa between different
  superoutbursts.  A period of 0.06455 d was used to draw this figure.
  Approximate cycle counts ($E$) after the appearance of the
  superhumps were used.
  }
  \label{fig:bcumacomp}
\end{figure}

   A comparison of $O-C$ diagrams between different superoutbursts
is shown in figure \ref{fig:bcumacomp}.  The duration of the stage B
was shorter in the 2003 superoutburst, corresponding to the maximum
brightness of the outbursts (11.1 mag for 2000 and 12.2 mag for 2003).

\begin{table}
\caption{Superhump maxima of BC UMa (2000).}\label{tab:bcumaoc2000}
\begin{center}

\end{center}
\end{table}

\begin{table}
\caption{Superhump maxima of BC UMa (2003).}\label{tab:bcumaoc2003}
\begin{center}
%
\end{center}
\end{table}

\addtocounter{table}{-1}
\begin{table}
\caption{Superhump maxima of BC UMa (2003). (continued)}
\begin{center}
%
\end{center}
\end{table}

\subsection{BZ Ursae Majoris}\label{sec:bzuma}\label{obj:bzuma}

\begin{figure}
  \begin{center}
    \FigureFile(88mm,110mm){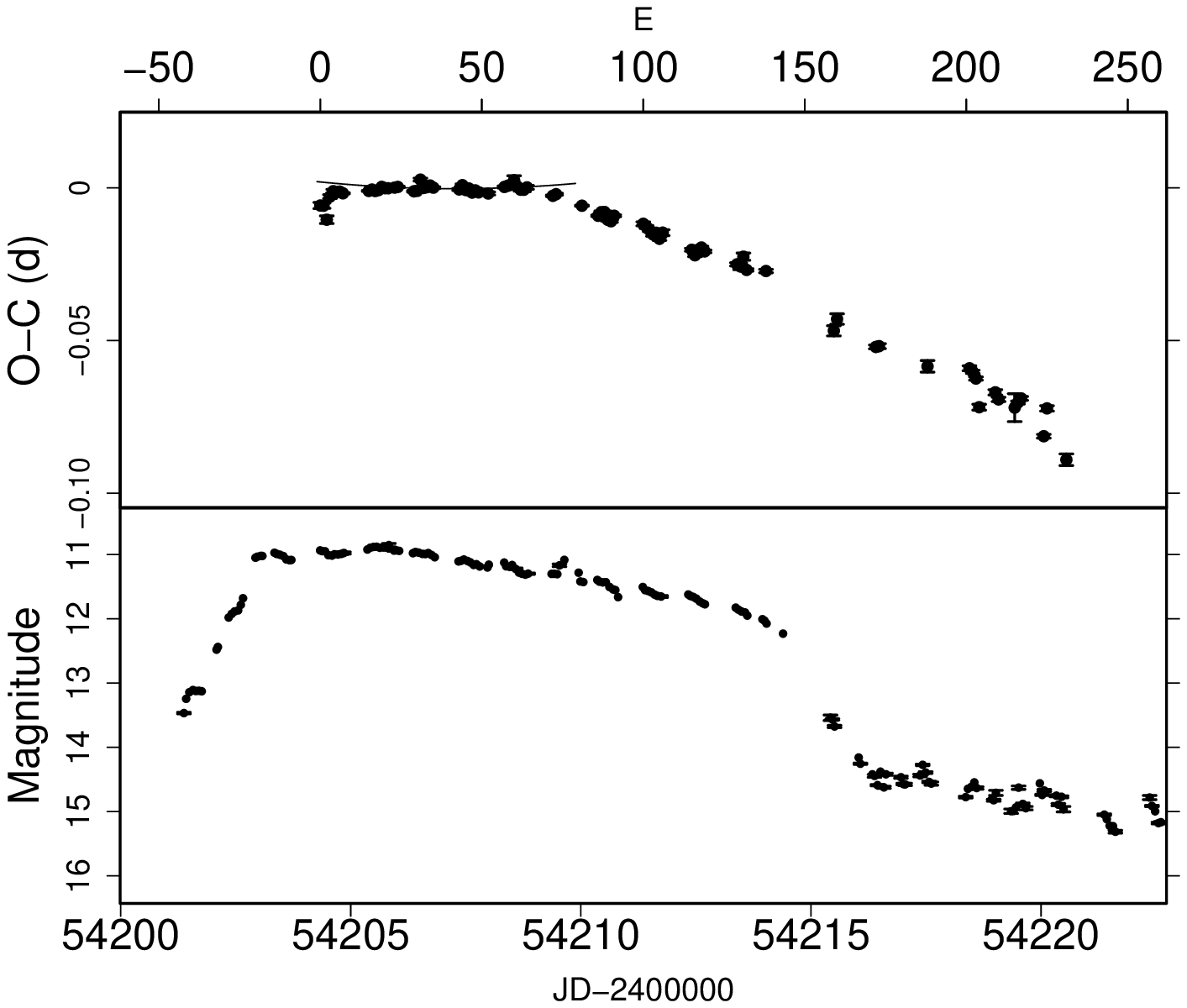}
  \end{center}
  \caption{$O-C$ of superhumps BZ UMa (2007).
  (Upper): $O-C$ diagram.  The values of $O-C$'s are different from
  those listed in table \ref{tab:j0824oc2007} and were calculated from
  a linear fit for the times of superhumps for $19 \le E \le 64$.
  The curve represents a quadratic fit with $P_{\rm dot}$
  = $+3.2 \times 10^{-5}$.
  (Lower): Light curve.  The rise of the superoutburst was very slow,
  apparently accompanied by a stagnation phase (BJD 2454201.5--2454201.8).
  There was a relatively rapid fading probably corresponding to a
  precursor outburst (BJD 2454203.3--2454203.7)}
  \label{fig:bzumaoc}
\end{figure}

   Although BZ UMa had long been suspected to be an SU UMa-type dwarf
nova, no definite superoutbursts were recorded before 2007
(\cite{jur94bzuma}; \cite{rin90bzuma}).
The first-ever recorded superoutburst occurred in 2007.
The times of superhump maxima during this superoutburst are listed
in table \ref{tab:bzumaoc2007}.  We included hump maxima during the
post-superoutburst stage, which will be discussed later.
During the first night of the observation ($E \le 4$), we observed
the growing stage of superhumps.  The superhump period was almost
constant for $19 \le E \le 64$ with
$P_{\rm dot}$ = $+3.6(3.3) \times 10^{-5}$.  We regard this as
the stage B.  An discontinuous transition to a shorter period
(stage C, $72 \le E \le 138$) occurred.  The mean periods for the stages
B and C were 0.07018(1) d and 0.06979(1) d, 3.3 \% and 2.6 \% longer than
the orbital period, respectively.  The superhump period further experienced
a discontinuous shortening after $E=138$ to 0.06968(5) d, 2.4 \% longer than
the orbital period.
BZ UMa is a rare SU UMa-type object around $P_{\rm SH} = $0.07 d without
a distinct segment having a positive $P_{\rm dot}$.  This feature may be
related to the extreme rarity of its superoutbursts.  Furthermore,
the present superoutburst was accompanied by a slow rise before superhumps
grew, suggesting that the outburst was an ``inside-out'' type
(vsnet-alert 9300), rarely met in SU UMa-type dwarf novae.
There was also a suggestion of the presence of a precursor-type
outburst (figure \ref{fig:bzumaoc}).
These features possibly suggest that the 3:1 resonance is difficult to
achieve in this system, and the superhumps were critically excited during
this superoutburst, likely leaving little mass beyond the 3:1 resonance.

   After $E = 167$ (around the start of the post-superoutburst stage),
secondary hump maxima were sometimes present, which later became stronger
than the original hump maxima.  The times of these secondary humps are
listed in table \ref{tab:bzumaoc2007b}, giving $O-C$'s using the same
ephemeris as in the earlier maxima.  These data indicated that
post-superoutburst superhumps persisted at least for $\sim$150 cycles,
or $\sim$10 d.

\begin{table}
\caption{Superhump maxima of BZ UMa (2007).}\label{tab:bzumaoc2007}
\begin{center}

\end{center}
\end{table}

\addtocounter{table}{-1}
\begin{table}
\caption{Superhump maxima of BZ UMa (2007). (continued)}
\begin{center}
%
\end{center}
\end{table}

\begin{table}
\caption{Secondary Maxima of BZ UMa (2007).}\label{tab:bzumaoc2007b}
\begin{center}
%
\end{center}
\end{table}

\subsection{CI Ursae Majoris}\label{obj:ciuma}

   CI UMa was discovered as a dwarf nova by \citet{gor72ciuma}.
\citet{nog97ciuma} observed the 1995 superoutburst and reported
the superhump period.  This observation was not long enough to determine
the period derivative.  Although \citet{nog97ciuma} suggested a supercycle
of $\sim$ 140 d based on the shortest interval between apparent superoutbursts
\citep{kol79cpdraciuma}, recent observations suggest that superoutbursts
occur less regularly.

\begin{figure}
  \begin{center}
    \FigureFile(88mm,70mm){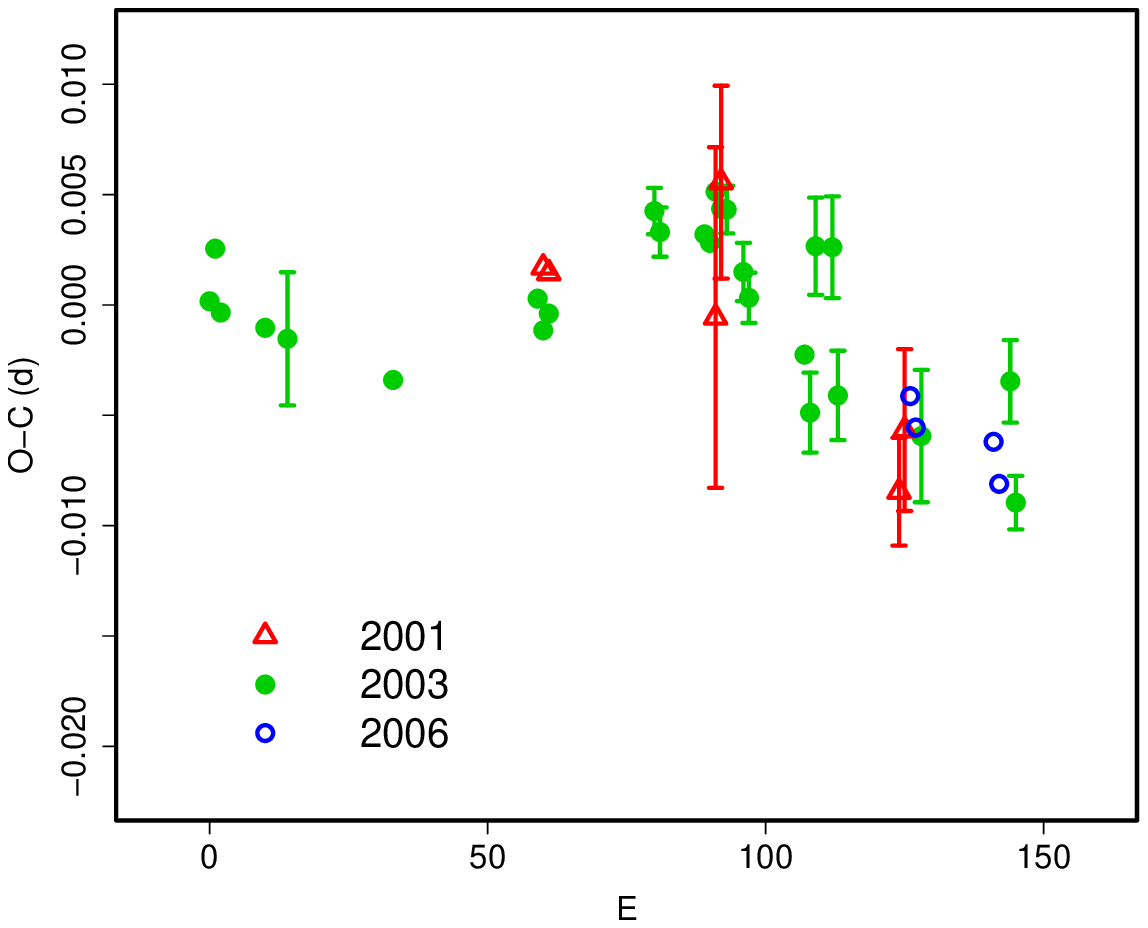}
  \end{center}
  \caption{Comparison of $O-C$ diagrams of CI UMa between different
  superoutbursts.  A period of 0.06264 d was used to draw this figure.
  Since the start of the outbursts were not clearly defined, the $O-C$
  diagrams were shifted to best match the 2003 one.
  }
  \label{fig:ciumacomp}
\end{figure}

   We further observed the 2001, 2003 and 2006 superoutburst
(tables \ref{tab:ciumaoc2001}, \ref{tab:ciumaoc2003},
\ref{tab:ciumaoc2006}).
The 2001 observation probably covered only the later part of the
superoutburst and likely recorded a stage B--C transition.
The 2003 $O-C$ diagram showed a clear stage B--C transition
(cf. figure \ref{fig:octrans}).
We determined $P_{\rm dot}$ = $+6.4(1.2) \times 10^{-5}$ for $E \le 93$.
The 2006 observation recorded the terminal stage of the superoutburst
(stage C superhumps).
A combined $O-C$ diagram is presented in figure \ref{fig:ciumacomp}.

\begin{table}
\caption{Superhump maxima of CI UMa (2001).}\label{tab:ciumaoc2001}
\begin{center}

\end{center}
\end{table}

\begin{table}
\caption{Superhump maxima of CI UMa (2003).}\label{tab:ciumaoc2003}
\begin{center}
%
\end{center}
\end{table}

\begin{table}
\caption{Superhump maxima of CI UMa (2006).}\label{tab:ciumaoc2006}
\begin{center}
%
\end{center}
\end{table}

\subsection{CY Ursae Majoris}\label{obj:cyuma}

   \citet{har95cyuma} observed the 1995 superoutburst and reported
a global $P_{\rm dot}$ of $-5.8 \times 10^{-5}$.  Their $O-C$ diagram,
however, also can be interpreted as a transition from a longer to
a shorter period (stage B--C transition) during the late stage of
the superoutburst.
Using the earlier part ($E \le 73$) of their table of superhump maxima,
we obtained $P_{\rm dot}$ = $+2.7(1.0) \times 10^{-5}$.

\begin{figure}
  \begin{center}
    \FigureFile(88mm,70mm){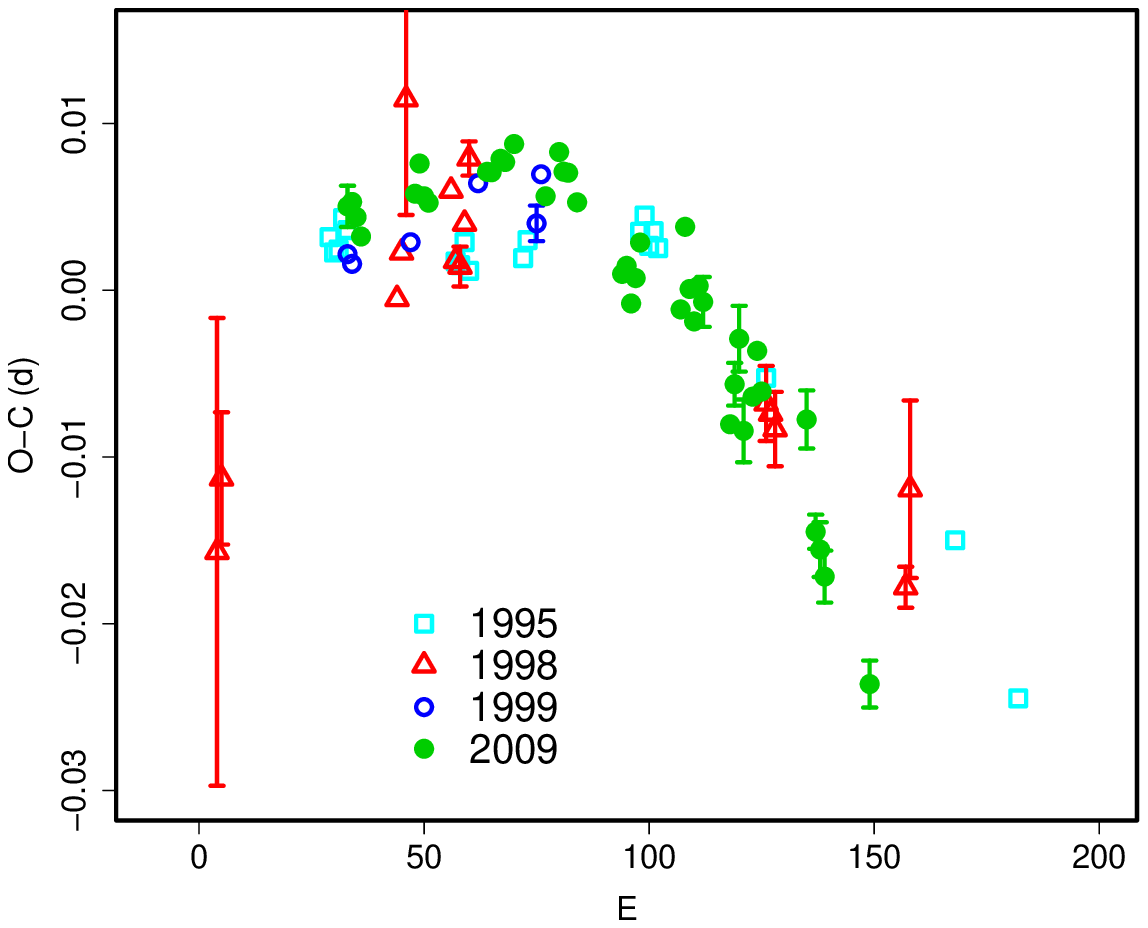}
  \end{center}
  \caption{Comparison of $O-C$ diagrams of CY UMa between different
  superoutbursts.  A period of 0.07212 d was used to draw this figure.
  Approximate cycle counts ($E$) after the start of the
  superoutburst were used.
  }
  \label{fig:cyumacomp}
\end{figure}

   We analyzed the 1998 AAVSO data and found a clear stage B--C
transition (table \ref{tab:cyumaoc1998}).  The parameters are given
in table \ref{tab:perlist}.
Our 1999 observation \citep{kat99cyuma} did not show a clear tendency of
a period decrease, probably because of the insufficient data coverage
(table \ref{tab:cyumaoc1999}).
The 2009 superoutburst was well-observed during the middle-to-late
stage (table \ref{tab:cyumaoc2009}).  A clear stage B--C transition
was recorded.
A comparison of $O-C$ diagrams between different superoutbursts
is shown in figure \ref{fig:cyumacomp}.  There was a possible slight
difference in behavior during the stage B between different superoutbursts.
Observations at early epochs of superoutbursts are wanted.

\begin{table}
\caption{Superhump maxima of CY UMa (1998).}\label{tab:cyumaoc1998}
\begin{center}

\end{center}
\end{table}

\begin{table}
\caption{Superhump maxima of CY UMa (1999).}\label{tab:cyumaoc1999}
\begin{center}
%
\end{center}
\end{table}

\begin{table}
\caption{Superhump maxima of CY UMa (2009).}\label{tab:cyumaoc2009}
\begin{center}
%
\end{center}
\end{table}

\subsection{DV Ursae Majoris}\label{obj:dvuma}

   This eclipsing SU UMa-type dwarf nova has been well documented
(\cite{pat00dvuma}, \cite{nog01dvuma}).  Relatively large negative
period derivatives have been reported.  We summarize our observations
of five superoutbursts (1997, 1999, 2002, 2005 and 2007).
The maxima were determined from observations
outside the eclipses, as described in V2051 Oph.

   The times of superhump maxima for the 1997 superoutburst
(table \ref{tab:dvumaoc1997}) were determined by using a combination of
the AAVSO and data in \citet{nog01dvuma}.
We also incorporated times of superhump maxima reported in
\citet{pat00dvuma}.  Although the AAVSO data we analyzed were included
in \citet{pat00dvuma}, we presented our new determinations because
\citet{pat00dvuma} gave epochs only to 0.001 d.  Since the mean difference
between our measurements and those by \citet{pat00dvuma} was negligible
(0.0010(9) d), we did not make a systematic correction between them.
The combined result showed negative $O-C$'s for the earliest stage
(stage A, $E \le 5$), followed
by a segment of relatively constant period (stage B, $7 \le E \le 79$),
then by a transition to a shorter period (stage C, $104 \le E \le 184$).
The mean superhump periods of the stages B and C
were 0.08878(4) d and 0.08840(3) d, respectively.  The $P_{\rm dot}$
for the stage B was $-0.9(4.0) \times 10^{-5}$.

   During The 1999 superoutburst (table \ref{tab:dvumaoc1999}),
there was a discontinuous shortening (stage B to C)
of the period after $E = 80$ as in the 1997 superoutburst.
The mean periods before and after this transitions were
0.08893(3) d and 0.08836(8) d, respectively.  The $P_{\rm dot}$
before the transition was $-4.7(3.4) \times 10^{-5}$.
The 2002 superoutburst showed a similar pattern of $O-C$ variation
(table \ref{tab:dvumaoc2002}), although the observations were rather
sparse.

\begin{figure}
  \begin{center}
    \FigureFile(88mm,70mm){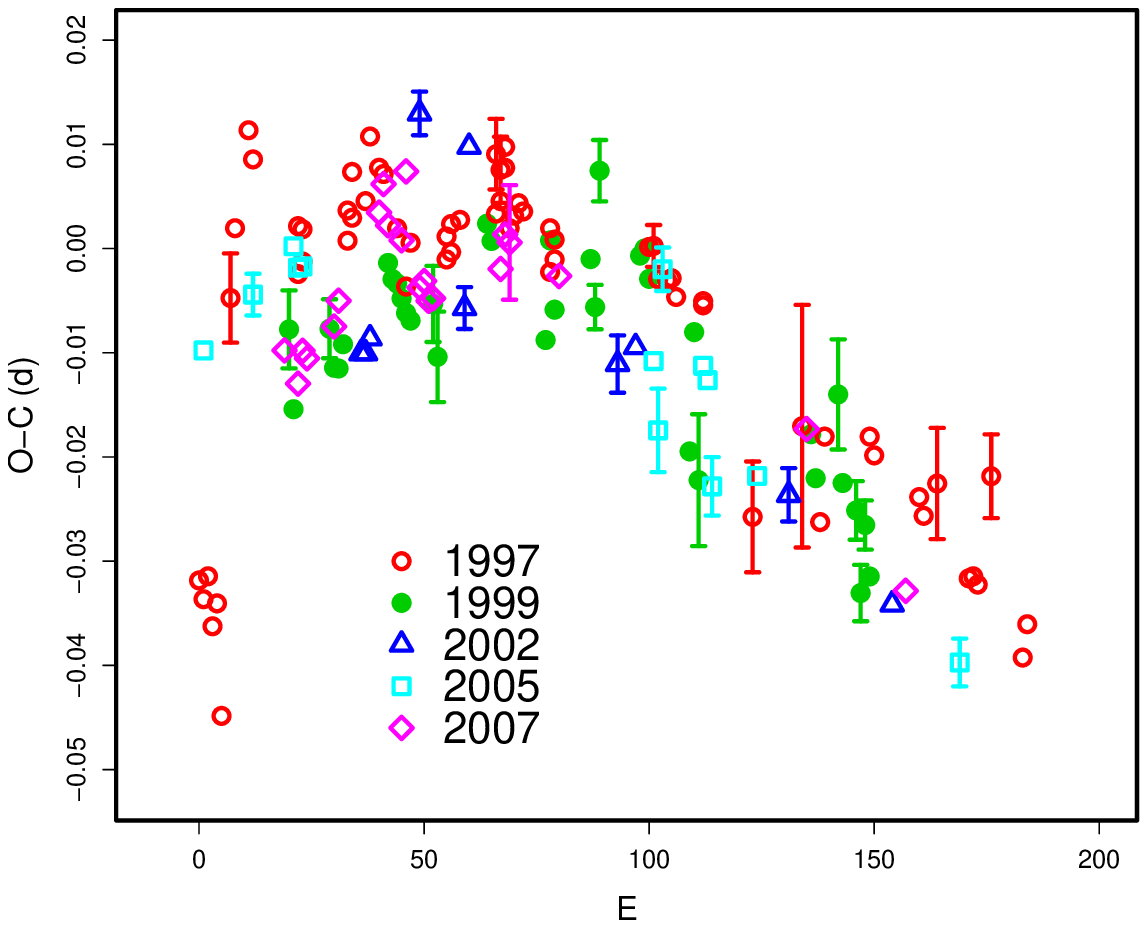}
  \end{center}
  \caption{Comparison of $O-C$ diagrams of DV UMa between different
  superoutbursts.  The $O-C$'s were calculated against a period of
  0.0888 d.  Approximate cycle counts ($E$) after the start of the
  superoutburst were used.
  }
  \label{fig:dvumacomp}
\end{figure}

   The 2005 and 2007 observations well
covered the growing stage of superhumps (tables \ref{tab:dvumaoc2005}
and \ref{tab:dvumaoc2007}).
As in other systems and as in 1997 superoutburst, the $O-C$'s of this
evolutionary stage were negative.
Regarding the 2007 superoutburst, we can determine
$P_{\rm dot}$ = $-1.7(1.8) \times 10^{-5}$ after this evolutionary stage,
corresponding to the stage B of the 1997 superoutburst.
In summary, we did not find strong difference in the behavior of period
variation between different superoutbursts (figure \ref{fig:dvumacomp}).

\begin{table}
\caption{Superhump maxima of DV UMa (1997).}\label{tab:dvumaoc1997}
\begin{center}

\end{center}
\end{table}

\addtocounter{table}{-1}
\begin{table}
\caption{Superhump maxima of DV UMa (1997). (continued)}
\begin{center}
%
\end{center}
\end{table}

\begin{table}
\caption{Superhump maxima of DV UMa (1999).}\label{tab:dvumaoc1999}
\begin{center}
%
\end{center}
\end{table}

\begin{table}
\caption{Superhump maxima of DV UMa (2002).}\label{tab:dvumaoc2002}
\begin{center}
%
\end{center}
\end{table}

\begin{table}
\caption{Superhump maxima of DV UMa (2005).}\label{tab:dvumaoc2005}
\begin{center}
%
\end{center}
\end{table}

\subsection{ER Ursae Majoris}\label{obj:eruma}

   The times of superhump maxima used to draw figure \ref{fig:erumahumpall}
are listed in tables \ref{tab:erumaoc1995-1} and \ref{tab:erumaoc1995-2}.

\begin{table}
\caption{Superhump maxima (1) of ER UMa (1995).}\label{tab:erumaoc1995-1}
\begin{center}
%
\end{center}
\end{table}

\begin{table}
\caption{Superhump maxima (2) of ER UMa (1995).}\label{tab:erumaoc1995-2}
\begin{center}
%
\end{center}
\end{table}

\subsection{IY Ursae Majoris}\label{obj:iyuma}

   This eclipsing SU UMa-type dwarf nova has been well documented
(\cite{uem00iyuma}; \cite{pat00iyuma}).  The superhump maxima
were determined from observations outside the eclipses, as described
in V2051 Oph.

   \citet{pat00iyuma} reported a ``normal'' negative period derivative.
We combined the reported times of superhump maxima with ours,
by adding a systematic difference of 0.0028 d (presumably due to
the difference in the procedure of determination of maxima) to
the times of \citet{pat00iyuma}.  The resultant times are listed in
table \ref{tab:iyumaoc2000}.  We restricted the analysis to the interval
before the rapid fading started, i.e. excluding times of late superhumps.
The $O-C$ diagram was complex and was different from the one in
\citet{pat00iyuma}, in that the present diagram clearly showed
a transition from a longer period to a stable period around $E = 23$
(stage A--B transition).
The main difference in appearance between \citet{pat00iyuma} and ours
was thus caused by the lack of early-stage superhumps in \citet{pat00iyuma}.

   The $P_{\rm dot}$ during the later interval was much closer
to zero than the global $P_{\rm dot}$ reported in \citet{pat00iyuma}.
The behavior after $E = 106$ was slightly different between ours
and \citet{pat00iyuma}.  Our data suggested a shortening of the
period while \citet{pat00iyuma} showed a steady increase.
This may have been caused by the increasing signal of late superhumps,
which predominated in later epochs, during the observation of
\citet{pat00iyuma}.  Excluding $E < 23$ and $E \ge 106$,
we obtained $P_{\rm dot}$ = $-1.8(2.2) \times 10^{-5}$.

   We also analyzed the 2002, 2004 and 2006 superoutbursts
(table \ref{tab:iyumaoc2002}, \ref{tab:iyumaoc2004} and
\ref{tab:iyumaoc2006}).
The 2002 observation covered the middle-to-late part of the outburst.
There was an apparent discontinuous transition to a shorter
period around $E = 137$.  Due to the gap in the observation, we could
not significantly determine the $P_{\rm dot}$ before this transition.
This 2004 observation covered the middle part to the latter half of
the outburst.
Although the initial stage of the 2006 superoutburst
was observed, the superhump maxima incidentally fell amid the eclipses.
We excluded most of the first two nights for calculating times of
superhump maxima.
The superhump profile at this stage was probably double-peaked.
Such a feature may have reflected the growing stage of the superhumps
and needs to be investigated in future superoutbursts.
The $O-C$'s after $E > 221$ apparently showed a phase shift attributable
to traditional late superhumps, as in the 2000 superoutburst.
The periods given in table \ref{tab:perlist} were determined
by excluding the maximum $E=176$.

   The 2007 superoutburst was caught by chance at $V=14$.  Judging from
the superhump maxima (table \ref{tab:iyumaoc2007}), there was a clear
decrease in the superhump period.  In conjunction with the faintness,
we probably observed the late stage of a superoutburst associated
with a stage B--C transition.  The nominal value
$P_{\rm dot}$ = $-16.0(6.5) \times 10^{-5}$ would not be a good
representative period derivative.

   The 2009 superoutburst was well-observed during the middle-to-late
stages (table \ref{tab:iyumaoc2009}).  The $O-C$ diagram clearly depicts
the presence of stages B and C.  The first ($E=0$ epoch probably corresponds
to the stage A.  Although orbital humps emerged after the rapid decline
from the superoutburst plateau, no prominent traditional late superhumps
were recorded (cf. the 2000 superoutburst, \cite{pat00iyuma}).

\begin{figure}
  \begin{center}
    \FigureFile(88mm,70mm){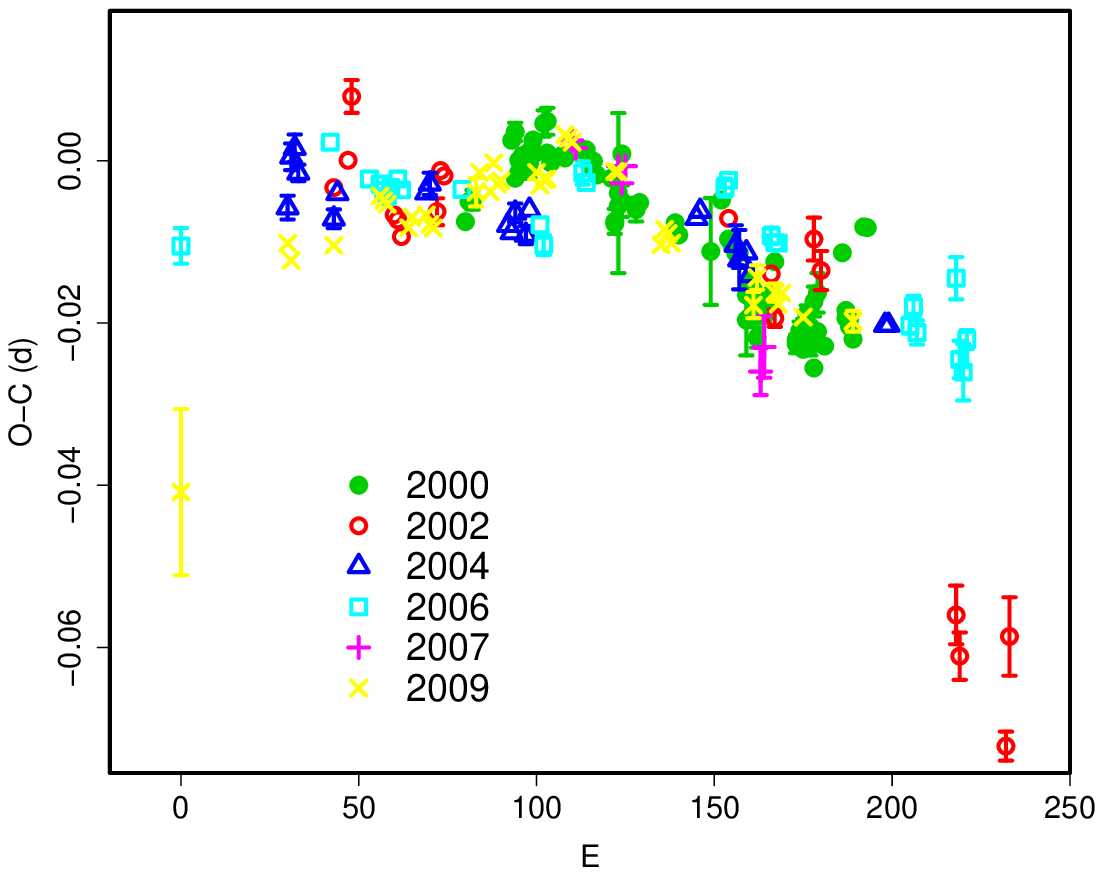}
  \end{center}
  \caption{Comparison of $O-C$ diagrams of IY UMa between different
  superoutbursts.  A period of 0.07610 d was used to draw this figure.
  Approximate cycle counts ($E$) after the start of the
  superoutburst were used.
  }
  \label{fig:iyumacomp}
\end{figure}

   A combined $O-C$ diagram drawn from all the superoutburst is
presented in figure \ref{fig:iyumacomp}.
The combined diagram appears to show stage B
lasting for $\sim$ 120 cycles with a positive $P_{\rm dot}$.
The duration of the stage B is compatible that during the 2009
superoutburst, although the behavior of the 2009 $O-C$ looks slightly
different from others during its early stage.
The likely presence of a positive $P_{\rm dot}$, then
would suggest the similarity to NSV 4838 \citep{ima09nsv4838}.
The lack of a positive $P_{\rm dot}$ in individual superoutbursts
may have been a result from the deficiency of observations around
the end of the stage B.  Some of superoutbursts seem to show
stage C superhumps while others tend to show humps resembling
traditional late superhumps.  Future observations of this object
at these epochs will be particularly important.

\begin{table}
\caption{Superhump maxima of IY UMa (2000).}\label{tab:iyumaoc2000}
\begin{center}

\end{center}
\end{table}

\addtocounter{table}{-1}
\begin{table}
\caption{Superhump maxima of IY UMa (2000) (continued).}
\begin{center}
%
\end{center}
\end{table}

\begin{table}
\caption{Superhump maxima of IY UMa (2002).}\label{tab:iyumaoc2002}
\begin{center}
%
\end{center}
\end{table}

\begin{table}
\caption{Superhump maxima of IY UMa (2004).}\label{tab:iyumaoc2004}
\begin{center}
%
\end{center}
\end{table}

\begin{table}
\caption{Superhump maxima of IY UMa (2006).}\label{tab:iyumaoc2006}
\begin{center}
%
\end{center}
\end{table}

\subsection{KS Ursae Majoris}\label{obj:ksuma}

   KS UMa (=SBS1017$+$533) was originally discovered as an emission-line
object \citep{bal97SBS2spec}.  In 1998, the object was found to be
in outburst during a spectroscopic survey (P. Garnavich, vsnet-alert 1441).
T. Vanmunster reported the detection of superhumps with a period of
0.069(1) d during this outburst (CVC 161, also in vsnet-alert 1448).
\citet{haz99ksuma} surveyed historical outbursts.  \citet{jia00RASSCV}
also selected this CV from the ROSAT all-sky survey.

   \citet{ole03ksuma} reported on the period variation of superhumps
in KS UMa.  We had more extensive data on the same superoutburst,
notably covering the earlier stage than in \citet{ole03ksuma}.
Table \ref{tab:ksumaoc2003} presents the combined list of times of
superhump maxima, after adding a systematic difference of 0.003 d to
\citet{ole03ksuma}.
The entire data now clearly show a sharp transition from a longer period
in the early stage (before $E = 15$), stabilized segment with a slightly
positive $P_{\rm dot}$, and followed by a sharp transition to
a shorter period after $E = 95$.  The pattern of period change can
be reasonably interpreted as stages A--C.
The negative $P_{\rm dot}$ in \citet{ole03ksuma} was a result of
the fit to the stages B and C together.
Our data yielded $P_{\rm dot}$ = $+2.2(1.1) \times 10^{-5}$ for the stage B.

   We also observed the 2007 superoutburst (table \ref{tab:ksumaoc2007}).
The observation covered the middle-to-late plateau stage.  Excluding
the last point (taken during the rapid fading stage), we obtained
$P_{\rm dot}$ = $+1.5(1.9) \times 10^{-5}$, probably corresponding to
the stage B of the 2003 superoutburst.

\begin{figure}
  \begin{center}
    \FigureFile(88mm,70mm){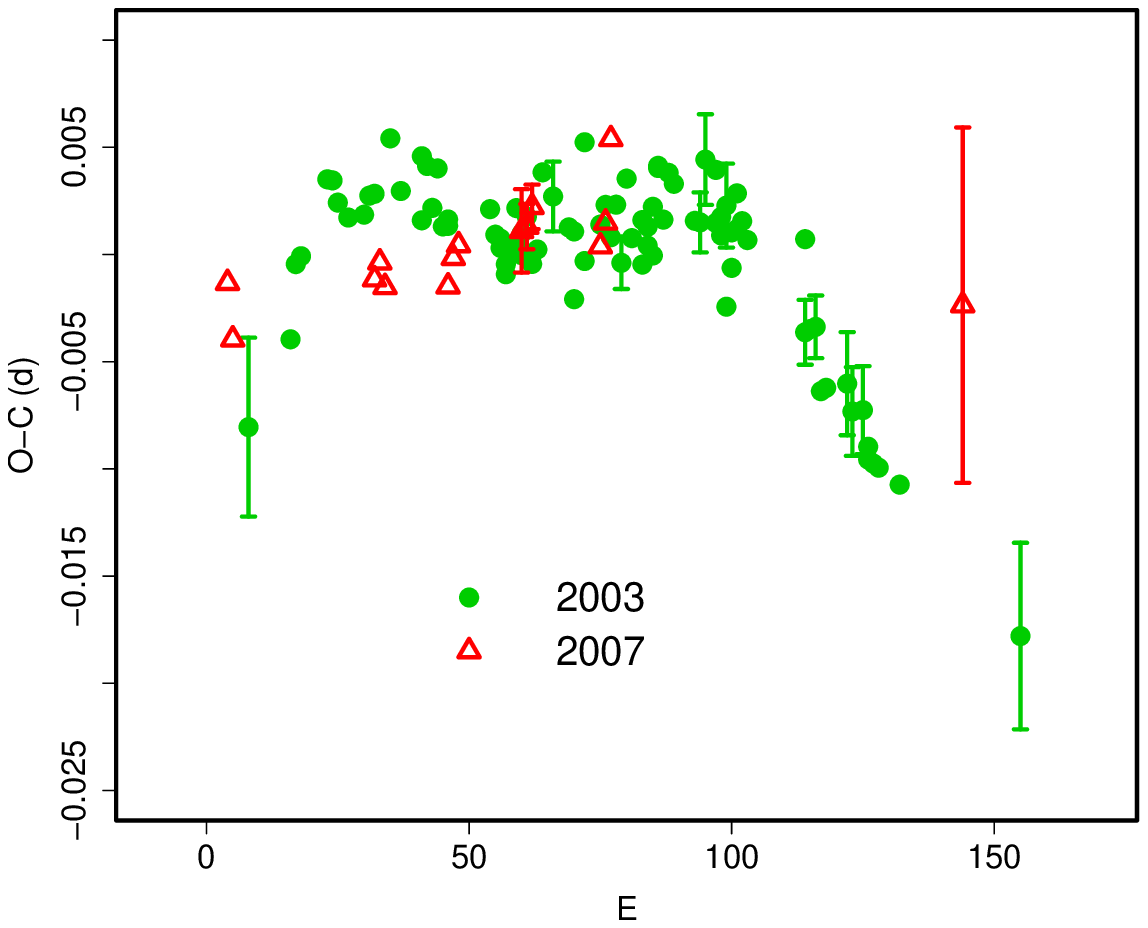}
  \end{center}
  \caption{Comparison of $O-C$ diagrams of KS UMa between different
  superoutbursts.  A period of 0.07019 d was used to draw this figure.
  Approximate cycle counts ($E$) after the start of the
  superoutburst were used.
  }
  \label{fig:ksumacomp}
\end{figure}

   The $O-C$ behavior was slightly different between the 2003 and 2007
superoutbursts (figure \ref{fig:ksumacomp}).  This may have been a result
of a longer duration of the 2003 superoutburst than the 2007 one.

\begin{table}
\caption{Superhump maxima of KS UMa (2003).}\label{tab:ksumaoc2003}
\begin{center}

\end{center}
\end{table}

\addtocounter{table}{-1}
\begin{table}
\caption{Superhump maxima of KS UMa (2003) (continued).}
\begin{center}
%
\end{center}
\end{table}

\begin{table}
\caption{Superhump maxima of KS UMa (2007).}\label{tab:ksumaoc2007}
\begin{center}
%
\end{center}
\end{table}

\subsection{KV Ursae Majoris}\label{sec:kvuma}\label{obj:kvuma}

   This object is a BHXT.  The times of superhump maxima, a reanalysis
of \citet{uem02j1118}, used for drawing figure \ref{fig:kvumaoc}
(subsection \ref{sec:BHXT}) are listed in table \ref{tab:kvumaoc2000}.

   The $O-C$ diagram was composed of three stages as in SU UMa-type
dwarf novae: stage A ($E \le 124$) with a mean $P_{\rm SH}$ = 0.17082(7) d,
stage B for $124 \le E \le 348$ (mean $P_{\rm SH}$ ($P_1$) = 0.17056(3) d
and $P_{\rm dot}$ = $+0.9(0.6) \times 10^{-5}$)
and stage C for $E \ge 238$ (mean $P_{\rm SH}$ ($P_2$) = 0.17038(3) d).
The global $P_{\rm dot}$ was $-0.43(0.05) \times 10^{-5}$.

\begin{table}
\caption{Superhump maxima of KV UMa (2000).}\label{tab:kvumaoc2000}
\begin{center}

\end{center}
\end{table}

\subsection{MR Ursae Majoris}\label{obj:mruma}

\begin{figure}
  \begin{center}
    \FigureFile(88mm,70mm){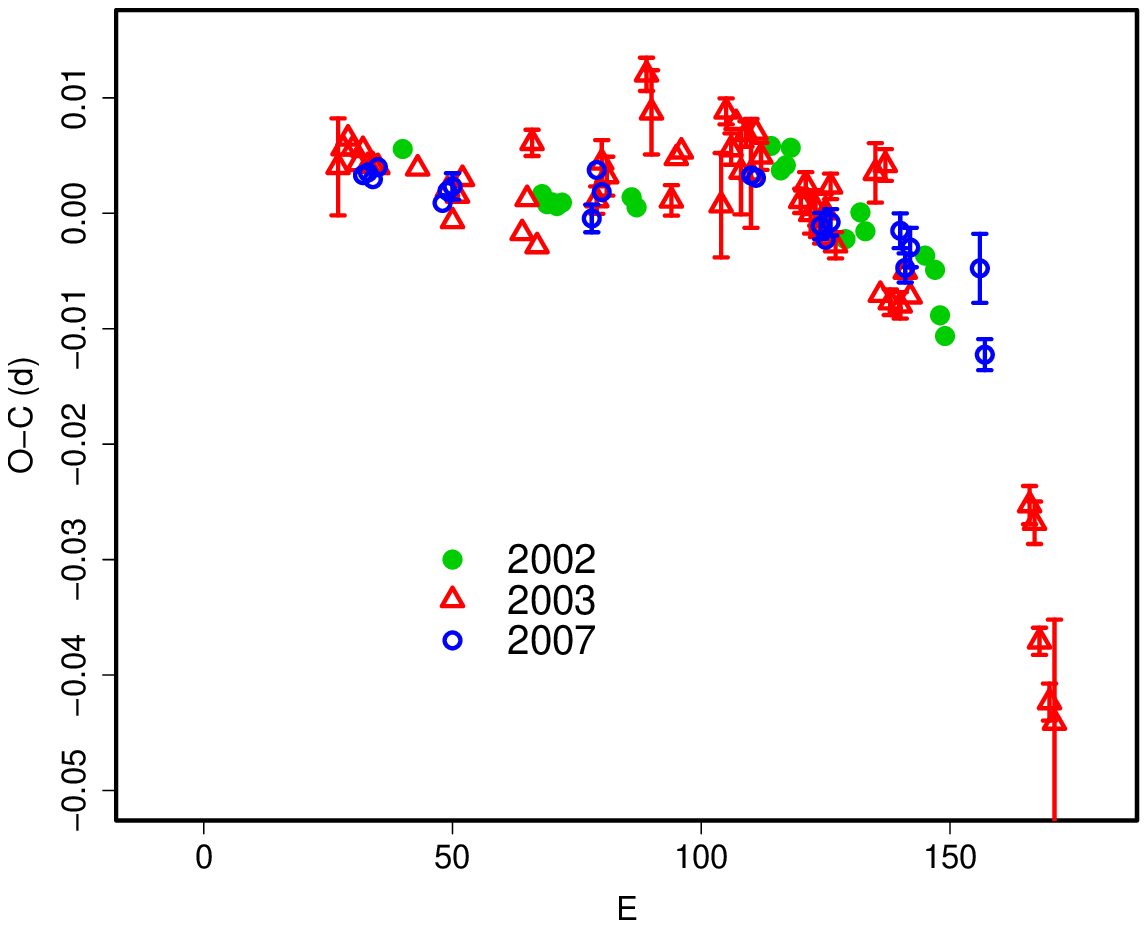}
  \end{center}
  \caption{Comparison of $O-C$ diagrams of MR UMa between different
  superoutbursts.  A period of 0.06512 d was used to draw this figure.
  Approximate cycle counts ($E$) after the start of the
  2007 superoutburst were used.  Since the starts of the 2002 and 2003
  superoutburst were not well constrained, we shifted the $O-C$ diagrams
  to best fit the 2007 one.
  }
  \label{fig:mrumacomp}
\end{figure}

   MR UMa = 1RXP J113123$+$4322.5 is an ROSAT-selected CV
\citep{wei97mruma}, which underwent the first secure recorded
outburst in 2002 (vsnet-alert 7221).
We observed the middle-to-late stage of the 2002 superoutburst
(table \ref{tab:mrumaoc2002}, figure \ref{fig:octrans}).
The data clearly indicated a stage B--C transition around $E = 80$.
The $P_{\rm dot}$ of the stage B was $+9.3(1.2) \times 10^{-5}$.
The behavior was very similar during the 2003 and 2007 superoutbursts
(tables \ref{tab:mrumaoc2003}, \ref{tab:mrumaoc2007};
figure \ref{fig:mrumacomp}),
with $P_{\rm dot}$ = $+6.0(2.3) \times 10^{-5}$ (2003, $E \le 84$) and
$P_{\rm dot}$ = $+3.8(1.6) \times 10^{-5}$ (2007, $E \le 79$).
For more information of the 2003, 2004 and 2005 superoutbursts,
see \citet{tan07mruma}, although they did not distinguish different
stages of period evolution.

\begin{table}
\caption{Superhump maxima of MR UMa (2002).}\label{tab:mrumaoc2002}
\begin{center}

\end{center}
\end{table}

\begin{table}
\caption{Superhump maxima of MR UMa (2003).}\label{tab:mrumaoc2003}
\begin{center}
%
\end{center}
\end{table}

\begin{table}
\caption{Superhump maxima of MR UMa (2007).}\label{tab:mrumaoc2007}
\begin{center}
%
\end{center}
\end{table}

\subsection{CU Velorum}\label{obj:cuvel}

   Although CU Vel had long been known as an SU UMa-type dwarf nova
\citep{vog80suumastars}, the details of the reported superhump period
(0.0799 d, \cite{RitterCV3}) was not reported in a solid publication.
\citet{men96cuvel} reported an orbital period of 0.0785 d.

\begin{figure}
  \begin{center}
    \FigureFile(88mm,110mm){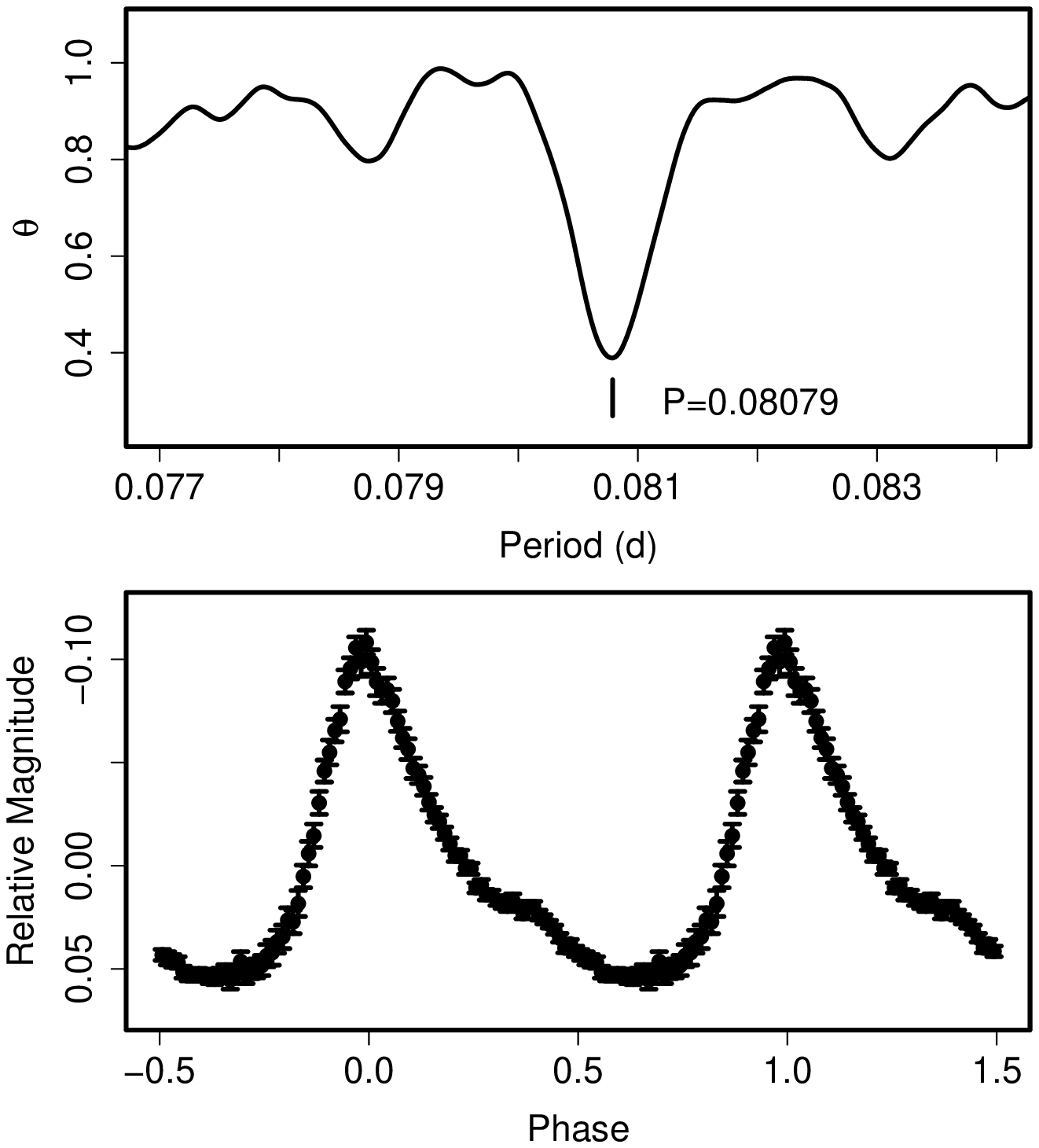}
  \end{center}
  \caption{Superhumps in CU Vel (2002). (Upper): PDM analysis.
     (Lower): Phase-averaged profile.}
  \label{fig:cuvelshpdm}
\end{figure}

   We observed the 2002 superoutburst.
The times of superhumps maxima are listed in table \ref{tab:cuveloc2002}.
The object clearly showed the stage A development with a longer period.
Excluding this epoch ($E = 0$), we obtained $P_{\rm dot}$ =
$-8.4(1.4) \times 10^{-5}$ for the stage B.
A PDM analysis yielded a mean superhump period of 0.080789(5) d
(figure \ref{fig:cuvelshpdm}).

\begin{table}
\caption{Superhump maxima of CU Vel (2002).}\label{tab:cuveloc2002}
\begin{center}
\begin{tabular}{ccccc}
\hline\hline
$E$ & max$^a$ & error & $O-C^b$ & $N^c$ \\
\hline
0 & 52620.2188 & 0.0003 & $-$0.0205 & 239 \\
22 & 52622.0161 & 0.0007 & $-$0.0012 & 20 \\
35 & 52623.0687 & 0.0002 & 0.0007 & 192 \\
36 & 52623.1495 & 0.0002 & 0.0006 & 408 \\
37 & 52623.2310 & 0.0003 & 0.0013 & 295 \\
49 & 52624.2007 & 0.0002 & 0.0011 & 146 \\
50 & 52624.2806 & 0.0004 & 0.0002 & 155 \\
51 & 52624.3687 & 0.0003 & 0.0075 & 124 \\
59 & 52625.0117 & 0.0003 & 0.0040 & 173 \\
60 & 52625.0935 & 0.0004 & 0.0049 & 220 \\
61 & 52625.1785 & 0.0003 & 0.0091 & 158 \\
72 & 52626.0623 & 0.0002 & 0.0038 & 304 \\
73 & 52626.1437 & 0.0002 & 0.0044 & 456 \\
74 & 52626.2257 & 0.0002 & 0.0056 & 506 \\
75 & 52626.3052 & 0.0003 & 0.0043 & 197 \\
97 & 52628.0777 & 0.0002 & $-$0.0013 & 111 \\
98 & 52628.1591 & 0.0002 & $-$0.0007 & 184 \\
109 & 52629.0445 & 0.0003 & $-$0.0044 & 281 \\
110 & 52629.1249 & 0.0002 & $-$0.0048 & 427 \\
111 & 52629.2062 & 0.0002 & $-$0.0043 & 301 \\
122 & 52630.0946 & 0.0006 & $-$0.0049 & 116 \\
123 & 52630.1752 & 0.0005 & $-$0.0051 & 113 \\
\hline
  \multicolumn{5}{l}{$^{a}$ BJD$-$2400000.} \\
  \multicolumn{5}{l}{$^{b}$ Against $max = 2452620.2393 + 0.080822 E$.} \\
  \multicolumn{5}{l}{$^{c}$ Number of points used to determine the maximum.} \\
\end{tabular}
\end{center}
\end{table}

\subsection{HS Virginis}\label{sec:hsvir}\label{obj:hsvir}

   We reanalyzed the data in \citet{kat98hsvir}.  Double-wave modulations
were observed on 1996 March 18 (BJD 2450161) during the fading stage from
the superoutburst plateau (the same feature was also recorded by
\cite{pat03suumas}).  These modulations were probably associated
with the manifestation of traditional late superhumps.  We listed
times of maxima of ordinary superhumps in table \ref{tab:hsviroc1996} and
secondary maxima in table \ref{tab:hsviroc1996-2}.  The agreement of
periods independently determined from these two sets strengthens the
identification of the latter as being traditional late superhumps.
Since \citet{kat98hsvir} did not take into account the present knowledge
in period variation and late superhumps, their period was contaminated
by these phenomena.  The mean $P_{\rm SH}$ for $23 \le E \le 99$ was
0.08006(3) d, giving a fractional period excess of 4.1 \%, slightly smaller
than the previous estimate.  The global $P_{\rm dot}$ was
$-18.3(3.8) \times 10^{-5}$, which is apparently affected by the stage A
evolution ($E \le 23$).

   The analysis of the 2008 superoutburst (table \ref{tab:hsviroc2008})
during its middle-to-late stage yielded a period of 0.08003(3) d,
in good agreement with the above analysis of the 1996 superoutburst.

\begin{table}
\caption{Superhump maxima of HS Vir (1996).}\label{tab:hsviroc1996}
\begin{center}
\begin{tabular}{ccccc}
\hline\hline
$E$ & max$^a$ & error & $O-C^b$ & $N^c$ \\
\hline
0 & 50153.3417 & 0.0006 & $-$0.0158 & 89 \\
12 & 50154.3201 & 0.0007 & 0.0002 & 88 \\
23 & 50155.2087 & 0.0018 & 0.0067 & 52 \\
35 & 50156.1718 & 0.0006 & 0.0073 & 124 \\
36 & 50156.2465 & 0.0008 & 0.0018 & 49 \\
37 & 50156.3324 & 0.0006 & 0.0074 & 73 \\
98 & 50161.2147 & 0.0020 & $-$0.0024 & 142 \\
99 & 50161.2922 & 0.0021 & $-$0.0052 & 125 \\
\hline
  \multicolumn{5}{l}{$^{a}$ BJD$-$2400000.} \\
  \multicolumn{5}{l}{$^{b}$ Against $max = 2450153.3575 + 0.080201 E$.} \\
  \multicolumn{5}{l}{$^{c}$ Number of points used to determine the maximum.} \\
\end{tabular}
\end{center}
\end{table}

\begin{table}
\caption{Secondary Maxima of HS Vir (1996).}\label{tab:hsviroc1996-2}
\begin{center}
\begin{tabular}{ccccc}
\hline\hline
$E$ & max$^a$ & error & $O-C^b$ & $N^c$ \\
\hline
0 & 50161.1697 & 0.0016 & $-$0.0003 & 86 \\
1 & 50161.2506 & 0.0014 & 0.0003 & 147 \\
26 & 50163.2594 & 0.0006 & $-$0.0000 & 95 \\
\hline
  \multicolumn{5}{l}{$^{a}$ BJD$-$2400000.} \\
  \multicolumn{5}{l}{$^{b}$ Against $max = 2450161.1700 + 0.080361 E$.} \\
  \multicolumn{5}{l}{$^{c}$ Number of points used to determine the maximum.} \\
\end{tabular}
\end{center}
\end{table}

\begin{table}
\caption{Superhump maxima of HS Vir (2008).}\label{tab:hsviroc2008}
\begin{center}
\begin{tabular}{ccccc}
\hline\hline
$E$ & max$^a$ & error & $O-C^b$ & $N^c$ \\
\hline
0 & 54619.0407 & 0.0006 & 0.0016 & 130 \\
11 & 54619.9184 & 0.0006 & $-$0.0010 & 106 \\
12 & 54619.9983 & 0.0005 & $-$0.0010 & 87 \\
62 & 54624.0011 & 0.0072 & 0.0004 & 66 \\
\hline
  \multicolumn{5}{l}{$^{a}$ BJD$-$2400000.} \\
  \multicolumn{5}{l}{$^{b}$ Against $max = 2454619.0390 + 0.080028 E$.} \\
  \multicolumn{5}{l}{$^{c}$ Number of points used to determine the maximum.} \\
\end{tabular}
\end{center}
\end{table}

\subsection{HV Virginis}\label{obj:hvvir}

   Analyses of superhumps of this WZ Sge-type dwarf nova have been
well documented (\cite{kat01hvvir}; \cite{ish03hvvir}).
We present our new observation of the 2008 superoutburst.
Only ordinary superhumps are treated here (table \ref{tab:hvviroc2008}).
The $O-C$ diagram resembles those of many systems with short superhump
periods, consisting of stages A--C (note, however, these stages
were preceded by a stage of early superhumps in this object).
The $P_{\rm dot}$ of the stage B was $+7.1(1.9) \times 10^{-5}$
($18 \le E \le 157$).  The value is in good agreement with those
obtained during previous superoutbursts:
$+7(1) \times 10^{-5}$ \citep{ish03hvvir} and
$+5.7(0.6) \times 10^{-5}$ \citep{kat01hvvir}.

\begin{figure}
  \begin{center}
    \FigureFile(88mm,70mm){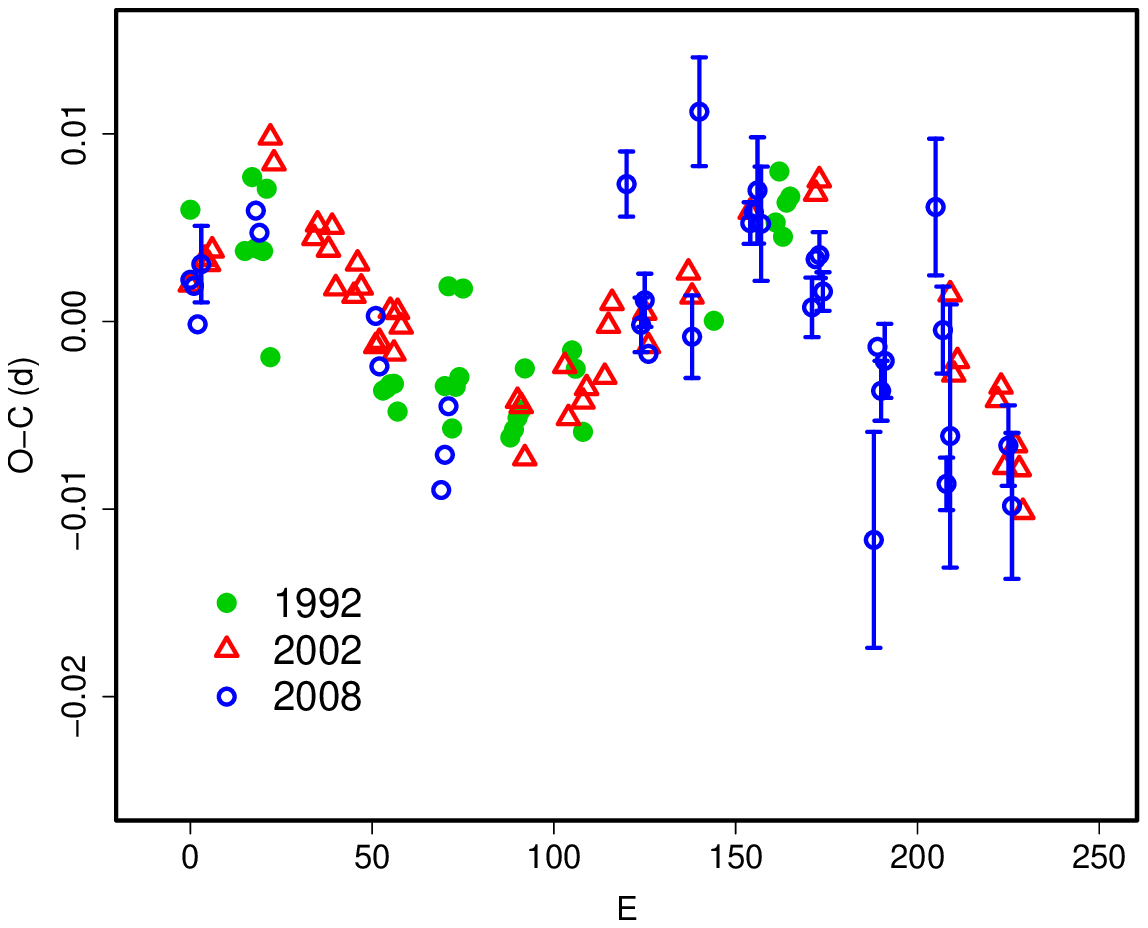}
  \end{center}
  \caption{Comparison of $O-C$ diagrams of HV Vir between different
  superoutbursts.  A period of 0.05828 d was used to draw this figure.
  Approximate cycle counts ($E$) after the appearance of the ordinary
  superhumps were used.
  }
  \label{fig:hvvircomp}
\end{figure}

   The $O-C$ diagrams after the appearance of ordinary superhumps were
similar between superoutburst (figure \ref{fig:hvvircomp}), although
the delay before the appearance of ordinary superhumps was shorter
in a fainter superoutburst in 2002 (subsection \ref{sec:wzsgedelay}).

\begin{table}
\caption{Superhump maxima of HV Vir (2008).}\label{tab:hvviroc2008}
\begin{center}
\begin{tabular}{ccccc}
\hline\hline
$E$ & max$^a$ & error & $O-C^b$ & $N^c$ \\
\hline
0 & 54517.1492 & 0.0005 & 0.0001 & 103 \\
1 & 54517.2071 & 0.0007 & $-$0.0003 & 146 \\
2 & 54517.2634 & 0.0009 & $-$0.0023 & 60 \\
3 & 54517.3248 & 0.0020 & 0.0009 & 61 \\
18 & 54518.2019 & 0.0003 & 0.0040 & 239 \\
19 & 54518.2590 & 0.0003 & 0.0029 & 219 \\
51 & 54520.1195 & 0.0008 & $-$0.0010 & 99 \\
52 & 54520.1751 & 0.0005 & $-$0.0037 & 103 \\
69 & 54521.1593 & 0.0007 & $-$0.0100 & 102 \\
70 & 54521.2194 & 0.0010 & $-$0.0081 & 92 \\
71 & 54521.2803 & 0.0009 & $-$0.0055 & 60 \\
120 & 54524.1479 & 0.0017 & 0.0072 & 103 \\
124 & 54524.3735 & 0.0015 & $-$0.0002 & 58 \\
125 & 54524.4331 & 0.0014 & 0.0011 & 65 \\
126 & 54524.4885 & 0.0010 & $-$0.0018 & 50 \\
138 & 54525.1888 & 0.0022 & $-$0.0006 & 389 \\
140 & 54525.3173 & 0.0029 & 0.0114 & 302 \\
154 & 54526.1273 & 0.0011 & 0.0057 & 96 \\
155 & 54526.1862 & 0.0008 & 0.0063 & 130 \\
156 & 54526.2456 & 0.0028 & 0.0075 & 55 \\
157 & 54526.3021 & 0.0030 & 0.0057 & 55 \\
171 & 54527.1136 & 0.0016 & 0.0015 & 36 \\
172 & 54527.1744 & 0.0005 & 0.0041 & 178 \\
173 & 54527.2329 & 0.0012 & 0.0043 & 135 \\
174 & 54527.2893 & 0.0010 & 0.0024 & 70 \\
188 & 54528.0919 & 0.0058 & $-$0.0106 & 32 \\
189 & 54528.1605 & 0.0009 & $-$0.0003 & 59 \\
190 & 54528.2164 & 0.0016 & $-$0.0026 & 60 \\
191 & 54528.2763 & 0.0020 & $-$0.0010 & 44 \\
205 & 54529.1004 & 0.0036 & 0.0074 & 37 \\
207 & 54529.2104 & 0.0023 & 0.0009 & 58 \\
208 & 54529.2605 & 0.0014 & $-$0.0073 & 60 \\
209 & 54529.3214 & 0.0070 & $-$0.0047 & 58 \\
225 & 54530.2533 & 0.0021 & $-$0.0050 & 38 \\
226 & 54530.3084 & 0.0039 & $-$0.0082 & 61 \\
\hline
  \multicolumn{5}{l}{$^{a}$ BJD$-$2400000.} \\
  \multicolumn{5}{l}{$^{b}$ Against $max = 2454517.1491 + 0.058263 E$.} \\
  \multicolumn{5}{l}{$^{c}$ Number of points used to determine the maximum.} \\
\end{tabular}
\end{center}
\end{table}

\subsection{OU Virginis}\label{obj:ouvir}

   OU Vir was a CV discovered through a survey for quasars \citep{ber92LBQS}.
\citet{van00ouvir} established that the object is an eclipsing SU UMa-type
dwarf nova, but their superhump period was rather poorly determined.
\citet{pat05SH} presented an analysis of the 2003 superoutburst
and reported a superhump period of 0.0751(1) d.  They did not give times
of superhump maxima.

\begin{figure}
  \begin{center}
    \FigureFile(88mm,110mm){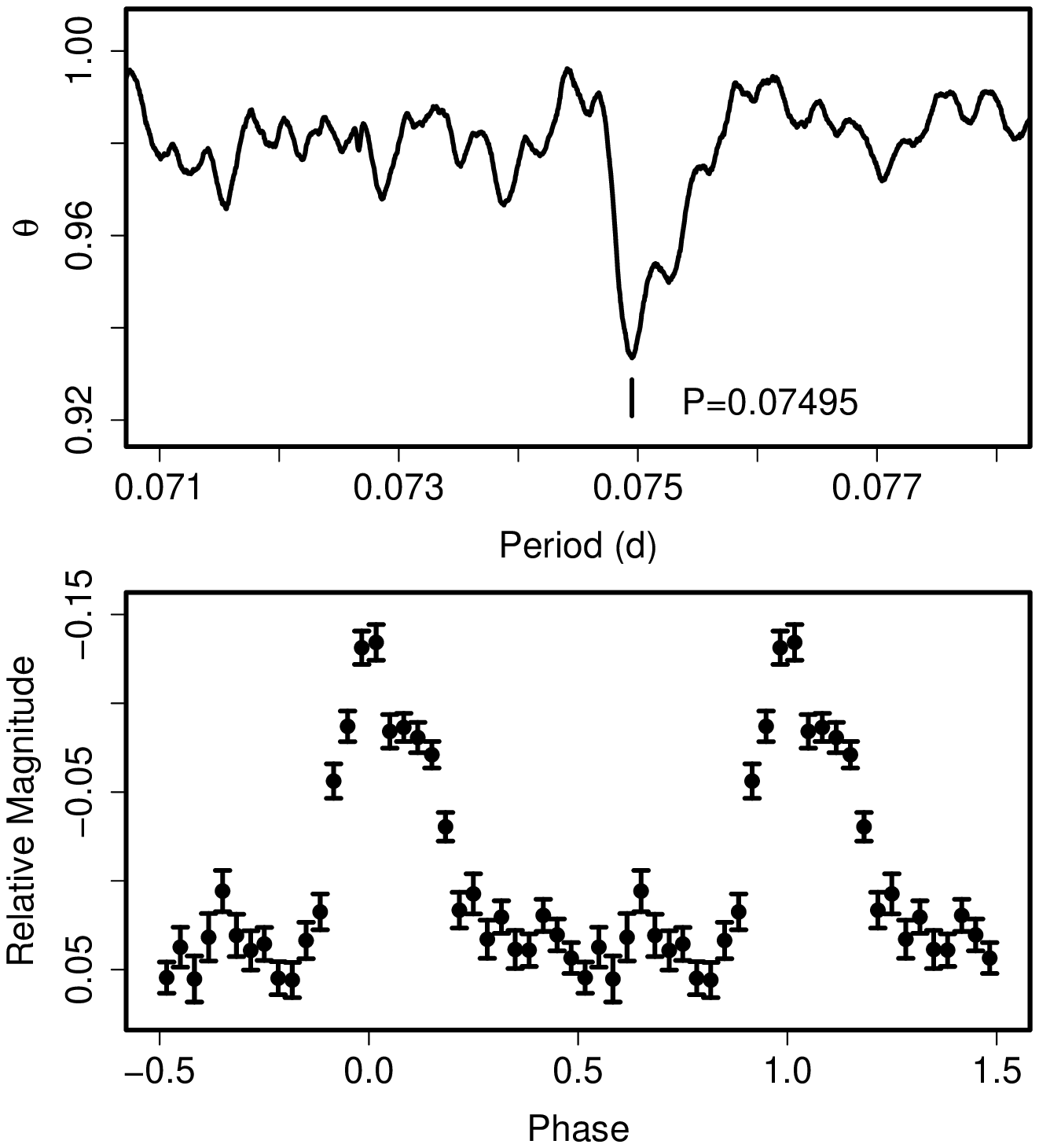}
  \end{center}
  \caption{Superhumps in OU Vir (2003). (Upper): PDM analysis.
     (Lower): Phase-averaged profile.}
  \label{fig:ouvirshpdm}
\end{figure}

   We present the analysis of the 2003 superoutburst, the data partly
overlapping those in \citet{pat05SH}.  Observations outside the eclipses,
as described in V2051 Oph, were used in analysis.
The mean superhump period with the PDM method was 0.074950(7) d
(figure \ref{fig:ouvirshpdm}).  The times of superhump maxima are
listed in table \ref{tab:ouviroc2003}.
The $O-C$'s showed a slight signature
of a discontinuous change around $E = 50$, but its nature remained
uncertain because of the relatively large scatter in the data.
Although we determined a global $P_{\rm dot}$ of $-1.8(0.6) \times 10^{-5}$,
this value apparently needs to be verified by a detailed future study
since eclipsing SU UMa-type dwarf novae are often associated
with more or less complexity in analysis.  The early stage of the
2008 superoutburst was also observed (table \ref{tab:ouviroc2008}).

\begin{table}
\caption{Superhump maxima of OU Vir (2003).}\label{tab:ouviroc2003}
\begin{center}

\end{center}
\end{table}

\begin{table}
\caption{Superhump maxima of OU Vir (2008).}\label{tab:ouviroc2008}
\begin{center}
%
\end{center}
\end{table}

\subsection{QZ Virginis}\label{obj:qzvir}

   We reanalyzed the data in \citet{kat97tleo}.  The refined times of
superhump maxima, together with those in \citet{lem93tleo}, are listed
in table \ref{tab:qzviroc1993}.  The earliest part ($E \le 1$) showed
large deviations from the nominal superhump period, as discussed in
\citet{kat97tleo}.  A strongly negative $O-C$ at $E = 9$ may be
interpreted as early development with a longer period (stage A).
Thanks to the improvement in determination of times of maxima,
it has now become evident that the segment $15 \le E \le 101$ showed
a positive $P_{\rm dot}$ (stage B).
Disregarding the discrepant points $E = 34$ and $E = 50$,
we obtained $P_{\rm dot}$ = $+7.0(1.4) \times 10^{-5}$.

\begin{figure}
  \begin{center}
    \FigureFile(88mm,70mm){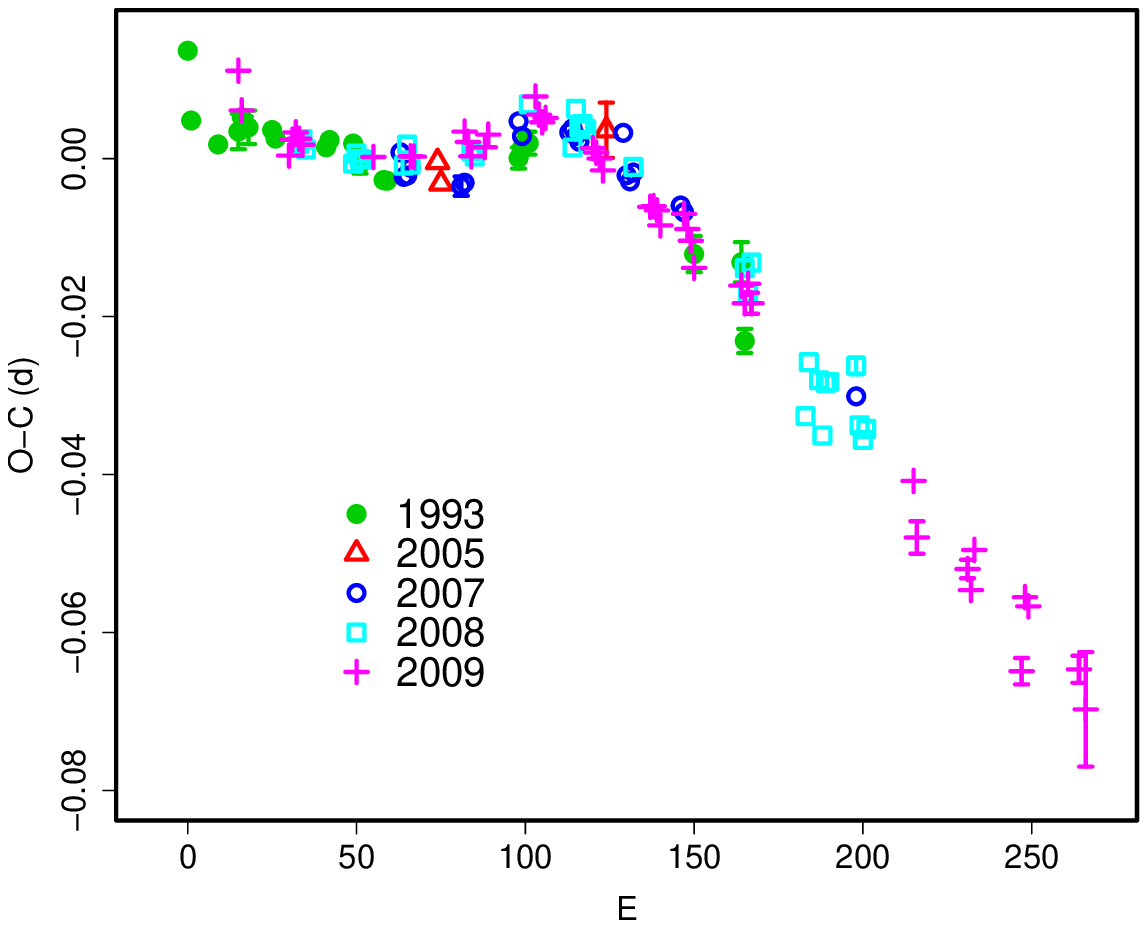}
  \end{center}
  \caption{Comparison of $O-C$ diagrams of QZ Vir between different
  superoutbursts.  A period of 0.06038 d was used to draw this figure.
  Approximate cycle counts ($E$) after the start of the
  superoutburst were used.  The start of the 2008 superoutburst was
  missed.  The $O-C$ analysis suggests that the superoutburst started
  two days before the initial detection.  The $O-C$ diagram was shifted
  by this value.
  }
  \label{fig:qzvircomp}
\end{figure}

   This period derivative and the overall behavior is similar to
those during the 2007 and 2008 superoutbursts (tables
\ref{tab:qzviroc2007}, \ref{tab:qzviroc2008};
figure \ref{fig:qzvircomp}).
The $P_{\rm dot}$'s for the corresponding segment
were $+4.5(7.6) \times 10^{-5}$ ($E \le 53$, 2007)
and $+4.7(1.9) \times 10^{-5}$ ($E \le 85$, 2008).
A fragmentary observation of the 2005 superoutburst is also given
(table \ref{tab:qzviroc2005}).
The negative $P_{\rm dot}$ in \citet{lem93tleo}
probably resulted from a stage A--B transition and sparse sampling.

   The 2009 superoutburst was particularly well observed
(table \ref{tab:qzviroc2009}).  This superoutburst was preceded by
a distinct precursor and followed by a rebrightening.
Despite the presence of a precursor, the $P_{\rm dot}$ during the
stage B ($E \le 91$) was positive with $+11.4(1.8) \times 10^{-5}$.
The stage C superhumps had a period of 0.06000(1) before $E=152$,
then the period slightly shortened to 0.05992(7) d.
These late-stage superhumps apparently endured during the period
of the rebrightening.

   Further detailed analysis will be presented in \citet{ohs09qzvir}.

\begin{table}
\caption{Superhump maxima of QZ Vir (1993).}\label{tab:qzviroc1993}
\begin{center}

\end{center}
\end{table}

\begin{table}
\caption{Superhump maxima of QZ Vir (2005).}\label{tab:qzviroc2005}
\begin{center}
%
\end{center}
\end{table}

\begin{table}
\caption{Superhump maxima of QZ Vir (2007).}\label{tab:qzviroc2007}
\begin{center}
%
\end{center}
\end{table}

\begin{table}
\caption{Superhump maxima of QZ Vir (2008).}\label{tab:qzviroc2008}
\begin{center}
%
\end{center}
\end{table}

\subsection{RX Volantis}\label{obj:rxvol}

\begin{figure}
  \begin{center}
    \FigureFile(88mm,110mm){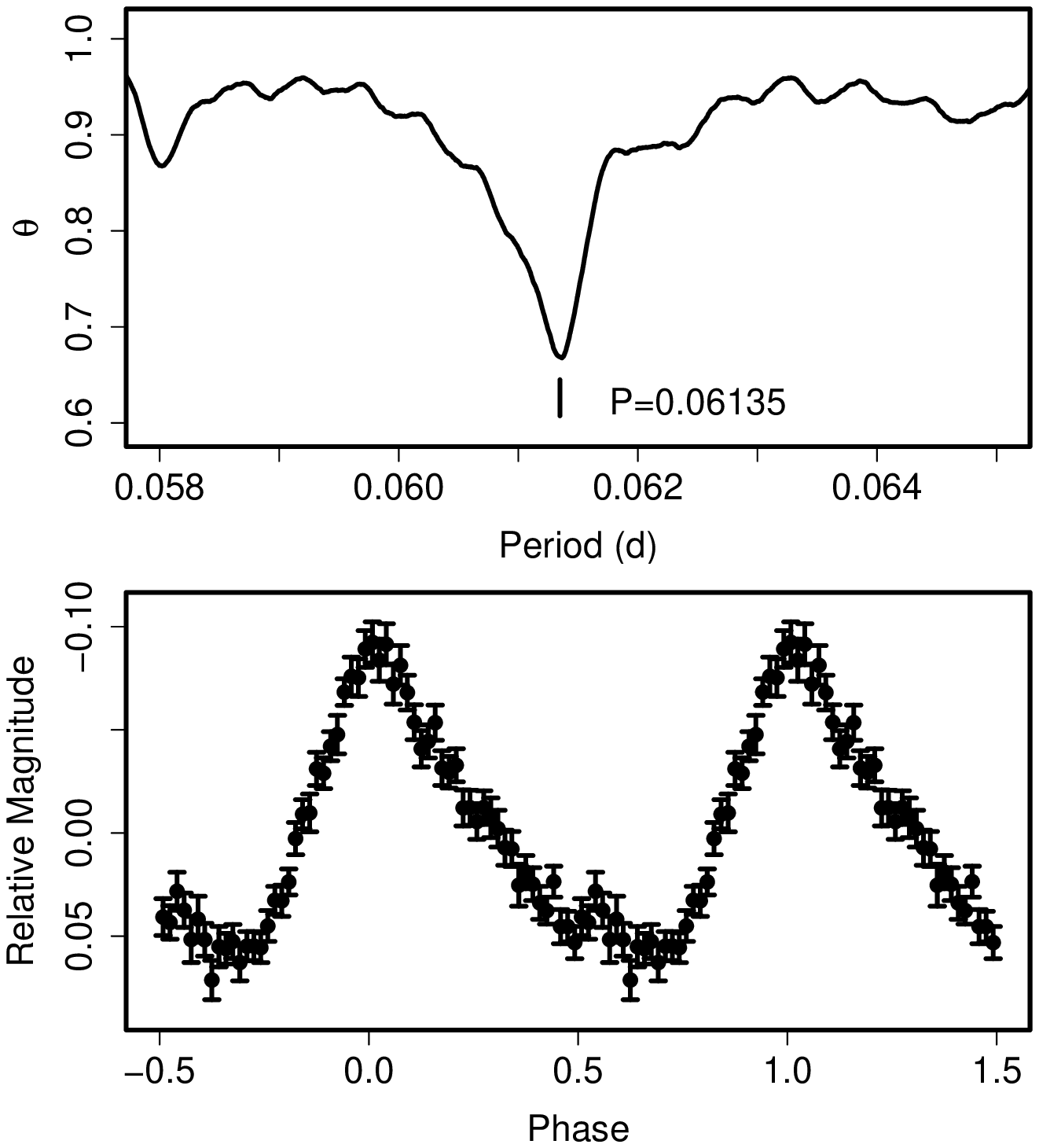}
  \end{center}
  \caption{Superhumps in RX Vol (2003). (Upper): PDM analysis.
     (Lower): Phase-averaged profile.}
  \label{fig:rxvolshpdm}
\end{figure}

   Although RX Vol was listed as a possible SU UMa-type dwarf nova
with a maximum of magnitude 16 \citep{GCVS}, little had been known
until 2003.
The first-ever outburst since the discovery, at an exceptional
brightness of 14.7, was reported on 2003 May 4
(R. Stubbings, vsnet-outburst 5482).  This outburst turned out
to be a superoutburst (vsnet-outburst 5502).
The mean superhump period with the PDM method was 0.061348(7) d
(figure \ref{fig:rxvolshpdm}).
The times of superhump maxima are listed in table \ref{tab:rxvoloc2003}.
The object clearly showed positive superhump derivative except for
the earliest part.  $P_{\rm dot}$ was $+5.8(0.8) \times 10^{-5}$ for
($E \ge 12$).  \citet{sch05rxvol} summarized the history of this
object and presented a spectrum in quiescence.

\begin{table}
\caption{Superhump maxima of RX Vol (2003).}\label{tab:rxvoloc2003}
\begin{center}
\begin{tabular}{ccccc}
\hline\hline
$E$ & max$^a$ & error & $O-C^b$ & $N^c$ \\
\hline
0 & 52764.0491 & 0.0009 & 0.0029 & 110 \\
1 & 52764.1074 & 0.0008 & $-$0.0001 & 100 \\
12 & 52764.7858 & 0.0005 & 0.0032 & 159 \\
13 & 52764.8466 & 0.0004 & 0.0026 & 270 \\
14 & 52764.9077 & 0.0004 & 0.0023 & 204 \\
15 & 52764.9688 & 0.0004 & 0.0021 & 204 \\
16 & 52765.0301 & 0.0004 & 0.0020 & 205 \\
19 & 52765.2143 & 0.0006 & 0.0022 & 62 \\
20 & 52765.2770 & 0.0007 & 0.0035 & 66 \\
21 & 52765.3368 & 0.0008 & 0.0019 & 70 \\
22 & 52765.3954 & 0.0009 & $-$0.0008 & 70 \\
31 & 52765.9479 & 0.0009 & $-$0.0006 & 161 \\
36 & 52766.2529 & 0.0006 & $-$0.0024 & 57 \\
37 & 52766.3164 & 0.0006 & $-$0.0003 & 61 \\
38 & 52766.3728 & 0.0012 & $-$0.0053 & 68 \\
52 & 52767.2333 & 0.0010 & $-$0.0039 & 72 \\
53 & 52767.2934 & 0.0012 & $-$0.0051 & 71 \\
54 & 52767.3553 & 0.0011 & $-$0.0046 & 58 \\
68 & 52768.2145 & 0.0010 & $-$0.0045 & 72 \\
69 & 52768.2757 & 0.0017 & $-$0.0046 & 71 \\
70 & 52768.3399 & 0.0012 & $-$0.0018 & 71 \\
85 & 52769.2619 & 0.0019 & $-$0.0003 & 71 \\
86 & 52769.3237 & 0.0016 & 0.0002 & 69 \\
101 & 52770.2446 & 0.0027 & 0.0006 & 70 \\
133 & 52772.2127 & 0.0013 & 0.0050 & 71 \\
134 & 52772.2747 & 0.0027 & 0.0057 & 71 \\
\hline
  \multicolumn{5}{l}{$^{a}$ BJD$-$2400000.} \\
  \multicolumn{5}{l}{$^{b}$ Against $max = 2452764.0462 + 0.061364 E$.} \\
  \multicolumn{5}{l}{$^{c}$ Number of points used to determine the maximum.} \\
\end{tabular}
\end{center}
\end{table}

\subsection{TY Vulpeculae}\label{obj:tyvul}

   \citet{kat99tyvul} suggested the SU UMa-type nature of this object
based on the observation of the 1999 September short outburst.
The SU UMa-type nature of TY Vul was established by Vanmunster et al.
(aavso-photometry message on 2003 December 7)\footnote{
  $<$http://www.aavso.org/pipermail/aavso-photometry/2003-December/000153.html$>$.
}, who reported a period of 0.0809(2) d.  We observed the same superoutburst
and obtained the times of superhump maxima after incorporating the
AAVSO data (table \ref{tab:tyvuloc2003}).
The period of 0.08048(7) d can satisfactorily expressed the maxima,
and the period was in agreement with the one by Vanmunster et al.
The resultant $O-C$ diagram showed a large negative period derivative
$P_{\rm dot}$ = $-14.8(3.0) \times 10^{-5}$ for the entire span of
observations.  This large variation can be attributed to a stage B--C
transition.  The parameters based on this interpretation are given
in table \ref{tab:perlist}.
The object may be similar to AX Cap and SDSS J1627 in the evolution
of the superhump period (see subsection \ref{sec:longp}).

\begin{table}
\caption{Superhump maxima of TY Vul (2003).}\label{tab:tyvuloc2003}
\begin{center}
\begin{tabular}{ccccc}
\hline\hline
$E$ & max$^a$ & error & $O-C^b$ & $N^c$ \\
\hline
0 & 52976.8787 & 0.0017 & $-$0.0213 & 115 \\
1 & 52976.9653 & 0.0036 & $-$0.0153 & 139 \\
7 & 52977.4546 & 0.0012 & $-$0.0088 & 45 \\
8 & 52977.5447 & 0.0008 & 0.0008 & 67 \\
9 & 52977.6266 & 0.0019 & 0.0022 & 37 \\
12 & 52977.8734 & 0.0009 & 0.0075 & 110 \\
13 & 52977.9471 & 0.0015 & 0.0007 & 148 \\
14 & 52978.0234 & 0.0037 & $-$0.0035 & 45 \\
42 & 52980.2944 & 0.0023 & 0.0140 & 112 \\
50 & 52980.9377 & 0.0024 & 0.0134 & 119 \\
51 & 52981.0194 & 0.0032 & 0.0147 & 72 \\
54 & 52981.2534 & 0.0011 & 0.0072 & 104 \\
55 & 52981.3292 & 0.0018 & 0.0025 & 97 \\
63 & 52981.9853 & 0.0044 & 0.0147 & 115 \\
67 & 52982.2998 & 0.0010 & 0.0073 & 89 \\
68 & 52982.3786 & 0.0019 & 0.0057 & 49 \\
70 & 52982.5368 & 0.0061 & 0.0029 & 61 \\
79 & 52983.2488 & 0.0006 & $-$0.0095 & 43 \\
80 & 52983.3292 & 0.0022 & $-$0.0096 & 43 \\
116 & 52986.2255 & 0.0014 & $-$0.0107 & 37 \\
120 & 52986.5434 & 0.0051 & $-$0.0148 & 20 \\
\hline
  \multicolumn{5}{l}{$^{a}$ BJD$-$2400000.} \\
  \multicolumn{5}{l}{$^{b}$ Against $max = 2452976.9001 + 0.080484 E$.} \\
  \multicolumn{5}{l}{$^{c}$ Number of points used to determine the maximum.} \\
\end{tabular}
\end{center}
\end{table}

\subsection{DO Vulpeculae}\label{obj:dovul}

   Although DO Vul had long been known as a dwarf nova \citep{baa28VS},
the identification was only recently known (\cite{ski97dovul};
\cite{hen01dovul}).

   Vanmunster reported the detection of superhumps with a period of
0.065 d during the 2005 outburst.

\begin{figure}
  \begin{center}
    \FigureFile(88mm,110mm){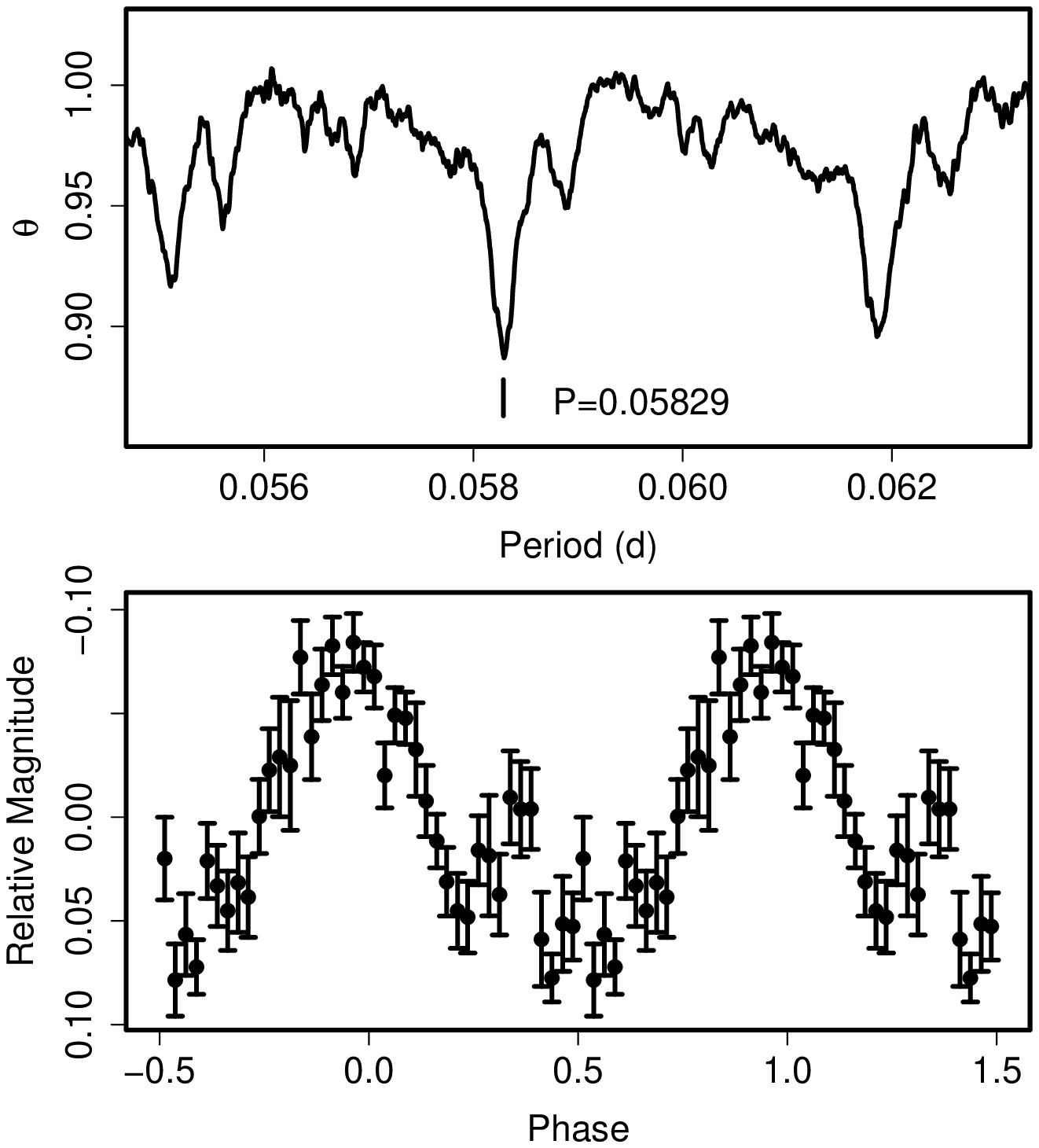}
  \end{center}
  \caption{Superhumps in DO Vul (2008). (Upper): PDM analysis.
     (Lower): Phase-averaged profile.}
  \label{fig:dovulshpdm}
\end{figure}

   Observations of the 2008 superoutburst yielded a mean period
of 0.058286(14) d (PDM analysis, figure \ref{fig:dovulshpdm})
and a $P_{\rm dot}$ of $+9.9(2.1) \times 10^{-5}$
(table \ref{tab:dovuloc2008}).

\begin{table}
\caption{Superhump maxima of DO Vul (2008).}\label{tab:dovuloc2008}
\begin{center}
\begin{tabular}{ccccc}
\hline\hline
$E$ & max$^a$ & error & $O-C^b$ & $N^c$ \\
\hline
0 & 54671.1058 & 0.0070 & 0.0170 & 83 \\
19 & 54672.1978 & 0.0060 & 0.0032 & 58 \\
33 & 54673.0010 & 0.0030 & $-$0.0085 & 76 \\
34 & 54673.0668 & 0.0008 & $-$0.0009 & 107 \\
50 & 54673.9948 & 0.0015 & $-$0.0041 & 67 \\
51 & 54674.0489 & 0.0003 & $-$0.0082 & 125 \\
52 & 54674.1111 & 0.0004 & $-$0.0043 & 96 \\
71 & 54675.2216 & 0.0027 & 0.0004 & 82 \\
120 & 54678.0700 & 0.0025 & $-$0.0032 & 29 \\
121 & 54678.1308 & 0.0045 & $-$0.0006 & 22 \\
137 & 54679.0580 & 0.0030 & $-$0.0046 & 120 \\
138 & 54679.1190 & 0.0015 & $-$0.0019 & 124 \\
155 & 54680.1195 & 0.0075 & 0.0092 & 84 \\
156 & 54680.1748 & 0.0018 & 0.0063 & 119 \\
\hline
  \multicolumn{5}{l}{$^{a}$ BJD$-$2400000.} \\
  \multicolumn{5}{l}{$^{b}$ Against $max = 2454671.0887 + 0.058204 E$.} \\
  \multicolumn{5}{l}{$^{c}$ Number of points used to determine the maximum.} \\
\end{tabular}
\end{center}
\end{table}

\subsection{NSV 4838}\label{obj:nsv4838}

   The times of superhump maxima during the 2005 and 2007 superoutbursts
are listed in tables \ref{tab:nsv4838oc2005} and \ref{tab:nsv4838oc2007}
The observations used here partly include the data in \citet{ima09nsv4838}.
The $P_{\rm dot}$ for the 2007 superoutburst ($0 \le E \le 101$, stage B)
was $+7.4(1.9) \times 10^{-5}$.  The 2005 superoutburst was apparently
observed during the stage C.  The period of 0.06960(3) d obtained by
the PDM analysis confirmed the $O-C$ analysis.  The object underwent
another superoutburst in 2009 February.

\begin{table}
\caption{Superhump maxima of NSV 4838 (2005).}\label{tab:nsv4838oc2005}
\begin{center}

\end{center}
\end{table}

\begin{table}
\caption{Superhump maxima of NSV 4838 (2007).}\label{tab:nsv4838oc2007}
\begin{center}
%
\end{center}
\end{table}

\subsection{NSV 5285}\label{obj:nsv5285}

   NSV 5285 was originally discovered as a blue eruptive object
that underwent an outburst at $B = 14.5$ \citep{kow75nsv5285}.
The object remained bright at least for five days and faded to
$B \sim 20$ thereafter.  \citet{kow75nsv5285} suggested that this
object is probably a quasar which underwent a 5.5-mag outburst.
\citet{dus08nsv5285cbet1574} detected an outbursting variable star,
which turned out to be identical with NSV 5285.  Subsequent photometric
observations established that this is an superoutburst of an SU UMa-type
dwarf nova (vsnet-alert 10726).  The times of superhump maxima are
given in table \ref{tab:nsv5285oc2008}.  The mean $P_{\rm SH}$
using the PDM method was 0.08082(3) d.

\begin{figure}
  \begin{center}
    \FigureFile(88mm,110mm){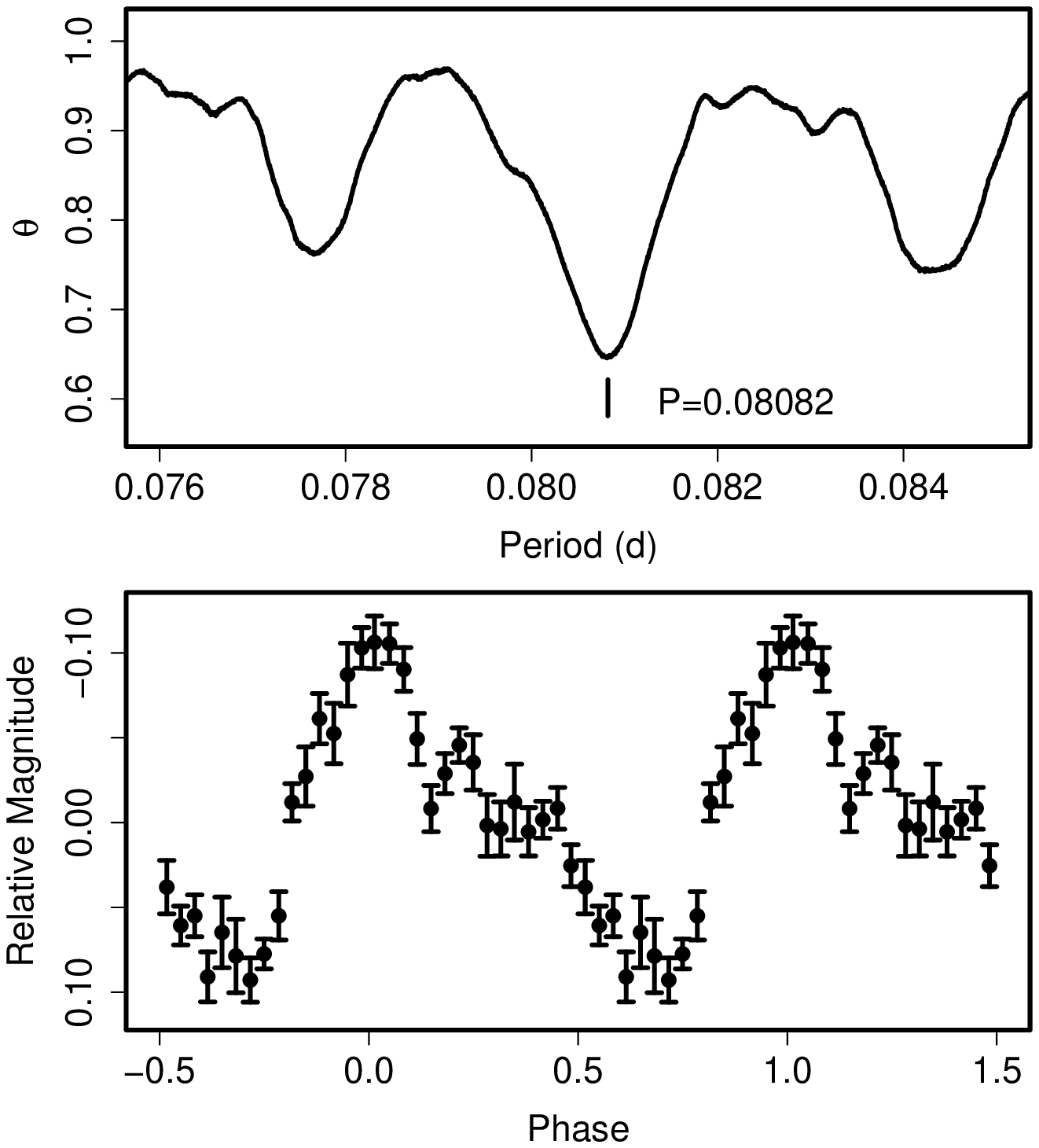}
  \end{center}
  \caption{Superhumps in NSV 5285 (2008). (Upper): PDM analysis.
     (Lower): Phase-averaged profile.}
  \label{fig:nsv5285shpdm}
\end{figure}

\begin{table}
\caption{Superhump maxima of NSV 5285 (2008).}\label{tab:nsv5285oc2008}
\begin{center}
\begin{tabular}{ccccc}
\hline\hline
$E$ & max$^a$ & error & $O-C^b$ & $N^c$ \\
\hline
0 & 54791.3043 & 0.0013 & 0.0019 & 52 \\
11 & 54792.2686 & 0.0008 & $-$0.0015 & 94 \\
12 & 54792.3566 & 0.0009 & $-$0.0014 & 97 \\
34 & 54794.2945 & 0.0008 & 0.0010 & 179 \\
\hline
  \multicolumn{5}{l}{$^{a}$ BJD$-$2400000.} \\
  \multicolumn{5}{l}{$^{b}$ Against $max = 2454791.3024 + 0.087973 E$.} \\
  \multicolumn{5}{l}{$^{c}$ Number of points used to determine the maximum.} \\
\end{tabular}
\end{center}
\end{table}

\subsection{NSV 14652}\label{obj:nsv14652}

   NSV 14652 was discovered by \citet{rei30VS} as a variable star
(AN 254.1930).  The object was positively recorded twice in 1901 and 1904.
The object was identified with a ROSAT source (vsnet-chat 3314).
T. Kinnunen detected the object in outburst on a Palomar Observatory
Sky Survey scan (cf. vsnet-alert 5203, 5205).

\begin{figure}
  \begin{center}
    \FigureFile(88mm,110mm){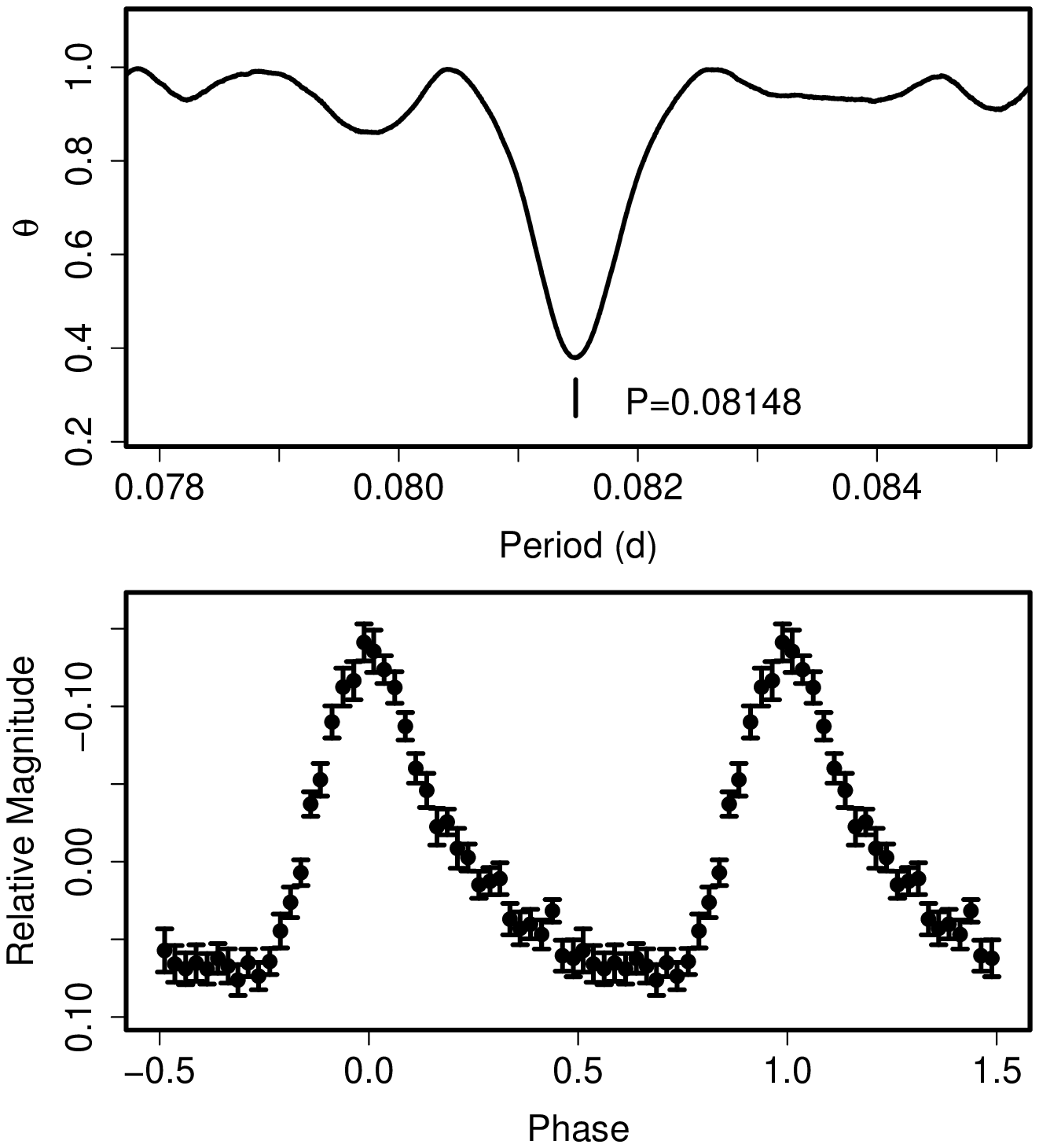}
  \end{center}
  \caption{Superhumps in NSV 14652 (2004). (Upper): PDM analysis.
     (Lower): Phase-averaged profile.}
  \label{fig:nsv14652shpdm}
\end{figure}

   We present times of superhump maxima during a superoutburst in 2004
September (table \ref{tab:nsv14652oc2004}).  The mean $P_{\rm SH}$
with the PDM method was 0.08148(1) d (figure \ref{fig:nsv14652shpdm}).
The $O-C$ diagram showed a stage B--C transition around $E=50$.
The $P_{\rm dot}$ during the stage B was close to zero,
$-3.0(3.6) \times 10^{-5}$.  The other parameters are listed in
table \ref{tab:perlist}.

\begin{table}
\caption{Superhump maxima of NSV 14652 (2004).}\label{tab:nsv14652oc2004}
\begin{center}
\begin{tabular}{ccccc}
\hline\hline
$E$ & max$^a$ & error & $O-C^b$ & $N^c$ \\
\hline
0 & 53251.4459 & 0.0006 & $-$0.0016 & 79 \\
1 & 53251.5282 & 0.0007 & $-$0.0008 & 82 \\
2 & 53251.6107 & 0.0010 & 0.0003 & 64 \\
12 & 53252.4245 & 0.0008 & $-$0.0005 & 69 \\
13 & 53252.5057 & 0.0016 & $-$0.0007 & 59 \\
36 & 53254.3814 & 0.0007 & 0.0015 & 68 \\
37 & 53254.4646 & 0.0008 & 0.0032 & 56 \\
38 & 53254.5438 & 0.0008 & 0.0010 & 52 \\
48 & 53255.3602 & 0.0011 & 0.0028 & 69 \\
49 & 53255.4397 & 0.0007 & 0.0008 & 67 \\
50 & 53255.5212 & 0.0008 & 0.0009 & 63 \\
60 & 53256.3308 & 0.0007 & $-$0.0041 & 53 \\
61 & 53256.4137 & 0.0015 & $-$0.0027 & 35 \\
\hline
  \multicolumn{5}{l}{$^{a}$ BJD$-$2400000.} \\
  \multicolumn{5}{l}{$^{b}$ Against $max = 2453251.4475 + 0.081457 E$.} \\
  \multicolumn{5}{l}{$^{c}$ Number of points used to determine the maximum.} \\
\end{tabular}
\end{center}
\end{table}

\subsection{1RXS J023238.8-371812}\label{sec:j0232}\label{obj:j0232}

\begin{figure}
  \begin{center}
    \FigureFile(88mm,110mm){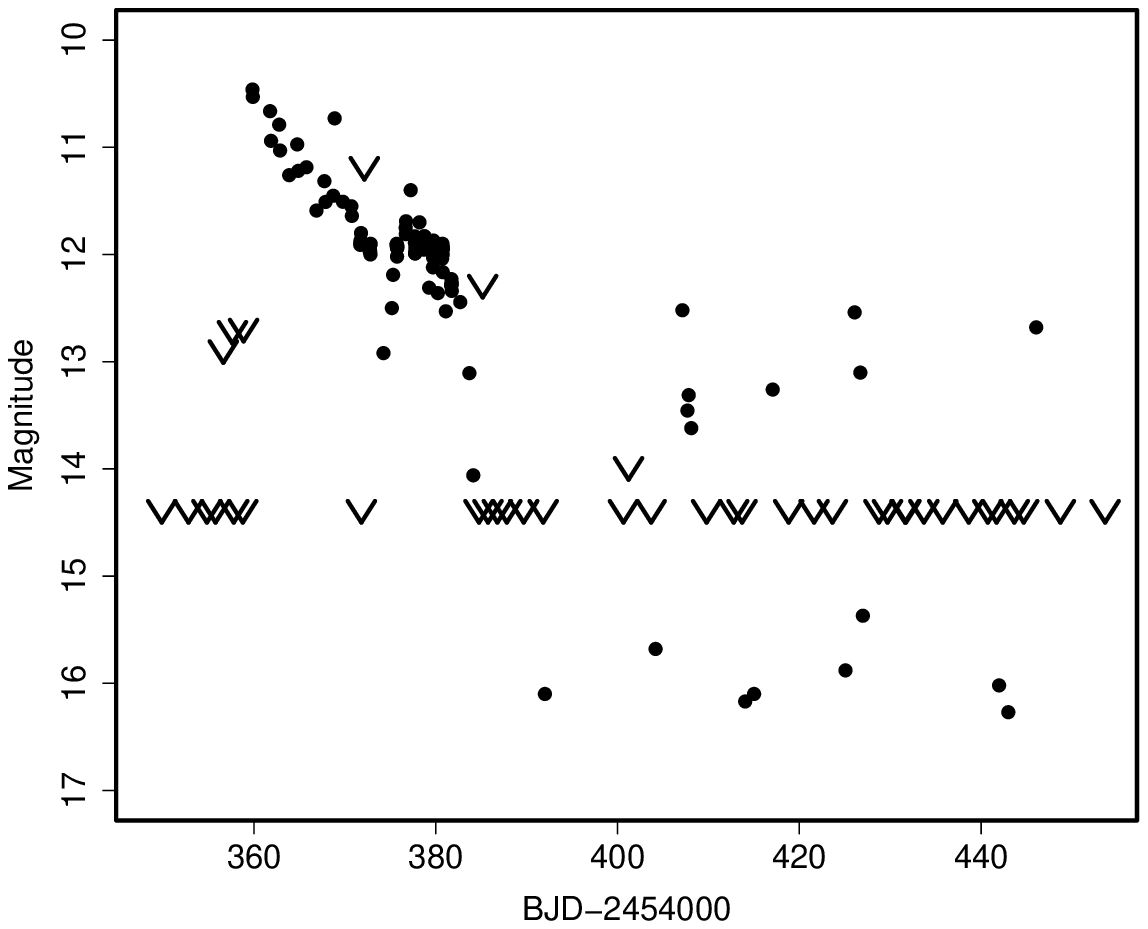}
  \end{center}
  \caption{Superoutburst of 1RXS J0232 in 2007.  The data are a combination
  of our observations, VSNET, ASAS-3 and ``Pi of the Sky'' observations.
  The ``V''-marks indicate upper limits.  There was a ``dip'' during the
  superoutburst plateau (around BJD 2454374).
  Four post-superoutburst rebrightenings were recorded.}
  \label{fig:j0232lc}
\end{figure}

   In 2007 October, K. Malek (``Pi of the Sky''\footnote{
   $<$http://grb.fuw.edu.pl/pi/index.html$>$.
}) reported a possible nova outburst (vsnet-alert 9622) close to
the location of 1RXS J023238.8$-$371812 (hereafter 1RXS J0232).
T. Kato suggested that the object
can be identified with a 6dF Galactic object and that it is most likely
a large-amplitude dwarf nova (vsnet-alert 9620).  This suggestion
was later confirmed by the detection of superhumps
(vsnet-alert 9634).  The superoutburst was unusual in that it had both
a ``dip'', characteristic to type-A superoutbursts
(subsection \ref{sec:wzsgeouttype}), during the superoutburst plateau and
four distinct post-superoutburst rebrightenings, characteristic to type-B
superoutbursts (figure \ref{fig:j0232lc}).

   The times of superhump maxima during the main superoutburst are
listed in table \ref{tab:j0232oc2007}.
The resultant $P_{\rm dot}$ was $-1.7(0.7) \times 10^{-5}$.
Since the only later portion of the superoutburst was observed,
this value may have been affected by a possible stage B--C transition.
The only small variation of the $P_{\rm SH}$, however,
may be associated with the extreme WZ Sge-type nature of this object.

\begin{figure}
  \begin{center}
    \FigureFile(88mm,110mm){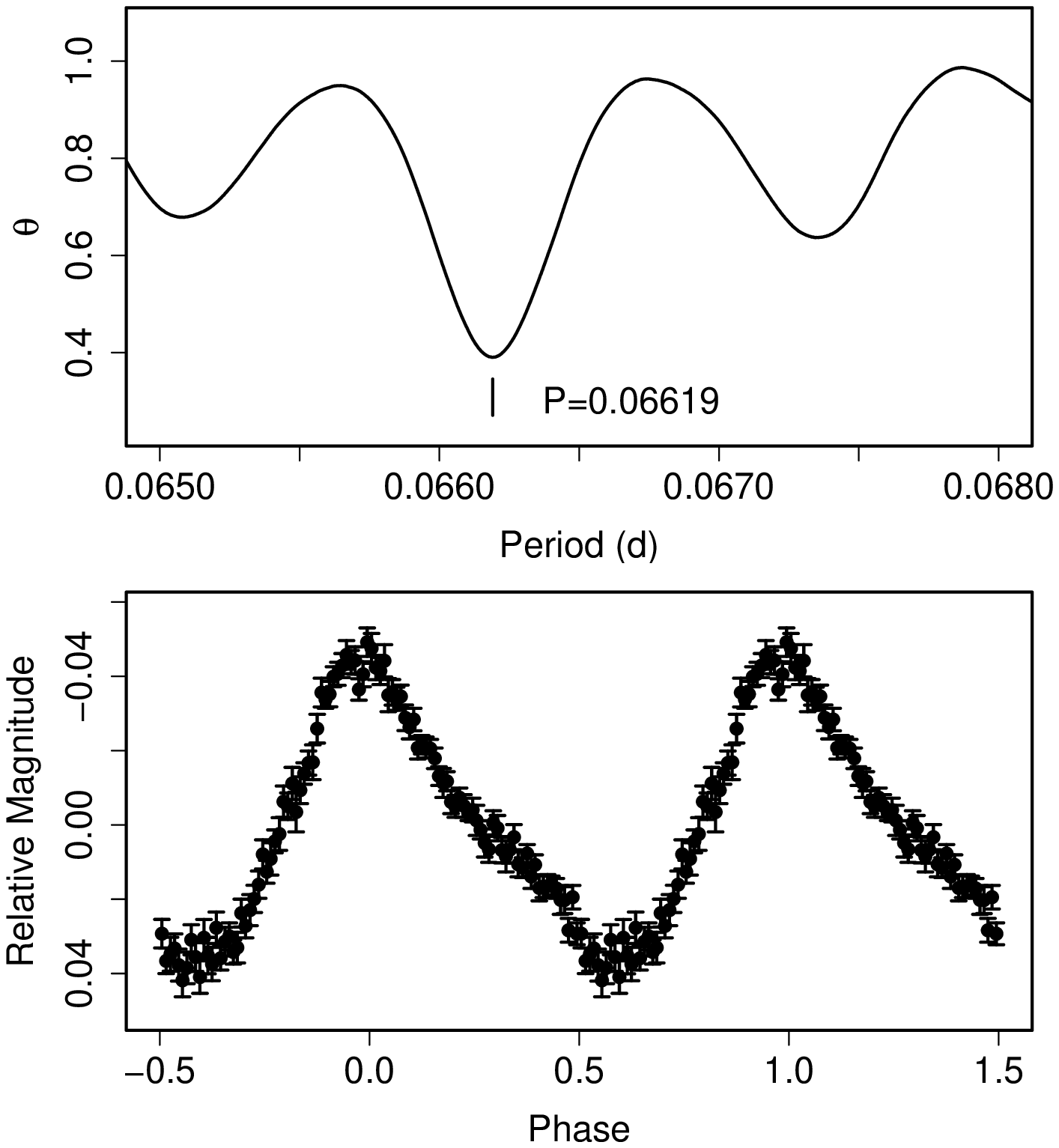}
  \end{center}
  \caption{Superhumps in 1RXS J0232 during the main superoutburst.
     (Upper): PDM analysis.
     (Lower): Phase-averaged profile.}
  \label{fig:j0232mainpdm}
\end{figure}

\begin{figure}
  \begin{center}
    \FigureFile(88mm,110mm){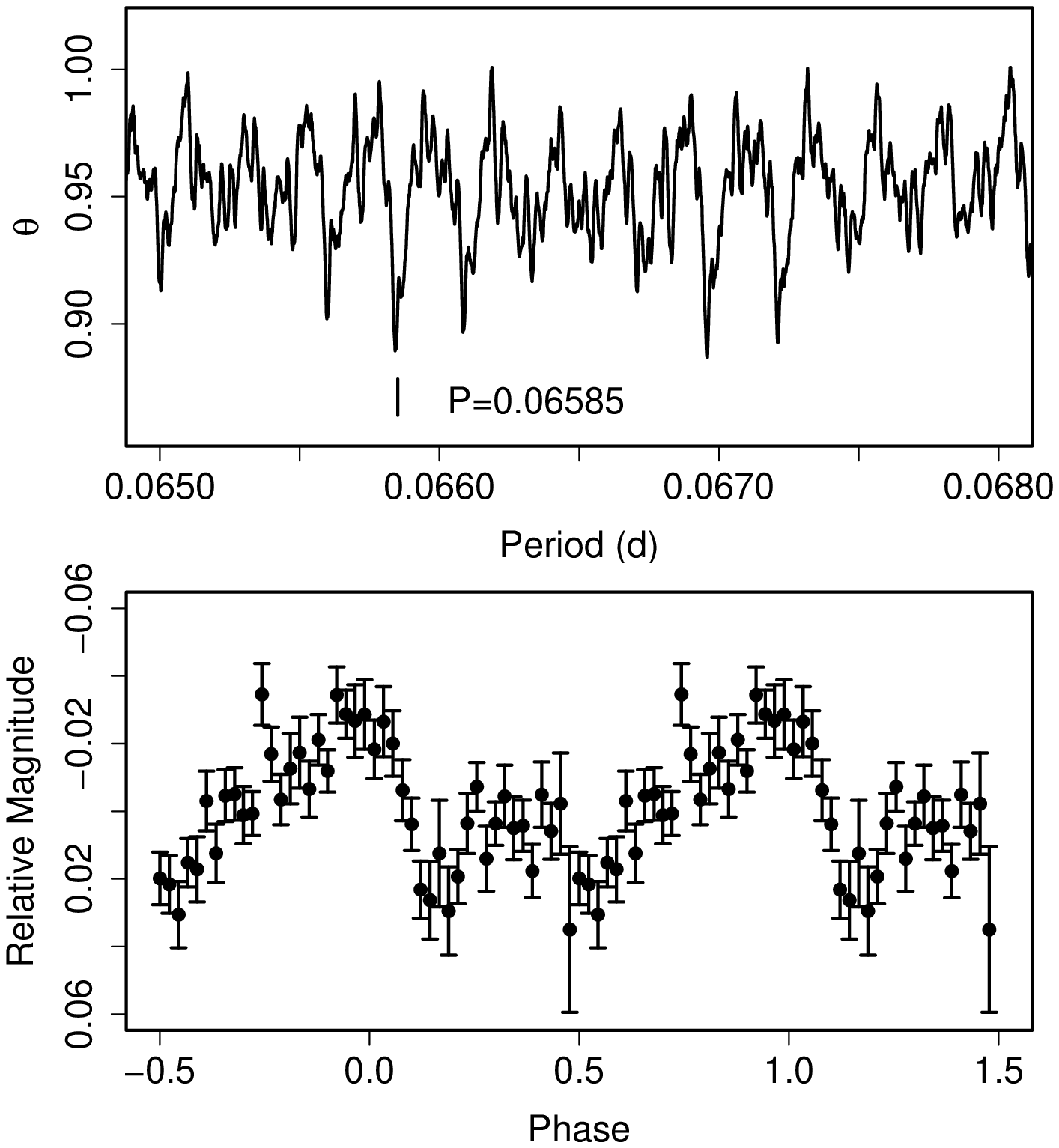}
  \end{center}
  \caption{Hump features in 1RXS J0232 during the rebrightening phase.
     (Upper): PDM analysis.  A period of 0.065850(4) is selected
     by a comparison with the superhump period (see text).
     Another potential period is 0.066963(4) d.
     (Lower): Phase-averaged profile at the period of 0.065850 d.}
  \label{fig:j0232rebpdm}
\end{figure}

   The period analyses and superhump profiles are presented in figures
\ref{fig:j0232mainpdm} and \ref{fig:j0232rebpdm}.  The analysis
during the rebrightening phase follows the same way as in SDSS J0804
\citep{kat09j0804}.  The mean $P_{\rm SH}$ during the main superoutburst
was 0.066191(4) d.  During the rebrightening phase, two candidate
periods were present: 0.066963(4) d and 0.065851(4) d.
Since the former period is 1.2 \% longer than the $P_{\rm SH}$ during
the main superoutburst, it appears to be slightly too long for a superhump
period at this stage (see subsection \ref{sec:latestage}).
The latter period, 0.5 \% shorter than the $P_{\rm SH}$, which
might represent the orbital period.
The phase-averaged profile also resembles that
of orbital humps rather than superhumps (cf. \cite{kat09j0804}).
If this identification of the period is confirmed, the small $\epsilon$
would place 1RXS J0232 similar to EG Cnc.  Since the periodicity can be
very complex during rebrightenings \citep{kat09j0804} and since the
coverage of the rebrightening phase was not sufficient, further
observations are needed to correctly identify the periods.

\begin{table}
\caption{Superhump maxima of 1RXS J0232 (2007).}\label{tab:j0232oc2007}
\begin{center}
\begin{tabular}{ccccc}
\hline\hline
$E$ & max$^a$ & error & $O-C^b$ & $N^c$ \\
\hline
0 & 54376.0443 & 0.0003 & $-$0.0020 & 191 \\
1 & 54376.1127 & 0.0002 & 0.0002 & 285 \\
46 & 54379.0904 & 0.0003 & 0.0005 & 66 \\
47 & 54379.1566 & 0.0003 & 0.0005 & 68 \\
48 & 54379.2231 & 0.0003 & 0.0009 & 65 \\
49 & 54379.2886 & 0.0004 & 0.0001 & 68 \\
50 & 54379.3550 & 0.0003 & 0.0004 & 69 \\
60 & 54380.0178 & 0.0004 & 0.0015 & 287 \\
61 & 54380.0835 & 0.0005 & 0.0010 & 337 \\
62 & 54380.1487 & 0.0004 & 0.0001 & 339 \\
63 & 54380.2149 & 0.0003 & 0.0001 & 68 \\
64 & 54380.2811 & 0.0003 & 0.0002 & 69 \\
65 & 54380.3467 & 0.0003 & $-$0.0004 & 69 \\
75 & 54381.0065 & 0.0005 & $-$0.0023 & 160 \\
106 & 54383.0589 & 0.0005 & $-$0.0010 & 152 \\
\hline
  \multicolumn{5}{l}{$^{a}$ BJD$-$2400000.} \\
  \multicolumn{5}{l}{$^{b}$ Against $max = 2454376.0463 + 0.066166 E$.} \\
  \multicolumn{5}{l}{$^{c}$ Number of points used to determine the maximum.} \\
\end{tabular}
\end{center}
\end{table}

\subsection{1RXS J042332$+$745300}\label{sec:j0423}\label{obj:j0423}

   1RXS J042332$+$745300 (=HS 0417$+$7445, hereafter 1RXS J0423) is
a CV \citep{wu01j0209j0423} selected from the ROSAT catalog and also
selected spectroscopically \citep{aun06HSCV}.  Although \citet{aun06HSCV}
detected superhumps during the 2001 superoutburst, the period was not
precisely determined.

   We observed the 2008 superoutburst and identified the correct
$P_{\rm SH}$.  Combined with the AAVSO observations, we obtained
a mean $P_{\rm SH}$ of 0.078320(6) d.
Among candidate $P_{\rm orb}$ given in \citet{aun06HSCV}, the period
of 0.07632 d best fits our $P_{\rm SH}$, and gives a fractional superhump
excess of 2.6 \%.
The times of superhump maxima are listed in table \ref{tab:j0423oc2008}.

   This outburst was associated with a precursor
($E \le 2$, figure \ref{fig:j0423oc}).
A longer period was observed for $E \le 30$ during the developmental
stage of superhumps.
This duration of stage A was thus rather unusually long.  Although there was
a slight indication of a stage B--C transition around $E=68$, the change
in the period was smaller than in other systems with similar
superhump periods.  The periods listed in table \ref{tab:perlist} are
based on this interpretation.
The presence of a precursor and the relatively short ($\sim$ 10 d) duration
of this superoutburst might signify a ``borderline'' superoutburst
as observed in BZ UMa in 2007 (subsection \ref{sec:bzuma}), which may be
responsible for the unusual development of superhumps.
Further investigation of this object is still needed.

   The light curve became double-humped during the post-superoutburst stage.
The three maxima of secondary humps ($E=153$, the first one of $E=166$,
and $E=167$) were excluded in the period analysis presented
in table \ref{tab:perlist}.

\begin{figure}
  \begin{center}
    \FigureFile(88mm,110mm){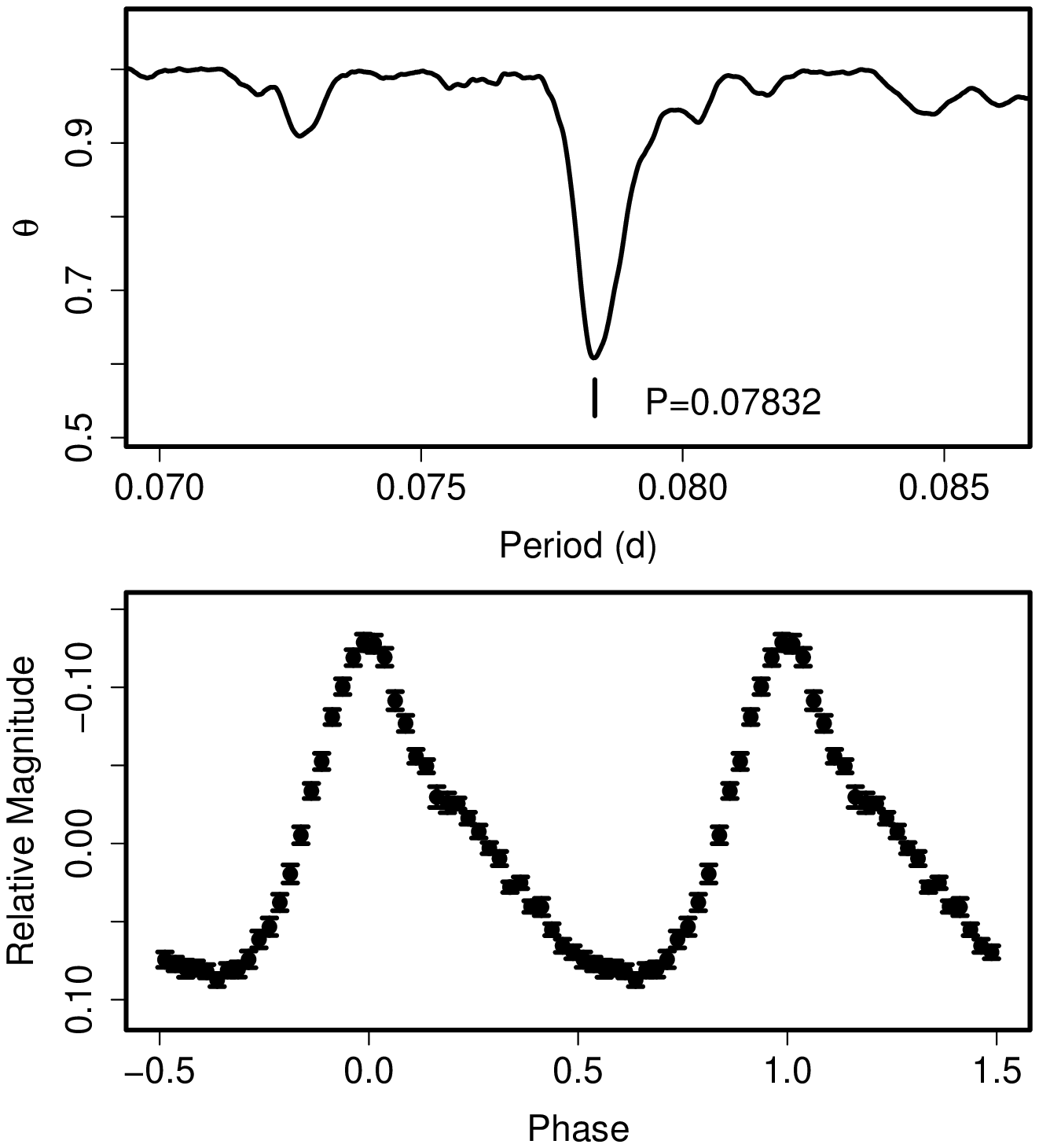}
  \end{center}
  \caption{Ordinary superhumps in 1RXS J0423 (2008). (Upper): PDM analysis.
     (Lower): Phase-averaged profile.}
  \label{fig:j0423shpdm}
\end{figure}

\begin{figure}
  \begin{center}
    \FigureFile(88mm,110mm){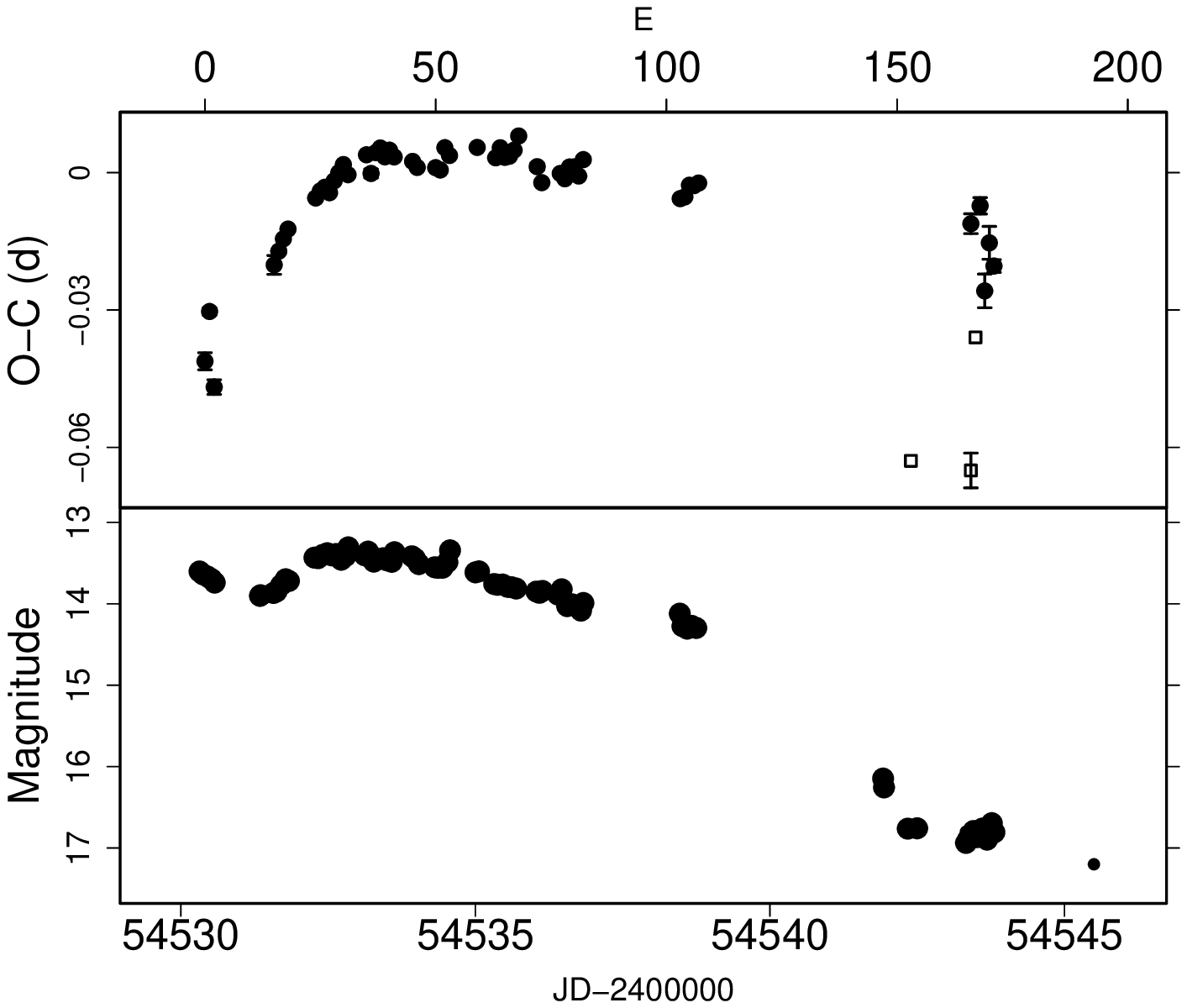}
  \end{center}
  \caption{$O-C$ of superhumps 1RXS J0423 (2008).
  (Upper): $O-C$ diagram.  The $O-C$ values were against the global
  mean period of 0.078320 d.  Open squares represent likely secondary
  hump maxima.
  (Lower): Light curve.  The last pre-outburst observation was
  on BJD 2454525.5 at magnitude 17.4.}
  \label{fig:j0423oc}
\end{figure}

\begin{table}
\caption{Superhump maxima of 1RXS J0423 (2008).}\label{tab:j0423oc2008}
\begin{center}

\end{center}
\end{table}

\subsection{1RXS J053234.9$+$624755}\label{obj:j0532}

\begin{figure}
  \begin{center}
    \FigureFile(88mm,110mm){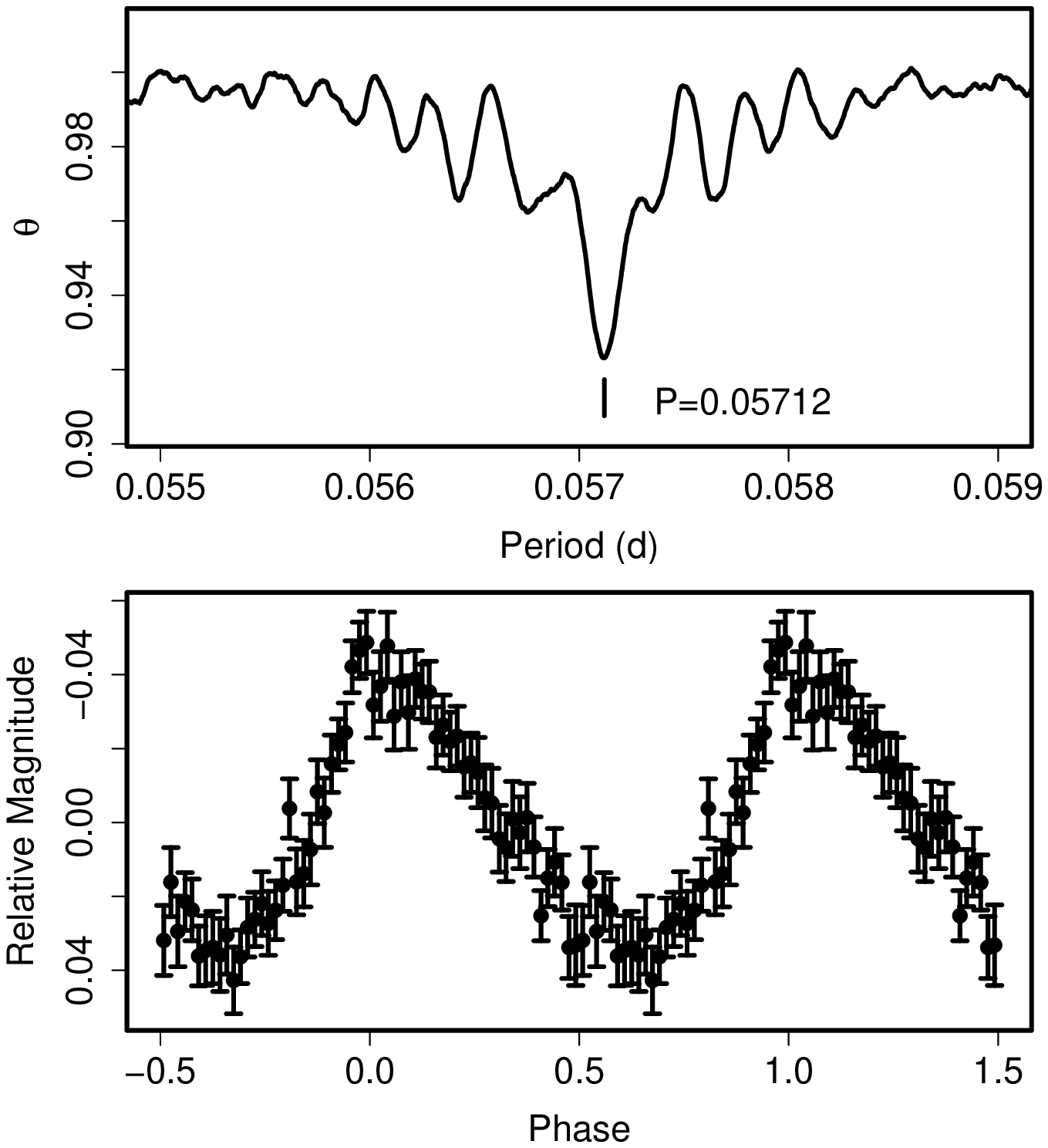}
  \end{center}
  \caption{Superhumps in 1RXS J0532 (2005). (Upper): PDM analysis.
     (Lower): Phase-averaged profile.}
  \label{fig:j0532shpdm}
\end{figure}

\begin{figure}
  \begin{center}
    \FigureFile(88mm,110mm){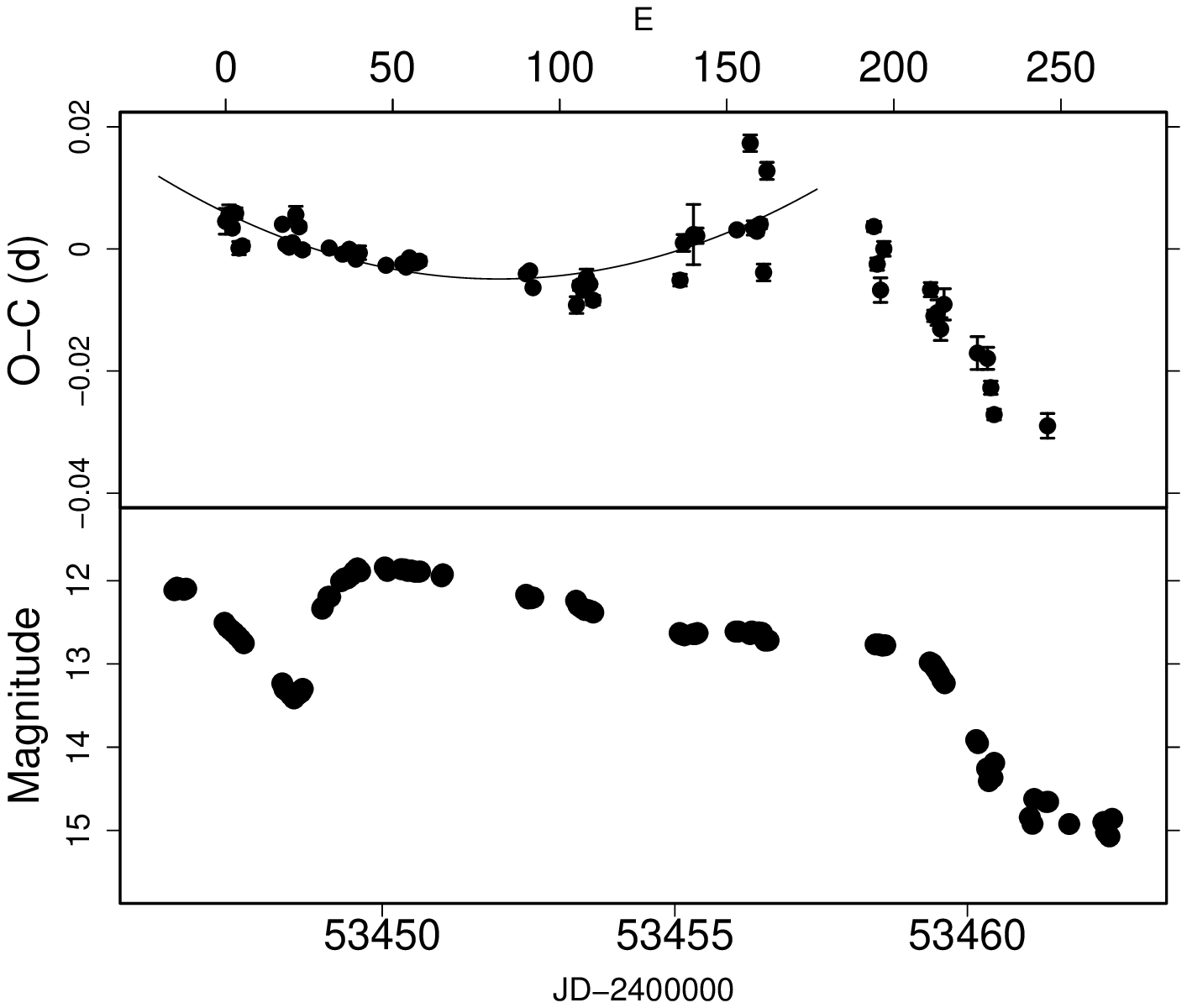}
  \end{center}
  \caption{$O-C$ of superhumps 1RXS J0532 (2005).
  (Upper): $O-C$ diagram.  The curve represents a quadratic fit to
  $E \le 162$.
  (Lower): Light curve.  The superoutburst was preceded by a precursor.
  }
  \label{fig:j05322005oc}
\end{figure}

\begin{figure}
  \begin{center}
    \FigureFile(88mm,70mm){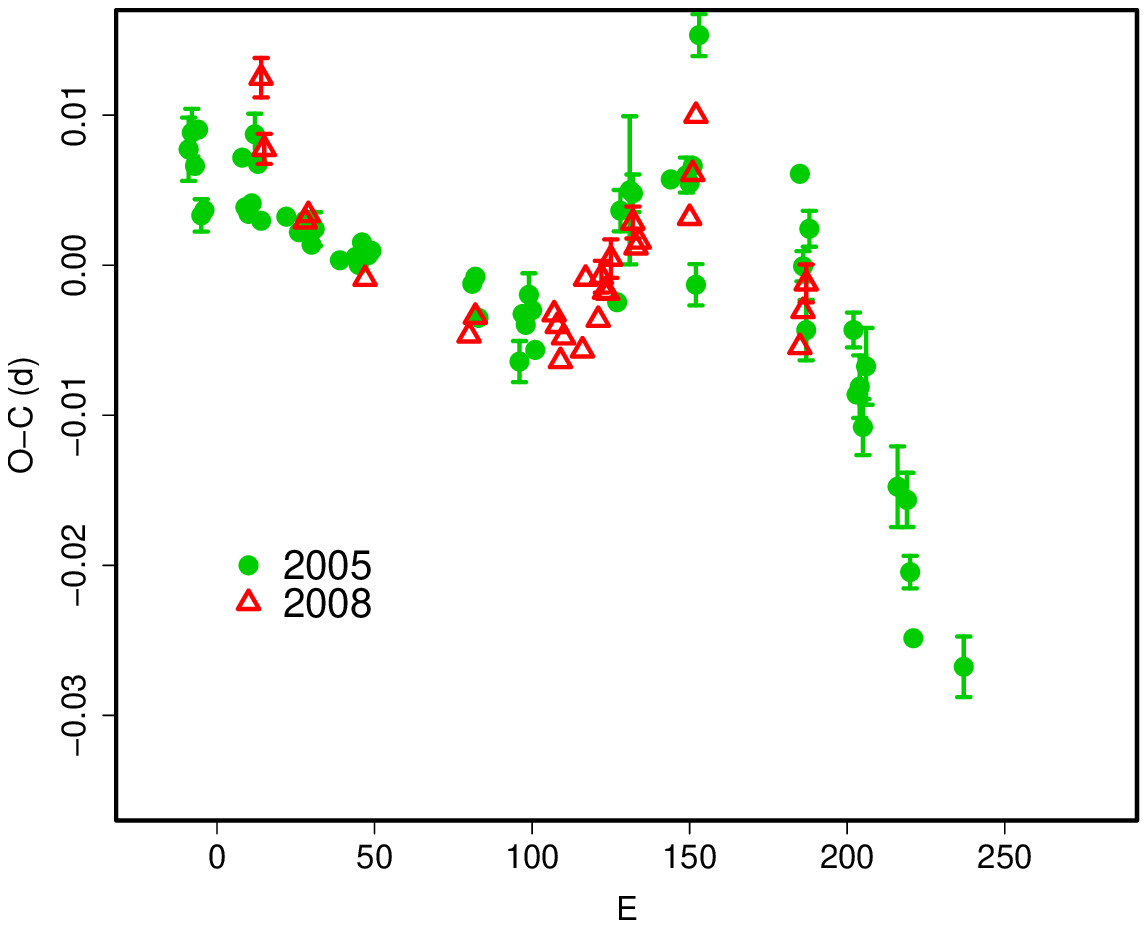}
  \end{center}
  \caption{Comparison of $O-C$ diagrams of 1RXS J0532 between different
  superoutbursts.  A period of 0.05716 d was used to draw this figure.
  Approximate cycle counts ($E$) after the start of the
  2008 superoutburst were used.  The $O-C$ diagram of the 2005 superoutburst
  bet fits the 2008 one by assuming an earlier development of the
  superhumps by $\sim$ 26 superhump cycles.
  }
  \label{fig:j0532comp}
\end{figure}

   This object (hereafter 1RXS J0532) was discovered as a dwarf nova
by \citet{ber05j0532}.  \citet{kap06j0532} provided a radial-velocity
study and yielded an orbital period of 0.05620(4) d.  (See also
\cite{kap06j0532} for the history of superhump observation).
\citet{par06j0532} observed the 2006 July superoutburst and reported
a period of 0.05707(12) d.  We report on the 2005 and 2008 superoutbursts.
The 2005 superoutburst (data from \cite{ima09j0532},
a combination of our data and the AAVSO observations)
showed a prominent precursor outburst
associated with superhumps.  This behavior was very similar to QZ Vir
(=T Leo) in 1993 \citep{kat97tleo}.
The mean superhump period with the PDM method was 0.057120(6) d
(figure \ref{fig:j0532shpdm}).
The times of superhump maxima are listed in table \ref{tab:j0532oc2005}.
The $O-C$ diagram showed a stage B--C transition
(figure \ref{fig:j05322005oc}).
The behavior in the period during the transition from the
precursor outburst was quite different from the one in QZ Vir in 1993
\citep{kat97tleo}.  The $P_{\rm dot}$ in the former interval
($E \le 162$) was $+5.7(0.8) \times 10^{-5}$.
The 2008 superoutburst was observed except for the late stage
(table \ref{tab:j0532oc2008}).  The $O-C$ behavior was similar to
that of the 2005 one, giving $P_{\rm dot}$ = $+10.2(0.8) \times 10^{-5}$
($E \le 138$).  Since there was no clear precursor at the onset of the
2008 superoutburst, the $P_{\rm dot}$ does not seem to show
very strong dependence on the presence of a precursor, on the contrary to
\citet{uem05tvcrv}.  The development of the superhumps during the 2005
superoutburst, however, may have been earlier by $\sim$ 26 superhump
cycles compared to the 2008 one (figure \ref{fig:j0532comp}).

\begin{table}
\caption{Superhump maxima of 1RXS J0532 (2005).}\label{tab:j0532oc2005}
\begin{center}

\end{center}
\end{table}

\addtocounter{table}{-1}
\begin{table}
\caption{Superhump maxima of 1RXS J0532 (2005) (continued).}
\begin{center}
%
\end{center}
\end{table}

\begin{table}
\caption{Superhump maxima of 1RXS J0532 (2008).}\label{tab:j0532oc2008}
\begin{center}
%
\end{center}
\end{table}

\subsection{2QZ J021927.9$-$304545}\label{obj:j0219}

   \citet{ima06j0219} reported the 2005 superoutburst of this object
(hereafter 2QZ J0219).  We also observed the 2009 superoutburst
during its middle-to-late stage.  The mean superhump period with
the PDM method was 0.08100(1) d, in good agreement with that of
stage C superhumps during the 2005 superoutburst.
The times of superhump maxima are listed in table \ref{tab:j0219oc2009}.

\begin{table}
\caption{Superhump maxima of 2QZ J0219 (2009)}\label{tab:j0219oc2009}
\begin{center}
%
\end{center}
\end{table}

\subsection{ASAS J002511$+$1217.2}\label{sec:asas0025}\label{obj:asas0025}

   ASAS J002511+1217.2 (hereafter ASAS J0025) is a dwarf nova discovered
by the ASAS-3 \citep{ASAS3} survey
(cf. \cite{pri04asas0025iauc8410}; for more information
see. e.g. \cite{gol05asas0025} and \cite{tem06asas0025}).

   \citet{gol05asas0025} presented a preliminary period analysis and
an $O-C$ diagram showing the presence of a variation in the superhump period.
\citet{tem06asas0025} claimed that the object belongs to WZ Sge-type
subclass based on their findings in the period variation and the presence
of a rebrightening.  The claim by \citet{tem06asas0025}, however,
led to a rather misguided conclusion because they compared the portions
of different stages (ordinary superhumps in ASAS J0025 and early superhumps
in WZ Sge), thereby resulting an inadequate period selection in drawing
the $O-C$ diagram.  We used combined data set used in
\citet{tem06asas0025} and ours, and determined superhump maxima
during the superoutburst plateau and subsequent rapid fading
(table \ref{tab:asas0025oc2004}).
The object showed a clear positive period
derivative before the terminal brightening (this agrees with
the general tendency in \cite{gol05asas0025}).  
The $P_{\rm dot}$ in this interval ($E \le 151$) was
$+8.7(0.4) \times 10^{-5}$.
The mean periods for the initial part ($E \le 30$) and the last part
($165 \le E \le 219$) were 0.05682(5) d and 0.05686(3) d, respectively,
while the mean period during the entire plateau was 0.057109(7) d.

   The superhumps in this object showed complex behavior
(see figure \ref{fig:asas0025humpall}).
After the termination of the main superoutburst, the superhumps became
doubly humped.
One the maxima (peak 1, dots in figure \ref{fig:asas0025humpall})
of these double waves, which are listed in table
\ref{tab:asas0025ochump2}, are on a smooth extension of the times of
maxima listed in table \ref{tab:asas0025oc2004} (filled circles
in figure \ref{fig:asas0025humpall}), but had a shorter period.
The other (peak 2, open squares in figure \ref{fig:asas0025humpall})
are on a smooth extension of the times of the maxima
during the post-rebrightening stage (table \ref{tab:asas0025ochump3}).
The mean periods of two components of the humps between the termination
of the main superoutburst and rebrightening were 0.056833(12) d (peak 1)
and 0.056829(21) d (peak 2), respectively.  These periods almost exactly
match the mean period during initial and last parts of the superoutburst
plateau.

The mean period of the
superhumps during the post-rebrightening stage (corresponding to
$347 \le E \le 661$) was 0.057000(6) d.  This period, longer than some
of observed (super)hump periods at earlier times, is unlikely the
orbital period.  Furthermore, the humps during the post-rebrightening
stage ($347 \le E \le 661$) appear to be on a smooth extension of the
superhumps at late stage of the superoutburst plateau ($165 \le E \le 219$).
The combined set of them yielded a mean period of 0.056995(3) d.
The stability of the period and phase for such a long interval
($166 \le E \le 661$, 28 d) is surprising.  These humps bear strong
resemblance to post-superoutburst superhumps in some of well-observed
WZ Sge-type dwarf novae
(\cite{kat08wzsgelateSH}; subsection \ref{sec:latestage}).
Following the same procedure as in \citet{kat08wzsgelateSH},
the mean period of these post-superoutburst superhumps
was found to be 0.3 \% longer than the superhump period near the onset
of the superoutburst (see discussion in \cite{kat08wzsgelateSH} for this
selection), which is close to the universal $\sim$ 0.5 \% excess described
in \citet{kat08wzsgelateSH}.

\begin{figure}
  \begin{center}
    \FigureFile(88mm,110mm){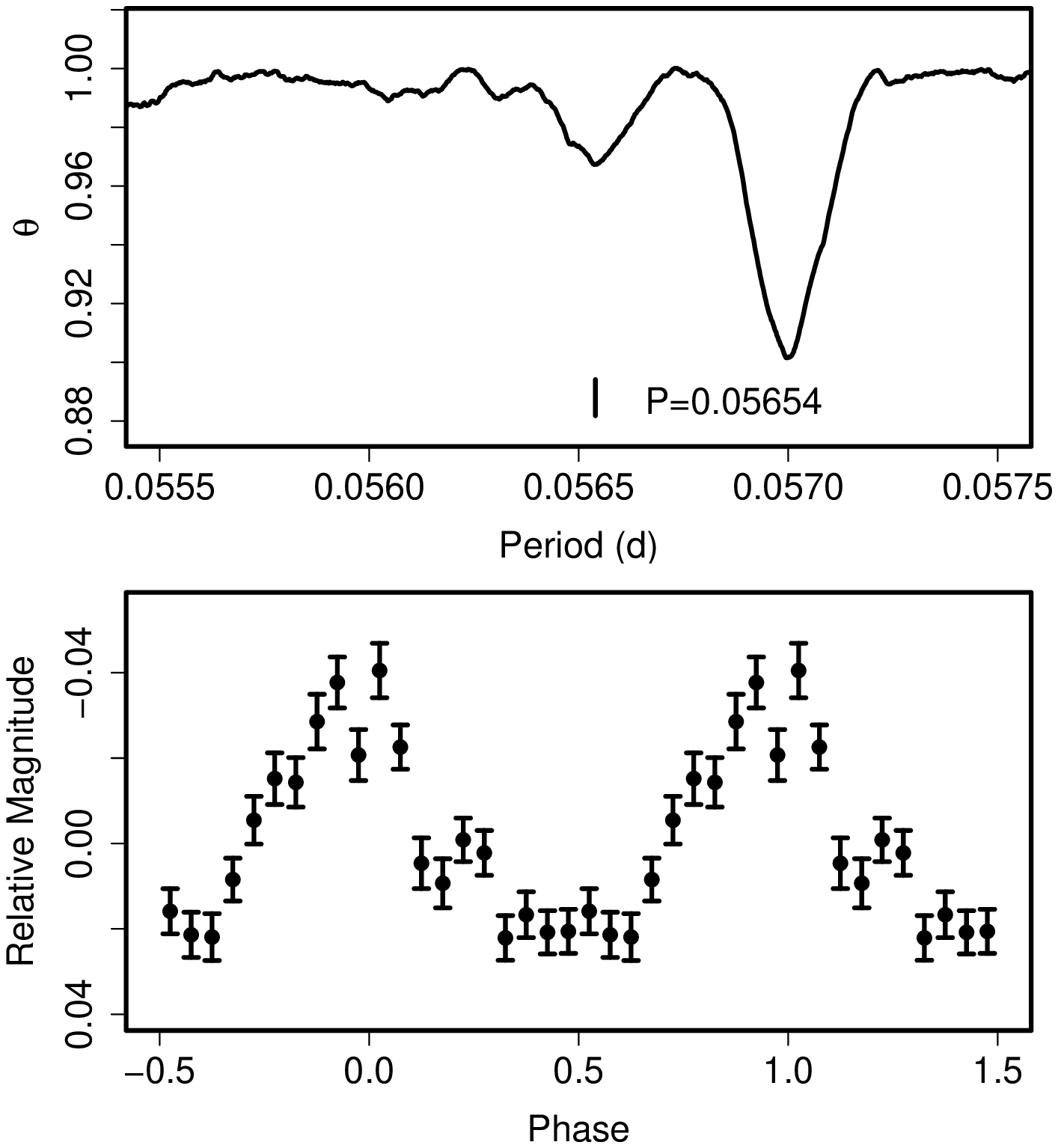}
  \end{center}
  \caption{Candidate orbital period after subtracting the superhump signal.
     (Upper): PDM analysis.  The tick denotes the candidate orbital period.
     The strong signal around $P=0.0570$ d is the residual superhump signal.
     (Lower): Phase-averaged profile.}
  \label{fig:asas0025orb}
\end{figure}

   We also performed a period analysis of the post-superoutburst stage
(BJD after 2453282) after subtracting fitted superhump signals
(figure \ref{fig:asas0025orb}).
The candidate $P_{\rm dot}$ was found with a period of 0.056540(3) d.
Although further spectroscopic confirmation is required, this period
gives $\epsilon$ of 1.0 \%.

\begin{table}
\caption{Superhump maxima of ASAS J0025 (2004).}\label{tab:asas0025oc2004}
\begin{center}

\end{center}
\end{table}

\addtocounter{table}{-1}
\begin{table}
\caption{Superhump maxima of ASAS J0025 (2004) (continued).}
\begin{center}
%
\end{center}
\end{table}

\addtocounter{table}{-1}
\begin{table}
\caption{Superhump maxima of ASAS J0025 (2004) (continued).}
\begin{center}
%
\end{center}
\end{table}

\begin{table}
\caption{Superhump maxima of ASAS J0025 (secondary).}\label{tab:asas0025ochump2}
\begin{center}
%
\end{center}
\end{table}

\begin{table}
\caption{Superhump maxima of ASAS J0025 (late stage).}\label{tab:asas0025ochump3}
\begin{center}
%
\end{center}
\end{table}

\addtocounter{table}{-1}
\begin{table}
\caption{Superhump maxima of ASAS J0025 (late stage, continued).}
\begin{center}
%
\end{center}
\end{table}

\addtocounter{table}{-1}
\begin{table}
\caption{Superhump maxima of ASAS J0025 (late stage, continued).}
\begin{center}
%
\end{center}
\end{table}

\subsection{ASAS J023322$-$1047.0}\label{obj:asas0233}

\begin{figure}
  \begin{center}
    \FigureFile(88mm,110mm){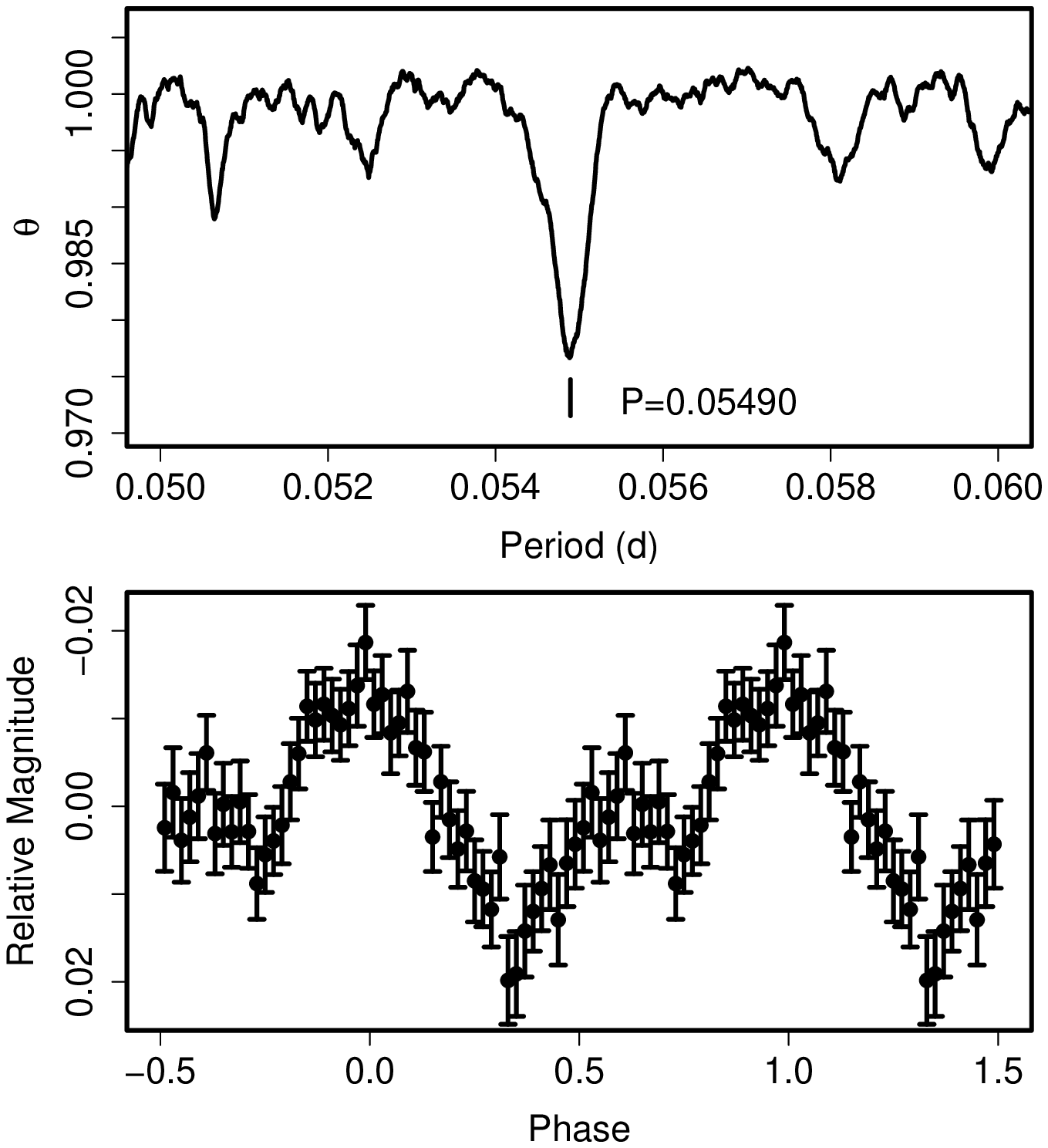}
  \end{center}
  \caption{Early superhumps in ASAS J0233 (2006). (Upper): PDM analysis.
     (Lower): Phase-averaged profile.}
  \label{fig:asas0233eshpdm}
\end{figure}

\begin{figure}
  \begin{center}
    \FigureFile(88mm,110mm){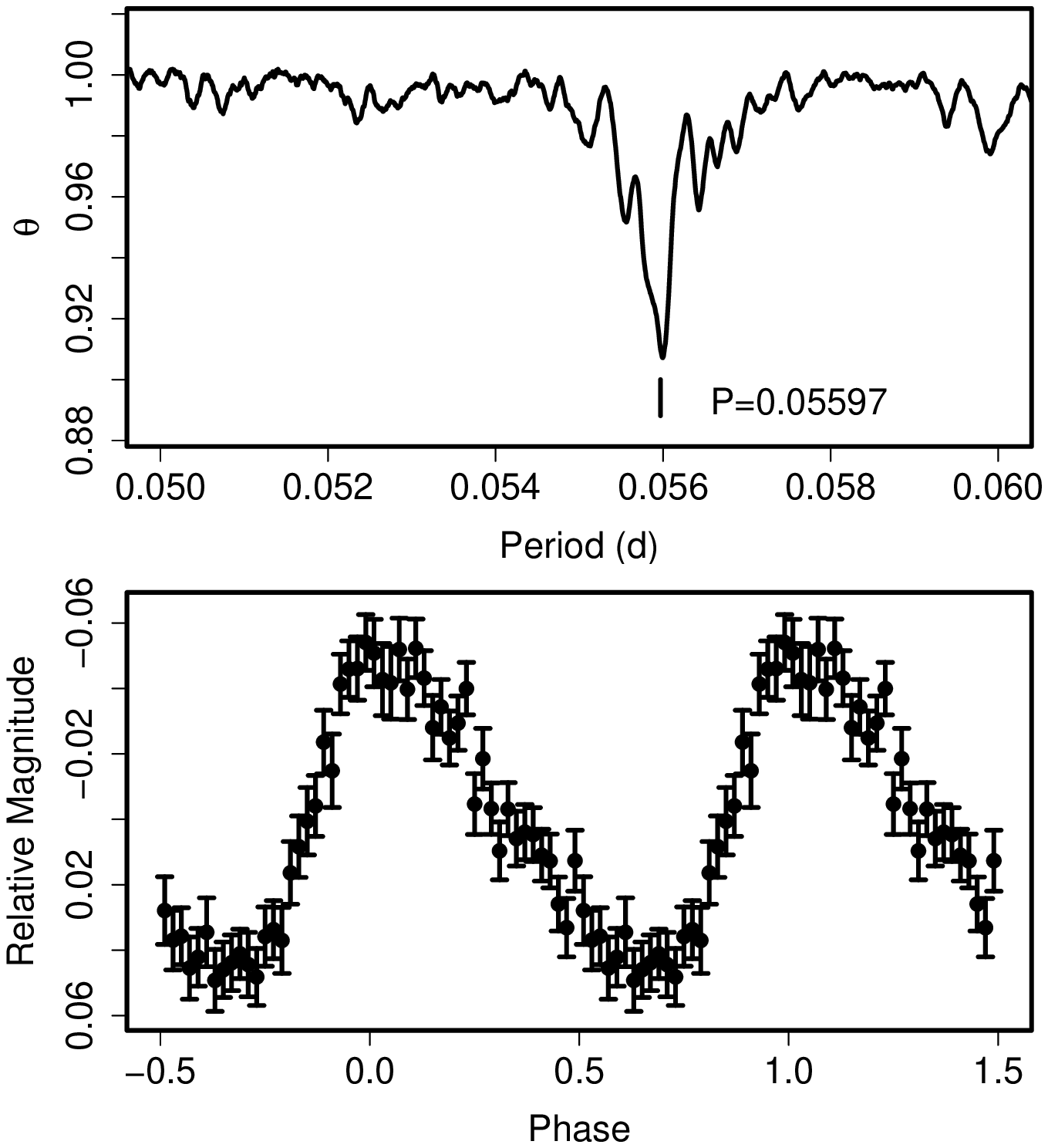}
  \end{center}
  \caption{Ordinary superhumps in ASAS J0233 (2006). (Upper): PDM analysis.
     (Lower): Phase-averaged profile.}
  \label{fig:asas0233shpdm}
\end{figure}

   ASAS J023322$-$1047.0 (hereafter ASAS J0233) is a dwarf nova detected
by ASAS-3 on 2006 January 20 ($V = 12.1$, vsnet-alert 8801).
Early superhumps were immediately detected (vsnet-alert 8815), and
ordinary superhumps developed eight days after the outburst detection
(vsnet-alert 8825).  \citet{van06asas0233asas1025} summarized this
outburst and reported period analyses.
The data was a combination of ours and the AAVSO data, as utilized in
\citet{van06asas0233asas1025}.
The mean periods of early and ordinary superhumps determined with
the PDM method were 0.054895(23) d (figure \ref{fig:asas0233eshpdm})
and 0.055970(9) d (figure \ref{fig:asas0233shpdm}), respectively.
The times of ordinary superhump maxima are listed in table
\ref{tab:asas0233oc2006}.  The $O-C$ diagram
(cf. figure \ref{fig:ocsamp}) clearly
demonstrates the presence of the early development stage (stage A, $E \le 2$),
stage B with a positive period derivative, and the final transition
to a shorter period (stage C).  The $P_{\rm dot}$ for the stage B
($7 \le E \le 216$) was $+4.9(0.5) \times 10^{-5}$.

\begin{table}
\caption{Superhump maxima of ASAS J0233 (2006).}\label{tab:asas0233oc2006}
\begin{center}

\end{center}
\end{table}

\subsection{ASAS J091858$-$2942.6 = Dwarf nova in Pyxis 2005}\label{obj:asas0918}

   This object (hereafter ASAS J0918) was independently discovered
by G. Pojmanski and K. Haseda \citep{poj05dnpyx}.
Follow-up spectroscopy revealed that the object
was not a nova, but a dwarf nova in outburst \citep{kaw05dnpyx}.
We undertook time-series photometry soon after the discovery announcement.

\begin{figure}
  \begin{center}
    \FigureFile(88mm,110mm){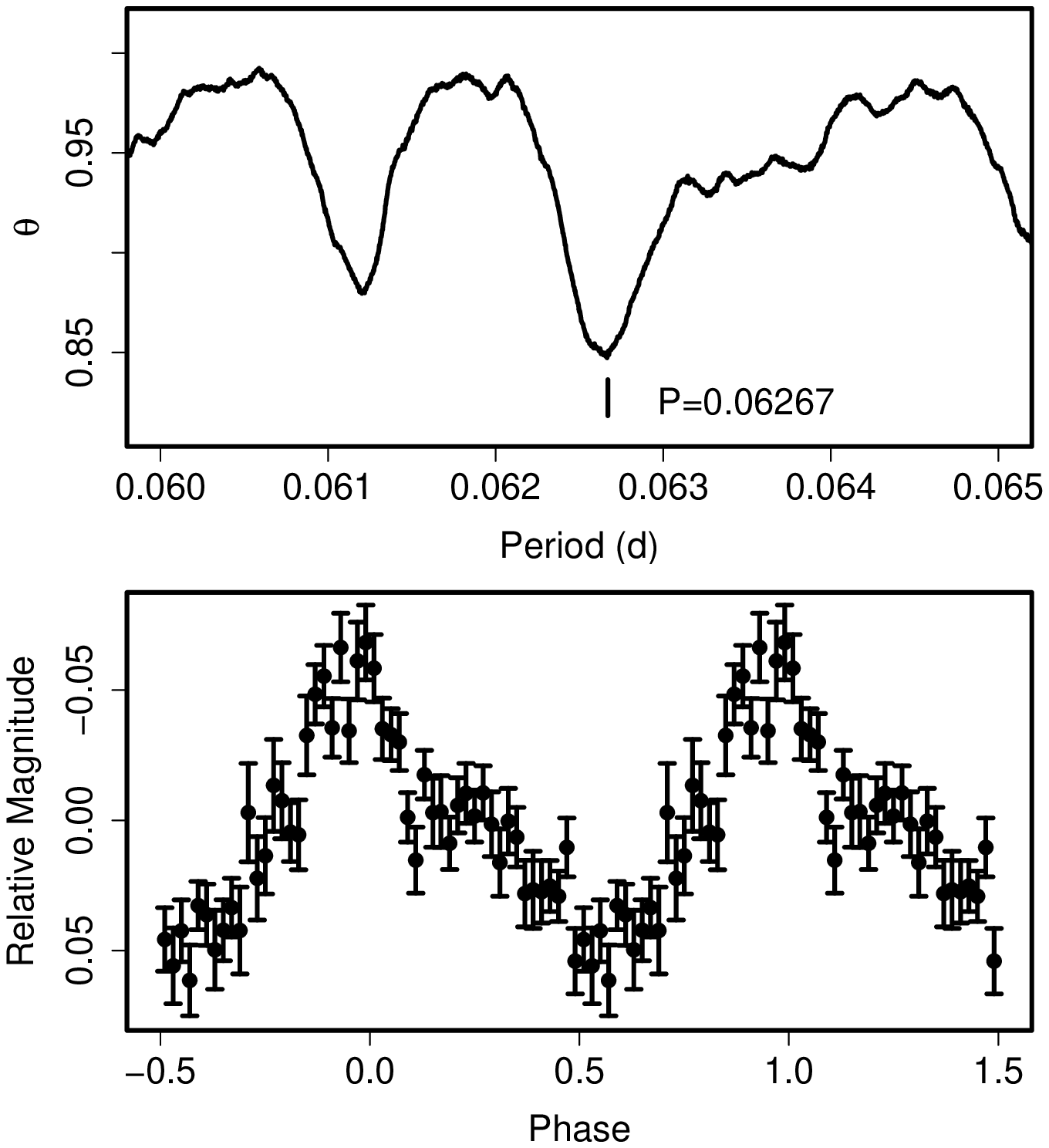}
  \end{center}
  \caption{Superhumps in ASAS J0918 (2005). (Upper): PDM analysis.
     (Lower): Phase-averaged profile.}
  \label{fig:asas0918shpdm}
\end{figure}

   A PDM analysis yielded a mean superhump period of 0.06267(2) d
(figure \ref{fig:asas0918shpdm}).
The times of superhump maxima are listed in table \ref{tab:dnpyxoc2005}.
Although we can derive a global $P_{\rm dot}$ = $-15.6(4.3) \times 10^{-5}$,
this value should not be regarded as the representative period derivative
of this system since the object showed a remarkable terminal rebrightening
before $E = 78$ and the observed $O-C$'s most likely reflected a shortening
of the superhump period between stage B and C.
The period derivative for the stage B was not significantly determined
from a short segment $E \le 32$.
Future observations starting from the early epoch of a superoutburst
are necessary to determine the period derivative, although continuous
monitoring by ASAS-3 has not detected any further outburst.

\begin{table}
\caption{Superhump maxima of ASAS J0918 (2005).}\label{tab:dnpyxoc2005}
\begin{center}
\begin{tabular}{ccccc}
\hline\hline
$E$ & max$^a$ & error & $O-C^b$ & $N^c$ \\
\hline
0 & 53448.0457 & 0.0011 & $-$0.0049 & 275 \\
1 & 53448.1088 & 0.0020 & $-$0.0044 & 171 \\
14 & 53448.9300 & 0.0021 & 0.0025 & 58 \\
15 & 53448.9939 & 0.0016 & 0.0037 & 91 \\
16 & 53449.0527 & 0.0041 & $-$0.0002 & 107 \\
17 & 53449.1128 & 0.0037 & $-$0.0027 & 93 \\
30 & 53449.9332 & 0.0014 & 0.0033 & 62 \\
31 & 53449.9968 & 0.0007 & 0.0043 & 84 \\
32 & 53450.0583 & 0.0007 & 0.0031 & 236 \\
78 & 53452.9311 & 0.0020 & $-$0.0055 & 63 \\
79 & 53453.0001 & 0.0013 & 0.0008 & 63 \\
\hline
  \multicolumn{5}{l}{$^{a}$ BJD$-$2400000.} \\
  \multicolumn{5}{l}{$^{b}$ Against $max = 2453448.0506 + 0.062642 E$.} \\
  \multicolumn{5}{l}{$^{c}$ Number of points used to determine the maximum.} \\
\end{tabular}
\end{center}
\end{table}

\subsection{ASAS J102522$-$1542.4}\label{obj:asas1025}

\begin{figure}
  \begin{center}
    \FigureFile(88mm,110mm){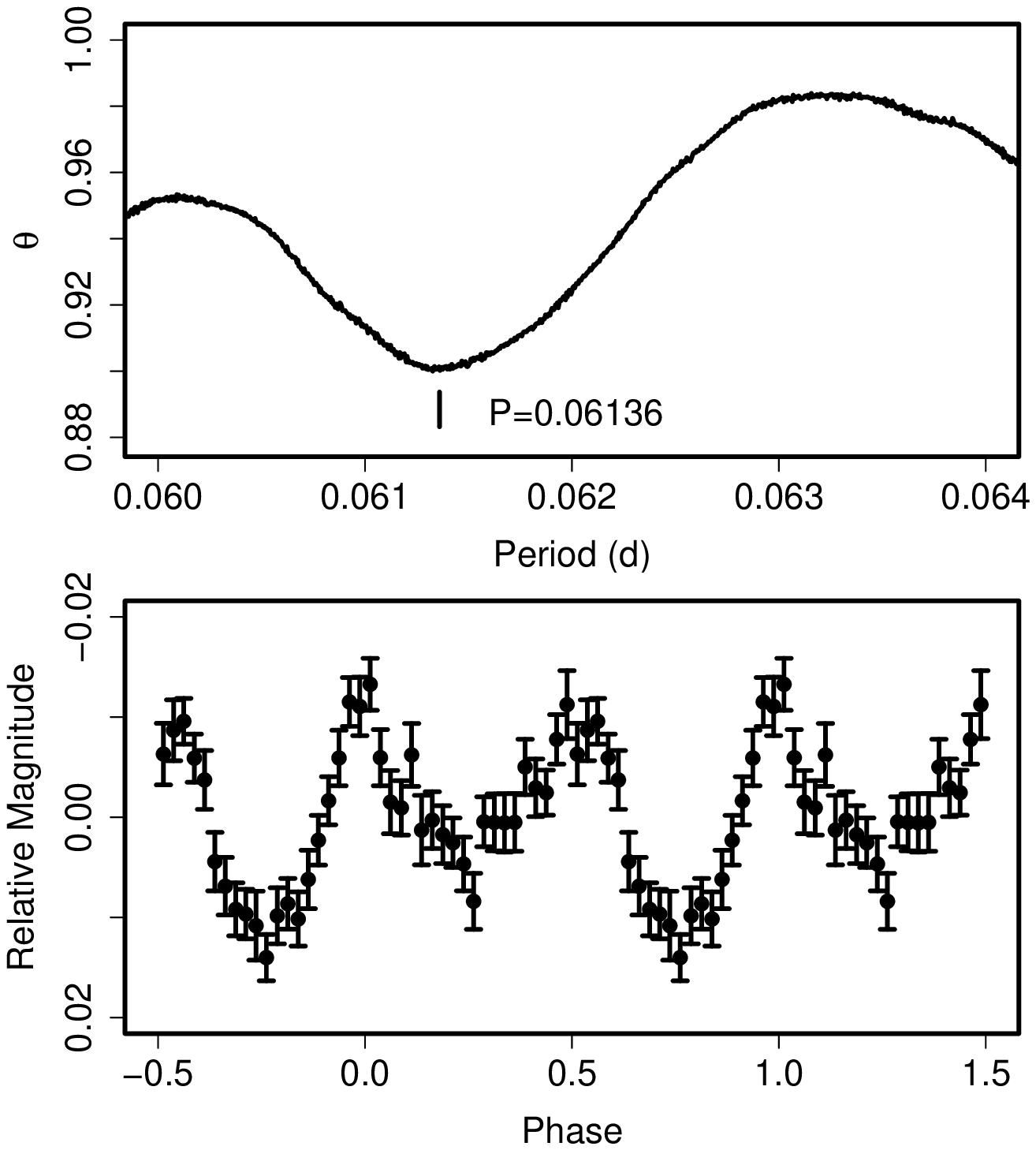}
  \end{center}
  \caption{Early superhumps in ASAS J1025 (2006). (Upper): PDM analysis.
     (Lower): Phase-averaged profile.}
  \label{fig:j1025eshpdm}
\end{figure}

\begin{figure}
  \begin{center}
    \FigureFile(88mm,110mm){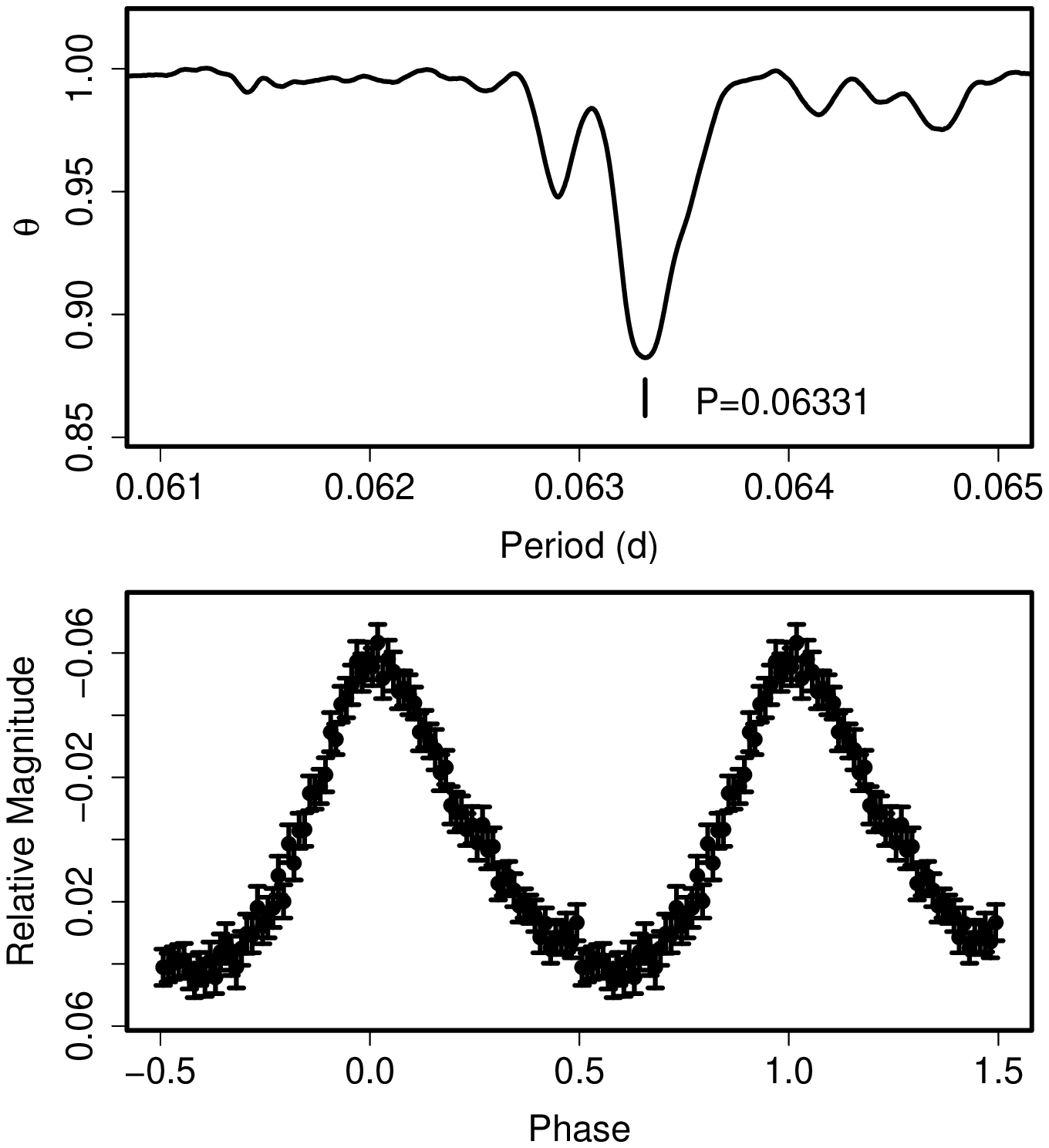}
  \end{center}
  \caption{Ordinary superhumps in ASAS J1025 (2006). (Upper): PDM analysis.
     (Lower): Phase-averaged profile.}
  \label{fig:j1025shpdm}
\end{figure}

   ASAS J102522$-$1542.4 (hereafter ASAS J1025) is a dwarf nova detected
by ASAS-3 on 2006 January 26 ($V = 12.2$, vsnet-alert 8821).
The detection of early superhumps (vsnet-alert 8824;
figure \ref{fig:j1025eshpdm}, period 0.06136(6) d) and ordinary
superhumps (vsnet-alert 8843; figure \ref{fig:j1025shpdm},
mean period 0.063314(5) d) led to a likely classification
as a WZ Sge-type dwarf nova.  \citet{van06asas0233asas1025} provided
a provisional analysis.

   The times of superhump maxima (excluding early superhumps)
are listed in table \ref{tab:asas1025oc2006}.
The $O-C$ diagram (cf. figure \ref{fig:ocsamp})
consisted of three stages A--C.  We obtained
$P_{\rm dot}$ = $+10.9(0.6) \times 10^{-5}$ (stage B, $27 \le E \le 142$).
The stage C superhumps persisted until the start of the rebrightening.

   The fractional superhump excess determined from the period of
early and ordinary superhumps was 3.2(1) \%, which is unusually large
for a WZ Sge-type dwarf nova.  Combined with the large $P_{\rm dot}$
and the short delay before ordinary superhumps emerged,
the object appears to be a ``borderline'' long-$P_{\rm SH}$ WZ Sge-like
dwarf nova similar to BC UMa \citep{pat03suumas} and
ASAS J1600 \citep{soe09asas1600}.
The exact identification of the $P_{\rm orb}$, however, should await
further observation because the period of early superhumps was determined
from a short baseline.

\begin{table}
\caption{Superhump maxima of ASAS J1025 (2006).}\label{tab:asas1025oc2006}
\begin{center}

\end{center}
\end{table}

\addtocounter{table}{-1}
\begin{table}
\caption{Superhump maxima of ASAS J1025 (2006) (continued).}
\begin{center}
%
\end{center}
\end{table}

\subsection{ASAS J153616$-$0839.1}\label{obj:asas1536}

\begin{figure}
  \begin{center}
    \FigureFile(88mm,70mm){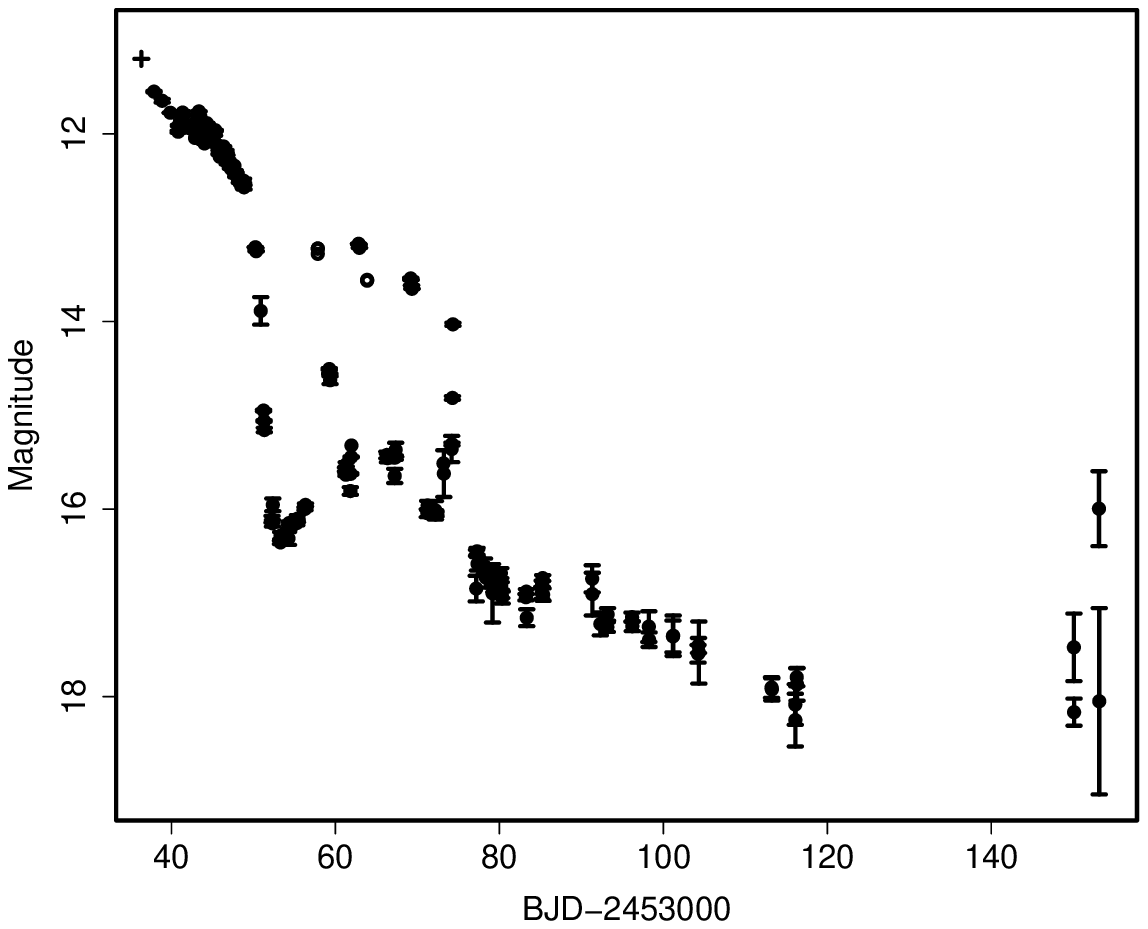}
  \end{center}
  \caption{Light curve of ASAS J1536 (2004).
     The filled circles, open circles and a cross represent CCD observations
     used here and ASAS-3 $V$ data, and Haseda's prediscovery photographic
     observation, respectively.}
  \label{fig:asas1536lc}
\end{figure}

   ASAS J153616$-$0839.1 (hereafter ASAS J1536) is a dwarf nova detected
by ASAS-3 on 2004 February 2 ($V = 11.54$).  A prediscovery observation by
K. Haseda on 2004 January 31 ($m_{\rm pg} = 11.2$) was reported
(vsnet-alert 7986, 7987; see also \cite{sch04asas1536}).
The object showed a relatively smooth fading until February 7, then
followed by a $\sim$ 0.2 mag rise associated with prominent superhumps.
The object underwent four post-superoutburst rebrightenings
(figure \ref{fig:asas1536lc}).

\begin{figure}
  \begin{center}
    \FigureFile(88mm,110mm){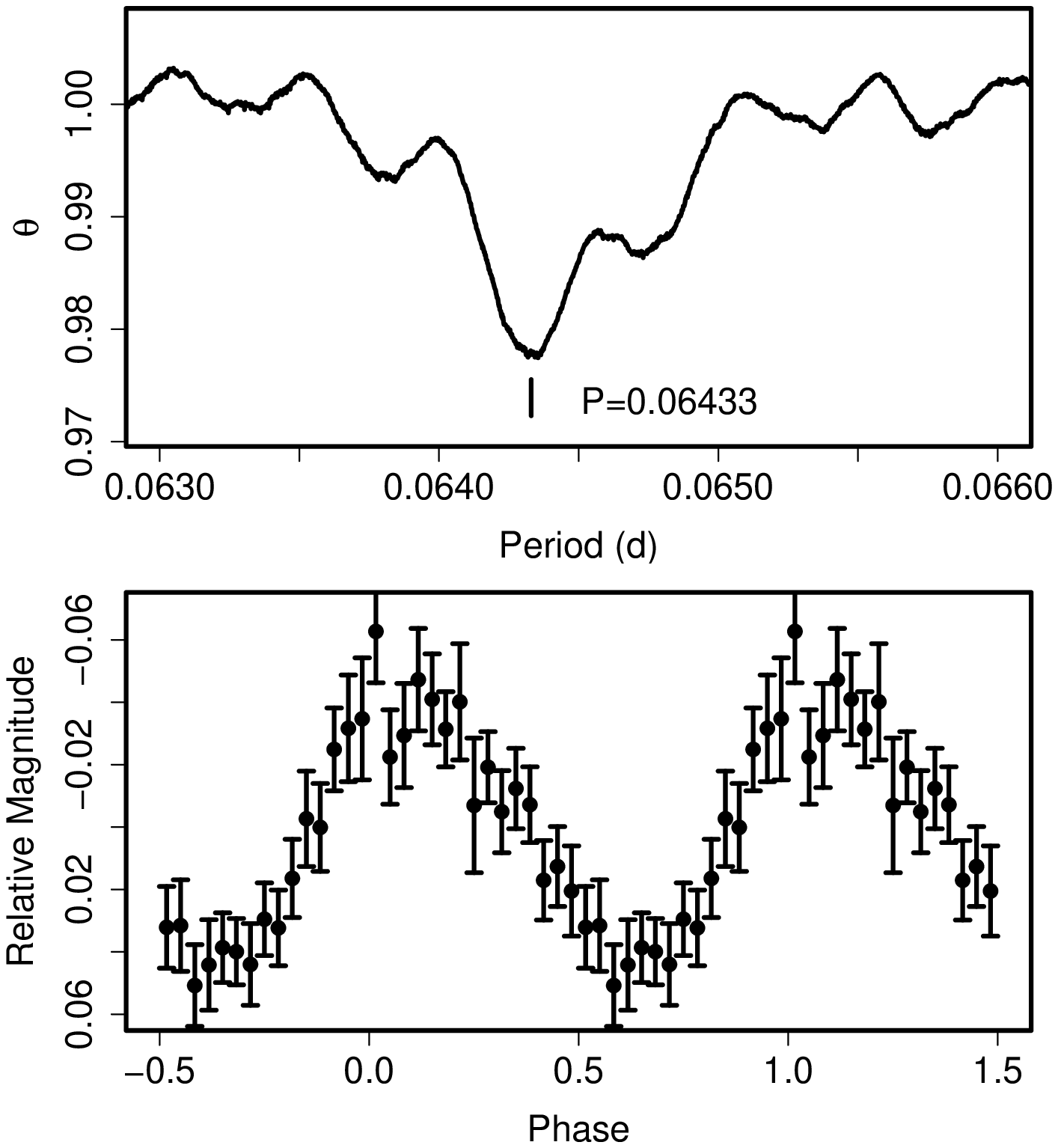}
  \end{center}
  \caption{Ordinary superhumps in ASAS J1536 (2004) after BJD 2453043
     (Upper): PDM analysis.
     (Lower): Phase-averaged profile.}
  \label{fig:asas1536shpdm}
\end{figure}

   The times of superhump maxima are listed in table \ref{tab:asas1536oc2004}.
There was a clear stage A--B transition around $E=30$.
The $P_{\rm dot}$ of the stage B was $+2.4(2.1) \times 10^{-5}$.
We know little information whether the object had already developed
superhumps or early superhumps before the start of our observation.
We, however, adopted this value as the representative period derivative
of this system since the object is likely a WZ Sge-type dwarf nova with
multiple rebrightenings and a rise associated with prominent superhumps
can be better interpreted as a signature of emergence of ordinary
superhumps (cf. \cite{pat98egcnc}).  Following this interpretation,
the epoch of our observation corresponds to the middle plateau stage of the
superoutburst rather than its final stage.  We present a representative
averaged light curve of superhumps (figure \ref{fig:asas1536shpdm}).

\begin{figure}
  \begin{center}
    \FigureFile(88mm,110mm){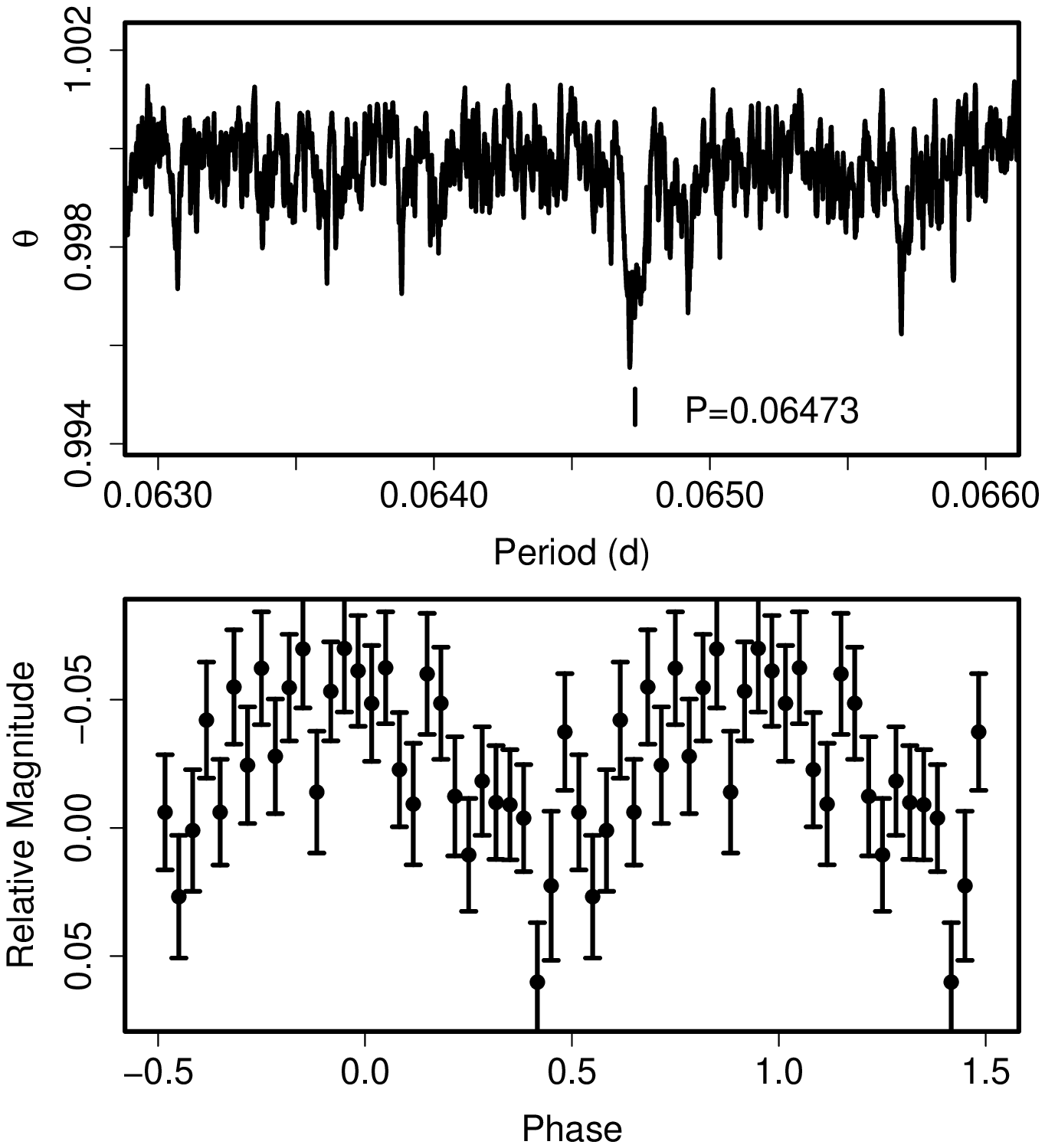}
  \end{center}
  \caption{Superhumps in ASAS J1536 (2004) during the rebrightenings
     and post-superoutburst stage.
     (Upper): PDM analysis.
     (Lower): Phase-averaged profile.}
  \label{fig:asas1536latepdm}
\end{figure}

   The object showed a weaker superhump signal during the rebrightening
and post-superoutburst stages (figure \ref{fig:asas1536latepdm}).
The period, 0.06473(1) d, appears to be longer than the $P_{\rm SH}$
during the superoutburst plateau, analogous to other WZ Sge-type
dwarf novae (subsection \ref{sec:latestage}).

\begin{table}
\caption{Superhump maxima of ASAS J1536 (2004).}\label{tab:asas1536oc2004}
\begin{center}
\begin{tabular}{ccccc}
\hline\hline
$E$ & max$^a$ & error & $O-C^b$ & $N^c$ \\
\hline
0 & 53041.3165 & 0.0019 & $-$0.0206 & 189 \\
15 & 53042.3057 & 0.0009 & $-$0.0017 & 200 \\
16 & 53042.3627 & 0.0024 & $-$0.0094 & 85 \\
30 & 53043.2831 & 0.0005 & 0.0053 & 89 \\
31 & 53043.3509 & 0.0006 & 0.0084 & 71 \\
39 & 53043.8620 & 0.0079 & 0.0020 & 10 \\
42 & 53044.0604 & 0.0005 & 0.0064 & 115 \\
43 & 53044.1220 & 0.0003 & 0.0032 & 106 \\
45 & 53044.2550 & 0.0003 & 0.0068 & 204 \\
46 & 53044.3178 & 0.0005 & 0.0050 & 205 \\
54 & 53044.8268 & 0.0026 & $-$0.0036 & 20 \\
58 & 53045.0914 & 0.0004 & 0.0024 & 133 \\
59 & 53045.1546 & 0.0003 & 0.0008 & 125 \\
61 & 53045.2840 & 0.0005 & 0.0008 & 205 \\
62 & 53045.3527 & 0.0006 & 0.0048 & 163 \\
69 & 53045.8043 & 0.0023 & 0.0036 & 16 \\
70 & 53045.8644 & 0.0073 & $-$0.0009 & 20 \\
73 & 53046.0600 & 0.0003 & 0.0005 & 125 \\
74 & 53046.1224 & 0.0006 & $-$0.0017 & 81 \\
76 & 53046.2592 & 0.0009 & 0.0057 & 191 \\
77 & 53046.3184 & 0.0008 & 0.0002 & 189 \\
78 & 53046.3816 & 0.0010 & $-$0.0013 & 57 \\
82 & 53046.6400 & 0.0006 & $-$0.0016 & 59 \\
84 & 53046.7845 & 0.0029 & 0.0134 & 16 \\
85 & 53046.8407 & 0.0086 & 0.0050 & 19 \\
89 & 53047.0923 & 0.0005 & $-$0.0022 & 106 \\
90 & 53047.1578 & 0.0003 & $-$0.0014 & 118 \\
92 & 53047.2875 & 0.0008 & $-$0.0011 & 219 \\
93 & 53047.3481 & 0.0009 & $-$0.0051 & 176 \\
98 & 53047.6728 & 0.0010 & $-$0.0039 & 42 \\
100 & 53047.8029 & 0.0049 & $-$0.0031 & 16 \\
101 & 53047.8652 & 0.0062 & $-$0.0055 & 18 \\
108 & 53048.3228 & 0.0007 & $-$0.0008 & 111 \\
115 & 53048.7757 & 0.0093 & $-$0.0007 & 13 \\
116 & 53048.8331 & 0.0169 & $-$0.0080 & 16 \\
138 & 53050.2659 & 0.0010 & 0.0017 & 151 \\
139 & 53050.3257 & 0.0015 & $-$0.0032 & 159 \\
\hline
  \multicolumn{5}{l}{$^{a}$ BJD$-$2400000.} \\
  \multicolumn{5}{l}{$^{b}$ Against $max = 2453041.3371 + 0.0646895 E$.} \\
  \multicolumn{5}{l}{$^{c}$ Number of points used to determine the maximum.} \\
\end{tabular}
\end{center}
\end{table}

\subsection{ASAS J160048$-$4846.2}\label{obj:asas1600}

   \citet{ima06asas1600} and \citet{soe09asas1600} reported a detailed report
of the 2005 superoutburst.  We further analyzed the data in combination
with the AAVSO observations.  The times of superhump maxima are
listed in table \ref{tab:asas1600oc2005}.  The result basically confirmed
the analysis by \citet{soe09asas1600}.  The maxima for $E \ge 243$
were humps observed during the rebrightening.  Since the $O-C$'s of these
humps were not on a smooth extension of the stage C superhumps, these
humps are less likely persisting superhumps.

\begin{table}
\caption{Superhump maxima of ASAS J1600 (2005).}\label{tab:asas1600oc2005}
\begin{center}

\end{center}
\end{table}

\addtocounter{table}{-1}
\begin{table}
\caption{Superhump maxima of ASAS J1600 (2005). (continued)}
\begin{center}
%
\end{center}
\end{table}

\subsection{CTCV J0549$-$4921}\label{obj:j0549}

   The identification of this object (hereafter CTCV J0549) as an SU UMa-type
dwarf nova was reported by \citet{ima08fltractcv0549}.  As in KK Tel
\citep{kat03v877arakktelpucma}, \citet{ima08fltractcv0549} failed to
identify the correct $P_{\rm SH}$ and $P_{\rm dot}$ due to the large
variation in $P_{\rm SH}$.  In table \ref{tab:j0549oc2006}, we listed
updated times of superhump maxima, measured from the data reported in
\citet{ima08fltractcv0549}.  Following the stage A period evolution
($E \le 1$), the $P_{\rm SH}$ varied strongly as in UV Gem and KK Tel.
The identified periods are given in table \ref{tab:perlist}.
After an examination of ASAS-3 light curve \citep{ASAS3}, we detected
a number of outbursts (table \ref{tab:j0549out}).  The object appears
to be more active than inferred by \citet{ima08fltractcv0549}.
The typical length of supercycle is 750--800 d.

\begin{table}
\caption{Superhump maxima of CTCV J0549 (2006).}\label{tab:j0549oc2006}
\begin{center}
\begin{tabular}{ccccc}
\hline\hline
$E$ & max$^a$ & error & $O-C^b$ & $N^c$ \\
\hline
0 & 53828.2560 & 0.0017 & $-$0.0251 & 188 \\
1 & 53828.3308 & 0.0027 & $-$0.0349 & 183 \\
23 & 53830.2389 & 0.0004 & 0.0126 & 105 \\
24 & 53830.3275 & 0.0005 & 0.0166 & 123 \\
35 & 53831.2610 & 0.0002 & 0.0197 & 193 \\
36 & 53831.3448 & 0.0004 & 0.0189 & 99 \\
47 & 53832.2610 & 0.0003 & 0.0048 & 189 \\
48 & 53832.3513 & 0.0010 & 0.0106 & 98 \\
118 & 53838.2499 & 0.0003 & $-$0.0111 & 190 \\
119 & 53838.3334 & 0.0008 & $-$0.0122 & 98 \\
\hline
  \multicolumn{5}{l}{$^{a}$ BJD$-$2400000.} \\
  \multicolumn{5}{l}{$^{b}$ Against $max = 2453828.2811 + 0.084575 E$.} \\
  \multicolumn{5}{l}{$^{c}$ Number of points used to determine the maximum.} \\
\end{tabular}
\end{center}
\end{table}

\begin{table}
\caption{Outbursts of CTCV J0549.}\label{tab:j0549out}
\begin{center}
\begin{tabular}{cccc}
\hline\hline
JD$-$2400000 & $V$ max & Duration (d) & Type \\
\hline
51952.5 & 13.5 & $>$9 & Super \\
52172.9 & 13.7 & 1 & Normal \\
52578.7 & 14.0 & 1 & Normal \\
53025.6 & 13.6 & $>$10 & Super \\
53489.5 & 13.8 & 1 & Normal \\
53740.6 & 13.6 & 2 & Normal \\
53813.7 & 15.0 & 1 & Normal \\
53830.5 & 13.3 & $>$6 & Super \\
54216.5 & 13.8 & 1 & Normal \\
54301.9 & 13.8 & 1 & Normal \\
54440.7 & 14.3 & 1 & Normal \\
54586.5 & 13.1 & $>$8 & Super \\
54705.9 & 14.4 & 1 & Normal \\
\hline
\end{tabular}
\end{center}
\end{table}

\subsection{Ha 0242$-$2802}\label{obj:ha0242}

\begin{figure}
  \begin{center}
    \FigureFile(88mm,110mm){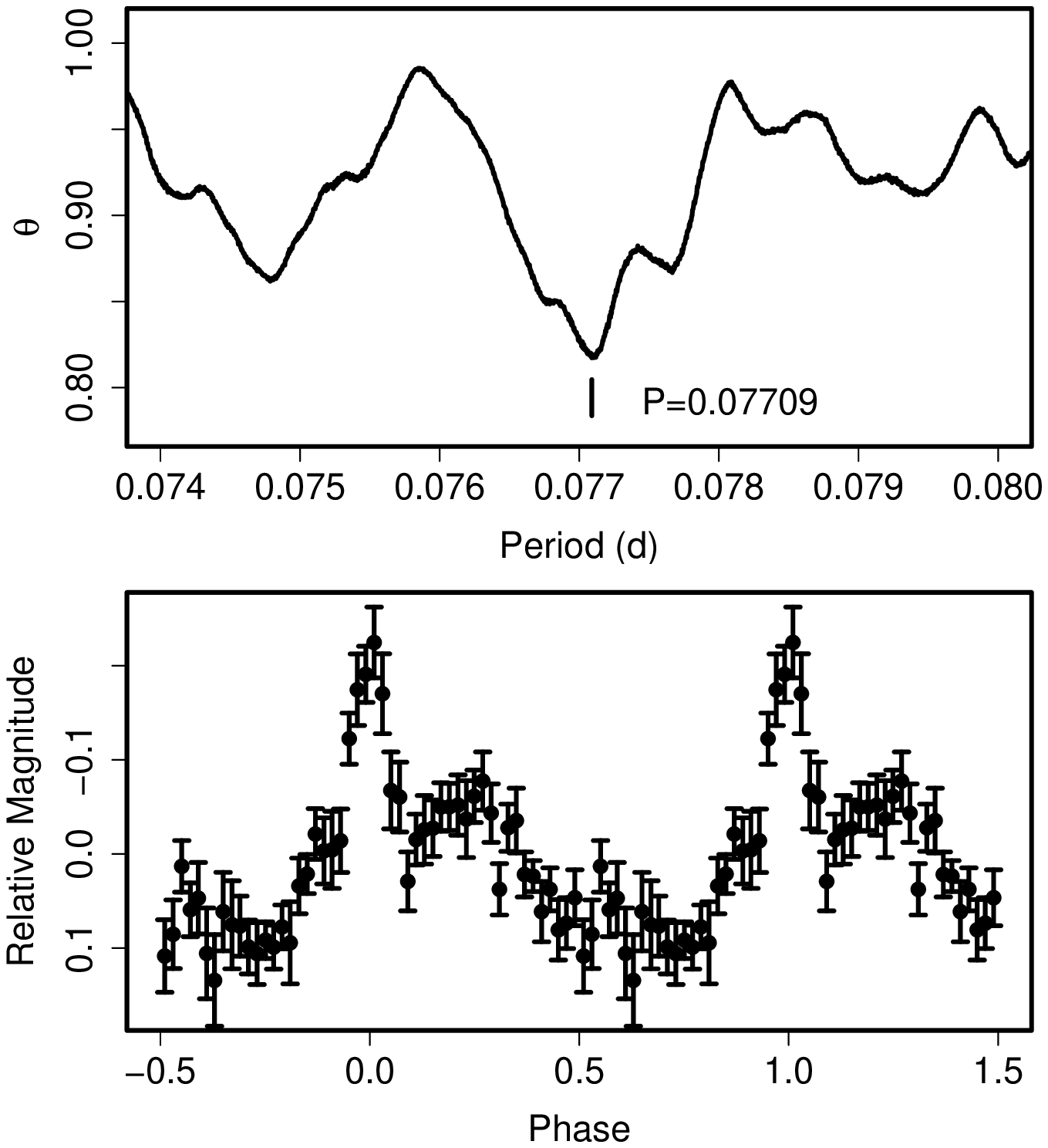}
  \end{center}
  \caption{Superhumps in Ha 0242 (2002). (Upper): PDM analysis.
     (Lower): Phase-averaged profile.}
  \label{fig:ha0242shpdm}
\end{figure}

   Ha 0242$-$2802 (hereafter Ha 0242) was discovered as a CV selected
by H$\alpha$ emission \citep{how02shortPCV}.  \citet{wou04CV4} presented
time-resolved photometry in quiescence and established its eclipsing
nature.  \citet{mas05ha0242} reported phase-resolved spectroscopy.
We observed the 2006 superoutburst and established its SU UMa-type
nature.  The times of superhump maxima, measured from observations
outside the eclipses, are listed in table \ref{tab:ha0242oc2006}.
Due to the overlapping eclipses, it is difficult to clearly determine
the period variation.  The mean $P_{\rm SH}$ with the PDM method
was 0.07709(2) d (figure \ref{fig:ha0242shpdm}), 3.3 \% longer than
the $P_{\rm orb}$ (updated using eclipse timings in \cite{kra06ha0242}).
This $P_{\rm SH}$ was adopted in table \ref{tab:perlist}.

\begin{table}
\caption{Superhump maxima of Ha 0242 (2006).}\label{tab:ha0242oc2006}
\begin{center}
\begin{tabular}{ccccc}
\hline\hline
$E$ & max$^a$ & error & $O-C^b$ & $N^c$ \\
\hline
0 & 53742.3224 & 0.0023 & $-$0.0047 & 102 \\
8 & 53742.9480 & 0.0028 & 0.0048 & 166 \\
9 & 53743.0219 & 0.0016 & 0.0017 & 128 \\
29 & 53744.5569 & 0.0041 & $-$0.0035 & 28 \\
30 & 53744.6442 & 0.0012 & 0.0068 & 47 \\
31 & 53744.7095 & 0.0022 & $-$0.0050 & 26 \\
42 & 53745.5612 & 0.0006 & $-$0.0005 & 53 \\
43 & 53745.6392 & 0.0005 & 0.0006 & 59 \\
\hline
  \multicolumn{5}{l}{$^{a}$ BJD$-$2400000.} \\
  \multicolumn{5}{l}{$^{b}$ Against $max = 2453742.3271 + 0.077013 E$.} \\
  \multicolumn{5}{l}{$^{c}$ Number of points used to determine the maximum.} \\
\end{tabular}
\end{center}
\end{table}

\subsection{SDSSp J013701.06$-$091234.9}\label{obj:j0137}

   This object (hereafter SDSS J0137) was extensively studied by
\citet{ima06j0137}.  We have reanalyzed the data and obtained improved
and newly measured times of superhump maxima (table \ref{tab:j0137oc2003}).
The $P_{\rm dot}$ for $E \le 98$ (before the remarkable period
shortening as described in \cite{ima06j0137})
was $+2.3(1.7) \times 10^{-5}$.

   The 2009 superoutburst was detected during its rising stage
(vsnet-alert 10994).  Only the stage C superhumps were recorded
(table \ref{tab:j0137oc2009}).  The mean superhump period with the
PDM method was 0.056443(8) d.  We adopted this value rather than the one
from the times of maxima because of fragmentary visibility of superhumps
due to the unfavorable seasonal condition.  The relatively low frequency
of superoutburst (once in three to five years) appears to be
confirmed.

\begin{table}
\caption{Superhump maxima of SDSS J0137 (2003--2004).}\label{tab:j0137oc2003}
\begin{center}

\end{center}
\end{table}

\begin{table}
\caption{Superhump maxima of SDSS J0137 (2009).}\label{tab:j0137oc2009}
\begin{center}
%
\end{center}
\end{table}

\subsection{SDSS J031051.66$-$075500.3}\label{obj:j0310}

\begin{figure}
  \begin{center}
    \FigureFile(88mm,110mm){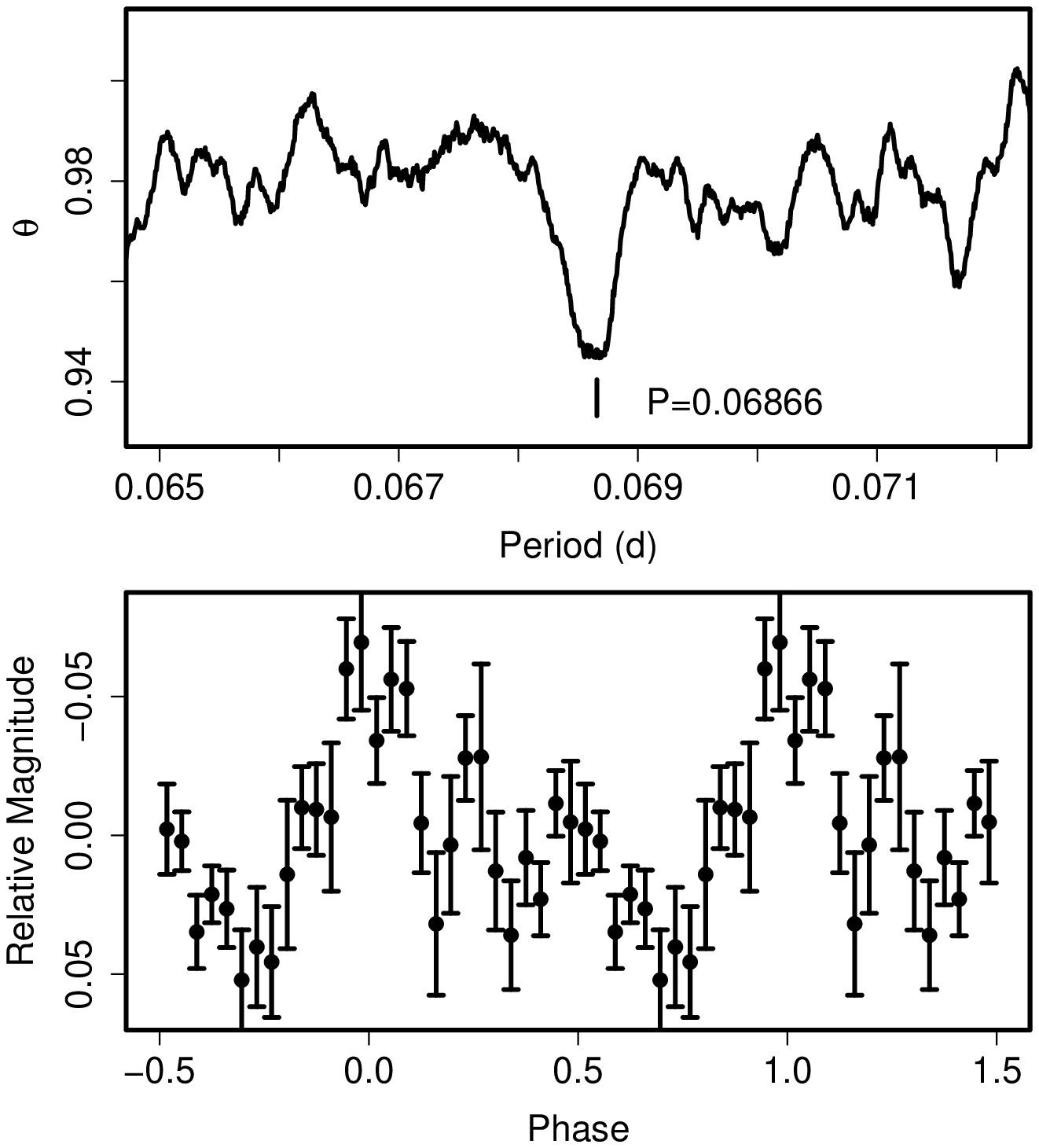}
  \end{center}
  \caption{Superhumps in SDSS J0310 (2004). (Upper): PDM analysis.
     (Lower): Phase-averaged profile.}
  \label{fig:j0310shpdm}
\end{figure}

   This object (hereafter SDSS J0310) is a CV selected
during the course of the SDSS \citep{szk03SDSSCV2}.
B. Monard detected an outburst in 2004 July and reported the presence
of superhumps (vsnet-alert 8236, 8239).  We analyzed the observation
of this superoutburst.  The best superhump period based on the first
three nights was 0.06866(6) d (figure \ref{fig:j0310shpdm}),
supporting the identification by D. Nogami (vsnet-alert 8240).
The times of superhump maxima based in this
period identification are listed in table \ref{tab:j0310oc2004}.
We obtained a global $P_{\rm dot}$ of
$+2.0(2.7) \times 10^{-5}$, which is probably a mixture of different stages
of period evolution.  The object underwent another superoutburst
in 2009 January--February (vsnet-alert 10995).
Further observations are absolutely needed to better qualify the period
evolution.

\begin{table}
\caption{Superhump maxima of SDSS J0310 (2004).}\label{tab:j0310oc2004}
\begin{center}
\begin{tabular}{ccccc}
\hline\hline
$E$ & max$^a$ & error & $O-C^b$ & $N^c$ \\
\hline
0 & 53198.5967 & 0.0009 & 0.0031 & 154 \\
1 & 53198.6602 & 0.0012 & $-$0.0020 & 98 \\
14 & 53199.5511 & 0.0086 & $-$0.0034 & 83 \\
15 & 53199.6279 & 0.0013 & 0.0048 & 155 \\
44 & 53201.6086 & 0.0023 & $-$0.0050 & 154 \\
49 & 53201.9664 & 0.0066 & 0.0097 & 73 \\
78 & 53203.9367 & 0.0019 & $-$0.0105 & 137 \\
160 & 53209.5801 & 0.0116 & 0.0047 & 87 \\
161 & 53209.6427 & 0.0084 & $-$0.0013 & 134 \\
\hline
  \multicolumn{5}{l}{$^{a}$ BJD$-$2400000.} \\
  \multicolumn{5}{l}{$^{b}$ Against $max = 2453198.5936 + 0.068637 E$.} \\
  \multicolumn{5}{l}{$^{c}$ Number of points used to determine the maximum.} \\
\end{tabular}
\end{center}
\end{table}

\subsection{SDSS J033449.86$-$071047.8}\label{obj:j0334}

\begin{figure}
  \begin{center}
    \FigureFile(88mm,110mm){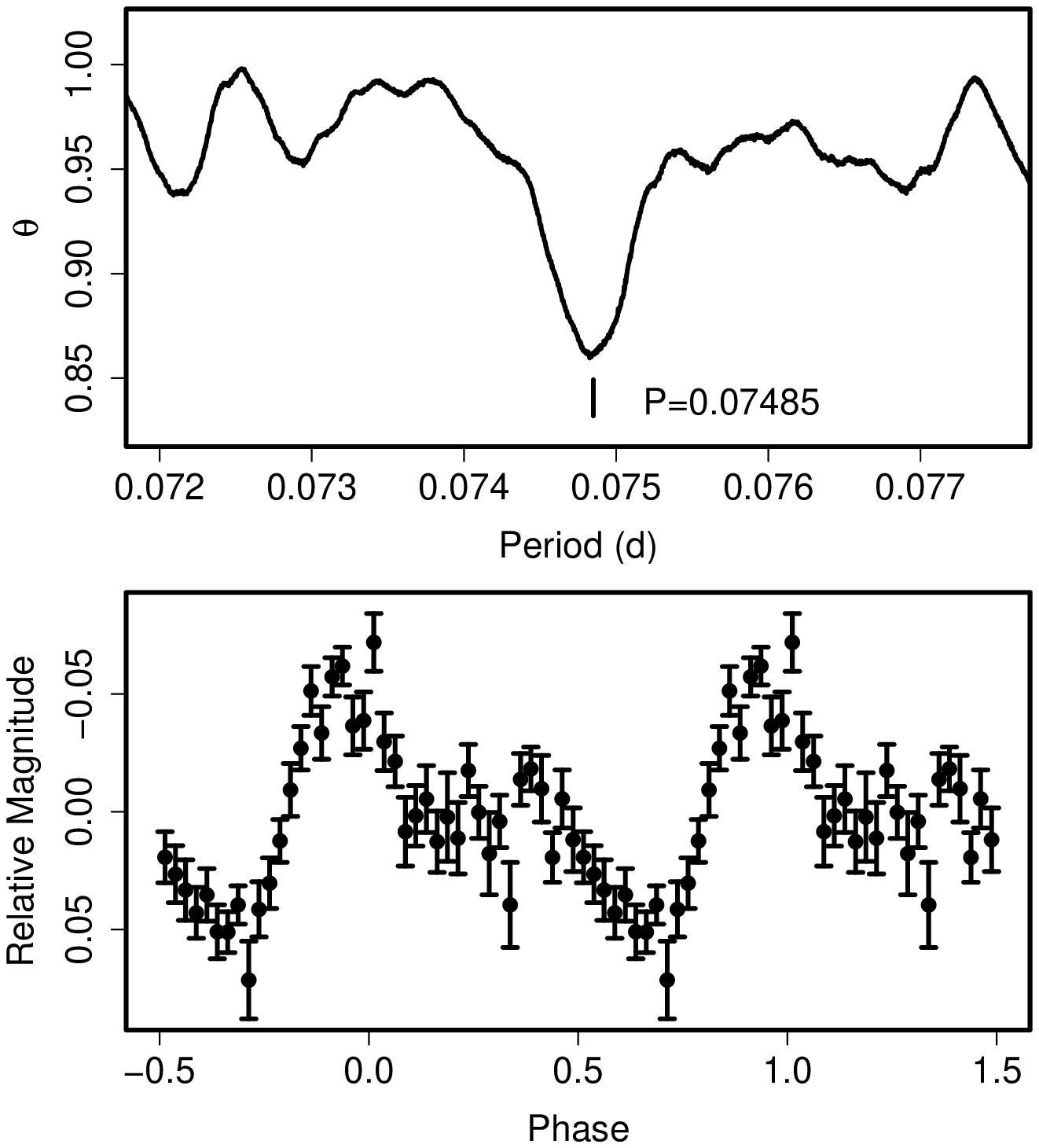}
  \end{center}
  \caption{Superhumps in SDSS J0334 (2009). (Upper): PDM analysis.
     (Lower): Phase-averaged profile.}
  \label{fig:j0334shpdm}
\end{figure}

   SDSS J033449.86$-$071047.8 (hereafter SDSS J0334) is a CV selected
during the course of the SDSS \citep{szk07SDSSCV6}, who reported
the classification as a dwarf nova and an orbital period of 0.079 d.
The 2009 outburst was detected by H. Maehara (vsnet-alert 10967).
The detection of superhumps qualified this object as an SU UMa-type
dwarf nova (vsnet-alert 10973).  The best superhump period determined
from the observations was 0.07485(3) d (figure \ref{fig:j0334shpdm}).
The times of superhump maxima are listed in table \ref{tab:j0334oc2009}.

\begin{table}
\caption{Superhump maxima of SDSS J0334 (2009).}\label{tab:j0334oc2009}
\begin{center}
\begin{tabular}{ccccc}
\hline\hline
$E$ & max$^a$ & error & $O-C^b$ & $N^c$ \\
\hline
0 & 54856.0144 & 0.0009 & $-$0.0041 & 160 \\
1 & 54856.0945 & 0.0024 & 0.0012 & 89 \\
12 & 54856.9133 & 0.0015 & $-$0.0025 & 73 \\
13 & 54856.9924 & 0.0014 & 0.0019 & 207 \\
14 & 54857.0689 & 0.0023 & 0.0036 & 127 \\
39 & 54858.9403 & 0.0014 & 0.0056 & 136 \\
40 & 54859.0062 & 0.0022 & $-$0.0032 & 166 \\
52 & 54859.9067 & 0.0024 & $-$0.0000 & 145 \\
53 & 54859.9821 & 0.0015 & 0.0006 & 189 \\
54 & 54860.0531 & 0.0037 & $-$0.0031 & 113 \\
\hline
  \multicolumn{5}{l}{$^{a}$ BJD$-$2400000.} \\
  \multicolumn{5}{l}{$^{b}$ Against $max = 2454856.0185 + 0.074773 E$.} \\
  \multicolumn{5}{l}{$^{c}$ Number of points used to determine the maximum.} \\
\end{tabular}
\end{center}
\end{table}

\subsection{SDSS J074640.62$+$173412.8}\label{obj:j0746}

\begin{figure}
  \begin{center}
    \FigureFile(88mm,110mm){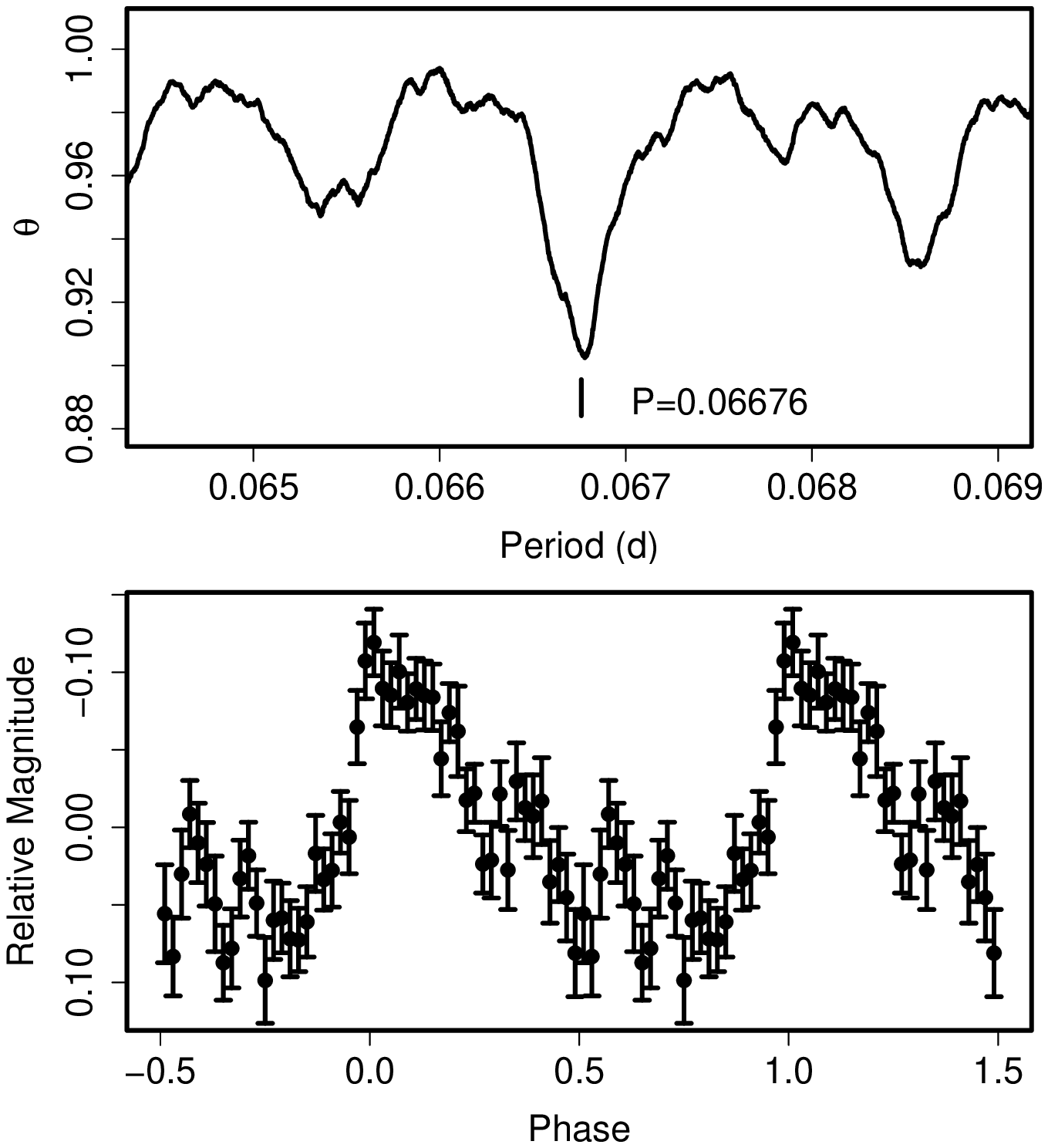}
  \end{center}
  \caption{Superhumps in SDSS J0746 (2009). (Upper): PDM analysis.
     (Lower): Phase-averaged profile.}
  \label{fig:j0746shpdm}
\end{figure}

   SDSS J074640.62$+$173412.8 (hereafter SDSS J0746) is a CV selected
during the course of the SDSS \citep{szk06SDSSCV5}, who suggested the
dwarf nova-type classification based on its variability.
J. Shears reported an outburst of this object in 2009 January
(cvnet-outburst 2949).  The detection of superhumps led to a classification
as an SU UMa-type dwarf nova (vsnet-alert 11069).
The mean superhump period with the PDM method was 0.066761(15) d
(figure \ref{fig:j0746shpdm}).
The times of superhump maxima are listed in table \ref{tab:j0746oc2009}.
There was a stage B--C transition around $E=78$.  Excluding $E=30$,
we obtained $P_{\rm dot}$ = $+9.3(2.5) \times 10^{-5}$, fairly common
for this $P_{\rm SH}$.

\begin{table}
\caption{Superhump maxima of SDSS J0746 (2009).}\label{tab:j0746oc2009}
\begin{center}
\begin{tabular}{ccccc}
\hline\hline
$E$ & max$^a$ & error & $O-C^b$ & $N^c$ \\
\hline
0 & 54874.9304 & 0.0011 & 0.0039 & 49 \\
1 & 54874.9954 & 0.0019 & 0.0021 & 70 \\
2 & 54875.0599 & 0.0015 & $-$0.0001 & 42 \\
30 & 54876.9158 & 0.0063 & $-$0.0127 & 42 \\
31 & 54876.9944 & 0.0026 & $-$0.0009 & 70 \\
32 & 54877.0607 & 0.0017 & $-$0.0013 & 257 \\
33 & 54877.1304 & 0.0015 & 0.0016 & 292 \\
34 & 54877.1928 & 0.0023 & $-$0.0027 & 119 \\
76 & 54880.0014 & 0.0027 & 0.0031 & 232 \\
77 & 54880.0711 & 0.0016 & 0.0061 & 314 \\
78 & 54880.1387 & 0.0020 & 0.0069 & 182 \\
90 & 54880.9367 & 0.0068 & 0.0042 & 64 \\
91 & 54880.9988 & 0.0033 & $-$0.0005 & 56 \\
92 & 54881.0658 & 0.0019 & $-$0.0003 & 42 \\
93 & 54881.1409 & 0.0018 & 0.0081 & 60 \\
94 & 54881.1859 & 0.0044 & $-$0.0136 & 71 \\
138 & 54884.1366 & 0.0039 & 0.0008 & 99 \\
139 & 54884.1979 & 0.0044 & $-$0.0046 & 74 \\
\hline
  \multicolumn{5}{l}{$^{a}$ BJD$-$2400000.} \\
  \multicolumn{5}{l}{$^{b}$ Against $max = 2454874.9265 + 0.066734 E$.} \\
  \multicolumn{5}{l}{$^{c}$ Number of points used to determine the maximum.} \\
\end{tabular}
\end{center}
\end{table}

\subsection{SDSS J081207.63$+$131824.4}\label{obj:j0812}

\begin{figure}
  \begin{center}
    \FigureFile(88mm,110mm){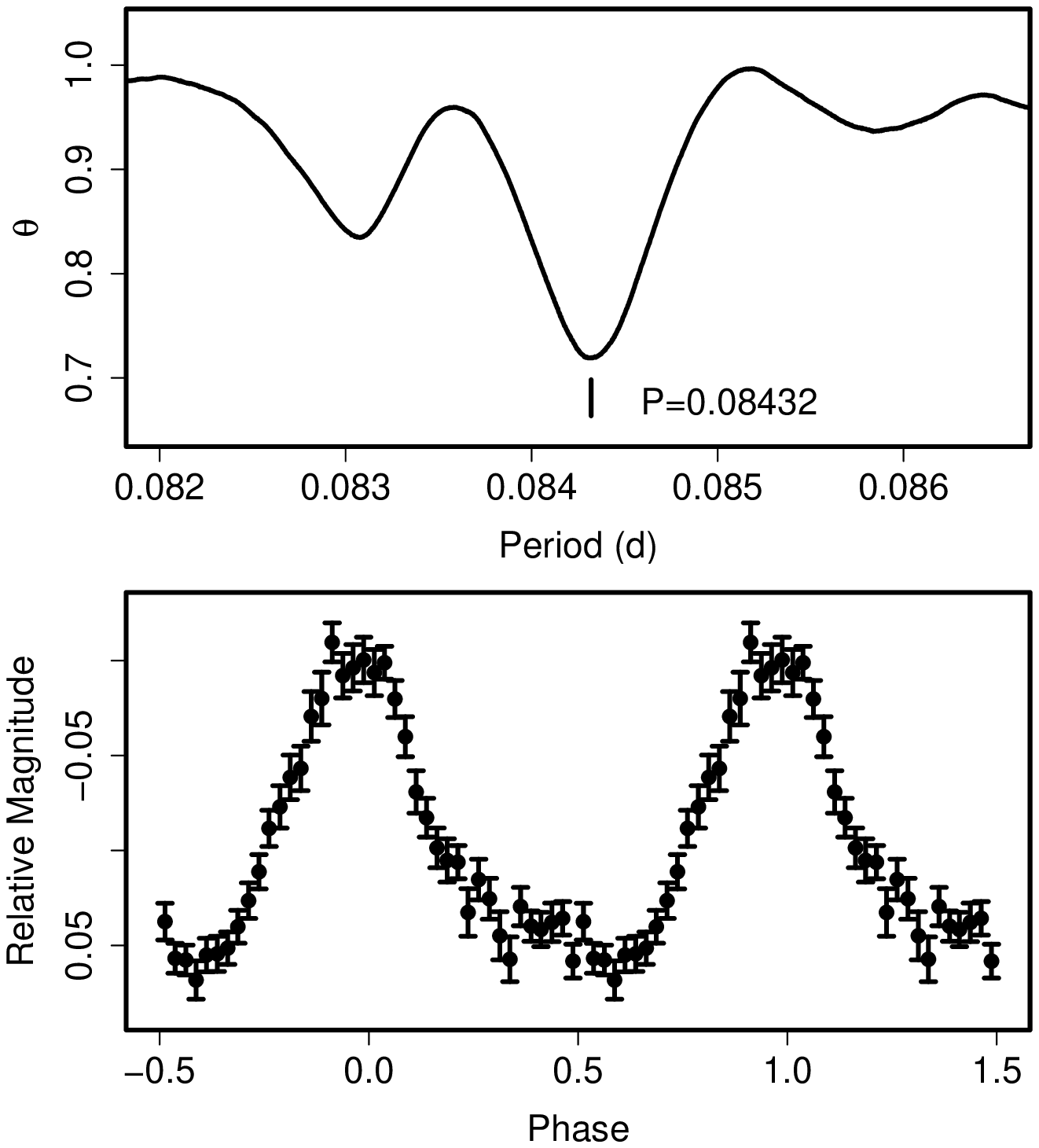}
  \end{center}
  \caption{Superhumps in SDSS J0812 (2008). (Upper): PDM analysis.
     (Lower): Phase-averaged profile.}
  \label{fig:j0812shpdm}
\end{figure}

   SDSS J081207.63$+$131824.4 (hereafter SDSS J0812) is a CV selected
during the course of the SDSS \citep{szk07SDSSCV6}.  The 2008 superoutburst
detected by K. Itagaki \citep{yam08j0812cbet1536} led to the classification
as a long-$P_{\rm orb}$ SU UMa-type dwarf nova.
The mean superhump period with the PDM method was 0.08432(1) d
(figure \ref{fig:j0812shpdm}).  The times of superhump
maxima are listed in table \ref{tab:j0812oc2008}.  We obtained a global
$P_{\rm dot}$ of $-24.0(5.2) \times 10^{-5}$, a value similar to
the one in UV Gem having a similar $P_{\rm SH}$.

\begin{table}
\caption{Superhump maxima of SDSS J0812 (2008).}\label{tab:j0812oc2008}
\begin{center}
\begin{tabular}{ccccc}
\hline\hline
$E$ & max$^a$ & error & $O-C^b$ & $N^c$ \\
\hline
0 & 54751.2800 & 0.0002 & $-$0.0146 & 333 \\
35 & 54754.2365 & 0.0036 & $-$0.0002 & 75 \\
36 & 54754.3234 & 0.0006 & 0.0027 & 154 \\
47 & 54755.2555 & 0.0004 & 0.0101 & 299 \\
48 & 54755.3309 & 0.0008 & 0.0014 & 188 \\
59 & 54756.2661 & 0.0015 & 0.0120 & 216 \\
60 & 54756.3402 & 0.0010 & 0.0020 & 120 \\
71 & 54757.2698 & 0.0008 & 0.0070 & 319 \\
83 & 54758.2727 & 0.0023 & 0.0012 & 88 \\
95 & 54759.2586 & 0.0018 & $-$0.0217 & 103 \\
\hline
  \multicolumn{5}{l}{$^{a}$ BJD$-$2400000.} \\
  \multicolumn{5}{l}{$^{b}$ Against $max = 2454751.2946 + 0.084059 E$.} \\
  \multicolumn{5}{l}{$^{c}$ Number of points used to determine the maximum.} \\
\end{tabular}
\end{center}
\end{table}

\subsection{SDSSp J082409.73$+$493124.4}\label{obj:j0824}

   SDSSp J082409.73$+$493124.4 (hereafter SDSS J0824) is a CV selected
during the course of the SDSS \citep{szk02SDSSCVs} (see \cite{boy08j0824}
for the history of observation).
\citet{boy08j0824} reported the detection of superhumps with a mean
period of 0.06954(5).  \citet{boy08j0824} interpreted an apparent phase
transition in the late course of the superoutburst as being the transition
to late superhumps.  Their data, however, had a gap in the middle stage
of the superoutburst.  Our own observations happened to fill the gap.
We used a combined data set by ours and
from the AAVSO database, the latter including the partial data in
\citet{boy08j0824}.  We used the data common to the AAVSO database
and those in \citet{boy08j0824} to determine the systematic difference
between our measurements and those by \citet{boy08j0824}.

\begin{figure}
  \begin{center}
    \FigureFile(88mm,110mm){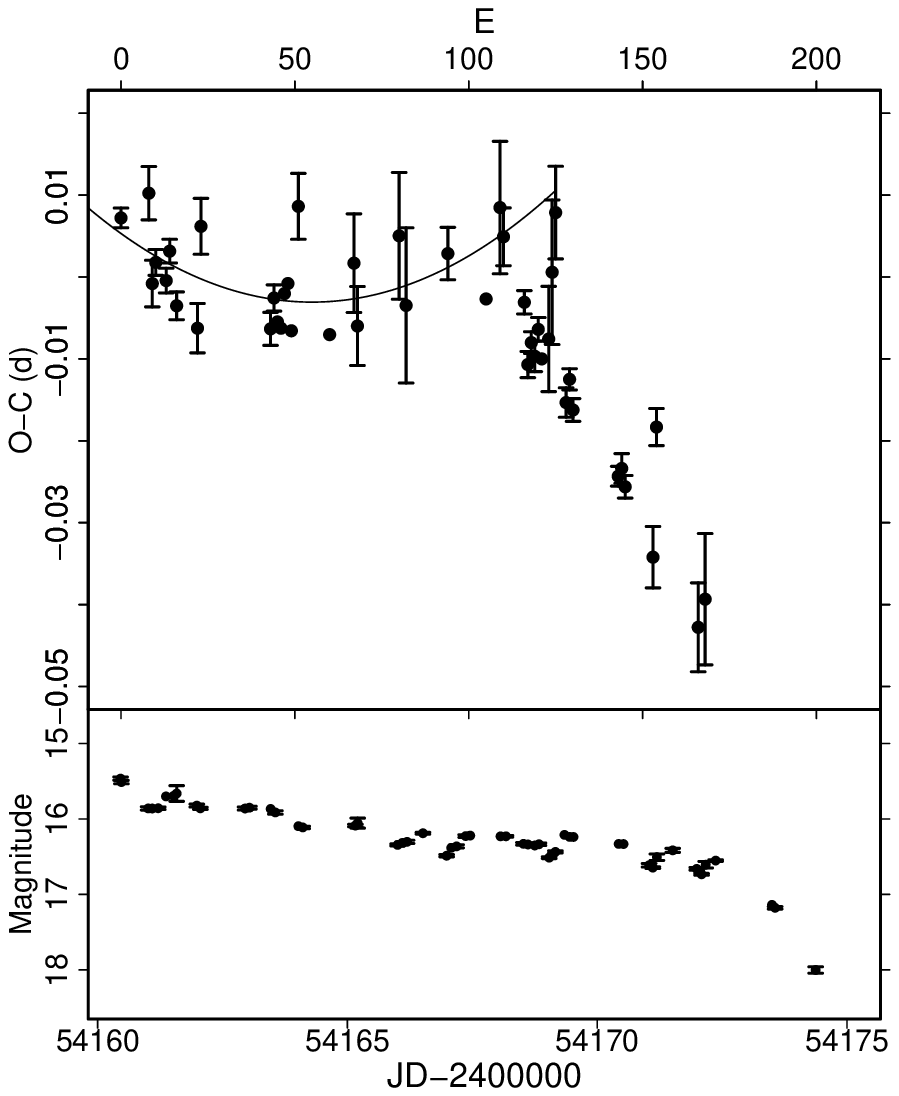}
  \end{center}
  \caption{$O-C$ of superhumps SDSS J0824.
  (Upper): $O-C$ diagram.  The values of $O-C$'s are different from
  those listed in table \ref{tab:j0824oc2007} and were calculated from
  a linear fit for the times of superhumps for $E \le 110$.
  The curve represents a quadratic fit with $P_{\rm dot}$
  = $+8.0 \times 10^{-5}$.
  (Lower): Light curve.}
  \label{fig:j0824reboc}
\end{figure}

   Table \ref{tab:j0824oc2007} lists combined times of superhump maxima,
after adding a systematic difference of 0.0035 d to \citet{boy08j0824}.
The entire data now clearly show a sharp transition from the stage B
with a slightly positive $P_{\rm dot}$ to the stage C after $E = 110$
(figure \ref{fig:j0824reboc}).
The phase discontinuity reported in \citet{boy08j0824} reflected this
period variation rather than a transition to late superhumps.
The $P_{\rm dot}$ of the first segment was $+8.0(2.5) \times 10^{-5}$.

   Another likely superoutburst was observed in 2007 December (J. Shears,
baavss-alert 1492), giving a supercycle length of $\sim$ 300 d.

\begin{table}
\caption{Superhump maxima of SDSS J0824 (2007).}\label{tab:j0824oc2007}
\begin{center}

\end{center}
\end{table}

\subsection{SDSSp J083845.23$+$491055.5}\label{obj:j0838}

   SDSSp J083845.23$+$491055.5 (hereafter SDSS J0838) was discovered
as a CV having a typical spectrum of a dwarf novae \citep{szk02SDSSCVs}.
We analyzed the AAVSO data during the 2007 October superoutburst
(cf. baavss-alert 1383, 1386).
The times of superhump maxima are listed in table \ref{tab:j0838oc2007}.

\begin{figure}
  \begin{center}
    \FigureFile(88mm,110mm){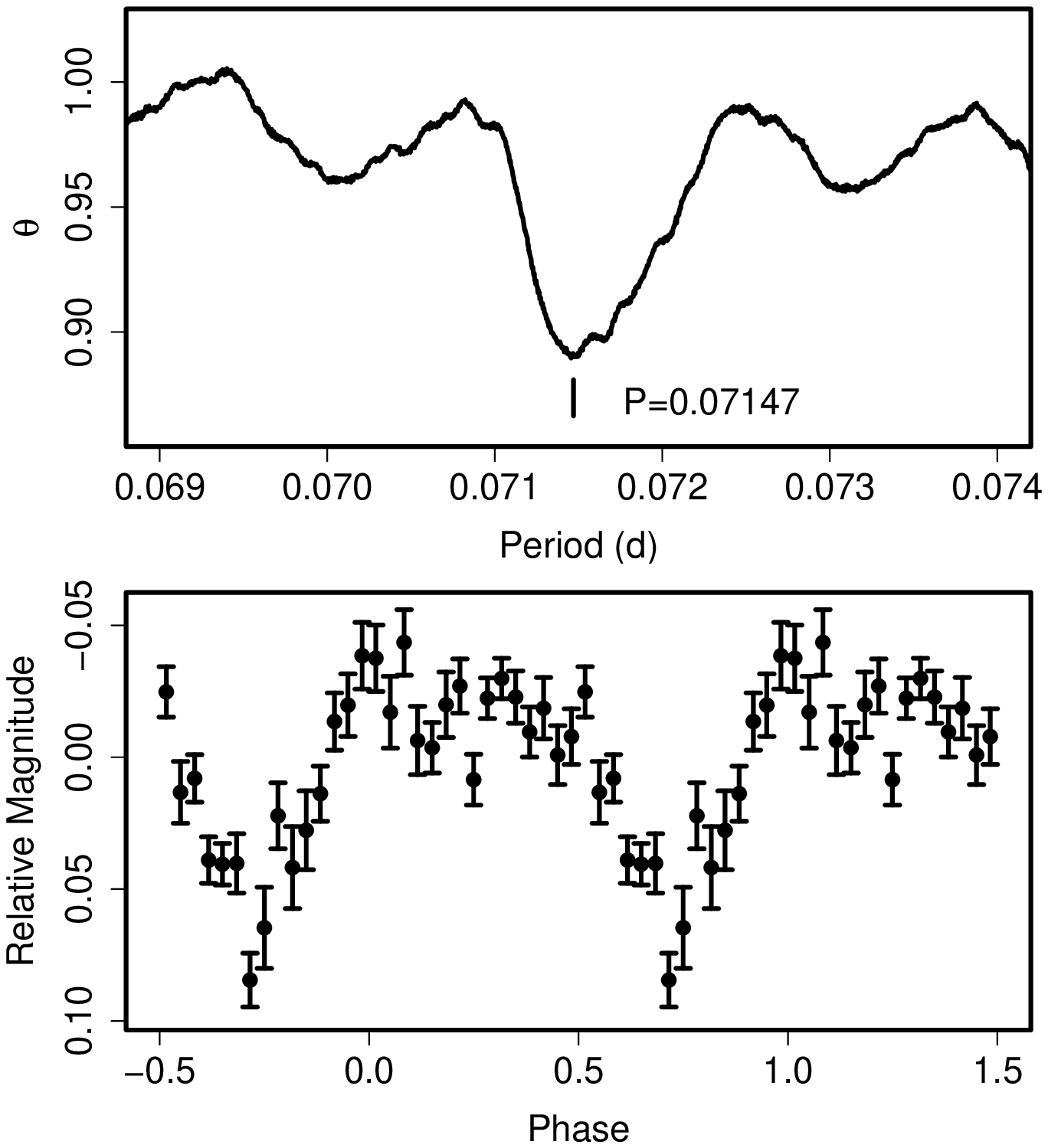}
  \end{center}
  \caption{Superhumps in SDSS J0838 (2009, late stage). (Upper): PDM analysis.
     (Lower): Phase-averaged profile.}
  \label{fig:j0838shpdm}
\end{figure}

   The object underwent another superoutburst in 2009 (baavss-alert
1944).  The observation of this superoutburst finally led to an
identification of the superhump period (vsnet-alert 11099).
The times of superhump maxima are listed in table \ref{tab:j0838oc2009}.
Although the identification of the cycle number between $E=2$ (stage B)
and $E=101$ was rather uncertain, the stage C superhumps with a mean
period of 0.07147(2) d (PDM method, figure \ref{fig:j0838shpdm}) were
eventually identified.

\begin{table}
\caption{Superhump maxima of SDSS J0838 (2007).}\label{tab:j0838oc2007}
\begin{center}
\begin{tabular}{ccccc}
\hline\hline
$E$ & max$^a$ & error & $O-C^b$ & $N^c$ \\
\hline
0 & 54396.5730 & 0.0008 & 0.0010 & 98 \\
1 & 54396.6432 & 0.0007 & $-$0.0020 & 101 \\
2 & 54396.7194 & 0.0008 & 0.0010 & 57 \\
\hline
  \multicolumn{5}{l}{$^{a}$ BJD$-$2400000.} \\
  \multicolumn{5}{l}{$^{b}$ Against $max = 2454396.5720 + 0.07316 E$.} \\
  \multicolumn{5}{l}{$^{c}$ Number of points used to determine the maximum.} \\
\end{tabular}
\end{center}
\end{table}

\begin{table}
\caption{Superhump maxima of SDSS J0838 (2009).}\label{tab:j0838oc2009}
\begin{center}
\begin{tabular}{ccccc}
\hline\hline
$E$ & max$^a$ & error & $O-C^b$ & $N^c$ \\
\hline
0 & 54884.1135 & 0.0002 & $-$0.0010 & 140 \\
1 & 54884.1829 & 0.0005 & $-$0.0034 & 69 \\
2 & 54884.2575 & 0.0003 & $-$0.0005 & 131 \\
101 & 54891.3654 & 0.0010 & 0.0059 & 51 \\
102 & 54891.4367 & 0.0009 & 0.0055 & 52 \\
103 & 54891.5065 & 0.0010 & 0.0036 & 49 \\
111 & 54892.0825 & 0.0010 & 0.0057 & 140 \\
112 & 54892.1485 & 0.0010 & $-$0.0000 & 151 \\
123 & 54892.9322 & 0.0045 & $-$0.0054 & 90 \\
129 & 54893.3651 & 0.0027 & $-$0.0029 & 51 \\
131 & 54893.5091 & 0.0022 & $-$0.0024 & 53 \\
152 & 54895.0228 & 0.0031 & 0.0049 & 126 \\
153 & 54895.0841 & 0.0038 & $-$0.0055 & 137 \\
154 & 54895.1732 & 0.0040 & 0.0118 & 146 \\
155 & 54895.2169 & 0.0062 & $-$0.0162 & 73 \\
\hline
  \multicolumn{5}{l}{$^{a}$ BJD$-$2400000.} \\
  \multicolumn{5}{l}{$^{b}$ Against $max = 2454884.1145 + 0.071732 E$.} \\
  \multicolumn{5}{l}{$^{c}$ Number of points used to determine the maximum.} \\
\end{tabular}
\end{center}
\end{table}

\subsection{SDSS J100515.39$+$191108.0}\label{obj:j1005}

\begin{figure}
  \begin{center}
    \FigureFile(88mm,110mm){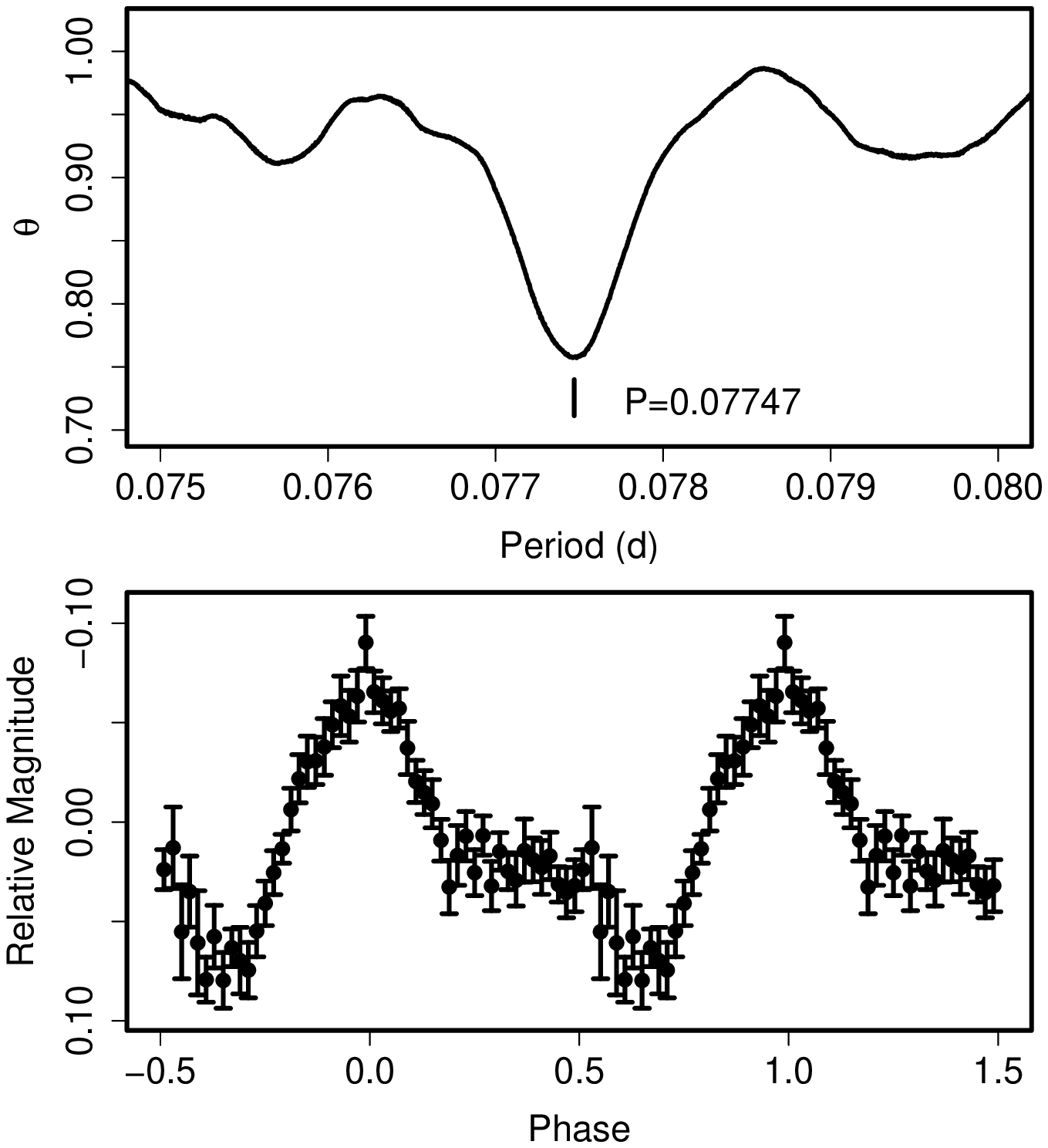}
  \end{center}
  \caption{Superhumps in SDSS J1005 (2009). (Upper): PDM analysis.
     (Lower): Phase-averaged profile.}
  \label{fig:j1005shpdm}
\end{figure}

   The 2009 outburst of this object (hereafter SDSS J1005) was reported
S. Brady (cvnet-outburst 2859), which later turned out to be a
superoutburst.  We analyzed the available data of this outburst during
its late stage and obtained a mean superhump period of 0.07747(2) d
with the PDM method (figure \ref{fig:j1005shpdm}).
The times of superhump maxima are listed in table \ref{tab:j1005oc2009}.
These superhumps most likely correspond to stage C superhumps.
J. Pietz reported a period of 0.0779 d (cvnet-outburst 2866).

\begin{table}
\caption{Superhump maxima of SDSS J1005 (2009).}\label{tab:j1005oc2009}
\begin{center}
\begin{tabular}{ccccc}
\hline\hline
$E$ & max$^a$ & error & $O-C^b$ & $N^c$ \\
\hline
0 & 54838.5679 & 0.0004 & $-$0.0016 & 106 \\
1 & 54838.6471 & 0.0005 & 0.0002 & 140 \\
10 & 54839.3442 & 0.0003 & 0.0007 & 166 \\
16 & 54839.8075 & 0.0008 & $-$0.0005 & 51 \\
17 & 54839.8855 & 0.0004 & 0.0001 & 80 \\
18 & 54839.9622 & 0.0006 & $-$0.0006 & 40 \\
36 & 54841.3606 & 0.0006 & 0.0046 & 112 \\
41 & 54841.7439 & 0.0018 & 0.0009 & 54 \\
42 & 54841.8201 & 0.0007 & $-$0.0003 & 81 \\
43 & 54841.8990 & 0.0007 & 0.0011 & 67 \\
46 & 54842.1303 & 0.0054 & 0.0003 & 56 \\
47 & 54842.2034 & 0.0014 & $-$0.0041 & 142 \\
48 & 54842.2891 & 0.0025 & 0.0043 & 149 \\
49 & 54842.3570 & 0.0025 & $-$0.0052 & 115 \\
\hline
  \multicolumn{5}{l}{$^{a}$ BJD$-$2400000.} \\
  \multicolumn{5}{l}{$^{b}$ Against $max = 2454838.5695 + 0.077404 E$.} \\
  \multicolumn{5}{l}{$^{c}$ Number of points used to determine the maximum.} \\
\end{tabular}
\end{center}
\end{table}

\subsection{SDSS J110014.72$+$131552.1}\label{obj:j1100}

   SDSS J110014.72$+$131552.1 (hereafter SDSS J1100) was selected as
a CV during the course of the SDSS \citep{szk06SDSSCV5}.
During the 2009 outburst detected by the CRTS (vsnet-alert 11188),
superhumps were detected (vsnet-alert 11198, 11202).
Although the short baseline of the observations makes alias selection
slightly ambiguous, we present the $O-C$'s based on the period of
0.06757(2) d (PDM method).

\begin{figure}
  \begin{center}
    \FigureFile(88mm,110mm){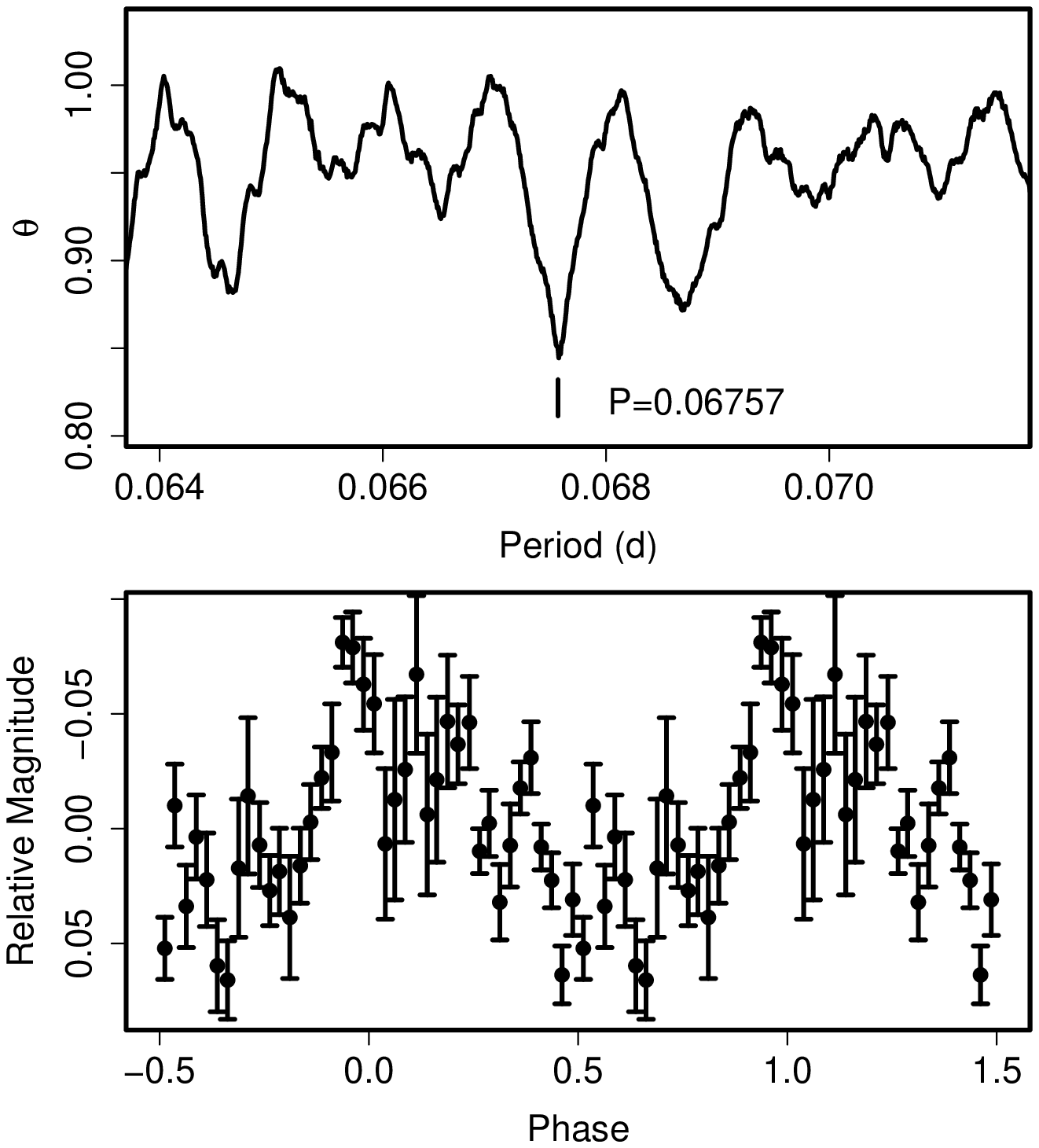}
  \end{center}
  \caption{Superhumps in SDSS J1100 (2009). (Upper): PDM analysis.
     (Lower): Phase-averaged profile.}
  \label{fig:j1100shpdm}
\end{figure}

\begin{table}
\caption{Superhump maxima of SDSS J1100 (2009).}\label{tab:j1100oc2009}
\begin{center}
\begin{tabular}{ccccc}
\hline\hline
$E$ & max$^a$ & error & $O-C^b$ & $N^c$ \\
\hline
0 & 54940.1283 & 0.0015 & $-$0.0020 & 122 \\
15 & 54941.1459 & 0.0045 & 0.0026 & 71 \\
63 & 54944.3870 & 0.0014 & 0.0024 & 108 \\
64 & 54944.4492 & 0.0027 & $-$0.0030 & 61 \\
\hline
  \multicolumn{5}{l}{$^{a}$ BJD$-$2400000.} \\
  \multicolumn{5}{l}{$^{b}$ Against $max = 2454940.1303 + 0.067529 E$.} \\
  \multicolumn{5}{l}{$^{c}$ Number of points used to determine the maximum.} \\
\end{tabular}
\end{center}
\end{table}

\subsection{SDSS J122740.83$+$513925.0}\label{obj:j1227}

   SDSS J122740.83$+$513925.0 (hereafter SDSS J1227) was selected as
a high-inclination CV during the course of the SDSS \citep{szk04SDSSCV3}.
\citet{lit08eclCV} reported parameters of eclipses.  \citet{she08j1227}
reported the detection of superhumps and discussed on the variation of
eclipses during the 2007 superoutburst.  Using the times of eclipses
published in \citet{lit08eclCV} and \citet{she08j1227}, we obtained
the following updated ephemeris (equation \ref{equ:j1227ecl}).

\begin{equation}
{\rm Min(BJD)} = 2453796.2478(4) + 0.06295835(5) E
\label{equ:j1227ecl}.
\end{equation}

   We analyzed the combined data set with ours, AAVSO data, and data
extracted from figures in \citet{she08j1227} which were not included
in ours nor in the AAVSO data.
The times of superhump maxima are listed in
table \ref{tab:j1227oc2007}.  The first night of the observation
either corresponded to the stage A or the complex profile disturbed
the $O-C$'s.  The period appears almost constant for the interval
$33 \le E \le 124$, with a mean $P_{\rm SH}$ of 0.064552(21) d
and $P_{\rm dot}$ = $+2.8(2.5) \times 10^{-5}$.  This $P_{\rm dot}$
appears rather unusual for this $P_{\rm SH}$.  The positive $O-C$'s for
$126 \le E \le 129$ may reflect the terminal stage of the stage B,
when the $P_{\rm SH}$ usually lengthens.  This identification seems
to be supported by the apparent increase of the amplitudes of
superhumps at this epoch.  Using the entire interval for
$33 \le E \le 129$, we obtained a mean $P_{\rm SH}$ of 0.064593(22) d
and $P_{\rm dot}$ = $+6.1(2.1) \times 10^{-5}$.  We adopted these
values in table \ref{tab:perlist}.
The fractional superhump excesses for these periods are
2.5 \% and 2.6 \%, respectively.

\begin{table}
\caption{Superhump maxima of SDSS J1227 (2007).}\label{tab:j1227oc2007}
\begin{center}
\begin{tabular}{ccccc}
\hline\hline
$E$ & max$^a$ & error & $O-C^b$ & $N^c$ \\
\hline
0 & 54256.4358 & 0.0012 & $-$0.0237 & -- \\
1 & 54256.5064 & 0.0021 & $-$0.0178 & -- \\
33 & 54258.6073 & 0.0004 & 0.0114 & 61 \\
34 & 54258.6739 & 0.0003 & 0.0133 & 60 \\
35 & 54258.7354 & 0.0008 & 0.0100 & 58 \\
40 & 54259.0611 & 0.0010 & 0.0120 & 104 \\
61 & 54260.4153 & 0.0032 & 0.0066 & 57 \\
62 & 54260.4762 & 0.0006 & 0.0028 & 37 \\
63 & 54260.5428 & 0.0004 & 0.0047 & 100 \\
77 & 54261.4417 & 0.0012 & $-$0.0028 & 103 \\
78 & 54261.5105 & 0.0007 & 0.0012 & 98 \\
80 & 54261.6421 & 0.0008 & 0.0034 & 31 \\
81 & 54261.7047 & 0.0010 & 0.0012 & 33 \\
82 & 54261.7706 & 0.0009 & 0.0024 & 35 \\
86 & 54262.0316 & 0.0009 & 0.0044 & 49 \\
101 & 54262.9979 & 0.0011 & $-$0.0004 & 81 \\
102 & 54263.0632 & 0.0014 & 0.0002 & 114 \\
111 & 54263.6429 & 0.0016 & $-$0.0028 & 34 \\
112 & 54263.7045 & 0.0016 & $-$0.0059 & 35 \\
113 & 54263.7691 & 0.0024 & $-$0.0060 & 35 \\
114 & 54263.8314 & 0.0034 & $-$0.0085 & 35 \\
123 & 54264.4155 & 0.0026 & $-$0.0070 & 74 \\
124 & 54264.4887 & 0.0006 & 0.0014 & 68 \\
126 & 54264.6177 & 0.0009 & 0.0009 & 41 \\
127 & 54264.6816 & 0.0012 & 0.0001 & 39 \\
128 & 54264.7449 & 0.0010 & $-$0.0014 & 41 \\
129 & 54264.8114 & 0.0026 & 0.0004 & 40 \\
\hline
  \multicolumn{5}{l}{$^{a}$ BJD$-$2400000.} \\
  \multicolumn{5}{l}{$^{b}$ Against $max = 2454256.4595 + 0.064740 E$.} \\
  \multicolumn{5}{l}{$^{c}$ Number of points used to determine the maximum.} \\
\end{tabular}
\end{center}
\end{table}

\subsection{SDSS J152419.33$+$220920.0}\label{obj:j1524}

   SDSS J152419.33$+$220920.0 (hereafter SDSS J1524) was suggested to be
a high-inclination CV during the course of the SDSS \citep{szk09SDSSCV7}.
The 2009 outburst of this object was detected by the CRTS (vsnet-alert 11133).
Subsequent observations established the presence of superhumps and
eclipses (cvnet-outburst 3029).

   The times of eclipse minima, measured outside the eclipses as in
V2051 Oph, are listed in table \ref{tab:j1524ecl}.
The times for $E < 0$ were from the CRTS chance detections of eclipses.
The times of these epochs have typical uncertainties of 0.001--0.002 d
(approximately half duration of the eclipse).  The epochs for $E \ge 0$
were determined from time-resolved CCD observations; the typical uncertainty
of the determination is $\sim$ 0.001 d or less.  The resultant orbital
ephemeris is given in equation \ref{equ:j1524ecl}.

\begin{figure}
  \begin{center}
    \FigureFile(88mm,110mm){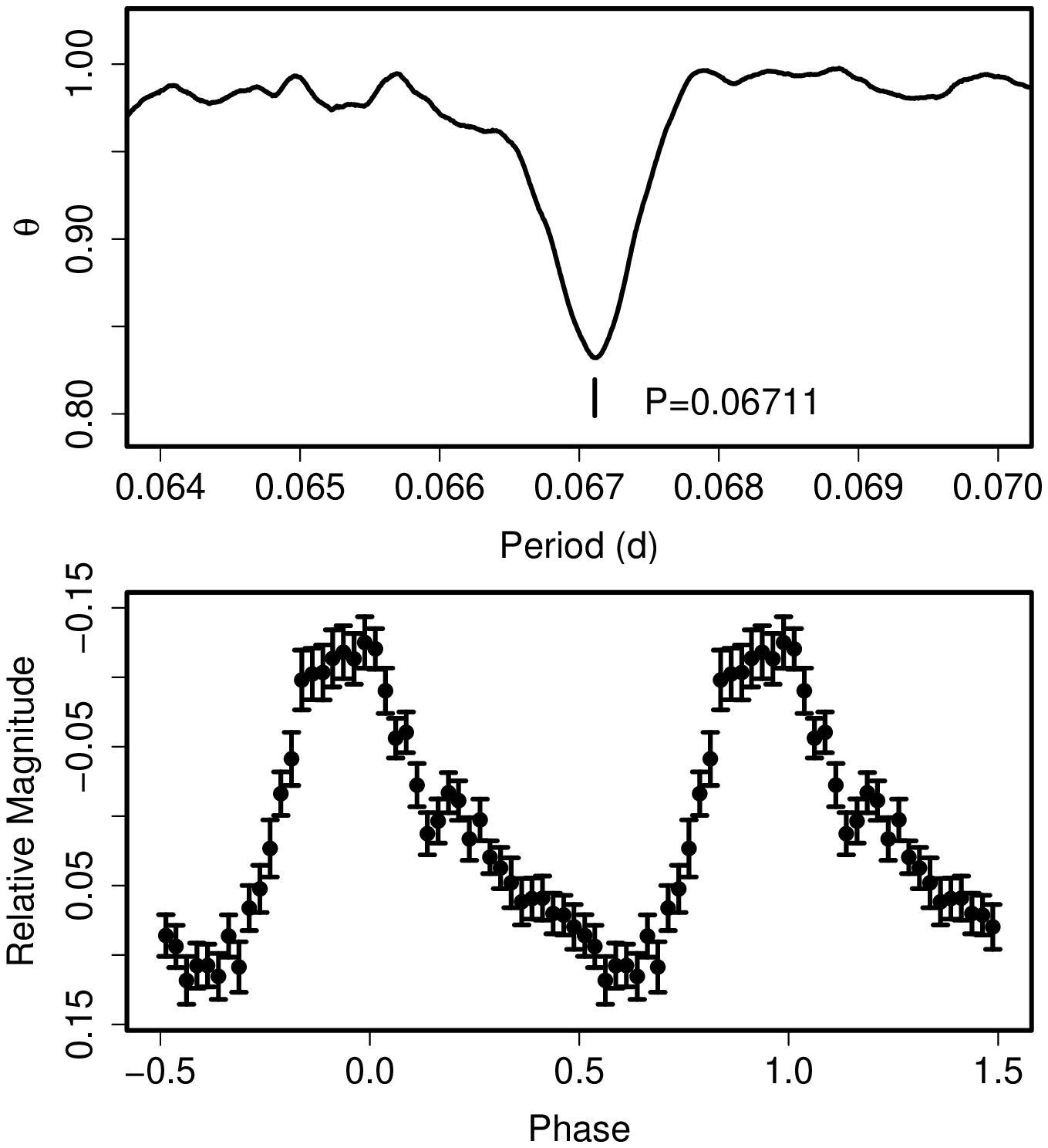}
  \end{center}
  \caption{Superhumps in SDSS J1524 (2009) before BJD 2454928.
     (Upper): PDM analysis.
     (Lower): Phase-averaged profile.}
  \label{fig:j1524shpdm}
\end{figure}

   The times of superhump maxima are listed in table \ref{tab:j1524oc2009}.
A stage B--C transition around $E=89$ was clearly detected.
The mean $P_{\rm SH}$ and $P_{\rm dot}$ during the stage B were
0.067111(14) d (PDM method, figure \ref{fig:j1524shpdm})
and $+8.2(2.6) \times 10^{-5}$, respectively.
The fractional superhump excess for $P_1$ was 2.7 \%.

\begin{equation}
{\rm Min(BJD)} = 2454921.5937(1) + 0.0653187(1)
\label{equ:j1524ecl}.
\end{equation}

\begin{table}
\caption{Eclipse Minima of SDSS J1524.}\label{tab:j1524ecl}
\begin{center}

\end{center}
\end{table}

\begin{table}
\caption{Superhump maxima of SDSS J1524 (2009).}\label{tab:j1524oc2009}
\begin{center}
%
\end{center}
\end{table}

\subsection{SDSS J155644.24$-$000950.2}\label{obj:j1556}

   SDSS J155644.24$-$000950.2 (hereafter SDSS J1556) was selected as
a dwarf nova during the course of the SDSS \citep{szk02SDSSCVs}.
\citet{wou04CV4} obtained 0.07408(1) d from quiescent orbital humps.
During the 2006 March outburst, H. Maehara reported the detection
of superhumps (vsnet-alert 9440).

\begin{figure}
  \begin{center}
    \FigureFile(88mm,110mm){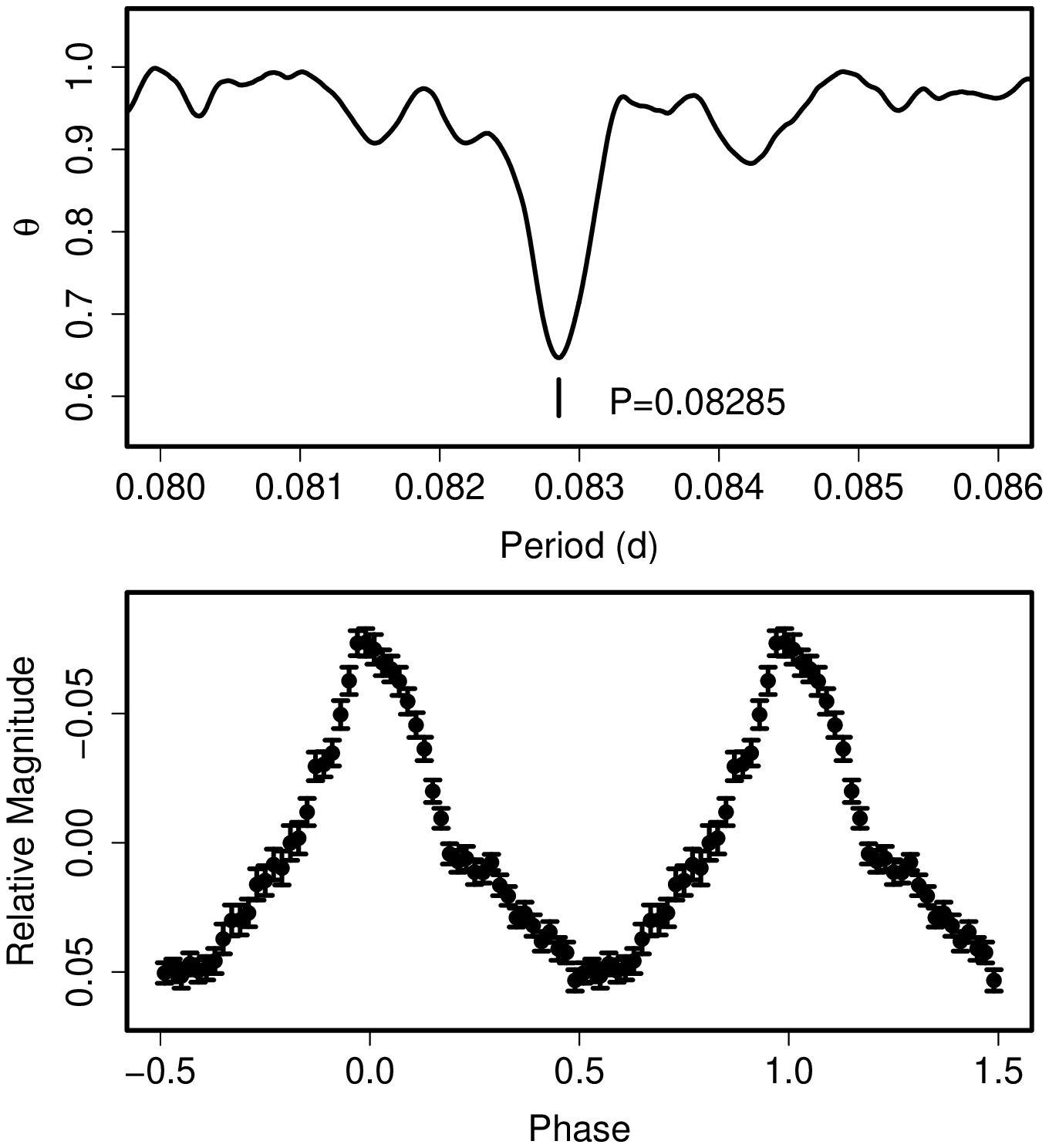}
  \end{center}
  \caption{Superhumps in SDSS J1556 (2007). (Upper): PDM analysis.
     (Lower): Phase-averaged profile.}
  \label{fig:j1556shpdm}
\end{figure}

   We observed the 2007 superoutburst.  A PDM analysis yielded a mean
superhump period of 0.082853(5) d (figure \ref{fig:j1556shpdm},
which corresponds to the longer one-day alias of \citet{wou04CV4}.
Both PDM analysis and superhump $O-C$ analyses
supported this alias selection.  Using the one-day alias period 0.08001(1)
calculated from \citet{wou04CV4}, we obtained a reasonable fractional
superhump excess of 3.6 \%.

   The times of superhump maxima are listed in table \ref{tab:j1556oc2007}.
The $O-C$ diagram (figure \ref{fig:ocsamp}) showed a strong decrease
in the superhump period.
The global $P_{\rm dot}$ was $-8.7(1.1) \times 10^{-5}$,
and was $-6.9(0.8) \times 10^{-5}$ excluding the initial stage of
development (stage A, $E \le 1$).
We consider the latter value as being the representative
period derivative.

   Details of these and other observations and discussion will be
presented in Maehara et al., in preparation.

\begin{table}
\caption{Superhump maxima of SDSS J1556 (2007).}\label{tab:j1556oc2007}
\begin{center}
\begin{tabular}{ccccc}
\hline\hline
$E$ & max$^a$ & error & $O-C^b$ & $N^c$ \\
\hline
0 & 54311.9945 & 0.0003 & $-$0.0114 & 204 \\
1 & 54312.0754 & 0.0005 & $-$0.0134 & 245 \\
12 & 54313.0013 & 0.0002 & 0.0010 & 256 \\
13 & 54313.0836 & 0.0003 & 0.0004 & 214 \\
24 & 54313.9997 & 0.0002 & 0.0050 & 234 \\
25 & 54314.0794 & 0.0004 & 0.0018 & 182 \\
61 & 54317.0704 & 0.0015 & 0.0096 & 54 \\
72 & 54317.9804 & 0.0005 & 0.0080 & 359 \\
73 & 54318.0629 & 0.0004 & 0.0077 & 311 \\
84 & 54318.9733 & 0.0002 & 0.0066 & 428 \\
85 & 54319.0570 & 0.0005 & 0.0074 & 360 \\
101 & 54320.3751 & 0.0004 & $-$0.0003 & 83 \\
121 & 54322.0285 & 0.0008 & $-$0.0043 & 304 \\
133 & 54323.0182 & 0.0008 & $-$0.0090 & 365 \\
145 & 54324.0123 & 0.0005 & $-$0.0093 & 378 \\
\hline
  \multicolumn{5}{l}{$^{a}$ BJD$-$2400000.} \\
  \multicolumn{5}{l}{$^{b}$ Against $max = 2454312.0059 + 0.082866 E$.} \\
  \multicolumn{5}{l}{$^{c}$ Number of points used to determine the maximum.} \\
\end{tabular}
\end{center}
\end{table}

\subsection{SDSS J162718.39$+$120435.0}\label{obj:j1627}

   The 2008 outburst of SDSS J162718.39$+$120435.0 (hereafter SDSS J1627) 
was detected by S. Brady (cvnet-outburst 2421), which was subsequently
proven to be a superoutburst (cvnet-outburst 2426).  The observations
presented here are a combination of \citet{she08j1627} and the
VSNET Collaboration.  The times of superhump maxima
(table \ref{tab:j1627oc2008}) indicated a long $P_{\rm SH}$ with
a strong global period variation of
$P_{\rm dot}$ = $-20.0(2.5) \times 10^{-5}$.
The $O-C$ diagram (figure \ref{fig:lp2})
was clearly composed of all stages A--C.  The abrupt
period change between stages B and C was also noted in \citet{she08j1627}.
The periods of each segments are listed in table \ref{tab:perlist}.

\begin{table}
\caption{Superhump maxima of SDSS J1627.}\label{tab:j1627oc2008}
\begin{center}
\begin{tabular}{cccc}
\hline\hline
$E$ & max$^a$ & error & $O-C^b$ \\
\hline
0 & 54617.7385 & 0.0016 & $-$0.0659 \\
2 & 54617.9384 & 0.0025 & $-$0.0843 \\
6 & 54618.4293 & 0.0047 & $-$0.0303 \\
7 & 54618.5422 & 0.0017 & $-$0.0266 \\
8 & 54618.6593 & 0.0011 & $-$0.0186 \\
9 & 54618.7619 & 0.0007 & $-$0.0252 \\
10 & 54618.8736 & 0.0004 & $-$0.0227 \\
15 & 54619.4453 & 0.0007 & 0.0029 \\
16 & 54619.5581 & 0.0006 & 0.0065 \\
17 & 54619.6670 & 0.0003 & 0.0063 \\
18 & 54619.7794 & 0.0003 & 0.0095 \\
19 & 54619.8891 & 0.0002 & 0.0100 \\
26 & 54620.6590 & 0.0002 & 0.0155 \\
27 & 54620.7679 & 0.0002 & 0.0151 \\
28 & 54620.8776 & 0.0003 & 0.0156 \\
33 & 54621.4239 & 0.0004 & 0.0159 \\
33 & 54621.4241 & 0.0004 & 0.0161 \\
34 & 54621.5389 & 0.0004 & 0.0218 \\
35 & 54621.6439 & 0.0011 & 0.0175 \\
36 & 54621.7584 & 0.0004 & 0.0228 \\
37 & 54621.8668 & 0.0003 & 0.0220 \\
38 & 54621.9815 & 0.0004 & 0.0275 \\
49 & 54623.1727 & 0.0005 & 0.0175 \\
50 & 54623.2884 & 0.0012 & 0.0239 \\
52 & 54623.5010 & 0.0005 & 0.0182 \\
54 & 54623.7199 & 0.0003 & 0.0187 \\
55 & 54623.8292 & 0.0003 & 0.0188 \\
56 & 54623.9345 & 0.0004 & 0.0149 \\
60 & 54624.3742 & 0.0006 & 0.0178 \\
71 & 54625.5686 & 0.0005 & 0.0110 \\
79 & 54626.4409 & 0.0005 & 0.0096 \\
80 & 54626.5494 & 0.0005 & 0.0090 \\
86 & 54627.1992 & 0.0008 & 0.0035 \\
98 & 54628.5042 & 0.0006 & $-$0.0019 \\
109 & 54629.7014 & 0.0005 & $-$0.0059 \\
110 & 54629.8139 & 0.0007 & $-$0.0026 \\
111 & 54629.9205 & 0.0010 & $-$0.0052 \\
117 & 54630.5779 & 0.0009 & $-$0.0030 \\
118 & 54630.6836 & 0.0013 & $-$0.0065 \\
119 & 54630.7940 & 0.0012 & $-$0.0054 \\
120 & 54630.9006 & 0.0010 & $-$0.0080 \\
127 & 54631.6609 & 0.0008 & $-$0.0121 \\
128 & 54631.7746 & 0.0010 & $-$0.0075 \\
149 & 54634.0514 & 0.0099 & $-$0.0239 \\
150 & 54634.1522 & 0.0082 & $-$0.0324 \\
\hline
  \multicolumn{4}{l}{$^{a}$ BJD$-$2400000.} \\
  \multicolumn{4}{l}{$^{b}$ Against $max = 2454617.8043 + 0.109202 E$.} \\
\end{tabular}
\end{center}
\end{table}

\subsection{SDSS J170213.26$+$322954.1}\label{sec:j1702}\label{obj:j1702}

\begin{figure}
  \begin{center}
    \FigureFile(88mm,110mm){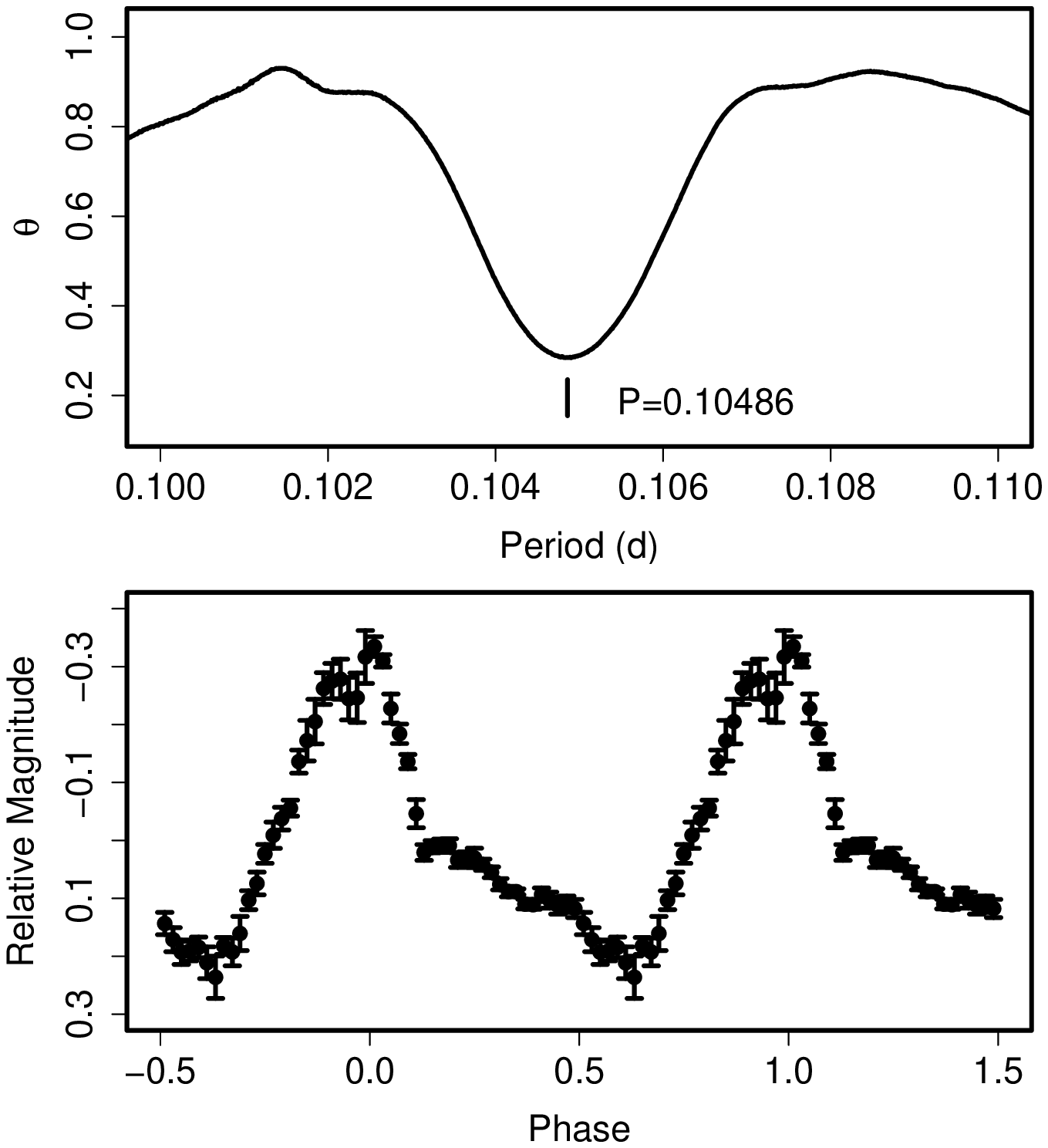}
  \end{center}
  \caption{Superhumps in SDSS J1702 before BJD 2453652. (Upper): PDM analysis.
     (Lower): Phase-averaged profile.}
  \label{fig:j1702pdma}
\end{figure}

\begin{figure}
  \begin{center}
    \FigureFile(88mm,110mm){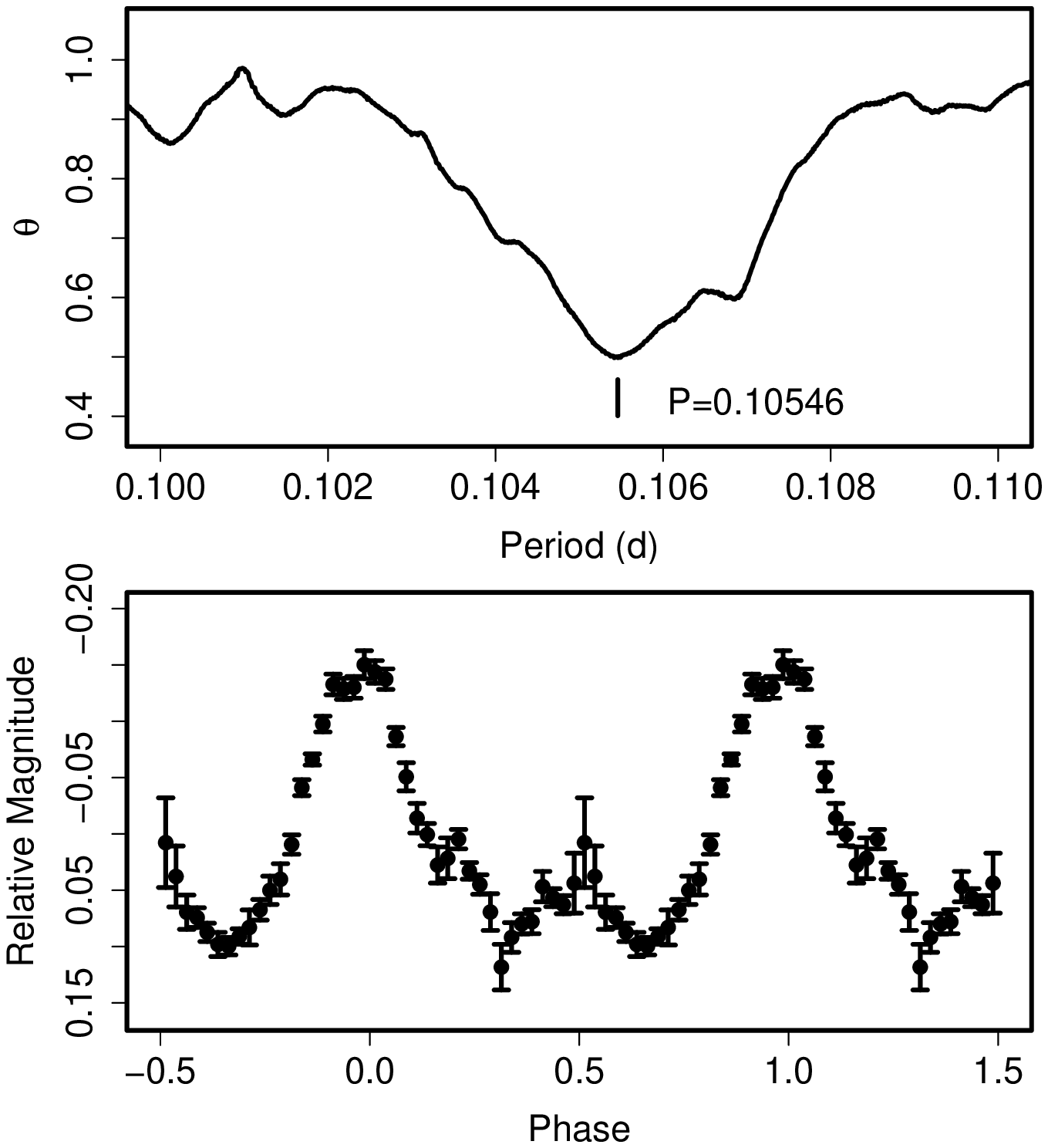}
  \end{center}
  \caption{Superhumps in SDSS J1702 after BJD 2453652. (Upper): PDM analysis.
     (Lower): Phase-averaged profile.}
  \label{fig:j1702pdmb}
\end{figure}

   This object (hereafter SDSS J1702) was discovered as a high-inclination
CV by \citet{szk04SDSSCV3}.  \citet{lit06j0702} identified this object
as an eclipsing CV in the period gap and suggested that it has an
evolved secondary.  \citet{boy06j1702} observed the 2005 superoutburst
of this object and established its SU UMa-type nature.  We used the
AAVSO data which includes the data used in \citet{boy06j1702}.
Using the eclipse ephemeris by \citet{boy06j1702}, we extracted the
times of superhump maxima outside the eclipses (table \ref{tab:j1702oc2005}).
Although our analysis basically confirmed the $P_{\rm SH}$, the periods
before $E \le 20$ and $E \ge 38$ appears to show a discontinuous change.
The mean periods determined with the PDM method were 0.10486(3) d
before BJD 2453652 (figure \ref{fig:j1702pdma})
and 0.10546(3) d after BJD 2453652 (figure \ref{fig:j1702pdmb}),
respectively.
Although the timing of $E=38$ maximum was affected by an eclipse,
the $O-C$ analysis also supports the same tendency.
These periods correspond to fractional superhump excesses of
4.8 \% and 5.4 \%, respectively.

It is very unusual for such a long $P_{\rm SH}$-system to show an
increase in the $P_{\rm SH}$ during the middle-to-late stage of
a superoutburst (cf. V725 Aql, subsection \ref{sec:v725aql}).
Although the effect of the overlapping orbital
variation can not be excluded, this object deserves further detailed
study for evolution of superhump periods.

\begin{table}
\caption{Superhump maxima of SDSS J1702 (2005).}\label{tab:j1702oc2005}
\begin{center}
\begin{tabular}{ccccc}
\hline\hline
$E$ & max$^a$ & error & $O-C^b$ & $N^c$ \\
\hline
0 & 53648.3842 & 0.0003 & 0.0045 & 152 \\
2 & 53648.5926 & 0.0018 & 0.0027 & 14 \\
9 & 53649.3267 & 0.0009 & 0.0014 & 215 \\
18 & 53650.2729 & 0.0002 & 0.0020 & 178 \\
19 & 53650.3820 & 0.0008 & 0.0060 & 93 \\
38 & 53652.3560 & 0.0033 & $-$0.0163 & 39 \\
47 & 53653.3115 & 0.0005 & $-$0.0063 & 263 \\
48 & 53653.4184 & 0.0008 & $-$0.0045 & 105 \\
56 & 53654.2611 & 0.0007 & $-$0.0023 & 146 \\
57 & 53654.3672 & 0.0006 & $-$0.0012 & 204 \\
66 & 53655.3154 & 0.0011 & 0.0014 & 225 \\
85 & 53657.3228 & 0.0038 & 0.0125 & 85 \\
\hline
  \multicolumn{5}{l}{$^{a}$ BJD$-$2400000.} \\
  \multicolumn{5}{l}{$^{b}$ Against $max = 2453648.3797 + 0.105066 E$.} \\
  \multicolumn{5}{l}{$^{c}$ Number of points used to determine the maximum.} \\
\end{tabular}
\end{center}
\end{table}

\subsection{SDSSp J173008.38$+$624754.7}\label{obj:j1730}

   SDSSp J173008.38+624754.7 (hereafter SDSS J1730) was selected as
a dwarf nova during the course of the SDSS \citep{szk02SDSSCVs}.
\citet{szk02SDSSCVs}
obtained an orbital period of 117(5) m (0.081(3) d) from radial-velocity
study, which made the object a good candidate for an SU UMa-type dwarf nova.

   We observed the 2001 October superoutburst, soon after the discovery
announcement of this object, 2002 February-March and 2004 March
superoutbursts.
We first analyzed the best sampled superoutburst in 2004
(table \ref{tab:j1730oc2004}).  The mean $P_{\rm SH}$ and the global
$P_{\rm dot}$ was 0.07948(2) d and $-7.7(3.5) \times 10^{-5}$.
This $P_{\rm dot}$ was likely from a stage B--C transition around
$E=10$.  The mean periods before and after this epoch were
0.08007(24) d and 0.07946(2) d, respectively.

The times of superhump maxima for the 2001 superoutburst are listed in
table \ref{tab:j1730oc2001}.
The last two superhumps ($E = 108$ and $E = 109$) have large
$O-C$'s.  These superhumps may have been traditional late superhumps,
or the period had largely changed before these observations.
We disregarded these maxima and obtained a mean
$P_{\rm SH}$ of 0.07941(10) d, which likely reflects stage C superhumps.

   The 2002 February-March superoutburst (table \ref{tab:j1730oc2002})
was probably observed during the stage C.  The mean $P_{\rm SH}$
was 0.07939(5) d, with an insignificant $P_{\rm dot}$ of
$+2.0(3.5) \times 10^{-5}$.

   We derived a mean supercycle of 109(1) d from the times of these four
superoutbursts and the 2002 September one.

   The variation of superhump period has generally been small
in this system.  In conjunction with the long superhump period,
the object resembles BF Ara and HV Aur.  The shortness of the cycle length
of normal outbursts (9--10 d) and supercycle also resembles
BF Ara (cf. \cite{kat03bfara}).
The lack of period variation, though, may be a result of the lack
of observations during the early stage (cf. the 2004 superoutburst).
This possibility should be resolved by future observations.

\begin{table}
\caption{Superhump maxima of SDSS J1730 (2004).}\label{tab:j1730oc2004}
\begin{center}

\end{center}
\end{table}

\begin{table}
\caption{Superhump maxima of SDSS J1730 (2001).}\label{tab:j1730oc2001}
\begin{center}
%
\end{center}
\end{table}

\begin{table}
\caption{Superhump maxima of SDSS J1730 (2002).}\label{tab:j1730oc2002}
\begin{center}
%
\end{center}
\end{table}

\subsection{SDSS J210014.12$+$004446.0}\label{obj:j2100}

   This object (hereafter SDSS J2100) was selected as a CV during the
course of the SDSS \citet{szk04SDSSCV3}.
\citet{tra05j2100} reported the detection of
superhumps with a period of 0.08746(8) d on two consecutive nights
during the 2003 superoutburst.
We observed the earliest stage of the 2007 superoutburst.  Assuming
that the first epoch observation was taken during the stage A
development, we assigned $E$ for superhumps (table \ref{tab:j2100oc2007}).
The mean period for $44 \le E \le 56$ was 0.08696(15) d.

\begin{table}
\caption{Superhump maxima of SDSS J2100 (2007).}\label{tab:j2100oc2007}
\begin{center}
\begin{tabular}{ccccc}
\hline\hline
$E$ & max$^a$ & error & $O-C^b$ & $N^c$ \\
\hline
0 & 54318.2574 & 0.0021 & $-$0.0027 & 180 \\
44 & 54322.1570 & 0.0038 & 0.0063 & 78 \\
45 & 54322.2459 & 0.0024 & 0.0068 & 46 \\
56 & 54323.2014 & 0.0033 & $-$0.0104 & 77 \\
\hline
  \multicolumn{5}{l}{$^{a}$ BJD$-$2400000.} \\
  \multicolumn{5}{l}{$^{b}$ Against $max = 2454318.2601 + 0.088423 E$.} \\
  \multicolumn{5}{l}{$^{c}$ Number of points used to determine the maximum.} \\
\end{tabular}
\end{center}
\end{table}

\subsection{SDSS J225831.18$-$094931.7}\label{obj:j2258}

   SDSS J225831.18$-$094931.7 (hereafter SDSS J2258) was selected as
a CV during the course of the SDSS \citep{szk03SDSSCV2}.
The SU UMa-type nature was established during
the 2004 June superoutburst (vsnet-alert 8162; the reported period of
0.045 d referred to a half of $P_{\rm SH}$).  During its superoutburst
in 2005 August, H. Maehara established a long $P_{\rm SH}$ of 0.083 d
(vsnet-campaign-dn 4489).

\begin{figure}
  \begin{center}
    \FigureFile(88mm,110mm){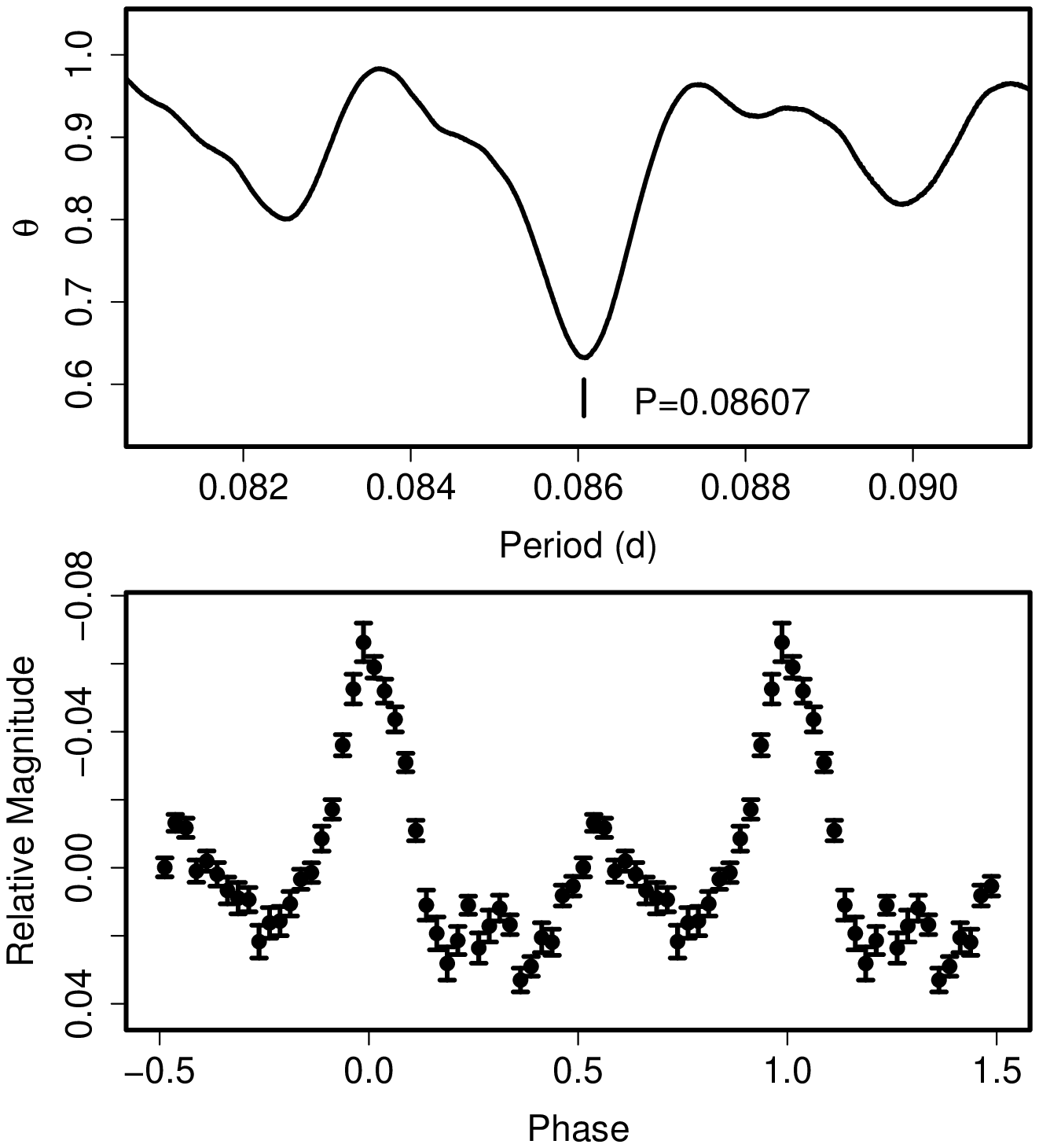}
  \end{center}
  \caption{Superhumps in SDSS J2258 (2004). (Upper): PDM analysis.
     (Lower): Phase-averaged profile.}
  \label{fig:j2258shpdm}
\end{figure}

   The times of superhump maxima during the 2008 superoutburst are
listed in table \ref{tab:j2258oc2008}.  This outburst was apparently
detected during its late stage, since the object already started fading
rapidly after six days.  The mean superhump period with the PDM method
was 0.08607(2) d (figure \ref{fig:j2258shpdm}), which most likely
refers to $P_2$, with an almost zero
$P_{\rm dot}$ of $+1.5(2.1) \times 10^{-5}$.
The maxima for $82 \le E \le 93$ refer to the post-superoutburst stage.
There was no apparent indication of a phase shift around the termination
of the superoutburst.

   The times of superhump maxima during the 2004 superoutburst
are also given (table \ref{tab:j2258oc2004}).  The 2004 superoutburst
was caught during its final stage.  The 2005 observation is omitted
because it was a single-night observation.

\begin{table}
\caption{Superhump maxima of SDSS J2258 (2004).}\label{tab:j2258oc2004}
\begin{center}
\begin{tabular}{ccccc}
\hline\hline
$E$ & max$^a$ & error & $O-C^b$ & $N^c$ \\
\hline
0 & 53159.5243 & 0.0004 & $-$0.0031 & 368 \\
1 & 53159.6148 & 0.0007 & 0.0015 & 211 \\
12 & 53160.5599 & 0.0006 & 0.0016 & 379 \\
13 & 53160.6464 & 0.0008 & 0.0023 & 290 \\
20 & 53161.2446 & 0.0023 & $-$0.0008 & 145 \\
23 & 53161.5019 & 0.0073 & $-$0.0012 & 224 \\
24 & 53161.5887 & 0.0012 & $-$0.0003 & 380 \\
\hline
  \multicolumn{5}{l}{$^{a}$ BJD$-$2400000.} \\
  \multicolumn{5}{l}{$^{b}$ Against $max = 2453159.5274 + 0.085900 E$.} \\
  \multicolumn{5}{l}{$^{c}$ Number of points used to determine the maximum.} \\
\end{tabular}
\end{center}
\end{table}

\begin{table}
\caption{Superhump maxima of SDSS J2258 (2008).}\label{tab:j2258oc2008}
\begin{center}
\begin{tabular}{ccccc}
\hline\hline
$E$ & max$^a$ & error & $O-C^b$ & $N^c$ \\
\hline
0 & 54788.9460 & 0.0004 & 0.0023 & 178 \\
1 & 54789.0298 & 0.0004 & $-$0.0001 & 260 \\
11 & 54789.8849 & 0.0019 & $-$0.0064 & 77 \\
12 & 54789.9817 & 0.0003 & 0.0042 & 389 \\
13 & 54790.0657 & 0.0004 & 0.0021 & 423 \\
23 & 54790.9250 & 0.0004 & $-$0.0000 & 309 \\
24 & 54791.0122 & 0.0003 & 0.0011 & 820 \\
25 & 54791.0971 & 0.0013 & $-$0.0002 & 286 \\
34 & 54791.8713 & 0.0023 & $-$0.0013 & 87 \\
35 & 54791.9586 & 0.0005 & $-$0.0001 & 260 \\
36 & 54792.0410 & 0.0010 & $-$0.0038 & 187 \\
46 & 54792.9068 & 0.0011 & 0.0005 & 122 \\
47 & 54792.9962 & 0.0012 & 0.0038 & 162 \\
58 & 54793.9362 & 0.0007 & $-$0.0037 & 289 \\
59 & 54794.0250 & 0.0016 & $-$0.0011 & 123 \\
82 & 54796.0086 & 0.0022 & 0.0013 & 64 \\
92 & 54796.8730 & 0.0034 & 0.0043 & 55 \\
93 & 54796.9520 & 0.0015 & $-$0.0029 & 81 \\
\hline
  \multicolumn{5}{l}{$^{a}$ BJD$-$2400000.} \\
  \multicolumn{5}{l}{$^{b}$ Against $max = 2454788.9438 + 0.086141 E$.} \\
  \multicolumn{5}{l}{$^{c}$ Number of points used to determine the maximum.} \\
\end{tabular}
\end{center}
\end{table}

\subsection{OT J004226.5$+$421537}\label{obj:j0042}

   This object (hereafter OT J0042) was discovered by K. Itagaki
as a possible nova in M31 which reached a peak magnitude of 14.5
around 2008 November 28.6 UT (=M31N 2008-11b, \cite{ita08j0042cbet1588}).
Multicolor photometry by S. Kiyota suggested that this object is
a foreground dwarf nova rather than a nova in M31 (vsnet-alert 10747).
The object was indeed spectroscopically confirmed as a dwarf nova
\citep{kas08j0042cbet1611}.

\begin{figure}
  \begin{center}
    \FigureFile(88mm,110mm){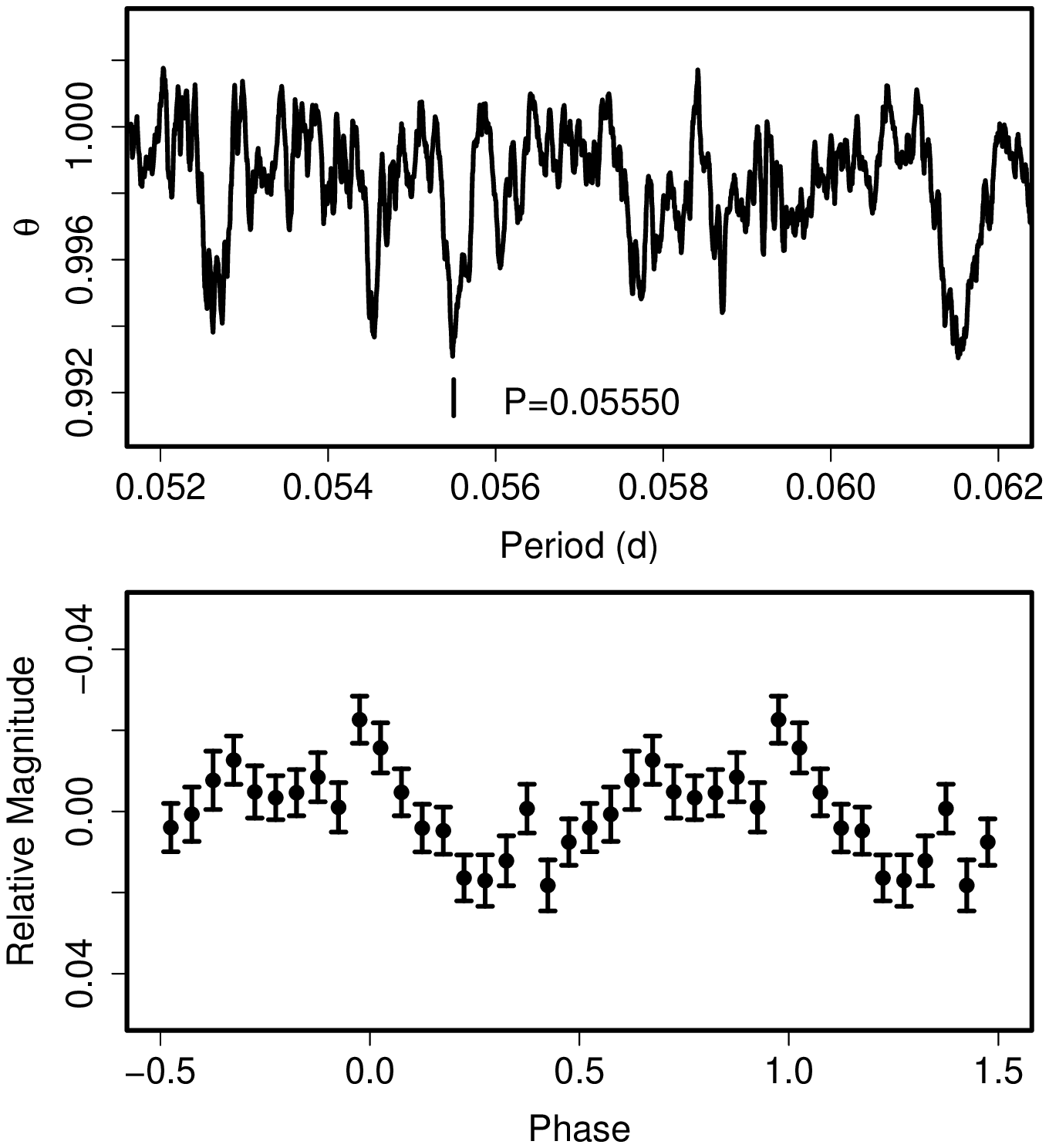}
  \end{center}
  \caption{Early superhumps in OT J0042 (2008). (Upper): PDM analysis.
     (Lower): Phase-averaged profile.}
  \label{fig:j0042eshpdm}
\end{figure}

   Until 2008 December 7, early superhumps were present (vsnet-alert
10747, 10763, 10786).  The mean period of early superhumps was
0.05550(2) d (figure \ref{fig:j0042eshpdm}).

\begin{figure}
  \begin{center}
    \FigureFile(88mm,110mm){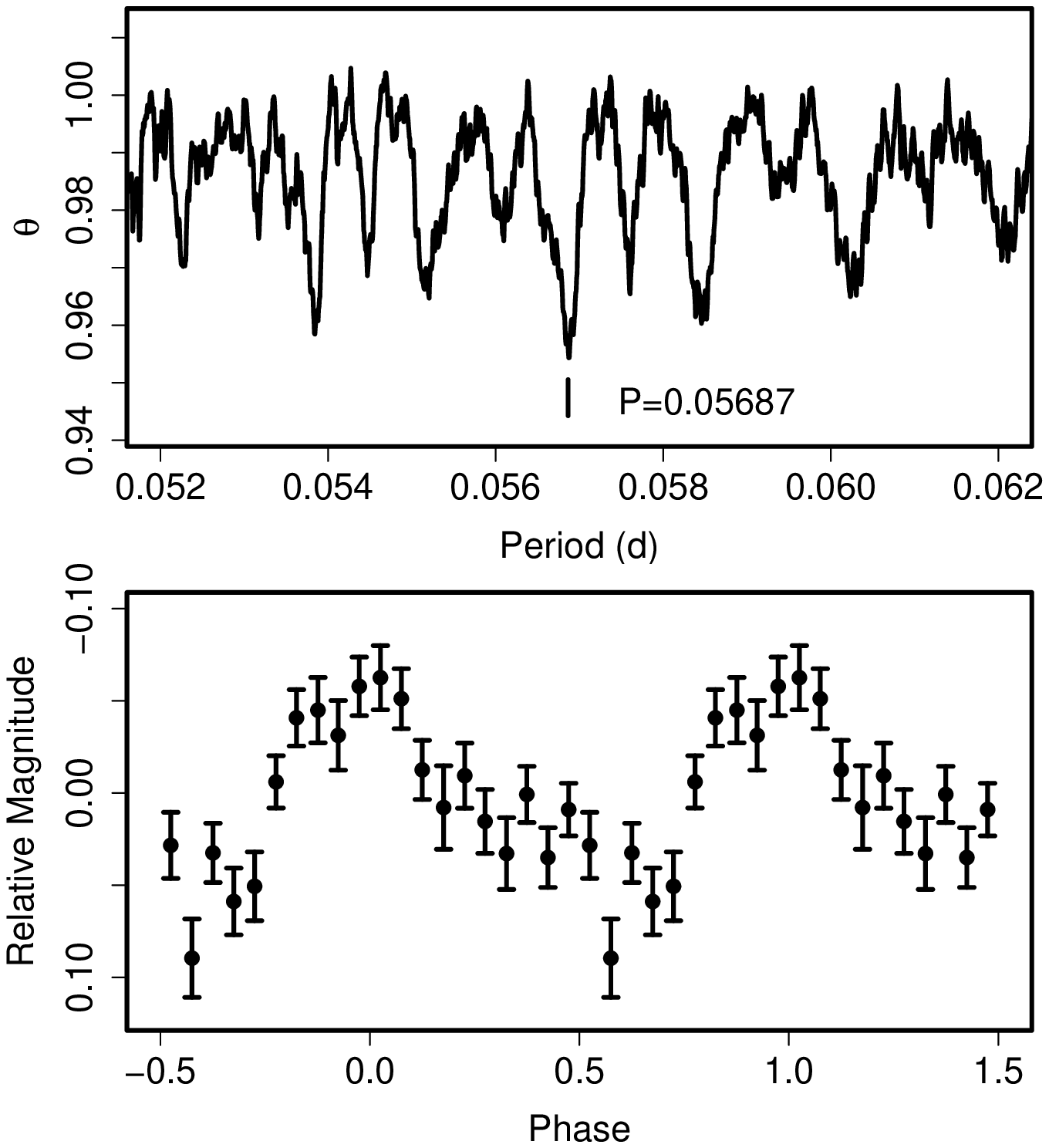}
  \end{center}
  \caption{Ordinary superhumps in OT J0042 (2008). (Upper): PDM analysis.
     (Lower): Phase-averaged profile.}
  \label{fig:j0042shpdm}
\end{figure}

   On December 10, ordinary superhump emerged (cvnet-outburst 2801,
vsnet-alert 10818).  The times of superhump maxima are listed in
table \ref{tab:j0042oc2008}.  The mean $P_{\rm SH}$ determined
with the PDM method was 0.05687(2) d (figure \ref{fig:j0042shpdm}).
The $P_{\rm dot}$ was slightly positive, $+4.0(1.8) \times 10^{-5}$.
The fractional superhump excess is 2.5(1) \%, which is slightly
large for a WZ Sge-type dwarf nova.  Since the amplitudes of
early superhumps and ordinary superhumps were low, these period
determinations may have been affected by non-ideal photometric
conditions and the fractional superhump excess needs to be treated
with caution.

\begin{table}
\caption{Superhump maxima of OT J0042 (2008).}\label{tab:j0042oc2008}
\begin{center}
\begin{tabular}{ccccc}
\hline\hline
$E$ & max$^a$ & error & $O-C^b$ & $N^c$ \\
\hline
0 & 54810.9649 & 0.0033 & 0.0052 & 43 \\
1 & 54811.0200 & 0.0016 & 0.0034 & 128 \\
2 & 54811.0747 & 0.0044 & 0.0013 & 98 \\
3 & 54811.1324 & 0.0034 & 0.0021 & 78 \\
4 & 54811.1963 & 0.0031 & 0.0091 & 43 \\
34 & 54812.8911 & 0.0023 & $-$0.0029 & 30 \\
35 & 54812.9431 & 0.0040 & $-$0.0077 & 28 \\
36 & 54813.0008 & 0.0063 & $-$0.0069 & 30 \\
37 & 54813.0675 & 0.0029 & 0.0029 & 40 \\
70 & 54814.9424 & 0.0034 & 0.0004 & 30 \\
71 & 54814.9984 & 0.0041 & $-$0.0005 & 95 \\
72 & 54815.0529 & 0.0017 & $-$0.0030 & 136 \\
73 & 54815.1110 & 0.0014 & $-$0.0017 & 119 \\
74 & 54815.1668 & 0.0032 & $-$0.0028 & 112 \\
89 & 54816.0244 & 0.0020 & 0.0014 & 67 \\
90 & 54816.0657 & 0.0023 & $-$0.0142 & 18 \\
158 & 54819.9633 & 0.0039 & 0.0148 & 30 \\
160 & 54820.0728 & 0.0027 & 0.0105 & 127 \\
161 & 54820.1130 & 0.0021 & $-$0.0061 & 112 \\
162 & 54820.1708 & 0.0087 & $-$0.0053 & 61 \\
\hline
  \multicolumn{5}{l}{$^{a}$ BJD$-$2400000.} \\
  \multicolumn{5}{l}{$^{b}$ Against $max = 2454810.9596 + 0.056892 E$.} \\
  \multicolumn{5}{l}{$^{c}$ Number of points used to determine the maximum.} \\
\end{tabular}
\end{center}
\end{table}

\subsection{OT J011306.7$+$215250}\label{obj:j0113}

   This object (=CSS080922:011307$+$215250, hereafter OT J0113)
was discovered by the Catalina Real-time Transient Survey
(CRTS, \cite{dra08atel1734}).\footnote{
   $<$http://nesssi.cacr.caltech.edu/catalina/$>$.
   For the information of the individual Catalina CVs, see
   $<$http://nesssi.cacr.caltech.edu/catalina/AllCV.html$>$.
}
H. Maehara detected superhumps and identified this object as
a long-$P_{\rm SH}$ SU UMa-type dwarf nova (vsnet-alert 10539).
The observation was performed during the last stage of the superoutburst
(table \ref{tab:j0113oc2008}).  The cycle count is based on period
determination in \citet{sha09v466andj0113}.

\begin{table}
\caption{Superhump maxima of OT J0113 (2008).}\label{tab:j0113oc2008}
\begin{center}
\begin{tabular}{ccccc}
\hline\hline
$E$ & max$^a$ & error & $O-C^b$ & $N^c$ \\
\hline
0 & 54732.1975 & 0.0007 & 0.0016 & 187 \\
21 & 54734.1751 & 0.0040 & $-$0.0017 & 93 \\
22 & 54734.2694 & 0.0032 & $-$0.0016 & 129 \\
43 & 54736.2535 & 0.0040 & 0.0016 & 51 \\
\hline
  \multicolumn{5}{l}{$^{a}$ BJD$-$2400000.} \\
  \multicolumn{5}{l}{$^{b}$ Against $max = 2454732.1959 + 0.094325 E$.} \\
  \multicolumn{5}{l}{$^{c}$ Number of points used to determine the maximum.} \\
\end{tabular}
\end{center}
\end{table}

\subsection{OT J021110.2$+$171624}\label{obj:j0211}

   This object (=CSS080130:021110$+$171624, hereafter OT J0211) was
discovered by the CRTS in 2008 January (\cite{djo08atel1416}; \cite{CRTS};
cvnet-discussion 1106). 
The detection of superhumps
led to a classification as an SU UMa-type dwarf nova
(cvnet-discussion 1109).  \citet{djo08atel1416} reported spectroscopic
confirmation as a CV.

\begin{figure}
  \begin{center}
    \FigureFile(88mm,110mm){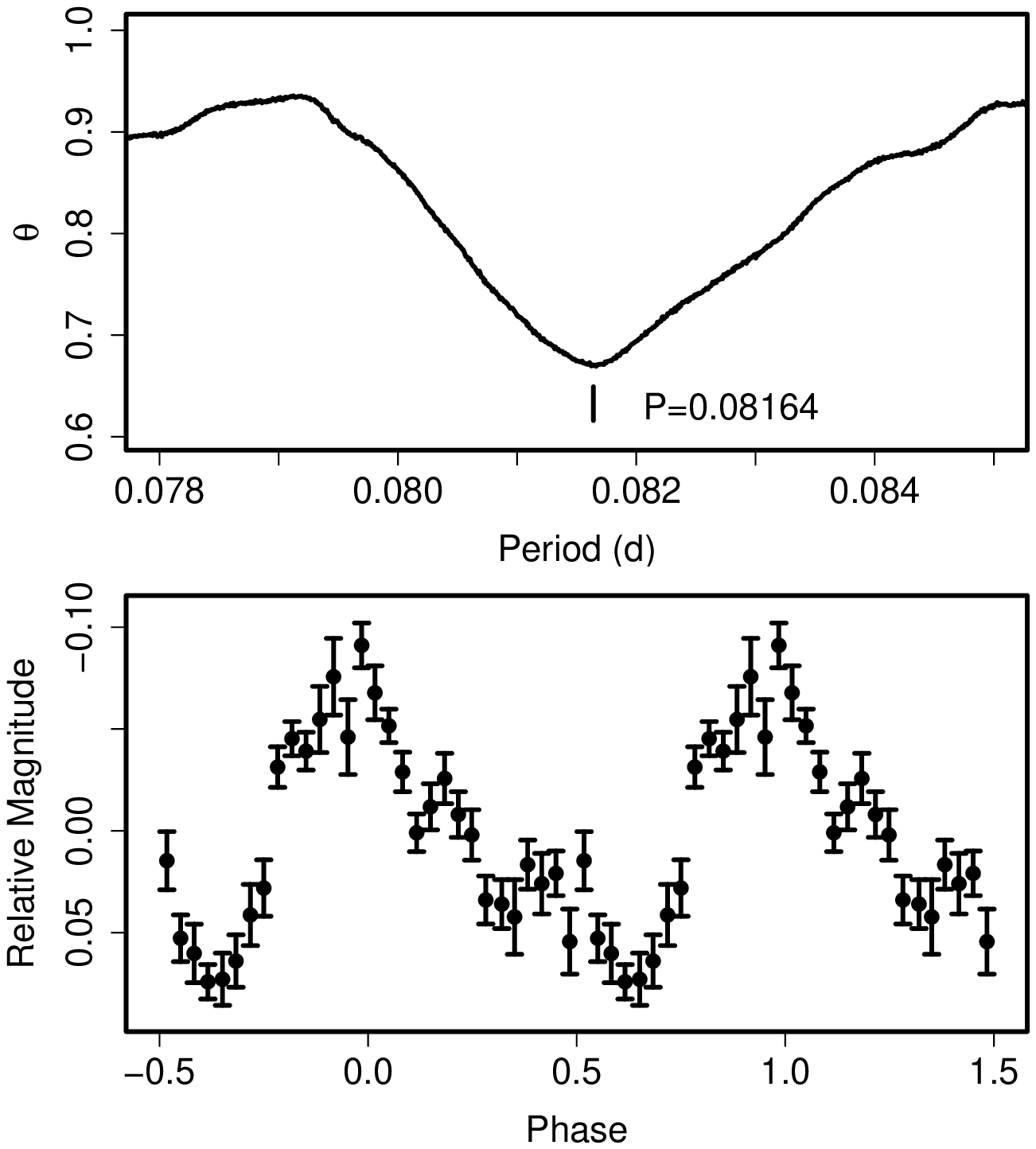}
  \end{center}
  \caption{Superhumps in OT J0211 (2008). (Upper): PDM analysis.
     (Lower): Phase-averaged profile.}
  \label{fig:j0211shpdm}
\end{figure}

   We observed the 2008 November superoutburst (vsnet-alert 10663)
and established the superhump period of 0.08164(6) d with the PDM method
(figure \ref{fig:j0211shpdm}).
The times of superhump maxima are listed in table \ref{tab:j0211oc2008}.
The object appears to have a relatively short supercycle of
$\sim$280 d, typical for an SU UMa-type dwarf nova with a long
$P_{\rm SH}$.

\begin{table}
\caption{Superhump maxima of OT J0211.}\label{tab:j0211oc2008}
\begin{center}
\begin{tabular}{ccccc}
\hline\hline
$E$ & max$^a$ & error & $O-C^b$ & $N^c$ \\
\hline
0 & 54775.0286 & 0.0046 & 0.0006 & 89 \\
1 & 54775.1098 & 0.0011 & 0.0001 & 129 \\
12 & 54776.0016 & 0.0037 & $-$0.0062 & 53 \\
13 & 54776.0932 & 0.0009 & 0.0038 & 227 \\
14 & 54776.1728 & 0.0013 & 0.0018 & 87 \\
\hline
  \multicolumn{5}{l}{$^{a}$ BJD$-$2400000.} \\
  \multicolumn{5}{l}{$^{b}$ Against $max = 2454775.0281 + 0.081643 E$.} \\
  \multicolumn{5}{l}{$^{c}$ Number of points used to determine the maximum.} \\
\end{tabular}
\end{center}
\end{table}

\subsection{OT J023839.1$+$355648}\label{sec:j0238}\label{obj:j0238}

\begin{figure}
  \begin{center}
    \FigureFile(88mm,110mm){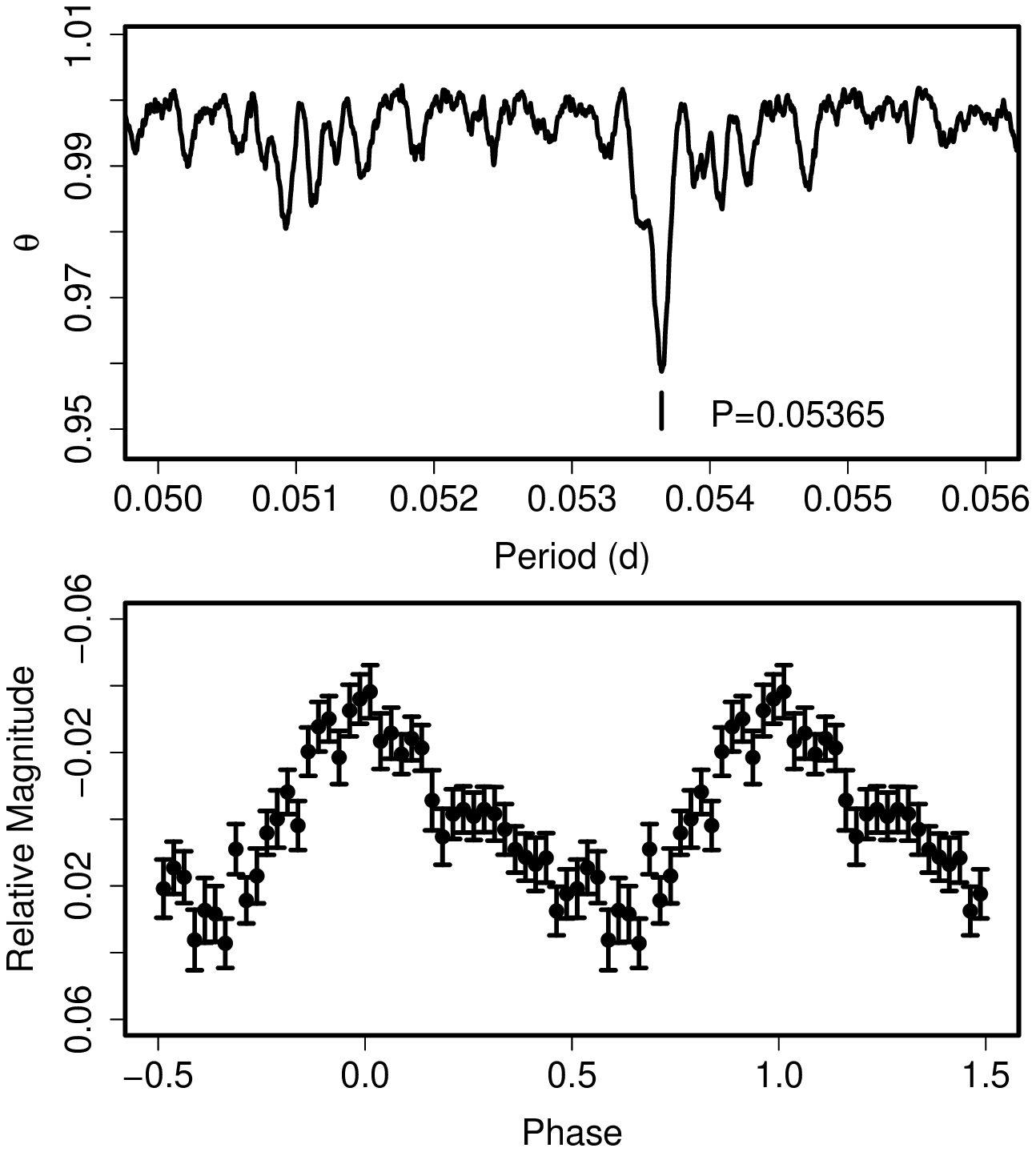}
  \end{center}
  \caption{Ordinary superhumps in OT J0238 (2008). (Upper): PDM analysis.
     (Lower): Phase-averaged profile.}
  \label{fig:j0238shpdm}
\end{figure}

\begin{figure}
  \begin{center}
    \FigureFile(88mm,110mm){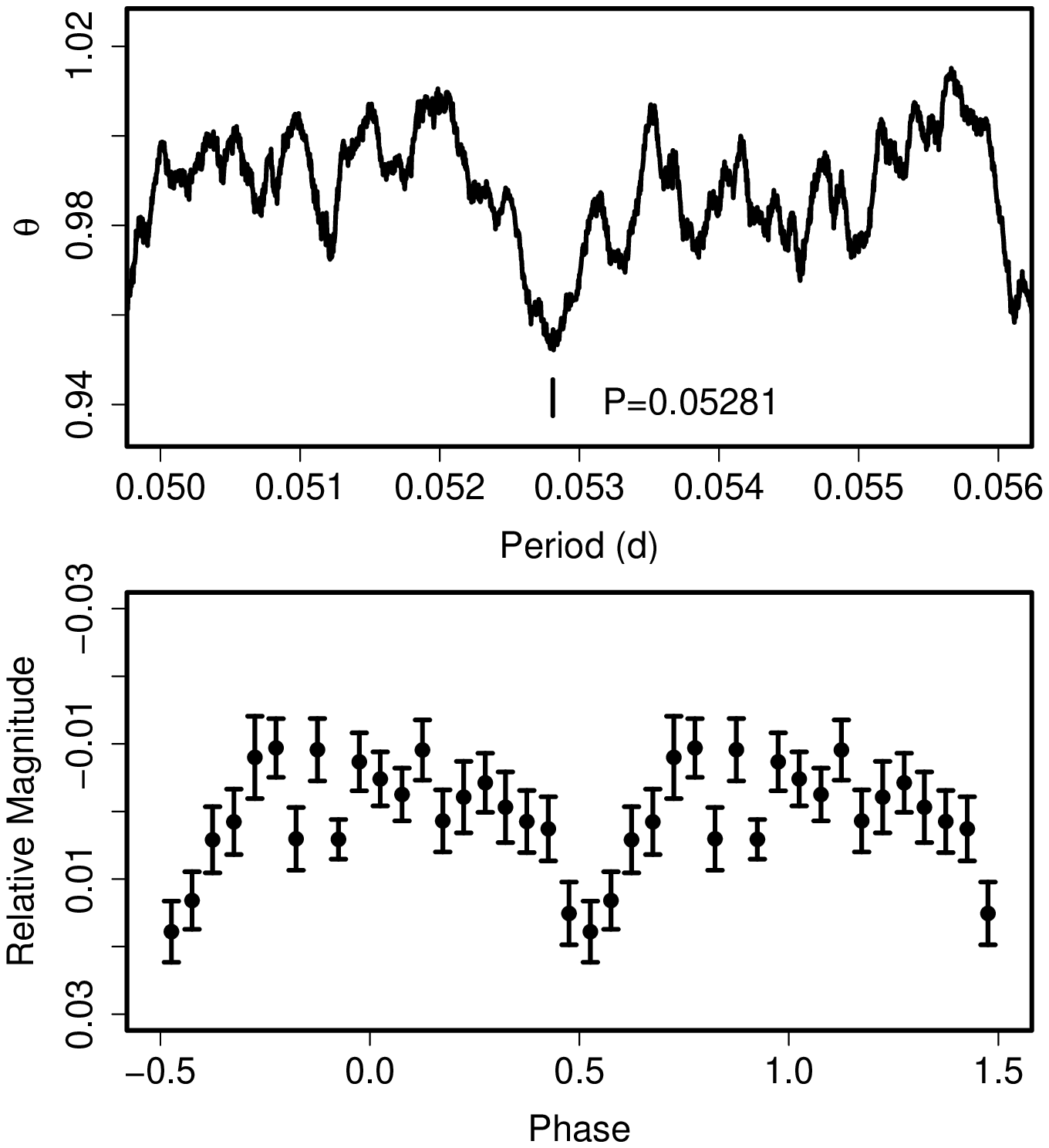}
  \end{center}
  \caption{Early superhumps in OT J0238 (2008) before BJD 2454769.5.
     (Upper): PDM analysis.
     The alias selection was based on $P_{\rm SH}$.
     (Lower): Phase-averaged profile.}
  \label{fig:j0238eshpdm}
\end{figure}

   This object (=CSS081026:023839$+$355648, hereafter OT J0238)
was discovered by the CRTS.
H. Maehara suggested that this object may be a WZ Sge-type dwarf nova
(vsnet-alert 10628).  Superhumps were later detected (vsnet-alert 10667,
figure \ref{fig:j0238shpdm}).
A reanalysis of the early data confirmed the presence of early superhumps
(vsnet-alert 10686), confirming the suggested
classification of the object as a WZ Sge-type dwarf nova with the
shortest known $P_{\rm SH}$.
\citet{shu08j0238} observed the same outburst and reported 
periods of 0.0531 d and 0.0537 d for early and ordinary superhumps.
We used the combined data set with ours and \citet{shu08j0238},
after selecting the best-quality segment, and refined the period
of early superhumps to be 0.05281(6) d (figure \ref{fig:j0238eshpdm}).

   The times of superhump maxima are listed in table \ref{tab:j0238oc2008}.
The $O-C$ diagram clearly consists of A--C stages.  The $P_{\rm dot}$
for the stage B ($67 \le E \le 350$, disregarding $E=347$)
was $+2.0(0.2) \times 10^{-5}$.  The duration of the stage A
(52 $P_{\rm SH}$ or longer) is longer than those of typical SU UMa-type
dwarf novae (20--30 $P_{\rm SH}$).  This might be a signature of slow
evolution of superhumps in this system.

   The details will be presented by Maehara et al., in preparation.

\begin{table}
\caption{Superhump maxima of OT J0238 (2008).}\label{tab:j0238oc2008}
\begin{center}

\end{center}
\end{table}

\addtocounter{table}{-1}
\begin{table}
\caption{Superhump maxima of OT J0238 (2008). (continued)}
\begin{center}
%
\end{center}
\end{table}

\subsection{OT J032912.3$+$125018}\label{obj:j0329}

\begin{figure}
  \begin{center}
    \FigureFile(88mm,110mm){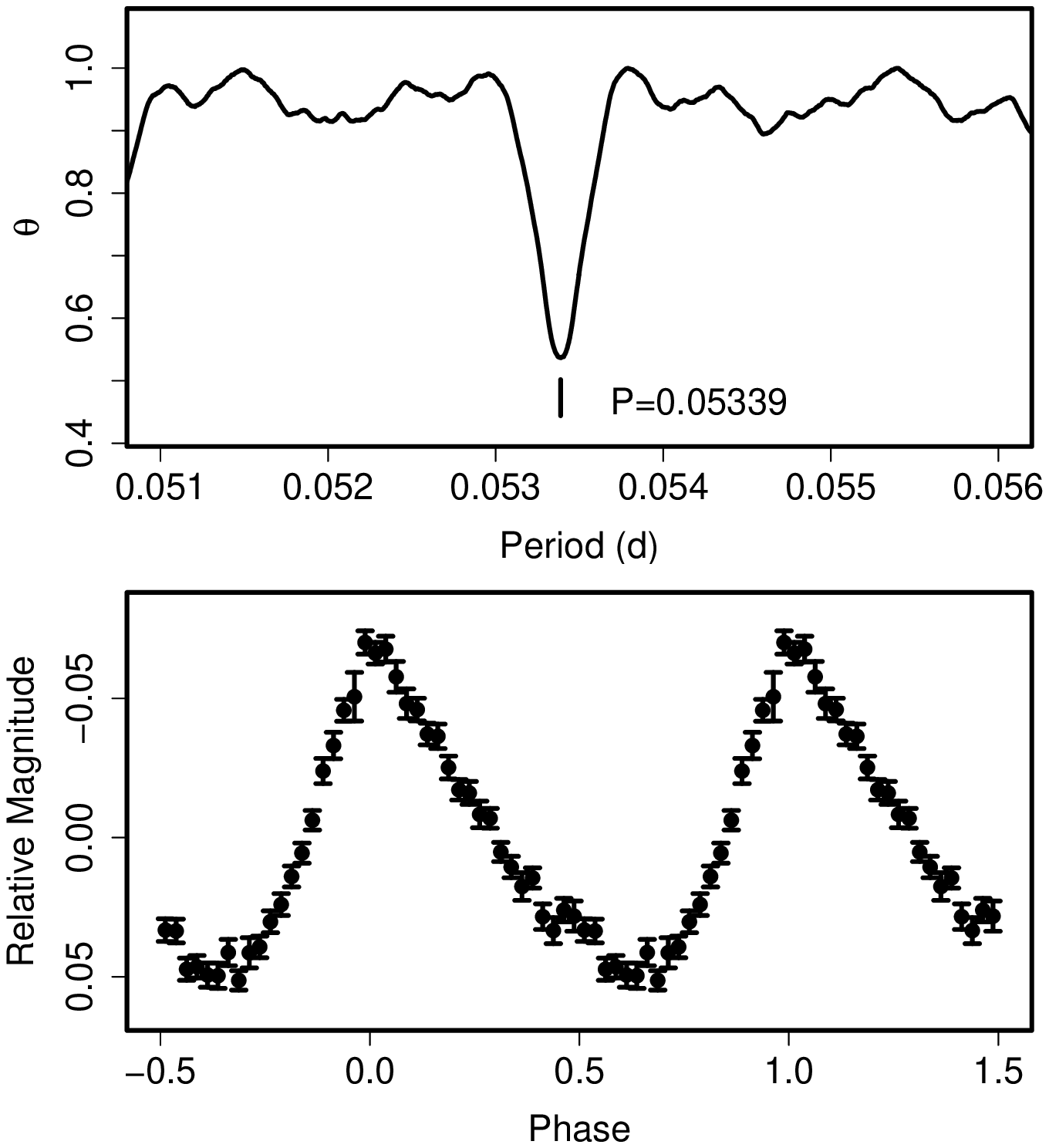}
  \end{center}
  \caption{Superhumps in OT J0329 (2006). (Upper): PDM analysis.
     (Lower): Phase-averaged profile.}
  \label{fig:j0329shpdm}
\end{figure}

\begin{figure}
  \begin{center}
    \FigureFile(88mm,90mm){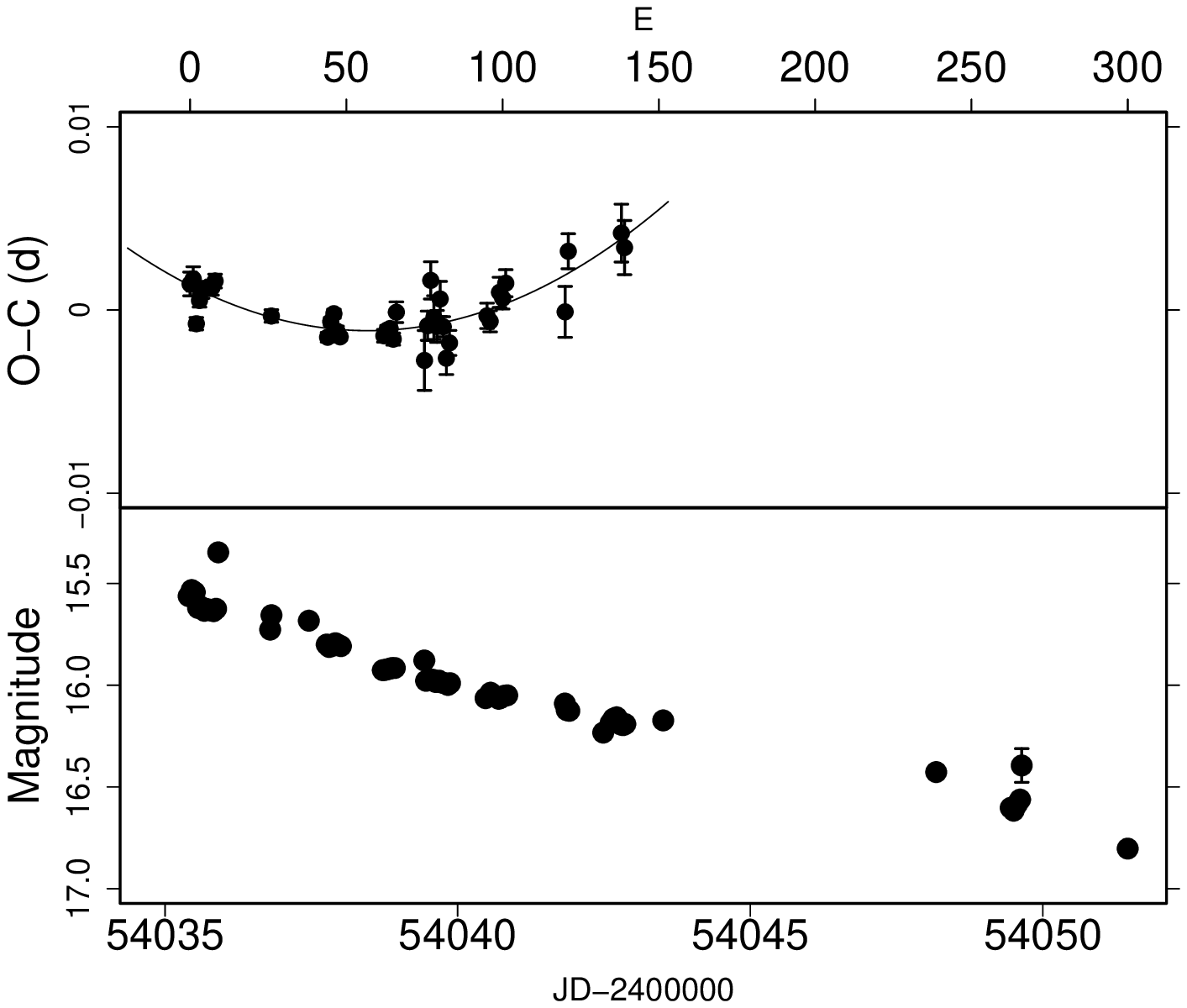}
  \end{center}
  \caption{$O-C$ of superhumps OT J0329 (2006).
  (Upper): $O-C$ diagram.  The $O-C$ values were against the mean period
  for the stage B ($E \le 139$, thin curve).  Late-stage humps with
  large errors were omitted.
  (Lower): Light curve.}
  \label{fig:j0329oc}
\end{figure}

   This object (also known as VS 0329$+$1250; hereafter OT J0329)
was discovered by \citet{skv06j0329cbet701}.
The detection of superhumps led to a classification as an SU UMa-type
dwarf nova \citep{waa06j0329cbet701}.
\citet{sha07j0329} reported a superhump period of 0.053394(7) d,
the shortest record at that time among ordinary SU UMa-type dwarf novae.
We used a combination of the photometric data by \citet{sha07j0329}
and AAVSO observations and obtained times of superhump maxima
(table \ref{tab:j0329oc2006}; the times for superhumps were systematically
different from those by \citet{sha07j0329} due to the difference in
the method for determining the maxima).
The mean $P_{\rm SH}$ with the PDM method was 0.053388(4) d
(figure \ref{fig:j0329shpdm}).
The $P_{\rm dot}$ was $+2.8(0.3) \times 10^{-5}$
($E \le 139$, figure \ref{fig:j0329oc}),
confirming the positive $P_{\rm dot}$ reported in \citet{sha07j0329}.
Although there appears to have been a transition to stage C after $E=139$,
we could not measure $P_2$ because of the lack of observations.

   According to the CRTS, this object
(=CSS081025:032912$+$125018) has a magnitude of 21 in quiescence and
experienced two further faint outbursts.  The relatively small outburst
amplitude for an extremely short $P_{\rm SH}$ and the presence of relatively
frequent (approximately once per year) outbursts, combined with
the relatively large $P_{\rm dot}$,
would place the object as a member of OT J0557
(group ``X'' in \cite{uem09j0557}, though the $P_{\rm dot}$ is larger
in OT J0557) rather than an extreme WZ Sge-type dwarf nova.

\begin{table}
\caption{Superhump maxima of OT J0329 (2006).}\label{tab:j0329oc2006}
\begin{center}

\end{center}
\end{table}

\subsection{OT J040659.8$+$005244}\label{obj:j0406}

\begin{figure}
  \begin{center}
    \FigureFile(88mm,110mm){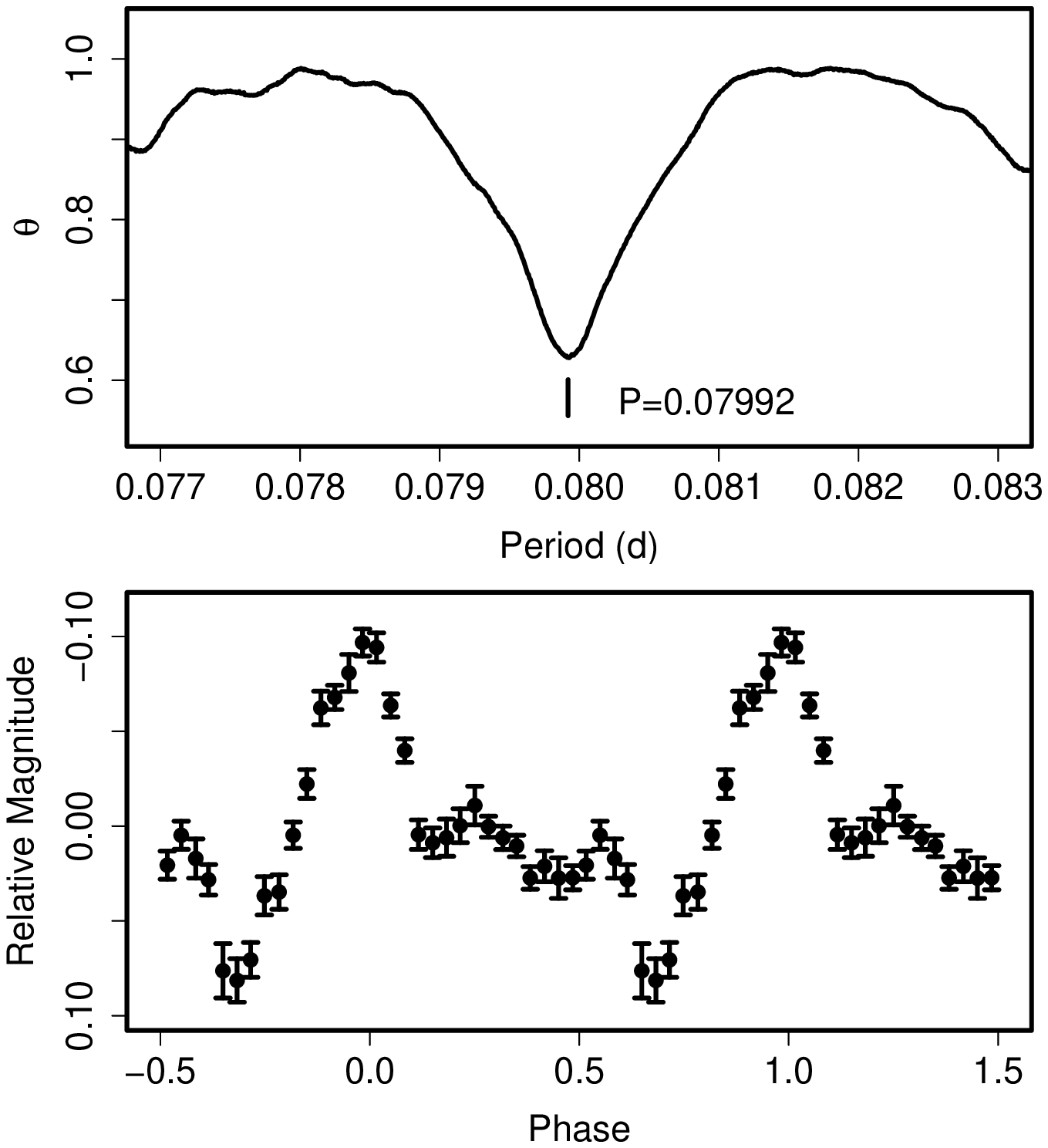}
  \end{center}
  \caption{Superhumps in OT J0406 (2008). (Upper): PDM analysis.
     (Lower): Phase-averaged profile.}
  \label{fig:j0406shpdm}
\end{figure}

   This object (hereafter OT J0406) was discovered by K. Itagaki
\citep{yam08j0406cbet1463}.  Subsequent observations confirmed the
SU UMa-type nature of the object (vsnet-alert 10422).
The mean superhump period with the PDM method was 0.07992(2) d
(figure \ref{fig:j0406shpdm}).
The times of superhump maxima are listed in table \ref{tab:j0406oc2008}.
The period was almost constant with $P_{\rm dot}$ = $+2.8(3.4) \times 10^{-5}$.
The outburst may have been detected during its late course, and
the lack of period variation may be attributed to stage C superhumps.

\begin{table}
\caption{Superhump maxima of OT J0406 (2008).}\label{tab:j0406oc2008}
\begin{center}
\begin{tabular}{ccccc}
\hline\hline
$E$ & max$^a$ & error & $O-C^b$ & $N^c$ \\
\hline
0 & 54687.3825 & 0.0004 & 0.0005 & 151 \\
11 & 54688.2619 & 0.0008 & 0.0005 & 350 \\
24 & 54689.2989 & 0.0009 & $-$0.0018 & 193 \\
36 & 54690.2605 & 0.0007 & 0.0005 & 167 \\
61 & 54692.2591 & 0.0013 & 0.0003 & 138 \\
\hline
  \multicolumn{5}{l}{$^{a}$ BJD$-$2400000.} \\
  \multicolumn{5}{l}{$^{b}$ Against $max = 2454687.3820 + 0.079947 E$.} \\
  \multicolumn{5}{l}{$^{c}$ Number of points used to determine the maximum.} \\
\end{tabular}
\end{center}
\end{table}

\subsection{OT J055718$+$683226}\label{obj:j0557}

   This object was discovered by \citet{klo06j0557cbet777} and was
extensively studied by \citet{uem09j0557}.  We present a supplementary
analysis using the combined data with \citet{uem09j0557} and the
AAVSO data (table \ref{tab:j0557oc2006}).
The $P_{\rm dot}$ for $E \le 110$ (stage B) was
$+9.0(2.1) \times 10^{-5}$.  The relatively large $P_{\rm dot}$
with a very short $P_{\rm SH}$ strengthens the similarity to V844 Her,
as suggested by \citet{uem09j0557}.

\begin{table}
\caption{Superhump maxima of OT J0557 (2006).}\label{tab:j0557oc2006}
\begin{center}

\end{center}
\end{table}

\subsection{OT J074727.6$+$065050}\label{obj:j0747}

\begin{figure}
  \begin{center}
    \FigureFile(88mm,70mm){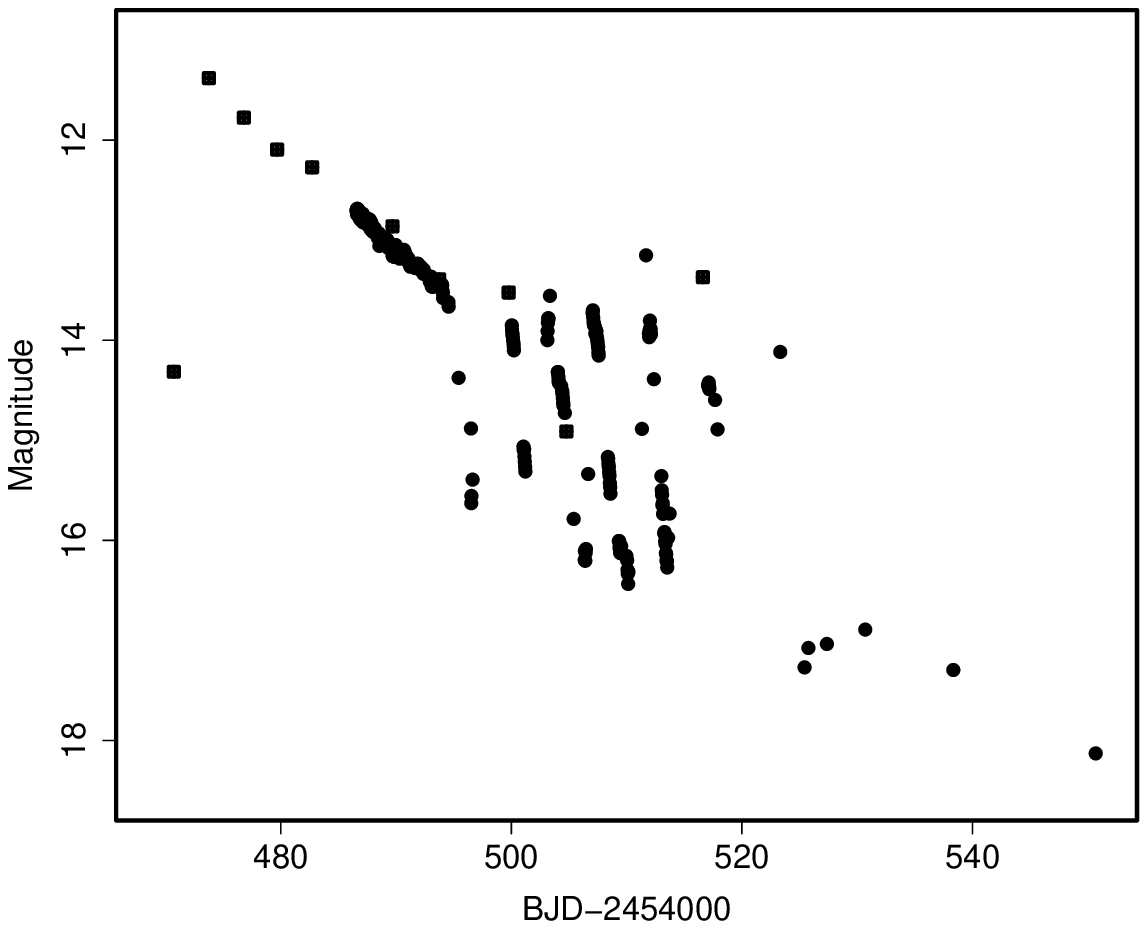}
  \end{center}
  \caption{Light curve of the 2008 superoutburst of OT J0747.
     The filled circles and filled squares represent CCD observations
     used here and ASAS-3 $V$ data, respectively.}
  \label{fig:j0747lc}
\end{figure}

   This object (hereafter OT J0747) was discovered by K. Itagaki
\citep{yam08j0747cbet1216}.  Soon after the discovery announcement
and spectroscopic confirmation, this object was proposed to be a
good candidate for a WZ Sge-type dwarf nova (vsnet-alert 9832).
The detection of superhumps and later repeated rebrightenings
confirmed this suggestion.  The outburst behavior was very similar
to those of EG Cnc or UZ Boo (figure \ref{fig:j0747lc}).
The post-superoutburst observations indicated that the final fading
was on a smooth extension of the quiescence during the rebrightening
phase, as in SDSS J0804 (for the implication, see \cite{kat09j0804}).

   The times of superhump maxima during the main superoutburst are
listed in table \ref{tab:j0747oc2008}.
The detection of the outburst was 11 d after the maximum ($V = 11.4$)
retrospectively measured with ASAS-3.  The stage of early superhumps
and early development of the ordinary superhumps were not recorded.
The $P_{\rm dot}$ during the plateau stage was $+4.0(0.8) \times 10^{-5}$
($E \le 109$).  \citet{she09j0747} reported $P_{\rm dot}$ of
$+4.4(0.9) \times 10^{-5}$ using a slightly different set of observations.

\begin{figure}
  \begin{center}
    \FigureFile(88mm,110mm){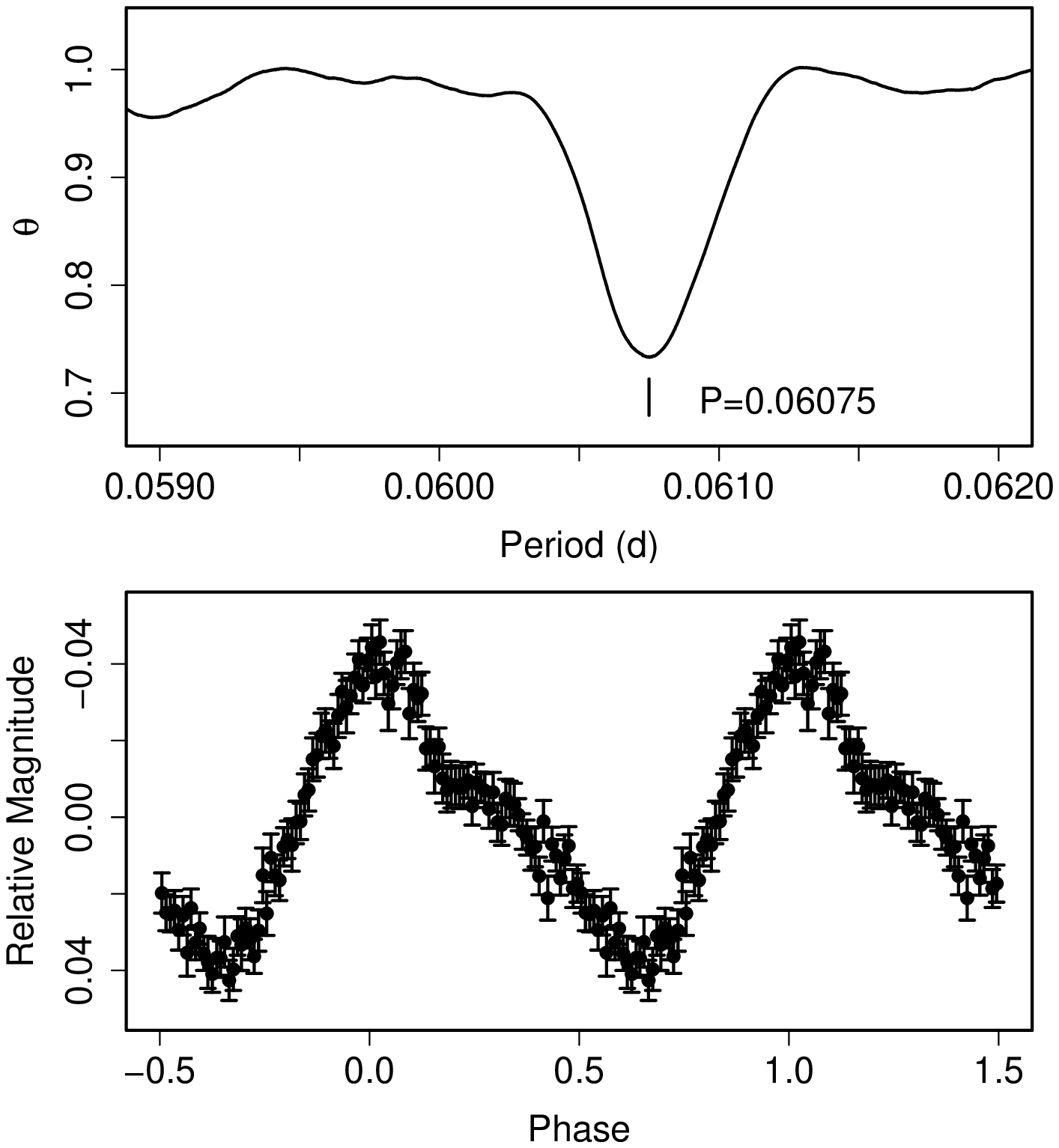}
  \end{center}
  \caption{Superhumps in OT J0747 during the superoutburst plateau
     (2008). (Upper): PDM analysis.
     (Lower): Phase-averaged profile.}
  \label{fig:j0747mainpdm}
\end{figure}

\begin{figure}
  \begin{center}
    \FigureFile(88mm,110mm){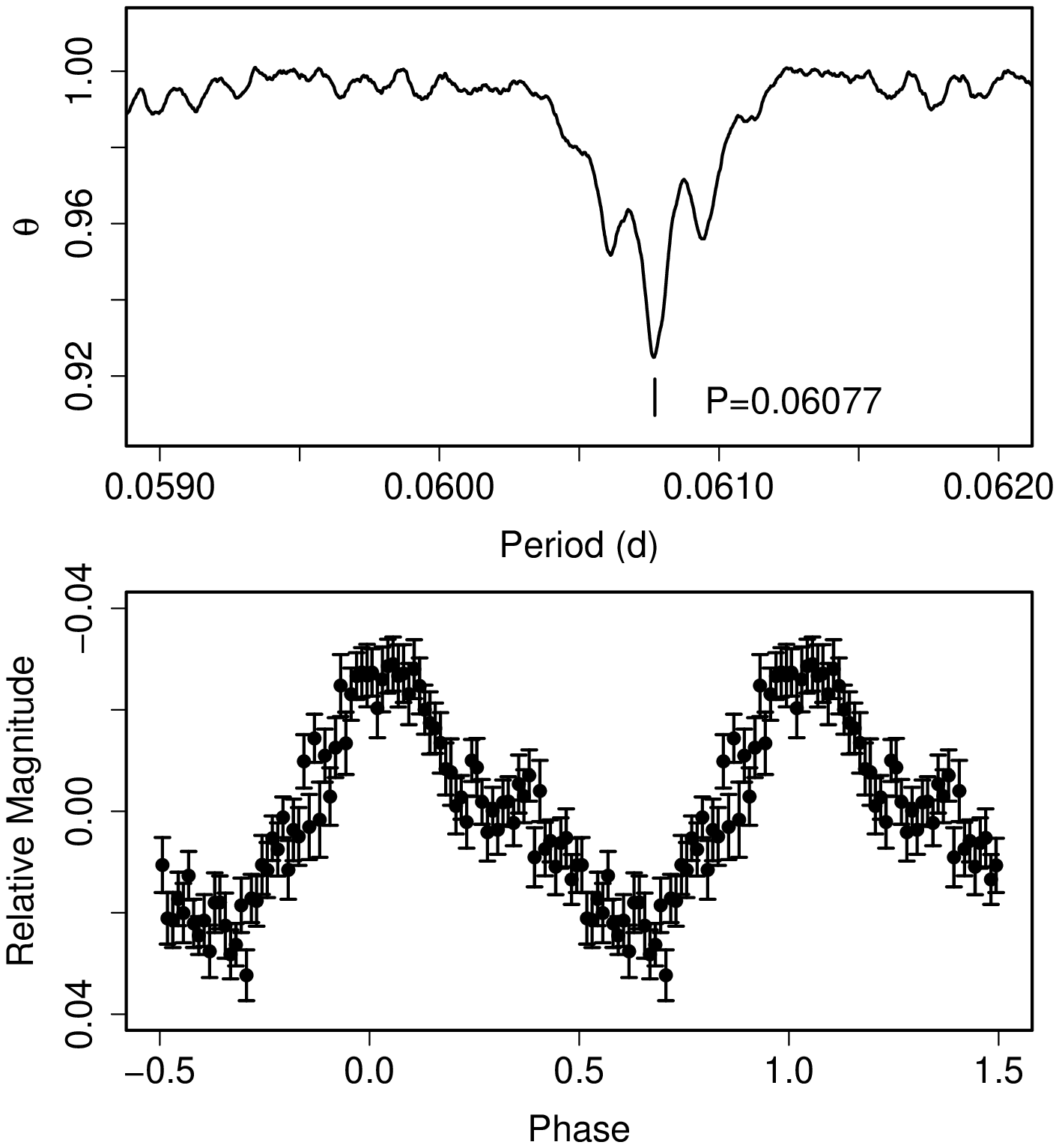}
  \end{center}
  \caption{Superhumps in OT J0747 during the rebrightening phase (2008).
     (Upper): PDM analysis.
     (Lower): Phase-averaged profile.}
  \label{fig:j0747rebpdm}
\end{figure}

   After removing the global trend of the outburst (the method is the
same as in \cite{kat09j0804}), PDM analyses yielded mean superhump
periods of 0.060750(7) d during the superoutburst
(figure \ref{fig:j0747mainpdm})
and 0.060771(3) d during the rebrightening phase
(figure \ref{fig:j0747rebpdm}).
The superhump period during the rebrightening phase is 0.3 \% longer
than that during the superoutburst plateau.
This behavior follows the general tendency in WZ Sge-type dwarf novae
(subsection \ref{sec:latestage}).

\begin{table}
\caption{Superhump maxima of OT J0747 (2008).}\label{tab:j0747oc2008}
\begin{center}

\end{center}
\end{table}

\subsection{OT J080714.2$+$113812}\label{obj:j0807}

\begin{figure}
  \begin{center}
    \FigureFile(88mm,90mm){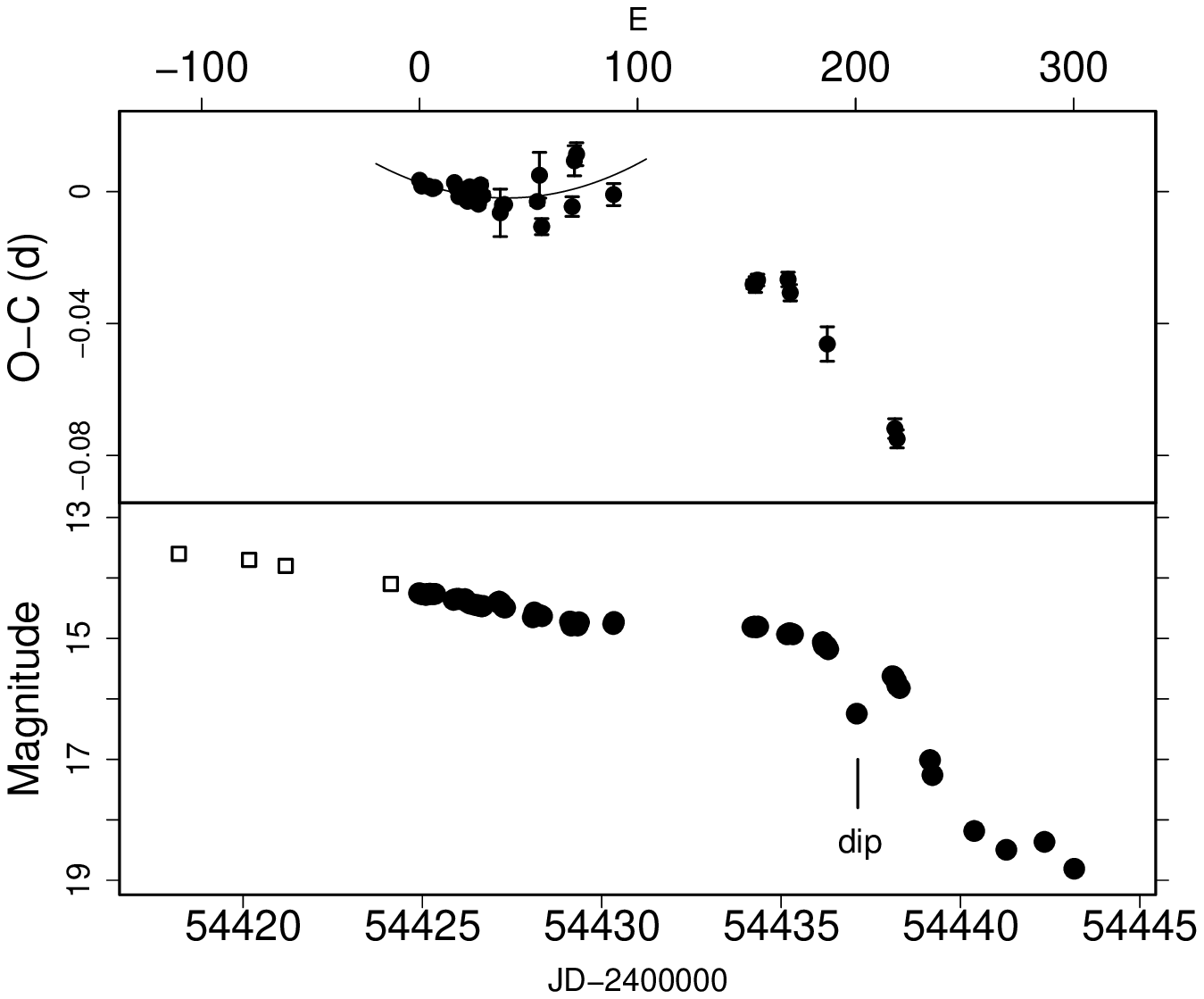}
  \end{center}
  \caption{$O-C$ of superhumps OT J0807 (2007).
  (Upper): $O-C$ diagram.  The $O-C$ values were against the mean period
  for the stage B ($E \le 89$, thin curve)
  (Lower): Light curve.  Large dots are our CCD observations and open
  squares are Itagaki's CCD observations.}
  \label{fig:j0807oc}
\end{figure}

\begin{figure}
  \begin{center}
    \FigureFile(88mm,110mm){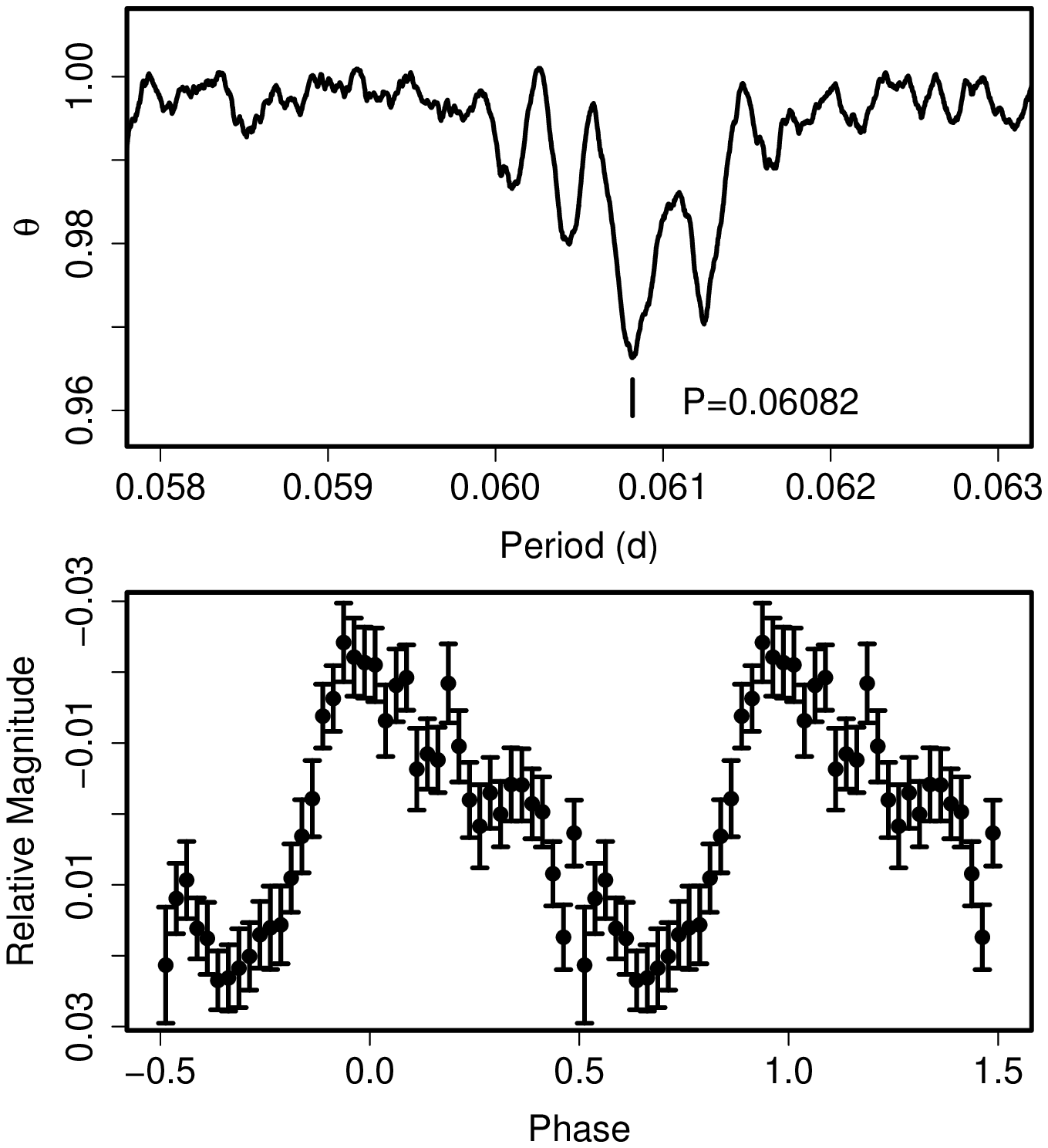}
  \end{center}
  \caption{Superhumps in OT J0807 (2007). (Upper): PDM analysis.
     (Lower): Phase-averaged profile.}
  \label{fig:j0807shpdm}
\end{figure}

   This object (hereafter OT J0807) was discovered by K. Itagaki
and was suggested to be a candidate WZ Sge-type dwarf nova
(vsnet-newvar 2602, vsnet-alert 9721, 9731).  The object was soon
confirmed to exhibit superhumps.  The outburst was associated with an
unusual rebrightening following a one-day dip near the termination
of the superoutburst (vsnet-alert 9745, 9746, figure \ref{fig:j0807oc}).
The mean superhump period with the PDM method was 0.060818(10) d
(figure \ref{fig:j0807shpdm}).
The times of superhump maxima are listed in table \ref{tab:j0807oc2007}.
Judging from the light curve and the variation of the amplitude of
superhumps, the outburst was probably detected during its
middle-to-late course.  The stage of early superhumps, if the object
is indeed a WZ Sge-type dwarf nova, and early development of
the ordinary superhumps thus were not recorded.  The table includes
the maxima during the rebrightening ($E = 218, 219$).
The break in the $O-C$ diagram most likely reflected a transition
to the stage C.
We determined a relatively large $P_{\rm dot}$ = $+9.5(4.8) \times 10^{-5}$
for the earlier phase ($E \le 89$).  This behavior of period variation
is similar to those observed in the stage B of short-period SU UMa-type
dwarf novae or some WZ Sge-type dwarf novae.
More detailed analysis will be reported by Maehara et al., in preparation.

\begin{table}
\caption{Superhump maxima of OT J0807 (2007).}\label{tab:j0807oc2007}
\begin{center}
\begin{tabular}{ccccc}
\hline\hline
$E$ & max$^a$ & error & $O-C^b$ & $N^c$ \\
\hline
0 & 54424.9170 & 0.0003 & $-$0.0049 & 113 \\
1 & 54424.9763 & 0.0003 & $-$0.0064 & 113 \\
4 & 54425.1594 & 0.0014 & $-$0.0056 & 90 \\
5 & 54425.2201 & 0.0004 & $-$0.0057 & 368 \\
6 & 54425.2808 & 0.0005 & $-$0.0058 & 373 \\
7 & 54425.3422 & 0.0005 & $-$0.0052 & 296 \\
16 & 54425.8931 & 0.0006 & $-$0.0012 & 111 \\
17 & 54425.9525 & 0.0009 & $-$0.0026 & 102 \\
18 & 54426.0110 & 0.0005 & $-$0.0049 & 108 \\
22 & 54426.2538 & 0.0012 & $-$0.0053 & 220 \\
23 & 54426.3192 & 0.0013 & $-$0.0006 & 116 \\
26 & 54426.5005 & 0.0014 & $-$0.0017 & 121 \\
27 & 54426.5581 & 0.0012 & $-$0.0048 & 123 \\
28 & 54426.6250 & 0.0013 & 0.0013 & 105 \\
29 & 54426.6828 & 0.0013 & $-$0.0017 & 116 \\
37 & 54427.1660 & 0.0072 & $-$0.0047 & 106 \\
38 & 54427.2294 & 0.0010 & $-$0.0021 & 66 \\
39 & 54427.2906 & 0.0006 & $-$0.0016 & 80 \\
54 & 54428.2073 & 0.0013 & 0.0034 & 321 \\
55 & 54428.2763 & 0.0069 & 0.0116 & 345 \\
56 & 54428.3218 & 0.0024 & $-$0.0037 & 267 \\
70 & 54429.1825 & 0.0030 & 0.0062 & 129 \\
71 & 54429.2575 & 0.0046 & 0.0204 & 131 \\
72 & 54429.3205 & 0.0035 & 0.0226 & 123 \\
89 & 54430.3462 & 0.0033 & 0.0150 & 37 \\
153 & 54434.2262 & 0.0013 & 0.0052 & 123 \\
154 & 54434.2872 & 0.0024 & 0.0054 & 123 \\
155 & 54434.3496 & 0.0018 & 0.0071 & 132 \\
169 & 54435.2044 & 0.0022 & 0.0111 & 147 \\
170 & 54435.2615 & 0.0025 & 0.0073 & 204 \\
187 & 54436.2838 & 0.0052 & $-$0.0036 & 54 \\
218 & 54438.1507 & 0.0030 & $-$0.0208 & 188 \\
219 & 54438.2086 & 0.0027 & $-$0.0237 & 175 \\
\hline
  \multicolumn{5}{l}{$^{a}$ BJD$-$2400000.} \\
  \multicolumn{5}{l}{$^{b}$ Against $max = 2454424.9219 + 0.060778 E$.} \\
  \multicolumn{5}{l}{$^{c}$ Number of points used to determine the maximum.} \\
\end{tabular}
\end{center}
\end{table}

\subsection{OT J081418.9$-$005022}\label{obj:j0814}

   This object (=CSS080409:081419$-$005022, hereafter OT J0814) was
discovered by the CRTS (\cite{dra08atel1479}; \cite{CRTS}; vsnet-alert 10038).
ASAS-3 detected a new outburst in 2008 October
(vsnet-alert 10594), during which superhumps were detected
(vsnet-alert 10603, 10630).  Due to the short visibility of the object,
it was difficult to uniquely determine $P_{\rm SH}$.  We adopted the
most likely period (0.0763 d) that best express all the recorded superhumps.
The times of superhump maxima are listed in table \ref{tab:j0814oc2008}.
There was likely a stage B--C transition.

\begin{table}
\caption{Superhump maxima of OT J0814 (2008).}\label{tab:j0814oc2008}
\begin{center}
\begin{tabular}{ccccc}
\hline\hline
$E$ & max$^a$ & error & $O-C^b$ & $N^c$ \\
\hline
0 & 54759.5713 & 0.0008 & $-$0.0041 & 116 \\
1 & 54759.6459 & 0.0009 & $-$0.0058 & 95 \\
79 & 54765.6153 & 0.0007 & 0.0148 & 149 \\
101 & 54767.2901 & 0.0066 & 0.0116 & 81 \\
141 & 54770.3126 & 0.0022 & $-$0.0166 & 91 \\
\hline
  \multicolumn{5}{l}{$^{a}$ BJD$-$2400000.} \\
  \multicolumn{5}{l}{$^{b}$ Against $max = 2454759.5754 + 0.076268 E$.} \\
  \multicolumn{5}{l}{$^{c}$ Number of points used to determine the maximum.} \\
\end{tabular}
\end{center}
\end{table}

\subsection{OT J084555.1$+$033930}\label{obj:j0845}

\begin{figure}
  \begin{center}
    \FigureFile(88mm,110mm){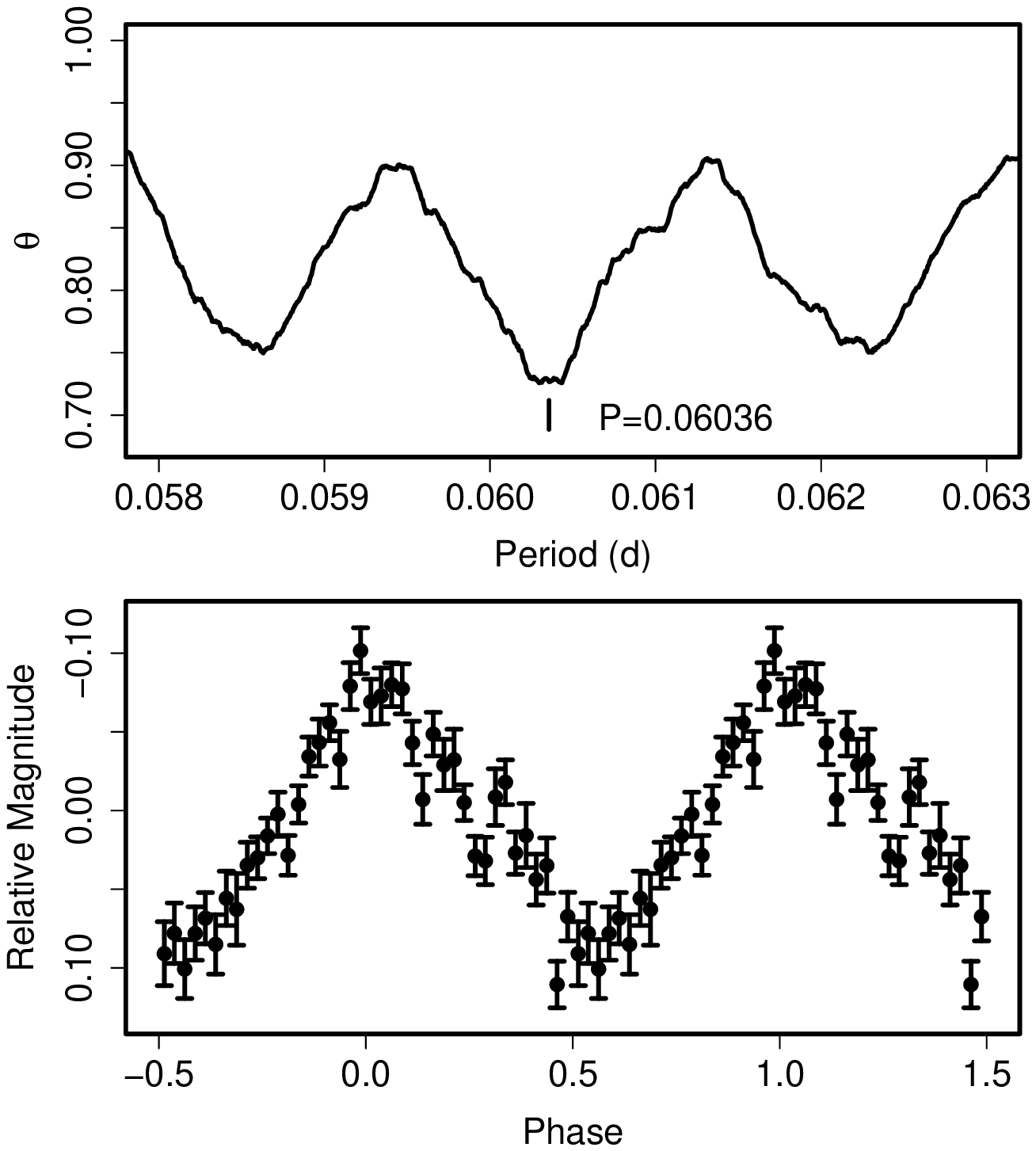}
  \end{center}
  \caption{Superhumps in OT J0845 (2008) after BJD 2454491.
     (Upper): PDM analysis.
     (Lower): Phase-averaged profile.}
  \label{fig:j0845shpdm}
\end{figure}

   This object (hereafter OT J0845) was discovered by K. Itagaki
(\cite{yam08j0845cbet1225}; \cite{hon08j0845cbet1229}).
The mean superhump period with the PDM method was 0.06036(2) d
(figure \ref{fig:j0845shpdm}, excluding the first night).
The times of superhump maxima are listed in table \ref{tab:j0845oc2008}.
The observation on the first night ($E = 0$) apparently caught the
evolutionary stage of superhumps (cf. vsnet-alert 9847).  We used
$E > 0$ data and obtained $P_{\rm dot}$ = $+6.7(3.4) \times 10^{-5}$.
The object is likely a large-amplitude
SU UMa-type dwarf nova rather than a typical WZ Sge-type star
(vsnet-alert 9852).

\begin{table}
\caption{Superhump maxima of OT J0845 (2008).}\label{tab:j0845oc2008}
\begin{center}
\begin{tabular}{ccccc}
\hline\hline
$E$ & max$^a$ & error & $O-C^b$ & $N^c$ \\
\hline
0 & 54487.1018 & 0.0032 & $-$0.0004 & 100 \\
66 & 54491.0961 & 0.0007 & 0.0023 & 208 \\
67 & 54491.1548 & 0.0006 & 0.0006 & 114 \\
68 & 54491.2181 & 0.0009 & 0.0034 & 110 \\
69 & 54491.2735 & 0.0015 & $-$0.0017 & 105 \\
99 & 54493.0894 & 0.0016 & $-$0.0001 & 168 \\
100 & 54493.1443 & 0.0013 & $-$0.0057 & 114 \\
167 & 54497.2036 & 0.0043 & 0.0016 & 52 \\
\hline
  \multicolumn{5}{l}{$^{a}$ BJD$-$2400000.} \\
  \multicolumn{5}{l}{$^{b}$ Against $max = 2454487.1022 + 0.060478 E$.} \\
  \multicolumn{5}{l}{$^{c}$ Number of points used to determine the maximum.} \\
\end{tabular}
\end{center}
\end{table}

\subsection{OT J090239.7$+$052501}\label{obj:j0902}

\begin{figure}
  \begin{center}
    \FigureFile(88mm,110mm){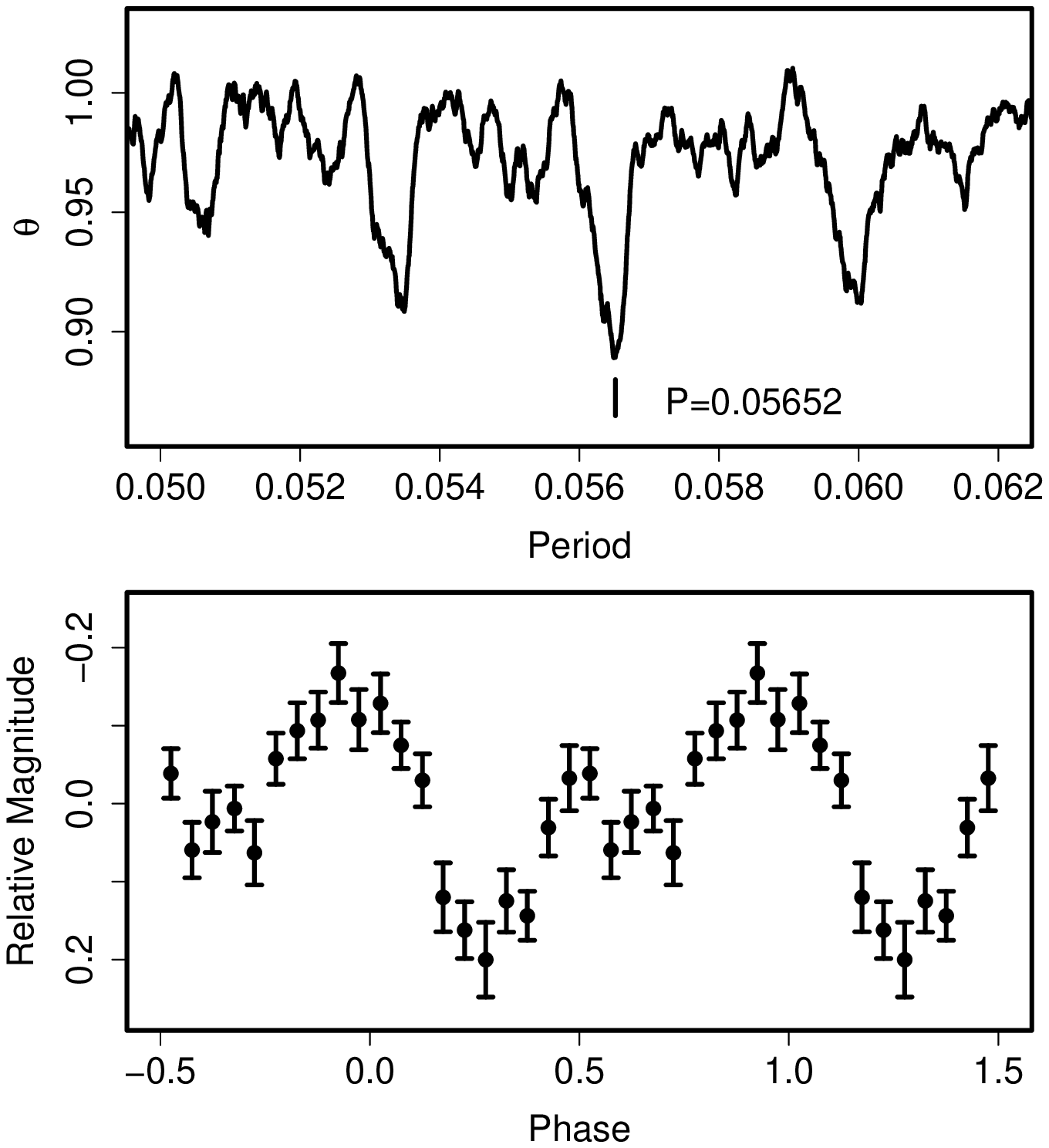}
  \end{center}
  \caption{Early superhumps in OT J0902 (2008). (Upper): PDM analysis.
     (Lower): Phase-averaged profile.}
  \label{fig:j0902eshpdm}
\end{figure}

   OT J090239.7$+$052501 (=CSS080304:090240$+$052501, hereafter OT J0902)
is a transient discovered by the CRTS \citet{CRTS}.
The object had a blue SDSS counterpart with $g = 23.17$, $g-r = +0.07$
(vsnet-alert 9945).  Spectroscopic observation of the outbursting
object revealed the presence of a broad He\textsc{II} emission lines
\citep{djo08j0902atel1411} which is suggestive of a WZ Sge-type
outburst in a high-inclination system (vsnet-alert 9948;
\cite{ima06tss0222}).  Early superhumps were subsequently detected
(vsnet-alert 9953, 9955, 9963).  The object was still in outburst
27 d after the outburst detection (vsnet-alert 10011).
Although we did not observe ordinary superhumps, we include this
object for improving the statistics of WZ Sge-type dwarf novae.
The mean period of early superhumps was 0.05652(3) d
(figure \ref{fig:j0902eshpdm}).
Uemura and Arai (vsnet-alert 9963) independently obtained the same
period.

\subsection{OT J102146.4$+$234926}\label{obj:j1021}

\begin{figure}
  \begin{center}
    \FigureFile(88mm,90mm){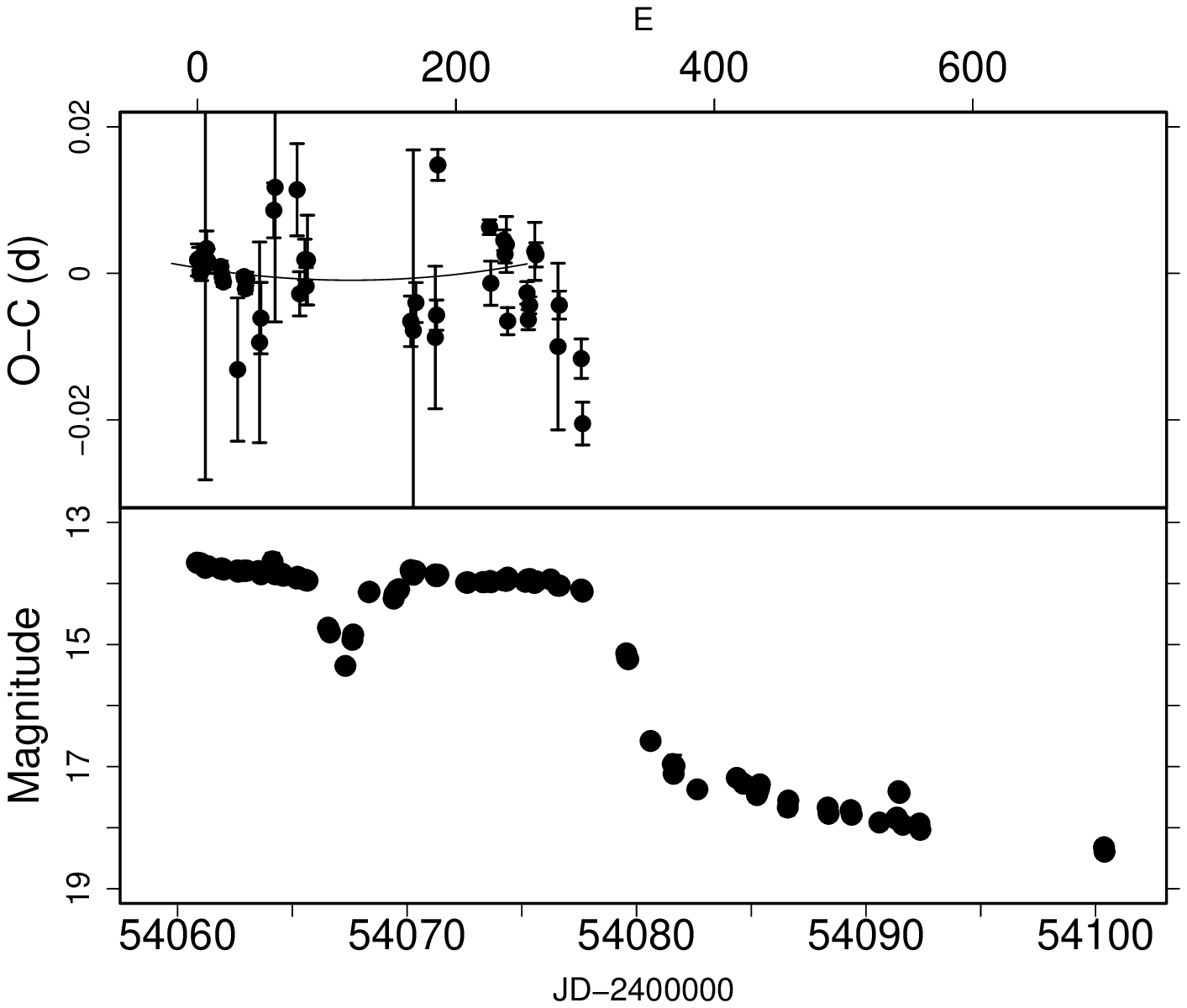}
  \end{center}
  \caption{$O-C$ of superhumps OT J1021 (2006).
  (Upper): $O-C$ diagram.  The $O-C$ values were against the mean period
  for the stage B ($E \le 240$, thin curve).
  (Lower): Light curve.}
  \label{fig:j1021oc}
\end{figure}

   This object (also called Var Leo 06, hereafter OT J1021) was discovered
by \citet{chr06j1021cbet746} in the course of the Catalina Sky Survey
(CSS).  \citet{gol07j1021} and \citet{uem08j1021} reported the detection
of superhumps and classified the object as a WZ Sge-type dwarf nova.
We reanalyzed the data for OT J1021 in \citet{uem08j1021} in combination
with the AAVSO data, and determined the superhump maxima
during the plateau stage and the rebrightening stage (table
\ref{tab:j1021oc2006}).  The maxima can be well expressed by a single
period of 0.056295(10) d without a phase shift (figure \ref{fig:j1021oc}).
This lack of a phase shift, as well as the smooth continuation of
the general fading trend before and after the ``dip'', the dip
phenomenon in this object can be better understood as a temporary
cooling of the disk, and the plateau stage of the main superoutburst
and the ``rebrightening'' comprise a continuous entity, rather than
the complete termination of a superoutburst and a newly triggered
superoutburst (see e.g. discussion for AL Com \cite{nog97alcom}).
A similar phenomenon was also observed in 1RXS J0232
(subsection \ref{sec:j0232}).

The $O-C$ apparently showed a break around $E = 240$ (corresponding
to a stage B--C transition), rather than a phase shift as
shown in \citet{uem08j1021}.  The mean period and $P_{\rm dot}$
for $E \le 240$ were 0.056312(12) d and $0.4(0.8) \times 10^{-5}$,
respectively.  The period after the transition was 0.056043(65),
which is probably identical to the newly appeared period of
0.055988(15) d during the fading tail \citep{uem08j1021}.
Although \citet{uem08j1021} attributed this period to a candidate
orbital period, the above behavior agrees with a transition to
a shorter superhump period, generally seen in SU UMa-type dwarf novae.

\begin{table}
\caption{Superhump maxima of OT J1021.}\label{tab:j1021oc2006}
\begin{center}

\end{center}
\end{table}

\subsection{OT J102637.0$+$475426}\label{obj:j1026}

\begin{figure}
  \begin{center}
    \FigureFile(88mm,110mm){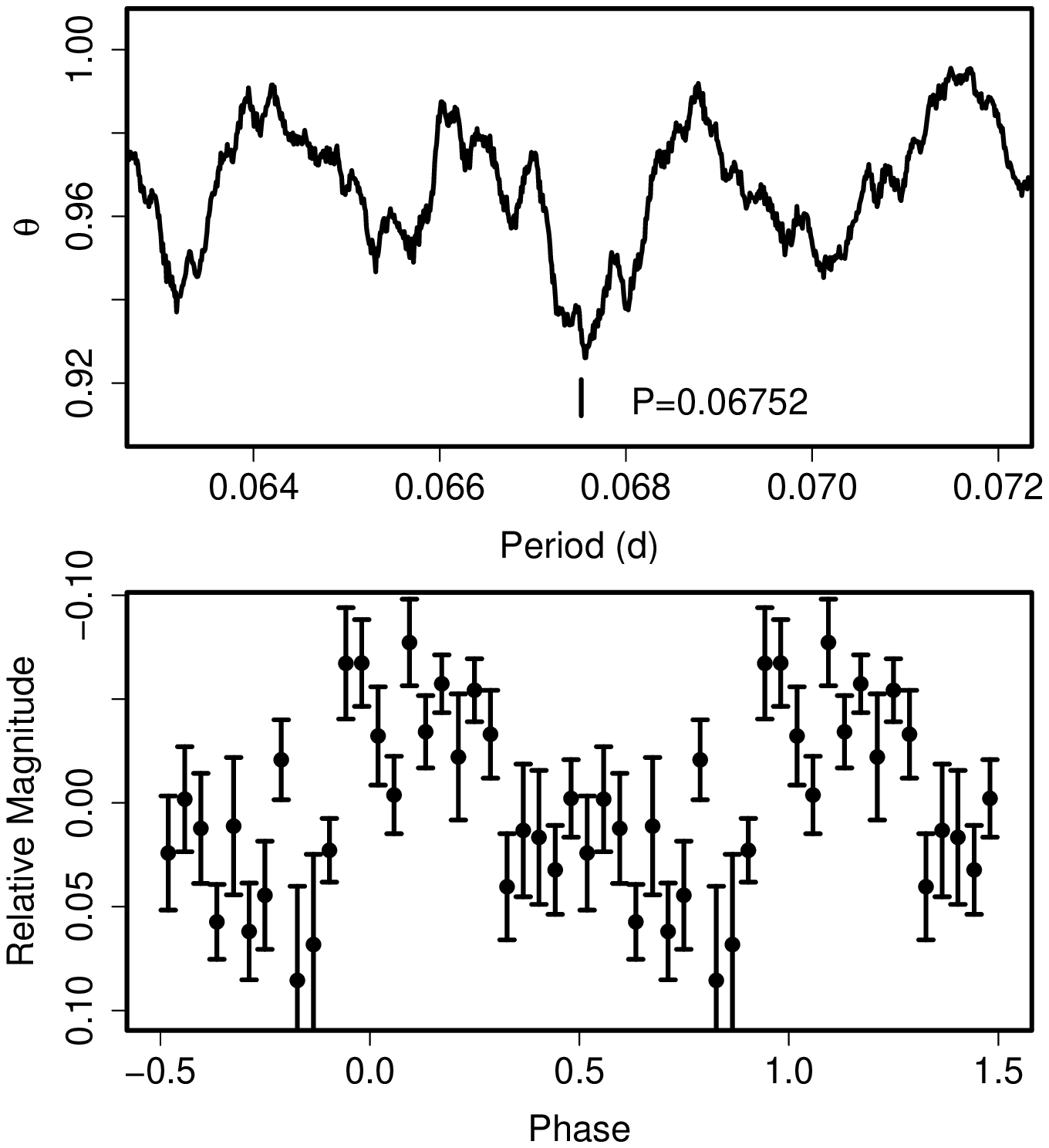}
  \end{center}
  \caption{Superhumps in OT J1026 (2009). (Upper): PDM analysis.
     (Lower): Phase-averaged profile.}
  \label{fig:j1026shpdm}
\end{figure}

   This object (hereafter OT J1026) was discovered by K. Itagaki
\citep{yam09j1026cbet1644}.  The SU UMa-type nature of this object was
immediately clarified (vsnet-alert 10882).  The object soon started fading,
indicating that the outburst was caught during its final stage.
A PDM analysis of the entire data set yielded a period of 0.06752(9) d
(figure \ref{fig:j1026shpdm}).
This period presumably corresponds to $P_2$.  The times of superhump
maxima are given in table \ref{tab:j1026oc2009}.

\begin{table}
\caption{Superhump maxima of OT J1026 (2009).}\label{tab:j1026oc2009}
\begin{center}
\begin{tabular}{ccccc}
\hline\hline
$E$ & max$^a$ & error & $O-C^b$ & $N^c$ \\
\hline
0 & 54835.1783 & 0.0137 & $-$0.0180 & 40 \\
1 & 54835.2703 & 0.0018 & 0.0064 & 72 \\
2 & 54835.3407 & 0.0026 & 0.0092 & 61 \\
28 & 54837.0918 & 0.0051 & 0.0034 & 28 \\
29 & 54837.1526 & 0.0035 & $-$0.0034 & 40 \\
30 & 54837.2300 & 0.0039 & 0.0065 & 67 \\
32 & 54837.3608 & 0.0019 & 0.0021 & 39 \\
46 & 54838.2987 & 0.0076 & $-$0.0061 & 71 \\
\hline
  \multicolumn{5}{l}{$^{a}$ BJD$-$2400000.} \\
  \multicolumn{5}{l}{$^{b}$ Against $max = 2454835.1963 + 0.067575 E$.} \\
  \multicolumn{5}{l}{$^{c}$ Number of points used to determine the maximum.} \\
\end{tabular}
\end{center}
\end{table}

\subsection{OT J102842.9$-$081927}\label{obj:j1028}

   This transient (=CSS090331:102843$-$081927, hereafter OT J1028)
was detected by the CRTS.  The object soon turned out to be
an ultrashort-period SU UMa-type dwarf nova (vsnet-alert 11149, 11158, 11164).
An unusual $V-J$ color was reported (vsnet-alert 11163).
A spectroscopic observation clarified its hydrogen-rich nature
(vsnet-alert 11166), suggesting that the object is similar to V485 Cen
and EI Psc.

\begin{figure}
  \begin{center}
    \FigureFile(88mm,110mm){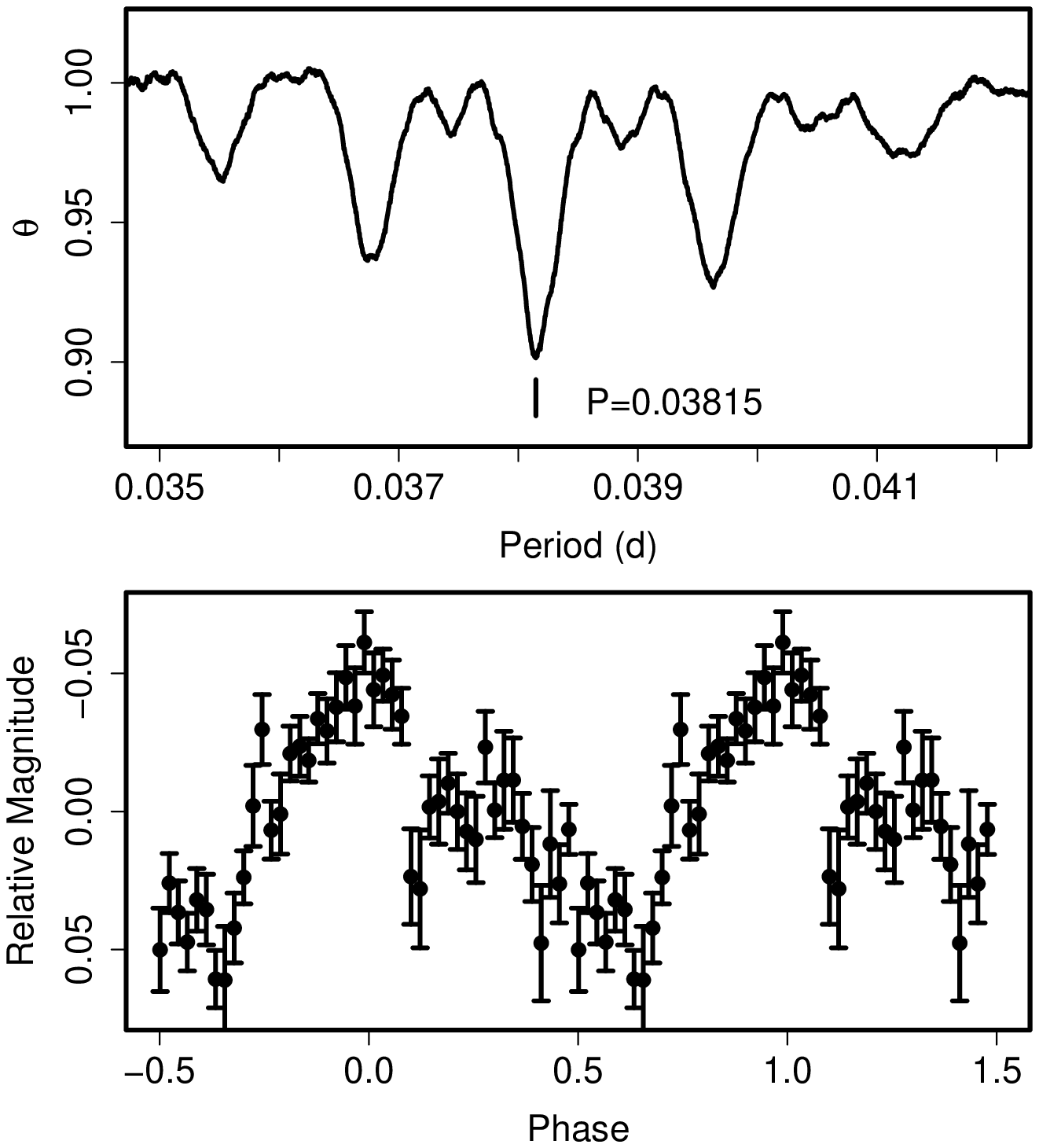}
  \end{center}
  \caption{Superhumps in OT J1028 (2009). (Upper): PDM analysis.
     (Lower): Phase-averaged profile.}
  \label{fig:j1028shpdm}
\end{figure}

   The times of superhump maxima are listed in table \ref{tab:j1028oc2009}.
The outburst was apparently observed during the relatively late stage
and the following decline phase.  Although we included times of maxima
after BJD 2454928 (decline phase) because of the continued detection
of the periodicity after the decline, this part of the data suffered from
the low signal-to-noise ratio.  We thus restricted to $E \le 59$ for
determining parameters, yielding a marginally positive
$P_{\rm dot}$ = $+11.6(8.5) \times 10^{-5}$.
A PDM analysis of the same interval yielded a period of 0.038147(14) d
(figure \ref{fig:j1028shpdm}).

\begin{table}
\caption{Superhump maxima of OT J1028 (2009).}\label{tab:j1028oc2009}
\begin{center}
\begin{tabular}{ccccc}
\hline\hline
$E$ & max$^a$ & error & $O-C^b$ & $N^c$ \\
\hline
0 & 54922.9883 & 0.0016 & 0.0015 & 56 \\
1 & 54923.0247 & 0.0009 & $-$0.0001 & 72 \\
2 & 54923.0621 & 0.0011 & $-$0.0009 & 72 \\
3 & 54923.0995 & 0.0010 & $-$0.0015 & 72 \\
4 & 54923.1380 & 0.0008 & $-$0.0012 & 72 \\
28 & 54924.0535 & 0.0021 & 0.0002 & 43 \\
29 & 54924.0918 & 0.0020 & 0.0005 & 119 \\
30 & 54924.1280 & 0.0010 & $-$0.0014 & 120 \\
31 & 54924.1655 & 0.0025 & $-$0.0020 & 119 \\
32 & 54924.2069 & 0.0017 & 0.0012 & 110 \\
52 & 54924.9632 & 0.0058 & $-$0.0042 & 42 \\
53 & 54925.0118 & 0.0038 & 0.0063 & 71 \\
54 & 54925.0446 & 0.0015 & 0.0010 & 71 \\
55 & 54925.0860 & 0.0009 & 0.0043 & 72 \\
56 & 54925.1237 & 0.0016 & 0.0039 & 71 \\
57 & 54925.1616 & 0.0014 & 0.0037 & 72 \\
58 & 54925.1983 & 0.0034 & 0.0023 & 70 \\
59 & 54925.2356 & 0.0008 & 0.0015 & 67 \\
134 & 54928.0912 & 0.0022 & 0.0005 & 62 \\
135 & 54928.1210 & 0.0051 & $-$0.0079 & 68 \\
159 & 54929.0377 & 0.0060 & $-$0.0053 & 50 \\
188 & 54930.1368 & 0.0054 & $-$0.0108 & 52 \\
189 & 54930.1747 & 0.0066 & $-$0.0110 & 66 \\
213 & 54931.0986 & 0.0017 & $-$0.0013 & 64 \\
214 & 54931.1529 & 0.0071 & 0.0150 & 46 \\
240 & 54932.1301 & 0.0032 & 0.0019 & 58 \\
423 & 54939.1023 & 0.0029 & 0.0037 & 44 \\
\hline
  \multicolumn{5}{l}{$^{a}$ BJD$-$2400000.} \\
  \multicolumn{5}{l}{$^{b}$ Against $max = 2454922.9868 + 0.038089 E$.} \\
  \multicolumn{5}{l}{$^{c}$ Number of points used to determine the maximum.} \\
\end{tabular}
\end{center}
\end{table}

\subsection{OT J111217.4$-$353829}\label{obj:j1112}

\begin{figure}
  \begin{center}
    \FigureFile(88mm,110mm){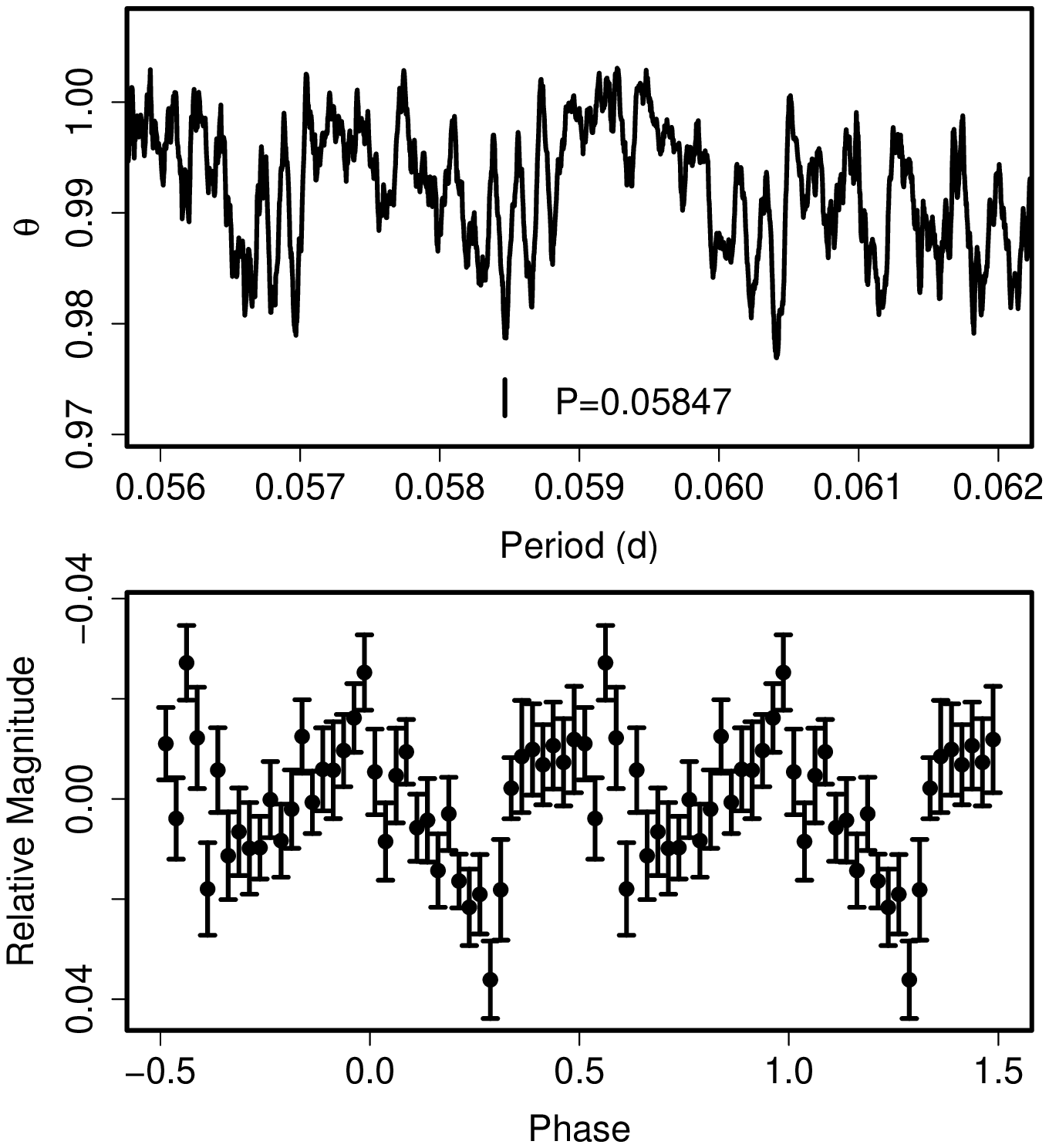}
  \end{center}
  \caption{Early superhumps in OT J1112 (2007). (Upper): PDM analysis.
     (Lower): Phase-averaged profile.}
  \label{fig:j1112eshpdm}
\end{figure}

\begin{figure}
  \begin{center}
    \FigureFile(88mm,110mm){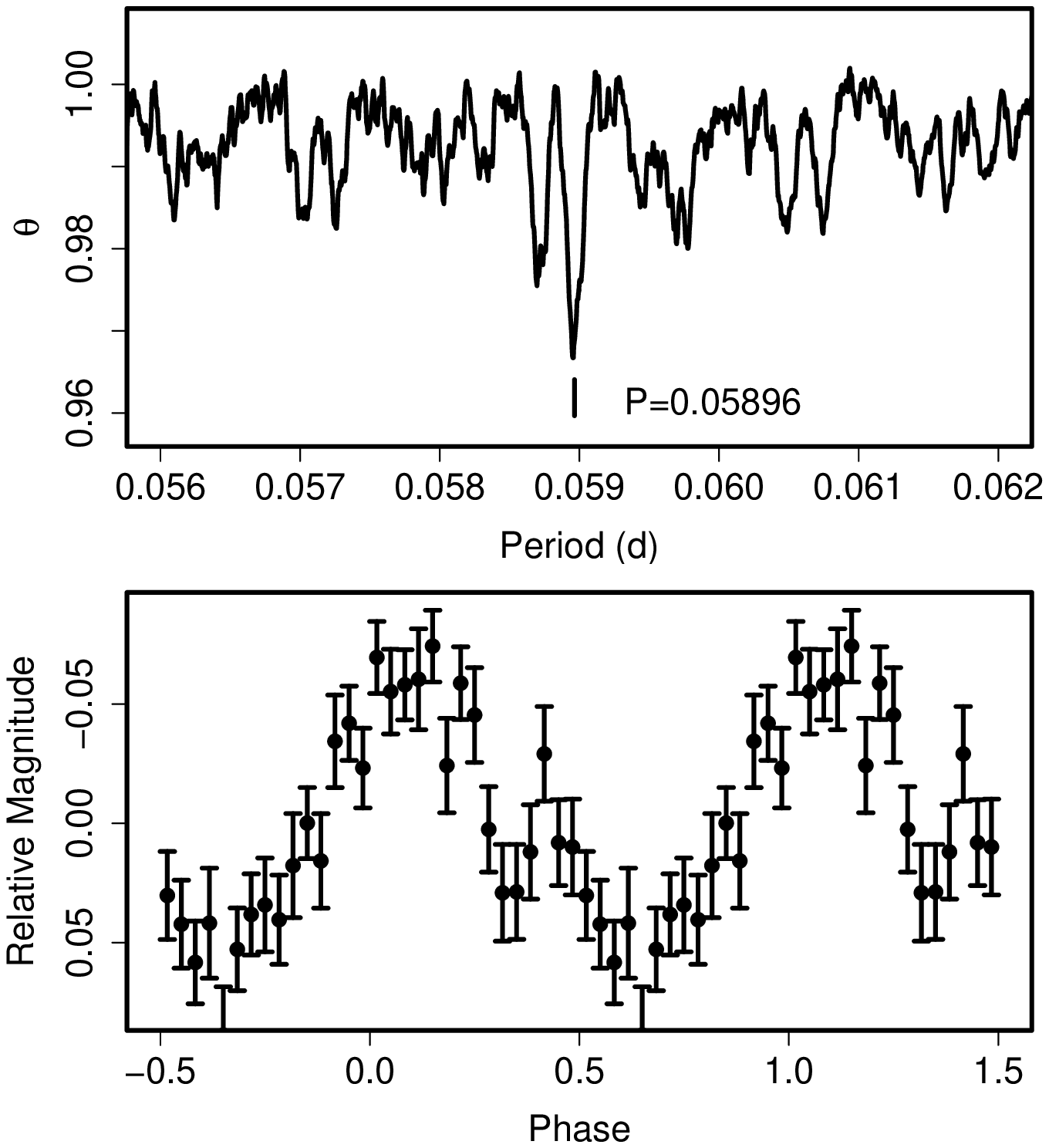}
  \end{center}
  \caption{Ordinary superhumps in OT J1112 (2007). (Upper): PDM analysis.
     (Lower): Phase-averaged profile.}
  \label{fig:j1112shpdm}
\end{figure}

   This object (hereafter OT J1112) was detected
by ``Pi of the Sky'' and its dwarf nova-type
nature was confirmed (vsnet-alert 9764, 9767, 9769, 9770, 9771).
The detection of early superhumps and ordinary superhumps led to
a classification of a typical WZ Sge-type dwarf nova (vsnet-alert
9775, 9806).  The presence of He\textsc{II} and C\textsc{IV}
emission lines in the spectrum was also very similar to WZ Sge
(vsnet-alert 9782).
The times of superhump maxima are listed are in table \ref{tab:j1112oc2007}.
The change in the superhump period was very small,
$P_{\rm dot}$ = $+0.5(0.3) \times 10^{-5}$, similar to WZ Sge
itself.  The mean periods of early and ordinary superhumps,
determined with the PDM method, were
0.05847(2) d (figure \ref{fig:j1112eshpdm}) and
0.058965(9) d (figure \ref{fig:j1112shpdm}),
respectively.  This $P_{\rm SH}$ is adopted in table \ref{tab:perlist}.
The fractional superhump excess was estimated to be 0.8(1) \%,
also very typical for a WZ Sge-type dwarf nova.
More detailed analysis will be presented in Maehara et al.,
in preparation.

\begin{table}
\caption{Superhump maxima of OT J1112 (2007--2008).}\label{tab:j1112oc2007}
\begin{center}
\begin{tabular}{ccccc}
\hline\hline
$E$ & max$^a$ & error & $O-C^b$ & $N^c$ \\
\hline
0 & 54475.3297 & 0.0027 & $-$0.0009 & 124 \\
16 & 54476.2778 & 0.0010 & 0.0030 & 281 \\
82 & 54480.1717 & 0.0008 & 0.0023 & 61 \\
83 & 54480.2255 & 0.0019 & $-$0.0029 & 96 \\
84 & 54480.2868 & 0.0009 & $-$0.0007 & 233 \\
85 & 54480.3474 & 0.0019 & 0.0009 & 303 \\
116 & 54482.1778 & 0.0010 & 0.0020 & 61 \\
117 & 54482.2350 & 0.0014 & 0.0002 & 61 \\
118 & 54482.2934 & 0.0007 & $-$0.0004 & 61 \\
119 & 54482.3517 & 0.0008 & $-$0.0011 & 51 \\
218 & 54488.1932 & 0.0080 & $-$0.0016 & 39 \\
219 & 54488.2539 & 0.0028 & 0.0001 & 61 \\
220 & 54488.3062 & 0.0014 & $-$0.0066 & 60 \\
221 & 54488.3695 & 0.0024 & $-$0.0023 & 33 \\
253 & 54490.2588 & 0.0019 & $-$0.0014 & 130 \\
254 & 54490.3219 & 0.0042 & 0.0028 & 112 \\
269 & 54491.2046 & 0.0021 & 0.0003 & 36 \\
270 & 54491.2656 & 0.0036 & 0.0023 & 108 \\
287 & 54492.2705 & 0.0014 & 0.0040 & 135 \\
\hline
  \multicolumn{5}{l}{$^{a}$ BJD$-$2400000.} \\
  \multicolumn{5}{l}{$^{b}$ Against $max = 2454475.3306 + 0.059010 E$.} \\
  \multicolumn{5}{l}{$^{c}$ Number of points used to determine the maximum.} \\
\end{tabular}
\end{center}
\end{table}

\subsection{OT J130030.3$+$115101}\label{obj:j1300}

\begin{figure}
  \begin{center}
    \FigureFile(88mm,110mm){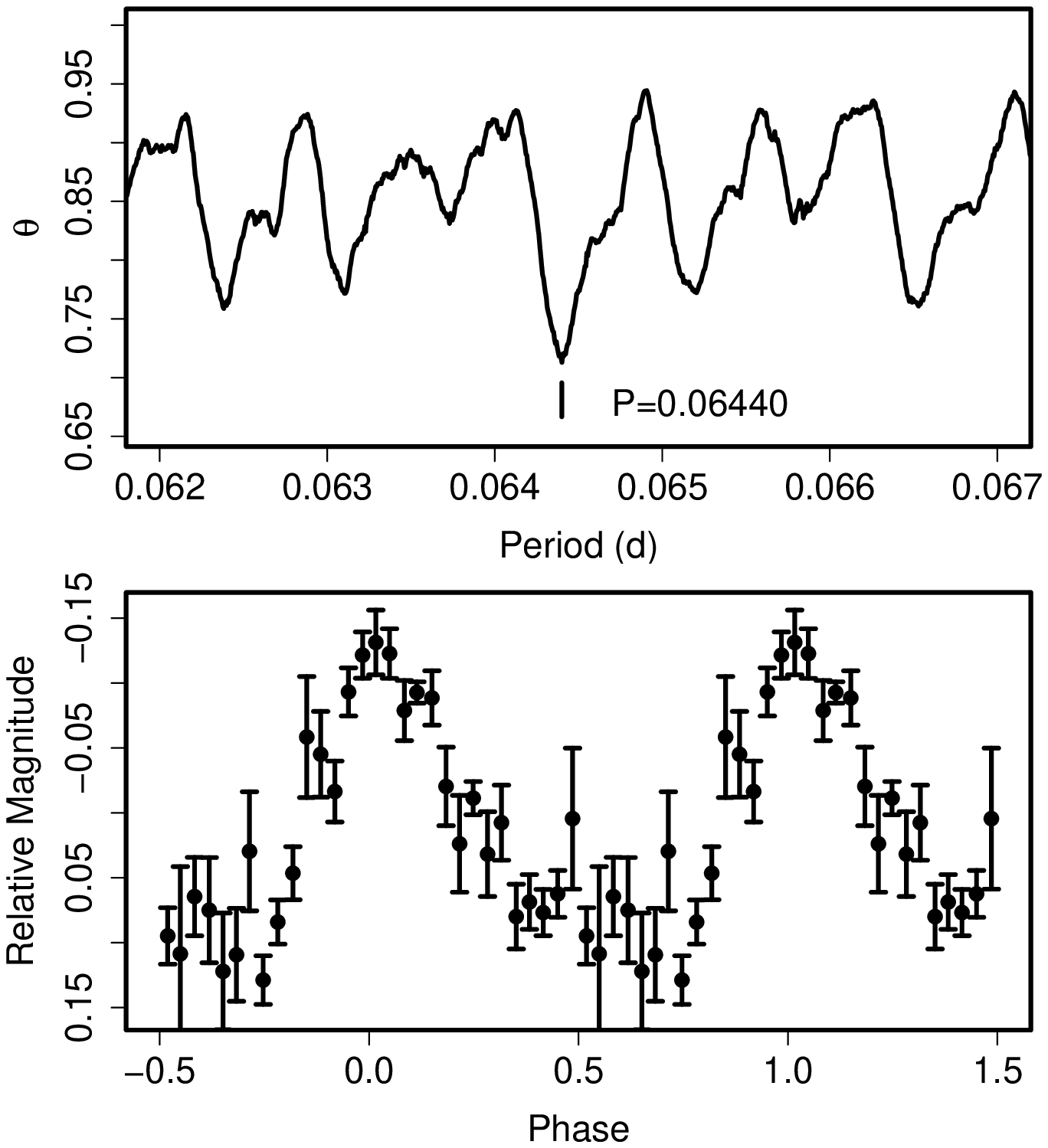}
  \end{center}
  \caption{Superhumps in OT J1300 (2008) after BJD 2454653.9.
     (Upper): PDM analysis.
     (Lower): Phase-averaged profile.}
  \label{fig:j1300shpdm}
\end{figure}

   This transient (=CSS080702:130030$+$115101, hereafter OT J1300)
was detected by the CRTS.
Independent detections by ASAS-3 suggested a superoutburst of an SU UMa-type
dwarf nova (vsnet-alert 10300).  Five days after the maximum, the object
showed superhumps (vsnet-alert 10311).
The mean superhump period with the PDM method was 0.06440(2) d
(figure \ref{fig:j1300shpdm})
The times of superhump maxima are listed in table \ref{tab:j1300oc2008}.
The epoch $E = 0$ corresponded
to a growing stage of superhumps.  Disregarding this epoch, we obtained
$P_{\rm dot}$ = $+14.4(1.5) \times 10^{-5}$.

\begin{table}
\caption{Superhump maxima of OT J1300 (2008).}\label{tab:j1300oc2008}
\begin{center}
\begin{tabular}{ccccc}
\hline\hline
$E$ & max$^a$ & error & $O-C^b$ & $N^c$ \\
\hline
0 & 54653.0315 & 0.0007 & $-$0.0016 & 120 \\
14 & 54653.9374 & 0.0012 & 0.0027 & 44 \\
15 & 54653.9998 & 0.0003 & 0.0007 & 108 \\
16 & 54654.0645 & 0.0008 & 0.0011 & 88 \\
77 & 54657.9847 & 0.0013 & $-$0.0069 & 51 \\
93 & 54659.0198 & 0.0039 & $-$0.0021 & 43 \\
108 & 54659.9915 & 0.0010 & 0.0037 & 80 \\
109 & 54660.0547 & 0.0014 & 0.0024 & 50 \\
\hline
  \multicolumn{5}{l}{$^{a}$ BJD$-$2400000.} \\
  \multicolumn{5}{l}{$^{b}$ Against $max = 2454653.0331 + 0.064396 E$.} \\
  \multicolumn{5}{l}{$^{c}$ Number of points used to determine the maximum.} \\
\end{tabular}
\end{center}
\end{table}

\subsection{OT J144011.0$+$494734}\label{obj:j1440}

\begin{figure}
  \begin{center}
    \FigureFile(88mm,110mm){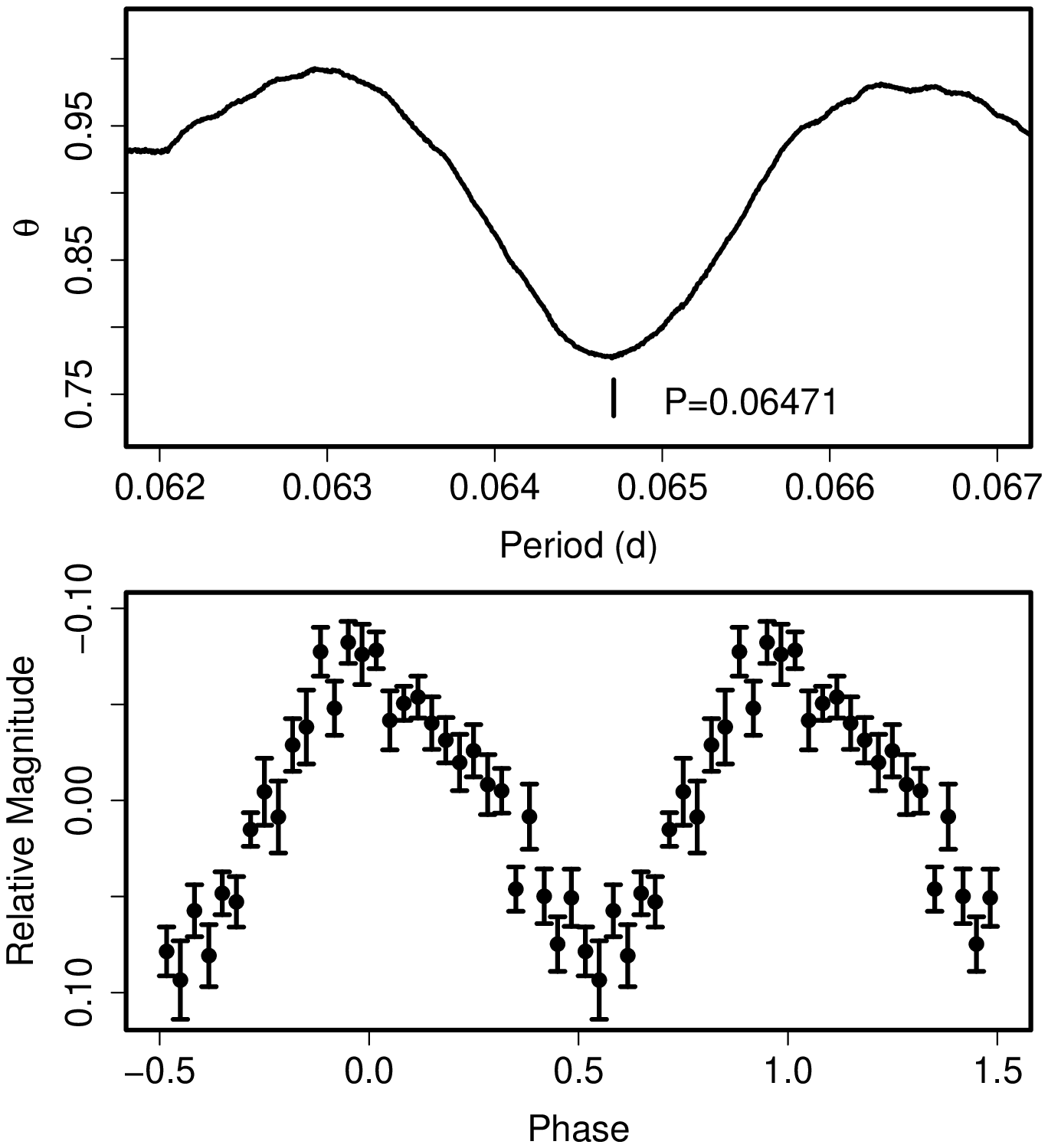}
  \end{center}
  \caption{Superhumps in OT J1440 (2009).
     (Upper): PDM analysis.
     (Lower): Phase-averaged profile.}
  \label{fig:j1440shpdm}
\end{figure}

   This transient (=CSS090530:144011$+$494734, hereafter OT J1440) was
detected by the CRTS.
The detection of superhumps confirmed the SU UMa-type classification
(vsnet-outburst 10297, vsnet-alert 11283).
The mean superhump period with the PDM method was 0.06471(5) d
(figure \ref{fig:j1440shpdm})
The times of superhump maxima are listed in table \ref{tab:j1440oc2009}.
Although there was a clear break in the $O-C$ diagram between $E=1$ and
$E=15$, it was unclear whether this break is attributed to stage A--B or
stage B--C transition.  We adopted the latter interpretation because the
period was almost constant after the break.

\begin{table}
\caption{Superhump maxima of OT J1440 (2009).}\label{tab:j1440oc2009}
\begin{center}
\begin{tabular}{ccccc}
\hline\hline
$E$ & max$^a$ & error & $O-C^b$ & $N^c$ \\
\hline
0 & 54983.0238 & 0.0031 & $-$0.0045 & 69 \\
1 & 54983.0935 & 0.0121 & 0.0003 & 64 \\
15 & 54984.0043 & 0.0016 & 0.0025 & 188 \\
16 & 54984.0689 & 0.0011 & 0.0023 & 224 \\
22 & 54984.4581 & 0.0005 & 0.0021 & 92 \\
23 & 54984.5209 & 0.0009 & $-$0.0001 & 101 \\
24 & 54984.5867 & 0.0006 & 0.0009 & 74 \\
37 & 54985.4313 & 0.0013 & 0.0018 & 73 \\
38 & 54985.4918 & 0.0007 & $-$0.0027 & 106 \\
39 & 54985.5566 & 0.0009 & $-$0.0027 & 104 \\
\hline
  \multicolumn{5}{l}{$^{a}$ BJD$-$2400000.} \\
  \multicolumn{5}{l}{$^{b}$ Against $max = 2454983.0283 + 0.064899 E$.} \\
  \multicolumn{5}{l}{$^{c}$ Number of points used to determine the maximum.} \\
\end{tabular}
\end{center}
\end{table}

\subsection{OT J144341.9$-$175550}\label{obj:j1443}

\begin{figure}
  \begin{center}
    \FigureFile(88mm,110mm){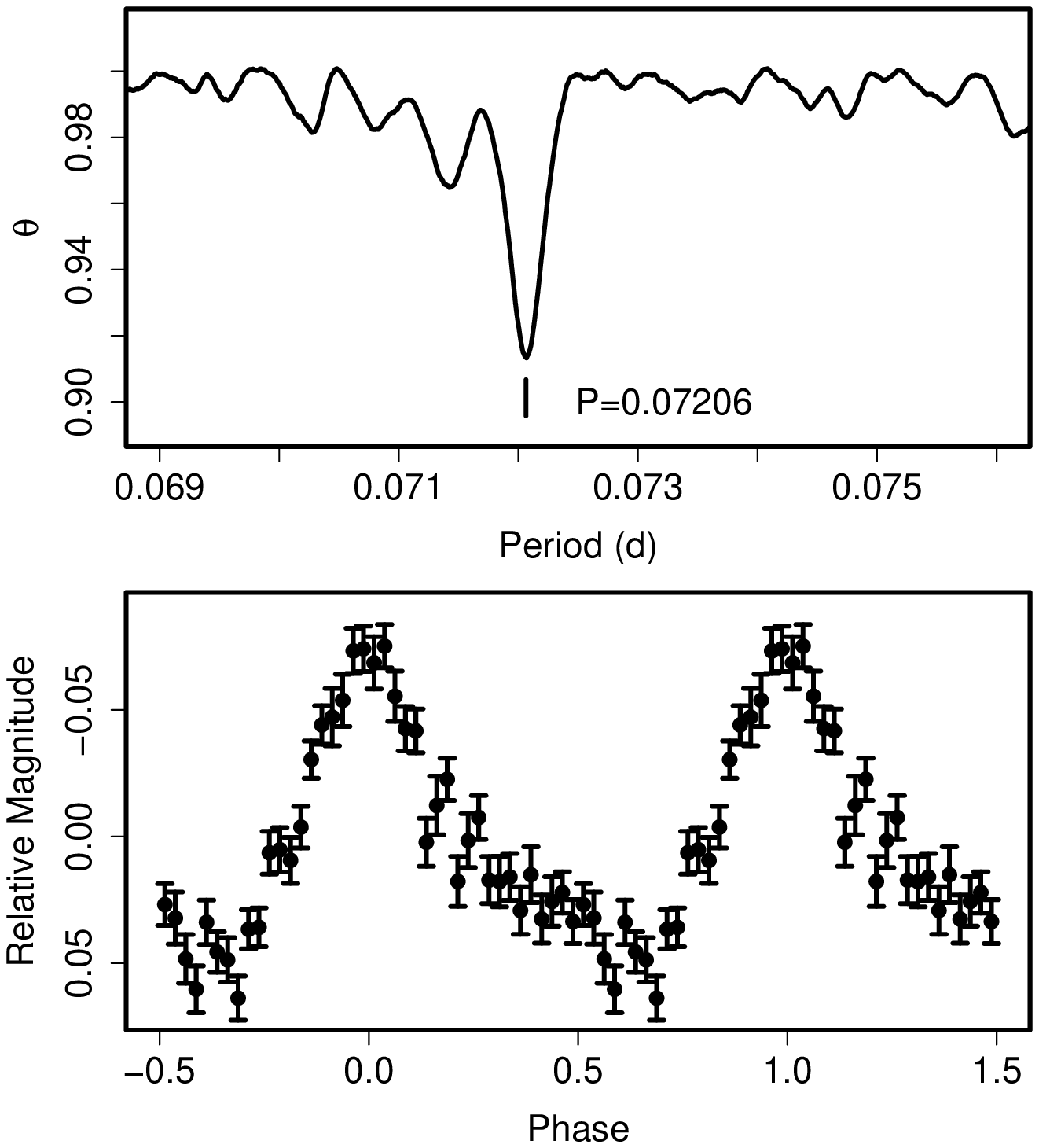}
  \end{center}
  \caption{Superhumps in OT J1443 (2009). (Upper): PDM analysis.
     (Lower): Phase-averaged profile.}
  \label{fig:j1443shpdm}
\end{figure}

   This transient (=CSS090418:144342$-$175550, hereafter OT J1443) was
detected by the CRTS.
The detection of superhumps confirmed the SU UMa-type classification
(vsnet-alert 11193, 11195, 11196, 11199, 11219).  The times of superhump
maxima are listed in table \ref{tab:j1443oc2009}.
Thanks to the early detection of the outburst, all stages A--C were
recorded.  The $P_{\rm dot}$ during the stage B was
$+11.0(1.3) \times 10^{-5}$ ($12 \le E \le 112$).
Other parameters are listed in table \ref{tab:perlist}.
The mean $P_{\rm SH}$ over the entire superoutburst was 0.072065(10) d
(PDM method, figure \ref{fig:j1443shpdm}).

\begin{table}
\caption{Superhump maxima of OT J1443 (2009).}\label{tab:j1443oc2009}
\begin{center}
\begin{tabular}{ccccc}
\hline\hline
$E$ & max$^a$ & error & $O-C^b$ & $N^c$ \\
\hline
0 & 54940.1823 & 0.0024 & $-$0.0308 & 134 \\
1 & 54940.2643 & 0.0019 & $-$0.0209 & 151 \\
12 & 54941.0808 & 0.0009 & 0.0025 & 42 \\
13 & 54941.1551 & 0.0009 & 0.0047 & 179 \\
14 & 54941.2256 & 0.0004 & 0.0031 & 75 \\
15 & 54941.2986 & 0.0009 & 0.0040 & 43 \\
26 & 54942.0921 & 0.0004 & 0.0044 & 58 \\
27 & 54942.1640 & 0.0004 & 0.0042 & 72 \\
28 & 54942.2359 & 0.0004 & 0.0040 & 74 \\
29 & 54942.3077 & 0.0003 & 0.0037 & 74 \\
30 & 54942.3790 & 0.0004 & 0.0029 & 60 \\
54 & 54944.1074 & 0.0004 & 0.0009 & 74 \\
55 & 54944.1785 & 0.0005 & $-$0.0001 & 75 \\
56 & 54944.2504 & 0.0004 & $-$0.0003 & 74 \\
57 & 54944.3236 & 0.0007 & 0.0008 & 48 \\
68 & 54945.1178 & 0.0006 & 0.0019 & 152 \\
69 & 54945.1862 & 0.0006 & $-$0.0017 & 261 \\
70 & 54945.2608 & 0.0021 & 0.0007 & 146 \\
95 & 54947.0691 & 0.0012 & 0.0066 & 119 \\
110 & 54948.1540 & 0.0005 & 0.0100 & 115 \\
111 & 54948.2340 & 0.0011 & 0.0178 & 120 \\
112 & 54948.3033 & 0.0018 & 0.0150 & 71 \\
123 & 54949.0852 & 0.0011 & 0.0038 & 141 \\
124 & 54949.1598 & 0.0010 & 0.0064 & 246 \\
125 & 54949.2346 & 0.0010 & 0.0090 & 184 \\
137 & 54950.0927 & 0.0010 & 0.0019 & 151 \\
138 & 54950.1625 & 0.0009 & $-$0.0003 & 314 \\
139 & 54950.2310 & 0.0015 & $-$0.0040 & 260 \\
151 & 54951.0961 & 0.0060 & $-$0.0040 & 319 \\
152 & 54951.1674 & 0.0009 & $-$0.0048 & 393 \\
153 & 54951.2305 & 0.0034 & $-$0.0139 & 209 \\
165 & 54952.1111 & 0.0063 & 0.0015 & 314 \\
166 & 54952.1698 & 0.0037 & $-$0.0119 & 242 \\
179 & 54953.1123 & 0.0021 & $-$0.0066 & 180 \\
180 & 54953.1803 & 0.0021 & $-$0.0107 & 285 \\
\hline
  \multicolumn{5}{l}{$^{a}$ BJD$-$2400000.} \\
  \multicolumn{5}{l}{$^{b}$ Against $max = 2454940.2131 + 0.072099 E$.} \\
  \multicolumn{5}{l}{$^{c}$ Number of points used to determine the maximum.} \\
\end{tabular}
\end{center}
\end{table}

\subsection{OT J163120.9$+$103134}\label{obj:j1631}

\begin{figure}
  \begin{center}
    \FigureFile(88mm,110mm){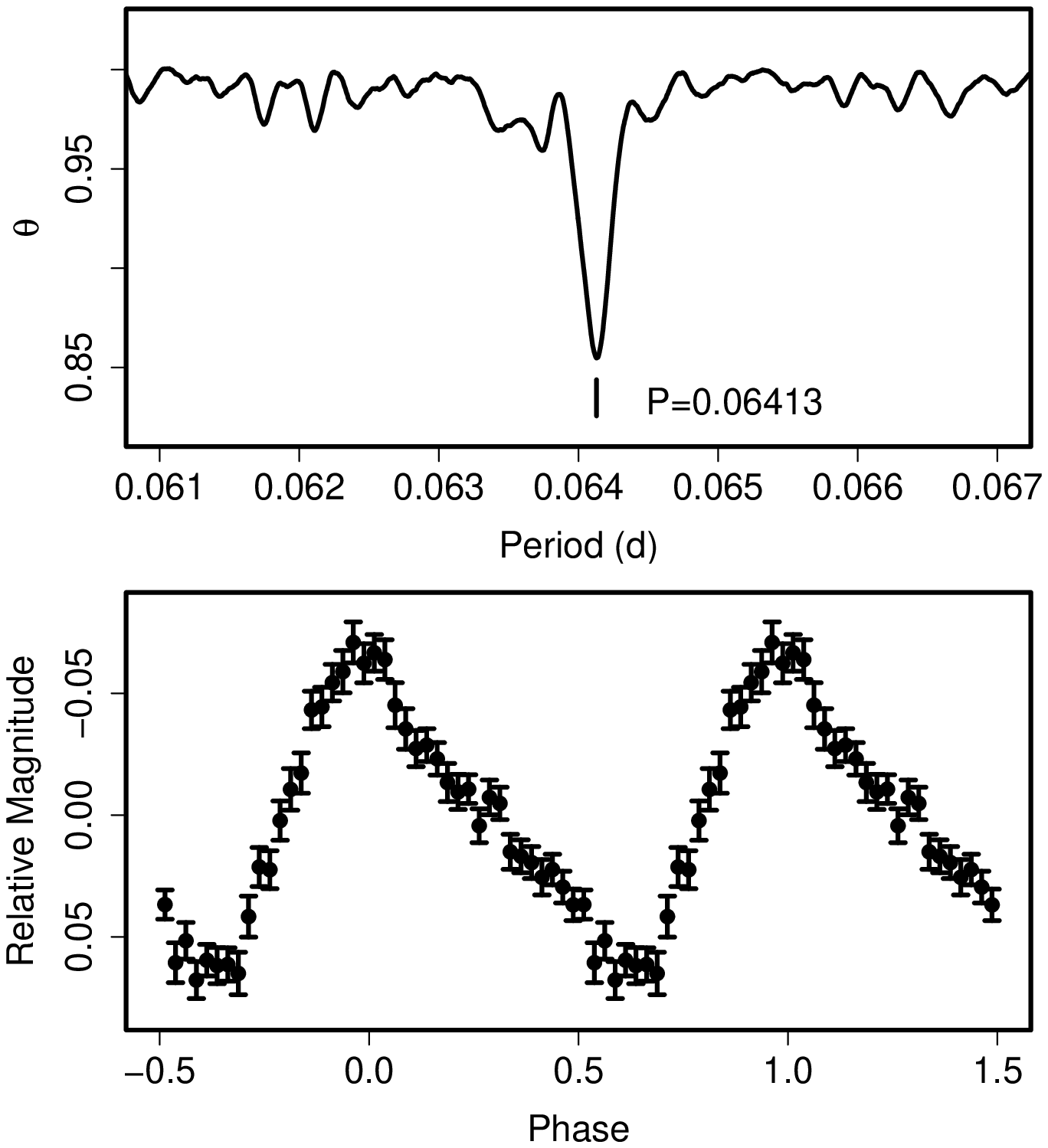}
  \end{center}
  \caption{Superhumps in OT J1631 (2008).
     (Upper): PDM analysis.
     (Lower): Phase-averaged profile.}
  \label{fig:j1631shpdm}
\end{figure}

   This transient (=CSS080505:163121$+$103134, hereafter OT J1631) was
discovered by the CRTS \citep{CRTS}.
Soon after the discovery announcement, past outbursts from ASAS-3 records
and the ROSAT identification were noticed (cvnet-discussion 1136,
vsnet-alert 10159).  The detection of superhump led to secure classification
of this object.  \citet{mah08atel1520} presented a spectroscopical
confirmation as a CV.
The mean superhump period with the PDM method was 0.064129(5) d
(figure \ref{fig:j1631shpdm}).
The times of superhump maxima are listed in
table \ref{tab:j1631oc2008}.  The $O-C$ diagram showed a clear positive
period derivative ($E \le 96$) before a transition to the stage C,
behavior typical for this superhump period (cf. figure \ref{fig:octrans}).
We obtained $P_{\rm dot}$ = $+12.5(1.3) \times 10^{-5}$ for the stage B.

\begin{table}
\caption{Superhump maxima of OT J1631 (2008).}\label{tab:j1631oc2008}
\begin{center}
\begin{tabular}{ccccc}
\hline\hline
$E$ & max$^a$ & error & $O-C^b$ & $N^c$ \\
\hline
0 & 54592.4052 & 0.0004 & 0.0062 & 247 \\
1 & 54592.4679 & 0.0005 & 0.0048 & 246 \\
2 & 54592.5331 & 0.0014 & 0.0059 & 135 \\
16 & 54593.4275 & 0.0004 & 0.0024 & 218 \\
17 & 54593.4905 & 0.0005 & 0.0012 & 200 \\
18 & 54593.5549 & 0.0006 & 0.0014 & 103 \\
27 & 54594.1302 & 0.0012 & $-$0.0005 & 66 \\
28 & 54594.1947 & 0.0012 & $-$0.0001 & 67 \\
31 & 54594.3860 & 0.0006 & $-$0.0013 & 184 \\
32 & 54594.4492 & 0.0005 & $-$0.0022 & 201 \\
33 & 54594.5102 & 0.0006 & $-$0.0054 & 156 \\
46 & 54595.3437 & 0.0009 & $-$0.0056 & 122 \\
47 & 54595.4096 & 0.0009 & $-$0.0039 & 204 \\
48 & 54595.4725 & 0.0011 & $-$0.0052 & 205 \\
63 & 54596.4348 & 0.0009 & $-$0.0049 & 84 \\
64 & 54596.5012 & 0.0012 & $-$0.0027 & 76 \\
65 & 54596.5619 & 0.0060 & $-$0.0061 & 52 \\
77 & 54597.3381 & 0.0012 & 0.0004 & 105 \\
78 & 54597.4012 & 0.0024 & $-$0.0006 & 182 \\
79 & 54597.4648 & 0.0011 & $-$0.0011 & 193 \\
80 & 54597.5281 & 0.0020 & $-$0.0020 & 175 \\
81 & 54597.5935 & 0.0010 & $-$0.0007 & 70 \\
89 & 54598.1132 & 0.0010 & 0.0058 & 126 \\
90 & 54598.1716 & 0.0083 & 0.0001 & 86 \\
91 & 54598.2398 & 0.0013 & 0.0042 & 137 \\
94 & 54598.4314 & 0.0011 & 0.0033 & 83 \\
95 & 54598.4930 & 0.0008 & 0.0008 & 82 \\
96 & 54598.5581 & 0.0025 & 0.0018 & 47 \\
109 & 54599.3918 & 0.0009 & 0.0017 & 72 \\
110 & 54599.4507 & 0.0009 & $-$0.0036 & 81 \\
125 & 54600.4145 & 0.0034 & $-$0.0019 & 76 \\
137 & 54601.1945 & 0.0030 & 0.0084 & 185 \\
138 & 54601.2497 & 0.0010 & $-$0.0005 & 135 \\
\hline
  \multicolumn{5}{l}{$^{a}$ BJD$-$2400000.} \\
  \multicolumn{5}{l}{$^{b}$ Against $max = 2454592.3989 + 0.064140 E$.} \\
  \multicolumn{5}{l}{$^{c}$ Number of points used to determine the maximum.} \\
\end{tabular}
\end{center}
\end{table}

\subsection{OT J191443.6$+$605214}\label{obj:j1914}

\begin{figure}
  \begin{center}
    \FigureFile(88mm,110mm){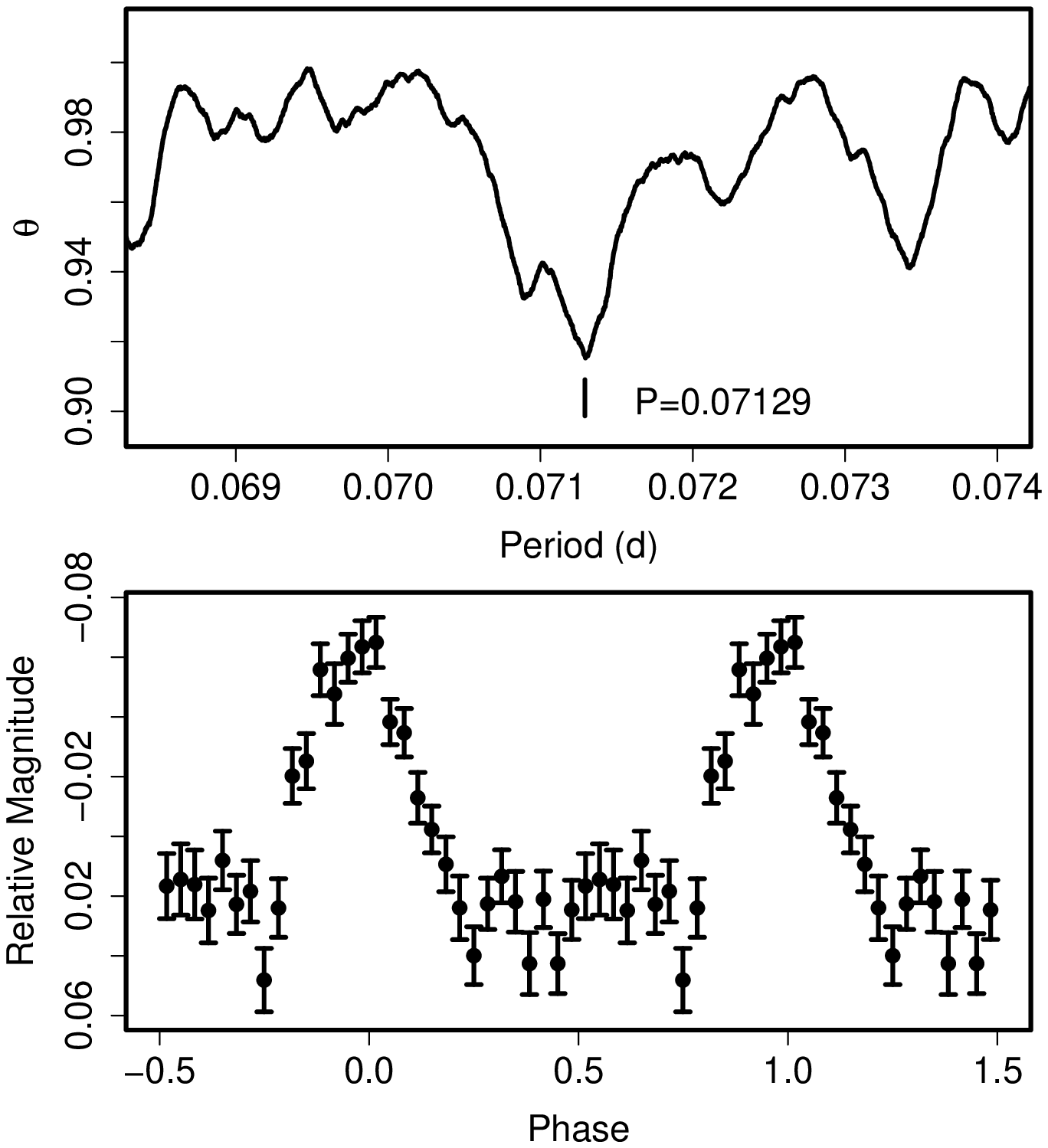}
  \end{center}
  \caption{Superhumps in OT J1914 (2008, plateau phase). (Upper): PDM analysis.
     (Lower): Phase-averaged profile.}
  \label{fig:j1914shpdm}
\end{figure}

   This transient (hereafter OT J1914) was detected by K. Itagaki
\citep{yam08j1914cbet1535}.  The SU UMa-type nature of this object
was soon established (vsnet-alert 10558).
The mean superhump period during the entire plateau phase was
0.071292(14) d (PDM method, figure \ref{fig:j1914shpdm}).
The times of superhump maxima are listed in table \ref{tab:j1914oc2008}.
A stage B--C transition was recorded around $E = 82$.
The mean $P_{\rm SH}$ and $P_{\rm dot}$ during the stage B were
0.07134(3) d and $+9.7(2.6) \times 10^{-5}$, respectively.
\citet{boy09j1914} reported $P_{\rm dot}$ = $+3.4(2.0) \times 10^{-5}$
using a slightly different treatment and data set.

\begin{table}
\caption{Superhump maxima of OT J1914.}\label{tab:j1914oc2008}
\begin{center}
\begin{tabular}{ccccc}
\hline\hline
$E$ & max$^a$ & error & $O-C^b$ & $N^c$ \\
\hline
0 & 54743.0965 & 0.0021 & $-$0.0093 & 128 \\
1 & 54743.1697 & 0.0023 & $-$0.0073 & 64 \\
40 & 54745.9467 & 0.0013 & $-$0.0050 & 152 \\
41 & 54746.0178 & 0.0012 & $-$0.0050 & 136 \\
68 & 54747.9442 & 0.0013 & 0.0003 & 291 \\
69 & 54748.0183 & 0.0010 & 0.0032 & 285 \\
70 & 54748.0888 & 0.0011 & 0.0026 & 215 \\
72 & 54748.2343 & 0.0006 & 0.0058 & 54 \\
73 & 54748.3078 & 0.0012 & 0.0082 & 23 \\
74 & 54748.3753 & 0.0016 & 0.0045 & 26 \\
82 & 54748.9491 & 0.0010 & 0.0091 & 203 \\
83 & 54749.0172 & 0.0015 & 0.0060 & 182 \\
84 & 54749.0870 & 0.0013 & 0.0048 & 147 \\
87 & 54749.3001 & 0.0019 & 0.0044 & 25 \\
88 & 54749.3682 & 0.0039 & 0.0013 & 26 \\
91 & 54749.5881 & 0.0012 & 0.0077 & 16 \\
100 & 54750.2198 & 0.0017 & $-$0.0008 & 28 \\
101 & 54750.2931 & 0.0006 & 0.0013 & 49 \\
105 & 54750.5809 & 0.0025 & 0.0045 & 15 \\
111 & 54751.0003 & 0.0021 & $-$0.0030 & 154 \\
112 & 54751.0633 & 0.0065 & $-$0.0111 & 31 \\
116 & 54751.3613 & 0.0014 & 0.0023 & 25 \\
117 & 54751.4338 & 0.0024 & 0.0036 & 25 \\
119 & 54751.5767 & 0.0030 & 0.0043 & 16 \\
124 & 54751.9109 & 0.0029 & $-$0.0173 & 54 \\
125 & 54752.0113 & 0.0058 & 0.0119 & 42 \\
126 & 54752.0664 & 0.0054 & $-$0.0041 & 29 \\
138 & 54752.9118 & 0.0041 & $-$0.0125 & 131 \\
139 & 54752.9948 & 0.0031 & $-$0.0007 & 302 \\
140 & 54753.0569 & 0.0030 & $-$0.0097 & 91 \\
\hline
  \multicolumn{5}{l}{$^{a}$ BJD$-$2400000.} \\
  \multicolumn{5}{l}{$^{b}$ Against $max = 2454743.1058 + 0.071148 E$.} \\
  \multicolumn{5}{l}{$^{c}$ Number of points used to determine the maximum.} \\
\end{tabular}
\end{center}
\end{table}

\subsection{OT J195951.3$+$224232}\label{obj:j1959}

   This object (also called Var Vul 05, hereafter OT J1959)
was discovered by J. Hanisch (vsnet-alert 8629; \cite{ren05j1959iauc8591}).
Subsequent observations
confirmed the presence of superhumps (cvnet-outburst 543, vsnet-alert
8640).  The large outburst amplitude ($\sim$ 8 mag, vsnet-alert 8654)
makes the object an excellent candidate for a WZ Sge-type dwarf nova.

   The object underwent another recorded outburst in 2008 April.\footnote{
$<$http://tech.groups.yahoo.com/group/VarVul05/message/98$>$.
}  The recurrence time may be an order of $\sim$ 1000 d.

\begin{figure}
  \begin{center}
    \FigureFile(88mm,110mm){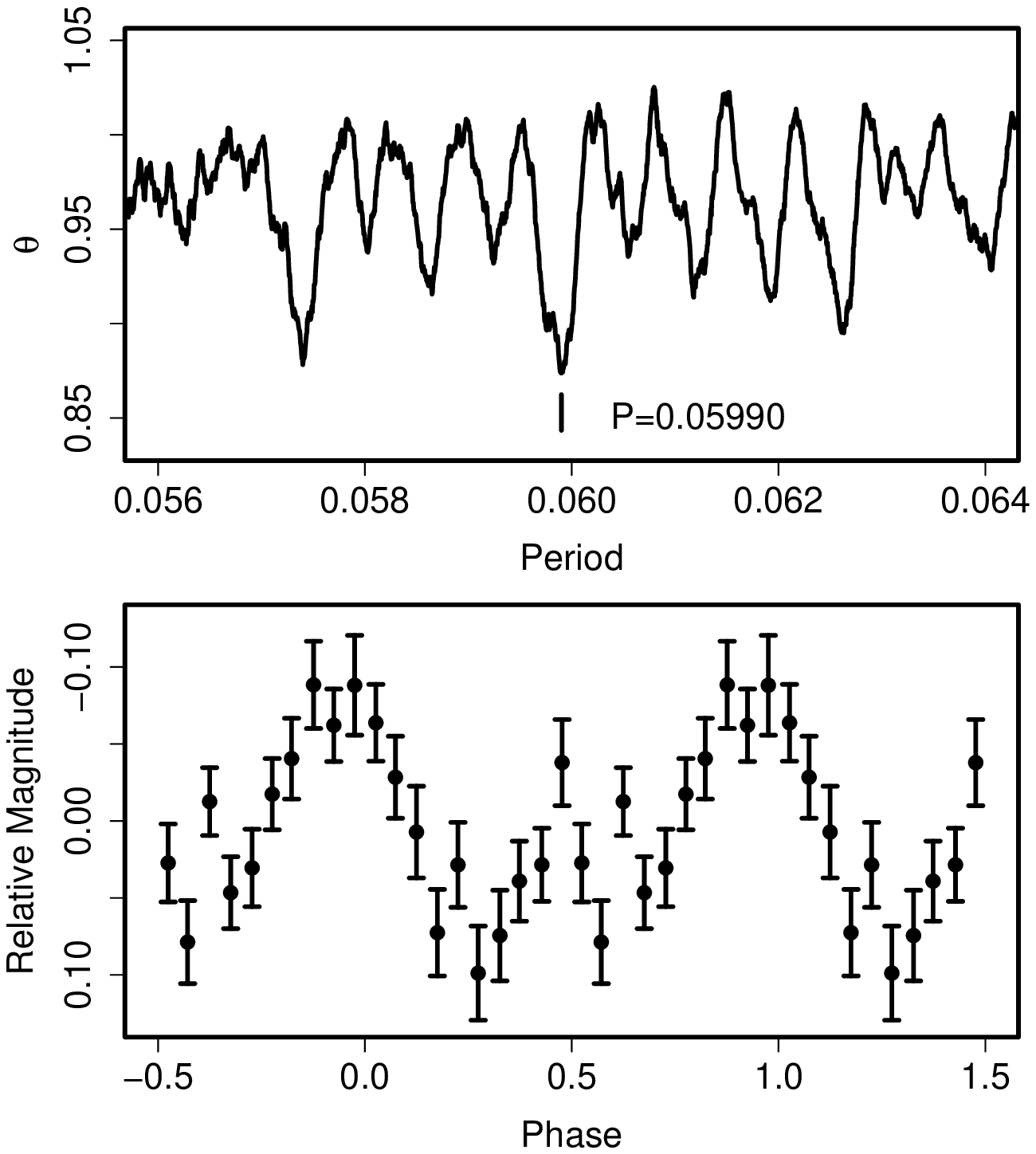}
  \end{center}
  \caption{Superhumps in OT J1959 (2005). (Upper): PDM analysis.
     (Lower): Phase-averaged profile.}
  \label{fig:j1959shpdm}
\end{figure}

We adopted a mean superhump period of 0.05990(3) d
(figure \ref{fig:j1959shpdm}).  Although there was some hint of
double-wave modulations suggesting early superhumps, the large amplitude
of the modulations and the epoch of the observation ($>$ 6 d after
the outburst detection) suggest the identification of these humps
as ordinary superhumps.
The times of superhump maxima are listed in table \ref{tab:j1959oc2005}.
The resultant $P_{\rm dot}$ is virtually zero, $-0.7(5.2) \times 10^{-5}$.

Although this object is provisionally listed as a WZ Sge-type
object based on its apparently large outburst amplitude
and the long outburst duration (table \ref{tab:wztab}),
this object might resemble a borderline object such as BC UMa
and RZ Leo.  Future detection of early superhumps and accurate determination
of $P_{\rm dot}$ are desired.

\begin{table}
\caption{Superhump maxima of OT J1959 (2005).}\label{tab:j1959oc2005}
\begin{center}
\begin{tabular}{ccccc}
\hline\hline
$E$ & max$^a$ & error & $O-C^b$ & $N^c$ \\
\hline
0 & 53606.3830 & 0.0012 & 0.0004 & 37 \\
1 & 53606.4448 & 0.0017 & 0.0024 & 26 \\
2 & 53606.4991 & 0.0023 & $-$0.0033 & 33 \\
44 & 53609.0202 & 0.0156 & 0.0012 & 42 \\
45 & 53609.0786 & 0.0020 & $-$0.0003 & 53 \\
68 & 53610.4518 & 0.0022 & $-$0.0052 & 36 \\
69 & 53610.5220 & 0.0058 & 0.0051 & 20 \\
93 & 53611.9546 & 0.0012 & $-$0.0003 & 96 \\
\hline
  \multicolumn{5}{l}{$^{a}$ BJD$-$2400000.} \\
  \multicolumn{5}{l}{$^{b}$ Against $max = 2453606.3825 + 0.059919 E$.} \\
  \multicolumn{5}{l}{$^{c}$ Number of points used to determine the maximum.} \\
\end{tabular}
\end{center}
\end{table}

\subsection{OT J213122.4$-$003937}\label{obj:j2131}

   This transient (hereafter OT J2131) was detected by K. Itagaki
\citep{yam08j1631cbet1631}.  Subsequent observations confirmed
the SU UMa-type nature of this object (vsnet-alert 10830).
Since the individual observations were not long enough, we could not
uniquely select the superhump period among one-day aliases
(e.g. 0.069 d, as in vsnet-alert 10830).
In table \ref{tab:j2131oc2008}, we list epochs based on the base period
of 0.06463(3) d, a candidate superhump period.
This selection of the alias needs to be verified by future observations.

\begin{table}
\caption{Superhump maxima of OT J2131 (2008).}\label{tab:j2131oc2008}
\begin{center}
\begin{tabular}{ccccc}
\hline\hline
$E$ & max$^a$ & error & $O-C^b$ & $N^c$ \\
\hline
0 & 54819.8932 & 0.0028 & $-$0.0020 & 179 \\
15 & 54820.8670 & 0.0024 & 0.0030 & 84 \\
16 & 54820.9282 & 0.0042 & $-$0.0004 & 119 \\
62 & 54823.8993 & 0.0011 & $-$0.0006 & 245 \\
\hline
  \multicolumn{5}{l}{$^{a}$ BJD$-$2400000.} \\
  \multicolumn{5}{l}{$^{b}$ Against $max = 2454819.8951 + 0.064593 E$.} \\
  \multicolumn{5}{l}{$^{c}$ Number of points used to determine the maximum.} \\
\end{tabular}
\end{center}
\end{table}

\subsection{OT J213701.8$+$071446}\label{obj:j2137}

\begin{figure}
  \begin{center}
    \FigureFile(88mm,110mm){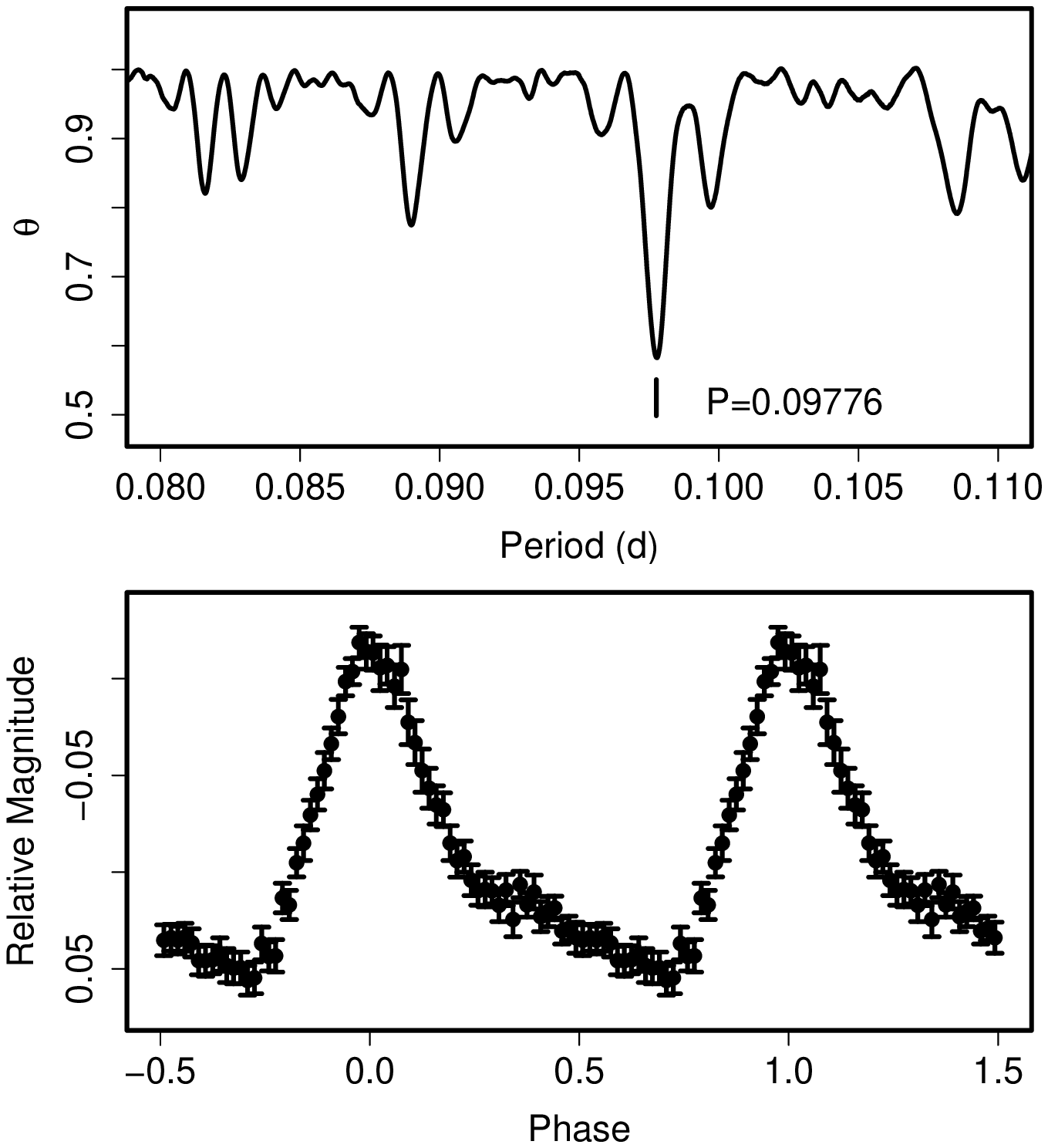}
  \end{center}
  \caption{Superhumps in OT J2137 (2008). (Upper): PDM analysis.
     (Lower): Phase-averaged profile.}
  \label{fig:j2137shpdm}
\end{figure}

   This transient (hereafter OT J2137) was detected by K. Itagaki
(vsnet-alert 10670, 10671).  The object was soon confirmed to be
an SU UMa-type dwarf nova in the period gap (vsnet-alert 10674, 10677).
The mean superhump period during the entire observation with the PDM
method was 0.097762(14) d (figure \ref{fig:j2137shpdm}).
The times of superhump maxima are listed in table \ref{tab:j2137oc2008}.
There was an apparent transition in the period between $E=0$ and $E=5$.
After $E=5$, the superhump period was almost constant
($P_{\rm SH}$ = 0.09768(3) d, $P_{\rm dot}$ = $+2.3(4.7) \times 10^{-5}$).
Since the object faded 6 d after this transition (vsnet-obs 62796),
we probably observed the stage C superhumps, which could explain the
lack of period variation.  The object underwent a rebrightening
(vsnet-alert 10708) 5 d after the fading.  Such a rebrightening is
rare in a long-$P_{\rm SH}$ system.  The object may resemble V725 Aql
in its period evolution of superhumps and in the presence of
a rebrightening (\cite{uem01v725aql}; subsection \ref{sec:v725aql}).

\begin{table}
\caption{Superhump maxima of OT J2137.}\label{tab:j2137oc2008}
\begin{center}

\end{center}
\end{table}

\subsection{TSS J022216.4$+$412260}\label{obj:j0222}

   The 2005 superoutburst of this WZ Sge-type dwarf nova was described
in \citet{ima06tss0222}.  We used the data used in \citet{ima06tss0222} and
determined times of superhump maxima during the plateau phase
(table \ref{tab:j0222oc2005}).
The superhumps were likely growing before $E = 37$.  We used the segment
later than this epoch and obtained
$P_{\rm dot}$ = $+2.2(1.5) \times 10^{-5}$.

\begin{table}
\caption{Superhump maxima of TSS J0222 (2005).}\label{tab:j0222oc2005}
\begin{center}
%
\end{center}
\end{table}

\section{Conclusion}\label{sec:conclusion}

   We systematically surveyed period variations of superhumps in
SU UMa-type dwarf novae based on newly obtained data and past
publications.  We found:

\begin{itemize}
\item In well-observed systems, the $O-C$ diagram of superhump maxima 
are usually composed of three distinct stages: early evolutionary
stage with a longer superhump period (stage A), middle stage with
systematically varying periods (stag B), and final stage with
a shorter superhump period (stage C).
\item During the stage B, the period derivative is strongly correlated
to the orbital period, or, more likely, to the mass ratio of the system.
Previously reported anomalously large period derivatives in EI Psc and
V485 Cen were not confirmed.
\item Upon transition to stage C, the superhump period generally decreases
by 0.5--1.0 \%.
\item We generally did not find strong evidence that period derivatives
vary between different superoutburst of the same object.  No apparent
correlation with the presence of a precursor outburst was recorded.
\item The superhump period at the start of stage B is close to that
in the stage C.  The fractional superhump excesses of these periods
are strongly correlated to the orbital period, or the mass ratio.
This period is slightly shorter than that expected for the precession
rate of single-particle dynamical 3:1 resonance.
\item In systems with positive period derivatives, the maximum period
at the end of stage B has a limit correlated to the mass ratio.
We interpret that the lengthening of the period is a result of outward
propagation of the eccentricity wave and this upper limit of the period
corresponds to the radius near the tidal truncation.
\item We interpreted that stage C superhumps are rejuvenized excitation
of 3:1 resonance when the superhumps in the outer disk is effectively
quenched.
\item Traditional phase reversal in ``late superhumps'' was not
recorded in many systems.  We suggested that some of these observations
misinterpreted stage C superhumps.
\item In some systems, particularly WZ Sge-type dwarf novae and analogous
systems, long-enduring superhump signals were recorded
during the post-superoutburst stage.  The $O-C$ analysis suggests that
these superhumps evolved from superhumps in the stages B or C.
The periods of these persisting superhumps are usually longer than the
periods of superhumps during the main superoutburst by $0.2-0.5$ \%.
\item The period variation in systems with long superhump periods
 vary from system to system.  Some systems show a very large decrease
in the superhump period.  While some systems show a stepwise decrease
as in short-period systems, some systems show a more continuous change.
\item Some long-period systems apparently lack period variations,
and there is even a hint of positive period derivatives in systems
with very infrequent outbursts.  The superoutbursts in these systems
resemble those of short-period systems in the frequent presence of
a rebrightening.
\item The positive period derivatives appears to be confirmed in
ER UMa-type dwarf novae.  In ER UMa itself, the stage C superhumps
seem to appear earlier than in other SU UMa-type dwarf nova accompanied
by a phase $\sim$0.5 offset.
\item In WZ Sge-type dwarf novae, period derivatives are an excellent
function of the fractional superhump excess or the mass-ratio.
\item In WZ Sge-type dwarf novae, the type of rebrightening is correlated
with the period variation.  Superoutbursts with multiple rebrightenings
or with a long-lasting rebrightening tend to have smaller period
derivatives while superoutbursts with a single rebrightening tend to
have larger period derivatives.
\item The superhumps of at least one outburst of a black-hole X-ray binary
(KV UMa) exhibited the same evolutionary sequence as in SU UMa-type
dwarf novae, although the degree of period variation was an order
of magnitude smaller.
\item We refined the empirical relations between the fractional superhump
excess and the mass ratio, and the fractional superhump excess and
the superhump period.
\end{itemize}

   The present survey has clarified the relation between general behavior of
period variation of superhumps and the system mass ratio (or the superhump
period).  Although this would seem to indicate that SU UMa-type dwarf
novae are ``single parameter systems'' regarding the period variation
of superhumps, the difference in behavior between different objects
with nearly equal superhump periods or mass ratios is much larger than
the variation within the same system.  This suggests the presence of
a mechanism causing diversity in different systems; questions whether
this diversity is related to outburst characteristics, or to the condition
of the accretion disk, need to be answered by future investigations.
There have also been indications of unusual development of superhumps in
several systems, making future observations of superhumps in even
well-observed objects still attractive.  The early emergence of the
stage C superhumps in ER UMa-type dwarf novae and some other systems,
and the superhumps in WZ Sge-type dwarf novae, particularly the late-stage
humps and transient enhancement of orbital humps, are still poorly understood.
The study presents an alternative idea to the traditional picture of
decreasing superhump period due to the shrinkage of the accretion disk
from the radius of the 3:1 resonance, which anticipates novel theoretical
progress in understanding the superhump phenomenon.

\vskip 3mm

This work was supported by the Grant-in-Aid for the Global COE Program
``The Next Generation of Physics, Spun from Universality and Emergence''
from the Ministry of Education, Culture, Sports, Science and Technology
(MEXT) of Japan.
This work was partly supported by a Grant-in-Aid from the 
Ministry of Education, Culture, Sports, Science and Technology 
of Japan (19740104).
Part of this work is supported by a Research Fellowship of
the Japan Society for the Promotion of Science for Young Scientists (AI).
The authors are grateful to observers of VSNET Collaboration and
VSOLJ observers who supplied vital data.  We also benefited from
the data by Martin Nicholson, Achim Sucker, Paulo Cacella,
T. Kryachko and his colleagues, Doug West, Oksana I. Dudka,
and Masayuki Moriyama.
We acknowledge with thanks the variable star
observations from the AAVSO International Database contributed by
observers worldwide and used in this research.
This work is deeply indebted to outburst detections and announcement
by a number of variable star observers worldwide, including participants of
CVNET, BAA VSS alert and AVSON networks. 
We are grateful to Allen W. Shafter for providing observations of
OT J0329, and Artur Rutkowski for providing the times of superhump
maxima for DI UMa.
The CCD operation of the Bronberg Observatory is partly sponsored by
the Center for Backyard Astrophysics.
The CCD operation by Peter Nelson is on loan from the AAVSO,
funded by the Curry Foundation.
P. Schmeer's observations were made with the Iowa Robotic Observatory,
and he wishes to thank Robert Mutel and his students.
We are grateful to the Catalina Real-time Transient Survey
team for making their real-time
detection of transient objects available to the public, and providing
times of eclipses for SDSS J1524.

\end{document}